\documentclass[aps,prd,12pt,nofootinbib,longbibliography]{revtex4-1}

\newif\ifcombine
\newif\ifarxiv

\usepackage{graphicx}
\usepackage[font=small,format=plain,labelfont=bf,textfont=it,justification=RaggedRight,singlelinecheck=false]{caption}
\usepackage{subcaption}
\usepackage{grffile}
\usepackage{amsmath}
\usepackage{mathtools}
\usepackage{amssymb}
\usepackage{xfrac}
\usepackage{fancyhdr}
\usepackage{enumerate}
\usepackage{indentfirst}
\usepackage[hang,flushmargin]{footmisc} 
\usepackage{bibentry}
\usepackage{natbib}
\usepackage{xspace}
\usepackage{wrapfig}
\usepackage{geometry}
\usepackage{verbatim}
\usepackage{lineno}
\usepackage{color}
\usepackage{cancel}
\usepackage{placeins}
\usepackage{booktabs}
\usepackage{multirow}
\usepackage{enumitem}
\usepackage{tabularx}
\usepackage{pdfpages}
\usepackage{titlesec}
\usepackage{titletoc}
\usepackage[
  pdfusetitle,
  pdfkeywords={SBN, Fermilab, neutrino, oscillation, sterile},
  bookmarks,bookmarksopen,
  bookmarksnumbered,
  pdfpagelabels,
  colorlinks,linkcolor=blue,citecolor=red,urlcolor=blue,
  pdfview={Fit},
]
{hyperref}

\graphicspath{{Figures/}}

\makeatletter
\def\l@subsubsection#1#2{}
\makeatother

\def\thepart{\arabic{part}}

\def\p@section{\thepart\.}

\makeatletter
\@addtoreset{section}{part}
\makeatother

\titleformat{\part}[display]{\bfseries\Large\itshape}{}{-3ex}
{\centering\rule{0.8\textwidth}{0.3pt}\\*\vspace{0.75ex}\partname~\thepart:~}
[\vspace{-0.5ex}\rule{0.8\textwidth}{0.3pt}]

\titlecontents{part}[0em]
{\centering\normalsize\bfseries\protect\addvspace{15pt}}
{}{\textcolor{blue}{\partname}~\thecontentslabel}{}

\titlecontents{section}[0em]
{\normalsize\protect\addvspace{10pt}}
{\textcolor{blue}{\thecontentslabel.~}}{}
{\titlerule*[1pc]{.}\contentspage}

\titlecontents{subsection}[0.8em]
{\normalsize\protect\addvspace{0.8pt}}
{\textcolor{blue}{\thecontentslabel.~}}{}
{\titlerule*[1pc]{.}\contentspage}

\newcommand{\proposalTitle}{\begin{center} \Large \bf A Proposal for a Three Detector \\ Short-Baseline Neutrino Oscillation Program \\ in the Fermilab Booster Neutrino Beam \end{center}}

\newcommand{\lartpc}{\text{LAr-TPC}\xspace}
\newcommand{\lartpcs}{\text{LAr-TPCs}\xspace}
\newcommand{\uboone}{\text{MicroBooNE}\xspace}
\newcommand{\MB}{MiniBooNE\xspace}
\newcommand{\uB}{MicroBooNE\xspace}
\newcommand{\SB}{SciBooNE\xspace}
\newcommand{\larnd}{LAr1-ND\xspace}
\newcommand{\icarus}{ICARUS-T600\xspace}
\newcommand{\nova}{\text{NO\ensuremath{\nu}A}\xspace}

\newcommand{\pot}{\ensuremath{6.6\times 10^{20}}\xspace}
\newcommand{\potuB}{\ensuremath{13.2\times 10^{20}}\xspace}
\newcommand{\apach}{5,632}
\newcommand{\detch}{11,264}
\newcommand{\lntwo}{\text{LN}\ensuremath{_{2}}\xspace}

\newcommand{\numu}{\ensuremath{\nu_{\mu}}\xspace}
\newcommand{\nue}{\ensuremath{\nu_{e}}\xspace}
\newcommand{\nutau}{\ensuremath{\nu_{\tau}}\xspace}
\newcommand{\numubar}{\ensuremath{\bar\nu_{\mu}}\xspace}
\newcommand{\nuebar}{\ensuremath{\bar\nu_{e}}\xspace}

\newcommand{\nufour}{\ensuremath{\nu_{4}}\xspace}

\newcommand{\nualpha}{\ensuremath{\nu_{\alpha}}\xspace}
\newcommand{\nubeta}{\ensuremath{\nu_{\beta}}\xspace}
\newcommand{\pizero}{\ensuremath{\pi^{0}}\xspace}

\newcommand{\numunue}{\ensuremath{\numu \rightarrow \nue}\xspace}
\newcommand{\numunuebar}{\ensuremath{\bar{\nu}_{\mu} \rightarrow \bar{\nu}_{e}}\xspace}
\newcommand{\numunueboth}{\ensuremath{\overset{\text{\tiny (}-\text{\tiny )}}{\numu} \rightarrow \overset{\text{\tiny (}-\text{\tiny )}}{\nue}\xspace}}
\newcommand{\numudis}{\ensuremath{\numu \rightarrow \nu_{x}}\xspace}
\newcommand{\nuedis}{\ensuremath{\nue \rightarrow \nu_{x}}\xspace}
\newcommand{\numunumu}{\ensuremath{\numu \rightarrow \numu}\xspace}
\newcommand{\dmsq}{\ensuremath{\Delta m^2}\xspace}

\newcommand{\sinth}{\ensuremath{\sin^22\theta}\xspace}
\newcommand{\Enu}{\ensuremath{E_{\nu}}\xspace}

\newcommand{\lenu}{\ensuremath{L/E_{\nu}}\xspace}

\newcommand{\GeV}{\ensuremath{\mbox{GeV}}\xspace}
\newcommand{\MeV}{\ensuremath{\mbox{MeV}}\xspace}

\newcommand{\eVsq}{\ensuremath{\mbox{eV}^2}\xspace}

\newcommand{\us}{\ensuremath{\mu s}\xspace}

\begin{document}

\combinetrue
\arxivtrue

\pagestyle{empty}

\proposalTitle
\bigskip

\begin{figure}[h]
\centering
\setlength{\fboxsep}{0pt}
\setlength{\fboxrule}{1pt}
\fbox{\includegraphics[width=0.95\textwidth,trim=20mm 0mm 0mm 0mm,clip]{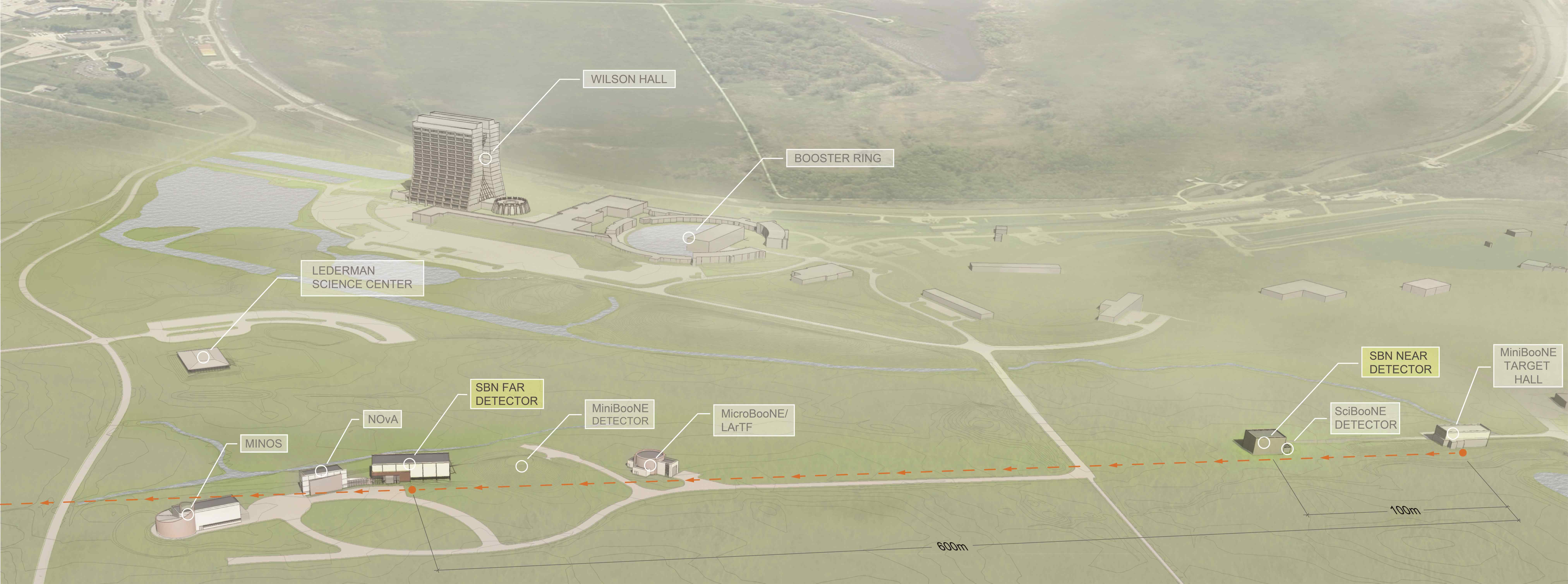}}
\end{figure}

%\begin{center}
%{\large
%DRAFT \\
%\bigskip
%\today
%\proposaldate
%}\end{center}
 
\clearpage

%\includepdf[pages={1}]{SBN_Proposal__Icarus_Author_List.pdf}
%\includepdf[pages={1}]{SBN_Proposal__LAr1-ND_Author_List.pdf}
%\includepdf[pages={1}]{SBN_Proposal__MicroBooNE_Author_List.pdf}
%\includepdf[pages={1}]{SBN_Proposal__Fermilab_Author_List.pdf}


# The ICARUS-WA104 Collaboration

M. Antonello[16], B. Baibussinov[31], V. Bellini[5], P. Benetti[32], S. Bertolucci[6], H. Bilokon[15], F. Boffelli[32], M. Bonesini[17], J. Bremer[6], E. Calligarich[32], S. Centro[31], A.G. Cocco[19], A. Dermenev[20], A. Falcone[32], C. Farnese[31], A. Fava[31], A. Ferrari[6], D. Gibin[31], S. Gninenko[20], N. Golubev[20], A. Guglielmi[31], A. Ivashkin[20], M. Kirsanov[20], J. Kisiel[38], U. Kose[6], F. Mammoliti[5], G. Mannocchi[15], A. Menegolli[32], G. Meng[31], D. Mladenov[6], C. Montanari[32], M. Nessi[6], M. Nicoletto[31], F. Noto[6], P. Picchi[15], F. Pietropaolo[31], P. Płoński[42], R. Potenza[5], A. Rappoldi[32], G.L. Raselli[32], M. Rossella[32], C. Rubbia*[6,11,16], P. Sala[18], A. Scaramelli[18], J. Sobczyk[44], M. Spanu[32], D. Stefan[18], R. Sulej[43], C.M. Sutera[5], M. Torti[32], F. Tortorici[5], F. Varanini[31], S. Ventura[31], C. Vignoli[16], T. Wachala[12], and A. Zani[32]

# The LAr1-ND Collaboration

C. Adams[45], C. Andreopoulos[23], A.M. Ankowski[41], J. Asaadi[40], L. Bagby[10], B. Baller[10], N. Barros[33], M. Bass[30], S. Bertolucci[6], M. Bishai[3], A. Bitadze[25], J. Bremer[6], L. Bugel[26], L. Camilleri[9], F. Cavanna[10,a], H. Chen[3], C. Chi[9], E. Church[10,45], D. Cianci[7], G.H. Collin[26], J.M. Conrad[26], G. De Geronimo[3], R. Dharmapalan[1], Z. Djurcic[1], A. Ereditato[2], J. Esquivel[40], J. Evans[25], B.T. Fleming[45], W.M. Foreman[7], J. Freestone[25], T. Gamble[37], G. Garvey[24], V. Genty[9], D. Göldi[2], E. Gramellini[45], H. Greenlee[10], R. Guenette[30], A. Hackenburg[45], R. Hänni[2], J. Ho[7], J. Howell[10], C. James[10], C.M. Jen[41], B.J.P. Jones[26], L.N. Kalousis[41], G. Karagiorgi[25], W. Ketchum[24], J. Klein[33], J. Klinger[37], U. Kose[6], I. Kreslo[2], V.A. Kudryavtsev[37], D. Lissauer[3], P. Lively[22], W.C. Louis[24], M. Lüthi[2], C. Mariani[41], K. Mavrokoridis[23], N. McCauley[23], N. McConkey[37], I. Mercer[22], T. Miao[10], G.B. Mills[24], D. Mladenov[6], D. Montanari[10], J. Moon[26], Z. Moss[26], S. Mufson[14], M. Nessi[6], B. Norris[10], F. Noto[6], J. Nowak[22], S. Pal[37], O. Palamara*[10,b], J. Pater[25], Z. Pavlovic[10], J. Perkin[37], G. Pulliam[40], X. Qian[3], L. Qiuguang[24], V. Radeka[3], R. Rameika[10], P.N. Ratoff[22], M. Richardson[37], C. Rudolf von Rohr[2], B. Russell[45], D.W. Schmitz*[7], M.H. Shaevitz[9], B. Sippach[9], M. Soderberg[40], S. Söldner-Rembold[25], J. Spitz[26], N. Spooner[37], T. Strauss[2], A.M. Szelc[45], C.E. Taylor[24], K. Terao[9], M. Thiesse[37], L. Thompson[37], M. Thomson[4], C. Thorn[3], M. Toups[26], C. Touramanis[23], R.G. Van de Water[24], M. Weber[2], D. Whittington[14], T. Wongjirad[26], B. Yu[3], G.P. Zeller[10], and J. Zennamo[7]

# The MicroBooNE Collaboration

R. Acciarri[10], C. Adams[45], R. An[13], A.M. Ankowski[41], J. Asaadi[40], L. Bagby[10], B. Baller[10], G. Barr[30], M. Bass[30], M. Bishai[3], A. Blake[4], T. Bolton[21], C. Bromberg[27], L. Bugel[26], L. Camilleri[9], D. Caratelli[9], B. Carls[10], F. Cavanna[10,a], H. Chen[3], E. Church[10,45], G.H. Collin[26], J.M. Conrad[26], M. Convery[39], S. Dytman[34], B. Eberly[39], A. Ereditato[2], J. Esquivel[40], B.T. Fleming*[45], W.M. Foreman[7], V. Genty[9], D. Göldi[2], S. Gollapinni[21], M. Graham[39], E. Gramellini[45], H. Greenlee[10], R. Grosso[8], R. Guenette[30], A. Hackenburg[45], O. Hen[26], J. Hewes[25], J. Ho[7], G. Horton-Smith[21], C. James[10], C.M. Jen[41], R.A. Johnson[8], B.J.P. Jones[26], J. Joshi[3], H. Jostlein[10], D. Kaleko[9], L.N. Kalousis[41], G. Karagiorgi[25], W. Ketchum[24], B. Kirby[3], M. Kirby[10], T. Kobilarcik[10], I. Kreslo[2], Y. Li[3], B. Littlejohn[13], D. Lissauer[3], S. Lockwitz[10], W.C. Louis[24], M. Lüthi[2], B. Lundberg[10], A. Marchionni[10], C. Mariani[41], J. Marshall[4], K. McDonald[35], V. Meddage[21], T. Miceli[28], G.B. Mills[24], J. Moon[26], M. Mooney[3], M.H. Moulai[26], N. Mu[9], R. Murrells[25], D. Naples[34], P. Nienaber[36], O. Palamara[10,b], V. Paolone[34], V. Papavassiliou[28], S. Pate[28], Z. Pavlovic[10], S. Pordes[10], G. Pulliam[40], X. Qian[3], J.L. Raaf[10], V. Radeka[3], R. Rameika[10], B. Rebel[10], L. Rochester[39], C. Rudolf von Rohr[2], B. Russell[45], D.W. Schmitz[7], A. Schukraft[10], W. Seligman[9], M.H. Shaevitz[9], M. Soderberg[40], S. Söldner-Rembold[25], J. Spitz[26], J. St. John[8], T. Strauss[2], A.M. Szelc[45], N. Tagg[29], K. Terao[9], M. Thomson[4], C. Thorn[3], M. Toups[26], Y. Tsai[39], T. Usher[39], R. Van de Water[24], M. Weber[2], S. Wolbers[10], T. Wongjirad[26], K. Woodruff[28], M. Xu[1], T. Yang[10], B. Yu[3], G.P. Zeller*[10], J. Zennamo[7], and C. Zhang[3]

# Additional Fermilab Contributors

W. Badgett[10], K. Biery[10], S.J. Brice[10], S. Dixon[10], M. Geynisman[10], C. Moore[10], E. Snider[10], and P. Wilson[10]

[1] Argonne National Laboratory, Lemont, IL, USA
[2] Laboratory for High Energy Physics, Universität Bern, Bern, Switzerland
[3] Brookhaven National Laboratory, Upton, NY, USA
[4] University of Cambridge, Cambridge, UK
[5] Department of Physics, Catania University, and INFN, Catania, Italy
[6] CERN, Geneva, Switzerland
[7] Enrico Fermi Institute, University of Chicago, Chicago, IL, USA
[8] University of Cincinnati, Cincinnati, OH, USA
[9] Nevis Labs, Columbia University, Irvington, NY, USA
[10] Fermi National Accelerator Laboratory, Batavia, IL, USA
[11] GSSI, Gran Sasso Science Institute, L'Aquila, Italy
[12] Henryk Niewodniczanski Institute of Nuclear Physics, Polish Academy of Science, Kraków, Poland
[13] Illinois Institute of Technology, Chicago, IL, USA
[14] Indiana University, Bloomington, IN, USA
[15] INFN LNF, Frascati (Roma), Italy
[16] INFN LNGS, Assergi (AQ), Italy
[17] INFN Milano Bicocca, Milano, Italy
[18] INFN Milano, Milano, Italy
[19] INFN Napoli, Napoli, Italy
[20] Institute for Nuclear Research of the Russian Academy of Sciences, Moscow, Russia
[21] Kansas State University, Manhattan, KS, USA
[22] Lancaster University, Lancaster, UK
[23] University of Liverpool, Liverpool, UK
[24] Los Alamos National Laboratory, Los Alamos, NM, USA
[25] University of Manchester, Manchester, UK
[26] Massachusetts Institute of Technology, Cambridge, MA, USA
[27] Michigan State University, East Lansing, MI, USA
[28] New Mexico State University, Las Cruces, NM, USA
[29] Otterbein University, Westerville, OH, USA
[30] University of Oxford, Oxford, UK
[31] Department of Physics and Astronomy, Padova University, and INFN, Padova, Italy
[32] Department of Physics, Pavia University, and INFN, Pavia, Italy
[33] University of Pennsylvania, Philadelphia, PA, USA
[34] University of Pittsburgh, Pittsburgh, PA, USA
[35] Princeton University, Princeton, NJ, USA
[36] Saint Mary's University of Minnesota, Winona, MN, USA
[37] University of Sheffield, Sheffield, UK
[38] Institute of Physics, University of Silesia, Katowice, Poland
[39] SLAC National Accelerator Laboratory, Menlo Park, CA, USA
[40] Syracuse University, Syracuse, NY, USA
[41] Center for Neutrino Physics, Virginia Tech, Blacksburg, VA, USA
[42] Institute for Radioelectronics, Warsaw University of Technology, Warsaw, Poland
[43] National Centre for Nuclear Research, Warsaw, Poland
[44] Institute of Theoretical Physics, Wroclaw University, Wroclaw, Poland
[45] Yale University, New Haven, CT, USA

*Spokespeople
[a] on leave of absence from University of L'Aquila and INFN, L'Aquila, Italy
[b] on leave of absence from INFN Gran Sasso Laboratories, Assergi (AQ), Italy


 
\ifcombine
\clearpage
\tableofcontents
\addtocontents{toc}{\protect\thispagestyle{empty}}
\clearpage
\setcounter{page}{0}
\else
\clearpage
\fi 
 
\pagestyle{fancy}
\rhead{\thepage}
\lhead{Introduction}
\cfoot{}

%%%%%%%%%%%%%%%%%%%%%%%%%%%%%%%%%%%%%%%%%%%%%%%%%%%%%%%%%%%%%%%%%%%%%%%%%%%%%%%%%%

\section*{Introduction}
\addcontentsline{toc}{section}{Introduction}

We propose a Short-Baseline Neutrino (SBN) physics program of three \lartpc detectors located along the Booster Neutrino Beam (BNB) at Fermilab.  This new SBN Program will deliver a rich and compelling physics opportunity, including the ability to resolve a class of experimental anomalies in neutrino physics and to perform the most sensitive search to date for sterile neutrinos at the eV mass-scale through both appearance and disappearance oscillation channels.  Additional physics of the SBN Program includes the study of neutrino-argon cross sections with millions of interactions using the well characterized neutrino fluxes of the BNB.  The SBN detectors will also record events from the off-axis flux of the NuMI neutrino beam with its higher electron neutrino content and different energy spectrum.  Finally, the SBN Program is an excellent opportunity to further develop this important technology for the future long-baseline neutrino program while utilizing its remarkable capabilities to explore one of the exciting open questions in neutrino physics today. 

The recent report of the Particle Physics Prioritization Panel (P5) specifically recommended a near-term, world-leading short-baseline experimental neutrino program with strong participation by the domestic and international neutrino physics communities working toward LBNF:
\begin{itemize}
\item P5 Recommendation \#12: In collaboration with international partners, develop a coherent short- and long-baseline neutrino program hosted at Fermilab.
\item P5 Recommendation \#15: Select and perform in the short term a set of small-scale short-baseline experiments that can conclusively address experimental hints of physics beyond the three-neutrino paradigm. Some of these experiments should use liquid argon to advance the technology and build the international community for LBNF at Fermilab.
\end{itemize}
This proposal outlines exactly such a program.  The SBN program brings together three \lartpc detectors built and operated by leading teams of scientists and engineers from Europe and the U.S.  The \icarus detector is the first successful large-scale \lartpc to be exposed to a neutrino beam and to this point the largest \lartpc for neutrino physics.  The \uboone detector is the largest \lartpc built in the U.S. and will have been operational for several years at the start of the three detector program. The new near detector, \larnd, is being developed by an international team with experience from ArgoNeuT, \uboone, and LBNE prototypes. The combination of these three detectors and associated collaborations represents a tremendous R\&D opportunity toward the future LBN program.  

At the January 2014 meeting of the Fermilab PAC, presentations were made by two collaborations to significantly enhance the physics capabilities of the Booster Neutrino Beam (BNB) with additional \lartpc detectors.  Both proposals were targeted at providing definitive measurements of the LSND and MiniBooNE anomalies.  The ICARUS collaboration proposed \cite{ICARUSPAC} a two detector experiment incorporating the existing T600 \lartpc located 700~m from the BNB as a far detector and a new T150 \lartpc located 150$\pm50$~m from the target as a near detector. The primary physics goal of the ICARUS proposal was the search for light sterile neutrinos. The idea to utilize multiple \lartpc detectors for a comprehensive test of neutrino anomalies was first put forth by the ICARUS collaboration at CERN as early as 2009~\cite{Baibussinov:2009tx,CRUBBIA2011,CRUBBIA345} and later extended to include the addition of magnetized spectrometer detectors~\cite{Bernardini:2011gy} by the NESSIE collaboration. The ICARUS-NESSIE proposal~\cite{Antonello:2012hf} required the construction of a new neutrino beam (CENF) from the 100 GeV proton SPS in the CERN north area, which has not since been realized.  

Also at the January 2014 Fermilab PAC meeting, the \larnd collaboration proposed \cite{LAr1-NDPAC} to install a new \lartpc based on LBNE-type technology 100~m from the BNB target in an existing enclosure that was constructed for the SciBooNE experiment.  The proposed \larnd detector, in concert with the \uboone experiment, would address the \MB neutrino mode anomaly and enable improved searches for oscillations. \larnd was seen as the next step in a phased short-baseline neutrino program at Fermilab. The full LAr1 detector, a 1~kton \lartpc previously presented in an LOI in 2012~\cite{lar1_loi}, could then, together with \larnd, definitively address the question of neutrino oscillations in the \dmsq $\approx$ 1~\eVsq region. LAr1 was not encouraged by P5, however, due to the high cost of a new detector of this scale.  

Following the recommendation of the PAC, the \uboone, \larnd, and ICARUS collaborations were asked by the Fermilab Director to propose a combined Short-Baseline Neutrino (SBN) program to address the short-baseline anomalies and search for sterile neutrinos.  The SBN Task Force was created to steer the activities of the collaborations required to create this proposal including a conceptual design for the program components.  The task force was charged jointly by the directorates of CERN, Fermilab, and INFN.  The task force consists of five members, one representing each of the three collaborations (\larnd, \uboone, ICARUS), one representing CERN, and the Fermilab SBN Program Coordinator.  This proposal is the result of this joint effort over a period of about nine months. The proposal is organized into six parts that are briefly summarized below. 

\begin{comment}
A set of five working groups were formed to address key issues:
\begin{enumerate}
\item Cosmic backgrounds,
\item Neutrino flux and systematics, 
\item Detector building configuration and siting,
\item Cryostat and cryogenic system design and integration,
\item BNB Improvements
\end{enumerate}
The first two working groups, each co-lead by members of LAr1-ND and ICARUS, are aimed at determining the optimal configuration of the experiments (e.g. baseline of near detector) based on analyses of the impact on the physics sensitivity.  The latter two working groups, each co-lead by a member of ICARUS and Fermilab staff, are aimed at refining experimental configurations for the purpose of establishing schedules and cost.  This report will summarize the current state of the work of these four groups and the plans for their future activities.  Although some conclusions have been reached, the working groups are expected to continue at least until a Conceptual Design Report has been completed.  The scope of the working groups activities or the number of working groups is likely to evolve as the work of defining the program progresses toward that design. 
\end{comment}

Part~1 describes the primary physics case for the SBN program: the search for sterile neutrinos and exploration of the LSND and MiniBooNE anomalies. This chapter describes the physics motivations including the current landscape of oscillation anomalies at the $\Delta m^2 \sim 1$~eV$^2$ scale. The physics sensitivity for \numu disappearance and \nue appearance are evaluated for the proposed three detector configuration.  The extensive evaluation of systematics includes the impact of neutrino interaction uncertainties and differences in the neutrino flux at the different detectors. The primary backgrounds to the \nue signal have been carefully considered in the sensitivities:
\begin{itemize}
\item Intrinsic \nue content of the beam,
\item Neutral current $\gamma$ production,
\item \numu charged current $\gamma$ production, 
\item Neutrino interactions in material surrounding the detectors,
\item Cosmogenic photons. 
\end{itemize}

Using data sets of \pot protons on target (P.O.T.) in the \larnd and T600 detectors plus \potuB P.O.T. in the \uboone detector, we conservatively estimate that a search for \numunue appearance can be performed covering with $\sim$5$\sigma$ sensitivity the LSND allowed (99\% C.L.) region.  This level of sensitivity is achieved using relatively simple event selection criteria that have undergone only a preliminary optimization. A more detailed analysis will likely yield even better results.  However, achieving this level of sensitivity will require careful control over cosmogenic backgrounds.  The external muon tagging/veto systems proposed in Parts~2~\&~3 of this proposal will be essential in ensuring an efficient independent method of identifying tracks in the TPCs associated with cosmogenic muons. Further, requiring precision timing of the already planned light detection systems will play an important role in rejecting cosmogenic photons. Since the measurement is statistically limited, an increase in the number of $\nu$ interactions provided by more protons on target and/or greater efficiency of the target/horn system will be highly desirable.  Part~5 of this proposal outlines a possible reconfiguration of the BNB with a two horn system.

Part~2 describes the conceptual design of the SBN near detector (\larnd) which will be located 110~m from the BNB target. The detector design draws extensively on the design of the LBNF far detector including cryostat technology, TPC design, and electronics.  Also described are the synergies between \larnd and the LBNF development. Potential benefits include expanded experience in construction of membrane cryostats, development of standardized cryogenic system modules, wire plane assembly techniques, testing of next generation cold electronics, and the development of scintillation light collection systems for \lartpcs. A system of scintillators external to the cryostat are proposed to provide a system for identifying tracks from cosmic rays. The near detector has been approved by Fermilab as a test experiment (T-1053).       

Part~3 describes the SBN far detector (\icarus), presently the largest physics oriented operational \lartpc detector \cite{ICARUS_jinst}, and its very successful performance during operations in the Gran Sasso laboratory (LNGS) on the CNGS beam. The T600 will be located 600~m from the BNB target.   In preparation for a thorough refurbishing, the T600 was transported  from LNGS to CERN at the end of 2014. The light detection system will be improved with many additional photo-multipliers to provide better timing resolution and spatial segmentation of the light signals.  These improvements are essential to handling the much higher flux of cosmic rays through the detector with operation near the surface on the BNB.  Like the near detector, counters surrounding the sensitive LAr-TPC volume are proposed to provide a system for identifying tracks from cosmic rays. These cosmic tagger systems provide a clear opportunity for shared development within the SBN program.

Part~4 describes the conceptual design of infrastructure needed for the \larnd and T600 detectors.  This includes construction of two new detector enclosures on the beamline at  110~m and 600~m from the BNB target for the near and far detectors, respectively.  Each of the detectors will be located on axis with the BNB.  The design of the new detector cryostats and cryogenic systems are also described.  The new cryogenics infrastructure is being developed by a joint team of engineers at CERN, Fermilab, and INFN taking advantage of opportunities for common solutions for the two detectors.  While a completely new LAr filtration and circulation system is needed for the near detector, the LAr systems used at LNGS for the T600 will be reused. A possible model for a common DAQ software platform for the detectors is described. Similarly, a model is described for common reconstruction and analysis tool development based on the presently working algorithms from the \icarus experience and the LArSoft platform currently in use by ArgoNeuT, \uboone, and the LBNE 35~ton prototype.      

Part~5 describes potential improvements to the Booster Neutrino Beamline.  A modest reconfiguration of the beamline leading up to target and horn could provide sufficient space to convert the beamline from a single horn to a two horn system.  An optimization based on a fast simulation demonstrates that the \numu rate can be approximately doubled relative to the existing \MB horn. A more detailed study is required to take this pre-conceptual design to a conceptual design that could be used to estimate cost and schedule to create a two horn configuration.  Such an improvement would be an extremely valuable addition to the program providing headroom on the statistical power of the measurements.  We propose that a detailed study of the cost and schedule for conversion to a two horn system be initiated immediately. This should include the cost of new horns, new or refurbished power supplies capable of 20~Hz operation, and necessary work for reconfiguration of the incoming beamline and of the collimator. 

Part~6 describes the organization of the SBN program and the schedule for completion of the detectors. A set of high level milestones is shown that has installation of both the near and far detectors during 2017 culminating in the three detector configuration ready for beam data-taking in the spring of 2018.  Funding for the program is expected from a combination of US DOE, US NSF, and international in-kind contributions. The overhauling, improvements, and transport of the \icarus is a major contribution of INFN and of CERN in terms of equipment and of associated funding to the realization of the present program.  Commitments are already in place for funding from CERN, INFN, and UK-STFC. Funding is being sought from CH-NSF and discussions are in progress with other international partners.  Organization of the program draws upon the very successful model of the LHC experiments at CERN.  Under the proposed structure, the program will be monitored by an oversight committee organized by Fermilab on behalf of the international partners.  \clearpage 

\pagestyle{empty}

\proposalTitle

\setcounter{part}{0}
\part{Physics Program}

%\begin{center}
%{\large
%DRAFT \\
%\bigskip
%\today
%\proposaldate 
%}\end{center}

\ifcombine
\clearpage
\else
\clearpage
\tableofcontents
\addtocontents{toc}{\protect\thispagestyle{empty}}
\clearpage
\setcounter{page}{0}
\fi

\pagestyle{fancy}
\rhead{I-\thepage}
\lhead{SBN Physics Program}
\cfoot{}

%%%%%%%%%%%%%%%%%%%%%%%%%%%%%%%%%%%%%%%%%%%%%%%%%%%%%%%%%%%%%%%%%%%%%%%%
% Introductory Material
%%%%%%%%%%%%%%%%%%%%%%%%%%%%%%%%%%%%%%%%%%%%%%%%%%%%%%%%%%%%%%%%%%%%%%%%

\section{Overview of the SBN Experimental Program}

The future short-baseline experimental configuration is proposed to include three Liquid Argon Time Projection Chamber detectors (\lartpcs) located on-axis in the Booster Neutrino Beam (BNB) as summarized in Table~\ref{tab:detectors}.  The near detector (\larnd) will be located in a new building directly downstream of the existing \SB enclosure 110~m from the BNB target.  The \uboone detector, which is currently in the final stages of installation, is located in the Liquid Argon Test Facility (LArTF) at 470~m.  The far detector (the improved \icarus) will be located in a new building 600~m from the BNB target and between \MB and the NOvA near detector surface building. The detector locations were chosen to optimize sensitivity to neutrino oscillations and minimize the impact of flux systematic uncertainties as reported in \cite{JulyPAC}. 

\begin{figure}[h]
\centering
\setlength{\fboxsep}{0pt}%
\setlength{\fboxrule}{0.5pt}%
\fbox{\includegraphics[height=0.45\textheight, trim=0mm 0mm 0mm 0mm, clip]{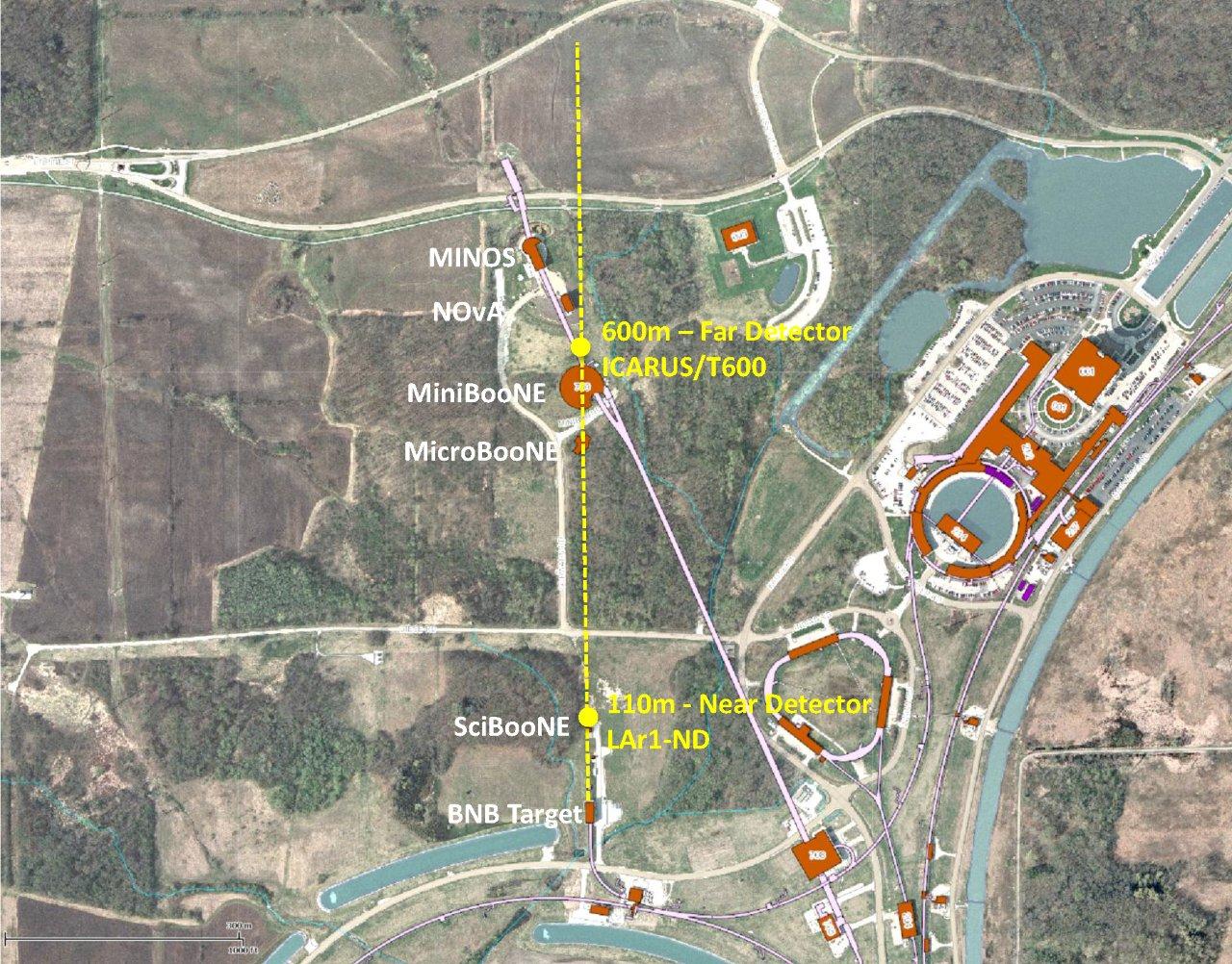}}
\caption{Map of the Fermilab neutrino beamline area showing the axis of the BNB (yellow dashed line) and approximate locations of the SBN detectors at 110~m, 470~m, and 600~m. The pink line indicates the axis of the NuMI neutrino beam for reference.}
\label{fig:SBNmap}
\end{figure}

Figure~\ref{fig:SBNmap} shows the locations of the detectors superimposed on an aerial view of the Fermilab neutrino experimental area. The following Sections briefly describe the attributes of the three detectors; more detailed descriptions are provided in dedicated Design Reports submitted with this proposal (see Part II and Part III).  Initial physics studies are based on current BNB fluxes, however, studies are on-going to determine what changes could be made to the target and horn systems to re-optimize for \lartpc detectors and increase event rates per proton on target (see Part V).  

\begin{table}[h]
\centering
\begin{tabular}{|c|c|c|c|}
\hline
Detector	& Distance from BNB Target	& LAr Total Mass & LAr Active Mass \\
\hline \hline
\larnd 		&   110~m  	&  220 t	& 112 t \\ \hline
\uboone 	&   470~m 	&  170 t 	& 89 t 	\\ \hline
\icarus		&   600~m	&  760 t  	& 476 t \\ \hline
\end{tabular}
\caption{Summary of the SBN detector locations and masses.}
\label{tab:detectors}
\end{table}

\subsection{The Booster Neutrino Beam}
\label{sec:BNB}

\begin{figure}[htp]
\centering
\setlength{\fboxsep}{0pt}%
\setlength{\fboxrule}{0.5pt}%
\mbox{\includegraphics[width=0.39\textwidth]{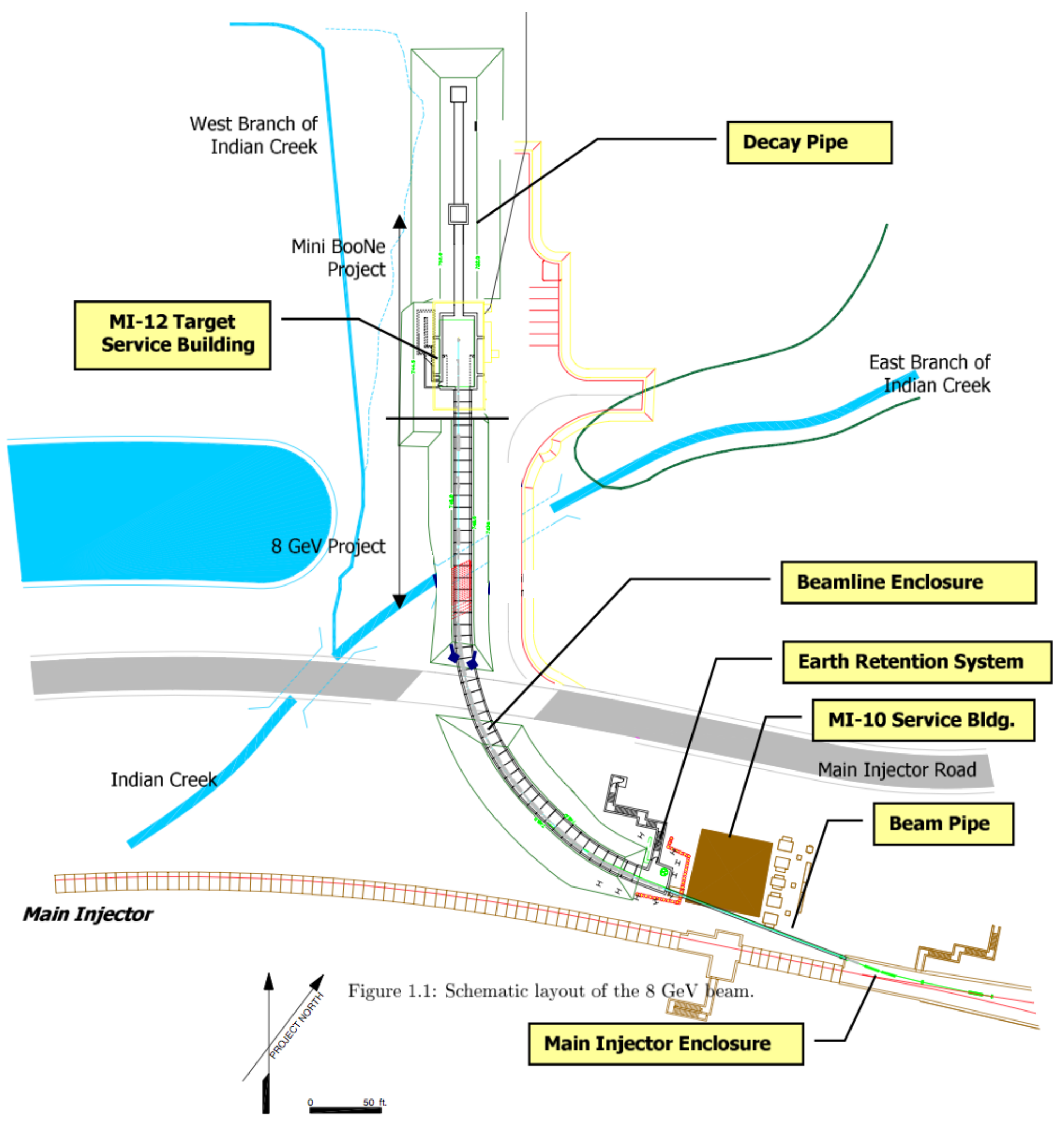}
\raisebox{10mm}{\fbox{\includegraphics[width=0.6\textwidth, trim=0mm 0mm 0mm 0mm]{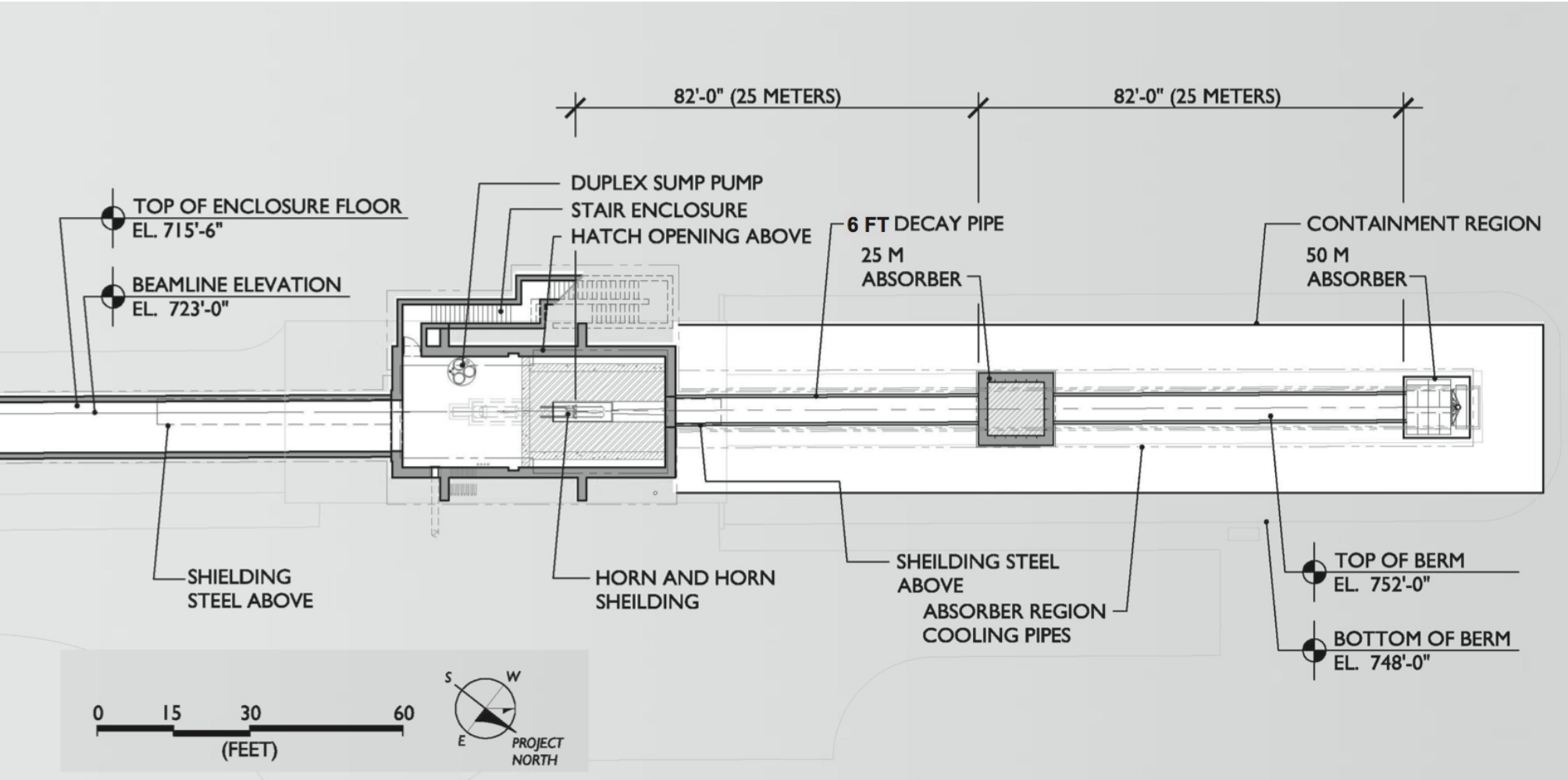}}}}
\caption{Schematic drawings of the Booster Neutrino Beamline including the 8 GeV extraction line, target hall and decay region.} 
\label{fig:BNB}
\end{figure}

The Booster Neutrino Beam is created by extracting protons from the Booster accelerator at 8~GeV kinetic energy (8.89~GeV$/c$ momentum) and impacting them on a 1.7$\lambda$ beryllium (Be) target to produce a secondary beam of hadrons, mainly pions. Charged secondaries are focused by a single toroidal aluminum alloy focusing horn that surrounds the target.  The horn is supplied with 174~kA in 143~$\mu$s pulses coincident with proton delivery. The horn can be pulsed with either polarity, thus focusing either positives or negatives and de-focusing the other. Focused mesons are allowed to propagate down a 50~m long, 0.91~m radius air-filled tunnel where the majority will decay to produce muon and electron neutrinos. The remainder are absorbed into a concrete and steel absorber at the end of the 50~m decay region. Suspended above the decay region at 25~m are concrete and steel plates which can be deployed to reduce the available decay length, thus systematically altering the neutrino fluxes.  A schematic of the BNB target station and decay region is shown in Figure~\ref{fig:BNB}. See Refs. \cite{BNB_8GeV,BNB_Beam} for technical design reports on the 8 GeV extraction line and the Booster Neutrino Beam.  

The timing structure of the delivered proton beam is an important aspect for the physics program.  The Booster spill length is 1.6~$\mu s$ with nominally $\sim5 \times 10^{12}$ protons per spill delivered to the beryllium target. The main Booster RF is operated at 52.8~MHz, with some 81 buckets filled out of 84.  The beam is extracted into the BNB using a fast-rising kicker that extracts all of the particles in a single turn. The resulting structure is a series of 81 bunches of protons each $\sim$2~ns wide and 19~ns apart. While the operating rate of the Booster is 15~Hz, the maximum allowable average spill delivery rate to the BNB is 5~Hz, set by the design of the horn and its power supply.  

The BNB has already successfully and stably operated for 12 years in both neutrino and anti-neutrino modes. The fluxes are well understood thanks to a detailed simulation~\cite{MiniFlux} developed by the \MB Collaboration and the availability of dedicated hadron production data for 8.9 GeV/c $p$+Be interactions collected at the HARP experiment at CERN~\cite{HARP,Schmitz:2008zz}. Systematic uncertainties associated with the beam have also been characterized in a detailed way as seen in Refs.~\cite{MiniFlux,MB_QE} with a total error of $\sim$9\% at the peak of the \numu flux and larger in the low and high energy regions.   

\begin{figure}[htp]\centering
\mbox{\includegraphics[width=0.333\textwidth,trim = 0mm 60mm 18mm 60mm, clip]{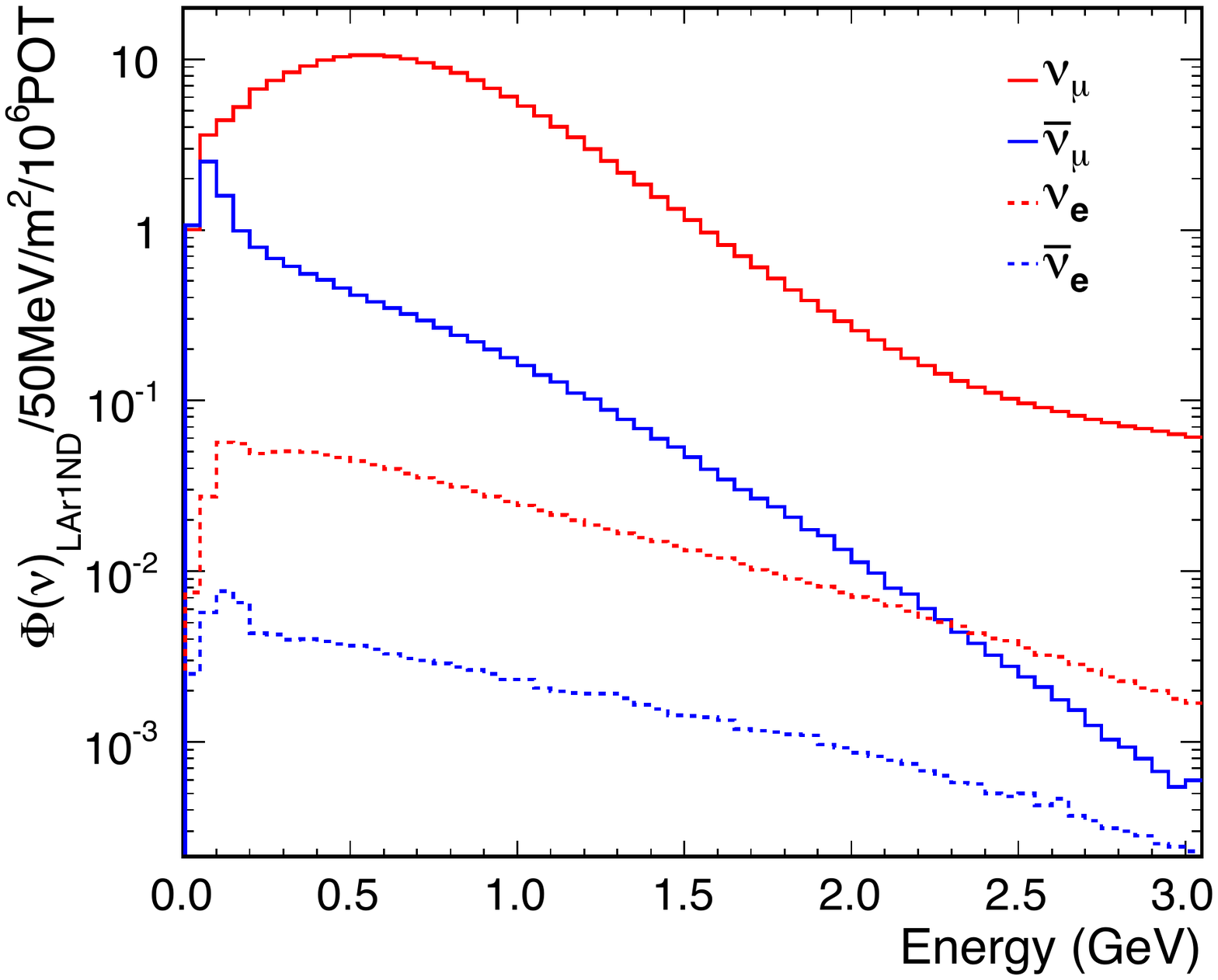}
\includegraphics[width=0.333\textwidth,trim = 0mm 60mm 18mm 60mm, clip]{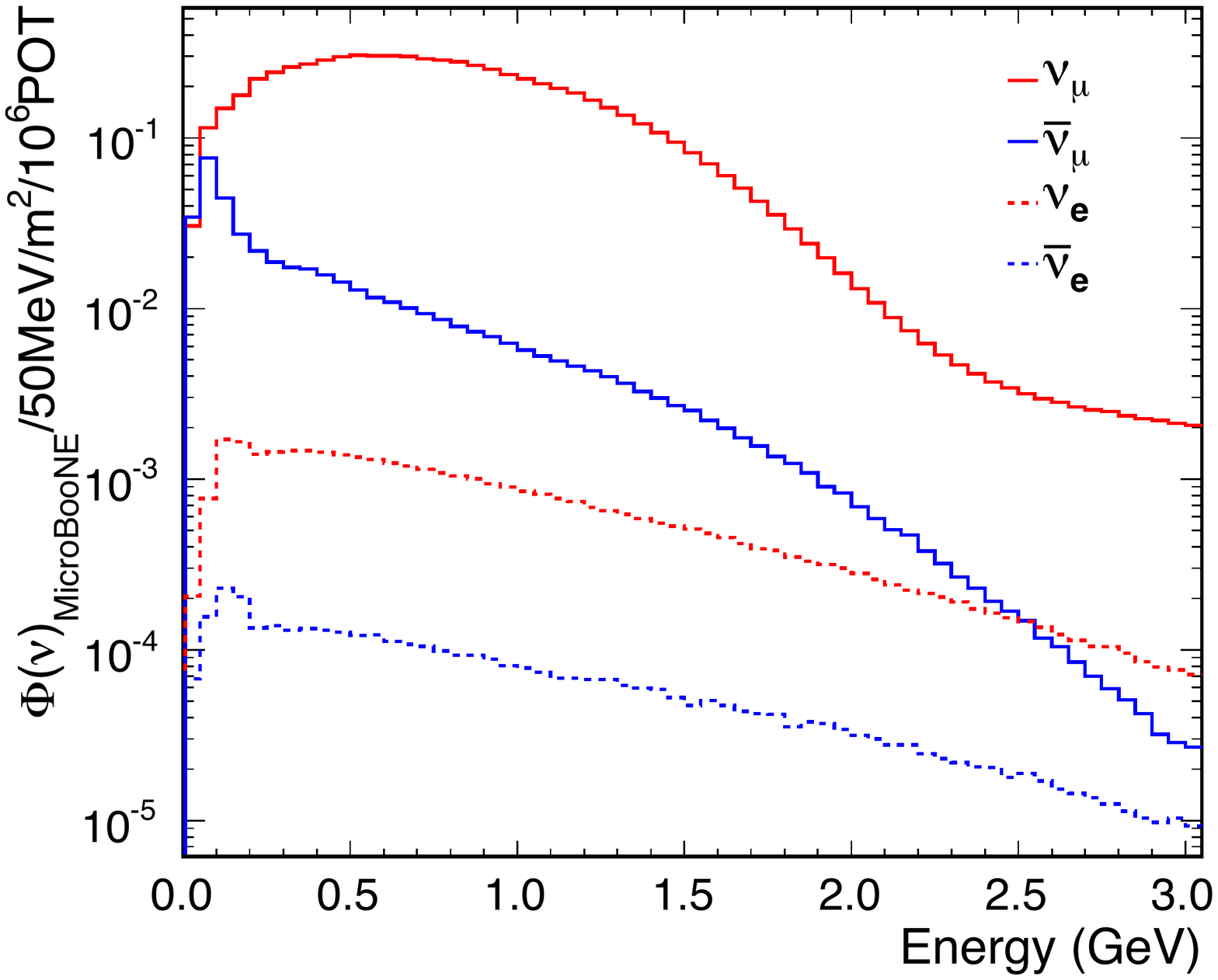}
\includegraphics[width=0.333\textwidth,trim = 0mm 60mm 18mm 60mm, clip]{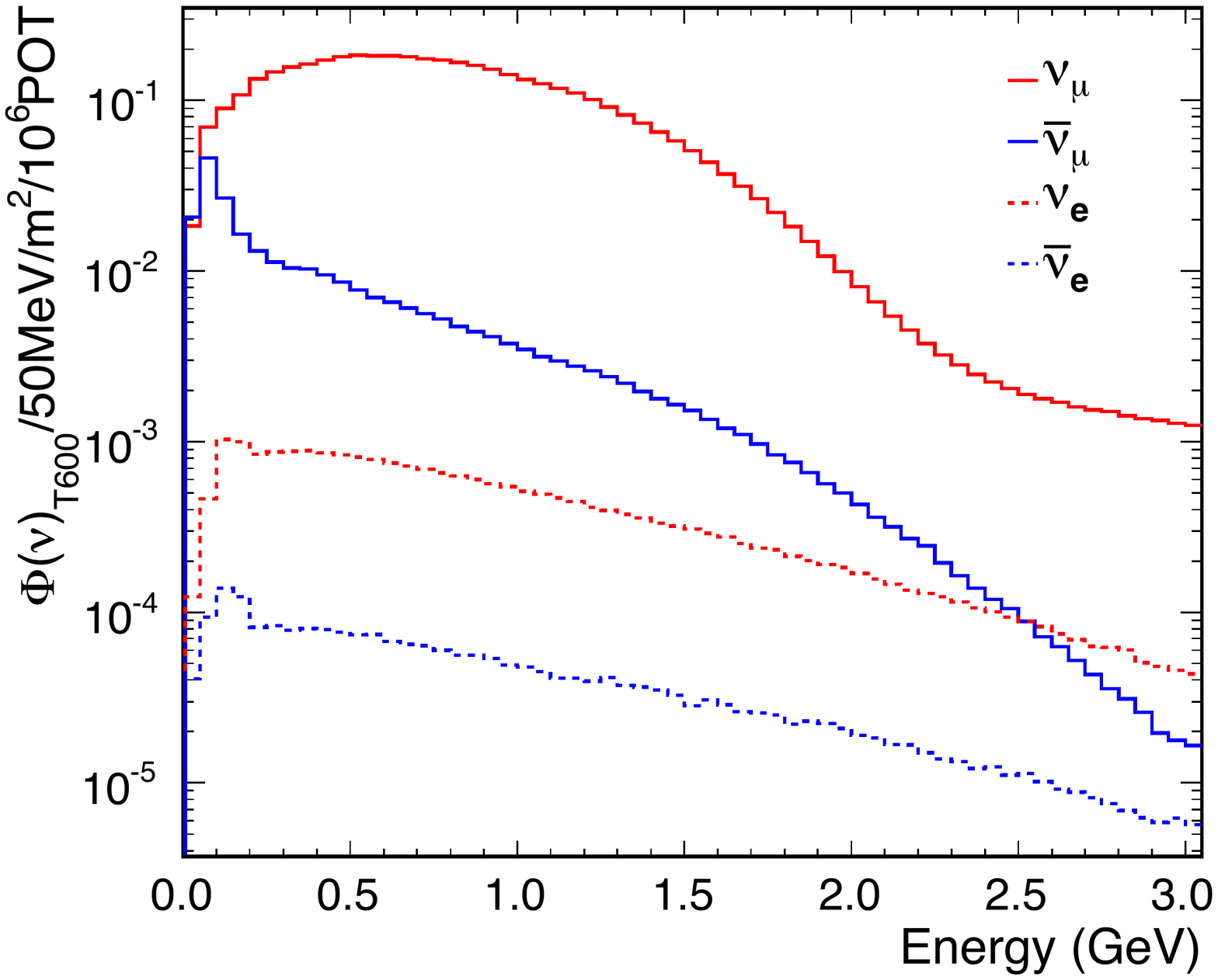}}
\mbox{\includegraphics[width=0.333\textwidth,trim = 0mm 0mm 10mm 0mm, clip]{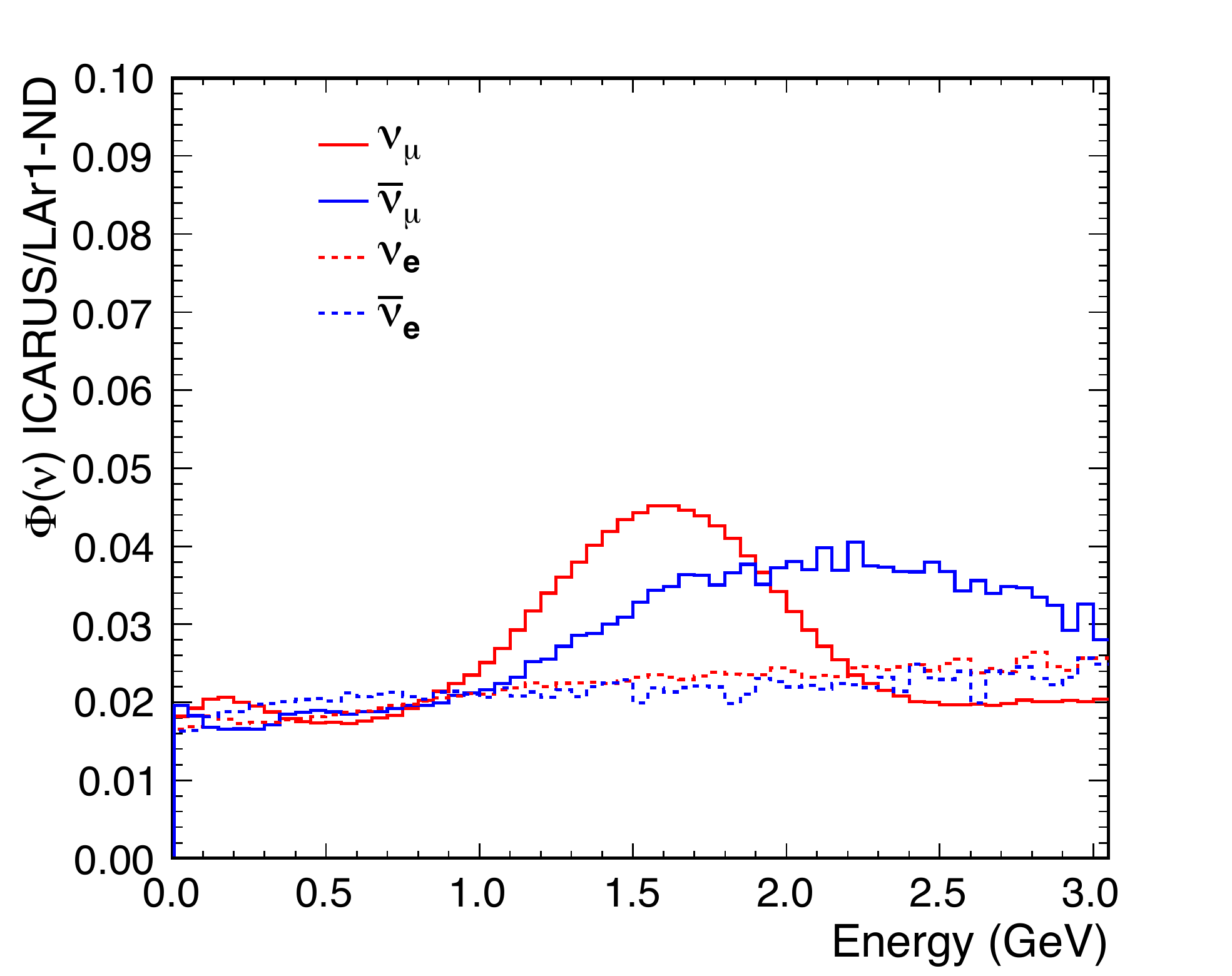}\quad
\includegraphics[width=0.333\textwidth,trim = 0mm 0mm 10mm 0mm, clip]{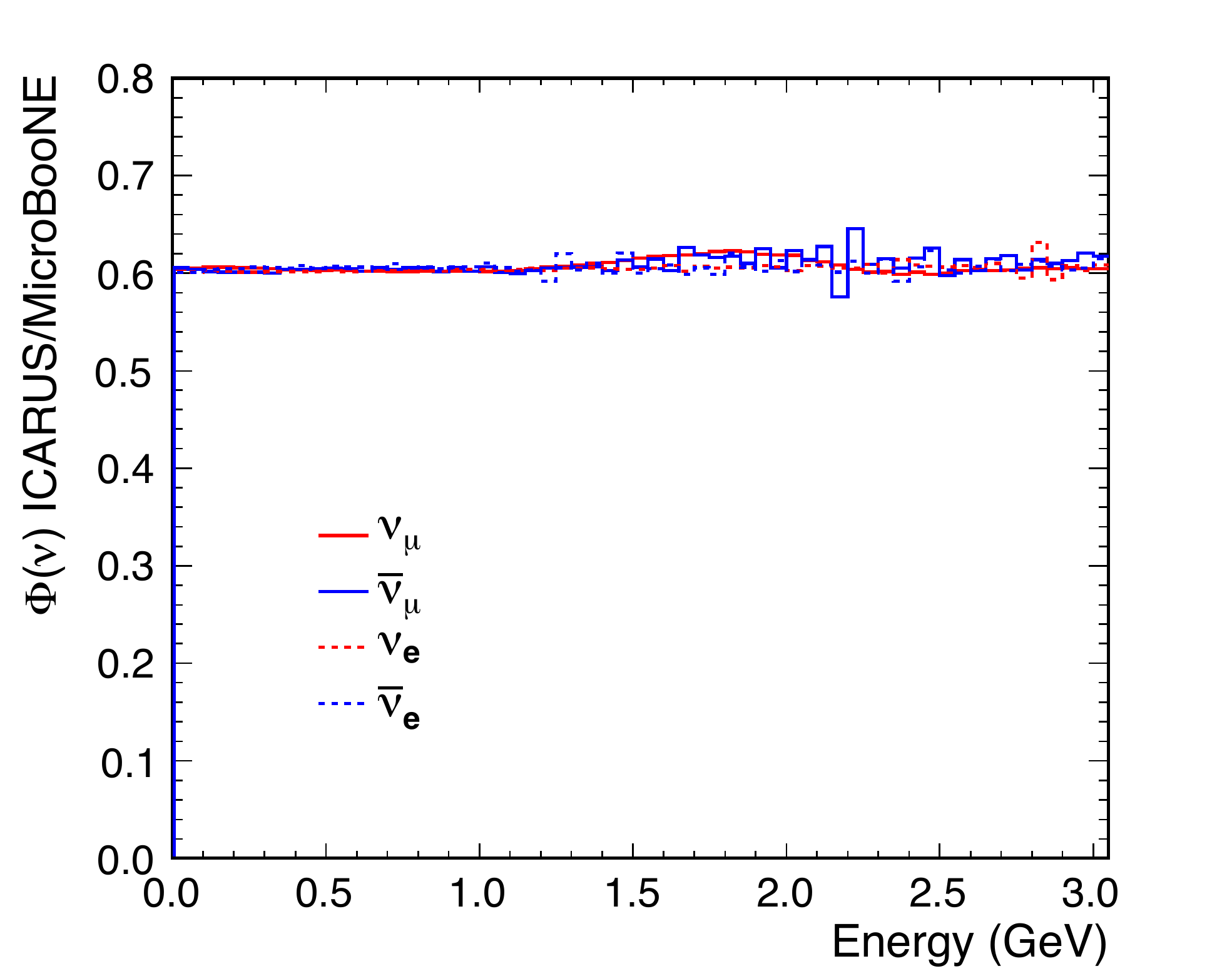}
}
\caption{(Top) The Booster Neutrino Beam flux at the three SBN detectors: (left) \larnd, (center) \uboone, and (right) \icarus. (Bottom) Ratio of the fluxes for each neutrino species between ICARUS and \larnd (left) and between ICARUS and \uboone (right). Fluxes at the far detectors fall off faster than $1/r^2$ when compared to the 110~m location and the \numu/\numubar spectra are harder due to the restricted solid angle at the far locations. These effects and associated systematic uncertainties are fully considered in the analysis. The far detector locations are clearly in the $1/r^2$ regime with $470^2/600^2 = 0.61$. } 
\label{fig:BNB_flux}
\end{figure}

The neutrino fluxes observed at the three SBN detector locations are shown in Figure~\ref{fig:BNB_flux}. Note the rate in the near detector is 20-30 times higher than at the \uboone and ICARUS locations. Also, one sees the \numu spectrum is slightly harder at the far locations as a result of the narrower solid angle viewed by the far detector.  We'll see later, however, that this does not introduce a significant systematic in oscillation searches. The shapes of the \nue/\nuebar fluxes are more similar.  The composition of the flux in neutrino mode (focusing positive hadrons) is energy dependent, but is dominated by \numu ($\sim$93.6\%), followed by \numubar ($\sim$5.9\%), with an intrinsic \nue/\nuebar contamination at the level of 0.5\% at energies below 1.5~GeV. The majority of the \numu flux originates from pion decay in flight ($\pi^{+}\rightarrow\mu^{+}+\numu$) except above $\sim$2~GeV where charged kaon decay is the largest contributor. A substantial portion of the intrinsic \nue flux, 51\%, originates from the pion $\rightarrow$ muon decay chain ($\pi^{+}\rightarrow\mu^{+}\rightarrow e^{+}+\nue+\numu$) with the remaining portion from $K^+$ and $K^0$ decays.

\subsection{The Detector Systems: \uB, \larnd, \icarus}

\subsubsection*{\uB}
\label{sec:uB}

\uboone is currently in the final stages of construction and will be commissioned at the end of 2014.  The experiment will measure neutrino interactions in argon for multiple reaction channels and investigate the source of the currently unexplained excess of low energy electromagnetic events observed by \MB. \uboone also incorporates several important R\&D features: the use of a non-evacuated cryostat, passive insulation of the cryostat and cryogenics, cold (in liquid) electronics, a long 2.56 meter drift distance, and a novel UV laser calibration system~\cite{ub-laser}. To accomplish these goals, the \uboone detector is a 170 ton total mass (89 ton active mass) liquid argon TPC contained within a conventional cryostat~\cite{ub-tdr}. The active region of the TPC is a rectangular volume of dimensions 2.33~m $\times$ 2.56~m $\times$ 10.37~m.  The TPC cathode plane forms the vertical boundary of the active volume on the left side of the detector when viewed along the neutrino beam direction (beam left). The \uboone TPC design allows ionization electrons from charged particle tracks in the active liquid argon volume to drift up to 2.56 meters to a three-plane wire chamber. Three readout planes, spaced by 3 mm, form the beam-right side of the detector, with 3,456 Y wires arrayed vertically and 2,400 U and 2,400 V wires oriented at $\pm 60$ degrees with respect to vertical.  An array of 32 PMTs are mounted behind the wire planes on the beam right side of the detector to collect prompt scintillation light produced in the argon~\cite{ub-pmts}.

\uboone is approved to receive an exposure of $6.6\times10^{20}$ protons on target in neutrino running mode from the BNB. It will also record interactions from an off-axis component of the NuMI neutrino beam. During \uboone running, the BNB will be operated in the same configuration that successfully delivered neutrino and anti-neutrino beam to \MB for more than a decade, thereby significantly reducing systematic uncertainties in the comparison of \uboone data with that from \MB.

As of the writing of this document, construction of the \uboone TPC has been completed and on June 23, 2014, the \uboone vessel was moved to the 
%Liquid Argon Test Facility
LArTF, a new Fermilab enclosure just upstream of the \MB detector hall. Final installation and detector commissioning has begun. \uboone is on schedule to begin taking neutrino data in early 2015.

\subsubsection*{\larnd}
\label{sec:lar1nd}

\begin{figure}[t]
\centering
\setlength{\fboxsep}{0pt}%
\setlength{\fboxrule}{0.6pt}%
\mbox{\fbox{\includegraphics[height=0.21\textheight]{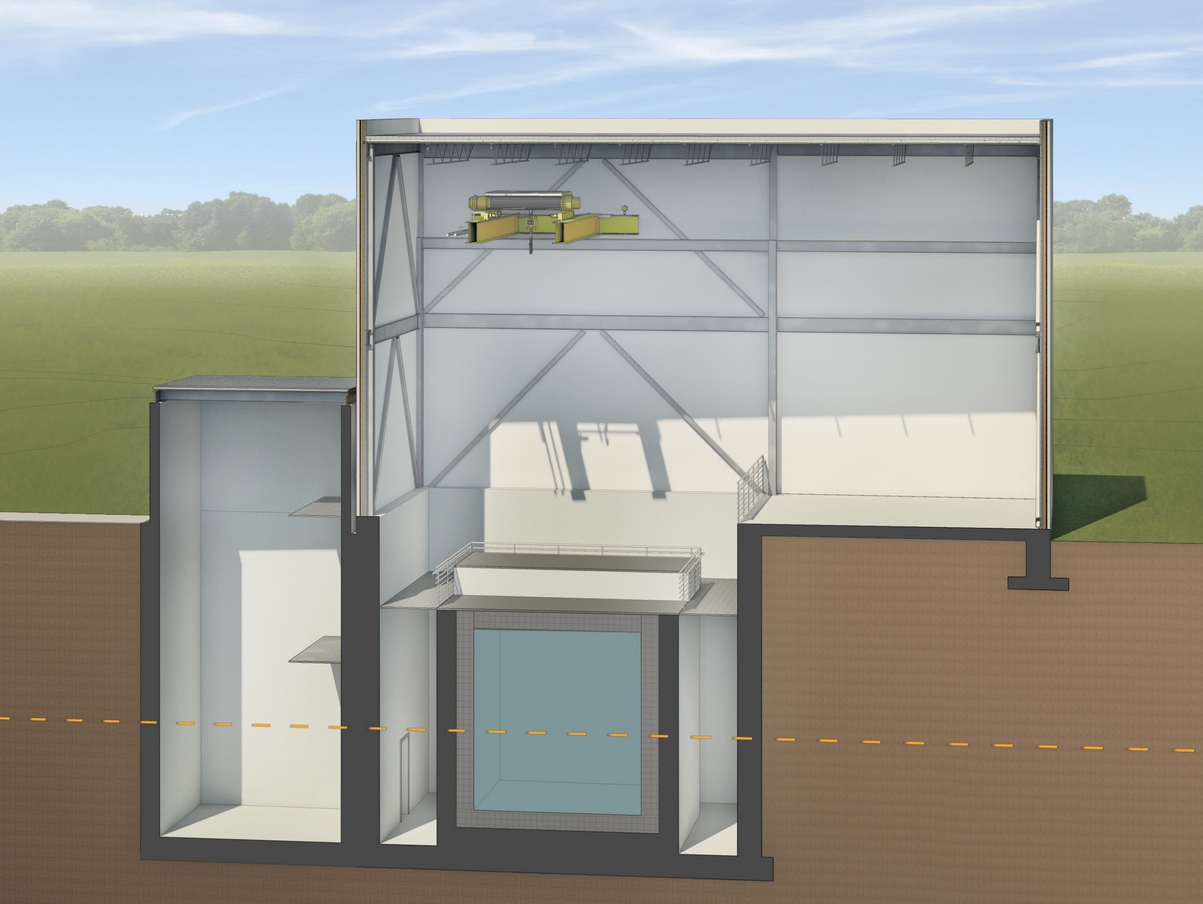}} \quad
\includegraphics[height=0.21\textheight, trim=0mm 0mm 0mm 0mm, clip]{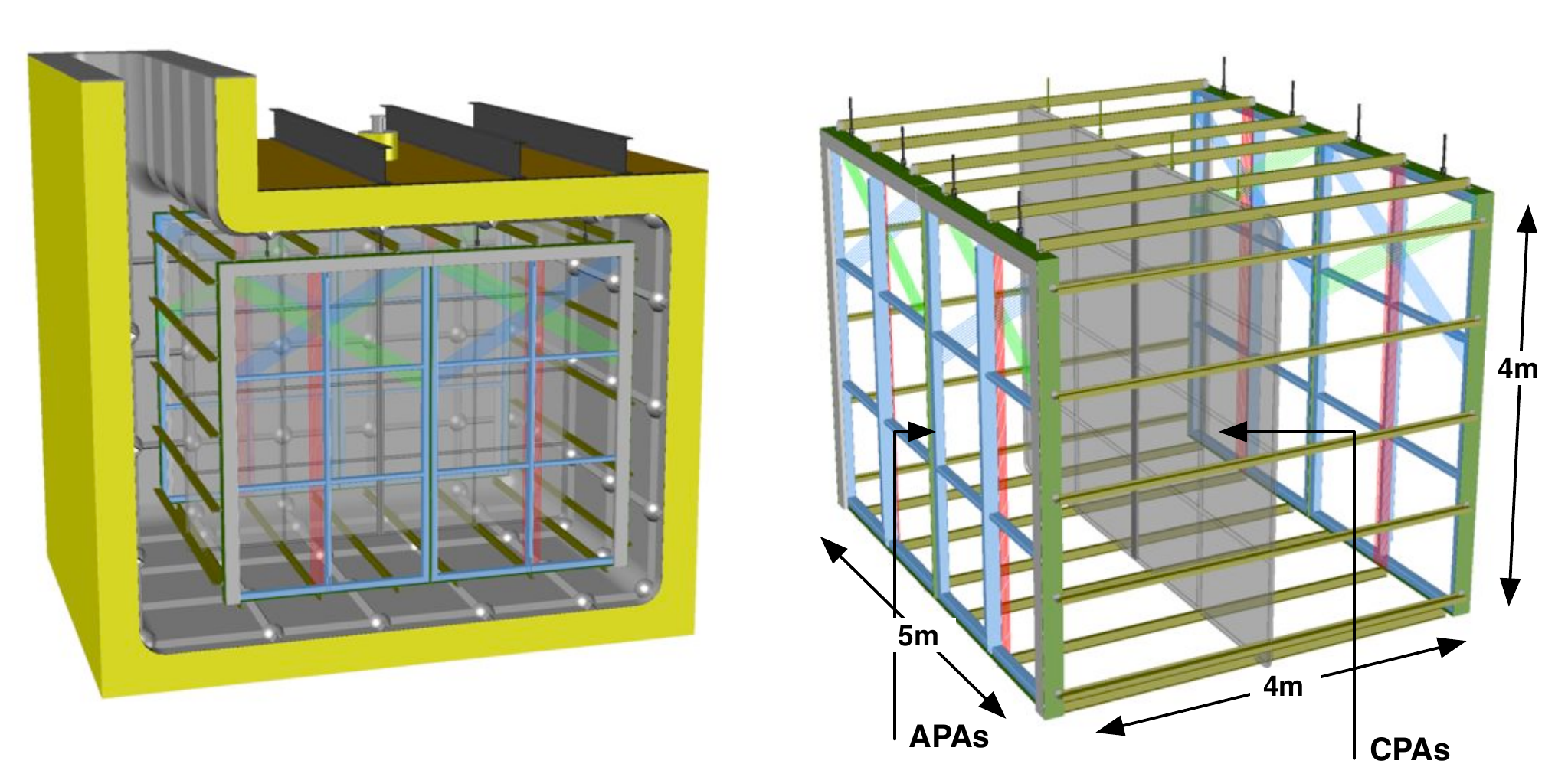}}
\caption{\small \it  (Left) The \larnd detector building concept.  The neutrino beam center is indicated by the orange dashed line and enters from the left.  (Right) The \larnd TPC conceptual design.}
\label{fig:larnd}
\end{figure}

The design of the Liquid Argon Near Detector, or \larnd\cite{LAr1-NDPAC}, builds on many years of \lartpc detector R\&D and experience from design and construction of the \icarus, ArgoNeuT, \uboone, and LBNF detectors. The basic concept is to construct a membrane-style cryostat in a new on-axis enclosure adjacent to and directly downstream of the existing \SB hall. The membrane cryostat will house a CPA (Cathode Plane Assembly) and four APAs (Anode Plane Assemblies) to read out ionization electron signals.  The active TPC volume is 4.0~m (width) $\times$ 4.0~m (height) $\times$ 5.0~m (length, beam direction), containing 112 tons of liquid argon.  Figure \ref{fig:larnd} shows the state of the conceptual design for the Near Detector building and the \larnd TPC.

The two APAs located near the beam-left and beam-right walls of the cryostat will each hold 3 planes of wires with 3~mm wire spacing. The APAs use the same wire bonding method developed for the LBNF APAs, but without the continuous helical wrapping to avoid ambiguity in track reconstruction.  Along the common edge of neighboring APAs, the U \& V wires are electrically ``jumped''.  TPC signals are then read out with banks of cold electronics boards at the top and two outer vertical sides of each detector half. The total number of readout channels is 2,816 per APA (11,264 in the entire detector). The CPA has the same dimensions as the APAs and is centered between them. It is made of a stainless-steel framework, with an array of stainless-steel sheets mounted over the frame openings. Each pair of facing CPA and APA hence forms an electron-drift region. The open sides between each APA and the CPA are surrounded by 4 FCAs (Field Cage Assemblies), constructed from FR4 printed circuit panels with parallel copper strips, to create a uniform drift field. The drift distance between each APA and the CPA is 2~m such that the cathode plane will need to be biased at -100 kV for a nominal 500~V/cm field. The \larnd design will additionally include a light collection system for detecting scintillation light produced in the argon volume. 

Overall, the design philosophy of the \larnd detector is to serve as a prototype for LBNF that functions as a physics experiment.  While the present conceptual design described here is an excellent test of LBNF detector systems sited in a neutrino beam, the \larnd Collaboration is exploring innovations in this design and the opportunity to further test them in a running experiment.

\subsubsection*{\icarus}
\label{sec:icarus}

The \icarus detector previously installed in the underground INFN-LNGS Gran Sasso Laboratory has been the first large-mass \lartpc operating as a continuously sensitive general purpose observatory. The successful operation of the \icarus \lartpc demonstrates the enormous potential of this detection technique, addressing a wide physics program with the simultaneous exposure to the CNGS neutrino beam and cosmic-rays~\cite{ICARUS_jinst}. 

The \icarus detector consists of two large identical modules with internal dimensions $3.6 \times 3.9 \times 19.6$ m$^3$ filled with 
$\sim 760$ tons of ultra-pure liquid argon, surrounded by a common thermal insulation ~\cite{ICARUS_jinst, ICARUS_NIM}. Each module houses two TPCs separated by a common central cathode for an active volume of $3.2 \times 2.96 \times 18.0$ m$^3$. A uniform electric field (E$_D$ = 500 V/cm) is applied to the drift volume. The reliable operation of the high-voltage system has been extensively tested in the \icarus up to about twice the operating voltage (150~kV, corresponding to E$_D$ = 1 kV/cm). Each TPC is made of three parallel wire planes, 3~mm apart, with 3 mm pitch, facing the drift path (1.5~m) and with wires oriented at $0^0$, $\pm 60^0$ with respect to the horizontal direction, respectively. Globally, 53,248 wires with length up to 9 m are installed in the detector. A three-dimensional image of the ionizing event is reconstructed combining the wire coordinate on each plane at a given drift time with $\sim$1~mm$^3$ resolution over the whole active volume (340~m$^3$ corresponding to 476~tons). 

\begin{figure}[t]
\centering
\setlength{\fboxsep}{0pt}%
\setlength{\fboxrule}{0.6pt}%
\mbox{\fbox{\includegraphics[height=0.18\textheight]{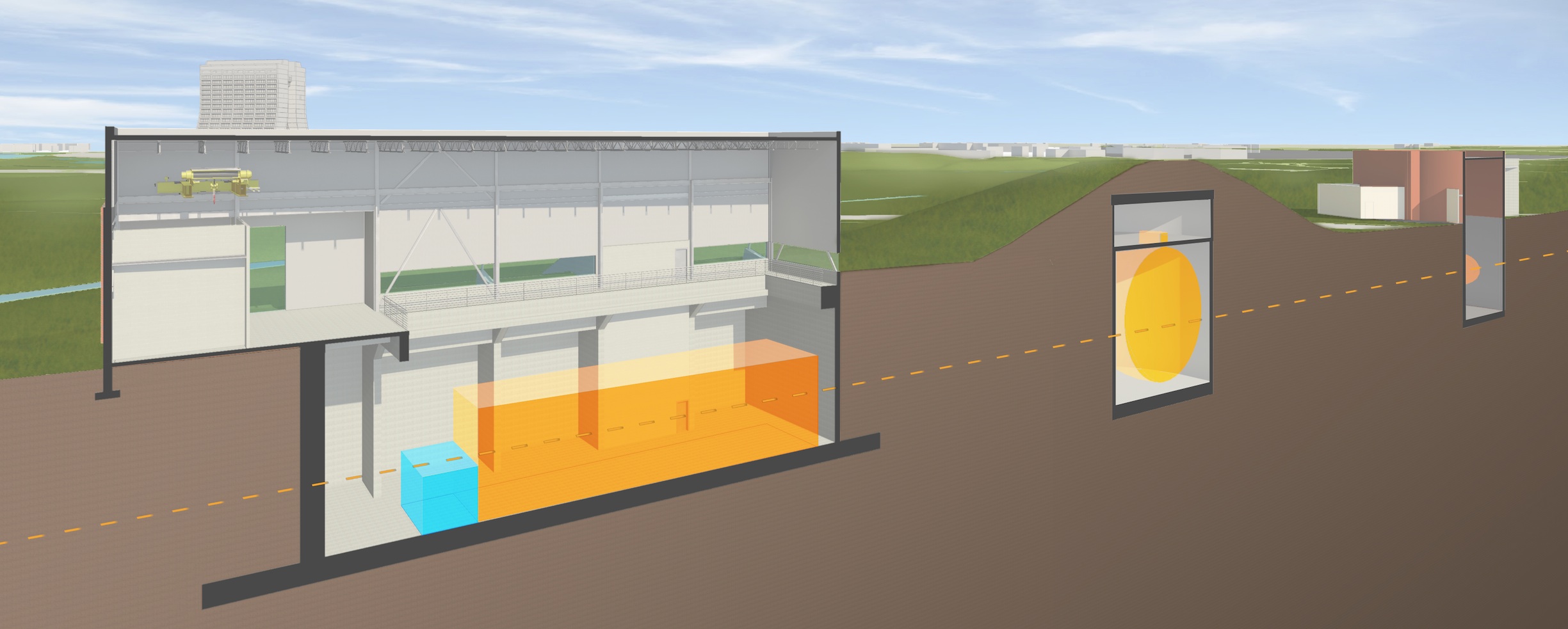}}
\fbox{\includegraphics[height=0.18\textheight, trim = 5mm 210mm 5mm 5mm, clip]{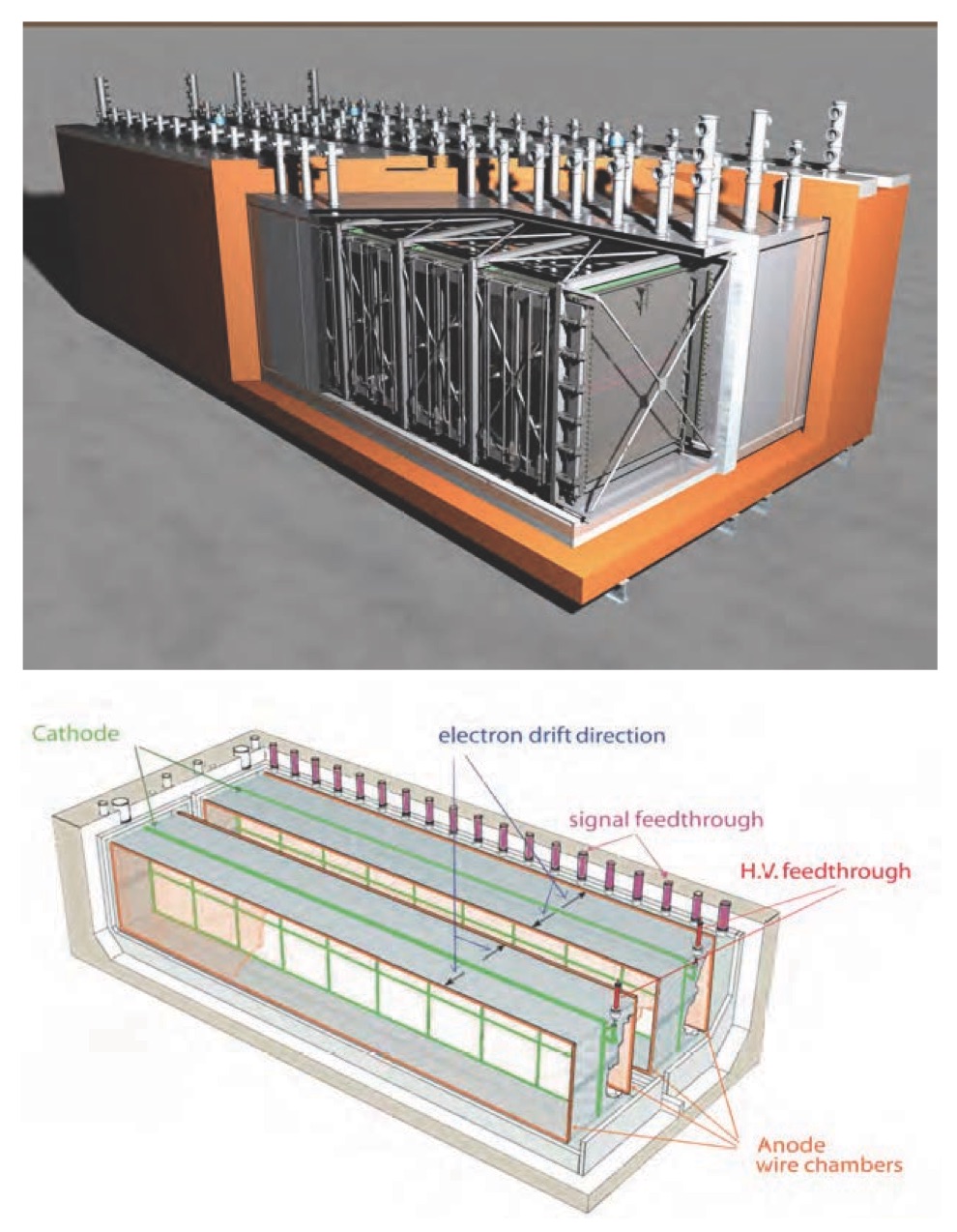}}}
\caption{\small \it  (Left) The \icarus detector building concept.  The neutrino beam center is indicated by the orange dashed line and enters from the right.  The existing \MB and \uB buildings are also shown. (Right) \icarus detector schematic showing both modules and the common insulation surrounding the detector.}
\label{fig:icarus}
\end{figure}

The \icarus detector has been moved to CERN for an overhauling preserving most of the existing operational equipment, while upgrading some components with up-to-date technology in view of its future near surface operation. The
refurbishing program, described in detail in Part~3 and Part~4, has been endorsed by a dedicated MoU between INFN and CERN. This mainly includes: 

\begin{itemize}
 \item realization of new vessels for LAr containment and new thermal insulation; %, based on similar technology as foreseen for LBNF and the near detector;
 \item implementation of an improved  light collection system, to allow a more precise event localization and the disentangling of the background induced by cosmic rays;
 \item although the present electronics would be perfectly adequate for the SBN program, several reasons exist for its substitution with a more modern version that preserves the general architecture with more updated components. A possible solution already at prototype level is described in Part~3. The final solution is under evaluation and cost sharing and responsibilities will be object of a special addendum of the MoU.
  
\end{itemize}

Moreover an anti-coincidence system, common to the SBN detectors, will be constructed to automatically tag cosmic rays crossing the LAr active volume. 

For what concerns the maintenance and the adaptation of the cryogenic systems to the new experimental layout at FNAL, this activity will be carried-out under the supervision of the ICARUS Collaboration with a major involvement of CERN.  

The detector is expected to be transported to FNAL at the beginning of 2017. Installation and operation at Fermilab will require significant involvement of Fermilab technical personnel. All of the above mentioned activities will also bring considerable value as R\&D for a future long-baseline neutrino facility
based on LAr.

\subsection{SBL Neutrino Anomalies and the Physics of Sterile Neutrinos}
\label{sec:Steriles}

Experimental observations of neutrino oscillations have established a picture consistent with the mixing of three neutrino flavors ($\nu_e$, $\nu_\mu$, $\nu_\tau$) with three mass eigenstates ($\nu_1$, $\nu_2$, $\nu_3$) whose mass differences turn out to be relatively small, with $\dmsq_{31} \simeq 2.4 \times 10^{-3}$ eV$^2$ and $\Delta m_{21} \simeq 7.5 \times 10^{-5}$ eV$^2$~\cite{PDG-2014}.  However, in recent years, several experimental ``anomalies" have been reported which, if experimentally confirmed, could be hinting at the presence of additional neutrino states with larger mass-squared differences participating in the mixing \cite{Abazajian:2012ys}. 

Two distinct classes of anomalies pointing at additional physics beyond the Standard Model in the neutrino sector have been reported, namely \emph{a}) the apparent  disappearance signal in low energy electron anti-neutrinos from nuclear reactors beyond the expected $\theta_{13}$ effect \cite{reactor} (the ``reactor anomaly") and from Mega-Curie radioactive electron neutrino sources in the Gallium experiments \cite{Gallex,Sage} originally designed to detect solar neutrinos (the ``Gallium anomaly"), and \emph{b}) evidence for an electron-like excess in interactions coming from muon neutrinos and anti-neutrinos from particle accelerators \cite{lsnd,mbosc1,mbosc2,mbcomb} (the ``LSND and \MB anomalies"). None of these results can be described by oscillations between the three Standard Model neutrinos and, therefore, could be suggesting important new physics with the possible existence of at least one fourth non-standard neutrino state, driving neutrino oscillations at a small distance, with typically $\Delta m_{new}^2 \geq 0.1 $ eV$^2$.

The ``reactor anomaly" refers to the deficit of electron anti-neutrinos observed in numerous detectors a few meters away from nuclear reactors compared to the predicted rates, with $R_{\text{avg}} = N_{\text{obs}}/N_{\text{pred}}= 0.927 \pm 0.023$ \cite{reactor}. The reference spectra take advantage of an evaluation of inverse beta decay cross sections impacting the neutron lifetime and account for long-lived radioisotopes accumulating in reactors \cite{reactorspectra1, reactorspectra2}. Recent updates have changed the predictions slightly giving a ratio $R_{\text{avg}} = 0.938 \pm 0.023$, a 2.7$\sigma$ deviations from unity \cite{lasserre_appec}. Moreover, some lack of knowledge of the reactor neutrino fluxes is still remaining and a detailed treatment of forbidden transitions in the reactor spectra computation  may result in a few percent increase of systematic uncertainties \cite{reactorspectra4}. A similar indication for electron neutrino disappearance has been recorded by the SAGE and GALLEX solar neutrino experiments measuring the calibration signal produced by intense k-capture sources of $^{51}$Cr and $^{37}$Ar. The combined ratio between the detected and the predicted neutrino rates from the sources is $R = 0.86 \pm 0.05$, again about 2.7 standard deviations from $R = 1$ \cite{Gallex,Sage}. Both of these deficits of low energy electron neutrinos over very short baselines could be explained through $\nue$ disappearance due to oscillations at $\dmsq\geq 1$ eV$^2$.

The LSND experiment \cite{lsnd} at Los Alamos National Laboratory used a decay-at-rest pion beam to produce muon anti-neutrinos between 20-53~MeV about 30~m from a liquid scintillator-based detector  where \nuebar could be detected through inverse beta decay (IBD) on carbon, $\nuebar p \rightarrow e^{+}n$. After 5 years of data taking $89.7 \pm 22.4 \pm 6.0$ \nuebar candidate events were observed above backgrounds, corresponding to 3.8$\sigma$ evidence for $\overline{\nu}_{\mu} \rightarrow \overline{\nu}_{e}$ oscillations  \cite{lsnd} occurring at a \dmsq in the 1 \eVsq region. This signal, therefore, cannot be accommodated with the three Standard Model neutrinos, and like the other short-baseline hints for oscillations at \lenu $\sim $1~m/\MeV, implies new physics.

\begin{figure}
\centering
\mbox{ \includegraphics[width=0.37\textwidth, trim=1cm -1.5cm 0cm 0cm]{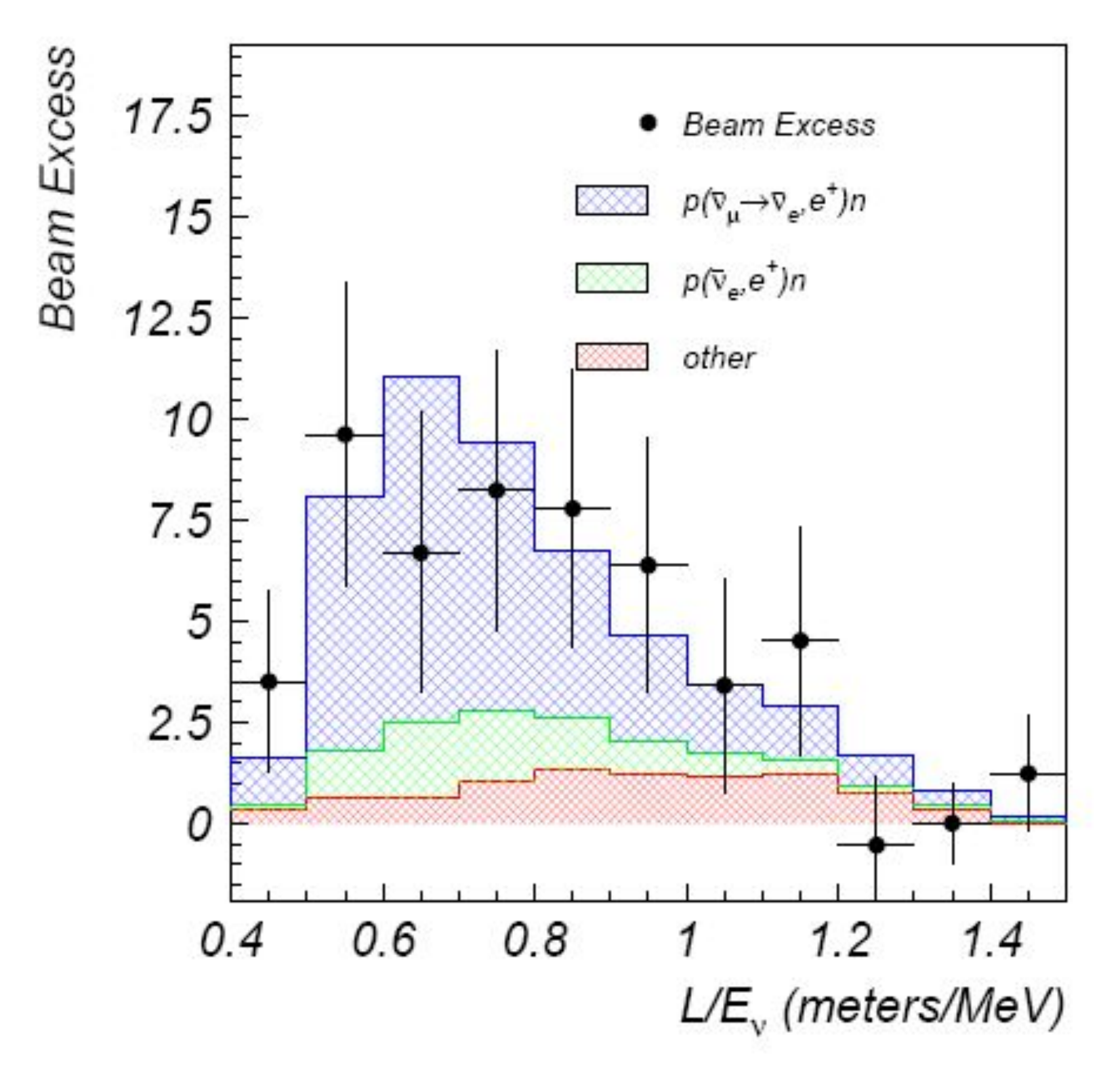} \quad
\includegraphics[width=0.6\textwidth, trim=1cm 1.2cm 1cm 1cm]{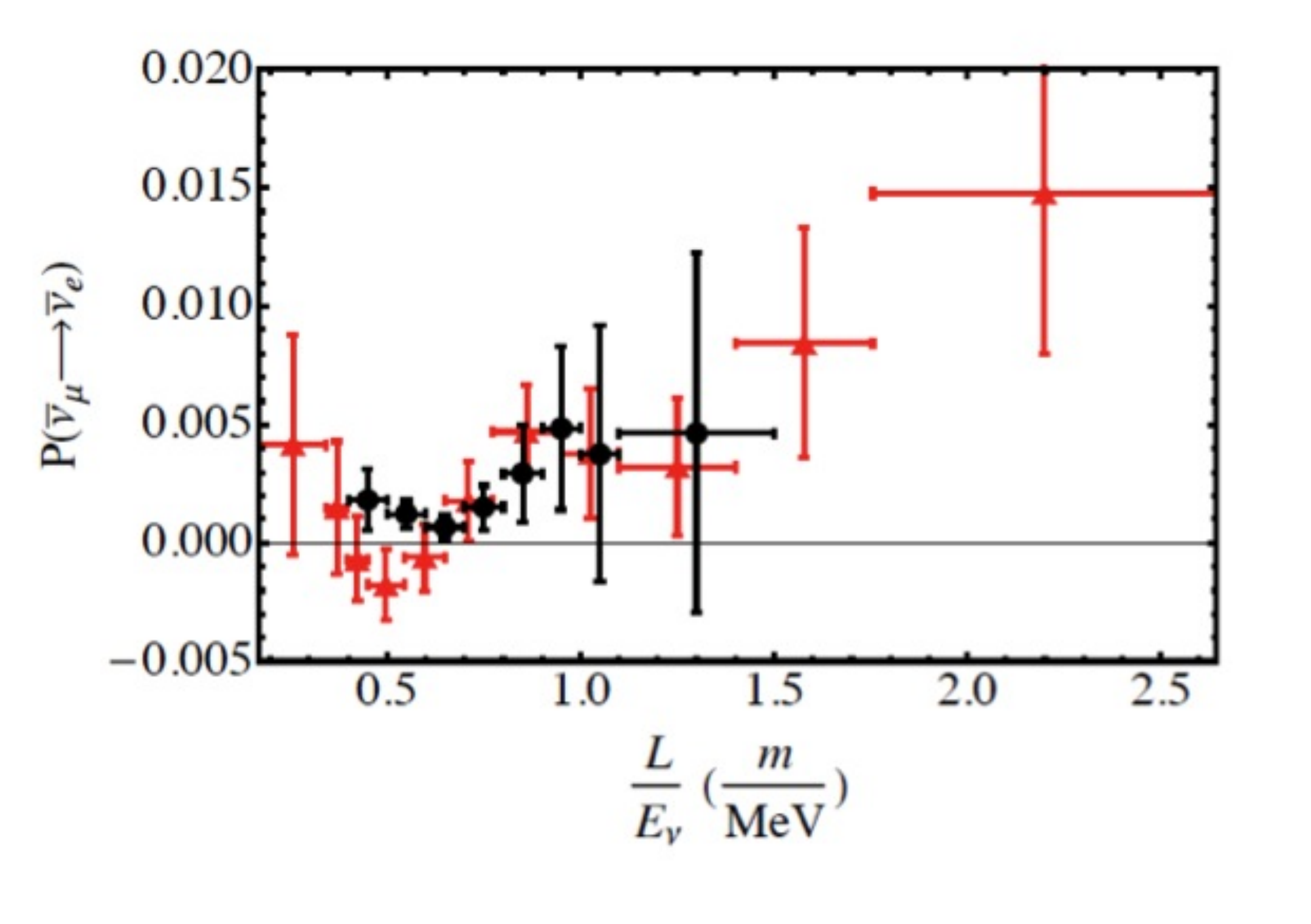}}
\caption{Left: Excess of electron neutrino candidate events observed by the LSND experiment \cite{lsnd}.  Right: Oscillation probability as a function of \lenu if the excess candidate events are assumed to be due to \numunuebar transitions using \MB (red) and LSND (black) data.}
\label{fig:mbooneLSND}
\end{figure}

\begin{figure}[h]
\centering
\mbox{\includegraphics[width=0.5\textwidth]{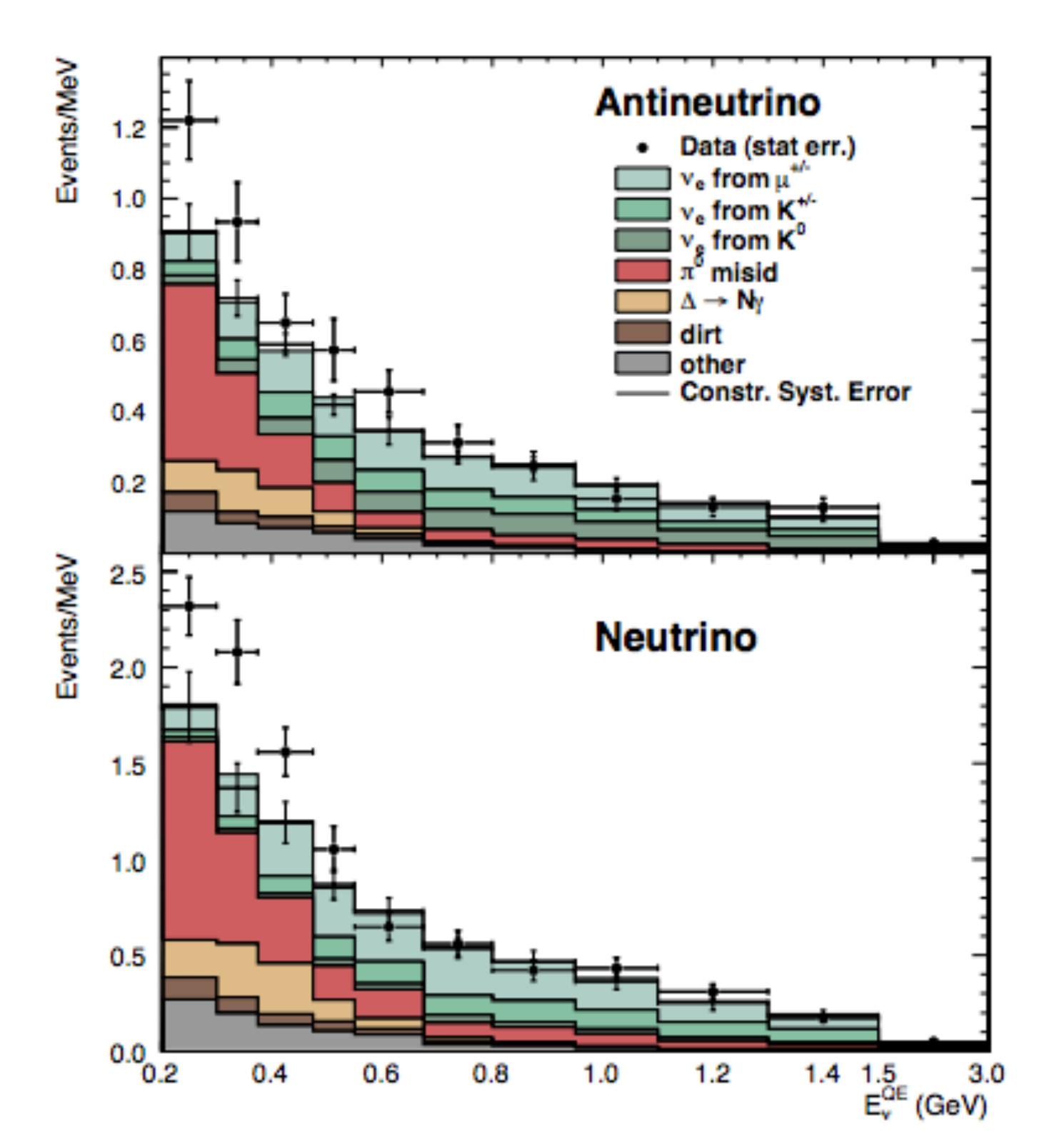}
\includegraphics[width=0.5\textwidth]{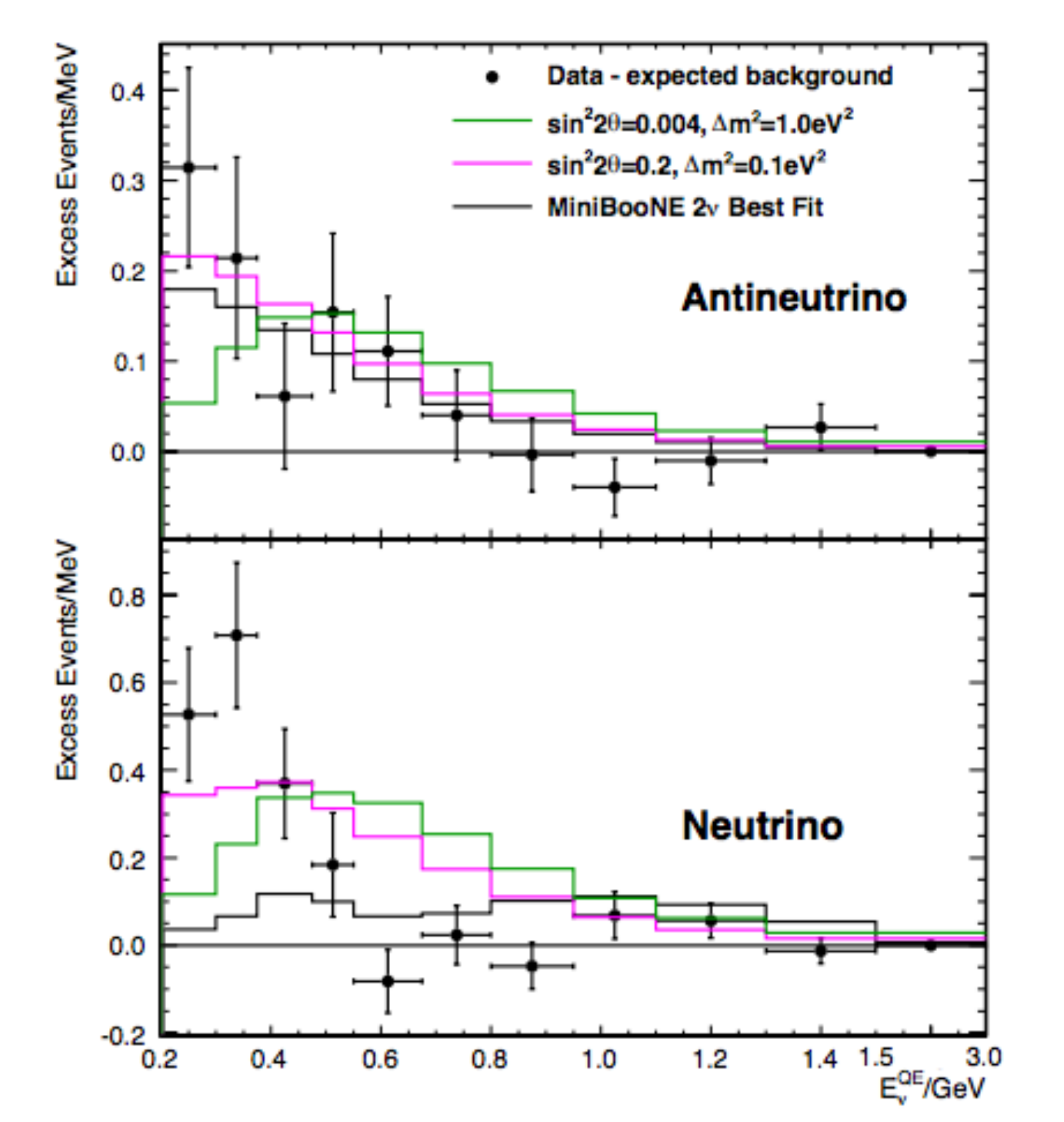}}
\caption{Left: \nuebar (top) and \nue (bottom) candidate events and predicted backgrounds showing the observed excesses in the \MB data. Right: background subtracted event rates in the \MB anti-neutrino (top) and neutrino (bottom) data~\cite{mbcomb}. $E^{QE}_{\nu}$ refers to the reconstructed neutrino event energy, where a quasi-elastic interaction is assumed in the reconstruction.}
\label{fig:mboone}
\end{figure}

The \MB experiment at Fermilab measured neutrino interactions 540~m from the target of the Booster Neutrino Beam (BNB), a predominantly muon neutrino beam peaking at 700~\MeV. Muon and electron neutrinos are identified in charged-current interactions by the characteristic signatures of Cherenkov rings for muons and electrons. In a ten year data set including both neutrino and anti-neutrino running \cite{mbosc1, mbosc2, mbcomb, mbosc3}, \MB has observed a 3.4$\sigma$ signal excess of \nue candidates in neutrino mode ($162.0 \pm 47.8$ electromagnetic events) and a 2.8$\sigma$ excess of \nuebar candidates in anti-neutrino mode ($78.4 \pm 28.5$ electromagnetic events) as shown in Figure~\ref{fig:mboone}. Figure~\ref{fig:mbooneLSND} compares the $L/E_{\nu}$ dependence of the \MB anti-neutrino events to the excess observed at LSND.  The excess events can be electrons or single photons since these are indistinguishable in \MB's Cherenkov imaging detector.  \uboone will address this question at the same baseline as \MB by utilizing the added capability to separately identify electrons and photons.  
%by applying the \lartpc technology to separately identify electrons and photons.  

\begin{figure}[h]
\centering
\hspace*{-5mm}
\mbox{
\raisebox{1mm}{\includegraphics[height=0.235\textheight,trim=0mm 0mm 5mm 17mm, clip]{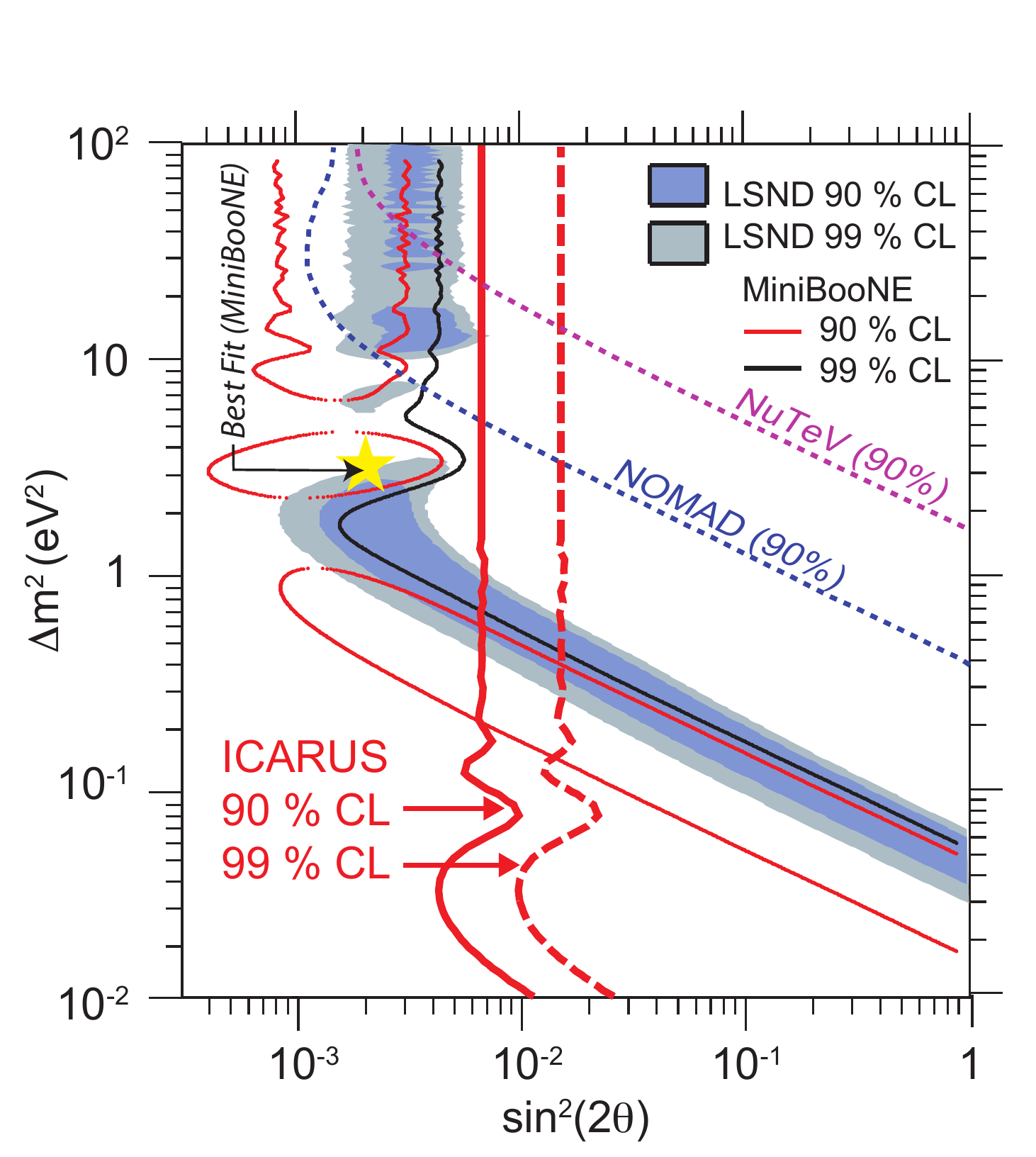}}
\raisebox{0.5mm}{\includegraphics[height=0.24\textheight,trim=0mm 0mm 5mm 0mm, clip]{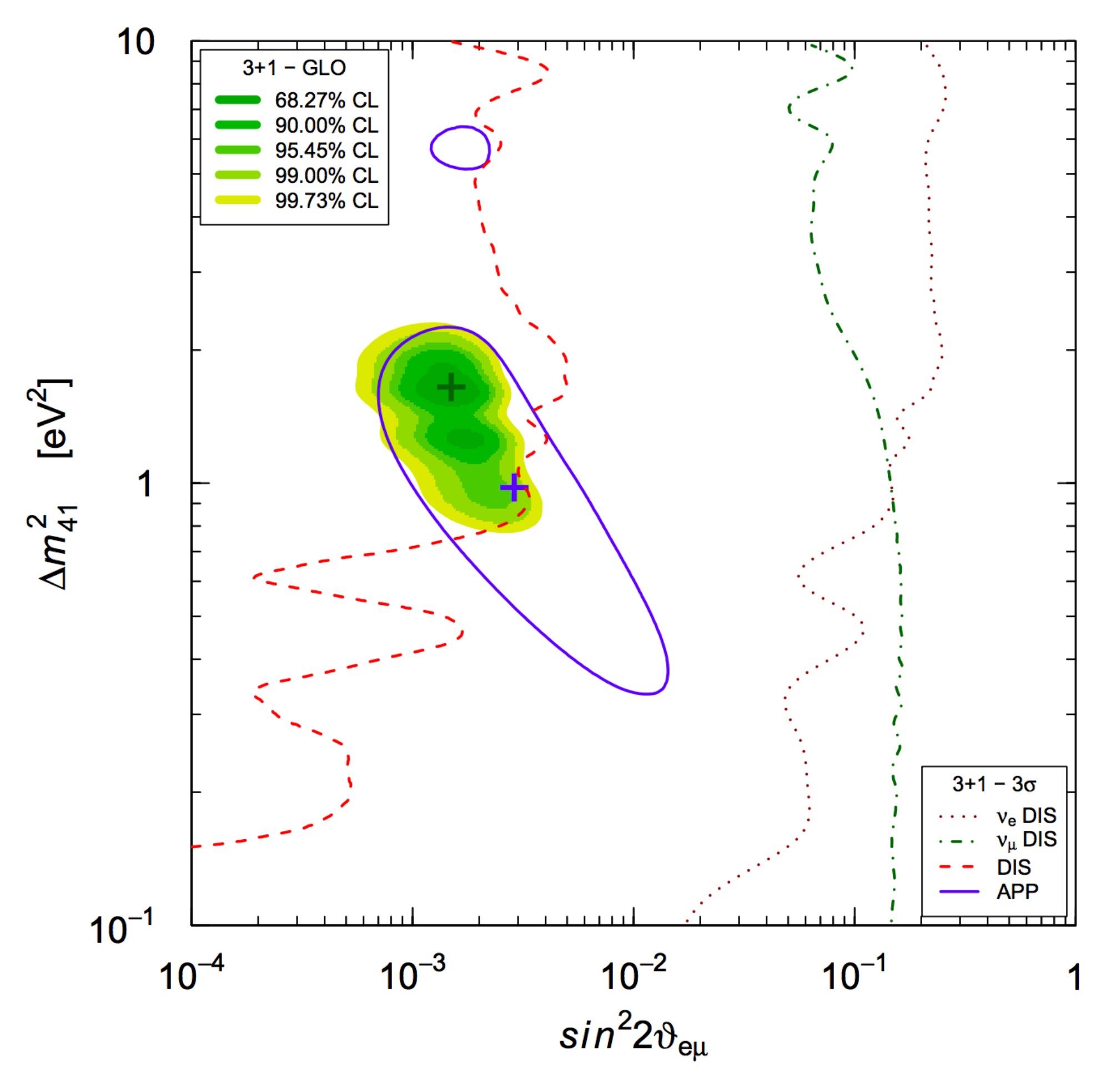}}
\includegraphics[height=0.24\textheight,trim=0mm 0mm 5mm 0mm, clip]{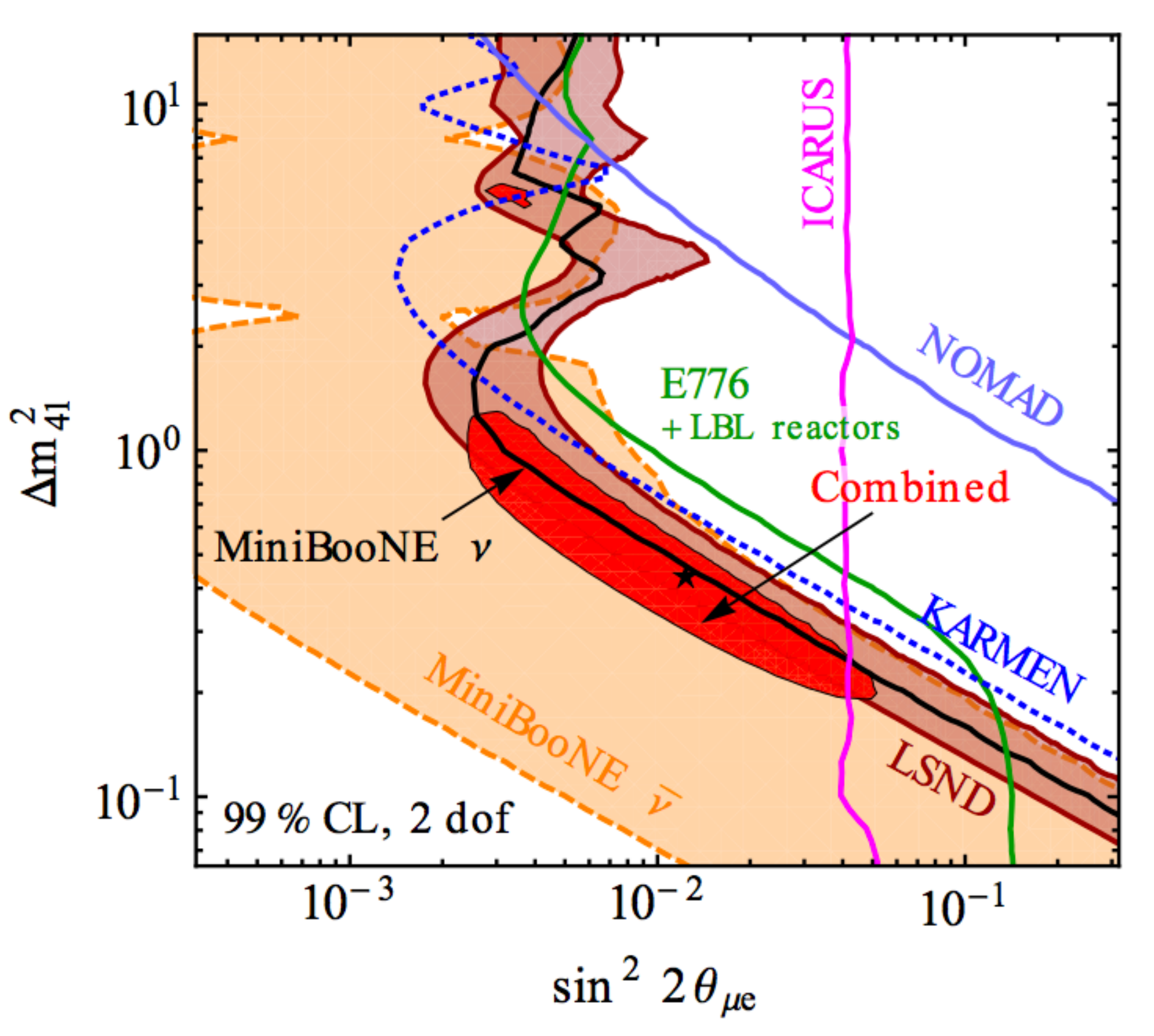}}
\caption{(Left) The main published experimental results sensitive to \numunue at large \dmsq \cite{lsnd,mbosc1,mbosc2,mbcomb,karmen,ccfr,nutev,nomad} including the present ICARUS limit \cite{ICARUS_EPJ-2} from the run in Gran Sasso.  Global analysis of short-baseline neutrino results from Giunti et al. \cite{laveder} (center) and Kopp et al. \cite{kopp} (right). The blue closed contour on the left and the red solid area on the right are the allowed parameter regions for \numunueboth appearance data and both indicate preferred $\dmsq_{41}$ values in the $\sim$[0.2--2] eV$^\text{2}$ range.}
\label{fig:sterile_fits}
\end{figure}

The most common interpretation of this collection of data is evidence for the existence of one or more additional, mostly ``sterile'' neutrino states with masses at or below the few eV range. The minimal model consists of a hierarchical 3+1 neutrino mixing, acting as a perturbation of the standard three-neutrino model dominated by the three  \nue, \numu and \nutau active neutrinos with only small contributions from sterile flavors.  The new sterile neutrino would mainly be composed  of a heavy neutrino $\nu_4$ with mass $m_4$ such that the new $\dmsq = \dmsq_{41}$ and $m_1,\; m_2,\; m_3 \ll m_4$ with $\dmsq_{41} \approx [0.1-10]~\eVsq$. 

In the 3+1 minimal extension to the Standard Model, the effective \nue appearance and \numu disappearance probabilities are described by:
\begin{equation}
P^{3+1}_{\nualpha \rightarrow \nubeta} = \delta_{\alpha\beta} - 4\left|U_{\alpha 4}\right|^2\left(\delta_{\alpha\beta} - \left|U_{\beta 4}\right|^2\right) \sin^2\left(\frac{\Delta m^2_{41}L}{4\Enu}\right)
\label{eq:oscprob}
\end{equation}
\noindent where $U_{ij}$ are elements of the now 4$\times$4 mixing matrix and $L$ is the travel distance of the neutrino of energy \Enu.  The interpretation of both the LSND and \MB anomalies in terms of light sterile neutrino oscillations requires mixing of the sterile  neutrino with both electron and muon neutrinos. Constraints on sterile neutrino mixing from \numu and neutral-current disappearance data are also available~\cite{SciBooNE_MiniBooNE,MiniBooNE_disappearance,MINOS_CCdis,MINOS_NCdis,CDHS}. An explanation of all the available observations in terms of oscillations suffers from significant tension between appearance and disappearance data, particularly due to the absence of \numu disappearance in the $\Delta m^2\sim 1$ \eVsq region. Many global analyses of experimental results have been performed fitting to models including one or more sterile neutrinos.  Figure \ref{fig:sterile_fits} shows two recent examples~\cite{kopp, laveder} of fits to a 3+1 model which indicate similar allowed parameter regions in the $\dmsq_{41} \approx [0.2-2] ~\eVsq$ range when considering available \nue/\nuebar appearance data.  Later, in Section \ref{sec:Analysis}, we will compare SBN sensitivity predictions to the original LSND allowed region and the allowed parameter space in the global data fit from Kopp et al. \cite{kopp} (the red combined region from Figure \ref{fig:sterile_fits}, right) and Giunti et al. \cite{laveder} (the green combined region from Figure \ref{fig:sterile_fits}, center).
%Later, in Section \ref{sec:osc}, we will compare %SBN sensitivity predictions to the original LSND %allowed region and the allowed parameter space in %the global data fit from Kopp et al. \cite{kopp} %(the red combined region from Fig. %\ref{fig:sterile_fits}, right).      

An important contribution to the sterile neutrino search has already been made using the \icarus detector running in the underground INFN-LNGS Gran Sasso Laboratory and exposed to the CERN to Gran Sasso (CNGS) neutrino beam~\cite{ICARUS_jinst}. Although not testing fully the relevant space of oscillation parameters, ICARUS results, corroborated by the OPERA experiment \cite{OPERA}, limit the window for the LSND anomaly to a narrow region around $\dmsq \sim 0.5$ \eVsq and $\sinth \sim 0.005$ \cite{ICARUS_EPJ,ICARUS_EPJ-2}.  In this region, there is overall agreement between the present ICARUS limit, the limit from the KARMEN experiment \cite{karmen}, and the positive signals of LSND and \MB.

%%%%%%%%%%%%%%%%%%%%%%%%%%%%%%%%%%%%%%%%%%%%%%%%%%%%%%%%%%%%%%%%%%%%%%%
%%%%%%%%%%%%%%%%%%%%%%%%%%%%%%%%%%%%%%%%%%%%%%%%%%%%%%%%%%%%%%%%%%%%%%%

\subsection{The Current Experimental Landscape}

Given the importance of a sterile neutrino discovery, it is clear that the existing anomalies must be explored further by repeating the existing measurements in an effective way capable of addressing the oscillation hypothesis and many experiments are setting out to explore it \cite{lasserre}.

New reactor experiments searching for oscillations with $L/\Enu \sim 1$ m/MeV are in preparation aiming to detect an oscillation pattern imprinted in the energy distribution of events.  Experimentally the detection technique relies on the IBD reaction, $\nuebar p \rightarrow e^{+}n$, where the positron carries out the \nuebar energy and tagging the neutron provides a discriminant signature against backgrounds. The backgrounds from radioactive contaminants or induced by the reactor core and by cosmic rays can partially be suppressed through passive shielding while the remaining contribution can be measured in-situ at the analysis stage. The Nucifer experiment \cite{nucifer} at the Osiris nuclear reactor in Saclay could provide first new constraints by 2015. The Stereo experiment \cite{stereo} will be constructed next to the ILL reactor in Grenoble, France. The DANSS \cite{danss} and Neutrino4 \cite{neutrino4} experiments are under construction in Russia and should provide first data in 2015. Finally, comprehensive projects for searching for sterile neutrinos at reactors in China \cite{china} and the US \cite{us} are currently under study. All these experiments are designed to test the space of oscillation parameters deduced from the interpretation of the reactor anti-neutrino deficits.

New projects aiming to search for evidence of oscillations using neutrinos from intense radioactive sources have also been proposed. The SOX experiment~\cite{SOX} will perform such a measurement with a 10~MCi $^{51}$Cr source deployed at 8.25~m  from the center of the Borexino detector in 2017. At Baksan a 3~MCi $^{51}$Cr source could be placed at the center of a target, containing 50~tons of liquid metallic gallium divided into two areas, an inner 8~ton zone and an outer 42~ton zone. The ratio of the two measured capture rates to its expectation could signify an oscillation. This is a well-proven technique free of backgrounds, developed for the SAGE solar neutrino experiment. The CeLAND and CeSOX projects plan to use 100~kCi of $^{144}$Ce in KamLAND~\cite{PBqSource, CeLAND} and Borexino~\cite{PBqSource, SOX} to produce an intense anti-neutrino flux which can be detected through the inverse beta decay process. The goal is to deploy the $^{144}$Ce radioisotope about 10~m away from the detector center and to search for an oscillating pattern in both event spatial and energy distributions that would determine neutrino mass differences and mixing angles unambiguously. The CeSOX experiment could take data as early as the end of 2015 at LNGS with Borexino.

A new neutrino, $\nu_{4}$, heavier than the three active neutrinos should be detected in the KATRIN experiment~\cite{KATRIN}. The detector aims as measuring precisely the high energy tail of the tritium $\beta$-decay spectrum by combining an intense molecular tritium source with an integrating high-resolution spectrometer reaching a 200 meV sensitivity on the effective electron neutrino mass at 90\% C.L. The detection principle for a new sterile neutrino state is to search for a distortion at the high energy endpoint of the electron spectrum of tritium $\beta$-decay, since its shape is \emph{a priori} very precisely understood. The KATRIN experiment can probe part of the current allowed region of the reactor anti-neutrino anomaly, especially for $\Delta m_{new}^2  > 1~\eVsq$, with 3 years of data-taking~\cite{Formaggio2011, Esmaili2012}. First results are expected in 2016.

As a long term project, a huge statistics of \nuebar $\to$ \numubar from the $\beta$-decay of $^{8}$Li could be obtained through the development of a high-power low energy cyclotron. The IsoDAR project~\cite{Aberle2013} proposes to place such a device underground in the Kamioka mine to search for an oscillation pattern in the KamLAND detector. This would be a disappearance experiment directly testing both the reactor and the gallium anomalies starting from a well known  \numubar spectrum.

The OscSNS project~\cite{Elnimr2013} proposes to locate an 800-ton gadolinium-doped scintillator detector 60 m away from the Spallation Neutron Source (SNS) at the Oak Ridge National Laboratory in order to directly test the LSND results. This kind of facility has the advantage of producing a well-understood source of electron and muon neutrinos from $\pi^+$ and $\mu^+$ decays-at-rest. The main search channel would be the appearance of \nuebar, taking advantage of the low duty factor of SNS to reduce cosmic induced backgrounds.

\begin{figure}[t]
\centering
\includegraphics[width=1\textwidth]{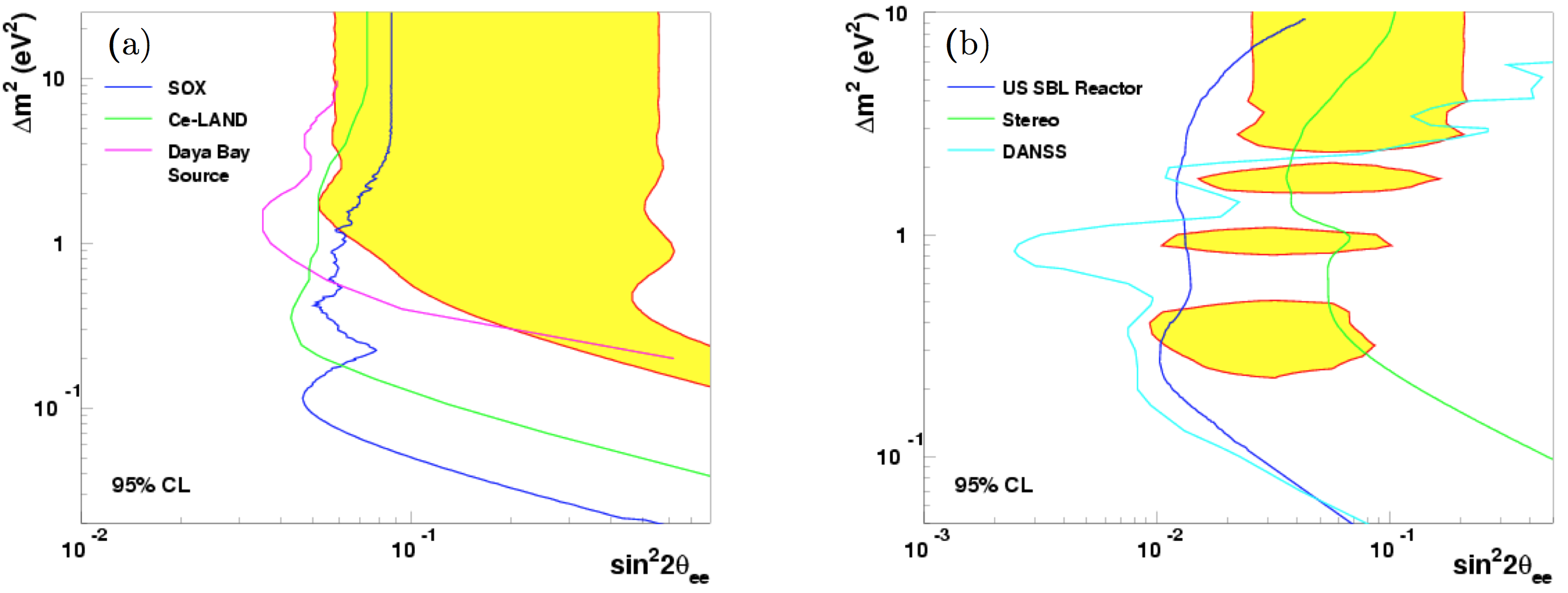}
\caption{Expected sensitivity curves at 95\% C.L. for proposed neutrino experiments with radioactive sources (a) and reactors (b) with the global fits to the existing gallium and reactor data (yellow regions) \cite{reactorspectra3}.}
\label{fig:reactor_source}
\end{figure}

A precision sterile neutrino search has been proposed with a clean and well-understood beam of \nue and \numubar produced in a low energy neutrino factory by the decay of stored muons both at CERN~\cite{Adey2013} and Fermilab~\cite{Kyberd2013} by the nuSTORM project. Such a neutrino beam could be used to probe both appearance and disappearance processes including the golden channel of \numu appearance in a muon-free electron neutrino beam, which is not possible in a meson decay-in-flight beam.
 
However, considering the present experimental scenario, an accelerator-based neutrino beam facility provides the best opportunity for a rich oscillation research program with a single experiment, where the existence of an oscillation signal in \nue appearance and disappearance modes as well as \numu disappearance can be simultaneously investigated. Neutrino or anti-neutrino beams can be produced in the same experiment and, at accelerator beam energies, both charged-current and neutral-current channels can be explored.  This is the approach of the short-baseline neutrino oscillation program on the FNAL Booster Neutrino Beam proposed here. \uboone is blazing the trail on the BNB with liquid argon technology now, but the challenge of predicting absolute neutrino fluxes in accelerator beam experiments and the large uncertainties associated with neutrino-nucleus interactions, strongly motivate the use of multiple detectors at different baselines to reduce systematic uncertainties in the search for oscillations. The anomalous short-baseline results discussed in Section \ref{sec:Steriles} may be hinting at neutrinos oscillating with an amplitude 10 to 100 times \emph{smaller} than the $\theta_{13}$ signals in experiments like Daya Bay, T2K, or MINOS, all multiple detector experiments. The Fermilab SBN Program, using detectors at different distances from the BNB source, will cover at high confidence level the entirety of the sterile neutrino parameter space suggested by the anomalies.
 
Finally, the observed set of anomalous results in neutrino physics call for conclusive new experiments capable of exploring the indicated parameter regions in a definitive way and to clarify the possible existence of eV-scale sterile neutrinos.  The accelerator-based short-baseline program presented in this proposal is the only means of testing the sterile neutrino picture through multiple channels in a single beam.

%\clearpage

%%%%%%%%%%%%%%%%%%%%%%%%%%%%%%%%%%%%%%%%%%%%%%%%%%%%%%%%%%%%%%%%%%%%%%%%
% SBN Sensitivities
%%%%%%%%%%%%%%%%%%%%%%%%%%%%%%%%%%%%%%%%%%%%%%%%%%%%%%%%%%%%%%%%%%%%%%%%

\section{SBN Oscillation Searches}
\label{sec:osc}

%The SBN program of three \lartpc detectors along the Fermilab Booster Neutrino Beam delivers a rich and diverse physics opportunity.  %Neutrino-argon cross sections can be studied first in \uboone and later with even higher statistics in \larnd using the well characterized neutrino fluxes of the BNB \cite{MiniFlux}.  \uboone and ICARUS will also record samples of events from the off-axis NuMI beam~\cite{MiniBooNE_NuMI}. The source of the \MB electromagnetic event excess will be directly checked in the same beam using the \lartpc technology in order to separate $e^{\pm}$ from single $\gamma$ interactions.  
Multiple \lartpc detectors at different baselines along the BNB will allow a very sensitive search for high-\dmsq neutrino oscillations in multiple channels.  These searches constitute the flagship measurements of the SBN program, and so we dedicate this section to a careful and detailed description of the sensitivity analysis for \numunue appearance and \numudis disappearance.  

This section is organized into subsections as follows. In Section \ref{sec:Analysis} we provide a mathematical description of the analysis methods used to calculate the sensitivities.  In Section \ref{sec:EventRates} we describe the procedures for selecting events for the \numu and \nue analyses and characterize in-detector intrinsic beam-related backgrounds to each.  In Sections \ref{sec:Flux} and \ref{sec:XSec} we present the systematic uncertainties impacting these predictions related to the neutrino fluxes and neutrino interaction model, with particular emphasis on the correlations between different detector locations that enable the increased sensitivity of a multi-detector experiment. Section \ref{sec:DetSyst} discusses detector related systematic uncertainties. Section \ref{sec:Dirt} deals with out-of-detector but beam-induced backgrounds; these include neutrino interactions in the earth surrounding each detector building, hence we often refer to this category as ``dirt'' events, though interactions in the the building, cryostat, and inactive argon surrounding the TPC which deposit energy in the detector are all included.  In \ref{sec:Cosmics} we discuss cosmogenic backgrounds and the strategies to mitigate them.  Both the dirt and cosmogenic backgrounds only affect the \nue analysis.  Finally, we bring it all together and present the oscillation sensitivities of the SBN program to \numunue appearance and \numudis disappearance in Sections \ref{sec:SBN_nue} and \ref{sec:SBN_numu}, respectively.

\subsection{Analysis Methods}
\label{sec:Analysis}

The sensitivity of the SBN program will be demonstrated using the commonly assumed framework of three active and one sterile neutrino, or a ``3+1 model'', as our baseline for evaluation.  Of course, other models could be assumed, such as those with multiple sterile states, but this choice provides a straight-forward way to compare to previous experimental results as well as to global data fits that were analyzed using the 3+1 model\footnote{Of course, what we would like to know is the general ability of the experiment to observe either an excess or a deficit relative to the expectation in the absence of any oscillation.  In a sense, the 3+1 sensitivity contains this information, but for many different possible distributions of the signal events across the observed energy spectrum.}. In the 3+1 model, the effective oscillation probabilities are described by Eq. \ref{eq:oscprob}, reproduced here explicitly for \nue appearance (\numunue) and \numu disappearance (\numunumu):

\begin{equation}
P^{3+1}_{\numu \rightarrow \nue} = \sin^2 2\theta_{\mu e} 
\sin^2\left(\frac{\Delta m^2 L}{4\Enu}\right) \hspace{15mm}
P^{3+1}_{\numu \rightarrow \numu} = 1 - \sin^2 2\theta_{\mu \mu} \sin^2\left(\frac{\Delta m^2 L}{4\Enu}\right)
\nonumber
\end{equation}

\noindent with $L$ the propagation length of the neutrino and \Enu the neutrino energy, $\sin^2 2\theta_{\mu e} \equiv 4|U_{\mu 4}U_{e4}|^2$ is an effective mixing amplitude that depends on the amount of mixing of both \numu and \nue with mass state \nufour, and $\sin^2 2\theta_{\mu\mu} \equiv 4|U_{\mu 4}|^2(1 - |U_{\mu 4}|^2)$ only depends on the amount of  \numu--\nufour mixing.  In our standard picture, any observation of \nue appearance due to oscillations must be accompanied by some amount of \numu disappearance as well as for the similar \nue disappearance.

\begin{figure}[t]
%\centering
\setlength{\fboxsep}{0pt}%
\setlength{\fboxrule}{0.5pt}%
\mbox{
\fbox{
\includegraphics[height=0.12\textheight,width=0.17\textheight,trim=0mm 0mm 10mm 0mm, clip]{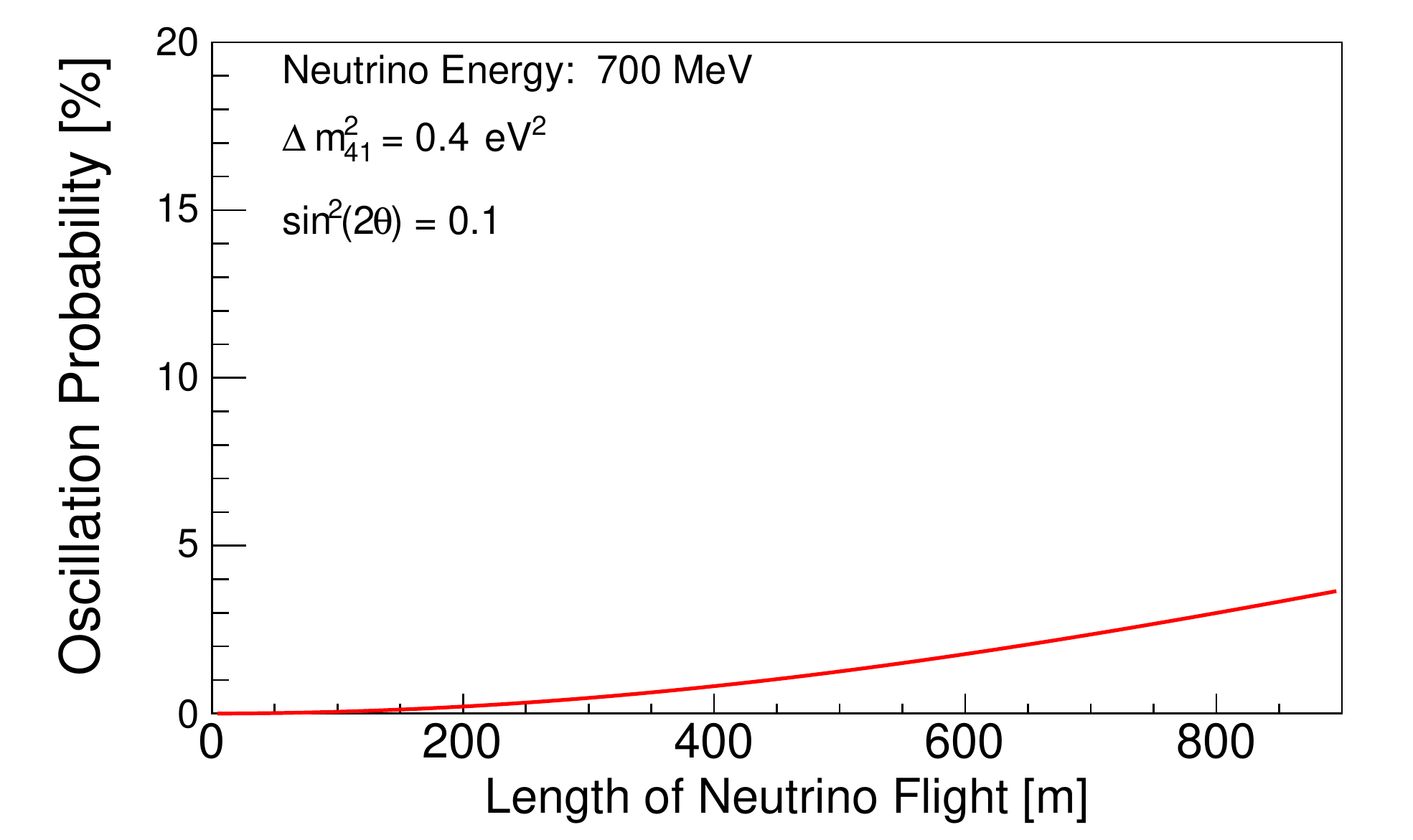}
\includegraphics[height=0.12\textheight,width=0.17\textheight,trim=0mm 0mm 5mm 0mm, clip]{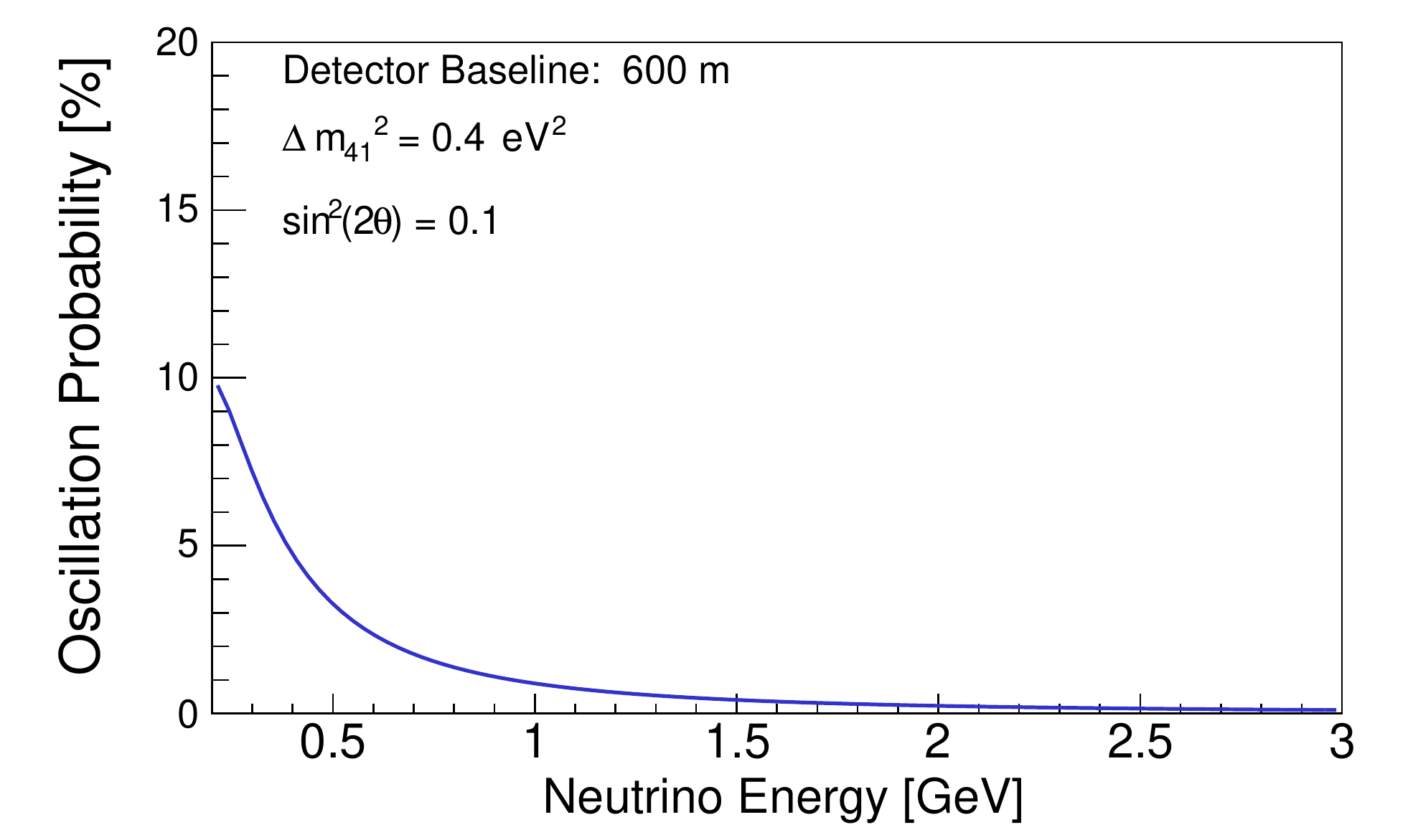}}
%\hspace{1pt}
\fbox{
\includegraphics[height=0.12\textheight,width=0.17\textheight,trim=0mm 0mm 10mm 0mm, clip]{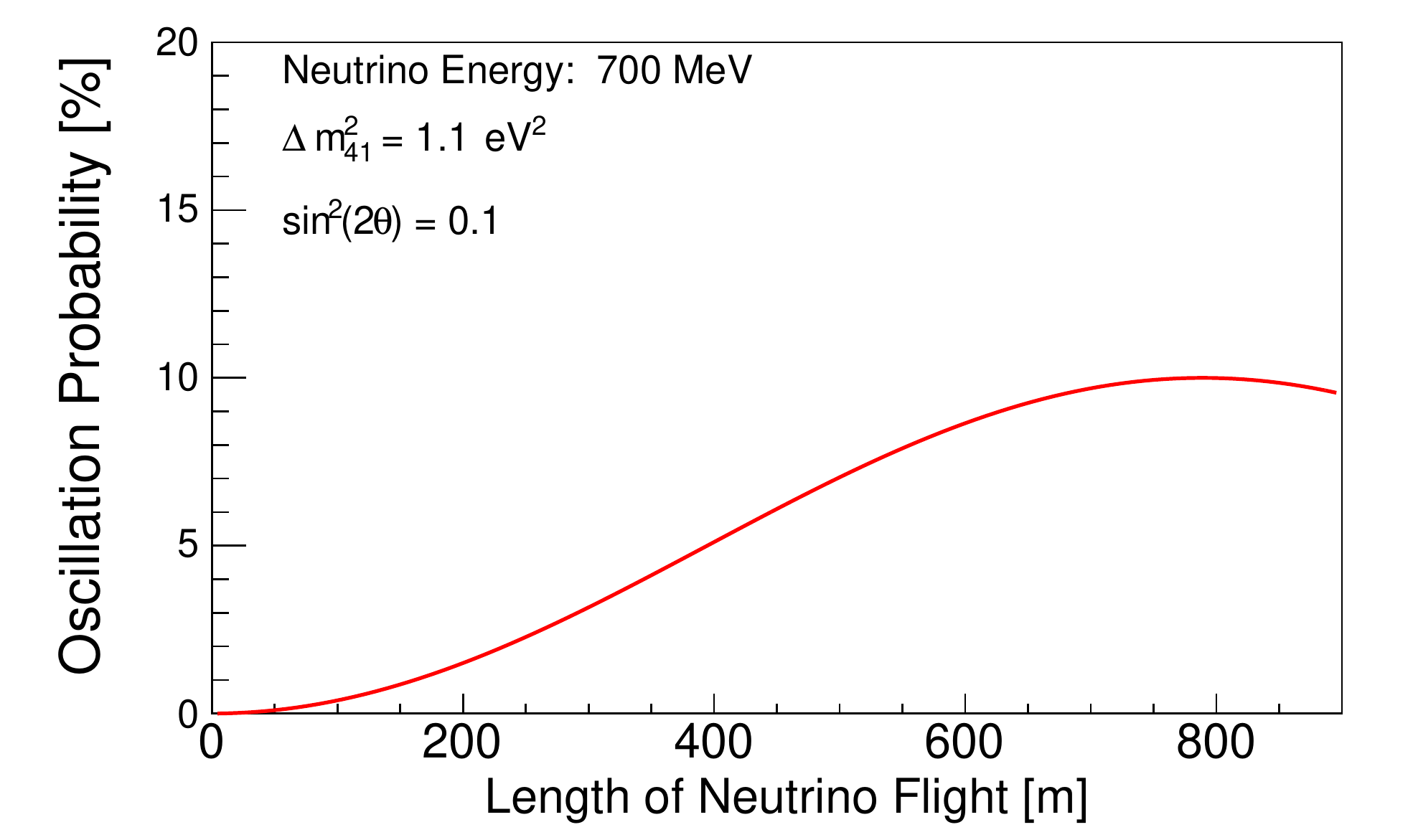}
\includegraphics[height=0.12\textheight,width=0.17\textheight,trim=0mm 0mm 5mm 0mm, clip]{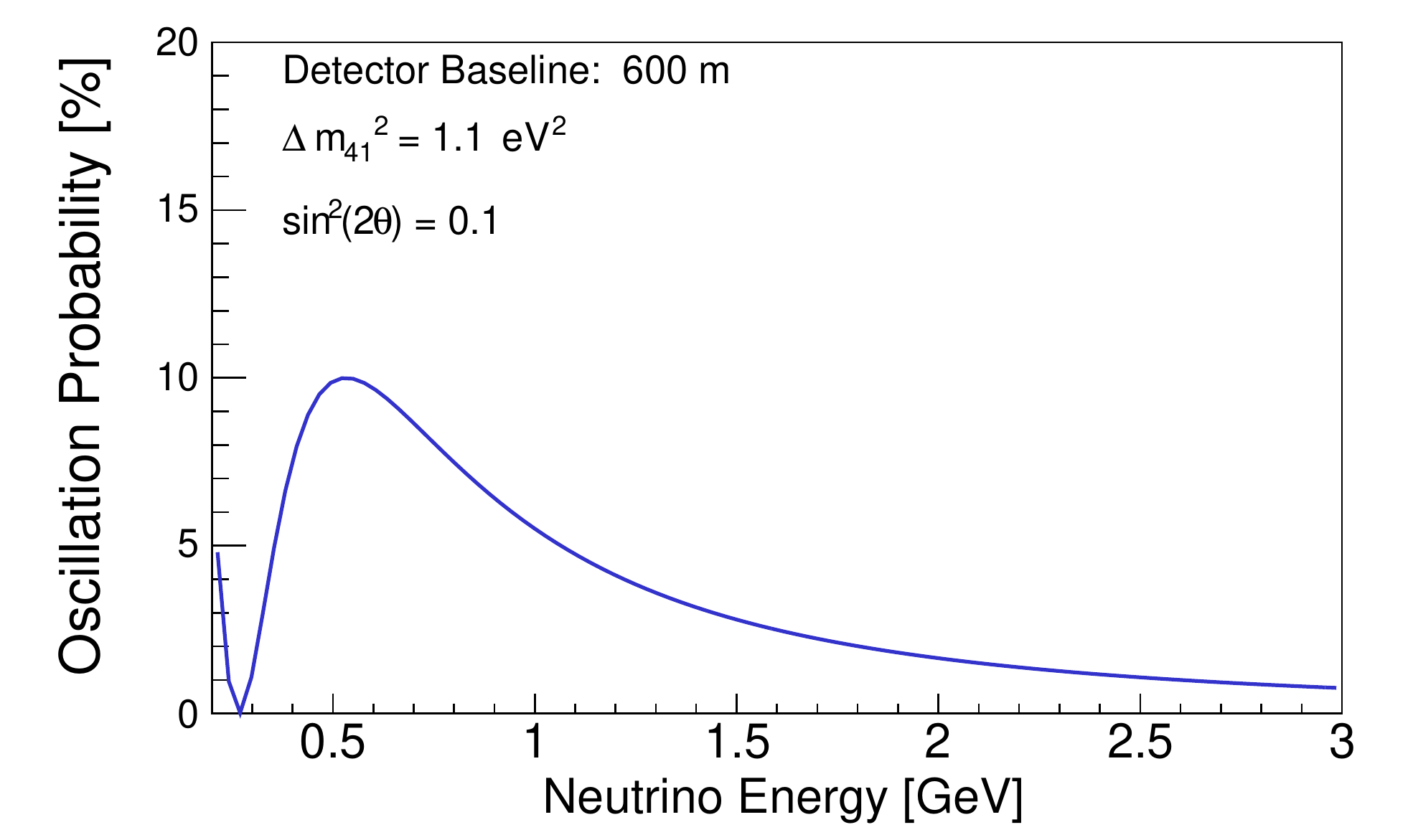}}}
\mbox{
\fbox{
\includegraphics[height=0.12\textheight,width=0.17\textheight,trim=0mm 0mm 10mm 0mm, clip]{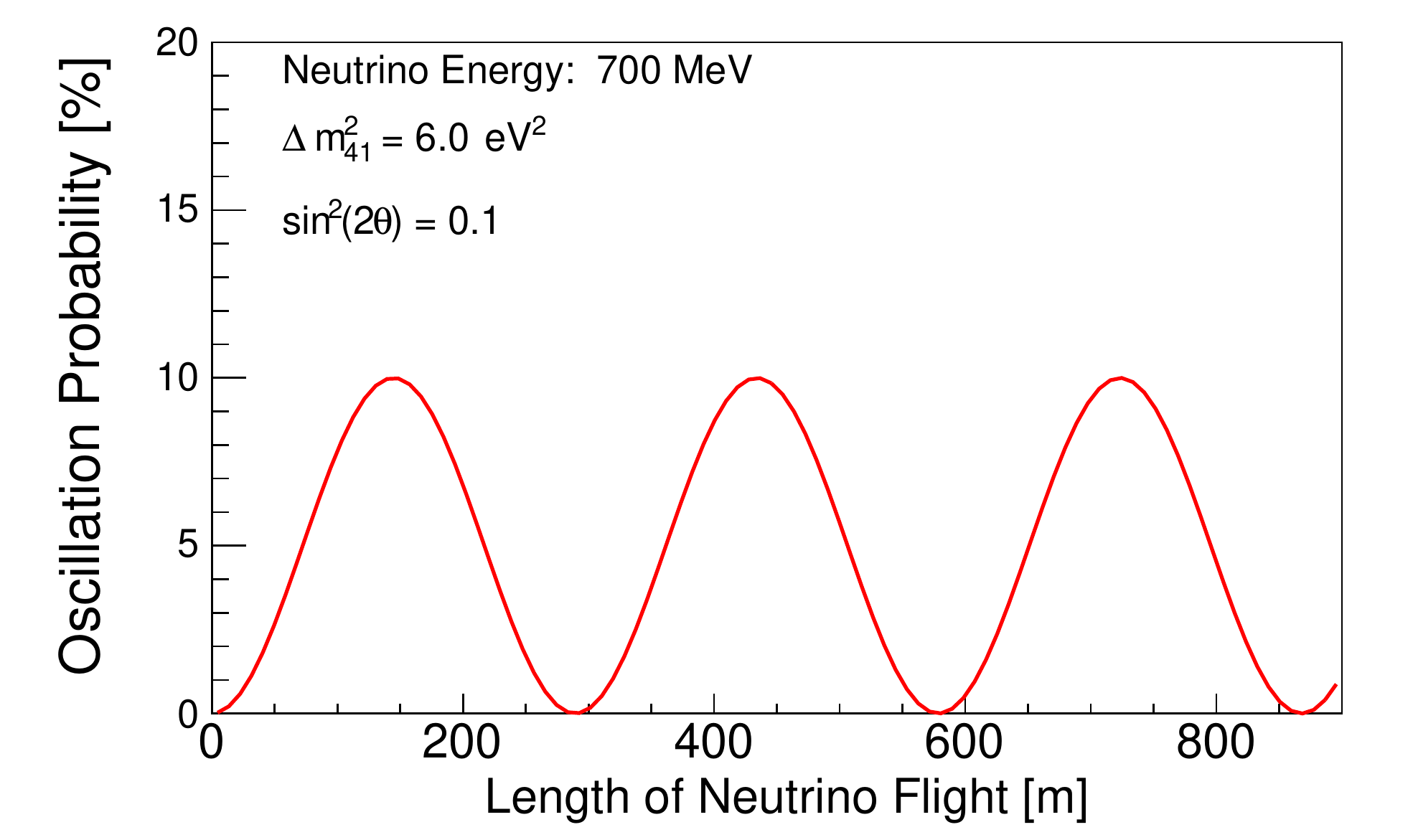}
\includegraphics[height=0.12\textheight,width=0.17\textheight,trim=0mm 0mm 5mm 0mm, clip]{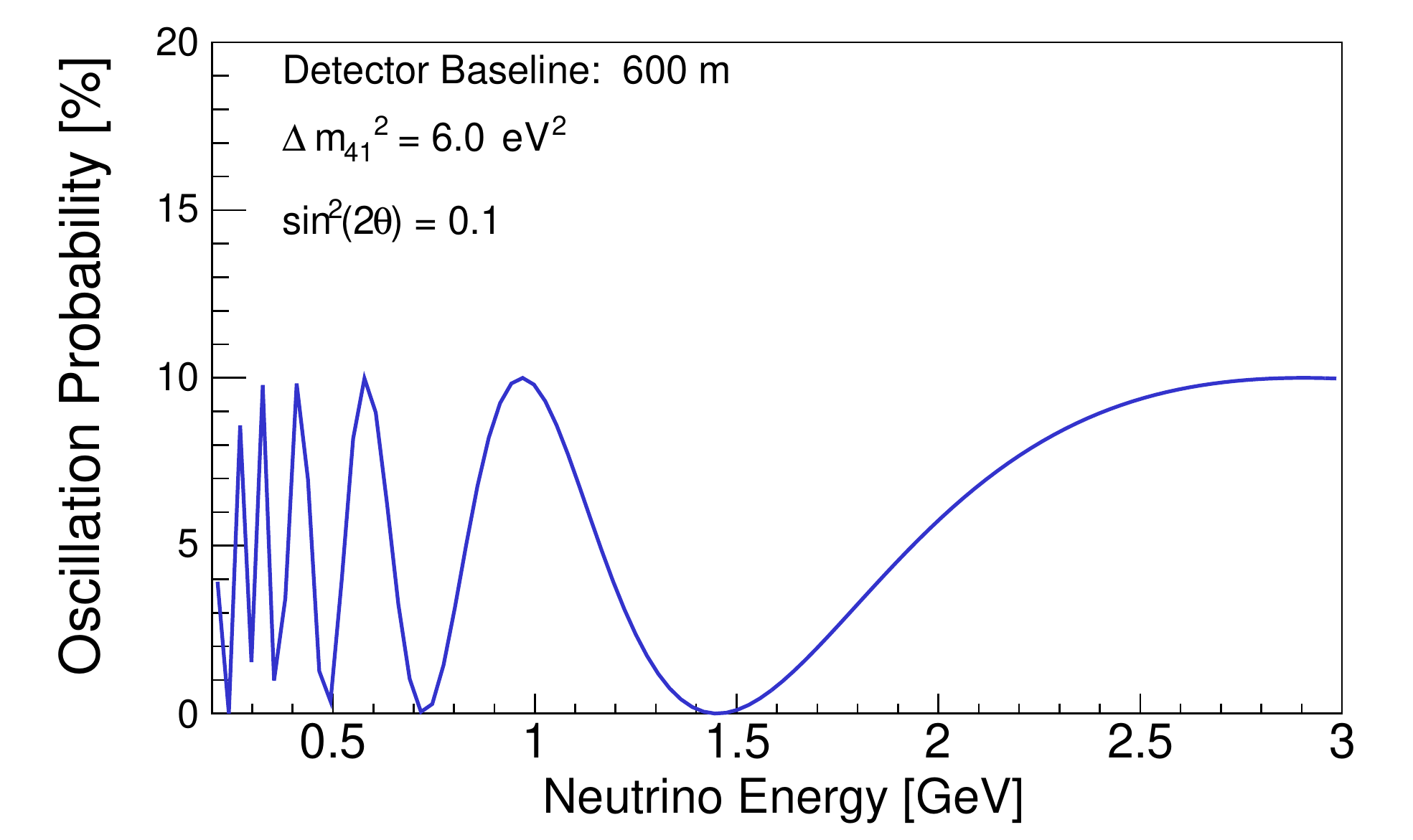}}
%\hspace{1pt}
\fbox{
\includegraphics[height=0.12\textheight,width=0.17\textheight,trim=0mm 0mm 10mm 0mm, clip]{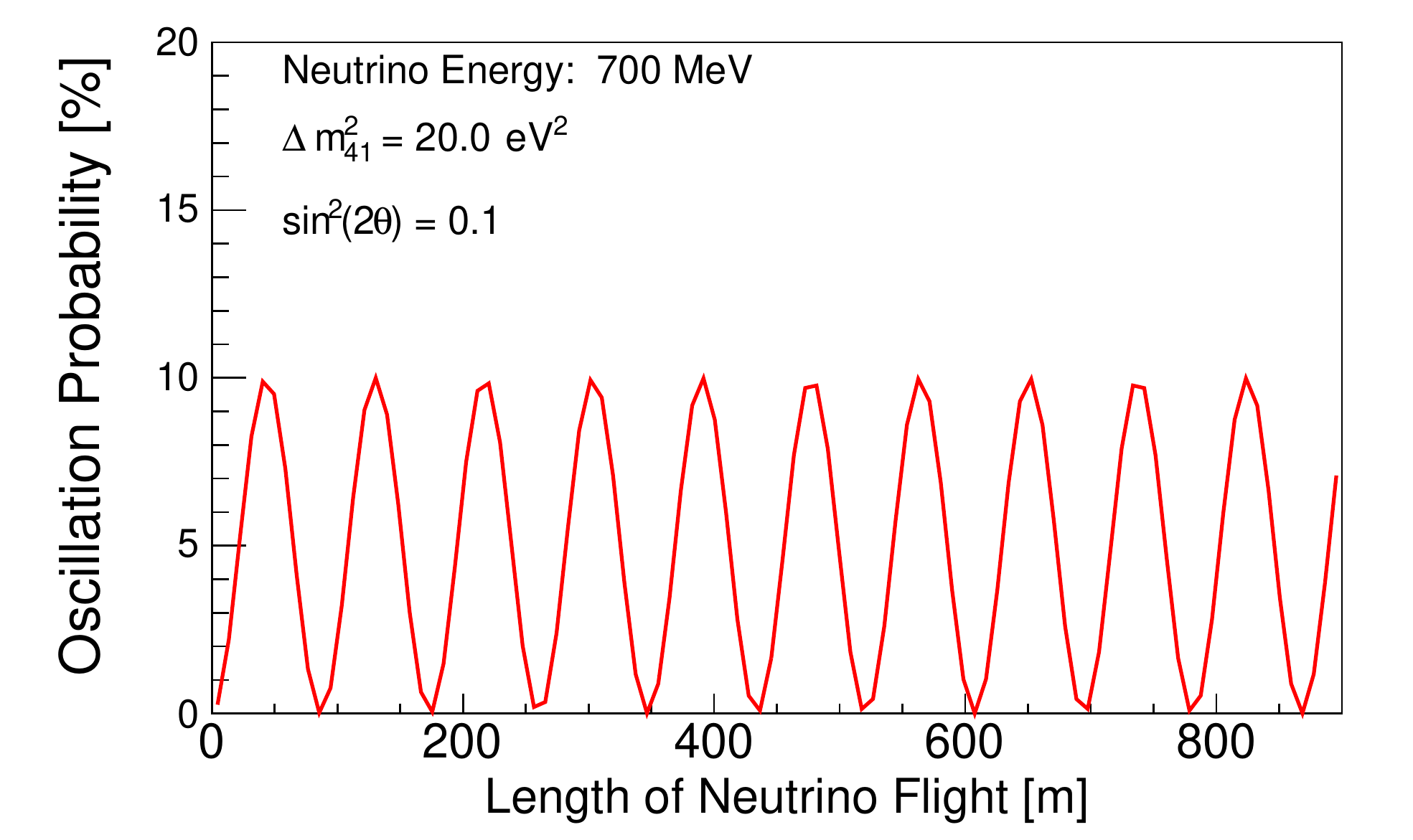}
\includegraphics[height=0.12\textheight,width=0.17\textheight,trim=0mm 0mm 5mm 0mm, clip]{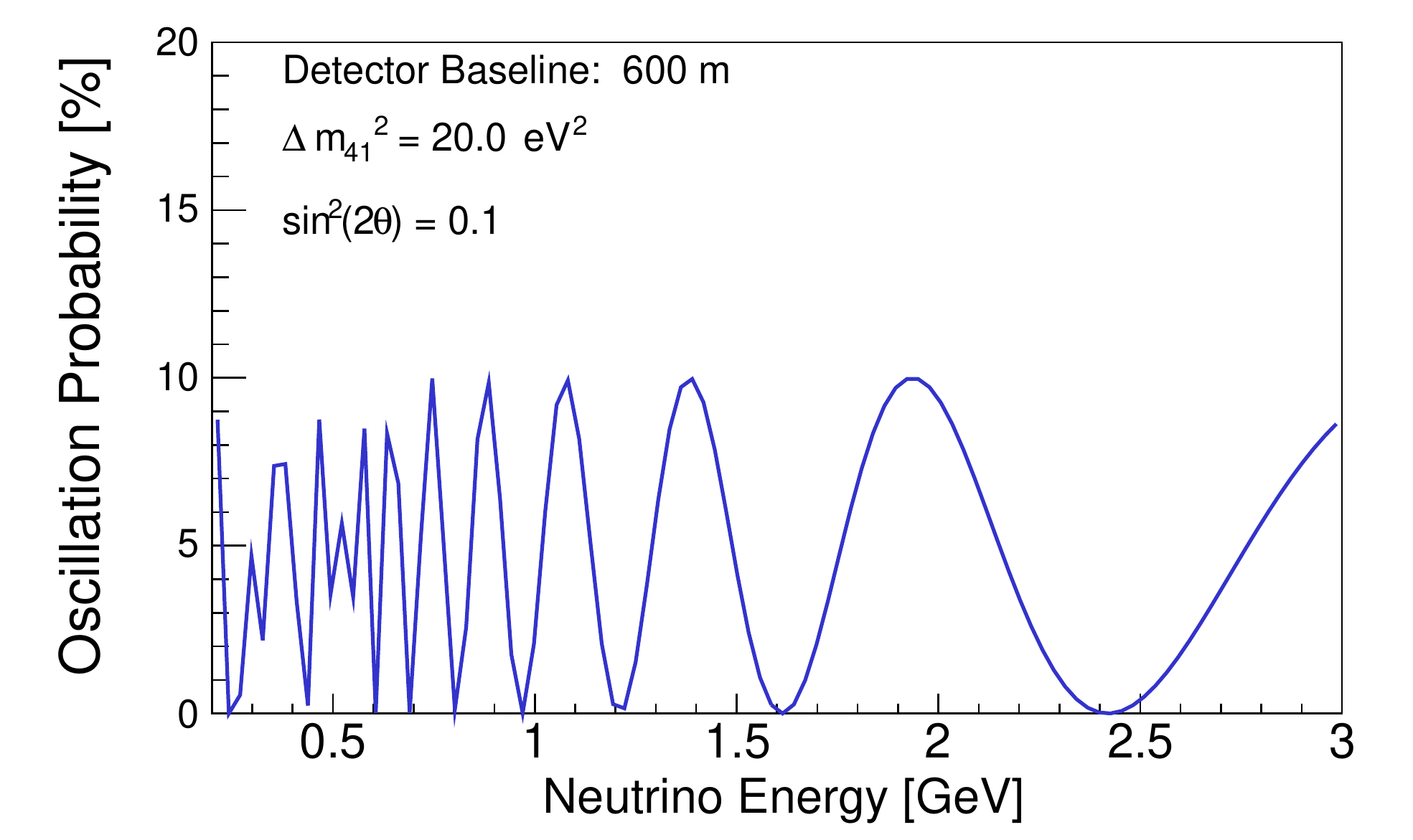}}}
\caption{Illustrations of the oscillation probability at SBN for four different values of \dmsq and \sinth = 0.1 in a 3+1 sterile neutrino model: \dmsq = 0.4 \eVsq (upper left), 1.1 \eVsq (upper right), 6 \eVsq (lower left), and 20 \eVsq (lower right). In each panel, the left red curve shows the evolution of the probability with \emph{distance} at a fixed energy (\Enu = 700~MeV).  The right blue curve shows the probability versus \emph{energy} at a fixed location (600~m, the \icarus location).}
\label{fig:oscprobs}
\end{figure}

Figure~\ref{fig:oscprobs} illustrates the shape of the oscillation probability in the SBN experiments for four different possible values of \dmsq (\sinth = 0.1).  The red curves show the evolution of the oscillation probability with distance for a fixed neutrino energy, \Enu = 700~MeV, while the blue curves demonstrate the oscillation probability across the full BNB neutrino energy range at the far detector location, 600~m.  From the top row (0.4 \eVsq and 1.1 \eVsq), one can clearly see why the sensitivity increases with \dmsq up to and a little beyond 1 \eVsq as the oscillation probability at 600~m increases but also shifts toward the peak of the BNB flux.  Also, note that the level of signal at the near detector location (110~m) is very small, making the near detector measurement an excellent constraint on the intrinsic beam content.  For \dmsq much larger than 1 \eVsq, as we see in the bottom row (6 \eVsq and 20 \eVsq), the oscillation wavelength becomes short compared to the 600~m baseline.  As a function of energy in \emph{all} detectors, the oscillations are rapid in neutrino energy and one observes an overall excess (or deficit) at all energies equal to half the value of \sinth. Therefore, at high \dmsq, the near detector is also contaminated with signal and absolute normalization uncertainties become important in determining the sensitivity.             

The sensitivity is calculated by computing a $\chi^2$ surface in the ($\dmsq_{41}$, $\sin^22\theta$) oscillation parameter plane according to:

\begin{equation}
\hspace*{-5mm} 
\chi^{2}(\Delta m_{41}^2, \sin^2{2\theta})= \sum_{i,j}\left[N^{null}_{i} - N^{osc}_{i}(\Delta m_{41}^2, \sin^2{2\theta})\right]\left(E_{ij}\right)^{-1}\left[N^{null}_{j} - N^{osc}_{j}(\Delta m_{41}^2, \sin^2{2\theta})\right]
\label{eq:chi}
\end{equation} 

\noindent where $N^{null}_{i}$ is the expected event distribution in the absence of oscillations and $N^{osc}_{i}(\Delta m_{41}^2, \sin^2{2\theta})$ is the event prediction for an oscillation signal determined by Eq. \ref{eq:oscprob} with mass splitting $\Delta m_{41}^2$ and amplitude $\sin^2{2\theta}$. The labels $i$ and $j$ indicate bins of reconstructed neutrino energy.  Uncertainties, both statistical and systematic, are encoded in the covariance matrix, $E_{ij}$. From this surface, sensitivity contours at different confidence levels (C.L.)\footnote{$\Delta\chi^2_{90}=1.64, \Delta\chi^2_{3\sigma}=7.74$, and $\Delta\chi^2_{5\sigma}=23.40$ corresponding to a one-sided, one degree of freedom $\Delta\chi^2$ cut.} can be identified based on the $\chi^2$ values relative to the overall minimum value.   We devote the next five Sections to describing how we estimate the background event vectors $N^{null}_{i}$ and the covariance matrices $E_{ij}$ for the \nue appearance and \numu disappearance analyses.  
%To build the final $\chi^2$ we need to have a selection of events which pass our analysis cuts (see Sec.~\ref{sec:EventRates}) and a complete covariance matrix.

The total systematic covariance matrix is a combination of independent matrices constructed for each of the systematic uncertainties considered:

\begin{equation}
E^{\text{syst}} = E^{\text{flux}}+E^{\text{cross section}}+E^{\text{cosmic bkgd}}+E^{\text{dirt bkgd}}+E^{\text{detector}}
\label{eq:TotCov}
\end{equation} 

\noindent and $E^{total} = E^{\text{stat}}+E^{\text{syst}}$ where $E^{\text{stat}}$ is the completely uncorrelated statistical error matrix, $E^{stat}_{ii} = N_{ii}^{null}$.   
%$E^{\text{flux}}$ is matrix which includes the uncertainties associated with the neutrino beam flux (see Sec.~\ref{sec:Flux}). $E^{\text{cross section}}$ encodes the uncertainties related to all the relevant neutrino cross section uncertainties (see Sec.~\ref{sec:XSec}). $E^{\text{cosmic bkgd}}$ accounts for the uncertainty in the cosmic background estimation (see Sec.~\ref{sec:Cosmics}). $E^{\text{dirt bkgd}}$ provides the uncertainties for the estimation of backgrounds associated with neutrino interactions outside of the detector (see Sec.~\ref{sec:Dirt}). Finally, $E^{\text{detector}}$ accounts for the detector based systematic uncertainties which will affect the measurements.  
%We will rely on measurements to constrain both the cosmic and ``dirt'' backgrounds such that the matrices will be diagonal and based on the statistics of the sample used to constrain the backgrounds. 
%and thus we will use a reweighting method to construct the covariance matrices associated with these uncertainties.
%The covariance matrix for the flux and cross section backgrounds will encode all the experimental uncertainties along with their correlations, both between detectors and between energy bins within the same detector. 
%For the flux and neutrino cross section systematic uncertainties, $\alpha$, The matrices are built using:
The flux and neutrino cross section covariance matrices are calculated using detailed Monte Carlo simulations based on GEANT4 and the GENIE neutrino event generator, respectively. Reweighting techniques are used to construct possible variations on the event distributions due to uncertainties on the underlying parameters in the models.  $\mathcal{N}$ such ``universes'' can be combined to construct the covariance matrix:    

\begin{equation}
E_{ij} =\frac{1}{\mathcal{N}} \sum^{\mathcal{N}}_{m=1} [N_{\text{CV}}^{i} - N_{m}^{i}] \times [N_{\text{CV}}^{j} - N_{m}^{j}], 
\label{eq:Ematrix}
\end{equation}

\noindent where $i$ and $j$ correspond to neutrino energy bins \emph{across all three detectors}, $N_{\text{CV}}$ is the number of entries in each energy bin of the nominal event distribution, and $N_{m}$ is the number of entries in the $m^{th}$ ``universe''. $E_{ij}$ is the total covariance matrix, sometimes called the total error matrix, with matrix element units of (events)$^2$.  The fractional covariance matrix is generally a more useful result and is defined as

\begin{equation}
F_{ij} = \frac{E_{ij}}{N_{\text{CV}}^{i}N_{\text{CV}}^{j}}.
\end{equation}

\noindent From $E_{ij}$ can also be extracted the correlation matrix, 

\begin{equation}
\rho_{ij} = \frac{E_{ij}}{\sqrt{E_{ii}}\sqrt{E_{jj}}}~~~~~~~~[-1 \leq \rho \leq 1], 
\end{equation}

\noindent where $\rho_{ij}$ describes the level of correlation between bins $i$ and $j$ of the neutrino energy distributions.   

The flux and cross section error matrices have been constructed according to Eq. \ref{eq:Ematrix}, while the cosmic background and dirt background error matrices are constructed differently as will be explained in the relevant Sections below.

%This provides the absolute uncertainty of the $\alpha$ systematic along the diagonal of the matrix, while the off diagonal entries correspond to the correlations between the various energy bins.

\subsection{\nue and \numu Signal Selection}
\label{sec:EventRates}

We begin with a discussion of beam-induced neutrino interactions within the TPC active volumes that are selected when isolating \nue and \numu charged-current events samples for analysis.

%%%%%%%%%%%%%%%%%%%%%%%%%%%%%%%%%%%%%%%%%%%%%%%%%%%%%%%%%%%%%%%%%%%%%%%%%%%%%%%%%%%%%%%%%
\subsubsection*{Electron Neutrino Charged-Current Candidates}

Electron neutrino event candidates include intrinsic \nue charged-current (CC) interactions as well as other beam-related (mostly \numu-induced) mis-identification backgrounds.  The event selection algorithms are given below and are applied identically to all three detectors in the analysis. A full GEANT simulation of GENIE produced neutrino interactions in argon is used and selections are made based on predicted event kinematics. As a cross-check, neutrino interactions in the \icarus detector have been also independently simulated using FLUKA\cite{Fluka1,Fluka2,Flukaneutrino}, and consistent results were found.  %The results from the different simulations have been found in good agreement in terms of event topologies and detection efficiency.
The efficiencies applied to different event types are based on inputs from other simulation results, hand-scanning studies of both simulated and real events in different detectors, and analysis results from \lartpc experiments (e.g. ICARUS, ArgoNeuT).

\begin{figure}[tp]
\centering
\hspace*{-3mm}
\mbox{
\includegraphics[width=0.333\textwidth,height=0.24\textwidth,trim = 11mm 0mm 6mm 15mm, clip]{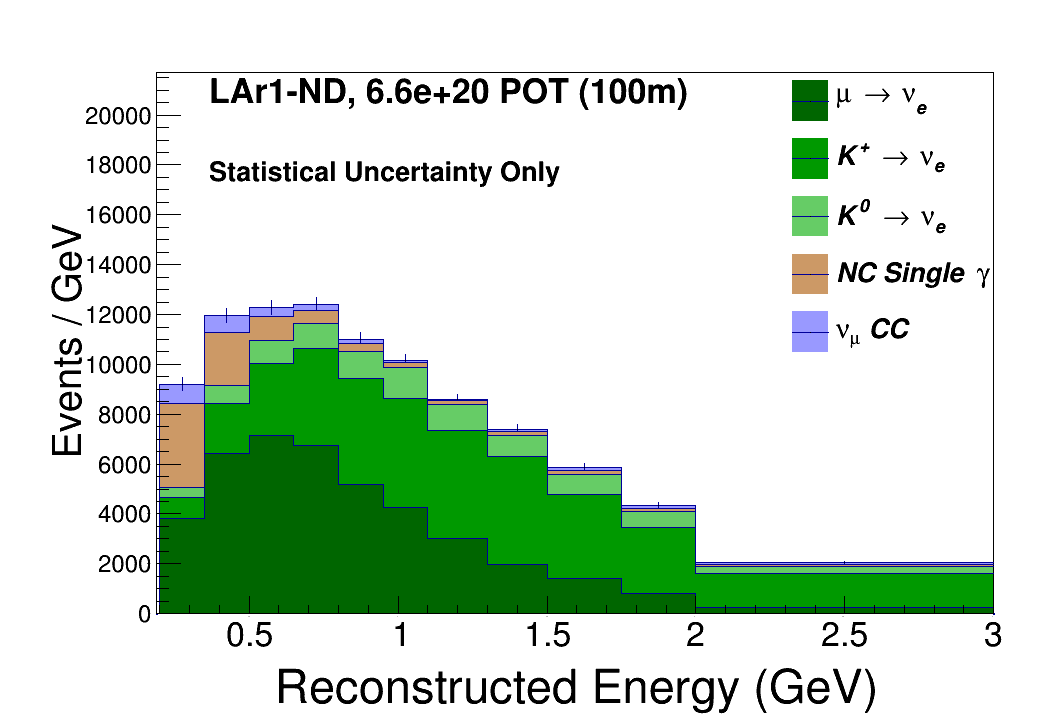}
\includegraphics[width=0.333\textwidth,height=0.24\textwidth,trim = 11mm 0mm 6mm 15mm, clip]{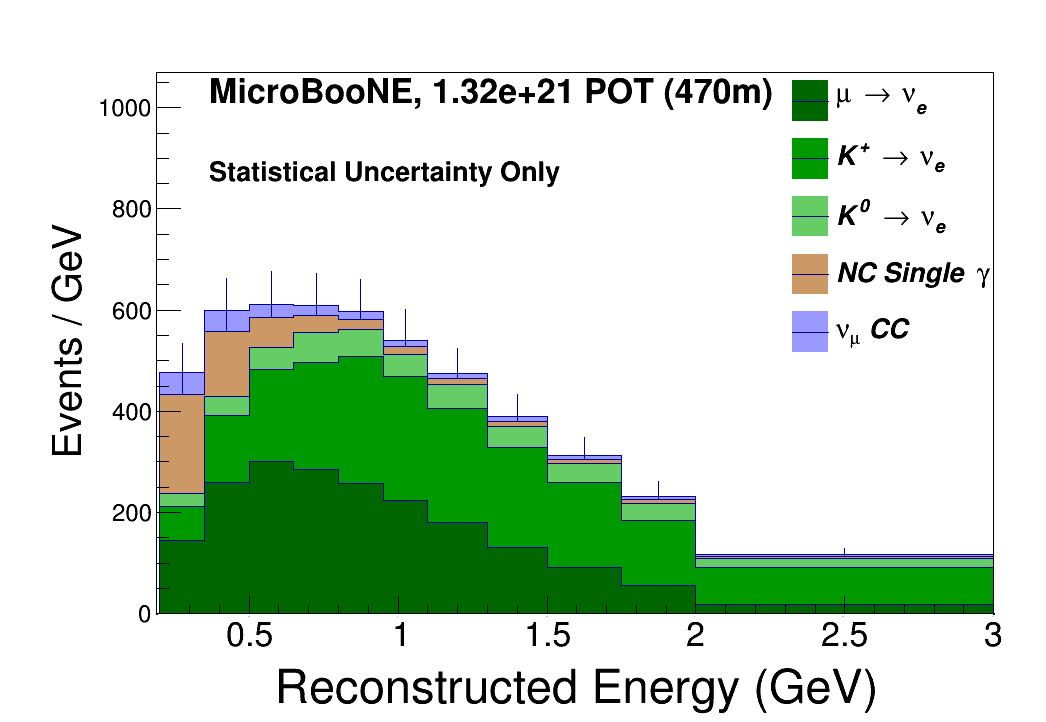}
\includegraphics[width=0.333\textwidth,height=0.24\textwidth,trim = 11mm 0mm 6mm 15mm, clip]{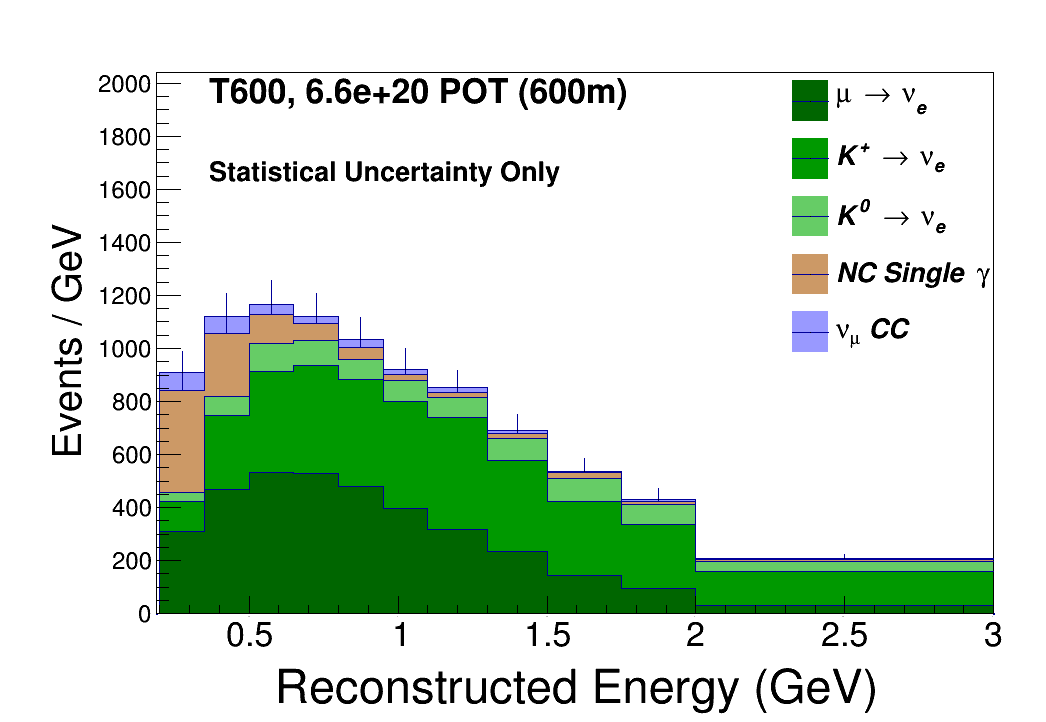}
}
\caption{Beam-related electron neutrino charged-current candidate events in \larnd (left), \uboone (center), and \icarus (right).  Statistical uncertainties only are shown. Data exposures are indicated on the plots and assume inclusion of the full \uboone data set.} 
\label{fig:nu_e_intrinsic}
\end{figure}

\begin{enumerate}

\item {\bf Intrinsic/Signal} $\mathbf{\nue}$ {\bf CC :} \nue charged-current interactions producing an electron with $E_{e}>200~\MeV$ are accepted with an assumed 80\% identification efficiency (after fiducial volume selection) in our baseline sensitivity analysis.  The 200~\MeV shower threshold is applied to ensure good event reconstruction and identification.  The simulation estimates this requirement sacrifices $\sim$30\% of the events in the 200-350~\MeV reconstructed neutrino energy bin and less than 5\% above 350~MeV.  It must be noted, however, that the threshold for analysis of events in LAr should be well below this and lower energy events will be studied in the SBN experiments. The 80\% efficiency is informed by hand-scanning exercises of simulated events in \lartpcs and significant effort is currently on-going to verify this performance with automated reconstruction algorithms. Stricter requirements on \nue CC event selection have been discussed in the context of rejecting cosmogenic backgrounds (such as requiring hadronic activity at the vertex, a clear indicator of a $\nu+N$ interaction), but other handles on cosmogenic event rejection will likely deem this unnecessary (see Section \ref{sec:Cosmics}). Also, selection efficiencies can depend on specific detector performance parameters. %, such as the signal to noise ratio in the different TPC views.  
For instance, scanning exercises in the ICARUS detector indicate that the efficiency for recognizing isolated electron showers after the vertex is reduced by $\sim$12\% if only one 2-D view (collection) out of three is available for a complete event reconstruction (e.g. due to low signal-to-noise in the induction views). It will be important to carefully monitor such effects. Selected intrinsic \nue CC candidates are shown in the green histograms in Figure \ref{fig:nu_e_intrinsic}.

\item {\bf NC $\mathbf{\gamma}$ production :} Photons creating a shower above the 200~\MeV selection threshold can fake the \nue CC signature described above. For example, neutral-current interactions with any number of \pizero in the final state or radiative resonance decays are sources of such $\gamma$'s. These events are analyzed according to the following criteria:

\begin{itemize}  

\item Second photon cut: If the second photon from a \pizero decay (with $E_{\gamma}>100~\MeV$ as an observation threshold) converts within the TPC active volume, the event is rejected.

\item Conversion gap cut: If the neutrino interaction is inside the active volume and produces more than 50~MeV of charged hadronic activity at the vertex, then the vertex is deemed visible.  With a visible vertex, if all photon showers convert more than 3~cm from that vertex, the event is rejected. 

\item $dE/dx$ cut: For events passing the previous two cuts, a 94\% photon rejection rate is applied, corresponding to the expected power of separating $e/\gamma$ showers in the \lartpc using the energy deposited in the first few centimeters of an electromagnetic shower. 

\end{itemize}

Beam-related photon backgrounds are shown as the orange histograms in Figure \ref{fig:nu_e_intrinsic} labeled ``NC Single $\gamma$''.

%\item {\bf NC $\mathbf{\gamma}$ production :} Other neutral-current interactions resulting in photons in the final state (not from \pizero decays) are also considered if the photon converts within the fiducial volume.  Similar to NC \pizero events,

%\begin{itemize}  

%\item Conversion gap cut: If the neutrino interaction is inside the active volume and produces more than 50~MeV of charged hadronic activity at the vertex, then the vertex is deemed visible.  With a visible vertex, if all photon showers convert more than 3~cm from that vertex, the event is rejected. 

%\item $dE/dx$ cut: For events passing the previous cut, again the 94\% rejection rate from $dE/dx$ is applied. 

%\end{itemize}

\item {\bf $\mathbf{\numu}$ CC :} \numu charged-current interactions with an identified primary electromagnetic (e.m.) shower within the fiducial volume could also be mis-identified as \nue interactions if the muon is not identified. Minimum ionizing tracks longer than 1~m in BNB events are essentially all muons, so events with $L_{\mu} \geq 1$~m are rejected.  Events with $L_{\mu} < 1$~m and a single e.m. shower attached to the CC event vertex could be identified as a $\mu+\gamma$ (\numu CC) or $\pi+e$ (\nue CC) final state.  We, therefore, check for the presence of candidate e.m. showers in \numu CC interactions following the same criteria as for NC $\gamma$ events described above, and if not rejected we retain the event as a background for the \nue CC sample.  These are represented by the blue histograms in Figure \ref{fig:nu_e_intrinsic}.        
%Careful hand-scans of \uboone and ICARUS simulated events have indicated this rate is $\leq$ 0.1\%, so we use the upper limit of 0.1\% as the \numu CC mis-identification rate in this analysis.

\item {\bf Neutrino Electron Scattering :} Neutrinos can scatter off an orbiting electron in an atom, ejecting the electron at high energy.  Experimentally, the signature is a very forward going electron and nothing else in the event, which mimics a \nue charged current interaction and will be selected with the same efficiency.  However, the $\nu+e$ cross section is very low and so forms a secondary background.  These are too small to be seen in Figure \ref{fig:nu_e_intrinsic} but are included in the analysis.   

\end{enumerate} 

For estimating these background rates, the full GEANT simulation of events is of fundamental importance.  By analyzing the conversion points of photons instead of just the true neutrino interaction vertex, we accurately account for acceptance effects in the differently shaped detectors.  Because the $e/\gamma$ separation is performed entirely with the first few centimeters of a shower, differences in total shower containment do not affect the assumption that the photon identification efficiency should be the same in each detector.  

To simulate calorimetric energy reconstruction, the incoming neutrino energy in each Monte Carlo event is estimated by summing the energy of the lepton (or the $\gamma$ faking an electron)  and all charged hadrons above observation thresholds present in the final state. This approach is used in the analysis of both \nue and \numu charged-current events described next. It should be noted that this method is one possible approach to estimating the neutrino energy. The liquid argon TPC technology enables a full calorimetric reconstruction, but other methods can be used as well, such as isolating charged-current quasi-elastic (CCQE) events and assuming QE kinematics. The ability to apply complementary approaches to event identification and energy reconstruction will provide valuable cross checks of the measurements performed.  The stacked beam-related backgrounds to the \nue analysis are summarized in Figure~\ref{fig:nu_e_intrinsic} as a function of the calorimetric reconstructed energy for each of the SBN detectors.  Event totals for each background class are tabulated later in Section~\ref{sec:SBN_nue} in Table~\ref{tab:nueevents}.

%%%%%%%%%%%%%%%%%%%%%%%%%%%%%%%%%%%%%%%%%%%%%%%%%%%%%%%%%%%%%%%%%%%%%%%%%%%%%%%%%%%%%%%%%%%%%%%%%%%%%%%%
\subsubsection*{Muon Neutrino Charged-Current Candidates}

\begin{figure}[htp]
\centering
\mbox{\includegraphics[width=0.333\textwidth,trim = 0mm 0mm 5mm 0mm, clip]{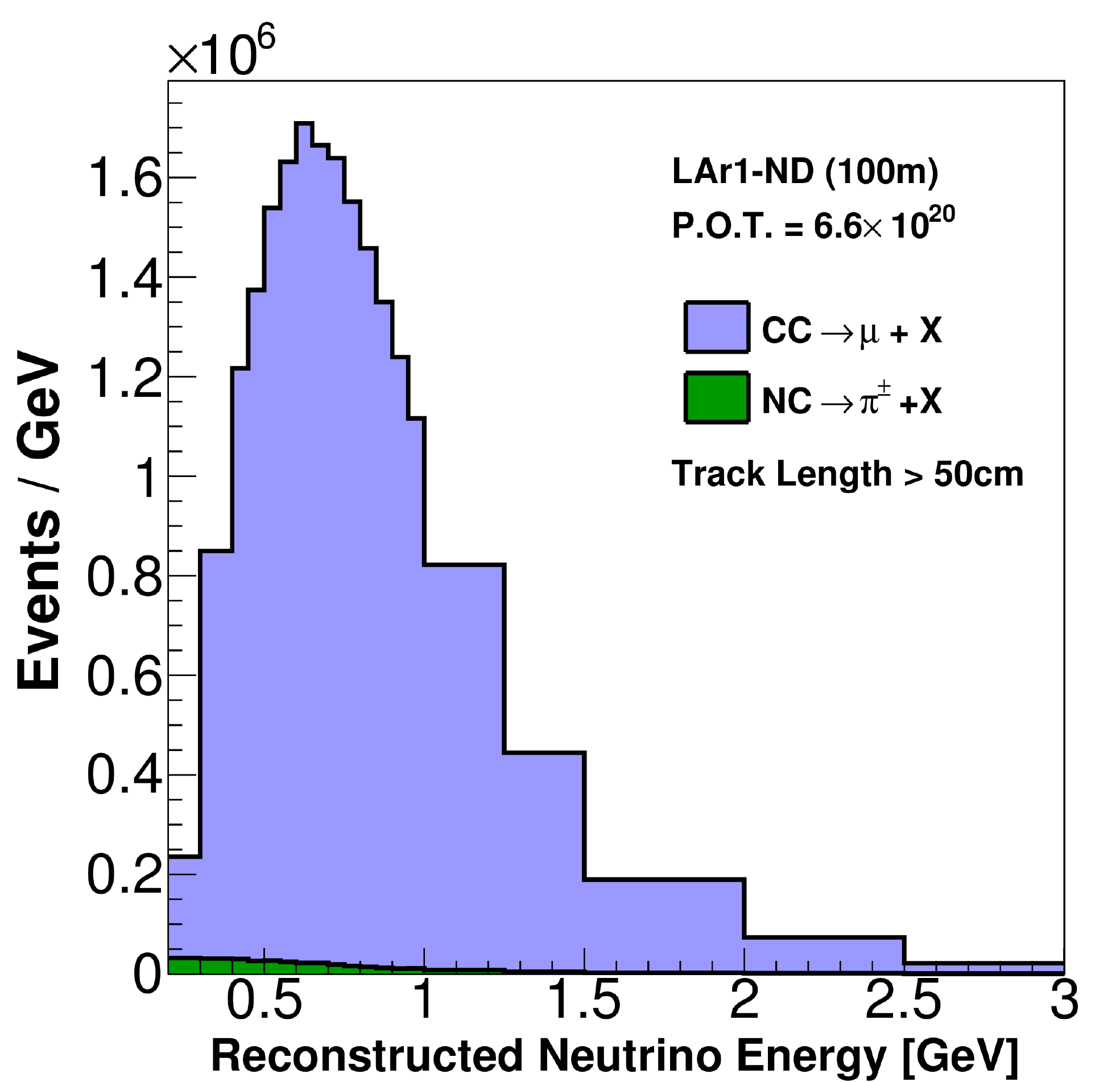}
\includegraphics[width=0.333\textwidth,trim = 0mm 0mm 5mm 0mm, clip]{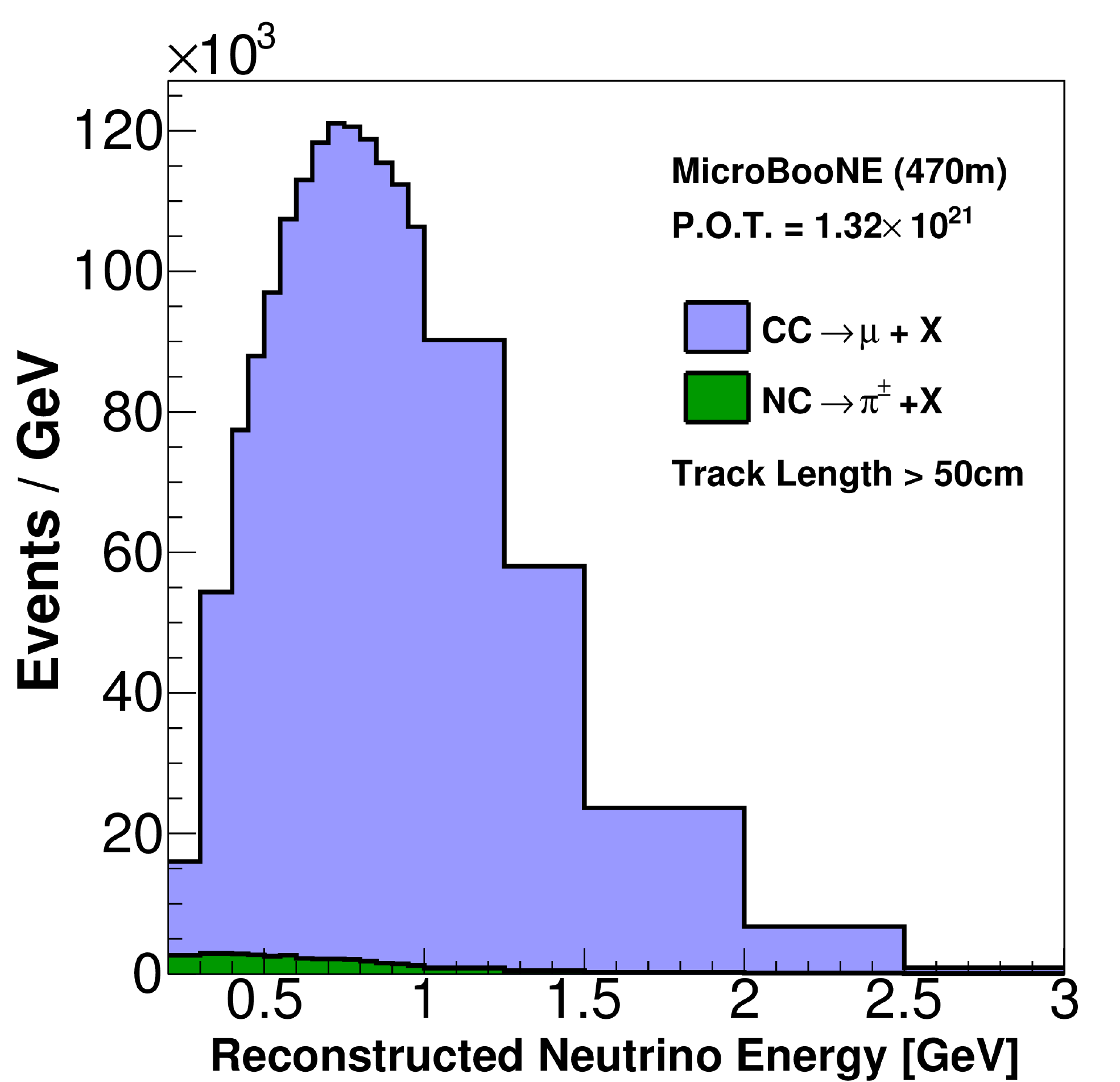}
\includegraphics[width=0.333\textwidth,trim = 0mm 0mm 5mm 0mm, clip]{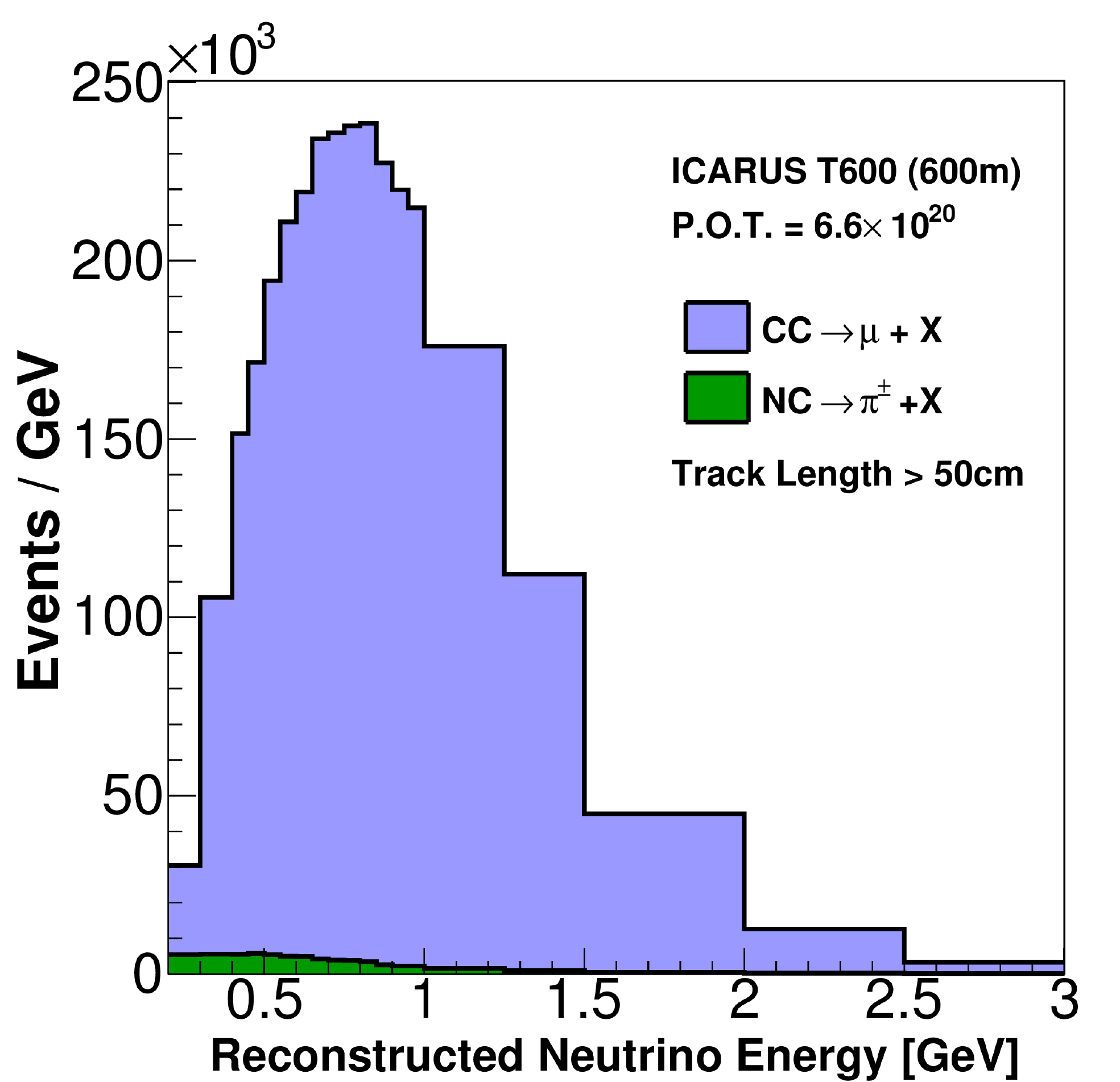}}
\caption{Selected muon neutrino charged-current inclusive candidate events in \larnd (left), \uboone (center), and the \icarus (right). Final state muon tracks that are fully contained in the TPC volume are required to travel greater than 50~cm.  Muons which exit the active detectors are required to travel $>1$~m before exiting. Statistical uncertainties only are shown. Data exposures are indicated on the plots and assume inclusion of the full \uboone data set.} 
\label{fig:nu_mu_intrinsic}
\end{figure}

Muon neutrino charged-current events are selected assuming an 80\% reconstruction and identification efficiency (after fiducial volume selection).  The only background contribution considered comes from neutral-current charged pion production, where the $\pi^{\pm}$ can be mistaken for a $\mu^{\pm}$.  Simulations show pion tracks produced in the BNB are short with most charged pions traveling less than half a meter in the liquid argon.  We therefore apply a simple cut requiring muon candidates that stop in the TPC active volume to be longer than 50~cm which minimizes the NC contamination in the \numu CC selection. The resulting contamination from NC events is shown in Figure~\ref{fig:nu_mu_intrinsic} and has a negligible impact on the oscillation sensitivity.  More sophisticated methods to separate pions and muons stopping in LAr are being explored, but this selection is sufficient for the current analysis.  

Measurement resolutions have been introduced for this analysis by smearing both the reconstructed muon energy and hadron energy in the event and $E_{\nu} = E_{\mu} + E_{\text{had-visible}}$. The smearing of the muon energy changes depending on if the muon is fully contained within the active volume or if it exits the active volume and the energy must be estimated via the multiple scattering of the track.  We require, therefore, all exiting tracks to have a minimum track length of 1~m in the active volume to enable this multiple scattering measurement with sufficient resolution.  The distributions of selected muon neutrino charged-current events in each detector are shown in Figure~\ref{fig:nu_mu_intrinsic}.

\subsection{Neutrino Flux Uncertainties}
\label{sec:Flux}

BNB neutrino flux predictions and related systematic uncertainties are assessed using a detailed Monte Carlo program developed by the \MB Collaboration \cite{MiniFlux}. In the simulation, charged pion production is constrained using dedicated 8~\GeV $p$+Be hadron production data from the HARP experiment~\cite{HARP} at CERN. 
%as well as data from the BNL E910 experiment that ran at similar energies~\cite{E910}. 
Neutral kaon production has been constrained by BNL E910 data~\cite{E910} and a measurement made at KEK by Abe et al.~\cite{KEK}. $K^{+}$ production uncertainties are set by measurements made with the \SB~\cite{SciBooNEkaon} detector when it ran in the BNB. In total, the BNB Monte Carlo treats systematic uncertainties related to the following sources:

\begin{itemize}
\renewcommand{\labelitemii}{$\circ$}
\item Primary production of $\pi^+$, $\pi^-$, $K^+$, $K^-$, and $K^0_L$ in $p$+Be collisions at 8~\GeV;
%\item $\pi^+$ production in primary p+Be collisions at 8 GeV. 
%\item $\pi^-$ production in primary p+Be collisions at 8 GeV. 
%\item $K^+$ production in primary p+Be collisions at 8 GeV. 
%\item $K^-$ production in primary p+Be collisions at 8 GeV. 
%\item $K^0_L$ production in primary p+Be collisions at 8 GeV.
\item Secondary interactions of $p$, $n$, $\pi^{\pm}$ in the beryllium target and aluminum horn;
%\item Secondary interactions in the steel, concrete, and other materials of the beamline
%Primary hadron production from non-target interactions
\item Beam focusing with the magnetic horn.
\end{itemize}

%\noindent Each of these components are constrained by data and has been investigated 
%thoroughly by the MiniBooNE collaboration in Ref.~\cite{MiniFlux}.  
%To evaluate these systematic uncertainties we will adopt the machinery 
%engineered by the MiniBooNE collaboration. 

%The primary pion production systematic uncertainties are constrained 
%using HARP data on a thin Be target impinged 8~GeV protons~\cite{HARP}. 
%The kaon production systematic uncertainties are constrained by BNL E910~\cite{E910} and KEK measurement~\cite{KEK}, while the $K^{+}$ constraint being mainly driven by SciBooNE measurements~\cite{SciBooNEkaon}. 

Primary hadron production uncertainties, whenever available, are taken directly from the measured cross sections which are used to constrain the Monte Carlo.  In particular, in the case of $\pi^{+}$ and $\pi^{-}$ production, the experimental uncertainties reported by the HARP experiment~\cite{HARP,Schmitz:2008zz}  are directly used to set the allowed variation within the beamline simulation.  

Secondary interaction uncertainties are also evaluated. Table \ref{tab:beamuni} summarizes allowed variations on hadron-Be and hadron-Al cross sections in the simulation. The total cross section, $\sigma_{\text{TOT}}$; the inelastic cross section, $\sigma_{\text{INE}}$; and the quasi-elastic cross sections, $\sigma_{\text{QEL}}$ are varied separately for nucleons and pions interacting with Be and Al. When we vary $\sigma_{\text{INE}}$ and $\sigma_{\text{QEL}}$ we fix the cross section of the other to hold the total cross section constant.   

\begin{table}[hb]
\begin{center}
\caption{\label{tab:beamuni} Cross section variations for systematic studies of secondary hadron interactions in the target and horn. For each hadron-nucleus
cross section type, the momentum-dependent cross section is offset by the amount shown~\cite{MiniFlux}.}
\begin{tabular}{lcccccc} \hline\hline
		 &  \multicolumn{2}{c}{ $\Delta\sigma_{\text{TOT}}$ (mb)}	&  \multicolumn{2}{c}{ $\Delta\sigma_{\text{INE}}$ (mb)}	&  \multicolumn{2}{c}{ $\Delta\sigma_{\text{QEL}}$ (mb)}	\\
		& Be & Al & Be & Al & Be & Al \\ \hline
$(p/n)$-(Be/Al)	   & $~~\pm 15\%$ & $~~\pm 25\%~~$ & $~~\pm 5\%$ & $~~\pm 10\%~~$ &  $~~\pm 20\%$ & $~~\pm 45\%~~$   \\
$\pi^{\pm}$-(Be/Al)& $~~\pm 11.9\%$ & $~~\pm 28.7\%~~$ 	& $~~\pm 10\%$ & $~~\pm 20\%~~$ &  $~~\pm 11.2\%$ & $~~\pm 25.9\%~~$ \\ \hline\hline
\end{tabular}
\end{center}
\end{table}

Beam focusing systematics include uncertainty on the magnitude of the horn current (174~$\pm$~1~kA) as well as skin depth effects describing where the current flows on the surfaces of the horn.  The skin depth effect allows the magnetic field to penetrate into the interior of the horn conductor which in turn creates a magnetic field within the conductor. This will lead to deflections of charged particles which traverse the conductor, especially higher energy particles which do not penetrate deeply into the horn conductor. The effect can be approximated by modeling an exponentially decreasing field to a depth of about 1.4~mm. To asses the systematic, the field is turned on and off, which leads to an energy dependent effect of 1 to 18\% for particles of $<1$~GeV to 2~GeV, respectively~\cite{MiniFlux}. 

We currently don't assess a systematic on hadron interactions with material downstream of the horn (including air, concrete, steel, etc.). These effects have been studied and found to contribute about 1\% (2\%) to the $\nu_{\mu}$ ($\nu_{e}$) fluxes, so even a large 50\% uncertainty would make a negligible contribution to the total errors. %When compared to the other systematic uncertainties this, fully correlated, uncertainty is a second order effect. 

%Secondary hadron interaction uncertainties are also studied. This includes both secondary interactions of the hadrons produced in the primary p+Be interaction, but also protons (and neutrons) interactions with the aluminum of the horn. To assess the systematic effect the cross sections of the hadrons on the horn/target media is varied. The total cross section, $\sigma_{\text{TOT}}$; the inelastic cross section, $\sigma_{\text{INE}}$; and the quasi-elastic cross sections, $\sigma_{\text{QEL}}$ are varied separately for nucleons and pions interacting with Be and Al. When we vary $\sigma_{\text{INE}}$ and $\sigma_{\text{QEL}}$ we fix the cross section of the other to hold the total cross section constant. Table~\ref{tab:beamuni} summarizes the variations which are allowed by the data, as described in Ref.~\cite{MiniFlux}.

Table \ref{tab:beamuni} reports the contributions of the underlying systematics to the integrated \numu and \nue fluxes along the BNB, revealing total normalization uncertainties of order 15\% on both absolute predictions.   %We can see that even with the dedicated pion production data, these errors still dominate, since they also contribute the overwhelming majority of the fluxes, and we are left with normalization uncertainties of order 15\%.     

\begin{table}[t]
\begin{center}
\caption{\label{tab:nusys}Variations in the total flux of
each neutrino species in neutrino mode due to the systematic uncertainties~\cite{MiniFlux}.}
\begin{tabular}{lrr}\hline\hline
Source of Uncertainty 		&$\nu_{\mu}$	&$\nu_{e}$		\\ \hline
$\pi^{+}$ production		& 14.7\%		&  9.3\%		\\
$\pi^{-}$ production		&  0.0\%		&  0.0\%		\\
$K^{+}$ production			&  0.9\%		& 11.5\%		\\
$K^{0}$ production			&  0.0\% 		&  2.1\%		\\
Horn field					&  2.2\%		&  0.6\%		\\
Nucleon cross sections		&  2.8\%		&  3.3\%		\\
Pion cross sections			&  1.2\%		&  0.8\%		\\ \hline\hline
%Total Uncertainty			&  15.2\%		&  15.3\%		\\\hline\hline
\end{tabular}
\end{center}
\end{table}

\begin{figure}[tp]\centering
\includegraphics[width=1.\textwidth,trim=0mm 0mm 0mm 0mm,clip]{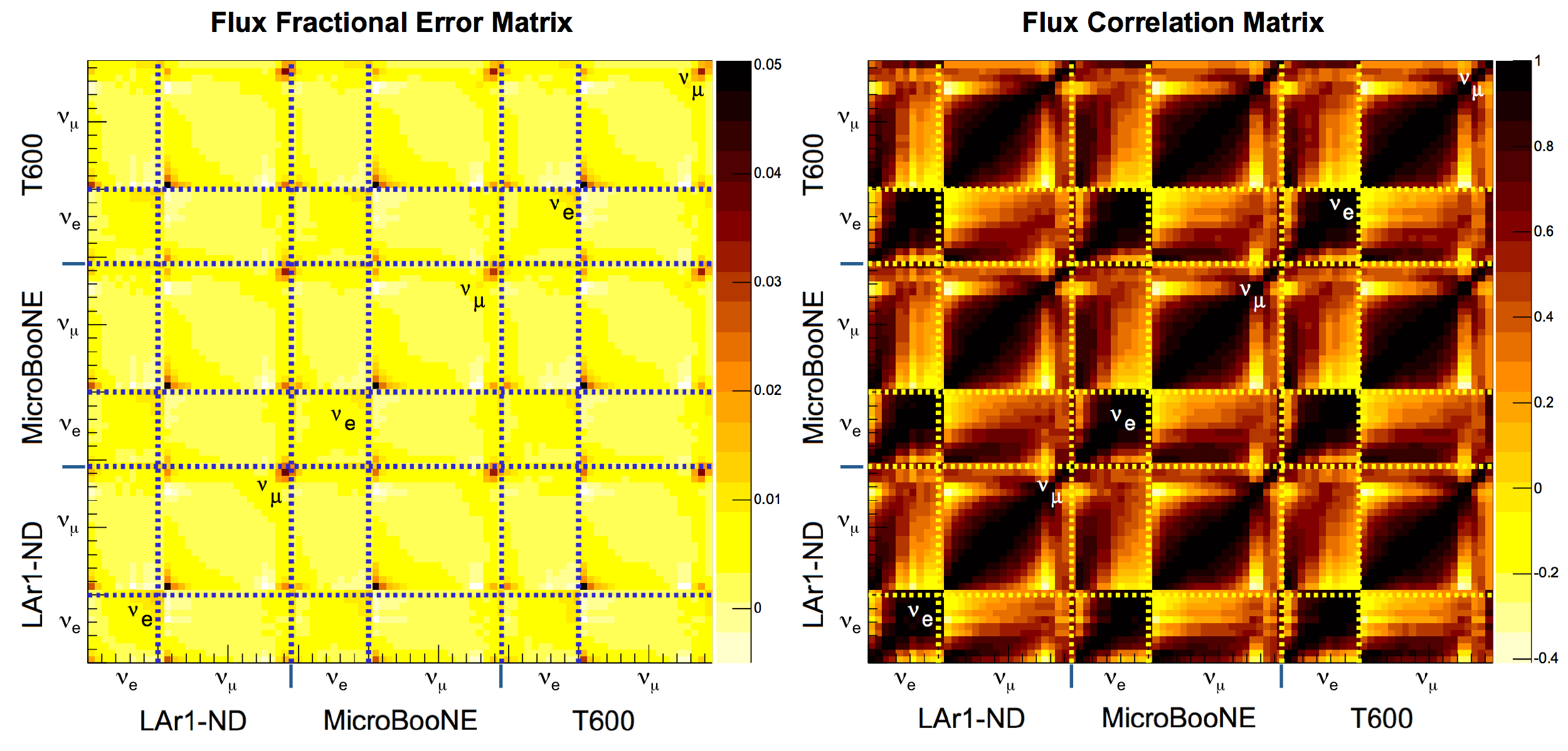}
\caption{The fractional flux covariance matrix (left) and correlation matrix (right) for the \nue and \numu charged-current reconstructed energy distributions.  Both the \nue events (11 energy bins from 0.2--3~\GeV) and \numu events (19 energy bins from 0.2--3~\GeV) at all three detector locations are represented; the dashed lines indicate the boundaries in the matrix.  For example, the lower left square marked ``\nue'' shows the fractional error (left) and correlations (right) within the reconstructed \nue CC event distribution in \larnd. In another example, the square four from the left and two from the bottom shows the correlations between the \numu CC event distributions in \larnd and \uboone.} 
\label{fig:fluxcorr}
\end{figure}

Using Eq. \ref{eq:Ematrix} we compute the covariance matrix for all the systematic variations in the flux model.  The fractional error matrix and flux correlation matrix are shown in Figure~\ref{fig:fluxcorr}.  We see that the event rates at different detector locations and for both \numu and \nue fluxes have large positive correlations.  These correlations are, of course, the key to SBN sensitivity.  The high statistics measurement made in the near detector, together with the high levels of correlation between the near and far locations will eliminate the large normalization uncertainty highlighted in Table \ref{tab:beamuni} when performing oscillation searches, a critical motivation for the multi-detector SBN configuration.

\subsection{Neutrino Interaction Uncertainties}
\label{sec:XSec}

Uncertainties in the neutrino interaction model 
%\footnote{including inclusive and exclusive differential cross sections off nucleons, models of the target nuclear medium, and final state interaction effects on produced particles before leaving the nucleus}
are the largest uncertainties affecting the normalization of events in the SBN detectors, but are expected to be highly correlated between detectors because of the use of the same target nucleus (argon).  Only through second order impacts of neutrino fluxes or differences in the geometric acceptance of events in the detectors can the correlations be different than 100\%. 

Neutrino interactions on argon are simulated using the GENIE~\cite{GENIE} neutrino event generator.  GENIE simulates each stage of the interaction including inclusive and exclusive differential cross sections off individual nucleons and the effects of the nuclear medium on final state particles as they propagate out of the target nucleus (final state interactions). Multi-nucleon correlation effects of the initial state are also a challenge in neutrino interactions and are not part of the present simulation.  This is true of other available Monte Carlo packages as well.  Incomplete modeling of nuclear effects can lead to biases in neutrino energy reconstruction and is a very active area of both experimental and theoretical research at the moment (see~\cite{Garvey:2014exa} and \cite{Alvarez-Ruso:2014bla} and the references therein). The data sets of the SBN \lartpc detectors will, in fact, be very valuable for studying these effects and improving simulations.  

GENIE does provide a built-in framework of event reweighting for evaluating systematic uncertainties and correlations in an analysis.  Table \ref{tab:xsec_enum} lists the uncertainties used for this analysis and their nominal percent variation at 1$\sigma$, according to the GENIE documentation.  This is a partial list of the available parameters within the GENIE framework, chosen here for their relevance to the SBN oscillation searches. The analysis does not currently include an estimate of uncertainties on final state interactions.  

\begin{figure}[h]
\centering
\mbox{\includegraphics[width=0.45\textwidth,trim = 0mm 0mm 20mm 20mm, clip]{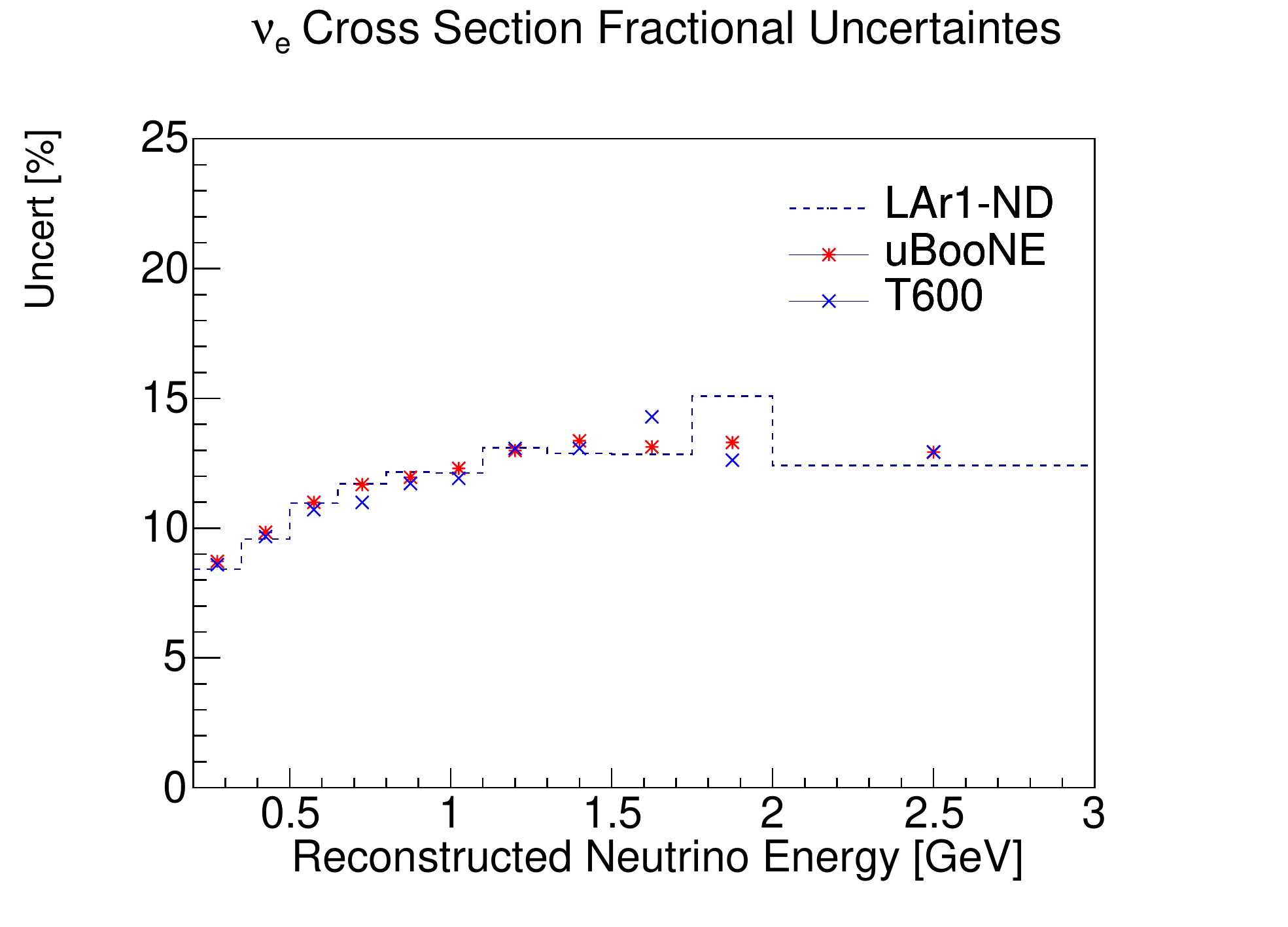}\quad\quad
\includegraphics[width=0.45\textwidth,trim = 0mm 0mm 20mm 20mm, clip]{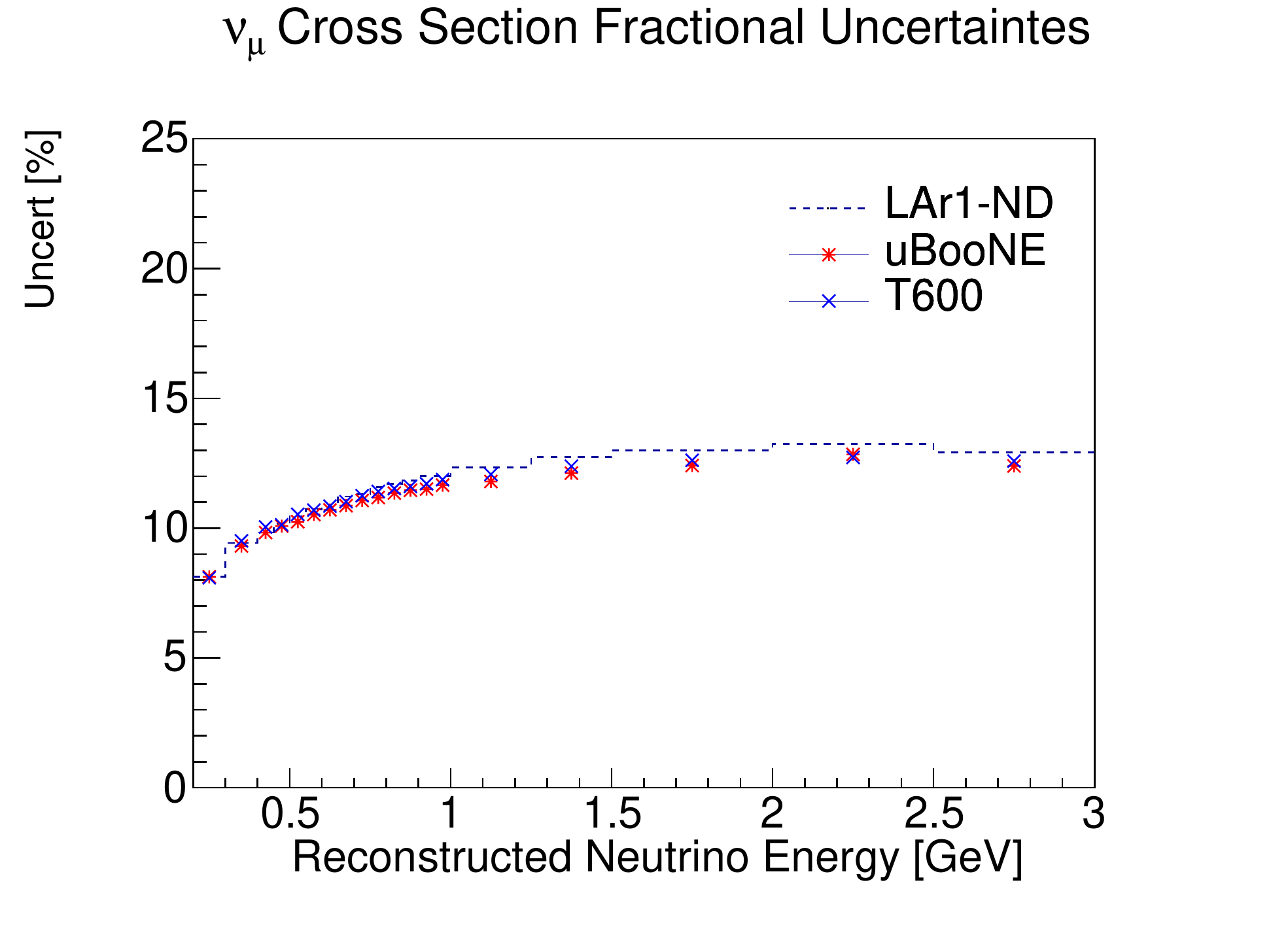}}
\mbox{\includegraphics[width=0.50\textwidth,trim=0mm 5mm 5mm 25mm,clip]{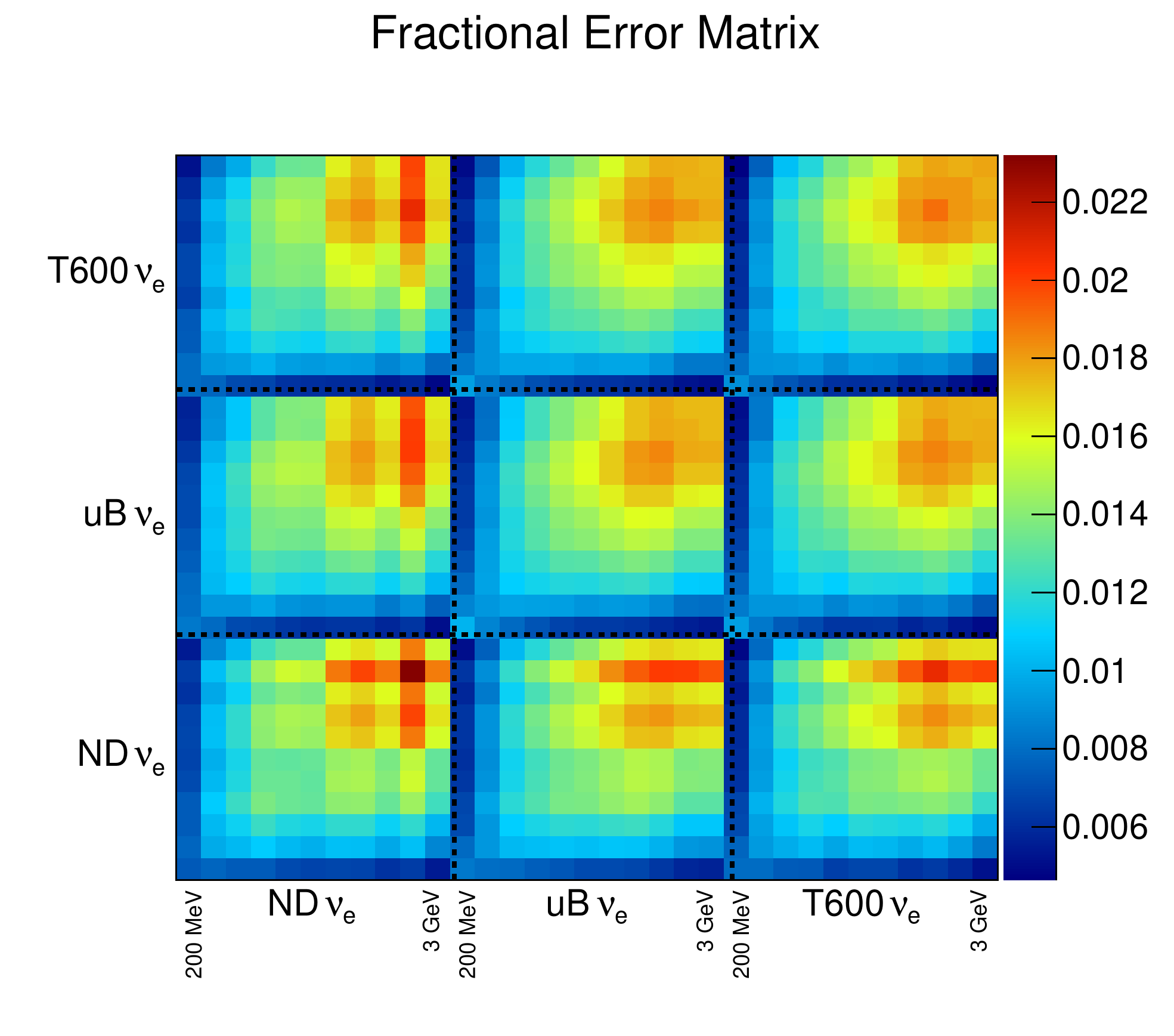}
\includegraphics[width=0.50\textwidth,trim=0mm 5mm 5mm 25mm,clip]{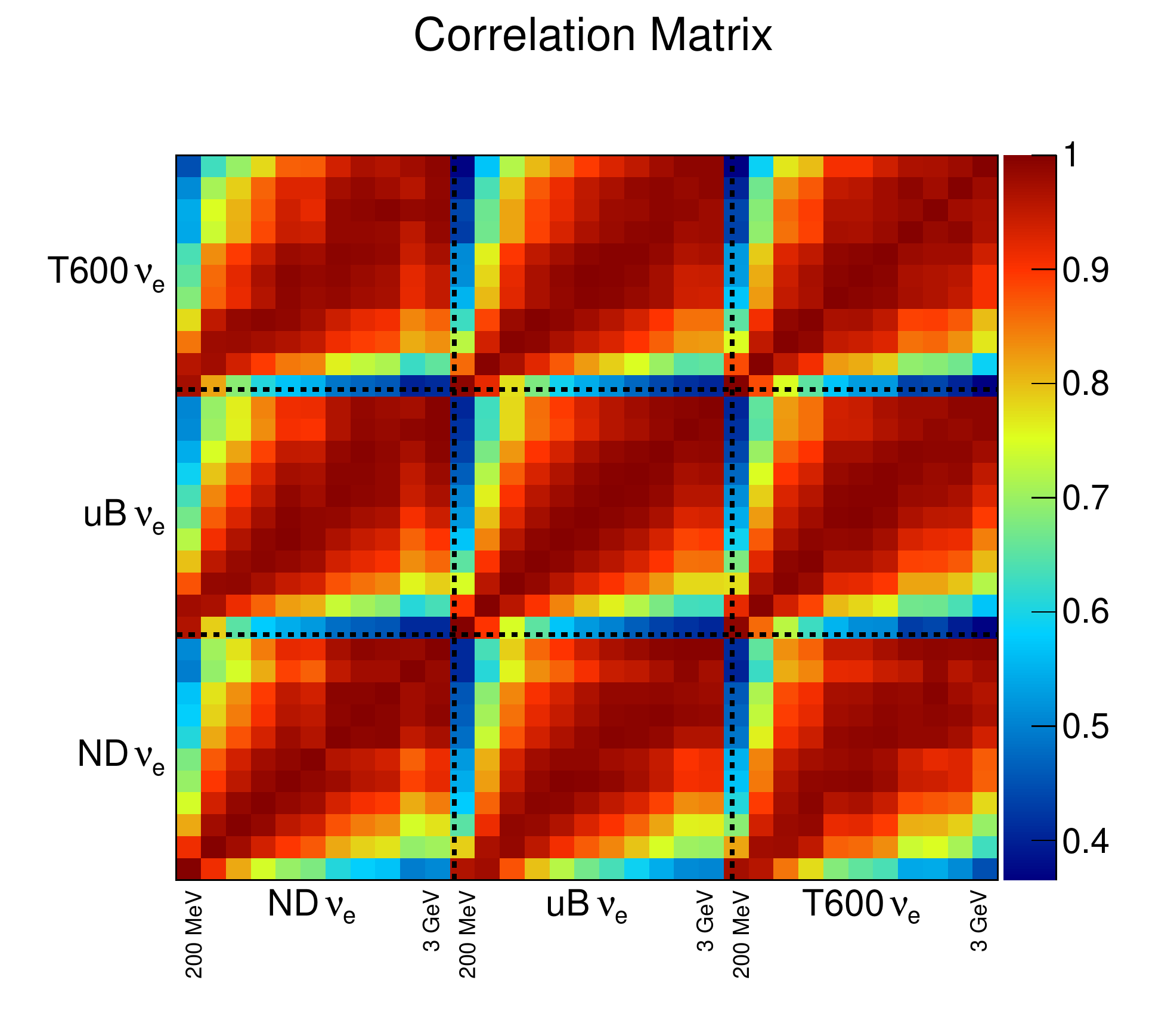}}
\caption{Absolute uncertainties on \nue (upper left) and \numu (upper right) event rates at each of the three SBN detectors due to neutrino cross section uncertainties. (Lower left) Fractional cross section covariance matrix, $E^{\text{cross section}}$, for \nue CC candidate events. (Lower right) The correlation matrix for \nue CC candidate events.  Inspection of the diagonal elements of the off-diagonal blocks shows the correlations between events in different detectors to be very near 1.0}
\label{fig:xsecUncerts}
\end{figure}

We simulated 250 different cross section ``universes'' in which each of the model parameters were varied at random from a Gaussian distribution with a 1$\sigma$ spread equal to the 1$\sigma$ uncertainty in the underlying physical quantity.  Much more detail is available from the GENIE manual, chapter 8~\cite{GENIE_users}, on both the underlying physical uncertainties and the methodology for propagating them to observed event distributions.  Figure \ref{fig:xsecUncerts} shows the RMS of the 250 simulated universes in the reconstructed neutrino energy bins used in the \nue and \numu analyses, indicating absolute neutrino interaction model uncertainties of 10--15\%.  From these variations, the cross section covariance matrix, $E^{\text{cross section}}$, is constructed using Eq. \ref{eq:Ematrix}.  Figure \ref{fig:xsecUncerts} shows the fractional covariance matrix and correlations for the \nue charged-current candidate events that were shown in Figure \ref{fig:nu_e_intrinsic}.  The off-diagonal blocks of the correlation matrix indicate the correlations between events in different detectors.  The diagonal elements within the off-diagonal blocks are the correlations between the same energy bins in different detectors and are seen to be near +1.0 in most cases.  

%On the other hand, the fluctuations due to cross section uncertainties are highly correlated between the three detectors.  As an illustration of this, Figure \ref{fig:xsecRatios} shows the ratio of event rates in the near detector to both far detectors (dark), as well as this same ratio in each of the ``universes'' (overlaid, gray).  While these ratios don't represent the input to any analyses, the do demonstrate that the ratio of event rates between detectors are highly correlated.  Note that the RMS of the ratio of event rates is at the 2\% level, compared to the 10\% to 15\% uncertainty in each of the individual event rates.

%Like the flux uncertainties, the computed cross section uncertainties can be built into a covariance matrix.  However, it is easier to see the underlying uncertainties and correlations by examining Figures \ref{fig:xsecFracUncert} and \ref{fig:xsecCorrelation}.  In Figure \ref{fig:xsecFracUncert}, the fractional covariance matrix is shown, while in Figure \ref{fig:xsecCorrelation} the correlations between the different detectors is evident.  In addition, there is still some correlation between the \nue and \numu samples, similar to the flux uncertainty correlations though for different underlying reasons.

\newcolumntype{L}[1]{>{\hsize=#1\hsize\raggedright\arraybackslash}X}%
\newcolumntype{R}[1]{>{\hsize=#1\hsize\raggedleft\arraybackslash}X}%
\newcolumntype{C}[1]{>{\hsize=#1\hsize\centering\arraybackslash}X}%

\begin{center}
\begin{table}[t]
  \begin{tabularx}{\textwidth}{L{.55} L{1.4} C{.55}}
    Parameter        & Description               & 1$\sigma$ Uncertainty (\%) \\  \hline
\rule{0pt}{4ex}
    $M_{A}^{CCQE} $  & Axial mass for CC quasi-elastic                   & -15\%+25\% \\
\rule{0pt}{3ex}
    $M_{A}^{CCRES}$  & Axial mass for CC resonance neutrino production   & $\pm$20\% \\
\rule{0pt}{3ex}
    $M_{A}^{NCRES}$  & Axial mass for NC resonance neutrino production   & $\pm$20\% \\
\rule{0pt}{3ex}
    $R_{bkg}^{\nu p, CC 1 \pi}$ & Non-resonance background in $\nu p, CC~1 \pi$ reactions. & $\pm$50\% \\
\rule{0pt}{3ex}
    $R_{bkg}^{\nu p, CC 2 \pi}$ & Non-resonance background in $\nu p, CC~2 \pi$ reactions. & $\pm$50\% \\
\rule{0pt}{3ex}
    $R_{bkg}^{\nu n, CC 1 \pi}$ & Non-resonance background in $\nu n, CC~1 \pi$ reactions. & $\pm$50\% \\
\rule{0pt}{3ex}
    $R_{bkg}^{\nu n, CC 2 \pi}$ & Non-resonance background in $\nu n, CC~2 \pi$ reactions. & $\pm$50\% \\
\rule{0pt}{3ex}
    $R_{bkg}^{\nu p, NC 1 \pi}$ & Non-resonance background in $\nu p, NC~1 \pi$ reactions. & $\pm$50\% \\
\rule{0pt}{3ex}
    $R_{bkg}^{\nu p, NC 2 \pi}$ & Non-resonance background in $\nu p, NC~2 \pi$ reactions. & $\pm$50\% \\
\rule{0pt}{3ex}
	$R_{bkg}^{\nu n, NC 1 \pi}$ & Non-resonance background in $\nu n, NC~1 \pi$ reactions. & $\pm$50\% \\
\rule{0pt}{3ex}
	$R_{bkg}^{\nu n, NC 2 \pi}$ & Non-resonance background in $\nu n, NC~2 \pi$ reactions. & $\pm$50\% \\
\rule{0pt}{3ex}
    NC    		  & Neutral current normalization & $\pm$25\% \\
\rule{0pt}{3ex}
    DIS-NuclMod   & DIS, nuclear model &  Model switch  \\ \hline
  \end{tabularx}
  \caption{Neutrino interaction model parameters and uncertainties. This information is reproduced here from the GENIE manual Section 8.1 \cite{GENIE} for convenience.}
  \label{tab:xsec_enum}
\end{table}
\end{center}

%It should be noted that the list of uncertainties probed is the result of a first pass of this analysis and does not represent the list of uncertainties that will be used in the final analysis of SBN data.  We expect that the list of uncertainties will grow to become more inclusive of all simulated interactions.  Also, this list of uncertainties does not include an estimate of uncertainties on final state interactions.

%In section \ref{sec:Analysis}, we described the methodology to take a series of weights for a neutrino interaction and convert it into a covariance matrix.  In the case of cross section uncertainties, each neutrino has been assigned a weight corresponding to a fluctuation of the above parameters (see Table \ref{tab:xsec_enum}).  By building an event rate distribution, in the same manner as in above sections, but modulating the weight of each neutrino by the cross section weight of a particular ``universe,'' we build a set of event rate distributions that represent the various possible outcomes due to uncertainties in cross section.  

%\begin{figure}[h]
%\centering
%\includegraphics[width=0.80 \textwidth]{nueRatio_470.pdf}
%\includegraphics[width=0.80 \textwidth]{nueRatio_600.pdf}
%\includegraphics[width=0.40 \textwidth]{joseph_placeholder.jpg}
%\caption{ [Left] Ratio of event rates to the near detector (dark) overlaid with the same ratio in each of the ``universes'' from which cross section weights were drawn.  [Right] The RMS on the ratio is suppressed compared to the absolute uncertainty on the event rates from Figure \ref{fig:xsecUncerts} due to correlations between the detectors. }
%\label{fig:xsecRatios}
%\end{figure}

%\begin{figure}[h]
%\centering
%\mbox{\includegraphics[width=0.50\textwidth,trim=0mm 5mm 5mm 25mm,clip]{nue_FracErrorMat.pdf}
%\includegraphics[width=0.50\textwidth,trim=0mm 5mm 5mm 25mm,clip]{nue_CorrelationMat.pdf}}
%\caption{(Left) Fractional cross section covariance matrix, $E^{\text{cross section}}$, for \nue CC candidate events. (Right) The correlation matrix for \nue CC candidate events.  Inspection of the diagonal elements of the off-diagonal blocks shows the correlations between events in different detectors to be very near 1.0}
%\label{fig:xsecMatrix_nues}
%\end{figure}

%\begin{figure}[h]
%\centering
%\includegraphics[width=0.80 \textwidth]{nue_CorrelationMat.pdf}
%\includegraphics[width=0.40 \textwidth]{joseph_placeholder.jpg}
%\caption{The correlation matrix due to cross section uncertainties.  As seen in the off diagonal entries, the correlations between detectors are large for this uncertainty.}
%\label{fig:xsecCorrelation}
%\end{figure}

\subsection{Detector Systematics}
\label{sec:DetSyst}

The response of the different detectors has to be known to a sufficient precision to  maximize the experimental sensitivity and avoid introducing artificial detector effects mimicking the sought for oscillation signal. In this respect, the adoption of the same detection technique for all the different detectors and of the same operation conditions, permits to virtually cancel out the impact of the detector response uncertainty on the final measurement. Possible second order effects can arise from differences in the details of the design and implementation of the various detectors. The most relevant physical parameters like the drift field and the TPC structure should be kept as close as possible. Detector systematic effects can be generated by differences between the near and the far detectors, for example:
\begin{itemize}
  \item The wire orientations in the TPCs;
  \item TPC readout electronics (shaping, sampling time, S/N ratio, general noise
           conditions affecting the identification/measurement efficiency);
  \item Residual differences in the electric drift field (absolute value and homogeneity);
  \item Residual differences in the detector calibrations including the light collection systems and the identification of off beam interactions by timing;
  \item LAr purity levels in the  detectors;
  \item Different drift lengths and space charge effects;
  \item Residual differences in background levels from dirt events and from cosmic rays including different coverage and efficiency of the cosmic tagging systems;
  \item Effects induced by the different event rates at the two sites, event selection and identification efficiency including the different aspect ratios of the near and far detectors.
\end{itemize} 

As an example, the impact of the different wire orientation on the electron identification  efficiency has been studied with a simulation of the primary electrons produced in \nue CC interactions of the beam, assuming in both detectors the electronic wire signal and noise level actually measured in the T600. 
The effect of the different collection wire orientation between \larnd and T600 turns out to be  negligible on the reconstructed $dE/dx$ distribution: a $\sim$0.1\% variation in the electron identification efficiency on the first 2 cm of the track is observed in the simulation, having fixed to the same value ($3.425$ MeV/cm) the maximum accepted $dE/dx$. The corresponding multiplicity of occupied collection wires is also affected by the different wire orientations. A $3 \%$ difference in the electron identification efficiency is expected when at least 3 collection wires are required for the measurement.  Conservatively, assuming to correct the angular dependence of the wire multiplicity to the 20\% level using the data themselves, a  residual $<1\%$ systematic effect in the selection efficiency is expected.
This effect would be made further negligible if the induction wire signals could also be exploited in the $dE/dx$ measurement. 

A further example are electric field distortions induced by the accumulation of positive argon ions in the TPC drift volumes from the high cosmic muon fluxes in the SBN detectors located at the surface. Since the distortions depend on the drift length, which are different among the detectors, a potential detector systematic uncertainty could arise. However, today only first estimates of the absolute size of such distortions exist and thus of their affect on tracks and reconstruction. The ICARUS technical run on the surface in Pavia did not see significant track distortions.  However, the effect could be more significant in \larnd and \uboone due to their longer drift distances. Different actions can be taken to reduce this effect. Tracks from the high rate of cosmic muons that create the effect can also be used to monitor the distortion in each detector.  For the near detector, as in \uboone, a laser calibration system is foreseen to provide information on the actual electric field (see Part II for more details). The laser and cosmic muon tracks allow to generate a correction to be applied in the event reconstruction. \uboone will provide important input on the scale of the effect and the performance of the reconstruction corrections. A possible hardware implementation could stabilize the electric field inside the sensitive volume of the \lartpc with the addition of widely spaced shaping wire planes at the voltage of the potentials of the field cage electrodes (see Part III for more details) thus reducing the space charge effect.

\begin{comment}

Possible electric field distortions due to the accumulation of positive argon ions in the TPC drift volumes from the high cosmic muon fluxes in the SBN detectors at the surface are being investigated. According to the ICARUS technical run on the surface in Pavia, the resulting track distortion could be at most a few mm over 1.5 m of drift. The effect could be more significant in \larnd and \uboone due to their longer drift distances. Different actions can be taken to reduce this effect to a negligible level. The high cosmic muon rate that creates the effect can also be used to monitor it and generate a correction to be applied in event reconstruction. For the near detector, as in \uboone, an additional laser calibration system is foreseen to provide a complementary calibration method (see Part II for more details). Another possibility could be to reduce the space charge effect by stabilizing the electric field inside the sensitive volume of the \lartpc with the addition of widely spaced shaping wire planes at the voltage of the potentials of the field cage electrodes (see Part III for more details).
\end{comment}

It should be noted that all the contributions listed above can be directly measured with the data, monitored during the experiment, and corrected for in the analysis, largely reducing their impact on the measurement. It has been estimated that an overall global detector systematic uncertainty in the 2--3\% range would preserve the experimental sensitivity. 
%and the capability to cover at the 5$\sigma$ level the LSND allowed parameter region. 
We assume this systematic level as a requirement for the detectors. 

\subsection{Beam-Induced ``Dirt'' Events}
\label{sec:Dirt}

Neutrinos from the BNB will interact in material surrounding the active detectors, including liquid argon outside of the TPC, the cryostat steel, structural elements or engineering support equipment in the detector hall, the building walls and floors, and the earth outside the detector enclosure.  These interactions can produce photons (through \pizero decay or other channels) which can enter the TPC and convert in the fiducial volume, potentially faking an electron signal.  While it turns out the majority of interactions producing this background occur relatively close to the detector volume, the moniker ``dirt'' events is kept in analogy to its use in \MB plots and publications. This description, however, will refer to any backgrounds generated by beam neutrino interactions \emph{occurring anywhere outside of the TPC active volume}.  We consider this background only for the \nue analysis as the out-of-detector contamination of the \numu charged-current sample is expected to be negligible.

%{\bf In-detector, in-beam} backgrounds associated with the beam have been studied in detail before, but until now we have not discussed the {\bf out-of-detector, in-beam}, or ``dirt'', backgrounds. Specifically, these are neutrino interactions which occur in the material surrounding the active TPC volume. All interactions which occur inside the active volume of the TPC are discussed in Sec.~\ref{}. 

%To model this background we use simulate all the dirt and material around the TPC volume and allow neutrino interactions in this material. To generate neutrino interactions we produce a spills worth of neutrinos and allow them to possibly interact. This means in a given spill there is the possibility that there can be more than one neutrino interaction. 

%These background will only affect the electron appearance analysis since any  muon or charged pion tracks which enter the TPC will cross the active volume buffer region which surrounds the fiducial volume and be vetoed. For the electron appearance analysis the main background contribution comes from photons (coming from $\pi^{0}\rightarrow\gamma\gamma$ decays, or other nuclear processes). These photons will be able to cross the active volume buffer and then convert inside the fiducial volume. 

\begin{figure}[t] 
\centering
\mbox{\includegraphics[height=0.26\textheight]{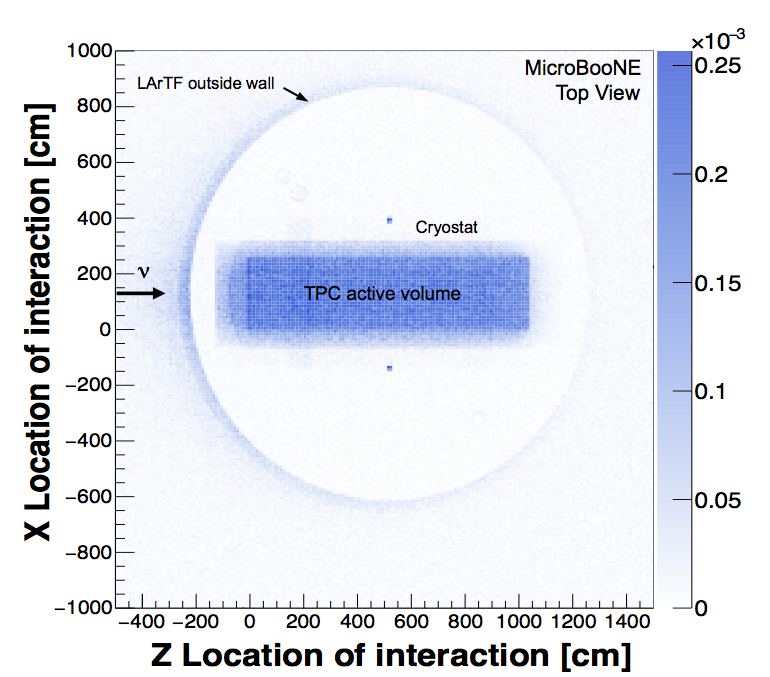}
\includegraphics[height=0.26\textheight]{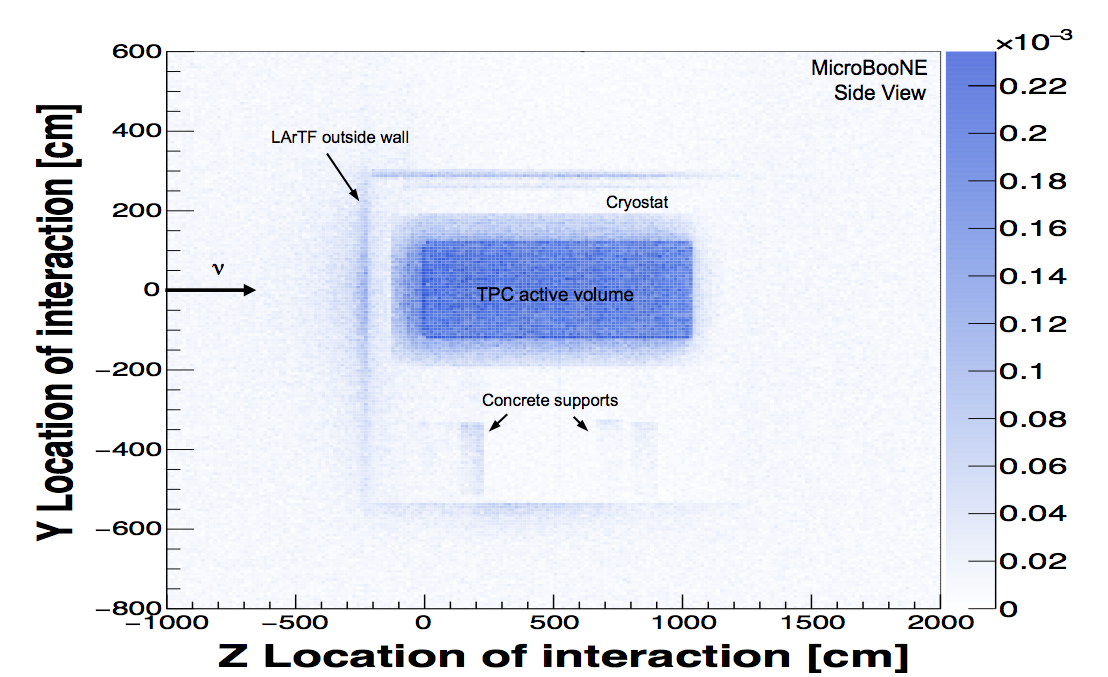}}
\caption{Location of interaction vertices for neutrinos which deposit any energy into the \uboone detector shown from above (left) and the side (right).}
\label{fig:dirt_int} 
\end{figure}

To estimate the dirt background, a Monte Carlo simulation is used which includes a realistic  geometry description of the material surrounding the detectors. Due to the large mass but small probability for any given interaction to create energy inside the detector, it is challenging to generate large statistics.  Substantial effort was put into generating a large Monte Carlo sample using the \uboone simulation where the geometry description is the most detailed.  Figure \ref{fig:dirt_int} shows the distribution of interaction vertices for BNB neutrinos which deposit any detectable energy into the \uboone detector. The walls of the LArTF building and the soil surrounding it are clearly visible.  In the right image, the concrete supports can be seen, but not the foam insulation saddles that sit between the supports and the cryostat.  The highest density of vertices is, of course, in the active volume of the detector. 
    
\begin{figure}[btp] 
\centering
\mbox{
\includegraphics[width=0.47\textwidth]{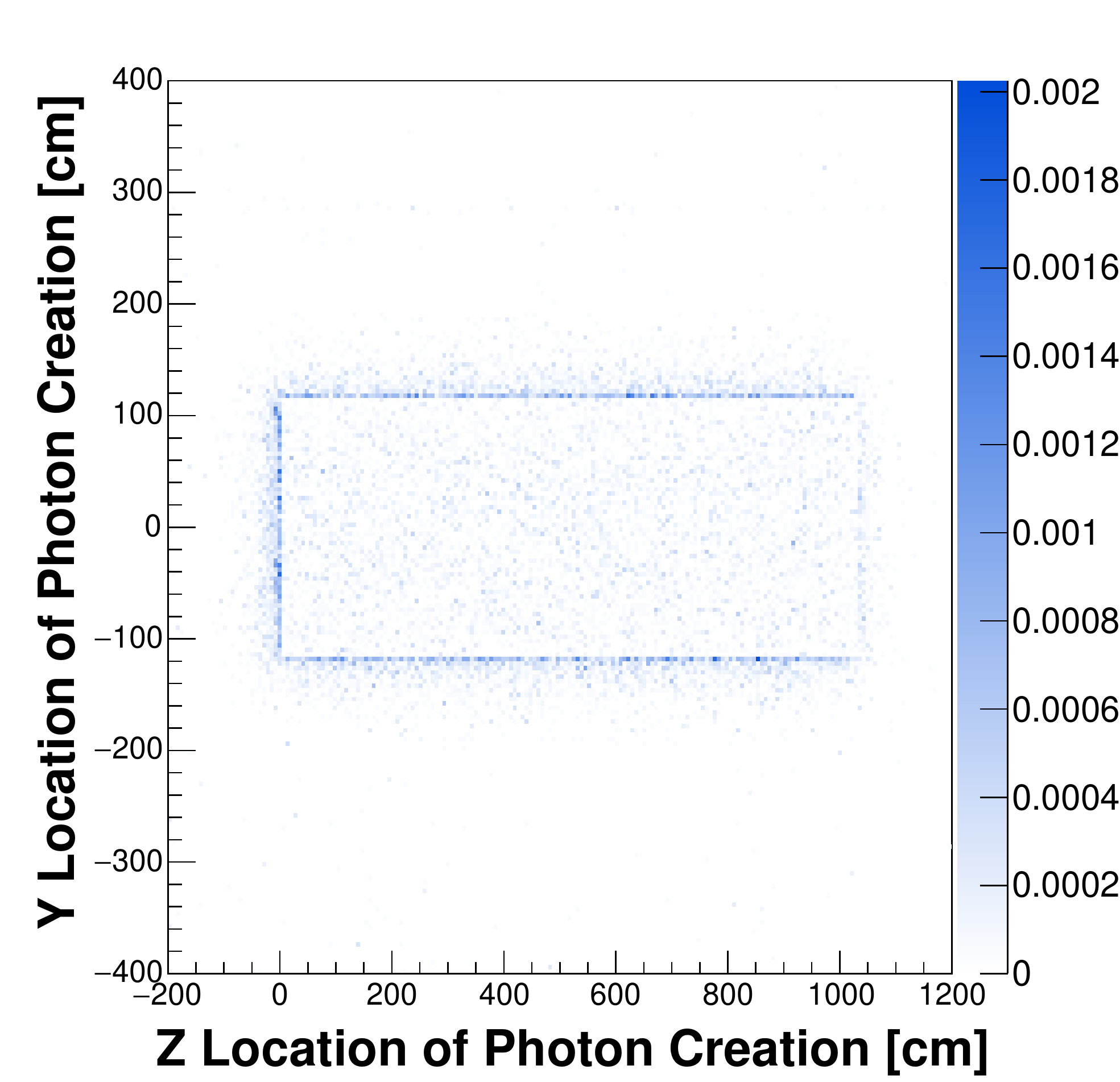}
\includegraphics[width=0.47\textwidth]{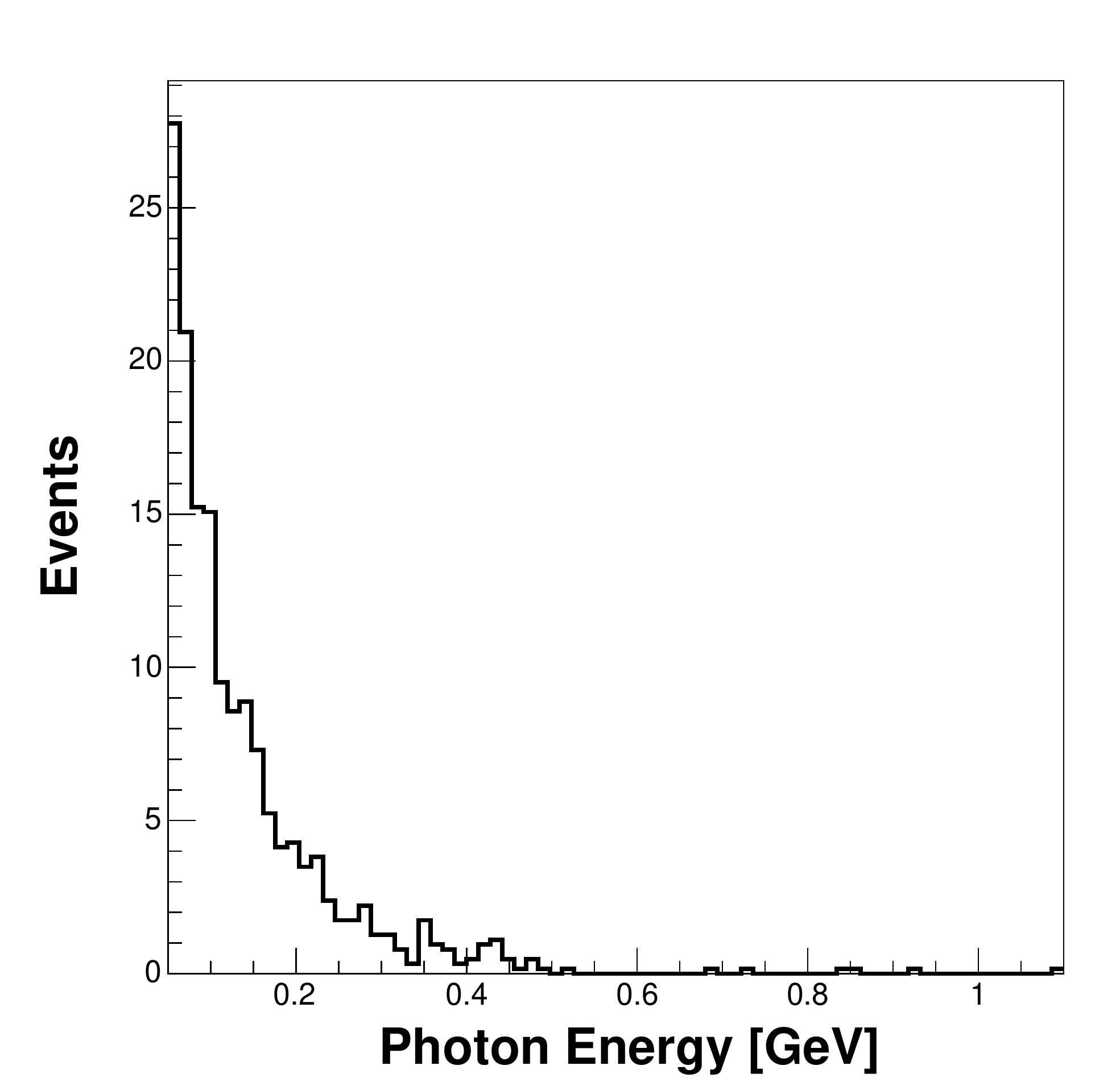}
}
\mbox{
\includegraphics[width=0.45\textwidth]{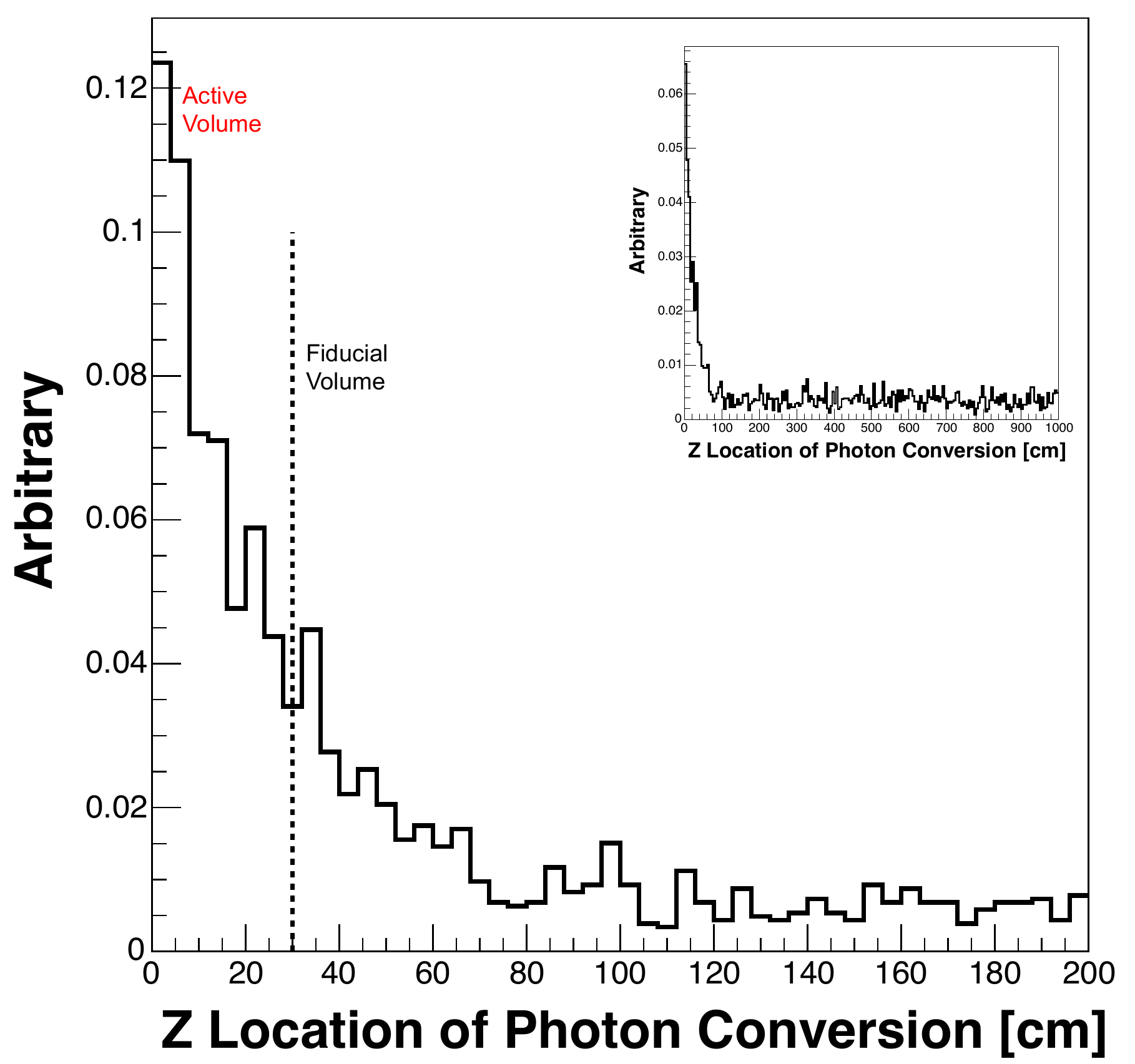}\quad\quad
\raisebox{-1mm}{\includegraphics[width=0.48\textwidth]{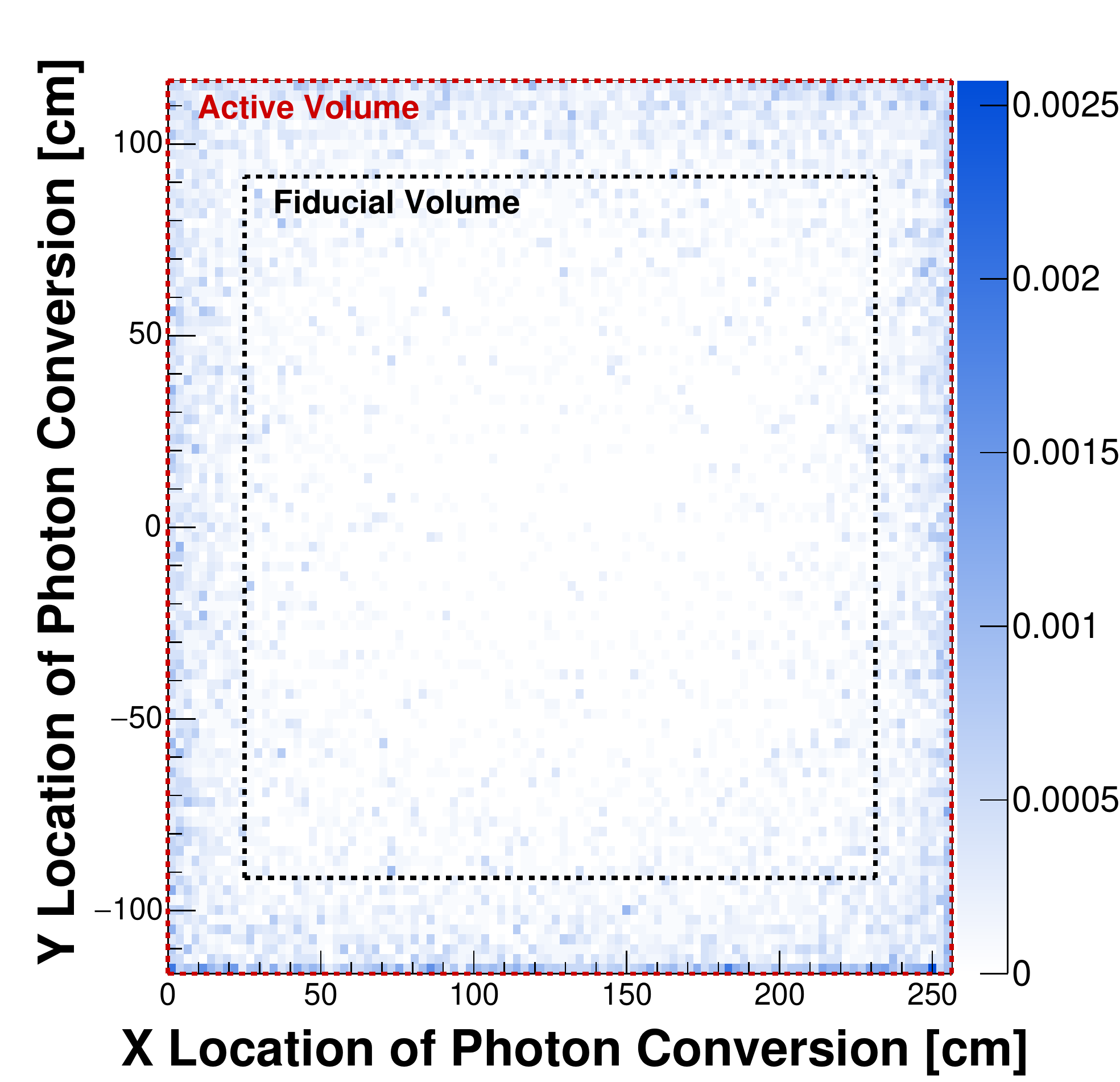}}
}
\caption{(Top left) Photon creation position in the $Y$--$Z$, side view, projection for photons which then convert inside the \uboone active volume, possibly faking a \nue CC interaction. (Top right) Energy of the photons that convert inside the \uboone detector but came from neutrino interactions outside of the detector active volume. (Bottom left) Photon conversion position inside the \uboone detector projected onto the $z$-axis ($z=0$ is the start of the TPC active volume; only first 200~cm shown). The vertical dashed line is 30~cm from the front of the TPC. (Bottom right) Photon conversion position in the $X$–-$Y$, front view, for photons which convert downstream of $z = 30~cm$ (plot boundary is the TPC active volume; fiducial volume for \nue analysis is indicated).}
\label{fig:dirt_photons} 
\end{figure}

From this sample, events with an interaction vertex outside of the active TPC volume but that generate a photon which converts inside the detector, are selected. Due to the short radiation length in liquid argon ($X_0 = 14$~cm), the argon volume surrounding the TPC inside the cryostat provides an effective shield for photons trying to enter from beyond the cryostat walls.  Most of the interactions capable of creating a photon inside the fiducial volume, therefore, tend to happen in this outer argon region.  This can be seen in the upper left panel of Figure \ref{fig:dirt_photons}.  The plot shows the creation point of all photons which then convert inside the \uboone active volume, and they clearly pile up in the region just beyond the active volume boundary.  Photons entering the detector are likely to interact within a few 10's of centimeters of the TPC boundary, providing a handle with which to minimize this background.  The lower panels of Figure \ref{fig:dirt_photons} show the photon conversion point within the active volume projected onto the $z$-axis (the beam direction) and the $x-y$ plane.  

This sample, as with all single photon shower backgrounds, is reduced by analyzing the $dE/dx$ at the start of the shower and rejecting 94\% of pair production interactions.  To further reduce out-of-detector dirt photons in the \nue analysis, we restrict the fiducial volume to an inner region of the detector 30~cm from the upstream and 25~cm from the side boundaries of the active TPC region, reducing the number of dirt background events by 80\% in \uboone. These fiducial volume boundaries are indicated in the figures for \uboone, but are used uniformly in all three detectors in the analysis. 

%\begin{figure}[btp] 
%\centering
%\mbox{
%\includegraphics[width=0.33\textwidth]{uB_dirt_ZY_Photon_creation.pdf}
%\includegraphics[width=0.33\textwidth]{uB_dirt_Z_Proj_conv_zoom.pdf}
%\includegraphics[width=0.33\textwidth]{ZY_Photon_creation_LAr1-ND.pdf}}
%\includegraphics[width=0.33\textwidth]{MicroBooNE_Dirt_Logy.pdf}}
%\caption{(Left) photon creation position in the $Y$--$Z$, side view, projection for photons which then convert inside the \uboone active volume, possibly faking a \nue CC interaction. (Center) Photon conversion position inside the \uboone detector projected onto the $z$-axis (only first 200~cm shown). The vertical dashed line is 30cm from the front of the TPC. (Right) Energy of the photons that convert inside the \uboone detector but came from neutrino interactions outside of the detector active volume.}
%\label{fig:dirt_photons} 
%\end{figure}

\begin{figure}[btp] 
\centering
\mbox{\includegraphics[width=0.45\textwidth]{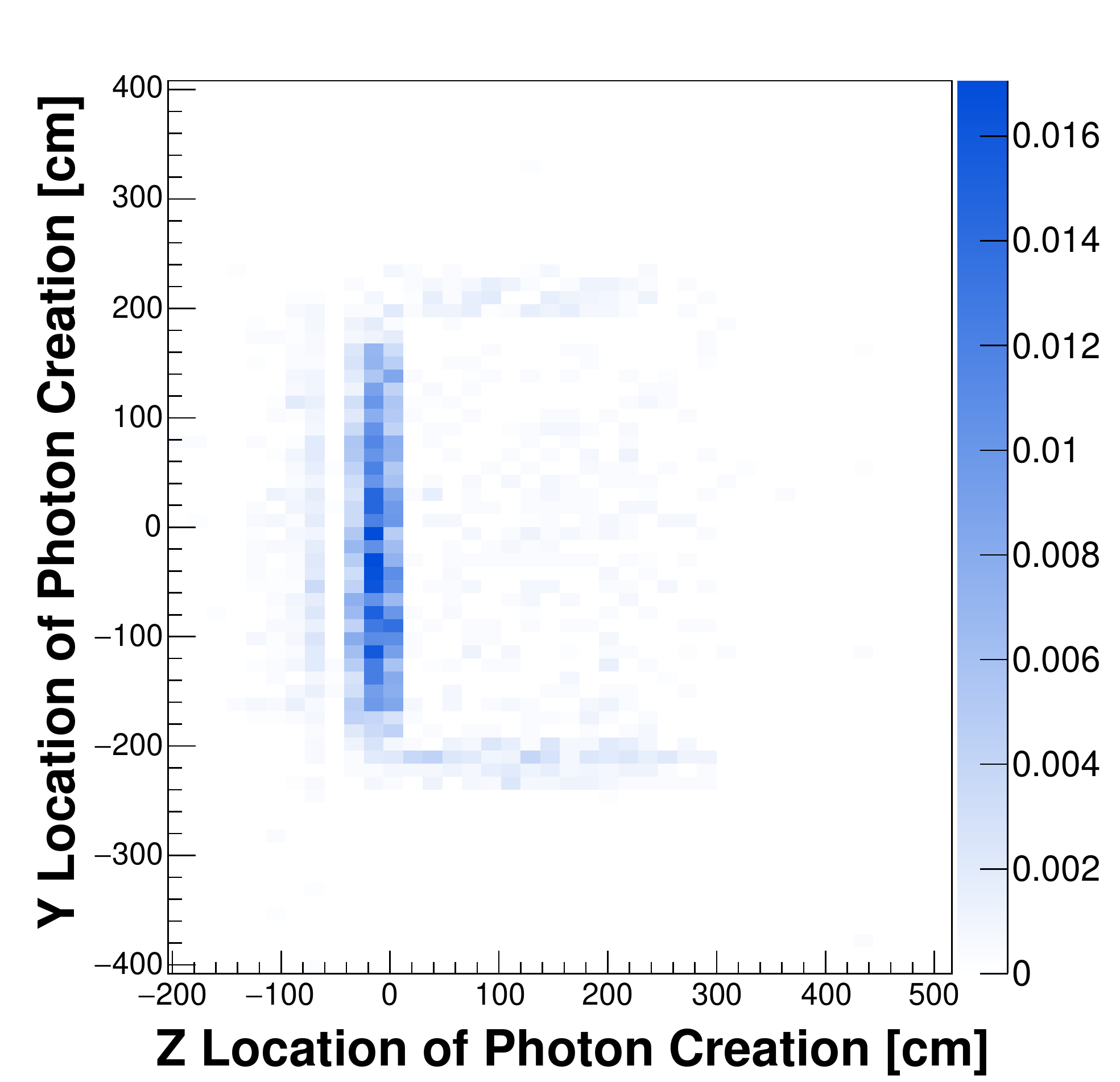}
\includegraphics[width=0.45\textwidth]{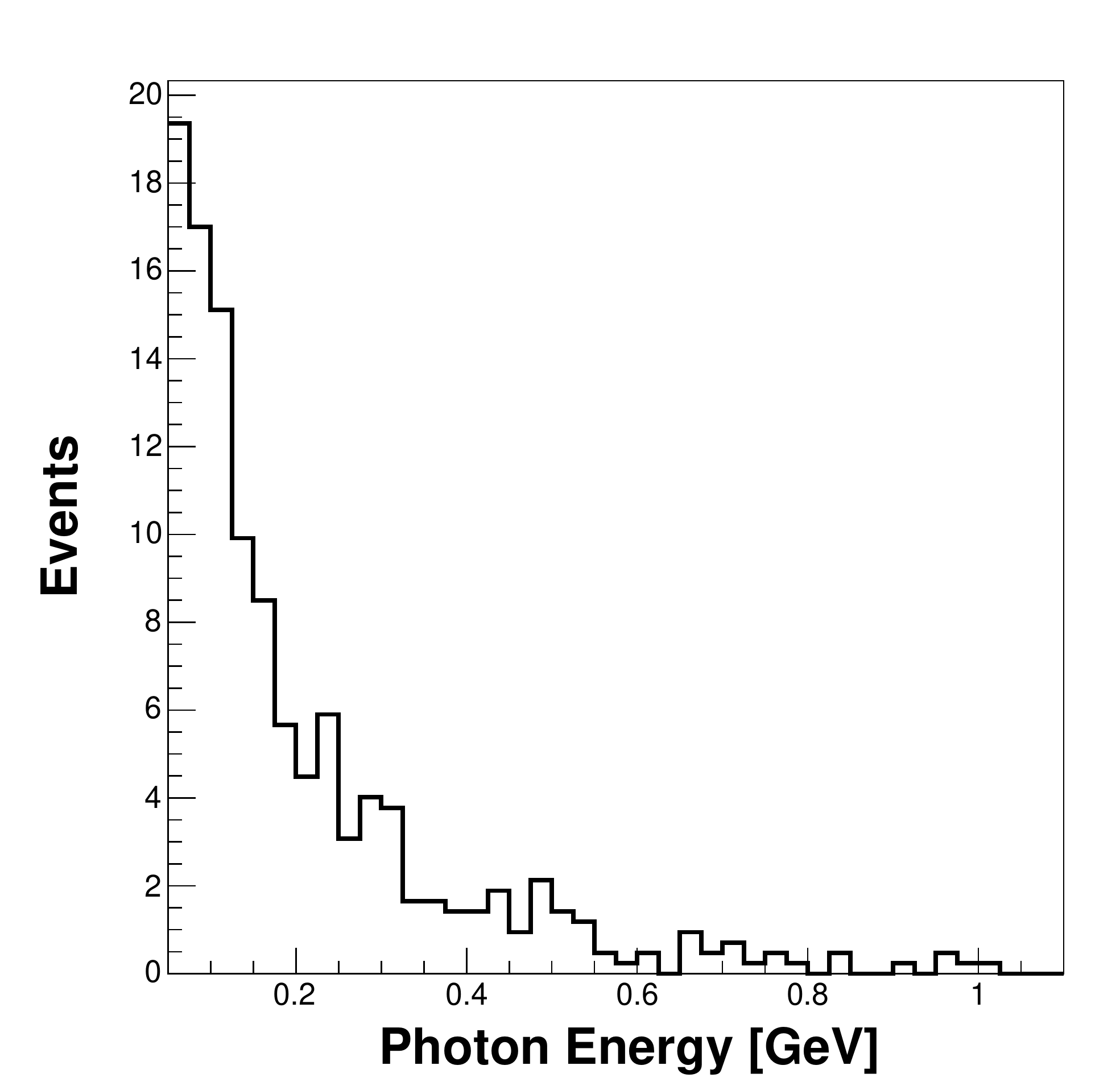}
}
\caption{(Left) Photon creation position in the $Y$--$Z$, side view, projection for photons which then convert inside the \larnd active volume. (Right) Energy of the photons that convert inside the \larnd detector but came from neutrino interactions outside of the detector active volume.}
\label{fig:dirt_photons_nd} 
\end{figure}

A similar Monte Carlo sample has been generated for the \larnd detector at 110~m. Figure \ref{fig:dirt_photons_nd} shows the creation point of all photons which then convert in the \larnd active volume.  While the dirt photons in \uboone come in from both the upstream face and the sides of the detector (see Figure \ref{fig:dirt_photons}), in \larnd they are more concentrated at the upstream face of the detector.  This difference is due to two factors, $i$) the neutrino flux at the \larnd location is still highly collimated so the event rate is peaked in the middle of the detector and falls off toward the detector sides, while it is uniform across the \uboone detector face, and $ii$) the amount of argon outside of the TPC in the square \larnd cryostat is less than the amount in the cylindrical \uboone cryostat.  

A dedicated simulation of out-of-detector interactions at the \icarus site has not been generated.  Instead, because the 470~m and 600~m locations are both in the region where the flux is wider than the detectors, we can use the \uboone predictions to scale to the far detector site and generate an estimate of the dirt background in ICARUS. We account for the different surface areas of the two detectors and scale the neutrino flux as $1/r^2$. To account for any differences in the background rate from photons entering the front vs. the sides of the detectors, we scale events in the beginning 50~cm of the \uboone detector separately from those further downstream: 

\begin{equation}
\setlength{\abovedisplayskip}{20pt}
\setlength{\belowdisplayskip}{20pt}
\small
N_{\text{dirt}}^{\text{T600}} =
\frac{470^2}{600^2}\times 2 \times
\left( \frac{\text{\footnotesize Front Area T300}}{\text{\footnotesize Front Area $\mu$BooNE}}N_{\text{dirt}}^{\mu B}(z\leq 50~\text{cm}) + 
\frac{\text{\footnotesize Side Area T300}}{\text{\footnotesize Side Area $\mu$BooNE}}N_{\text{dirt}}^{\mu B}(z>50~\text{cm}) \right)
\label{eq:dirt_scale}
\end{equation}

\noindent where \noindent $N_{\text{dirt}}^{\mu B}(z)$ is the number of dirt events predicted in \uboone and $z$ is the distance from the front of the active volume.  Table \ref{tab:dirt} provides the total number of dirt background events expected in each detector according to the simulations for \larnd and \uboone and using Eq. \ref{eq:dirt_scale} to estimate the rate in ICARUS. The scaling procedure adopted for the far detector has been checked with a  muon neutrino MC event sample. About 40 “dirt” events induced by neutrino interactions in the passive LAr volume behind the wire planes or in the top/bottom of the detector were found. This result is roughly in agreement with the previously described extrapolation if the events induced in the upstream and downstream passive LAr and in the other materials around the detector are neglected.

%when the fiducial volume boundary begins 30~cm from the upstream edge of the TPC, effectively creating an active buffer region in the front of the detector. 

%We find that a very simple, but effective way to substantially reduce the dirt background is to begin the fiducial volume boundary further away from the upstream edge of the TPC active volume.  
%We, therefore, have adopted a 30~cm active buffer upstream of the fiducial volume in each detector. %The suppression in from adopting a 30~cm buffer is largest in \larnd, as expected, but is significant also in \uboone.  

%\begin{table}[h]
%\begin{center}
%\caption{\label{tab:dirt}Estiamted background levels in the \nue charged-current sample from out-of-detector neutrino interactions in a \pot POT exposure. The total background is compared for a fiducial volume that begins 5~cm away from the upstream active detector boundary and one that begins 30~cm away.}
%\begin{tabular}{lcc}\hline\hline
%Detector 	& 5~cm front active buffer~~	& ~~30~cm front active buffer	\\ \hline
%\larnd		& 158					&  		\\
%\uboone		& 53 					&  		\\
%\icarus		& 48					&  		\\ \hline\hline
%\end{tabular}
%\end{center}
%\end{table}

\begin{table}[h]
\begin{center}
\caption{\label{tab:dirt}Estimated rates of photon induced showers from out-of-detector neutrino interactions faking a \nue CC interaction in a \pot POT exposure. A $dE/dx$ cut has been applied to reject 94\% of pair production events.}
\begin{tabular}{l c c c}\hline\hline
Detector 	& \multicolumn{3}{c}{Estimated Dirt Background Events (\pot POT)} \\ 
			&  ~~~~~$z \leq 50~cm$~~~~~  	&  ~~~$z > 50~cm$~~~	& Total \\ \hline
\larnd		& 26.2				&  17.0	 		&  43.2  \\
\uboone		& 2.38 	 			&  19.5			&  21.9  \\
\icarus		& 5.15				&  57.0			&  62.2  \\ \hline\hline
\end{tabular}
\end{center}
\end{table}

In the present analysis, we reduce the dirt background to manageable levels by restricting the fiducial volume used in the \nue analysis.  A more sophisticated approach has also been explored that would use the reconstructed shower direction in candidate events to project \emph{backwards} from the vertex and calculate the distance to the nearest TPC boundary in the backwards direction.  Cutting on this quantity on an event-by-event basis would allow us to further reduce the dirt backgrounds on all sides without sacrificing fiducial volume. This is referred to as the \emph{backwards-distance-to-wall} variable, and is not used in the current analysis.  

\subsubsection*{Error Matrix Contribution from Dirt Backgrounds}

Finally, we require an estimate of the error matrix associated with dirt backgrounds, $E^{\text{dirt}}$. The dirt background rate in each detector can be constrained with data using a sample of electromagnetic shower events near the TPC boundary, or showers where the reconstructed momentum is consistent with the particle having entered the detector.  This sample will be enhanced in dirt background events and can be used to validate the simulations.  At this time, we conservatively estimate a 15\% systematic uncertainty uncorrelated between detectors, but fully correlated within the energy spectrum in each detector.  This covariance matrix is constructed as

\begin{equation}
E^{\text{dirt}}_{ij} = \rho_{ij}(0.15\times N^{\text{dirt}}_{i})(0.15\times N^{\text{dirt}}_{j})
\label{eq:dirt}
\end{equation}

\noindent where $\rho_{ij}$ is 0 if $i$ and $j$ bins correspond to different detectors, and 1 if they correspond to the same detector.

\subsection{Cosmogenic Backgrounds}
\label{sec:Cosmics}

%An important background to consider, in particular   for the \nue appearance channel,  are cosmogenic events.  
Another important background to the \nue analysis is created by cosmogenic photons that generate electrons in the detector via Compton scattering or pair production interactions that are misidentified as a single electron. Photons are created either in the atmospheric shower (``primary photons'') or by cosmic muons propagating through the detector and nearby surrounding materials (``secondary photons'').  In the case of an un-shielded detector at the surface, the background to a \nue CC sample is mostly due to primary photons, but these can be easily absorbed by a few meters of earth or concrete shielding.  In simulations of the far detector, for example, a 3~m rock coverage reduces by a factor 400 the number of primary photons above 200~MeV in the active volume,  % and by a factor 4 the total number of photons, 
and secondary photons generated by muons passing through or very near the detectors becomes the dominant source of background. To further reduce the rate, we must identify cosmic showers through topological and timing information in the event.

In an ideal situation where precise timing information is known for every track or shower inside  the detector, only cosmogenic events in coincidence with the beam spill can contribute to the background. However, in a realistic situation, interactions occurring anytime within the acquisition time (which corresponds to the maximum electron drift time) may influence the data analysis, as will be explained below. Given the respective detector sizes, the maximum drift times are 1.28~ms in \larnd, 1.6~ms in \uboone and 0.96~ms in ICARUS, to be compared with the 1.6~$\mu$s duration of the beam spill from the BNB. Potential cosmogenic backgrounds can be categorized according to their time structure as:

\begin{enumerate}[label=Timing case \Alph*:,leftmargin=*]

\item Cosmogenic photon interacts in the detector in coincidence with the beam spill. %As already said, these constitute a direct background to the \nue appearance search. 

\item Cosmogenic photon interacts anywhere inside the drift time, and a \emph{different} cosmic event (muon or otherwise) is in the detector in coincidence with the beam spill. If the arrival time of the photon is poorly known, it could be mistaken for the in-spill event.

%\item  Cosmic photon interacts anywhere inside the drift time, with a trigger given by a neutrino interaction.  
%If the arrival time of the photon is poorly known, it could interfere with the reconstruction of a neutrino event in the fiducial volume.

\end{enumerate}

If not properly recognized, neutrino beam interactions occurring in LAr surrounding the TPC active volume or low energy neutral-current interactions that are not identified, can also provide a scintillation trigger in the beam spill leading to a situation similar to timing case B. This effect has been roughly estimated for the far detector, resulting in a very small additional contribution to the cosmogenic background due to the low neutrino interaction rate (Table~\ref{tab:cosmic_reduction}). Therefore, it is not currently included in the analysis.

% Due to the relatively low neutrino interaction rate, this effect is smaller and is not currently included in the analysis, but has been roughly estimated in the illustration for the far detector in Table~\ref{tab:cosmic_reduction}.

%at low energy occurring both in active volume and in the external LAr could mimic a cosmogenic trigger in time with the beam spill. In such a case cosmogenic photons inside the drift time are expected to contribute to the background escaping any mitigation effect from the muon tagging system and the precise time matching with the beam spill structure.

Key topological information includes the location of the photon within the detector (just as with dirt events, externally produced $\gamma$s will interact near the detector edges) and the proximity to the parent cosmic muon track in the case of secondary photons.   Therefore, we identify two main categories of event topology:

\begin{enumerate}[label=Topology \Roman*:,leftmargin=*]

\item Cosmogenic photon interacts inside the fiducial volume, and the parent muon also enters the TPC active volume.

\item Cosmogenic photon interacts inside the fiducial volume, but the photon originated from the atmospheric shower (a primary), the parent particle is not visible (e.g. neutrons), or the parent particle does not enter the TPC active volume (e.g. muon misses the active volume).

\end{enumerate}

%Below, we will consider various strategies, in addition to the electron/photon discrimination based on $dE/dx$ evaluation, to further reduce each of these background categories.

%It should be emphasized that the most important outcome at this time is to understand if cosmogenic backgrounds can be reduced to a level that oscillation signals will be observed with sufficient $S/\sqrt{B}$.  The absolute rate of events in the experiment will not introduce significant \emph{systematic} uncertainty because it will be measured with high precision using off-beam random event triggers. This is a critical aspect of the experiment, and designing the DAQ systems to record sufficient random triggers must be considered.  

\begin{figure}[t]
\centering
\mbox{\includegraphics[width=0.8\textwidth, trim=0mm 0mm 0mm 4mm, clip=true] {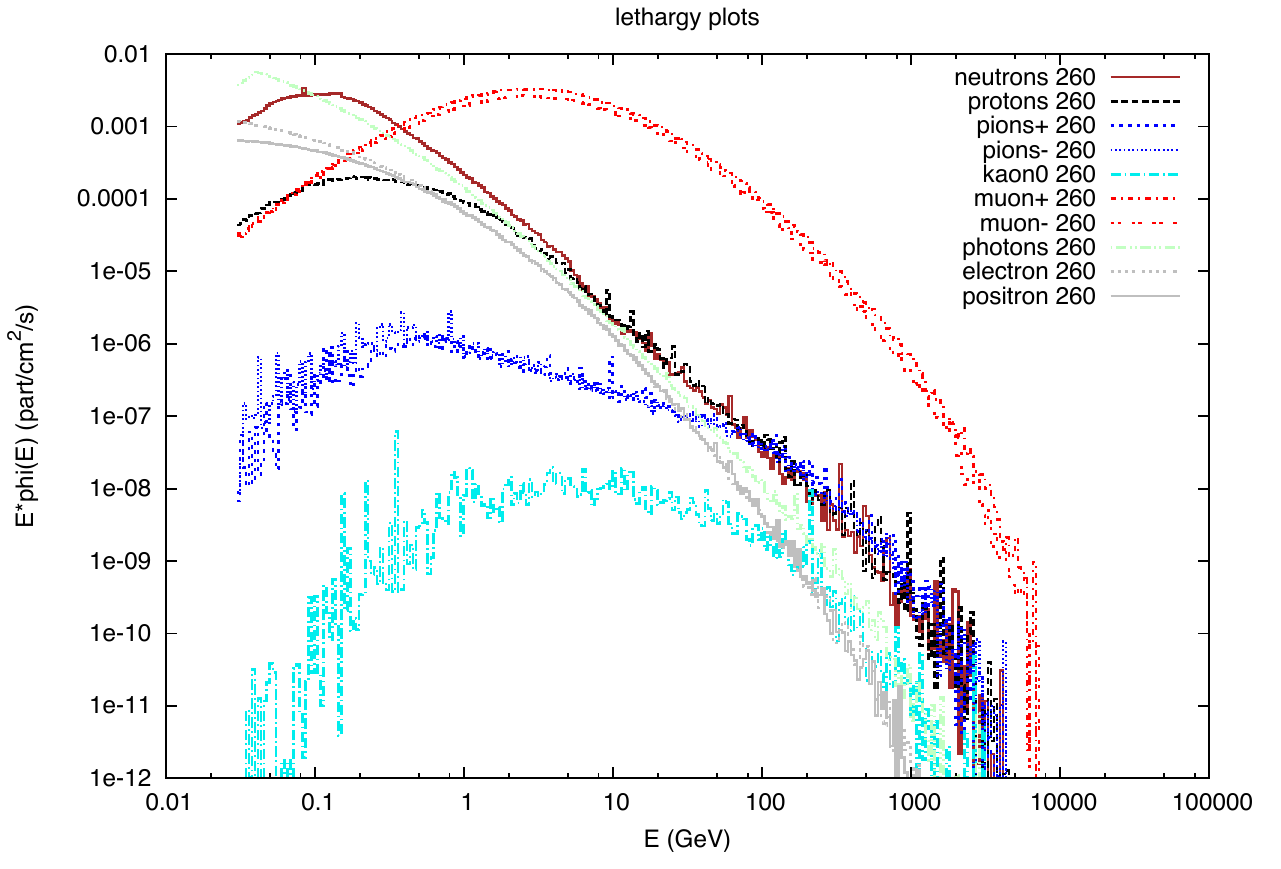}}
\caption{Particle fluxes in the atmosphere at 260~m elevation (FNAL is at 225~m) according to the FLUKA simulation.}
\label{fig:fluxes260}
\end{figure}

Estimation of the cosmogenic background rate requires a detailed simulation of the cosmic particle fluxes and their interactions in and around the detectors. As with the dirt backgrounds described in Section \ref{sec:Dirt}, a realistic geometry description and significant computational effort is required.  For the current analysis, independent simulations have been developed by the \uboone, \larnd, and ICARUS Collaborations. All future analysis of SBN data will, of course, be based on a common simulation, but the current development has provided some important opportunities for cross checks. We provide here brief descriptions of each simulation:

\begin{itemize}

\item ICARUS: The ICARUS simulation uses FLUKA~\cite{Fluka1,Fluka2} for both the cosmic ray showering and the particle transport to and inside the detector.  FLUKA is a multipurpose Monte Carlo code used for several years to simulate cosmic showers in the atmosphere. Examples of its performance can be found in the literature, for instance the simulated flux of muons at different depths in the atmosphere agrees with CAPRICE data within experimental errors~\cite{caprice}. Similar agreement~\cite{TesiSilvia} is obtained with the muon spectra measured by the L3 experiment, and predicted proton and lepton fluxes in the atmosphere are in very good agreement~\cite{AMS-fluka} with the AMS data.  The ICARUS simulation is the most complete in that it includes both proton and ion primary cosmic ray sources and generates all particle content in the showers. Primary neutrons, for example, are found to contribute about 10\% of electron-like events.  The energy spectra of different particle types predicted by the FLUKA simulations at 260~m above sea level (FNAL elevation is 225~m) are shown in Figure~\ref{fig:fluxes260}. The detector was simulated at the surface (not in a building) with and without 3~m of concrete overburden above the detector as mentioned above.  The default in this analysis is \emph{with} overburden.    

\item \uboone: The \uboone simulations are performed with the CRY cosmic-ray shower simulation \cite{CRY} as a primary particle generator and GEANT4 to transport particles into the \uboone detector.  The CRY package provides reasonable results in a fast and easy way, but has known limitations, such as the lack of a contribution from primary ions, a rigid binning structure that sacrifices some spectral details, and an under prediction of neutron, proton, electron, and $\gamma$ shower content. Comparisons have shown that results obtained with CRY+GEANT4 in \uboone and results scaled from the full FLUKA T600 simulation agree within a factor of two.  The detector geometry is the most detailed and includes the LArTF building and substantial infrastructure, so there is some shielding effects from the building and platforms above the detector.  The LArTF facility also has the ability to support concrete shielding blocks on the roof, but this is \emph{not} included in the present simulations. Studies are continuing by the \uboone Collaboration to determine if the additional shielding should be added for the upcoming physics run.          

\item \larnd: The \larnd cosmic muon flux is generated using Gaisser's parameterization \cite{gaisser}, with corrections for the Earth's curvature and the muon lifetime. The muon flux simulation is performed at the Fermilab latitude, and muons are propagated through the \larnd detector and building using GEANT4. The detector is simulated in a pit below grade but without additional shielding above the detector.  Only the muon component of the shower is included, but results from both the ICARUS and \uB simulations, which include all components, show that secondaries from muons are by far the dominant contribution to the background, and the exclusion of primary photons and hadrons in the simulation is nearly equivalent to simulating a detector with some overburden.       

\end{itemize}

To get a sense for the situation, it is instructive to first look at some basic numbers coming from the far detector simulation.  Cosmogenic interactions of all kinds depositing more than 100~\MeV of energy will occur in the T600 fiducial volume at $\sim$11~kHz, implying such an event inside the detector during 1 out of every 50 beam spills.  A \pot POT run represents approximately $1.32 \times 10^{8}$ spills at nominal intensity, corresponding to 211 seconds of beam-on time throughout the experiment. ICARUS will, therefore, see $2.5 \times 10^{6}$ cosmic events \emph{during the beam spill time} in the run. Further, $\sim$10 cosmic muon tracks will enter into the detector volume during the 0.96~ms drift time in each readout of the detector.   

\begin{figure}[t]
\centering
\mbox{\raisebox{0mm}{\includegraphics[width=0.47\textwidth, trim=0mm 0mm 20mm 30mm, clip=true]{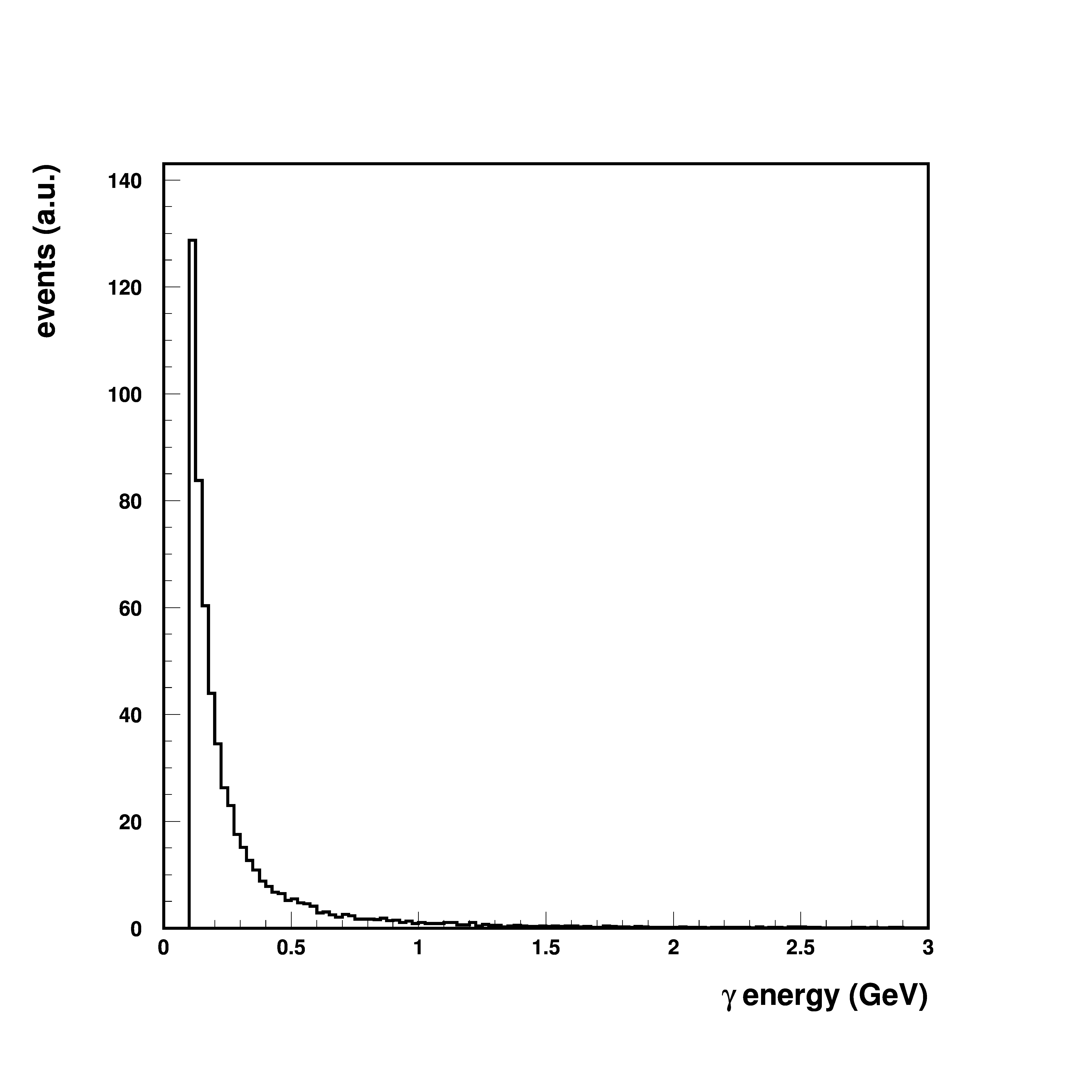}}
\includegraphics[width=0.52\textwidth, trim=0cm 0cm 0cm 1cm,clip=true]{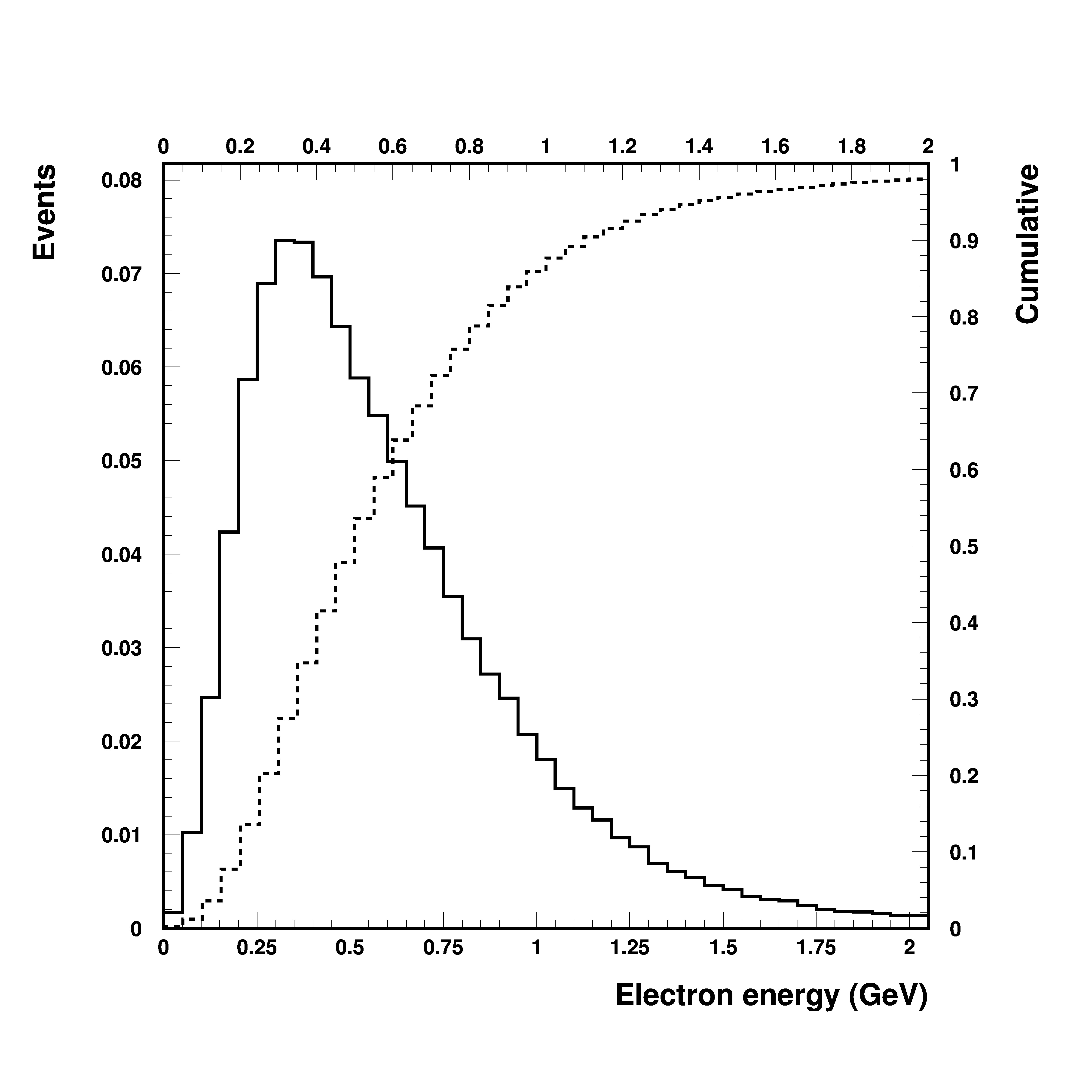}}
\caption{(Left) Energy distribution of cosmic background photons inside the far detector. (Right) Energy distribution of electrons produced in \nue interactions in the far detector.
%(Right) $\gamma+Ar$ cross sections showing the Compton (Incoherent Scattering) and pair production components versus photon energy (in MeV). Compton scattering contributes 3.7\% of the cross section at $E_{\gamma} = 200~\MeV$ and $<2\%$ above 400~\MeV.}    
}
\label{fig:eene}
\end{figure}

%\begin{figure}[h]
%\centering
%\includegraphics[width=0.4\textwidth]{argon_photon_cross_sections.pdf}
%\caption{Photon total cross section as a function of energy in argon.  Indicated are the contributions of pair production off a nuclear field or an electron field and incoherent scattering (Compton scattering off an electron) \cite{xcom}.}
%\label{fig:photonxsec}
%\end{figure}

Figure \ref{fig:eene} (left) shows the energy distribution of cosmogenic photons which interact in the TPC fiducial volume as calculated by far detector simulation (the others look similar, of course). The spectrum is steeply decreasing with energy.  For comparison, Figure \ref{fig:eene} (right) shows the energy distribution of the electrons produced in \nue charged-current interactions in the far detector by the BNB intrinsic \nue flux.  In the following, all background estimates are provided above an energy threshold of 200~MeV. %, corresponding to an efficiency loss of about 10% on signal events.  %photon total cross section in argon in this energy range and the contributions of pair production (off a nuclear or electron field) and incoherent scattering (Compton scattering off an electron) \cite{xcom}.    

Cosmogenic photon interaction rates have been estimated in the three detectors using the simulations described above and the results are detailed in Table~\ref{tab:bgrates}. In each detector, fiducial cuts as suggested by the beam dirt events analysis (Section \ref{sec:Dirt}) have been applied, namely 25~cm from the sides of the active volume, 30~cm from the upstream face, 50~cm from the downstream face, and 1.5~cm from the cathode when applicable (for the resulting fiducial masses see Table~\ref{tab:fv}).  Rates for both topology I and II events occurring within the beam spill (timing category A) are estimated directly from the simulations by scaling the time exposure represented in each Monte Carlo sample to 211 seconds.  Rows 1--4 of Table~\ref{tab:bgrates} give the raw rates for both Compton and pair producing photons inside the fiducial volume with and without a parent muon that enters the TPC active volume.  These numbers reveal several interesting features.  First, the ratio of row 1 to row 2 is $\sim$2\% in each case, which is consistent with the size of the Compton scattering cross section in this energy range.  Second, a comparison of rows 3--4 to 1--2 indicates that the likelihood of a photon converting in the fiducial volume where the parent muon completely misses the TPC is very small.  The 25~cm active buffer around the fiducial volume motivated by the dirt backgrounds is also very effective at absorbing cosmogenic photons entering the detector from outside. And as we will describe below, the presence of the parent muon in the TPC provides a strong handle for rejecting the photon shower as a beam-related event.  Finally, comparing the different columns of rows 1--4 does reveal some variability in the predicted cosmic photon rates in the three detectors.  Factors of 2-3 may be expected due to differences in the input simulations as described above.  The geometry of the detectors plays a role in the expected rates, as well.  For example, the probability that a crossing muon produces a photon in the detector will scale as the average muon track length in the detector, and \larnd has the largest average track length due to the detector's 4~m height. 

Rows 5--8 of Table \ref{tab:bgrates} present the number of events of timing category B and are calculated directly from rows 1--4. We assume, to first order, that the time signal during the beam spill is produced by a cosmic muon entering the detector.  Event category B is reducible if light signals in the argon are able to be correctly matched to the energy deposits that produce them, however, we initially assume this is not done.  The number of category B events, $N_B$, can then be calculated from the number of category A events, $N_A$, that were estimated directly from the simulation.  The scale factor ends up being $N_{\mu}^{\text{drift}}$, the average number of muons that enter the detector per readout during the full drift time:

\begingroup{
  \setlength{\abovedisplayskip}{20pt}
  \setlength{\belowdisplayskip}{20pt}
\begin{equation}
N_B = P_{\gamma}^{\text{drift}} \times P_{\mu}^{\text{spill}} =  
\left(N_A * \frac{t_{\text{drift}}}{t_{\text{spill}}}\right)\times \left(N_{\mu}^{\text{drift}}*\frac{t_{\text{spill}}}{t_{\text{drift}}}\right) = 
N_A * N_{\mu}^{\text{drift}}
\end{equation}
}

\noindent where $P_{\gamma}^{\text{drift}}$ is the probability of having a cosmogenic photon anywhere in the drift time and $P_{\mu}^{\text{spill}}$ is the probability that a muon crosses the detector during the beam spill time. Our simulations indicate that $N_{\mu}^{\text{drift-\larnd}} = 2.9$,  $N_{\mu}^{\text{drift-\uboone}} = 5.0$, and $N_{\mu}^{\text{drift-T300}} = 5.5$.  The T300 is the right unit for the ICARUS detector since each T300 module is an optically isolated element of the full T600 detector.  

\begin{table}[t]
\centering
\caption{Background rates, assuming 3 years of data taking for a total of
$6.6 \times 10^{20}$ protons on target, delivered in $1.32 \times
10^{8}$ beam spills equaling 211 seconds of beam time. Events with at least one photon shower above 200 MeV converting in the fiducial volume are counted in all the $\gamma$ entries. %, while the total number of events is counted for all energy depositions above 100 MeV in the active volume.}
}
\label{tab:bgrates}
% T600:0.960 msec drift, 
% 1.3210^8 beam triggers 
%211.2 sec beam time
\begin{tabular}{|c|l|c|c|c|c|c|}
\hline
\multicolumn{2}{|c|}{} & & & \multicolumn{3}{|c|}{\small $E_{\gamma}>200~\MeV$, Pair prod} \\
\multicolumn{2}{|c|}{Cosmic photon interaction description} & Timing & Topology & \multicolumn{3}{|c|}{\small $E_e>200~\MeV$, Compton} \\
\multicolumn{2}{|c|}{} & Cat. & Cat. & \multicolumn{1}{|c}{~\larnd~} & \multicolumn{1}{|c|}{~$\mu$BooNE~} & \multicolumn{1}{c|}{~ICARUS~} \\ \hline\hline
1 & $\gamma$  Compton in spill, primary $\mu$ enters AV & A & I
&887  	& 206  & 599\\
2 & $\gamma$  Pair prod in spill, primary $\mu$ enters AV & A & I
%&52337	& 11591 & 32000  \\
&52,300	& 11,600 & 32,000  \\
3 & $\gamma$  Compton in spill, primary misses AV & A & II
%& 0.95& 2.48 &  $<$4  \\
& $<$1 & $<$3 &  $<$4  \\
4 & $\gamma$  Pair prod in spill, primary misses AV & A & II
%& 55.3 & 82.0 & 11  \\ \hline
& 55 & 82 & 11  \\ \hline
5 & $\gamma$  Compton in drift, primary $\mu$ enters AV & B & I
%& 2254 & 1030 &  3300   \\
& 2,550 & 1,030 &  3,300   \\
6 & $\gamma$  Pair prod in drift, primary $\mu$ enters AV & B & I
%&132908 & 57956 & 176000  \\
&150,200 & 57,950 & 176,000  \\
7 & $\gamma$  Compton in drift, primary misses AV & B & II
%& 2.40 & 12.4 & $<$4  \\
& $<$3 & 12.4 & $<$4  \\
8 & $\gamma$  Pair prod in drift, primary misses AV & B & II
%&140.4 & 410 & 60  \\ \hline
&160 & 410 & 60  \\ \hline
\hline
\end{tabular}
\end{table}

Table \ref{tab:bgrates} represents the raw number of cosmogenic photons that interact within the fiducial volumes of each detector during the proposed run.  A number of strategies can be applied to reduce the cosmic backgrounds entering the \nue analysis sample. 
%which will have varying effectiveness depending on the background type. 
Below we list the strategies being considered.  Items 1--5 describe topology based cuts using TPC information only.  Items 6--8 use precise timing information to reject events that are not coincident with the neutrino beam or to eliminate TPC beam triggers that are contaminated by cosmic activity in the detector during the beam spill.  

\begin{enumerate}[label=\arabic*)]

\item $dE/dx$: Pair production events can be rejected with the reconstruction of $dE/dx$ in the initial part of the shower. Preliminary results show that only 6\% of pair conversions present a $dE/dx$ lower than 3.5 MeV/cm in the first 2.5~cm of the shower.
%The corresponding efficiency loss on \nue events is 10\%. 

\item Distance from the muon track: Figure~\ref{fig:dmuon} shows the distance of the cosmogenic photon conversion point from the parent muon track, whenever it also crosses the detector. Rejecting event candidates with a reconstructed vertex inside a cylindrical volume of 15~cm radius around each muon track rejects $>$99\% of the background photons above 200~\MeV. The resulting loss in fiducial volume for the \nue analysis ($\sum_{\mu}\pi R^2L_{\mu}$) is minimal, $\sim$1\% per event on average in the far detector considering all muons in one drift time in one module.  
% 0.7 comes from 0.136 ave per muon times 5.5 muons per T300 per drift

\item Clustering around muon tracks: Rather than a fixed cylindrical volume around tracks as in strategy 2), a variable volume cut around each muon/charged particle can be defined by the zone of connected electromagnetic activity. The ``connection'' is built by walking out from the primary track, clustering hits and gathering clusters together.  This appears to be a very effective cut, however its stability in different wire orientations and noise conditions has to be further established.

\item Activity at the vertex: Requiring the presence of another ionizing track from the vertex would reject all Compton events and a further fraction of the pair production events. 
%bringing the overall residual mis-identification to 3\%. 
However, the same selection on \nue events discards $\sim$25\% of the signal, making this a cut of last resort. %This is not implemented for the moment.
 
\item Backwards distance to the detector wall: This cut was introduced above in our discussion of dirt backgrounds to more efficiently identify showers from photons generated outside of the detector.  Using the reconstructed shower direction in candidate events, one can project back from the vertex and calculate the distance to the nearest TPC boundary in the backward direction.  Since the cosmogenic background is dominated by photons generated by muons inside the active volume, this cut has limited impact and needs further investigation before being applied.  
%a pointing distance of the shower from the detector wall can be calculated
%, as in figure \ref{fig:dwall}, 
%where pointing distance means backwards in the shower direction. This cut helps in identifying backgrounds generated outside the detector. 

\item Scintillation light: Precise event timing information is available through the detection of scintillation light in the liquid argon.  If light detector signals can be matched to the corresponding ionization signals with high efficiency, this would allow a large reduction of backgrounds falling into the timing category {\it B} introduced above. Studies are ongoing to characterize the matching performance and optimize the light collection systems in both \larnd and ICARUS.   
%by identifying the cosmogenic event that occurred within the beam spill time and rejecting the trigger.   A scintillation light system capable of associating different light pulses with different event segments would allow a drastic reduction of the timing category B backgrounds introduced above, and optimization of the light detection systems is progressing for all detectors. %For the time being, this possibility has not been included in the background rejection.

%detector readouts where a cosmic $\mu$ passed near the detector coincident with the beam spill and simply reject the data.  The estimated muon fluxes predict this approach would rejec        
%Note that crossing muons can be identified by the 3D reconstruction software, however the 3D reconstruction needs the information on the event time (the so-called t$_0$). The probability of an autoveto signal by charged particles generated in the neutrino interactions and escaping the \lartpcs has to be carefully evaluated. For a muon tagging system surrounding the detectors $\sim$10 percent of \numu and few percent of \nue events are expected to give signal.

\item  Proton beam spill time structure: Measurement of event times with $\sim$1-2 ns accuracy would enable the exploitation of the bunched beam structure within the spill ($\sim$2 ns wide bunches every 19 ns, see Section \ref{sec:BNB}), to reduce cosmic backgrounds 
%a factor of 3-4 
by rejecting events that occur between bunches. The possibility to reduce the number of bunches by a factor of 2-3, while keeping the same number of protons per spill, will also be investigated in order to further increase the rejection ability using precise timing.

\item Muon tagging: A powerful way to reduce cosmogenic backgrounds would be to employ a cosmic tagging system external to the TPC volume capable of independently measuring the position and time of entering charged tracks.  This information would greatly facilitate the reconstruction and identification of muon tracks in the TPC, leading to a reduction of both type {\it A} and {\it B} background categories.  In the simplest application of this information, an external tagging and tracking system with high (e.g. $>$95\%) coverage of the muon flux that creates potential backgrounds could be used to identify and reject detector readouts when a cosmic $\mu$ passes near the detector during the proton beam spill. Expected fluxes at the detector locations indicate this would reduce the beam data sets by roughly 1.5\%, 2\%, and 3\% at \larnd, \uboone, and ICARUS, respectively, while reducing the cosmic backgrounds in a very clean way.     

\end{enumerate}

\begin{figure}[t]
\centering
\mbox{\raisebox{1.3cm}{\includegraphics[width=0.42\textwidth, trim=0 0 0 0, clip=true]{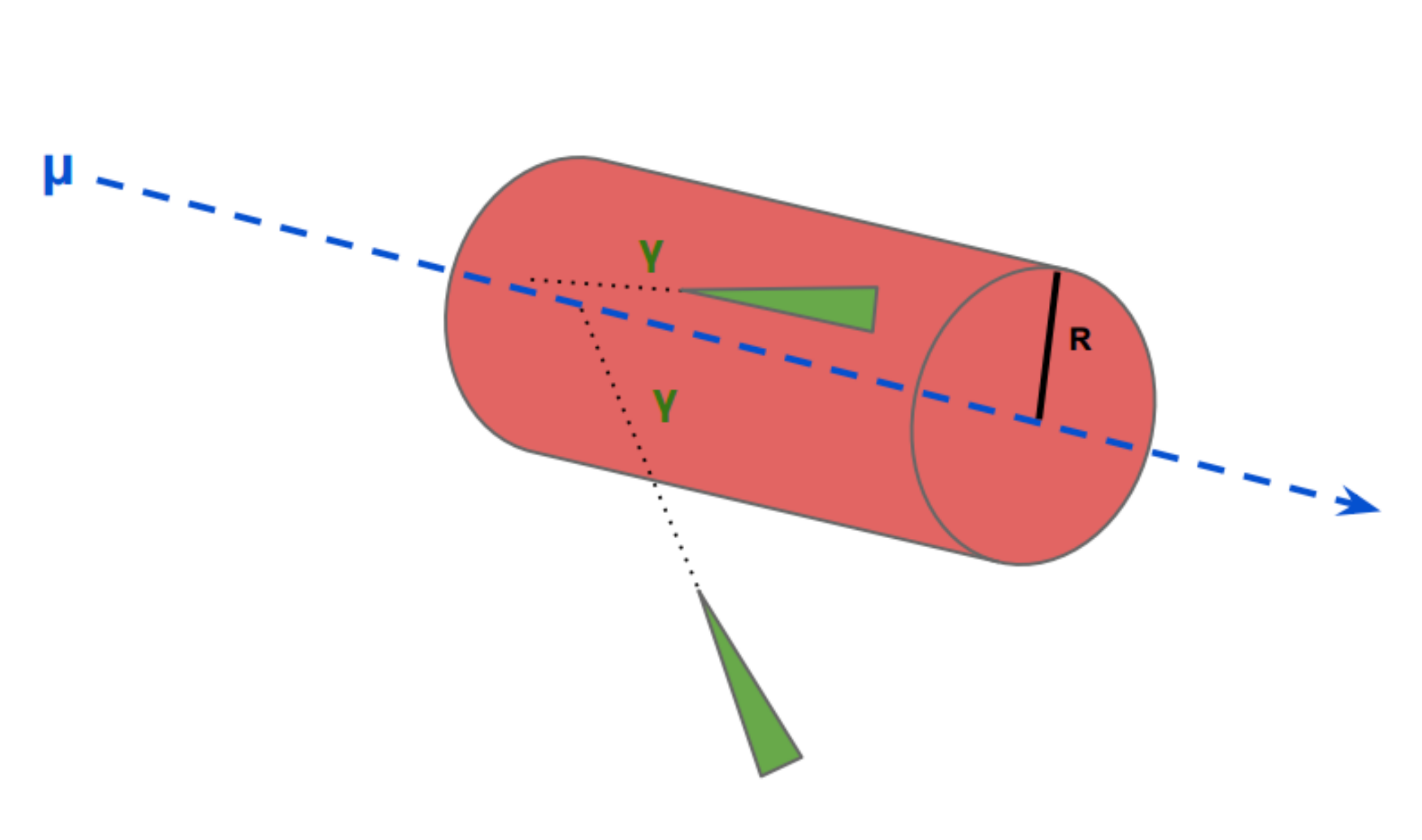}}\quad
\includegraphics[width=0.42\textwidth, trim=0 0 0 10mm, clip=true]{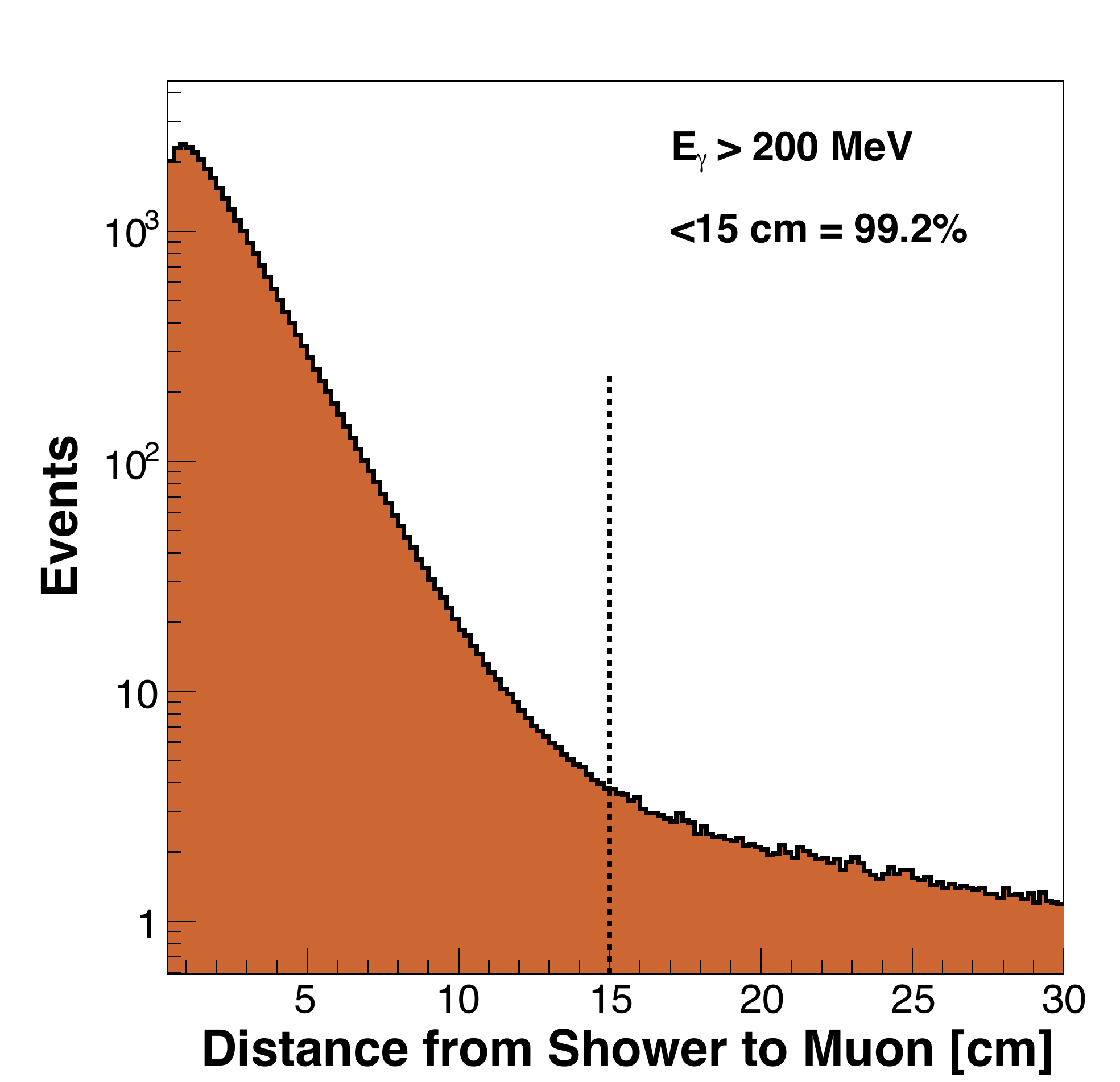}}
\caption{Shortest distance between the conversion point of cosmogenic photons and the parent muon track for photons above 200~\MeV.}
\label{fig:dmuon}
\end{figure}

Table \ref{tab:final_cosmic} illustrates the performance of topological cuts 1) and 2) applied to the Monte Carlo simulations.  In particular, photon showers within 15~cm  of the muon path are rejected and 94\% of $\gamma$ pair production showers are rejected corresponding to a $dE/dx > 3.5$~MeV/cm cut on the first 2.5~cm of the shower.  Remaining background levels in the three detectors (order 100 events) are summarized in Table~\ref{tab:final_cosmic}, which can be directly compared to Table \ref{tab:bgrates} before these cuts.  Also, listed for comparison is the expected numbers of intrinsic \nue CC events.
%and an example \nue signal corresponding to the best fit oscillation parameters from \cite{kopp}.  
In Section \ref{sec:SBN_nue}, we will present predicted event distributions when using these topological cuts, as well as illustrate the power of augmenting these with external muon tagging and timing selections. 

%Note that for the real data analysis, the cosmic background and the selection efficiencies will be measured with random triggers in between spills or in beam off periods.  

\begin{table}[h!]
\centering
\caption{Background rates, after topological cuts, assuming 3 years of data taking for a total of $6.6 \times 10^{20}$ protons on target, delivered in $1.32 \times 10^{8}$ beam spills equaling 211 seconds of beam time. The cuts that have been applied relative to Table~\ref{tab:bgrates} are (distance from the $\mu$ track) $<$ 15~cm and $dE/dx >$ 3.5~MeV/cm. %, while the total number of events is counted for all energy depositions above 100 MeV in the active volume.}
}
\label{tab:final_cosmic}
% T600:0.960 msec drift, 
% 1.3210^8 beam triggers 
%211.2 sec beam time
\begin{tabular}{|c|l|c|c|c|c|c|}
\hline
\multicolumn{2}{|c|}{} & & & \multicolumn{3}{|c|}{\small $E_{\gamma}>200~\MeV$, Pair prod} \\
%Total Cosmogenic Triggers ($E_{\text{dep}}>100~\MeV$) & & &  $2.5\cdot 10^6$  \\
%Total Cosmic Muons
%& & &  \\
%\hline\hline
\multicolumn{2}{|c|}{Interaction description} & Timing & Topology & \multicolumn{3}{|c|}{\small $E_e>200~\MeV$, Compton, \nue} \\
\multicolumn{2}{|c|}{} & Cat. & Cat. & \multicolumn{1}{|c}{~\larnd~} & \multicolumn{1}{|c|}{~$\mu$BooNE~} & \multicolumn{1}{c|}{~ICARUS~} \\ \hline\hline
1 & $\gamma$  Compton in spill, primary $\mu$ enters AV & A & I
& 8  	& $<$3  & $<$4  \\
2 & $\gamma$  Pair prod in spill, primary $\mu$ enters AV & A & I
%&52337	& 11591 & 32000  \\
&26	& 6 & 21  \\  %19
3 & $\gamma$  Compton in spill, primary misses AV & A & II
%& 0.95& 2.48 &  $<$4  \\
& $<$1 & $<$3 &  $<$4  \\
4 & $\gamma$  Pair prod in spill, primary misses AV & A & II
%& 55.3 & 82.0 & 11  \\ \hline
& $<$4 & 6 & $<$1  \\ \hline
5 & $\gamma$  Compton in drift, primary $\mu$ enters AV & B & I
%& 2254 & 1030 &  3300   \\
& 22 & 12 &  30  \\  %33
6 & $\gamma$  Pair prod in drift, primary $\mu$ enters AV & B & I
%&132908 & 57956 & 176000  \\
&74 & 29 & 113  \\
7 & $\gamma$  Compton in drift, primary misses AV & B & II
%& 2.40 & 12.4 & $<$4  \\
& $<$3 & 12 & $<$4  \\
8 & $\gamma$  Pair prod in drift, primary misses AV & B & II
%&140.4 & 410 & 60  \\ \hline
& 10 & 19 & $<$4  \\ \hline\hline
& Total Cosmogenic $\gamma$ backgrounds & & 
& 146 & 88 & 164\\
%Intrinsic \numu CC events
%& $2.2 \cdot 10^6$ & $1.2 \cdot 10^5$ & $5.4\cdot 10^5$  \\
& Intrinsic \nue CC & &
& 15,800 & 413 & 1,500  \\
%& Signal \nue CC (\dmsq = 0.43 \eVsq, \sinth = 0.013) & &
%& 140 & 84 & 615  \\
\hline
\end{tabular}
\end{table}

%\FloatBarrier

\subsubsection*{An Illustration of Cosmogenic Rate Reductions in ICARUS}

As an illustration of the detector capabilities in rejecting the cosmogenic background, the external muon tagging system and event matching to the proton spill time structure (introduced  in \cite {CRubbia-1,CRubbia-2}) can be applied in the first stages of the data selection in order to achieve an effective reduction of the data amount to be fully analyzed  (Table~\ref{tab:cosmic_reduction}).

From the previous calculations of the cosmic ray flux  impinging on the T600 detector (see Table~\ref{tab:bgrates}), the predicted number of triggers produced by cosmics inside the $1.6~\mu$s beam spill is globally $\sim 2.5 \times 10^6$ events.

The adoption of a full coverage external muon tagging system with a 95\% detection efficiency at each muon crossing can directly reduce the number of triggers produced by cosmic rays to $\sim 2.4 \times 10^4$ events. These resulting surving events come mainly from the $\sim 15 \%$ fraction of muons either coming to rest or decaying inside the detector since, in the case of double crossing of muons entering and exiting the \lartpc, the survival probability is considerably smaller ($\sim 0.25 \%$). As a consequence the cosmogenic background events of the ``Timing category A'' in  Table~\ref{tab:bgrates} associated to the primary crossing muons are suppressed down to $0.25 \%$.  Events of category B, triggered by the passing/stopping muons not identified by the tagging system, are instead reduced only to below 1 \% level ($0.96 \%$). 
As a result, the predicted fraction of $\gamma$-ray conversions per imaging picture is evaluated from the Timing Category A, Topology I and Timing Category B rescaled for the factors 0.0025 and 0.0096 respectively, is about 0.075.

Assuming a factor 3 of reduction can be achieved from the exploitation of the beam spill time structure, a total of 8020 events are retained, out of wich  $\sim 600$ contain a converting $\gamma$ with $E > 200$ MeV (event topology I). As described above, only $1 \%$ of converting $\gamma$'s accompanied by a visible muon will satisfy the requirement of a minimal distance of 15~cm of the photon conversion from the muons, leaving $\sim 6$ events.

In addition, from Table~\ref{tab:bgrates}, $<$4 $\gamma$-ray conversion ($E > 200$ MeV) events are expected without a visible muon in the TPC's  (event topology II) in time with the bunched beam structure and under the conservative assumption that only $50 \%$ of them are recognized by the tagging system. Therefore the surviving 9 event sample is further reduced to $\sim 1$ events 
by the reconstruction of the $dE/dx$ in the initial part of the shower.

If not properly recognized, neutrino beam interactions at low energy occurring both in active volume and in the external LAr could mimic a cosmogenic trigger in time with the beam spill. In such a case cosmogenic photons inside the drift time are expected to contribute to the background escaping any mitigation effect from the muon tagging system and the precise time matching with the beam spill structure. A rough conservative estimation of $\sim 3 \times 10^5$ events will result in 18 events satisfying the previous selection criteria, namely the requirement of the minimal 15 cm distance from muon and $dE/dx$ identification.

The resulting total $\sim 19$ background events could be further reduced to $\sim 5 $ events under the conservative assumption that the scintillation light system is capable to localize the triggering event within $\sim 4$ m along the beam direction.  

\begin{table}[t]
\caption{Expected background event reduction in the T600 detector exploiting the muon tagging system and the beam spill time structure.  Event topology I refers to events with a muon track crossing the active TPC volume and Event Topology II refers to events with no visible muon in the TPC. The contribution from the non-identified neutrino interactions is also added.}
\label{tab:cosmic_reduction}
\begin{tabular}{|p{0.72\linewidth}|cc|}
\hline
& \multicolumn{2}{c|}{Cosmic Background}\\
& \multicolumn{2}{c|}{Events} \\
\hline
Total cosmic events in beam spills (211 sec. total) 
		& \multicolumn{2}{c|}{$2.5 \times 10^6$} \\
Cosmic triggers after the tagging system  
		& \multicolumn{2}{c|}{$2.4 \times 10^4$ }\\
Surviving events after the spill structure exploitation  
		& \multicolumn{2}{c|}{8020} \\ \hline
& \multicolumn{2}{c|}{Event Topology} \\				  
&~~~~\underline{~~I~~}~~~~&~~~~\underline{~~II~~}~~~~  \\   
$\gamma$ conversions  &600 & 3 \\
After distance from muon cut (15 cm)    &6& 3\\ \hline
Remaining cosmogenic backgrounds after $dE/dx$ cut & \multicolumn{2}{c|}{1}\\
\hline
Remaining cosmogenic background in non-identified BNB $\nu$ interactions & \multicolumn{2}{c|}{18}\\
\hline
Total cosmogenic background after scintillation light exploitation & \multicolumn{2}{c|}{5}\\
\hline
\end{tabular}
\end{table}

The explicit request of absence of muon tagging in the event and the precise time matching with the bunched spill structure are expected to slightly reduce the \nue CC event acceptance by $\sim 3 \%$. 
%As already described, t
The 200 MeV electron energy threshold will result in a reduction  of $\sim 10 \%$ on the electron signal acceptance, while the corresponding reduction for a request of a minimal distance of the event vertex from the cosmic muon tracks is almost negligible, $0.7 \%$ on average.

%A similar cosmic background reduction level is expected for both the for \uboone and \larnd detectors. 

\subsubsection*{Error Matrix Contribution from Cosmogenic Backgrounds}

We now require an estimate of the error matrix associated with the predicted cosmogenic backgrounds, $E^{\text{cosmics}}$, to be used in the sensitivity analysis as discussed in Section~\ref{sec:Analysis}.

It is important to emphasize that the most important outcome at this time is to understand the approximate scale of the cosmogenic backgrounds in the experiment.  The exact rate does not introduce significant \emph{systematic} uncertainty because it will be measured with high precision using off-beam random event triggers. This is a critical aspect of the experiment, and designing the DAQ systems to record sufficient random triggers must be considered. For the sensitivity analysis, we construct the cosmic error matrix to account for the statistical uncertainty on the predicted sample, $E^{\text{cosmic}}_{ii} = N^{\text{cosmic}}_{i}$ and $E^{\text{cosmic}}_{ij} = 0$ for $i \neq j$.

%\begin{equation}
%E^{\text{cosmic}}_{ij} = \delta_{ij}(N^{\text{cosmic}}_{i})( N^{\text{cosmic}}_{j}).
%\label{eq:dirt}
%\end{equation}

The key thing to demonstrate at this time is that cosmic backgrounds can be reliably reduced to a level where oscillation signals can be observed over them with sufficient statistical significance.  The fact that the independent detector simulations have come together to produce predictions consistent within a factor of two gives confidence that the scale is correctly estimated.  Common simulations will, of course, be used in the actual analysis, but more importantly the background levels will be tightly constrained \emph{with data} in each detector.   

%, or $S/\sqrt{B}$ where $S$ is the number of signal events and $B$ is the total number of background events expected.     

\subsection{\numunue Appearance Sensitivity}
\label{sec:SBN_nue}

We are now ready to bring together the background predictions and uncertainty estimations detailed in the previous Sections to construct the experimental sensitivity to a \numunue oscillation signal.  Figure \ref{fig:nue_sigs} shows the full \nue background predictions in each detector, including intrinsic \nue beam events, neutral-current and \numu CC mis-IDs, out-of-detector beam related ``dirt'' backgrounds, and cosmogenic photon induced electromagnetic shower backgrounds. For comparison, a sample \numunue oscillation signal is also shown for each detector location corresponding to the best-fit parameters from the Kopp et al. analysis~\cite{kopp} of \dmsq = 0.43~\eVsq and \sinth = 0.013. 

On the left in Figure~\ref{fig:nue_sigs} is shown the result when using the topological cuts 1) and 2) described in Section~\ref{sec:Cosmics} to reduce cosmic backgrounds. This analysis, using $dE/dx$ information at the vertex and the 15 cm cylinder cut around crossing muons, demonstrates the power of TPC information alone in reducing these backgrounds, rejecting more than 99\% of cosmogenic photons when their parent muon is also visible in the TPC.  However, as can be seen in the figures, the cosmogenic backgrounds remain a large contribution to the analysis, particularly at low energies, and additional hardware-based systems that can initially reduce the data sample in a very clean way are considered important additions to guarantee the success of the experiment.    
  
\begin{figure}[th]
\centering
\mbox{\includegraphics[width=0.46\textwidth,trim=0mm 0mm 0mm 10mm,clip]{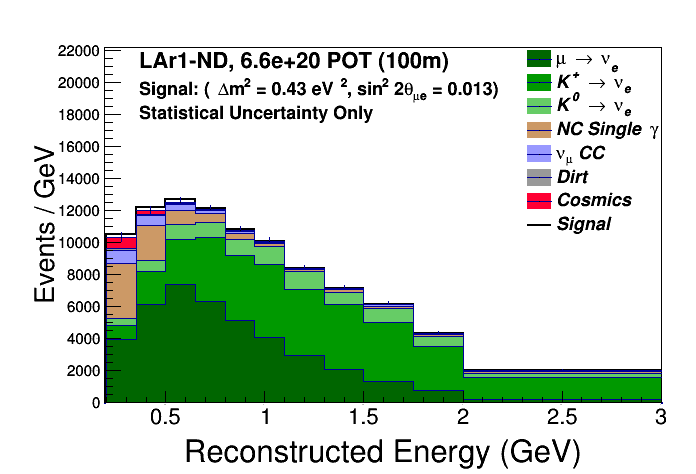}
\includegraphics[width=0.46\textwidth,trim=0mm 0mm 0mm 10mm,clip]{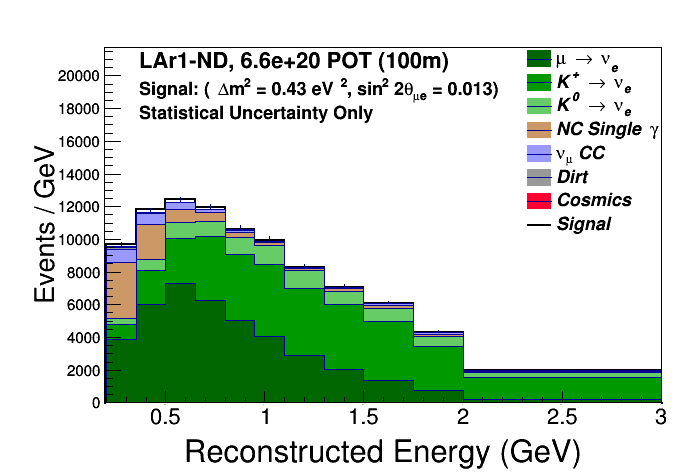}}
\mbox{\includegraphics[width=0.46\textwidth,trim=0mm 0mm 0mm 10mm,clip]{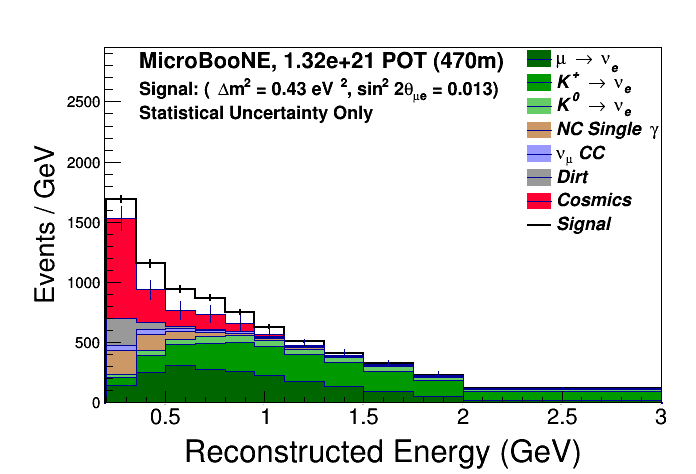}
\includegraphics[width=0.46\textwidth,trim=0mm 0mm 0mm 10mm,clip]{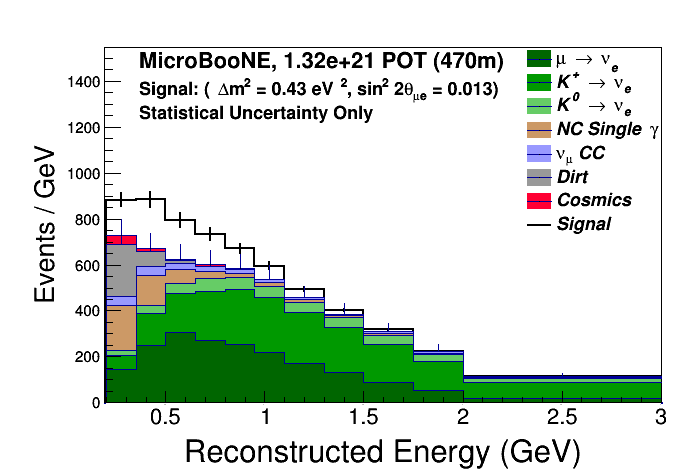}}
\mbox{\includegraphics[width=0.46\textwidth,trim=0mm 0mm 0mm 10mm,clip]{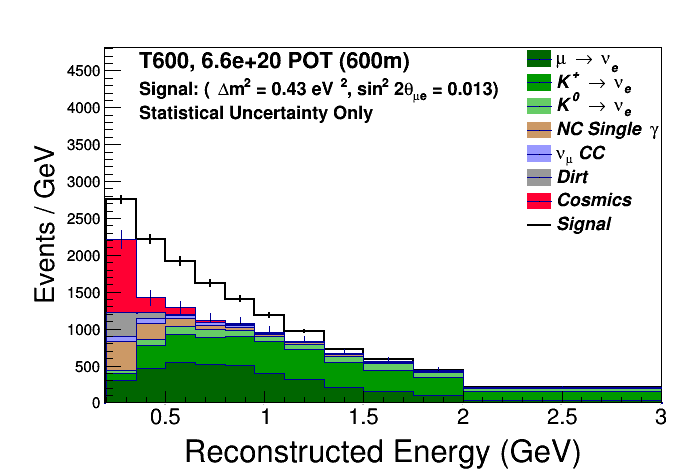}
\includegraphics[width=0.46\textwidth,trim=0mm 0mm 0mm 10mm,clip]{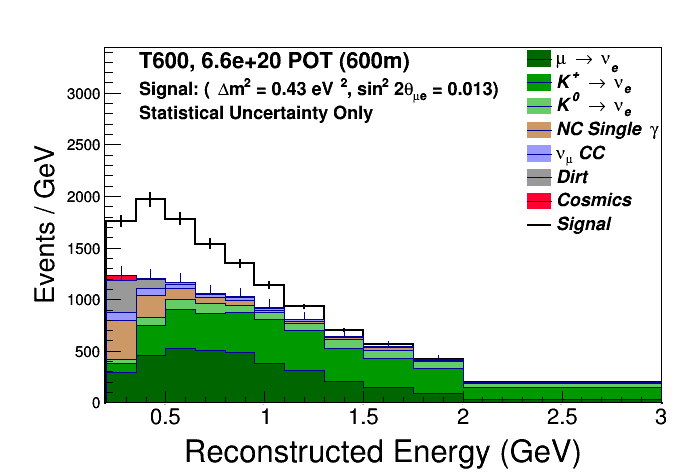}}
\caption{Electron neutrino charged-current candidate distributions in \larnd (top), \uboone (middle), and \icarus (bottom) shown as a function of reconstructed neutrino energy.  All backgrounds are shown.  In the left column, only muon proximity and $dE/dx$ cuts have been used to reject cosmogenic background sources.  In the right column, a combination of the internal light collection systems and external cosmic tagger systems at each detector are assumed to conservatively identify 95\% of the triggers with a cosmic muon in the beam spill time and those events are rejected.   Oscillation signal events for the best-fit oscillation parameters from Kopp et al. \cite{kopp} are indicated by the white histogram on top in each distribution.}
\label{fig:nue_sigs}
\end{figure}

The right column of Figure~\ref{fig:nue_sigs} demonstrates the potential improvement when employing additional hardware solutions such as  those introduced in Section~\ref{sec:Cosmics}.  Precise timing information, in particular, can augment the TPC data by rejecting triggers where the 1.6 $\mu s$ beam spill time is contaminated by a cosmic event in the detector. To generate the right hand distributions of Figure~\ref{fig:nue_sigs}, it is assumed that the combination of strategies 6--8 from Section~\ref{sec:Cosmics} are applied to remove 95\% of cosmogenic events in the first stages of data analysis, before entering into automated reconstruction and event selection algorithms. Given the dominance of muons passing very near the detectors as the source of cosmic backgrounds, most of this reduction should be straightforwardly achievable with a properly implemented external tracking system. Further rejection capabilities will come from precise event timing information from the internal scintillation light collection systems. The required level of rejection can be taken as a design requirement on these systems, determining the necessary coverage of the external tagging systems and the time resolution of the internal light collection, but this factor of 20 reduction is a fairly conservative estimate of the power of strategies 6--8 combined.   
 
Table \ref{tab:nueevents} lists the integrated event totals represented in the histograms of Figure~\ref{fig:nue_sigs}. The 20$\times$ reduction from additional cosmic tagging discussed above is indicated in parenthesis.  Vetoing of events with cosmic activity in the beam spill using timing results in a reduction of all beam related event categories of $\sim$1.5\%, 2\%, and 3\% in \larnd, \uboone, and ICARUS, respectively.  This reduction is not shown in the Table (for clarity) but is accounted for in Figure~\ref{fig:nue_sigs} (right) and in the final sensitivity.  One thing to note in Table \ref{tab:nueevents} is that the event counts listed for Dirt and Cosmogenic events are larger than those given in Sections \ref{sec:Dirt} and \ref{sec:Cosmics}. This is a result of energy smearing effects which are properly simulated in the final sensitivity analysis ($15\%/\sqrt{E}$), but not in the earlier stages of simulations where true energies were used to display the predictions.  The predicted background energy spectra are provided well below the 200~MeV cutoff value used in the analysis such that events can be properly smeared in both directions.  Because both backgrounds are steeply falling functions of photon energy, more events smear into the analysis range than smear out.  This is properly handled in the analysis and leads to an increase in event count relative to the earlier values which cut on generated photon energies.        

\begin{table}[t]
\begin{center}
\caption{\label{tab:nueevents}Event rates in the \nue charged-current candidate sample in the range 200--3000~\MeV reconstructed neutrino energy for \pot protons on target in \larnd and the \icarus and \potuB protons on target in \uboone. The numbers listed correspond to the application of topological cuts 1) \& 2) for reducing cosmogenic backgrounds. In parentheses are indicated the reduced cosmogenic background rate when a 95\% efficient time-based ID system is used to reject contaminated triggers.  Vetoing of these events results in a reduction of all beam related event categories of 1.5\%, 2\%, and 3\% in \larnd, \uboone, and ICARUS, respectively, which is not shown (for clarity) but is accounted for in Figure~\ref{fig:nue_sigs} and \ref{fig:nue_sens}.}
\begin{tabular}{l c c c}\hline\hline
	 									& ~\larnd~ 		& ~\uboone~ 	& ~\icarus~ \\ 
        & {\small ~\pot p.o.t.~}& {\small ~\potuB p.o.t.~} & {\small ~\pot p.o.t.~} \\ \hline 
$\mu \rightarrow \nue$					& 6,712			&  338			&  607  	\\
$K^{+} \rightarrow \nue$				& 7,333 		&  396			&  706  	\\
$K^{0} \rightarrow \nue$				& 1,786			&  94			&  180  	\\
NC $\pi^{0} \rightarrow \gamma\gamma$	& 1,356			&  81			&  149		\\
NC $\Delta \rightarrow \gamma$			& 87			&  5			&  9		\\
\numu CC								& 484			&  35			&  51		\\
Dirt events								& 44			&  47			&  67		\\
Cosmogenic events\footnote{These predictions exclude a small correction from the case where an unidentified neutrino interaction provides the scintillation trigger, as discussed in Section~\ref{sec:Cosmics}.}		
										& 170 (9)		&  220 (11)		&  204 (10)	\\ \hline
Signal (\dmsq = 0.43 \eVsq, \sinth = 0.013) \cite{kopp}	& 114	&  136		&  498	\\ \hline\hline
\end{tabular}
\end{center}
\end{table}

%Figure~\ref{fig:nue_sens} presents the experimental sensitivity of the proposed Fermilab SBN program to \numunue appearance signals. As discussed in Section~\ref{sec:Steriles}, a 3+1 model is used as the basis for quantifying the sensitivity, so we present the result in the (\dmsq, \sinth) plane and compare to the original LSND allowed region \cite{lsnd} and two recent 3+1 global data fit results \cite{laveder, kopp}.

Figure~\ref{fig:nue_sens} presents the experimental sensitivity of the proposed Fermilab SBN program to \numunue appearance signals in the (\dmsq, \sinth) plane compared to the original LSND allowed region \cite{lsnd}. 
%and two recent 3+1 global data fit results \cite{laveder, kopp}.
%Two sensitivities are shown, corresponding to the analysis with and without the additional 95\% cosmic background rejection coming from timing information described above. 
The sensitivity shown includes the additional 95\% cosmic background rejection coming from timing information described above and illustrated on the right in Figure~\ref{fig:nue_sigs}. We compare this to the case using only TPC topology cuts to identify cosmogenic events below.
%On the left is the sensitivity with only TPC topology cuts to identify cosmogenic events, and on the right is with the additional rejection.  The LSND 99\% C.L. allowed region is covered at almost the $5\sigma$ level in the case where only topological cuts are used. There is a clear improvement with the additional cosmogenic background rejection shown on the right with part of the LSND region covered at $>5\sigma$ level.  In both cases, one sees a clear reduction in the sensitivity at low \dmsq where oscillations occur at lower energies - exactly where cosmic backgrounds populate. 
The LSND 99\% C.L. allowed region is covered at the $\geq 5\sigma$ level above $\dmsq = 0.1~\eVsq$ and $> 4.5\sigma$ everywhere. Note that the region below $\dmsq = 0.1~\eVsq$ is already ruled out at more than $5\sigma$ by the previous results of ICARUS at Gran Sasso (see Figure~\ref{fig:sterile_fits}).
%There is a clear improvement with the additional cosmogenic background rejection shown on the right with part of the LSND region covered at $>5\sigma$ level.

%While the sensitivity seems fairly robust to this level of cosmic contamination, it is important to note that the proposed additional measures to mitigate these backgrounds further are strongly motivated. The low energy region is a particularly important region to explore, especially given the anomalies reported by the \MB experiment~\cite{mbosc2,mbcomb}. 
%While the 3+1 sensitivity seems fairly robust to this level of cosmic contamination, it is important to note that this is only in the context of a specific model and there are other motivations for taking additional measures to mitigate these backgrounds further as in the second case.  The low energy region is a particularly important region to explore, especially given the anomalies reported in that region by the \MB experiment~\cite{mbosc2,mbcomb}.

The sensitivity results presented in Figure~\ref{fig:nue_sens} incorporate all background sources and related uncertainties described in this proposal except detector related systematics as introduced in Section \ref{sec:DetSyst}.  Each of the rate predictions and other systematic uncertainties (i.e. flux and cross section) in the analysis are built using advanced, sophisticated simulation programs, while current estimates of detector related systematics come from hand scanning of events, empirical experience with these and other detectors, or toy Monte Carlo studies, and so are difficult to incorporate with the same sophistication at this time.  Instead, studies to investigate the level of uncorrelated detector systematics that can be tolerated while preserving the experimental sensitivity 
%and the capability to cover at the 5$\sigma$ level the LSND allowed parameter region 
have indicated that total uncertainties in the 2--3\% range are acceptable.  All studies performed to date suggest these can be well controlled for a multi-detector experiment, with individual studies coming in at $\leq1\%$ (see Section \ref{sec:DetSyst}).  

\begin{figure}[t]
\centering
%\mbox{\includegraphics[width=0.5\textwidth, trim=5mm 10mm 10mm 10mm, clip]{ND_100m_uB_T600_onaxis_nue_appearance_ecalo2_nu_vePhot0_05_gap3_fullCosmics_xsec_0_flux_6_dirt_cos_sensPlot.png}
\includegraphics[width=0.8\textwidth, trim=5mm 10mm 10mm 10mm, clip]{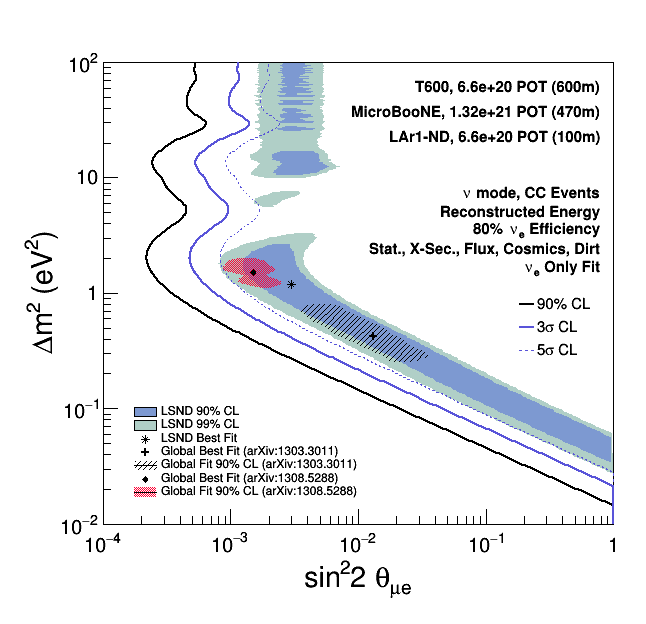}
%}
\caption{Sensitivity of the SBN Program to \numunue oscillation signals. All backgrounds and systematic uncertainties described in this proposal (except detector systematics, see text) are included.  The sensitivity shown corresponds to the event distributions on the right in Figure~\ref{fig:nue_sigs}, which includes the topological cuts on cosmic backgrounds and an additional 95\% rejection factor coming from an external cosmic tagging system and internal light collection system to reject cosmic rays arriving at the detector in time with the beam.}
\label{fig:nue_sens}
\end{figure}

In Figure~\ref{fig:nue_sens_compare}, we present the sensitivity in a different way that facilitates easier comparison between different results. Rather than displaying fixed confidence level contours ($90\%, 3\sigma, 5\sigma$) in the (\dmsq, \sinth) plane, we plot the significance with which the experiment covers the 99\% C.L. allowed region of the LSND experiment as a function of \dmsq.  The curves are extracted by asking what $\chi^2$ value the analysis produces at each point along the left edge of the 99\% C.L. LSND region.  The gray bands correspond to \dmsq ranges where LSND reports no allowed regions at 99\% C.L.

Two versions of this plot are shown in Figure~\ref{fig:nue_sens_compare}. The top presents the significance at which the LSND region would be covered for the different possible combinations of SBN detectors: \larnd+ \uboone only (blue), \larnd+ ICARUS only (black), and all three detectors in combination (red).  This presentation makes clear the contributions of the \uboone and \icarus detectors as far detectors in the oscillation search. The presence of the large mass added by the \icarus detector is imperative to achieving 5$\sigma$ coverage.  In addition, \uboone, by starting to run several years earlier, makes a valuable contribution particularly in the important 1~\eVsq region.

The bottom plot shows the full program sensitivity compared to the result when only topological information from the TPC is used to reject cosmogenic backgrounds as described above.  This is equivalent to comparing the cases depicted in the left and right columns of Figure~\ref{fig:nue_sigs}. The backgrounds at low energy impact the low-\dmsq region most and the ability to further suppress cosmic backgrounds through precise timing information clearly represents an improvement in the sensitivity at low \dmsq of about 0.75$\sigma$.   

We note that the selection criteria used here for the rejection of the dirt and cosmogenic backgrounds were chosen to illustrate the sensitivity of the proposed program in a conservative way.  For example, the fiducial cut on dirt backgrounds was chosen to aggressively remove this background but also reduces the fiducial volume (signal statistics) significantly.  A detailed optimization would likely result in a looser fiducial cut, allowing for an increase in signal statistics.  Similarly, a conservative rejection factor of 20 has been assumed for the combination of an external veto system and timing from the light detection systems.  Achieving a higher rejection factor from these systems would have significant positive impact on the program.

Finally, we note that this is a statistically limited measurement.  An increase in the number of neutrino interactions either through delivery of more protons to the target or a more efficient target and horn system would greatly benefit the program.  For this reason, we propose the further study of BNB improvements like those described in Part~V of this proposal. 

\begin{figure}[h!]
\centering
\hspace*{-0.03\textwidth}
\includegraphics[width=1.1\textwidth,trim=0mm 0mm 0mm 0mm,clip]{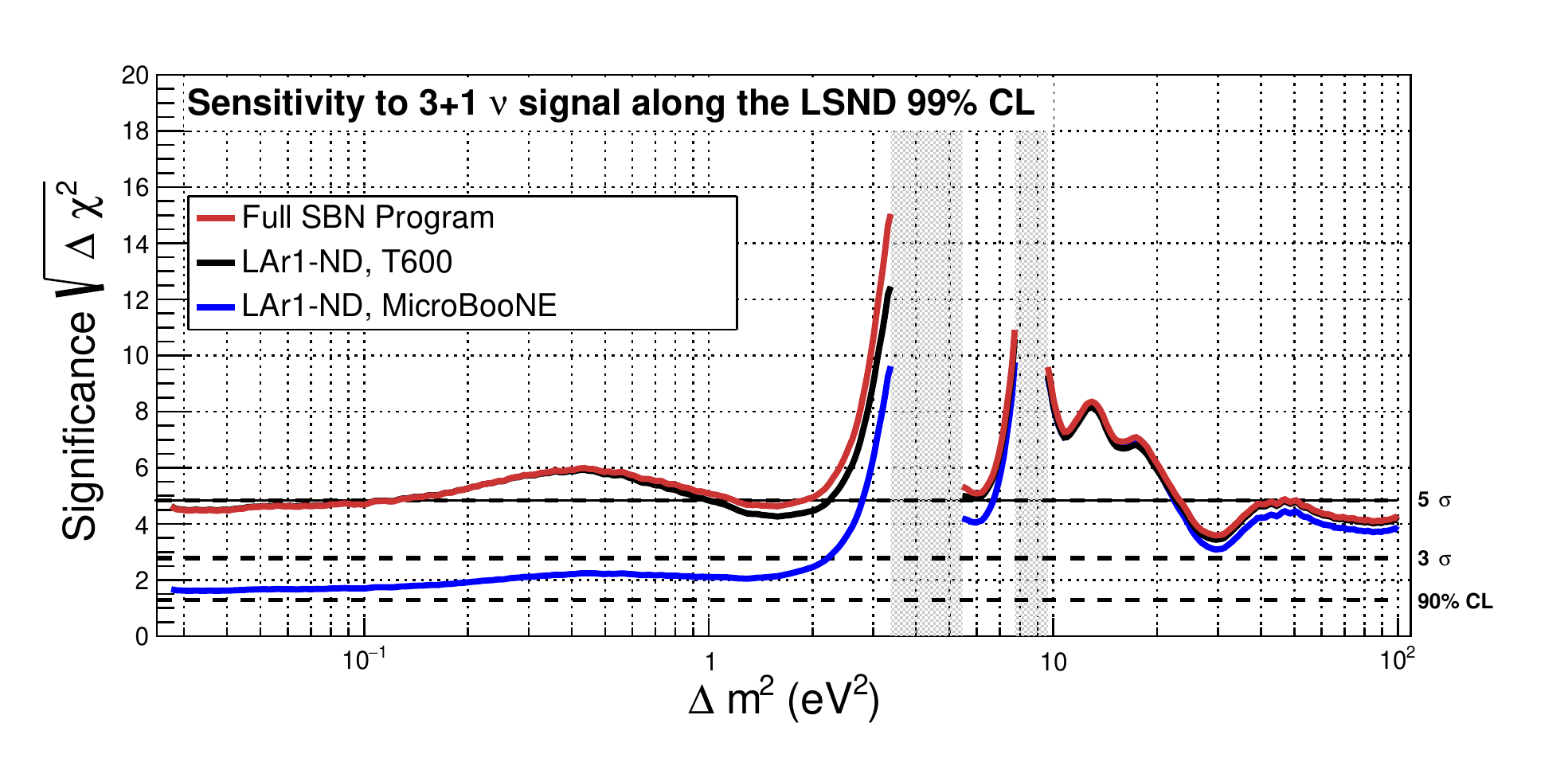}
\hspace*{-0.03\textwidth}
\includegraphics[width=1.1\textwidth,trim=0mm 0mm 0mm 0mm,clip]{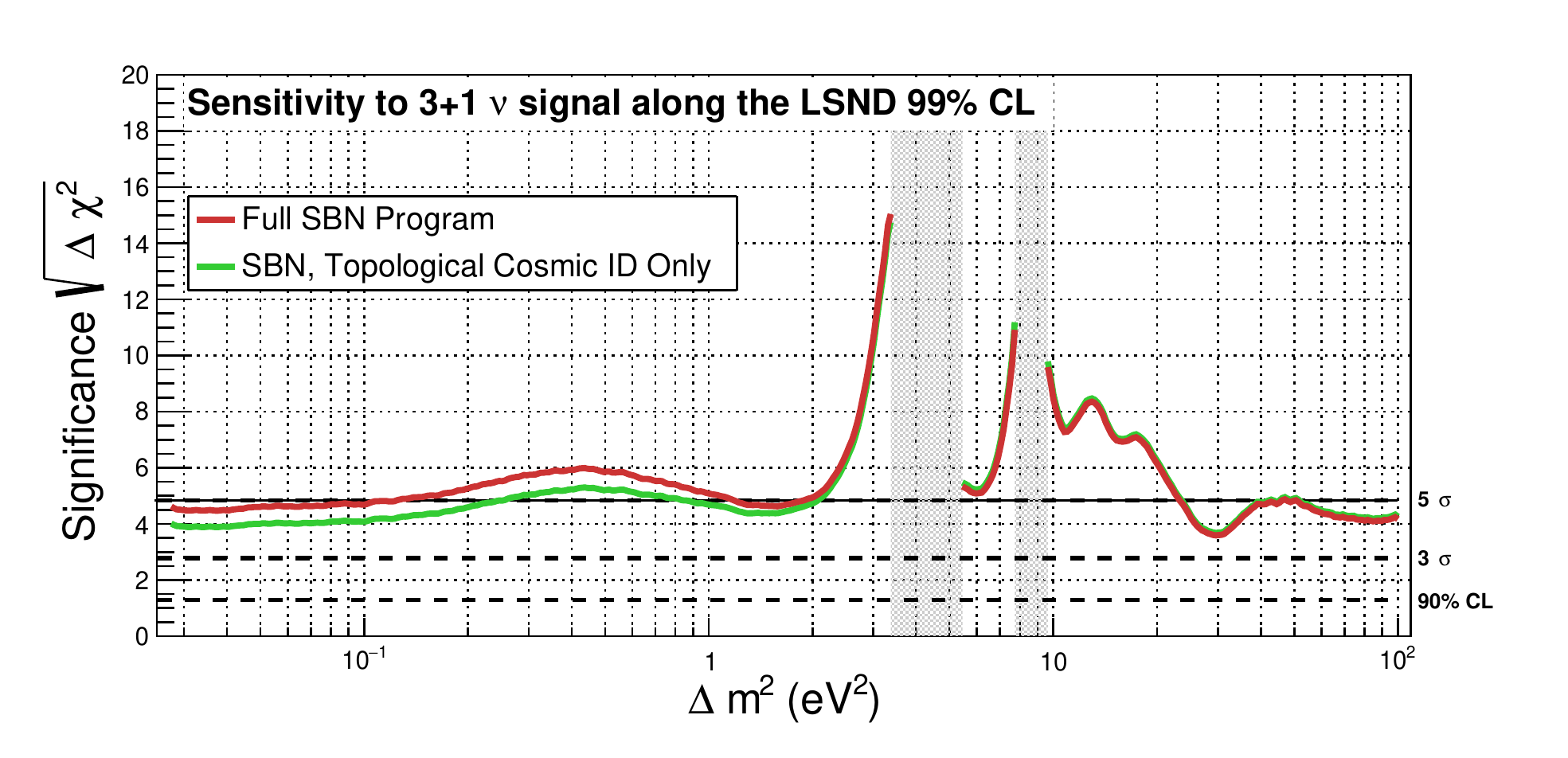}
\caption{Sensitivity comparisons for \numunue oscillations including all backgrounds and systematic uncertainties described in this proposal (except detector systematics, see text) assuming \pot protons on target in \larnd and the \icarus and \potuB protons on target in \uboone.  (Top) The three curves present the significance of coverage of the LSND 99\% allowed region (above) for the three different possible combinations of SBN detectors: \larnd+ \uboone only (blue), \larnd+ ICARUS only (black), and all three detectors (red). (Bottom) Comparison of the sensitivity with only topological cosmic background rejection and with additional suppression from timing information (see text).}
\label{fig:nue_sens_compare}
\end{figure}

\FloatBarrier

\subsection{\numudis Disappearance Sensitivity}
\label{sec:SBN_numu}

The \numu disappearance sensitivity for the SBN Program is also estimated.  The background evaluation is not as complete as for the \nue analysis, in particular possible contributions from dirt or cosmogenic sources are not considered, but they are expected to be small compared to the high \numu CC rate.  The critical aspects to this evaluation are the neutrino flux and interaction model uncertainties described in Sections \ref{sec:Flux} and \ref{sec:XSec}. The absolute flux and cross section uncertainties in any detector along the BNB are larger than 10\%, but the high correlations between the near detector and the \uboone/\icarus event samples along with the excellent statistical precision of the \larnd measurements will make the SBN program the most sensitive \numu disappearance experiment at \dmsq $\sim$1 \eVsq.

\begin{figure}[ht]
\centering
\mbox{\includegraphics[width=0.33\textwidth]{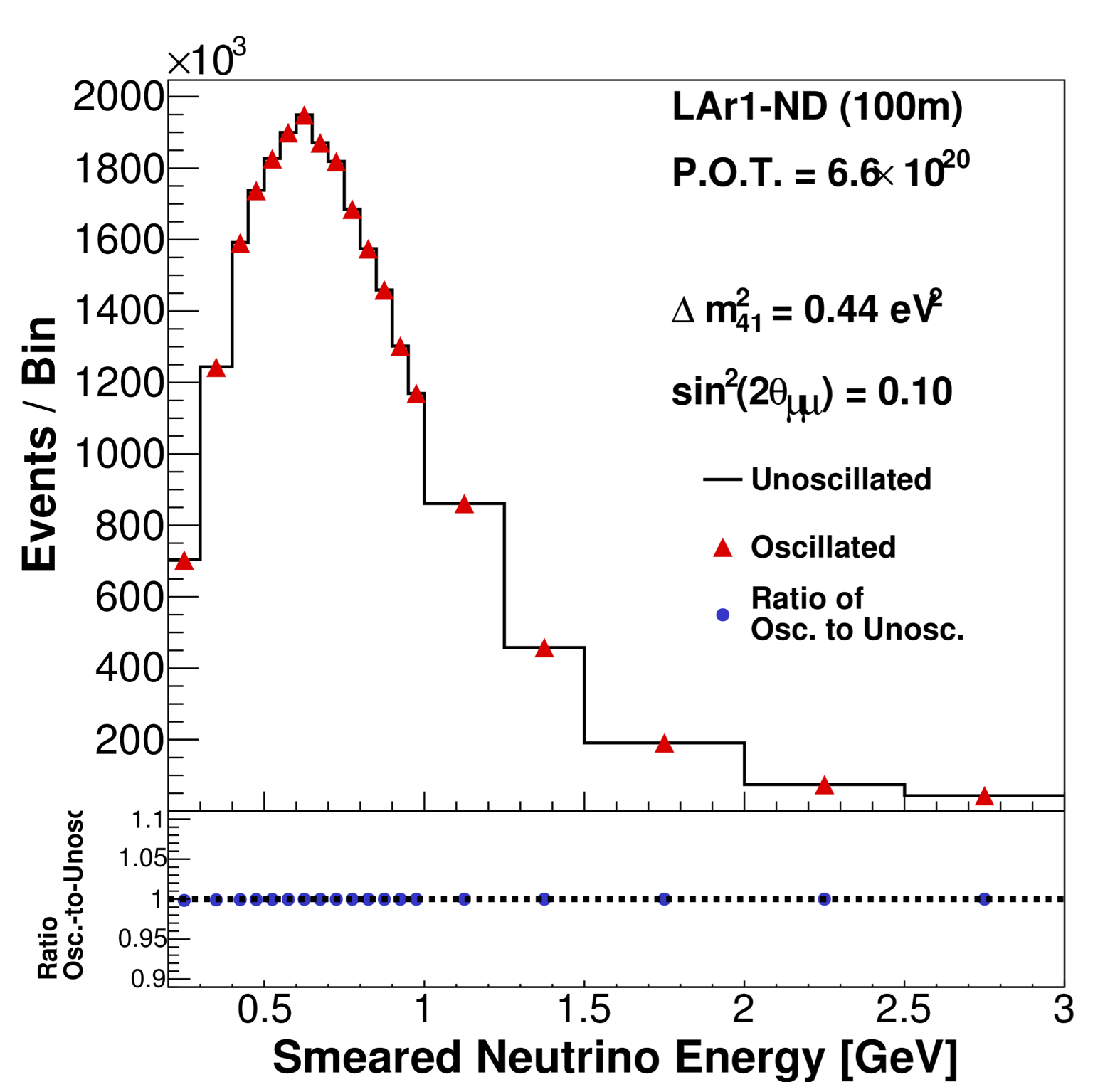}
\includegraphics[width=0.33\textwidth]{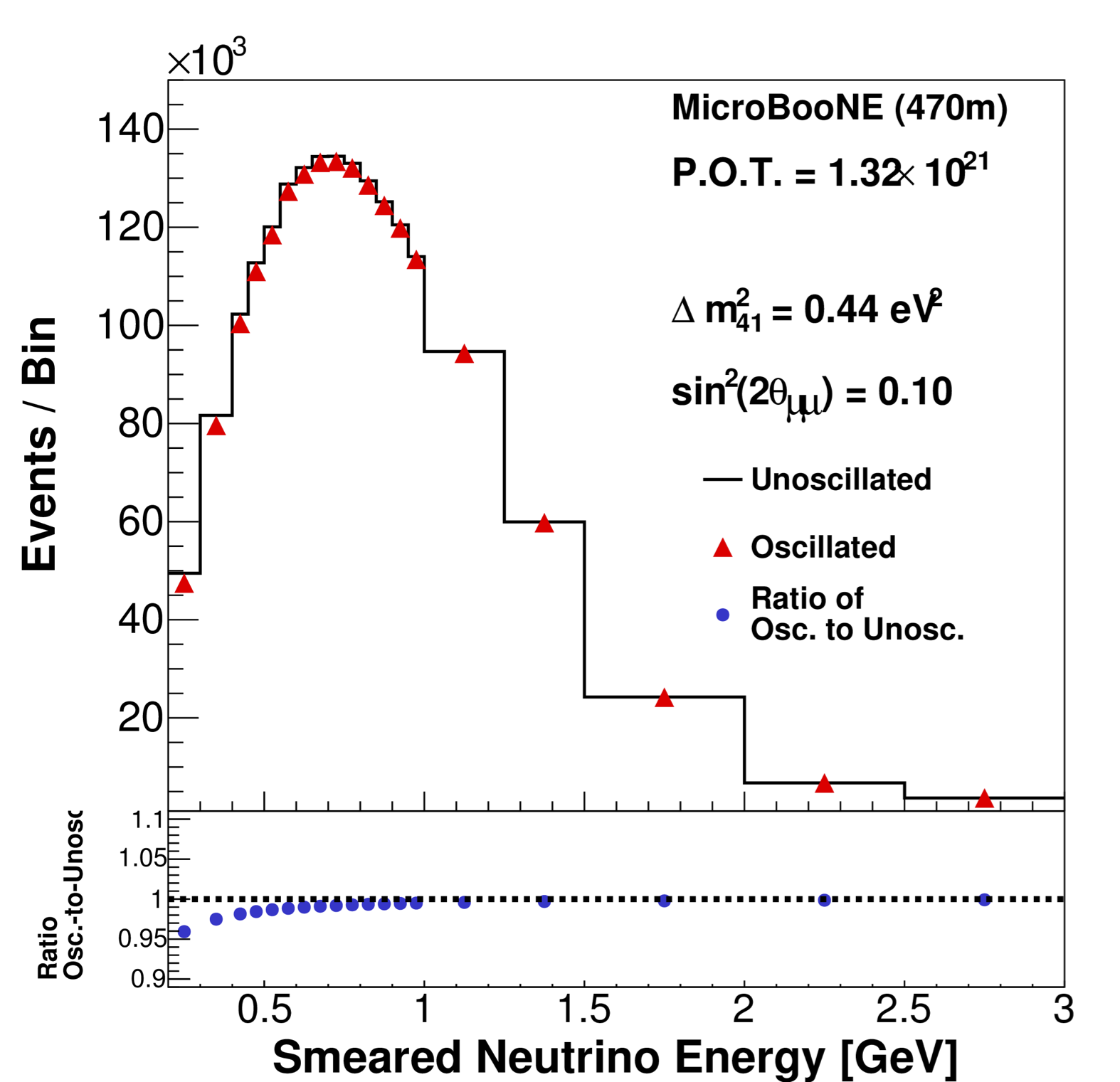}
\includegraphics[width=0.33\textwidth]{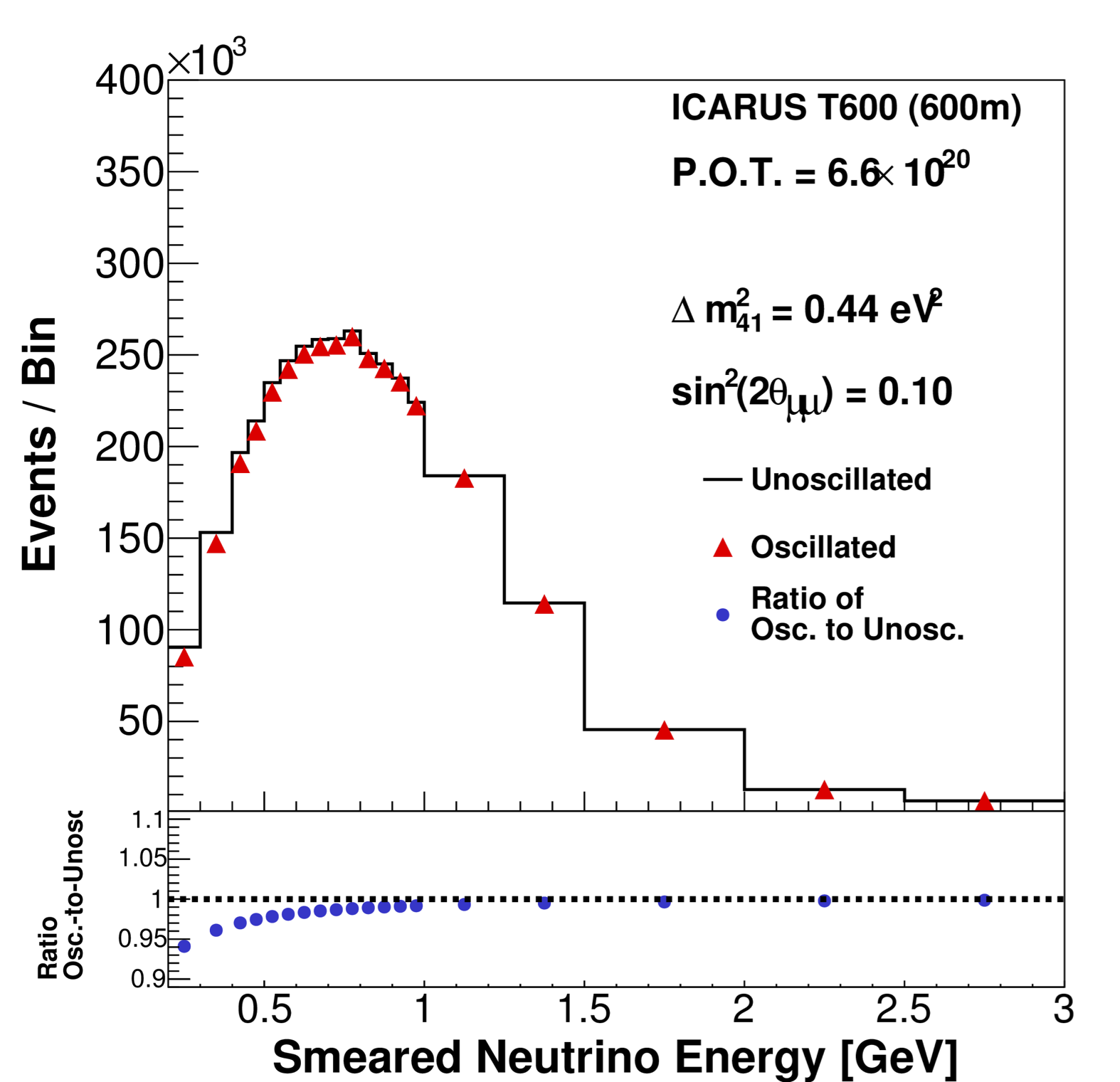}}
\mbox{\includegraphics[width=0.33\textwidth]{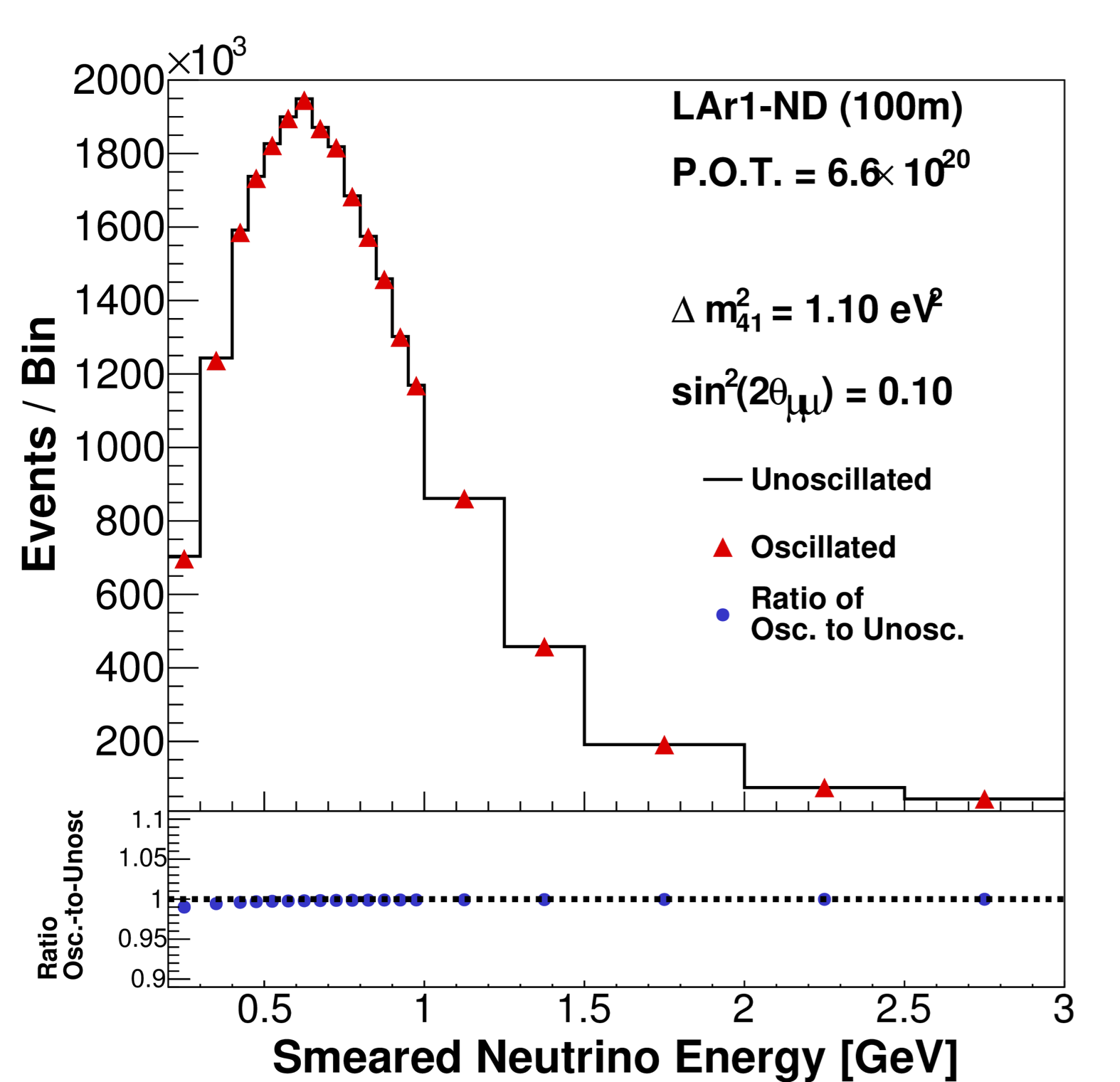}
\includegraphics[width=0.33\textwidth]{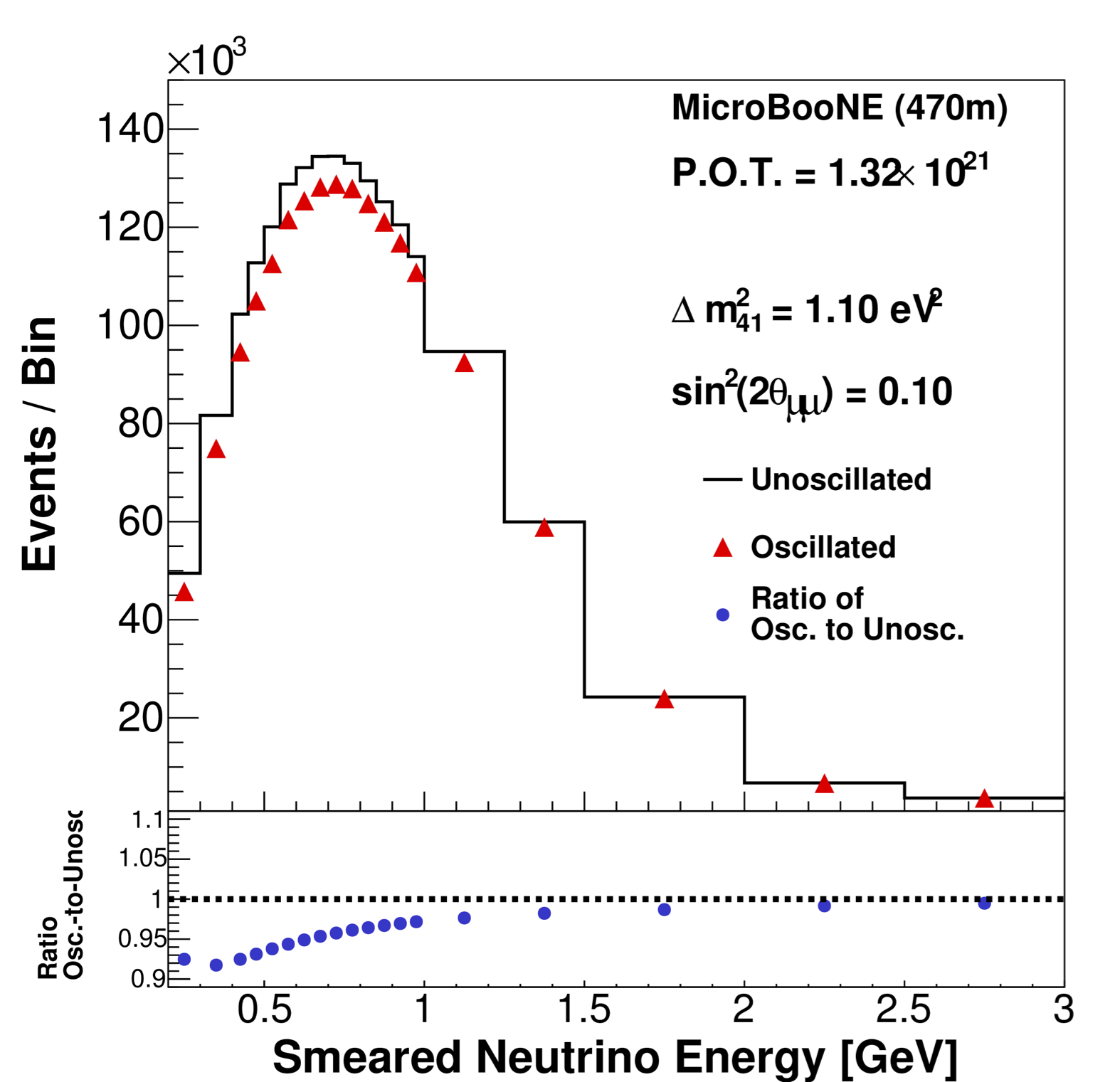}
\includegraphics[width=0.33\textwidth]{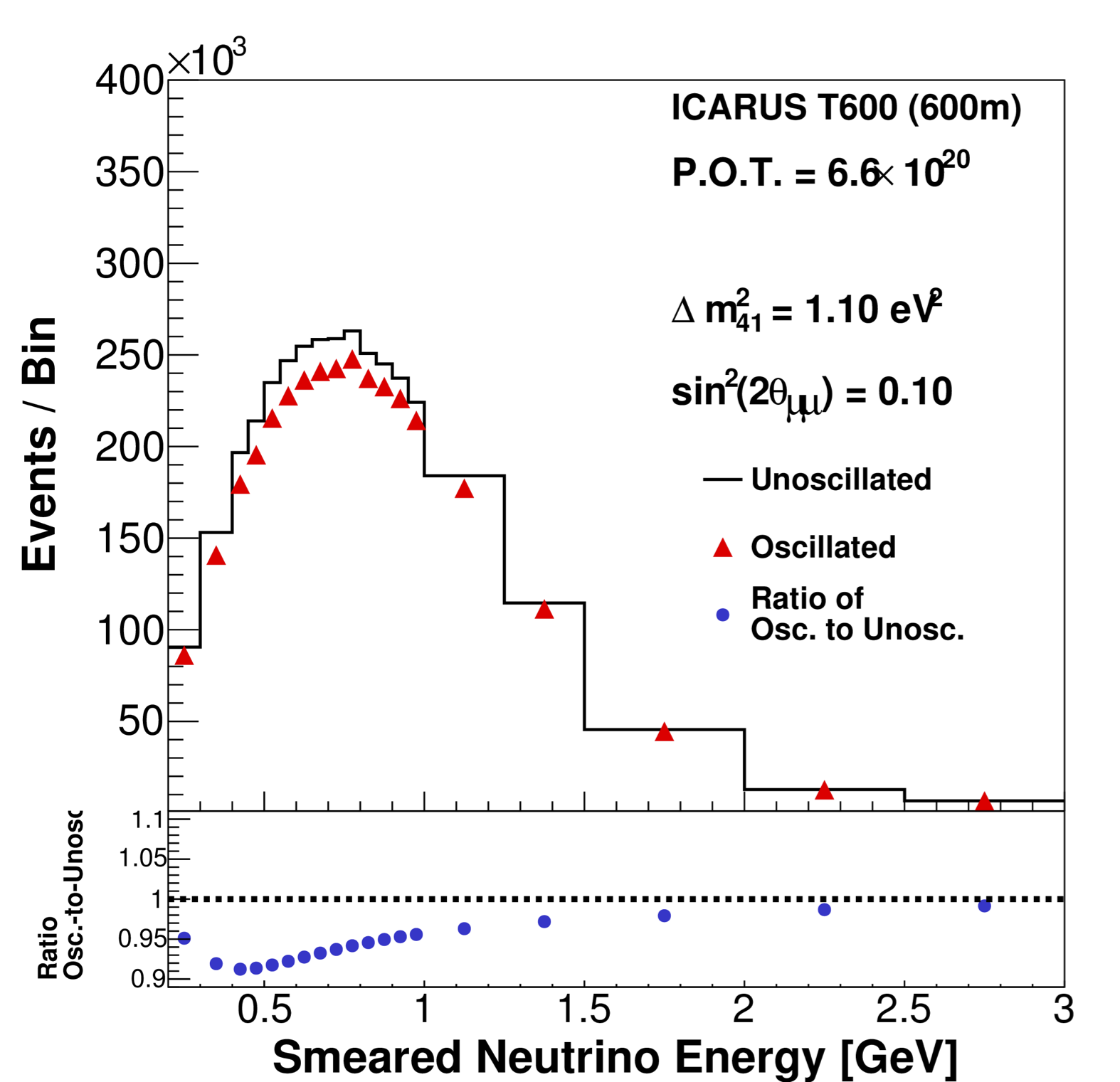}}
\caption{Examples of \numu disappearance signals in the SBN detectors for \dmsq = 0.44 \eVsq (top) and \dmsq = 1.1 \eVsq (bottom).}
\label{fig:numu_sigs}
\end{figure}

Figure \ref{fig:numu_sens} presents the \numu disappearance sensitivity assuming \pot protons on target exposure in \larnd and \icarus and \potuB protons on target in \uB.  The red curve is the 90\% confidence level limit set by the \SB and \MB joint analysis \cite{SciBooNE_MiniBooNE} and is to be compared to the solid black curve (also 90\% C.L.) for the LAr SBN program presented here.  SBN can extend the search for muon neutrino disappearance an order of magnitude  beyond the combined analysis of \SB and \MB.  Figure \ref{fig:numu_sigs} shows two examples of \numudis oscillation signals (for \dmsq~=~0.44~\eVsq and 1.1~\eVsq) in the three detectors for the exposures given above.  

The \numu disappearance measurement is a critical aspect of the SBN program and is needed to confirm a signal, if seen in \nue appearance, as oscillations.  A genuine \numunue appearance can also be accompanied by a disappearance of the intrinsic \nue beam component, since the three oscillation probabilities are related through a common mixing matrix. As an example, in the case of one additional sterile neutrino, $\sin^2 2 \theta_{\mu e} \leq 1/4 \sin^2 2 \theta_{\mu \mu} \cdot \sin^2 2 \theta_{e e}$, which is valid for small mixing angles. 

The ability to perform searches for oscillation signals in multiple channels is a major advantage for the FNAL SBN oscillation physics program.  By collecting the \numu and \nue event samples in the same experiment at the same time, correlations between the samples can be well understood and many systematics are common.  This implies that a simultaneous analysis of \nue CC and \numu CC events will be a very powerful way to explore oscillations and untangle the effects of \numunue, \numudis, and \nuedis, if they exist, in this mass-splitting range.

\begin{figure}[h]
\centering
\includegraphics[width=0.8\textwidth]{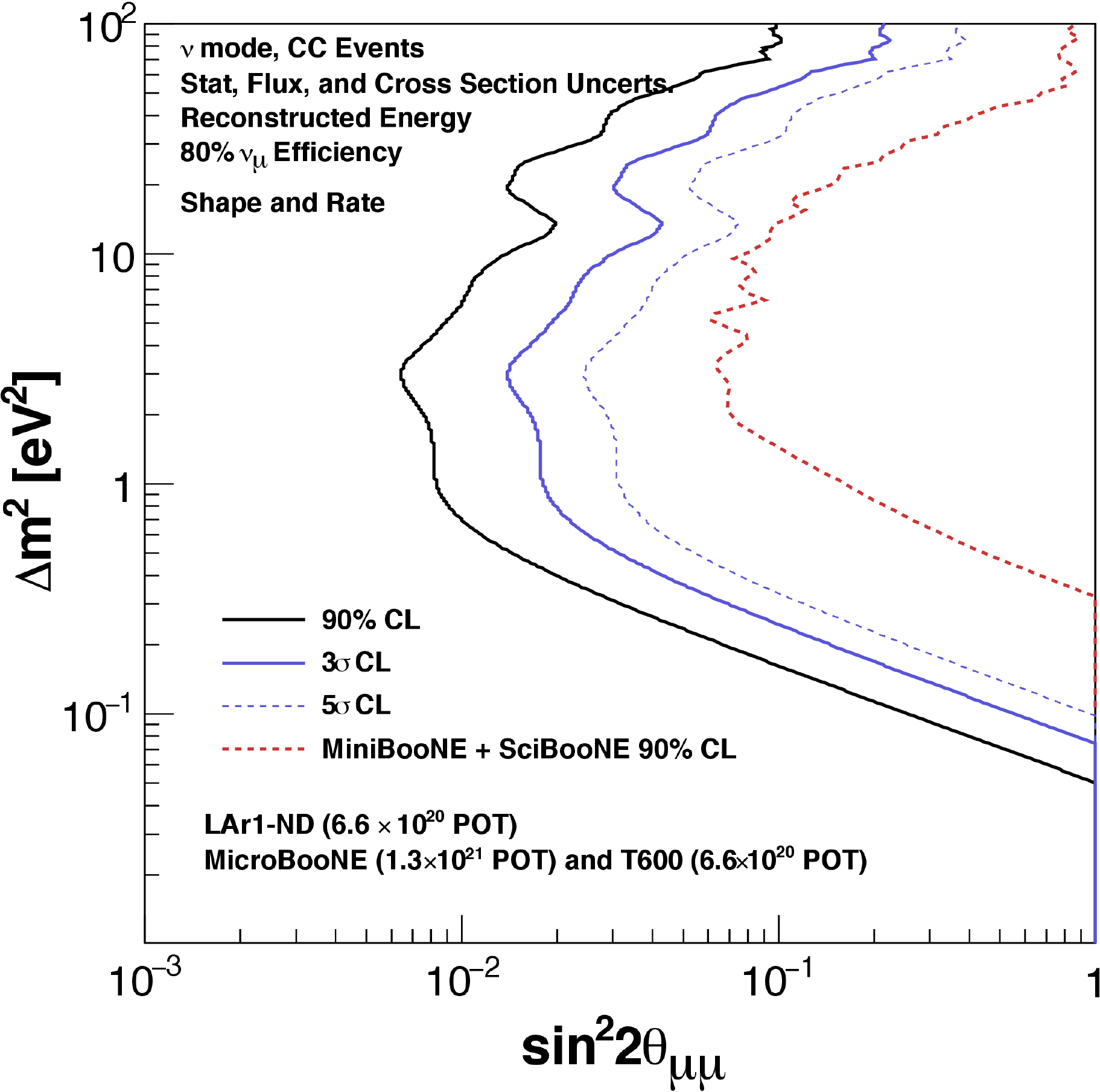}
\caption{Sensitivity prediction for the SBN program to \numudis oscillations including all backgrounds and systematic uncertainties described in this proposal (except detector systematics, see text). SBN can extend the search for muon neutrino disappearance an order of magnitude beyond the combined analysis of \SB and \MB.}
\label{fig:numu_sens}
\end{figure}

\FloatBarrier 

\section{Other SBN Physics}
\label{sec:otherphys}

The SBN program of three \lartpc detectors along the Fermilab Booster Neutrino Beam delivers a rich physics opportunity in addition to the oscillation searches detailed in Section~\ref{sec:osc}. We only briefly introduce some of them here.  In some cases, more details can be found in the individual detector Design Reports, Part II (\larnd) and Part III (\icarus) of this proposal.    %Neutrino-argon cross sections can be studied first in \uboone and later with even higher statistics in \larnd using the well characterized neutrino fluxes of the BNB \cite{MiniFlux}.  \uboone and ICARUS will also record samples of events from the off-axis NuMI beam~\cite{MiniBooNE_NuMI}. The source of the \MB electromagnetic event excess will be directly checked in the same beam using the \lartpc technology in order to separate $e^{\pm}$ from single $\gamma$ interactions.  
  
\subsection{Neutrino-Argon Interactions}

Precise neutrino-nucleus cross section measurements are a fundamental prerequisite for every neutrino oscillation experiment, including the future LAr long-baseline neutrino program.  In the GeV energy range, as a result of competitive physical processes and complicated nuclear effects, neutrino interactions on argon include a wide variety of final states. These can range from the emission of single or multiple nucleons to more complex topologies with multiple pions or other hadrons, all in addition to the leading lepton in charged-current events. Liquid argon TPC technology is particularly well suited to studying these interactions because of its excellent particle identification capability and calorimetric energy reconstruction down to very low thresholds.  

The SBN Program provides an ideal venue to conduct precision cross-section measurements in the few hundred MeV to few GeV energy range using the well characterized fluxes of the BNB~\cite{MiniFlux}. The detectors will collect neutrino samples with high statistics and will make the world's best measurements of \numu-Ar and \nue-Ar scattering.  \uboone will lead with initial measurements and pave the way for the T600 and \larnd to follow with even larger data sets for increased precision, including for rarer processes such as $\nu e$ scattering, strange particle production, multi-nucleon and multi-pion production, and coherent scattering with an argon nucleus.  Analysis of the different data sets will enable important cross-checks and each brings valuable additions to the physics reach.  The larger dimensions of the T600 will mean more complete containment of event final states including high energy muons and neutrons, leading to improved particle identification and energy reconstruction for some event classes.  \uboone and ICARUS will both also record large samples of events from the off-axis flux of the NuMI neutrino beam~\cite{MiniBooNE_NuMI} with its higher electron neutrino content and different energy spectrum.  \larnd, due to its proximity to the BNB source, will contribute the largest statistics, recording millions of neutrino interactions in a few year run.  In Part II of this proposal a table is provided that details the event rates for different event categories in the near detector (Table~I).

\subsection{Additional Searches} 

If short-baseline neutrino oscillations are observed in charged-current channels, then it will be possible to ``prove'' the existence of sterile neutrinos by searching for the disappearance of NC $\pi^0$ scattering events, $\nu_\mu Ar \rightarrow  \nu_\mu \pi^0 X$, in \uboone and \icarus relative to \larnd. As demonstrated by the \uboone experiment \cite{mbpi0}, this is a clean event sample with little background. Although the incident neutrino energy is not determined (the outgoing neutrino typically carries off most of the incident  neutrino energy), a decrease in this event rate in the \uboone and \icarus detectors would indicate the oscillation of active neutrino states into sterile neutrino states. %, $\nu_\mu \rightarrow \nu_s$.

SBN experiments will have good sensitivity to possible sterile neutrino decay. For the scenario of reference \cite{gninenko,gninenko2}, an active neutrino interacts by a neutral-current process inside the detector and produces a heavy neutrino (with a mass of a few hundred MeV) that quickly decays into a photon and a lighter neutrino. The signature, therefore, is an interaction vertex in the upstream portion of the detector and a single photon in the downstream portion of the detector, where the photon does not point back to the interaction vertex. With the superb spatial resolution and high event rates, this process can be searched for with good sensitivity.  The near detector has the advantage of high statistics, while \uboone and the T600 have longer volumes for observing the decay. 

SBN will also be able to search for sub-GeV dark matter (mass less than a few hundred MeV) \cite{wimps,wimps2,wimps3} by running in beam-dump mode, where the 8~\GeV proton beam is steered above the beryllium target and into the 50~m (or 25~m) downstream absorber. Beam-dump mode reduces the neutrino flux by a factor of $\sim$50, which makes the experiments more sensitive to low mass dark matter coming from $\pi^0$ and $\eta$ decay or from proton bremsstrahlung in the steel beam dump. The existence of low mass dark matter can then be inferred from the enhancement of neutral-current events relative to charged-current $\nu_\mu$ events compared to this ratio in normal beam-on-target running. Three different neutral-current channels can be studied: neutrino-electron elastic scattering, neutrino-nucleon elastic scattering, and neutral-current $\pi^0$ production ($\nu Ar \rightarrow \pi^0 X$). For each channel, low mass dark matter scattering on carbon will look just like neutral current neutrino scattering on carbon, and this will result in an enhancement of neutral current events. As a proof of principle, the MiniBooNE experiment just completed a one-year beam-dump run to search for low mass dark matter, and results are expected in 2015.

 \clearpage
\pagestyle{empty}

\proposalTitle

\setcounter{part}{1}
\part{\larnd Conceptual Design}

%\begin{center}
%{\large
%DRAFT \\
%\bigskip
%\proposaldate
%}\end{center}

\ifcombine
\clearpage
\else
\clearpage
\tableofcontents
\addtocontents{toc}{\protect\thispagestyle{empty}}
\clearpage
\setcounter{page}{0}
\fi

\pagestyle{fancy}
\rhead{II-\thepage}
\lhead{\larnd Conceptual Design}
\cfoot{}

%%%%%%%%%%%%%%%%%%%%%%%%%%%%%%%%%%%%%%%%%%%%%%%%%%%%%%%%

\section{Introduction}

The Fermilab Short Baseline Neutrino program includes the construction of a Liquid Argon Near Detector, \larnd, at 110~m from the Booster neutrino source in a new enclosure. Leveraging the advanced design work performed for LBNE and the very recent experience of the \uboone detector construction, the \larnd project has the potential to move forward quickly. As described in Part I, \larnd serves as the near detector in a three \lartpc experiment capable of definitively addressing existing anomalies in neutrino physics and making precision measurements of high-$\Delta m^2$ neutrino oscillations through both appearance and disappearance searches.

Due to the high event rate of neutrino interactions at the near location, significant physics output can be achieved with a relatively short run of the \larnd experiment. In addition to the physics program, \larnd, following the \uboone model, will have a development program serving as an engineering prototype for \lartpcs for long-baseline CP-violation searches in the future.

This Conceptual Design Report for the \larnd detector is organized as follows. Section~\ref{LAr1-NDPhysics} briefly reviews the \larnd stand alone physics program.  A short introduction to the \larnd detector design and dimensions is given in Section~\ref{LAr1-NDDetector}. Section~\ref{TPCDesign} presents the TPC design, while the TPC electronics, DAQ and trigger systems are described in Section~\ref{ElectronicsDAQTrigger}. Section~\ref{Laser} describes a UV laser-based field calibration system. Different options for the scintillation light collection system under consideration are described in  Section~\ref{LightDetectionSystem}. An external cosmic ray tagging system will complement the experiment, as reported in Section~\ref{CosmicRayTagging}. 

%Requirements in terms of cryogenics and building are summarized in Sect.~\ref{sec:cryogenics} and ~\ref{DetRequirements} respectively. Finally, Sect.\ref{CostSchedule} contains the cost estimate and the time scale for the \larnd experiment.

\section{Physics of \larnd}
\label{LAr1-NDPhysics}

While \larnd, in conjunction with \uboone and the \icarus, is a critical part of the oscillation physics program described in Part I, as a stand alone detector it enables a large number of relevant physics results. In this section we will discuss a sub-set of the physics measurements that can be performed with \larnd. These include studies of a possible \MB-like low energy excess of electromagnetic events that does not depend on the distance.  
%, an initial \numu disappearance search in combination with phase I \uboone data, 
%another analyses where the high rate and close location help provide insight. 

\subsection{\MB Low Energy Excess}
% A possible observation of a low energy excess signal by the \uboone collaboration in the 
% years leading up to the beginning of \larnd data taking, provides a possible new exciting opportunity for \larnd to explore.
% This excess, motivated by the previously observed low energy excess at \MB~\cite{mbosc}
% and being of electromagnetic nature, could be composed of either single-electron or single-photon events. 
% Looking for this excess and characterizing its nature is the main physics goal of \uboone. Truly understanding the source of this excess and in particular the underlying model responsible for these events may require additional information, such as a complementary event rate measurement with a near detector.

Looking for the low energy excess observed by the \MB experiment~\cite{mbosc} and characterizing its nature is the main physics goal of the \uboone experiment~\cite{uBProp}. This excess of electromagnetic events could be due to neutrino interactions with either a single-electron or single-photon in the final state.  
%Truly understanding the source of this excess and in particular the underlying model responsible for these events may require additional information, such as a complementary event rate measurement with a near detector. 
Observation of a low energy excess signal by \uboone in the years leading up to the beginning of \larnd data taking would immediately lead to the question of whether that excess is intrinsic to the beam or appears over the 470~m distance between source and detector.   

%A near detector will allow us to determine whether this excess depends upon the detector baseline, a signature of neutrino oscillations and other processes which are related to neutrino propagation, or whether the excess is baseline agnostic, which may suggest some new beam production mechanism as the excess source. Depending on the underlying model assumed to be creating the excess, the signal prediction at the near detector can vary considerably. For example, in the case where the excess is assumed to be due to sterile neutrino oscillations, a signal at the near detector would be generally reduced due to the very short baseline. The contribution that \larnd brings to such a scenario is a precise background rate determination, as covered in the SBN Physics Proposal~\cite{SBN-Physics}. 

%Here we instead focus on a model assumption where the \MB observed excess is created by single-electron events, and its rate scales as the total flux and the total electron neutrino charged-current inclusive cross-section. 

\larnd, at 110~m from the BNB target, can search for the same excess in a relatively short time.  Here we estimate the significance with which \larnd would observe \emph{the same} \MB-like electromagnetic excess ina the \nue candidate sample.  The methods used to estimate signal predictions for \larnd for the model described above follows those of a study performed by the \uboone Collaboration~\cite{dk_lowE}. Specifically, the excess of events observed from \MB is scaled to the \larnd predicted reconstructed rates. This is done accounting for the \MB reconstruction and particle identification efficiencies as reported in~\cite{MB_datarelease} to correct the raw excess event rates in \MB as a function of ($E_{lep}, \theta_{lep}$), and subsequently using the Monte Carlo predicted 2D matrix [$E_{\nu}^{true}, E_{lep}$] for charged current inclusive events on argon in order to properly correct for the cross-section dependence of Ar versus CH$_{2}$ and the flux ratio at \larnd with respect to \MB. Because only the lepton energy and angle are available in \MB data, we investigate the excess in \larnd as a function of the `lepton candidate' energy (after accounting for shower energy smearing). 

\begin{figure}[htp]\centering
\includegraphics[width=0.7\textwidth]{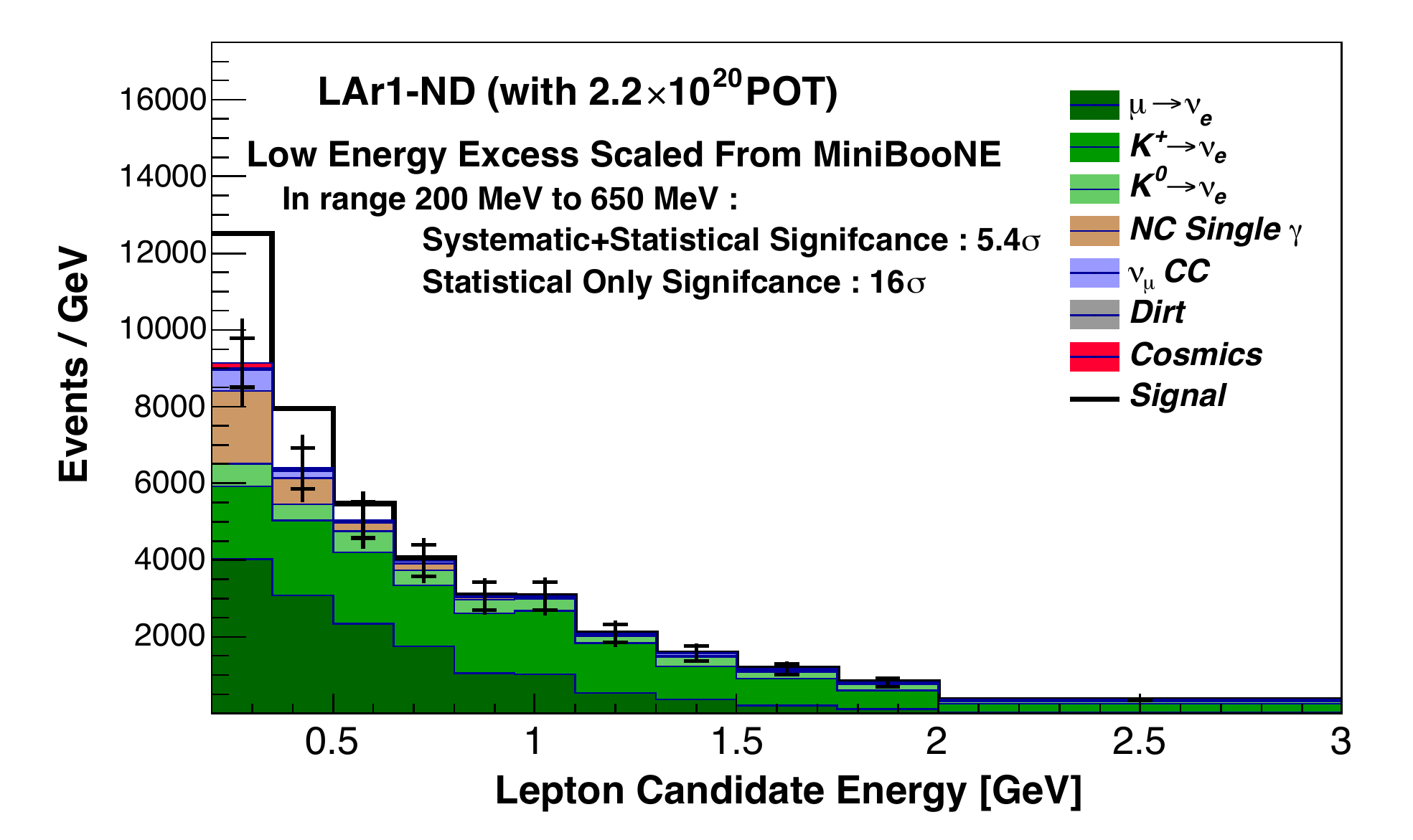}
\caption{Scaled \MB low energy excess events as a function of the lepton candidate energy in \larnd for an exposure of $2.2\times 10^{20}$ protons on target and using the same fiducial volume and backgrounds as the \nue appearance analysis described in Part I of the proposal. The signal prediction assumes the effect observed by \MB is electrons but is \emph{not distance dependent}. For the uncertainty on the background distribution the inner bars represent the statistical uncertainty and the outer bars represent the statistical + systematic uncertainties (see Part I).
\label{fig:LowEExcess}}
\end{figure}

Using the $\nu_e$ charged current inclusive backgrounds and systematic uncertainties discussed in Part I, we quantify the significance of a \MB-like excess in \larnd. In Fig.~\ref{fig:LowEExcess}, we report the excess events stacked on top of the expected backgrounds. In the 200--650~\MeV range in lepton energy, 803 excess events are expected, compared to a background of 3,177 events for an exposure of $2.2\times10^{20}$ protons on target. 
%To estimate the significance of this excess sitting above the given backgrounds we use the $\chi^{2}$ formulation presented in the SBN Physics Proposal, including the full set of systematic uncertainties. Using this we find that the significance of this excess 
This is a $5.4\sigma$ signal including both systematic and statistical uncertainties. Considering only statistical uncertainties, the excess sits $>10\sigma$ above the background. 
%Furthermore, using the 803 excess events we will be able to study the exclusive channels which produce this signal and study the event kinematics. 

% Note that there are likely many other exotic models that could be possibly be explored, including models which predict a photon excess, these have not been addressed yet but are actively being studied. 
If \uboone observes an excess of photons which are due to an as-yet unknown source of neutral current interactions producing single photons in the final state, \larnd at 110~m will immediately confirm that the excess is intrinsic to the beam (i.e. that it is due to some un-modeled neutral current interaction). Also, the event rate with which \larnd will be able to study these events will be more than one order of magnitude larger than \uboone. Such a sample will enable a measurement of this reaction with great precision and inform the development of cross section models in this energy range to include this process.

\subsection{Neutrino Cross Section Measurements}

%Precise cross section measurements are a fundamental prerequisite for every neutrino oscillation study. In the GeV energy range, as a result of competitive physical processes and complicated final state interactions and other nuclear effects, neutrino interactions on argon include a wide variety of final states. These can range from the emission of multiple nucleons to even more complex topologies with multiple pions, all in addition to the leading lepton in charged current event. Liquid argon TPC technology is particularly well suited to studying these interactions because of its excellent particle identification capability and calorimetric energy reconstruction down to very low thresholds.

As discussed in Part I, neutrino-nucleus interactions are critical to understand in neutrino oscillation experiments, including the future liquid argon long-baseline program.  \larnd provides an ideal venue to conduct precision cross section measurements in the GeV energy range. The experiment will collect enormous neutrino event samples and, continuing the studies done by \uboone and ICARUS, will make the world's highest statistics cross section measurements for many $\nu$-Ar scattering processes.

%Due to its location near the neutrino source (20-30 the flux at \uboone) and relatively large mass, LAr1-ND will make measurements of neutrino interactions with high statistics, as shown in 
Table~\ref{tab:events_xsec} shows the expected rates of $\nu_\mu$ and $\nu_e$ events separated into their main experimental topologies for an exposure of $6.6\times 10^{20}$ protons on target (POT). A novel approach based on the event categorization in terms of exclusive topologies can be used to analyze data and provide precise cross section measurements in many different $\nu_\mu$ and $\nu_e$ exclusive channels. Included for reference, we also show the classification by physical process from Monte Carlo truth information. 

The largest event sample corresponds to a \numu charged-current 0 pion final state, where there is an outgoing $\mu^-$, one or more recoil nucleons, and no outgoing pions or kaons. This cross section for scattering off nuclei largely depends on final state interactions and other nuclear effects and \larnd data will allow the study of nuclear effects in neutrino interactions in argon nuclei with high precision. 

% With the fine granularity of a LAr TPC, the experiment will be able to separate 
% neutral-current (NC) and charged-current (CC) interactions. 

\begin{table}[h!]
   \begin{center}
   \begin{tabular}{ l p{7cm} p{2cm} p{2cm} p{1.5cm}}
   {\bf Process} &  & {\bf No. }   & {\bf Events/} &{\bf Stat.   } \\
                 &  & {\bf Events} & {\bf  ton   } &{\bf Uncert. } \\
   \hline
    \hline
   \multicolumn{4}{c}{\emph{\numu ~ Events (By Final State Topology)}} \\
%  \hline
    CC Inclusive    & & 5,212,690 & 46,542 & 0.04\% \\
    CC 0 $\pi$~~~~~~~~~~~~ & $\numu N \rightarrow \mu + Np $  & 3,551,830 & 31,713   & 0.05\% \\
            & $~\cdot~\numu N \rightarrow \mu  + 0p  $        & 793,153   & 7,082    & 0.11\% \\
            & $~\cdot~\numu N \rightarrow \mu  + 1p  $        & 2,027,830 & 18,106   & 0.07\% \\
            & $~\cdot~\numu N \rightarrow \mu  + 2p  $        & 359,496   & 3,210    & 0.17\% \\
            & $~\cdot~\numu N \rightarrow \mu + \geq 3p  $    & 371,347   & 3,316    & 0.16\% \\
    CC 1 $\pi^{\pm}$      & $\numu N \rightarrow \mu + \mathrm{nucleons}~+ 1 \pi^{\pm}$      & 1,161,610 & 10,372 & 0.09\% \\
  CC $\geq$2$\pi^{\pm}$   & $\numu N \rightarrow \mu + \mathrm{nucleons}~+ \geq2\pi^{\pm}$   & 97,929    & 874    & 0.32\% \\
  CC $\geq$1$\pi^{0}$   & $\numu N \rightarrow \mu + \mathrm{nucleons}~+ \geq1\pi^{0}$       & 497,963   & 4,446  & 0.14\% \\
  \multicolumn{5}{c}{} \\
  NC Inclusive & & 1,988,110 & 17,751 & 0.07\% \\
  NC 0 $\pi$      & $\numu N \rightarrow \mathrm{nucleons} $     & 1,371,070 & 12,242 & 0.09\%  \\
    NC 1 $\pi^{\pm}$      & $\numu N \rightarrow \mathrm{nucleons} + 1\pi^{\pm}$       & 260,924 & 2,330 & 0.20\% \\
  NC $\geq$2$\pi^{\pm}$   & $\numu N \rightarrow \mathrm{nucleons}~+ \geq2\pi^{\pm}$   & 31,940  & 285   & 0.56\% \\
  NC $\geq$1$\pi^{0}$   & $\numu N \rightarrow \mathrm{nucleons}~+ \geq1\pi^{0}$       & 358,443 & 3,200 & 0.17\% \\
    \hline
  \hline
  \multicolumn{5}{c}{\emph{$\nu_{e}$   Events}} \\
   CC Inclusive   &    & 36798 & 329 & 0.52\% \\
   NC Inclusive   &    & 14351 & 128 & 0.83\% \\
   \hline
   \hline
   Total $\numu$ and $\nu_e$ Events &   & 7,251,948 & 64,750 & \\

    \multicolumn{5}{c}{} \\
    \hline
    \hline
    \multicolumn{5}{c}{\emph{\numu Events (By Physical Process)}} \\
%  \hline
    CC QE       & $\numu n \rightarrow \mu^{-} p $              & 3,122,600 & 27,880 & \\
    CC RES      & $\numu N \rightarrow \mu^{-} \pi N $          & 1,450,410 & 12,950 & \\
    CC DIS      & $\numu N \rightarrow \mu^{-} X$               & 542,516   & 4,844  & \\
    CC Coherent       & $\numu Ar \rightarrow \mu Ar +  \pi$    & 18,881    & 169    & \\
    \hline
    \hline

   \end{tabular}
   \caption{\small Estimated event rates using GENIE (v2.8) in the \larnd active volume (112 t) for a $6.6\times 10^{20}$ exposure. 
 In enumerating proton multiplicity, we assume an energy threshold on proton kinetic energy of 21~MeV. The $0 \pi $ topologies include any number of neutrons in the event.}
   \label{tab:events_xsec}
   \end{center}
\end{table}

In \larnd more than 2 million neutrino interactions will be collected per year in the full active volume (assuming $2.2\times 10^{20}$ POT), with 1.5 million \numu and 12,000 \nue charged current (CC) events. One year exposure of \larnd will provide an event sample 6-7 times larger than will be available in the full \uboone phase I run.

% Discussed below is
% a partial list of neutrino cross section measurements that will be made by LAr1-ND.

% \begin{itemize}
% \item{$\nu_\mu$ Charged-Current Quasi-Elastic Scattering}

Comparison of \nue CC and \numu CC cross sections is very important. By lepton universality, the cross sections should be the same after correcting for the outgoing charged-lepton mass. A difference in the cross sections would indicate a new process that violates lepton universality.

\larnd will also see several hundred $\nu_\mu e \rightarrow \nu_\mu e$ elastic scattering events in $6.6\times 10^{20}$ POT. These events are easily identified by an outgoing electron along the neutrino beam direction with $\cos \theta > 0.99$ and with no recoil nucleons. With this event sample, a measurement of $\sin^2\theta_W$ can be made at relatively low energy to be compared with the world average.

% \item{$\nu_\mu$ Nucleon Elastic Scattering}

We expect approximately a quarter million NC elastic scattering events identified by a single nucleon (proton or neutron) recoil track. If the recoil proton events can be cleanly separated from the recoil neutron events, then it may be possible to make a competitive measurement of $\Delta s$, the strange quark contribution to the proton spin.

% \item{$\nu_\mu C \rightarrow \nu_\mu \pi X$ Scattering}

% The cross section for neutral-current single $\pi$ scattering can also be measured
% by LAr1-ND. The cross section consists of two dominant contributions: coherent $\pi^0$
% and resonant $\pi^0$ and $\pi^+$ production. The resonant contribution occurs mostly 
% through the $\Delta$ resonance, while the coherent cross section per nucleon should 
% be enhanced by a factor of $\sim 3$ in argon relative to the cross section per nucleon 
% in carbon.
% \end{itemize}

Finally, by using the same neutrino beam that was used by the MiniBooNE experiment, we will be able to directly compare neutrino cross sections off carbon ($A=12$) and argon ($A=40$) targets and search for a nuclear dependence of the cross section.

\section{Overview of the \larnd Detector}
\label{LAr1-NDDetector}

The design of the \larnd detector~\cite{LAr1-NDPAC} builds on many years of LAr TPC detector R\&D and experience from design and construction of the \icarus, ArgoNeuT, \uboone, and LBNF detectors. The basic concept of the \larnd detector, based on LBNF-type technology, is to construct a membrane-style cryostat at 110~m from the Booster neutrino source in a new enclosure adjacent to and directly downstream of the existing SciBooNE hall. The membrane cryostat will house multiple cathode plane assembly (CPA) and anode plane assemblies (APAs) to read out ionization electron signals. The APAs located near the beam-left and beam-right walls of the cryostat will each hold 3 planes of wires with 3~mm wire spacing. The wire readout arrangement is identical to \uboone, with banks of cold electronics boards at the top and one vertical side of each APA. The total number of readout channels is \apach~per side (\detch~in the entire detector). The CPAs have the same dimensions as the APAs and are centered between them. Each pair of facing CPA and APA hence forms an electron-drift region. The open sides between each APA and the CPA are surrounded by 4 Field Cage Assembly (FCA) modules, constructed from FR4 printed circuit panels with parallel copper strips, to create a uniform drift field. The drift distance between each APA and the CPA is 2~m, such that the cathode plane will need to be biased at -100~kV to create an electric field of 500~V/cm. Accurate mapping of the electric field in the drift region will be performed through a UV laser-based calibration system. The active volume is 4.0~m (width) $\times$ 4.0~m (height) $\times$ 5.0~m (length), containing 112 tons of liquid argon. The \larnd design will additionally include a light collection system for the detection of scintillation light and the detector will be complemented by an external cosmic ray tagging system. In addition, we are looking into the possibility of placing shielding over the near detector should it be deemed necessary to reduce cosmogenic backgrounds. 

Overall, the design philosophy of the \larnd detector is to serve as a prototype for LBNF that functions as a physics experiment. While the present conceptual design described here is an excellent test of LBNF detector systems sited in a neutrino beam, the \larnd collaboration is exploring innovations in this design and the opportunity to further test them in a running experiment.

\subsection{Detector Dimensions}
\label{DetSizeOptmization}

The \larnd detector size has been optimized with respect to the preliminary design described in the \larnd proposal~\cite{LAr1-NDPAC}. In the original proposal the detector was designed to be located in the existing  experimental enclosure that previously housed the SciBooNE experiment, at 100 m from the BNB 
target. The dimensions of the detector were dictated by the size of the enclosure, leading to an active volume of 4.0~m (width) $\times$ 4.0~m (height) $\times$ 3.6~m (length), containing 82 tons of liquid argon. % this detector was large enough to enable a compelling physics program. 
Studies reported in Ref.~\cite{JulyPAC} indicated the advantage of locating the near detector in a new enclosure, directly downstream of the existing SciBooNE enclosure at 110 m. A new building opens the question of the detector dimensions and optimizing it for physics. Enlarging the dimensions in the transverse directions has not been considered in order to maintain the 2~m maximum drift length and the height of the detector (to avoid larger excavation costs). However, Monte Carlo studies of muon containment and photon background rejection as a function of the detector length in the beam direction have been performed. 

%The goal of these studies were to maximize the containment inside the detector of muons produced by charged current neutrino interactions, and reduce the single photon background, from $\pi^{0}$ for $\nu_e$ appearance analysis. The containment of photons produced from in-detector $\pi^0$ decays and the rejection of photons produced by out-of-detector $\pi^0$ decays have to be maximized.

%Studies with photons do not provide definitive insight into the optimal size of the detector.  
%A large active volume must be large enough to allow a large buffer (roughly 25~cm) around the fiducial volume to reduce the single photon background coming from dirt backgrounds, as described in reference~\cite{SBN-Physics}. 
Figure \ref{fig:muon_containment} shows the breakdown of the fate of muons produced in \numu CC interactions in a fixed fiducial volume as a function of the overall active detector length. 
%tudies of muons from \numu CC interactions show that increasing the length of the detector along the beam direction does provide significant gain in the containment of muons, as shown in Figure~\ref{fig:muon_containment}. 
%Increasing the length of the detector allows for more fully contained muons with respect to the shorter detector.
In a 4.0~m (width) $\times$ 4.0~m (height) $\times$ 3.6~m (length) detector, 53\% of muons are contained (so muon momentum can be measured through calorimetric reconstruction with very good accuracy), 35\% exit with a track longer than 1~m (so muon momentum is measured by multiple scattering, with less accuracy), and the remaining 12\% exit with a track shorter than 1~m (so muon momentum cannot be measured). Increasing the length of the detector in the direction of the beam to 5~m increases the fraction of fully contained muons to 62\%, a 17\% fractional increase.  The 11\% minimum on the fraction of exiting muons shown in the plot is due to tracks that leave the volume through the sides of the detector, and can be reduced by tightening the fiducial volume definition.  
%, 62\% of the muons are contained, 27\% exit with a track longer than 1~m, and 11\% exit with a track shorter than 1~m. 
%with further lengthening of the detector only slightly increases the fraction of contained muons  
%From these studies a detector length of 5~m was considered optimal, and is chosen for the detector design in the present CDR.

\begin{figure}
\centering
\includegraphics[width=0.5\textwidth]{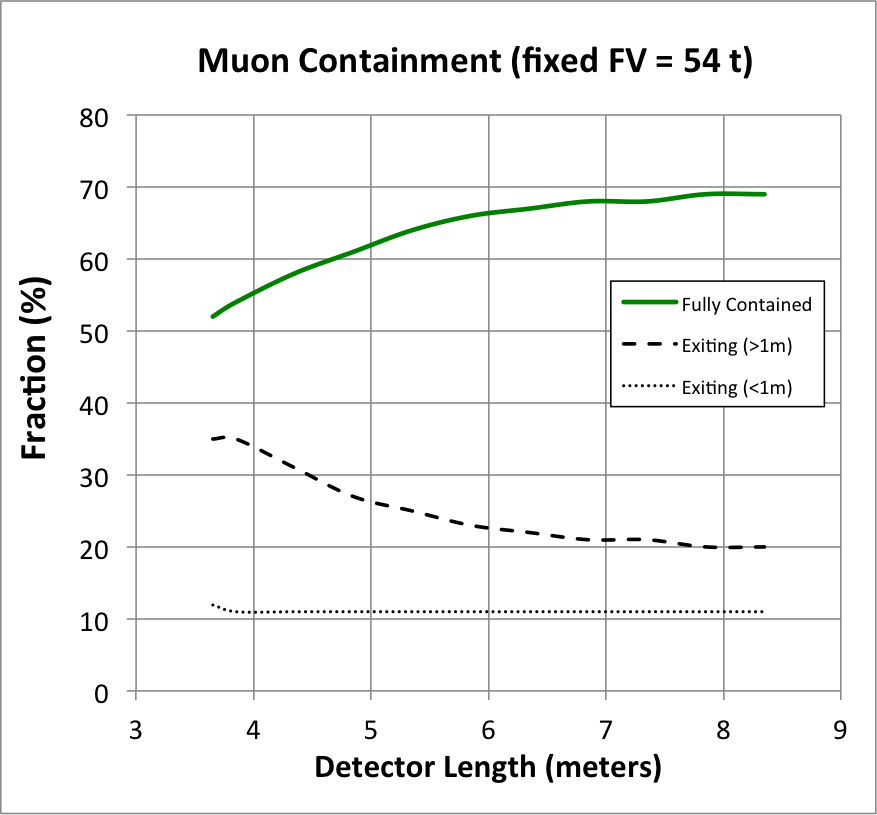}
\caption{\label{fig:muon_containment}Muon containment as a function of the detector length in the beam direction.}
\end{figure}

\section{TPC Design}
\label{TPCDesign}

The conceptual design for LAr1-ND is shown in Figure~\ref{fig:TPCcryo}a which shows the TPC housed inside a membrane-style cryostat. The LAr active volume is a rectangular parallelepiped with dimensions of 4~m vertically, 4~m horizontally, and 5~m along the beam direction.
The TPC consists of four anode plane assemblies~(APAs) and two central cathode plane assemblies~(CPAs), as indicated
 in Figure~\ref{fig:TPCcryo}b. The APAs and CPAs are large-scale elements with an area of 4~m~$\times$~2.5~m each. The overall dimensions of the individual APAs are restricted to be increments of  the top~(192~mm) and side~(222~mm) readout board dimensions.
The TPC is oriented such that the Booster neutrino beam passes perpendicular to the drift direction. The TPC key design parameters are summarized in Table~\ref{table:TPC_param}.

%%===========TPC parameters ==============
\begin{table}[b]
\caption[TPC Design Parameters.] {LAr1-ND TPC key design parameters.}

\begin{center}
\vspace{4mm}
{\footnotesize \begin{tabular}{lc}
\toprule \hline\hline
{\bf TPC Parameter} & {\bf Value} \\ \hline
\midrule
TPC active volume & 5~m~(L)~$\times$~4~m~(H)~$\times$~4~m~(W), 112 metric ton active LAr  mass\\ 
Number of TPC cells & 2 drift volumes, 2~m drift length in each\\
Maximum drift time & 1.28~ms\\
Anode Plane Assembly & 2.5~m~$\times$~4~m active area, with cold electronics mounted on 2 sides\\
Wire properties& 150~$\mu$m, CuBe \\
Wire planes &3 planes on each APA, U~\&~V at $\pm$60$^{\circ}$ to vertical (Y) \\
Cathode bias &-100~kV at 500~V/cm drift field \\
Number of Wires & 2816 channels/APA, 11264 wires total in TPC\\
Wire tension & 0.5~kg at room temperature\\ \hline\hline
\bottomrule

\end{tabular}}
\end{center}
\label{table:TPC_param}
\end{table}

\begin{figure}
\centering
\includegraphics[width=0.9\textwidth]{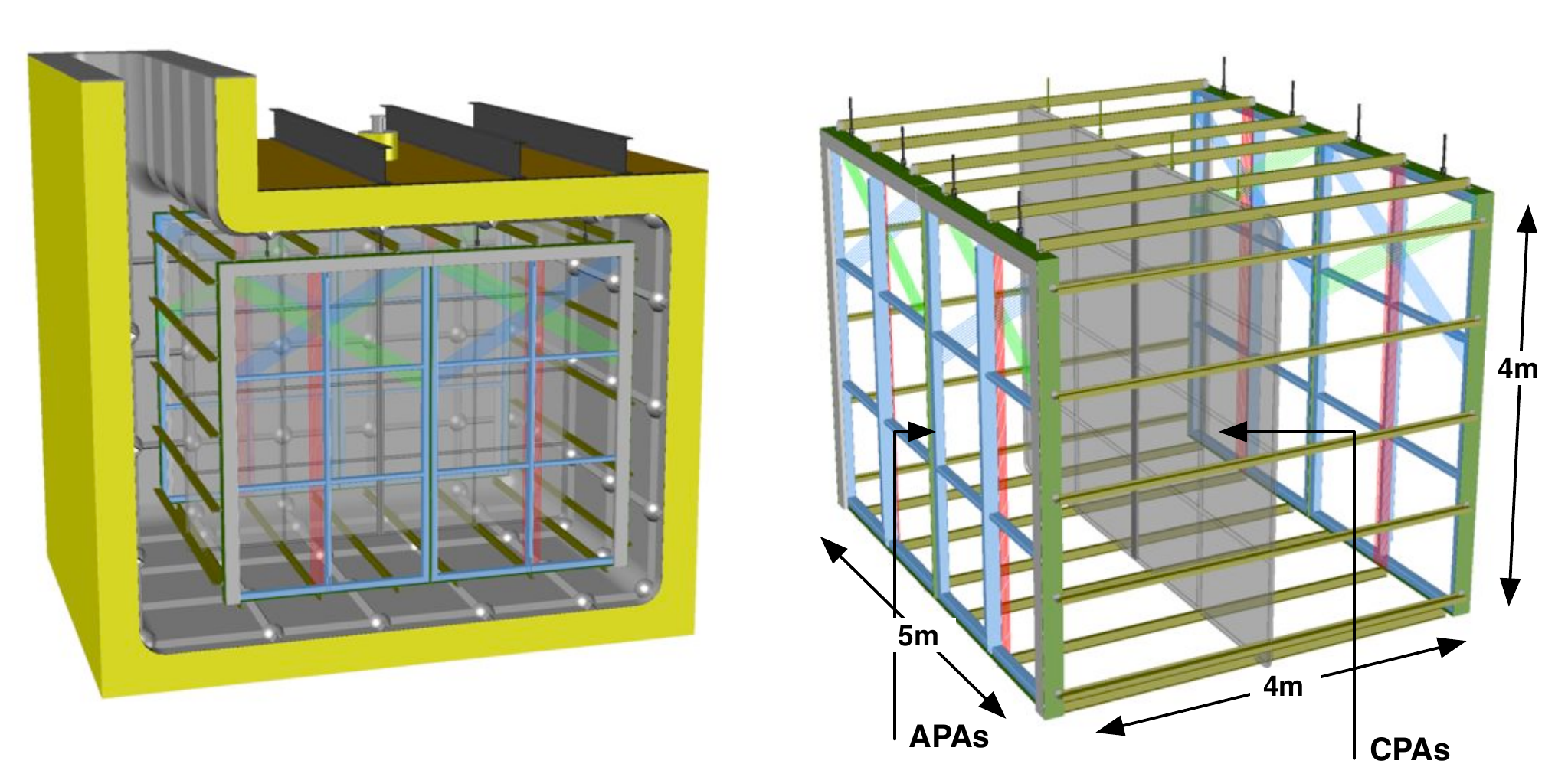}
\caption{\label{fig:TPCcryo}(Left) A conceptual design of the LAr1-ND. (Right) A model of the TPC, showing the four bridged APAs and the central CPAs.}
\end{figure}

The main requirements of the TPC are:

\begin{enumerate}
\item{The TPC volume is large enough to achieve the physics goals of the experiment.  The 4~m $\times$ 4~m dimensions of the TPC active volume in the transverse (perpendicular to beam) directions and 5~m dimension in the longitudinal (along beam) direction are determined based on studies of signal containment and background rejection (see Section~\ref{DetSizeOptmization}).}
\item{The 3~mm wire pitch is chosen, as in the MicroBooNE and \icarus detectors, to enable electron/photon separation to be achieved with identical efficiency.} 
% as described in section XXX.\textbf{update}.}
\item{The APAs are constructed in a manner that guarantees no wires will break during the operational life of the experiment.}
\item{The high voltage system and field cage provides a uniform and stable drift field in order to capably image the entire fiducial volume.}
\item{The electric field everywhere inside the cryostat must not exceed 40 kV/cm to prevent breakdown~\cite{Blatter2014}.}
\end{enumerate}

\subsection{The Anode Plane Assembly}
Each \larnd APA will consist of 3 planes (referred to by the direction they are oriented: Y, U and V) of 150~$\mu$m diameter Copper-Beryllium (CuBe) wires. The wire pitch and plane spacing is 3~mm, with the collection planes vertical (Y), and the two induction planes (U,V) each have wires at angles of $\pm$60$^{\circ}$ from the vertical. Bias voltages of approximately -200~V, 0~V, and +500~V will be applied to the  (U,V,Y) wire planes, respectively, to provide the 100$\%$ transparency condition necessary to allow all electrons to pass through the U and V planes and be collected by the Y plane.

All wires in the APAs are bonded mechanically with epoxy and terminated electrically with solder onto bonding boards made out of G10, which also provide connection to the readout electronics. The APA uses the same wire bonding method developed for the LBNF APAs, but without the continuous helical wrapping. Each wire will be tensioned at 0.5~kg  per wire when the APA is at room temperature.  The wires would acquire an additional 0.7~kg  if suddenly cooled to liquid argon temperatures while the support frame structure is still warm, therefore a controlled cooldown rate is needed.  Since CuBe has a nearly identical CTE as that of the stainless steel, the nominal wire tension will be restored once the entire APA is cooled down. This CuBe wire has a break load of approximately 3~kg at the LAr temperature, so the wire tension will be comfortably below this threshold.

 In order to minimize the cost of the readout electronics, each APA has cold readout electronics on two edges only.  The U wires of the two APA's observing the same drift volume are electrically connected at the joining edges via flexible jumper cables (see Figure~\ref{fig:APAsbridgedwires}). Similarly for the V wires. The installation/access of the jumpers should be done before the installation of the cold electronics to avoid ESD damage to the ASICs. The jumper cables use industry standard gold plated pin and socket connections, similar to all electrical interconnects between the cold electronics boards.  Although the jumpers may increase the risk of open circuit on the wire readout somewhat, its benefits in the reduction of readout channel count, and the APA size outweight the risks.  Nevertheless, one of our early design efforts will be the selection and evaluation of the interconnect components at the cryogenic temperatures. A special set of readout electronics may be needed to read out the joining edge of an APA during individual APA testing.

\begin{figure}
\centering
\includegraphics[width=0.6\textwidth]{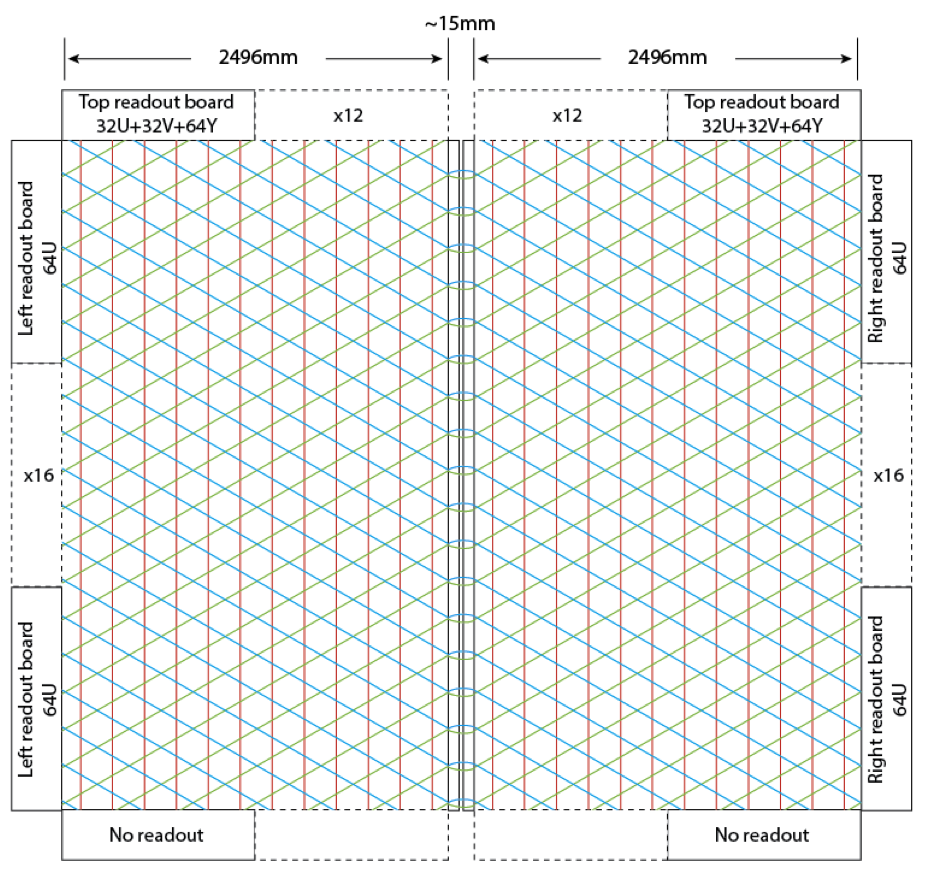}
\caption{\label{fig:APAsbridgedwires}A schematic of the bridged APAs.}
\end{figure}

In this design there is a gap of 15~mm between the two active apertures of the APAs which creates a ``dead" readout region.
To overcome this issue there is an option to insert a printed circuit board in the gap between the two APAs, and bias the circuit board strips with a voltage distribution such that the incoming electrons will be deflected away from this gap and land in the active region of the wire frame. 
In this field configuration, there will be no electron loss, but reconstructed inclined tracks will appear distorted at this gap. As this region has a fixed distortion, it can be easily mapped out and corrected (see Figure~\ref{fig:FieldShaping}).  This field shaping concept will be implemented at one section of the LBNE 35ton TPC and evaluated during the LBNE 35ton Phase II operation. 

\begin{figure}
\centering
\includegraphics[width=\textwidth]{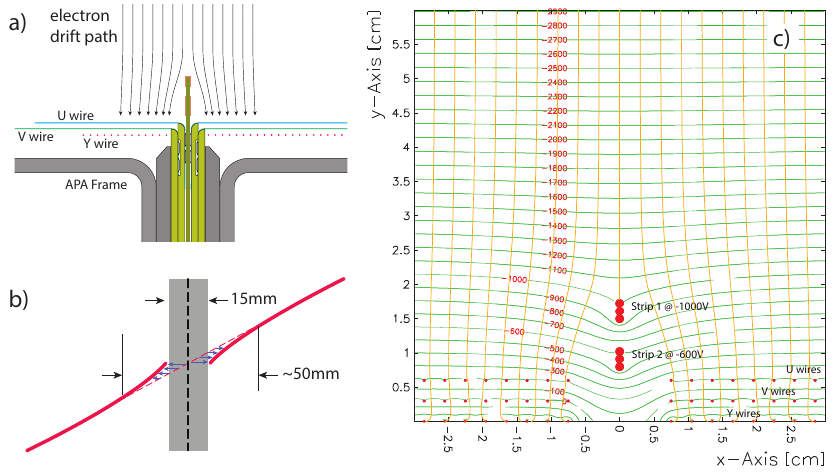}
\caption{\label{fig:FieldShaping}a) A concept to eliminate the dead region between the two APAs by adding a few properly biased electrodes at the center of the gap to divert the electrons to the nearby active regions. b) The distortion on the reconstructed tracks. c) A Garfield simulation of the electron drift lines in a two strip configuration.}
\end{figure}

Figure~\ref{fig:TPCCornerDetail1} is a view of an upper outer corner of an APA.  Three layers of wire bonding boards are stacked on the front face of the frame.  The wires are bonded to each board at the leading edge with epoxy, and then soldered to copper pads on the boards.
%(if the wire to be read out).  
Copper traces on the wire bonding boards bring the wire signal to the cold electronics boards mounted on the two outside edges of the frame.  Only the wires on the joining edge of the APA are bonded on ``LBNE'' style grooved boards (Figure~\ref{fig:FieldShaping}a, \ref{fig:TPCGapDetail}). This configuration minimizes the number of grooved boards which are labor intensive to fabricate, while maintaining a relatively small dead space between APAs.

High voltage capacitors and high value resistors are needed for each readout channel with a bias voltage (U \& Y).  These components can be integrated on the wire bonding boards  (MicroBooNE style), or mounted on intermediate CR boards between the wire bonding boards and the FEE boards (LBNE style).

The four  stainless steel APA frames are required to be flat and to have rigid tolerances. The distributed load on the frame is 
calculated to be 250~kg/m.  Since this load is applied to one side of the frame only, it has the tendency to bow the frame. This in turn could make the wire plane spacing non-uniform over the entire opening, resulting in different electron transparency. The transparency can be restored by over biasing the wires, but we should keep the flatness of the frame to better than $\sim$ 0.5~mm to avoid very high bias voltages. Since the APA frames are outside of the TPC's active volume, it is straightforward to design the frames to be stiff against such distortions.  Adjustment of the wire bonding boards can also be made at APA assembly time to further improve the wire plane precision over fabrication tolerances in the APA frames.

 A prototype LBNE type frame with dimensions 1.5~m$\times$ 0.5~m has been designed and fabricated at Sheffield University to prototype the manufacturing process, which minimizes distortions and leads to significant improvements in flatness. For the full LAr1-ND APA frame, a four-stage process is envisioned: i) secure the whole assembly to a purpose built fabrication jig during tack welding; ii) normalization of the structure (stress relief) to prevent buckling; iii) machine and drill all mounting points for additional components; and iv) chemical passivation to remove contaminants. 

\begin{figure}
\centering
\includegraphics[width=0.8\textwidth]{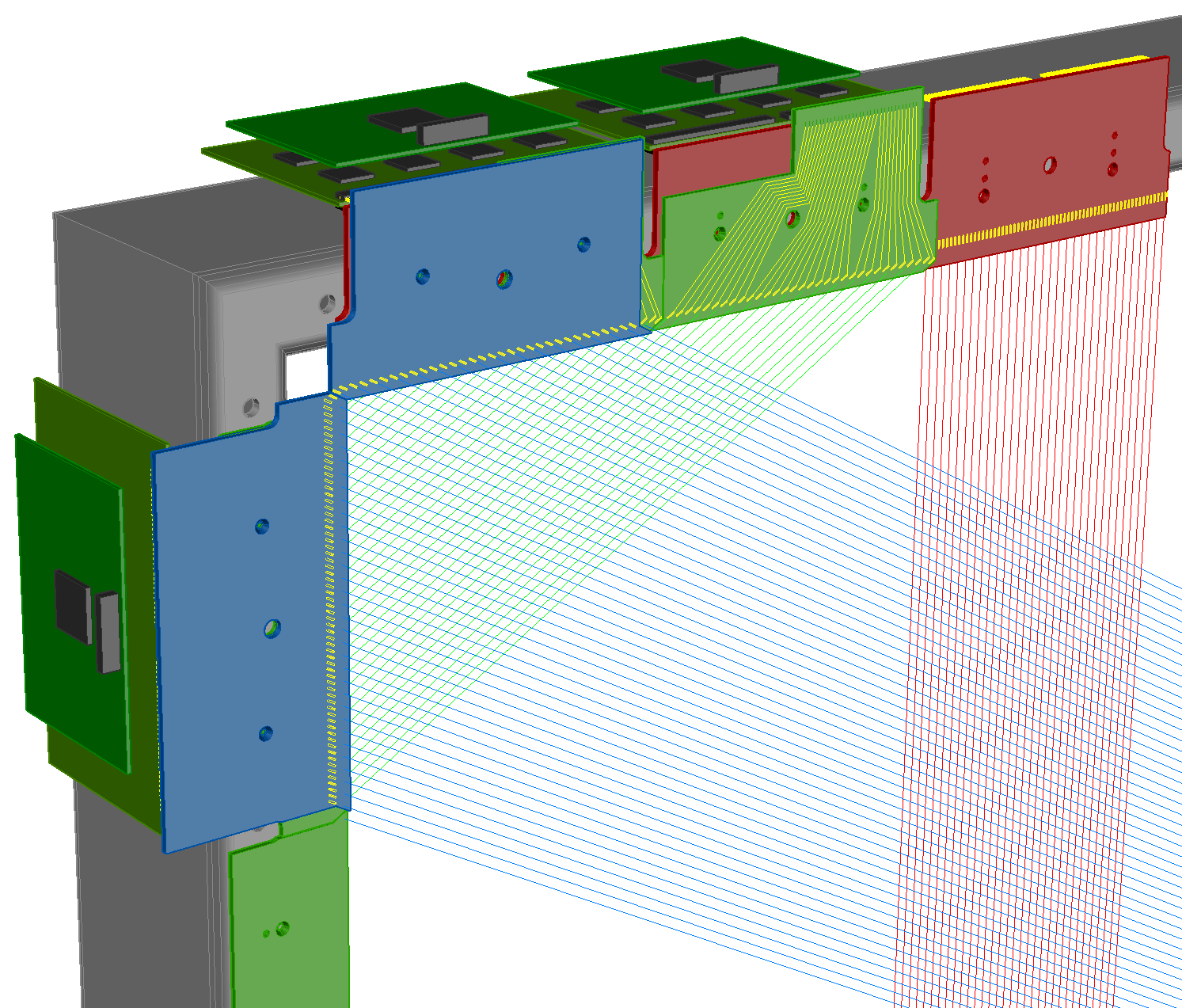}
\caption{\label{fig:TPCCornerDetail1}A conceptual design of an upper outer corner of an APA showing the different layers of wire bonding boards on the top end side of the APA frame.  The cold electronics boards are connected to these wire bonding boards.  Metal covers may be installed over the electronics boards to reduce noise pickup and contain the boiled argon bubble streams.}
\end{figure}

\begin{figure}
\centering
\includegraphics[width=0.8\textwidth]{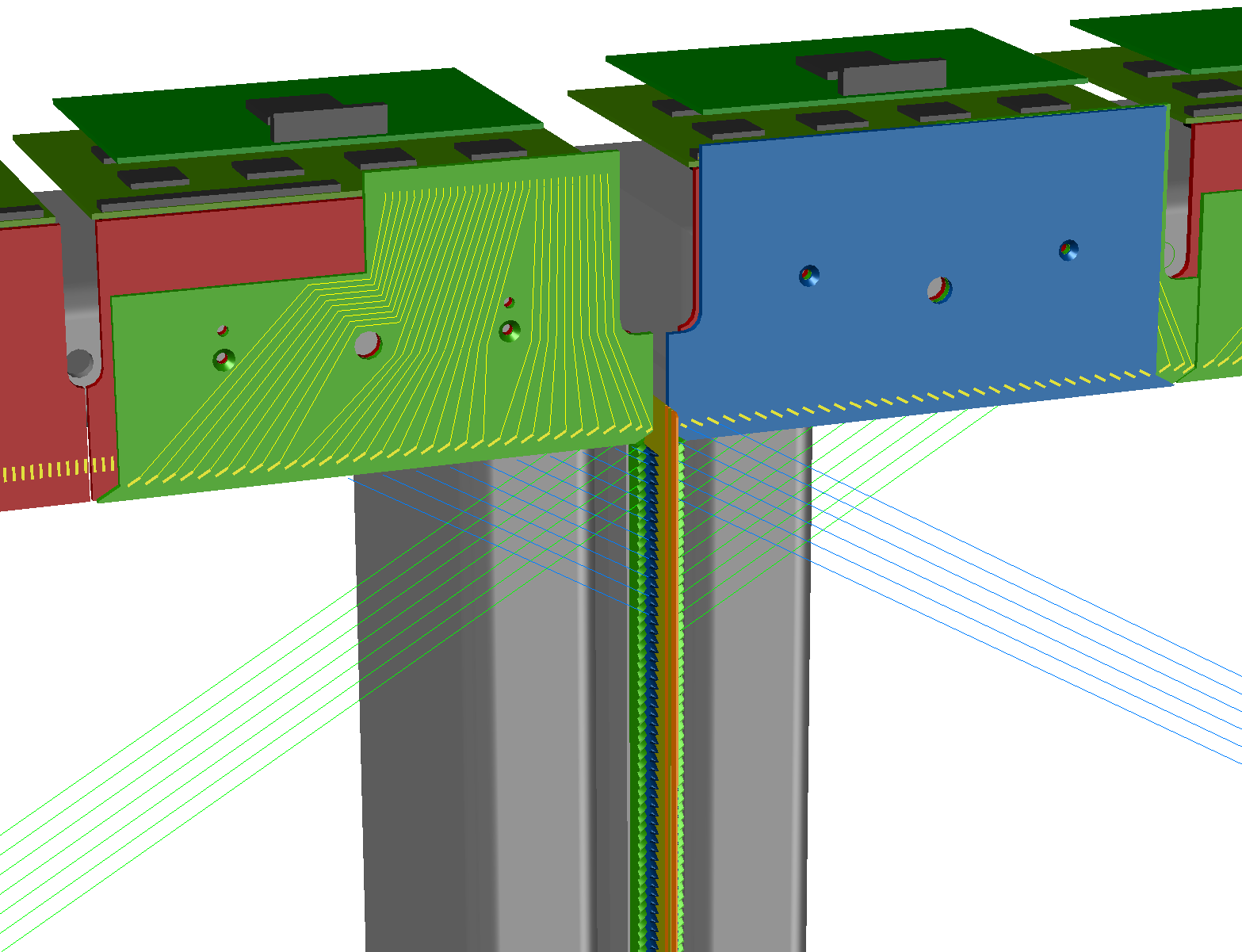}
\caption{\label{fig:TPCGapDetail} A closeup view of the gap between the two APAs showing a possible implementation of the electron diverter concept}
\end{figure}

\subsection{The Cathode Plane Assembly}
\label{CPA}

The CPA has the same dimensions as the APA and consists of a stainless-steel framework.
The surface of the CPA panels will be either a solid stainless steel sheet or a highly transparent  wire-mesh-plane. The requirements of the light collection system %(see Section~\ref{LCS_requirements}) 
will dictate which CPA surface is used.  For example, a double layer mesh cathode module enables TPB coated reflector foils to be mounted in between the mesh planes. This configuration allows the polymer foils to contract freely during cool-down.

During manufacture, all the CPA surfaces will be carefully polished in order to avoid any sharp edges that could lead to electrical discharge. If a transparent type CPA is chosen
the mesh will be tensioned and mounted between two steel frames in order to enclose the sharp edges of the mesh~(see Figure~\ref{fig:CPAzoom}). G10 mounting connectors will be pre-installed on the outer edges of the CPA to allow integration of the CPA and the Field Cage Assembly (FCA) modules. Finally, a HV cup will be integrated to allow connection of the HV feedthrough to the CPA.
 
 \begin{figure}
\centering
\includegraphics[width=0.5\textwidth]{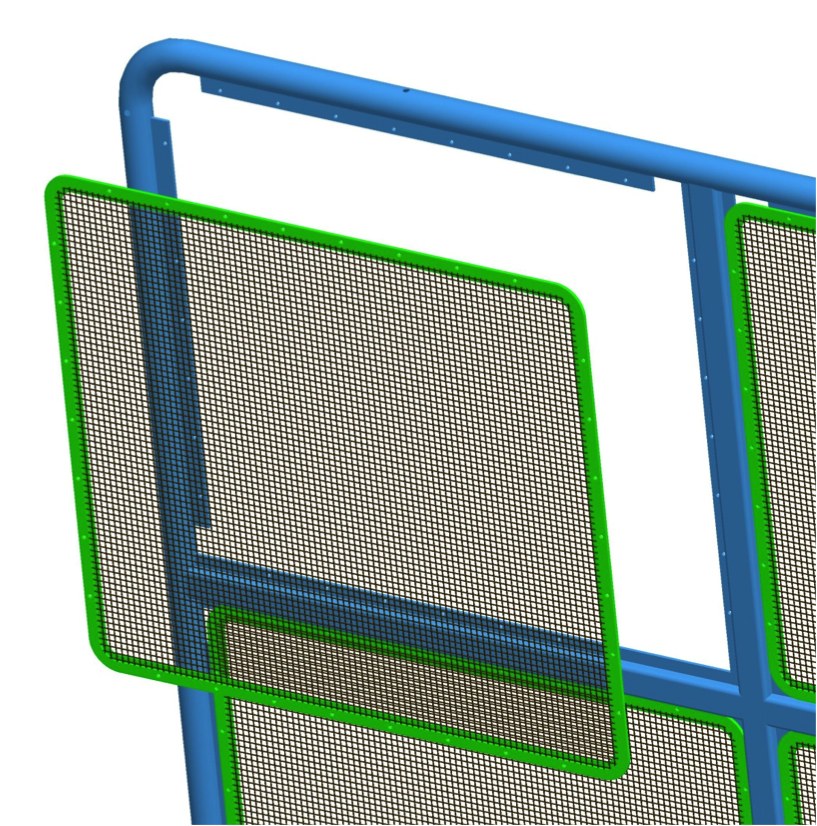}
\caption{\label{fig:CPAzoom}A model of the CPA corner, showing the mesh into the frame assembly.}
\end{figure}

\subsection{The High Voltage Feedthrough}

A single high voltage feedthrough is installed through a cryostat penetration to connect the high voltage power supply to the CPA. 
To provide the 500~V/cm drift field over the 2~m drift distance the CPA will sit at a potential of -100~kV.
To ensure safe and reliable operation, the HV feedthrough is constructed with a HV rating of no lower than 120~kV in liquid argon. 
The design of this feedthrough is based on the design of the ICARUS/ MicroBooNE feedthrough.
To avoid outgassing of impurities, the feedthrough uses a central stainless steel conducting core within an Ultra-High Molecular Weight Polyethylene (UHMW-PE) insulator surrounded by a stainless steel grounding sheath. UHMW-PE has a high dielectric strength, ensuring a compact feedthrough design, and its high-thermal expansion coefficient relative to stainless steel allows the vacuum seal to be cryo-fitted. The outer stainless steel sheath will terminate prior to the contact to the CPA to prevent electrical breakdown. As in the ICARUS/MicroBooNE design, the feedthrough will be removable from outside of the cryostat in the event that it needs to be replaced. A 150~kV power supply~(Glassman LX150R12)  in conjunction with a noise filter will be utilized as in MicroBooNE.

\subsection{The Field Cage}

The HV drift cage must provide a uniform 500~V/cm electric field over the TPC active volume to maintain linearity between drift distance and time of ionization. 
The field cage  gradually steps the voltage from the -100~kV applied to the CPA up to ground voltage.
It will be made from 1.6~mm double sided Cu clad FR4 PCBs, similar to the 35t LBNE prototype. The double sided Cu cladding effectively ensures that the
inter-strip capacitance is increased  thus minimizing over voltage conditions between the CPA and field cage strips in the event that the CPA or the field cage discharge to ground.
A photo of the 35t field cage panels is shown in Figure~\ref{fig:35tfieldcage}.
Slits between the Cu etched strips ensure good liquid argon flow.
Additionally, in order to accommodate the laser beam calibration system (see Section~\ref{Laser}) the side field cage walls will have two openings 50~mm ID in each drift volume at the half-height.
A resistor divider chain will supply the potential for the field cage electrodes. To reduce the field distortion caused by a possible resistor failure, four equal value 1~G$\Omega$ resistors, with each resistor rated at 5~kV and 1~W, will make parallel connections between neighboring electrodes. 
%\textbf{Zebra resistors?}.  
These resistors will be located on the inner wall of the TPC since the electric field in this volume is lower than that outside the TPC.  

Surge protection elements will be placed in parallel with the resistor divider chain to provide redundant protection to the resistors should a catastrophic voltage condition begin to arise in the chain.  This technique was instigated by the \uboone collaboration and shown to provide a reliable method of handling high-voltage breakdown issues~\cite{Asaadi2014}. 
%\textbf{refer to more papers}.  

\begin{figure}
\centering
\setlength{\fboxsep}{0pt}
\setlength{\fboxrule}{0.6pt}
\fbox{\includegraphics[width=0.7\textwidth]{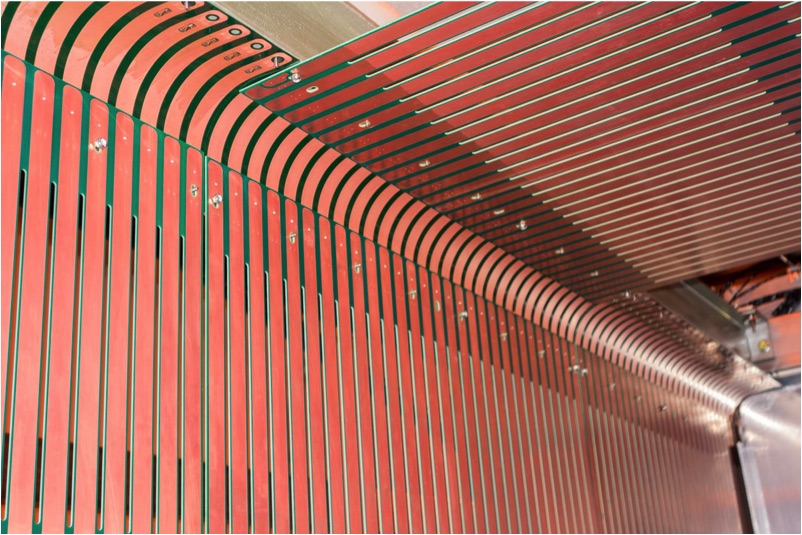}}
\caption{\label{fig:35tfieldcage}A photo of the 35 ton LBNE prototype field cage.}
\end{figure}

%\subsection{Prototyping}

%Build a 1/XXX size TPC

%\subsection*{Quality Assurance of APAs}

\subsection{Installation Procedure}
\label{TPC_installation}

There are two installation options for the TPC.
One option is to pass all TPC components through the cryostat chimney; this would require completion of the cryostat
prior to the TPC installation and the dimensions of all internal components would be restricted by the size of the chimney.
The second and more favorable option is to hang all the TPC assembly from a main lid before sealing. This option allows for parallel construction of the TPC and cryostat.
%\textbf{I don't understand....how is the roof of Fig. 1.a removable?}
However, this creates the necessity for the building housing the detector to be tall enough %(more than twice the height of the detector) 
and an adequate load capacity crane would be needed.

%\subsection{Alternatives Considered}

\subsection{Quality Control and Quality Assurance}
A QA program will be performed at Lancaster University to confirm the APAs constructed in the UK behave as expected when cooled to 77K using liquid nitrogen. These results will verify the design and manufacturing process of all APAs.
Cold tests will be carried out in a purpose-built $\sim$~3~m~$\times$~4~m~$\times$~0.2~m thermal cryo-vessel, large enough to contain the full-scale APA. The vessel will be constructed from stainless steel with polystyrene insulation. The vessel, with the APA inserted, will be cooled with gaseous nitrogen prior to filling with liquid nitrogen. The thermo-mechanical measurements will include: a survey of the bending and distortion of the APA frame structure and comparison with FEA calculations; vibration frequency based APA wire tension measurements; and resistance measurements to test electrical integrity. 

Travelers will be provided for all TPC components shipments detailing contents  and relevant instructions.

%Wire tension tests during APA construction...Travelers for all TPC components...etc...

\subsection{Risks}

Breakage of a single wire can jeopardize the detector's functionality, so a QA procedure will be followed to ensure the tension of the wires is never higher than that of their desired warm tension of 0.5~Kg, %($\textbf{update number}$),%
and wires are not subject to any kinking that could reduce their strength.  The tension of each wire installed on the APAs will be measured using a laser feedback system developed as part of the LBNE 35-ton project, and subsequently used by both MicroBooNE and LArIAT during TPC construction.  

Breakdowns in the liquid argon volume can produce over-voltage conditions across resistors in the field cage, which if damaged could produce distortions in the drift-field.  Surge protection devices, as developed by the MicroBooNE experiment, will be present mitigating the risk to the resistors.

%\subsection{ES\&H}

%Fermilab safety regulations consider CuBe as a hazardous material and as such any scrap will be deposited in dedicated waste containers.  

%Institutions that host detector construction are responsible for ES\&H at their institution. ES\&H guidance and oversight is provided by the LAr1-ND project.

%Safety procedures for work in membrane cryostat during construction...
\section{TPC Electronics, DAQ and Trigger}
\label{ElectronicsDAQTrigger}

\subsection{Introduction}
The aim of the TPC readout is to digitize, compress losslessly and record the TPC signals upon the reception of a variety of triggers such as neutrino beam and cosmic rays triggers generated by the light detection system as well as  external scintillation counters, calibration and random strobe triggers.  In order to fully reconstruct cosmic rays entering the drift space during an event drift time, data arriving from -1.28~msec before the trigger to +2.56~msec after the trigger (a total of 3 maximum drift times) will be recorded and compressed losslessly at each event. In a parallel stream, and as useful R\&D for LBNF, the readout will continuously record data, compress it and store it for a few hours awaiting for a possible supernova alert from the SNEWS network. 

The block diagram of a single TPC readout channel is outlined in Figure \ref{fig:LAr1-ND_Electronics}. The signal from each wire is pre-amplified and shaped by a CMOS analog front end ASIC, then digitized by a CMOS ADC ASIC inside the cryostat. The digitized signal is sent to an FPGA, which aggregates data from multiple ADC chips and multiplexes it to high speed serial links. The serial data is sent over cold cable through a feed-through to the warm interface board installed outside the cryostat on top of the signal feed-through. The warm interface board receives the electrical serial data from the cold electronics and converts it to optical signals for transmission over a fiber optical link to the TPC readout module, housed in a crate. Once the signal arrives at the TPC readout module, it is processed in an FPGA for compression, reduction, and storage. Processed data is buffered on board temporarily and then transmitted to DAQ PCs through the crate backplane and optical links. Data received on PCs is stored in hard drives for further analyses.

The design of the front end electronics, the TPC readout and the trigger is described in the next sections. A summary of the numbers of modules needed and of the spares as well as their distribution is reported in Table~\ref{tab:electronics_modules}.

\begin{figure}
\centering
\includegraphics[width=0.8\textwidth]{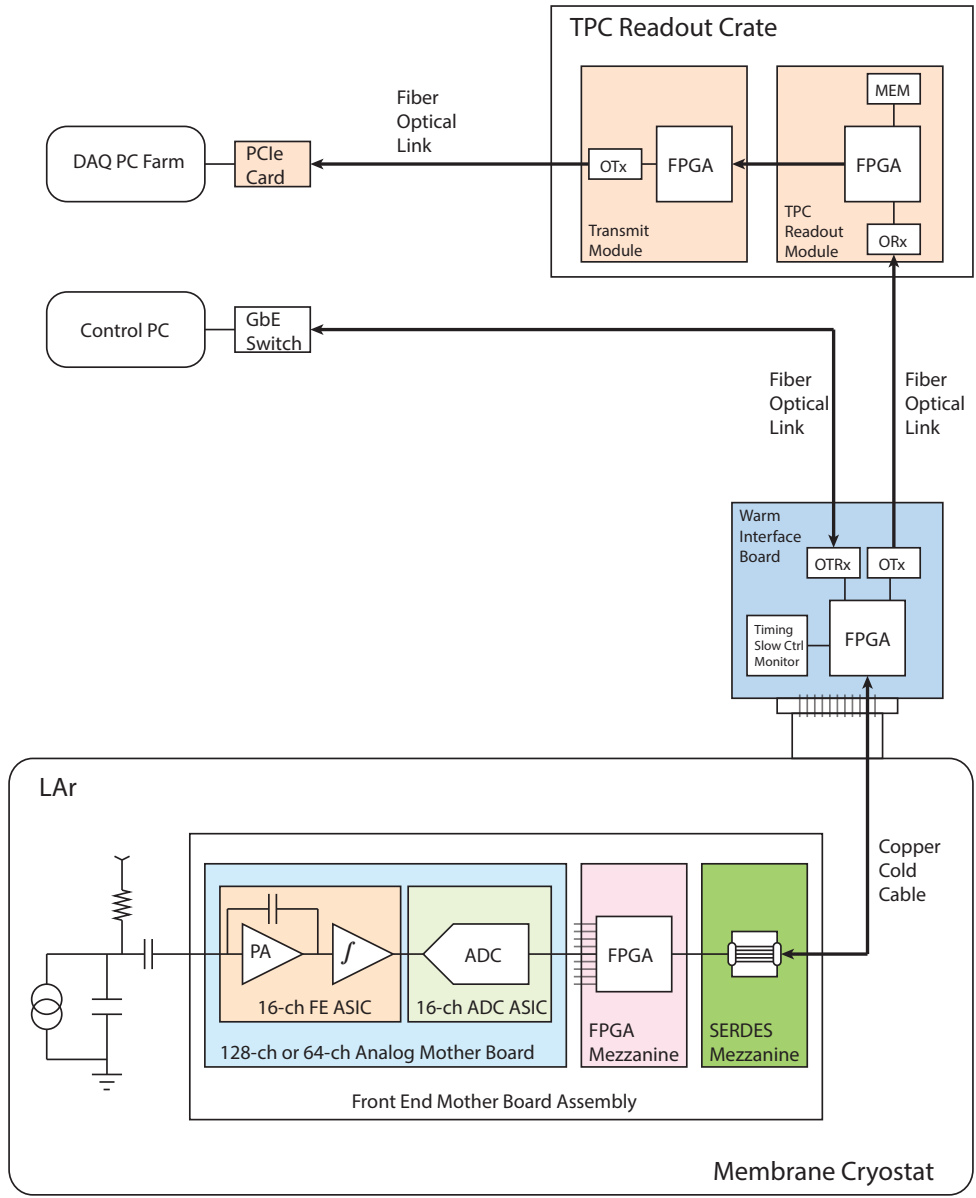}
\caption{\label{fig:LAr1-ND_Electronics}Information flow of a single TPC readout channel.}
\end{figure}

\subsection {The Front End Electronics}

The \larnd front end electronics is comprised of three parts: cold electronics, warm interface electronics and signal feed-through. The cold electronics will be installed on the TPC anode assembly and operated in LAr. The digitized detector signal will be sent to the warm interface electronics over cold cable. The warm interface electronics will be installed on the top signal feed-through assembly, and interface to both cold electronics and back end readout electronics and the DAQ system.

\begin{table}
{\footnotesize \begin{tabular}{cccccc}

%   \tablehead{1}{c}{b}{Module}
%   & \tablehead{1}{c}{b}{Channels\\ per Module}
%   & \tablehead{1}{c}{b}{Distribution}
%   & \tablehead{1}{c}{b}{Number of \\
% Modules needed}
%   & \tablehead{1}{c}{b}{Number of \\Spare Modules} 
% & \tablehead{1}{c}{b}{Total Number\\of Modules}\\

Module & Channels &  Distribution & Number of & Number of & Number \\
& per Module & & Modules needed & Spare Modules & of Modules \\
                
\hline
Front End Modules (FEM) & 64 wires/FEM& 16 FEM/crate  & 176 & 18 & 194\\
Crates &  &  & 11 & 2 & 13\\
Backplane  & & 1 Backplane/crate & 11 & 2 & 13 \\
XMIT  & & 1 XMIT/crate & 11 & 2 & 13\\
Crate Controller  (CC) & & 1 CC/crate & 11 & 2 & 13 \\
PCIe & & 3 PCIe/crate &  33 & 4 & 37 \\
Trigger module & &  & 1 & 2 & 3\\
 
\hline
Analog FE ASIC & 16 wires &  & 704 & 96 & 800\\
ADC ASIC & 16 wires &  & 704 & 96 & 800\\
128-ch Mother Board& 128 wires & & 52 & 6 & 58\\
64-ch Mother Board & 64 wires & & 72 & 8 & 80\\
FPGA Mezzanine & 128 or 64 wires & & 124 & 14 & 138\\
SERDES Mezzanine & 128 or 64 wires & & 124 & 14 & 138\\
Warm Interface Board & 704 wires & & 16 & 4 & 20\\
Service Board & 2,816 wires & & 4 & 2 & 6\\
Signal Feed-through & 2,816 wires & & 4 & 2 & 6\\
Cold Cable & 2,816 wires & & 4 & 2 & 6\\
\hline
\end{tabular}}
\caption{Numbers of modules needed and of spares as well as their distribution.
\label{tab:electronics_modules}}
\end{table}

\subsubsection*{Cold Electronics}

The \larnd TPC will have two APA modules,
%each module has 5,632 channels, total 11,264 readout channels.
on each side with \apach~channels, a total \detch~readout channels for the whole TPC. The large number of readout channels required to instrument the \larnd TPC motivates the use of CMOS ASICs for the electronics. Both analog FE ASIC and ADC ASIC, to a large extent, have already been developed for LBNE, and analog FE ASIC is being used in MicroBooNE. The entire front end electronics chain is immersed in the LAr and operates at 87 K to achieve an
optimum signal to noise ratio. It is composed of a 16-channel analog FE ASIC providing amplification and shaping, a 16-channel ADC ASIC implemented as a mixed-signal ASIC providing digitization, buffering and the first stage of multiplexing, a FPGA providing the second multiplexing stage, and voltage regulators. Analog FE ASICs, ADC ASICs plus a FPGA implementing multiplexer, clock, control and monitoring circuitry comprise a single 128-channel or 64-channel front end mother board assembly. The FPGA on each motherboard will transmit data out of the cryostat on twinax copper pairs running at multiple Gbit/s through a feedthrough to the warm interface electronics, and receive programming instructions and timing information from the warm interface electronics as well.

\begin{figure}
\centering
\includegraphics[width=0.8\textwidth]{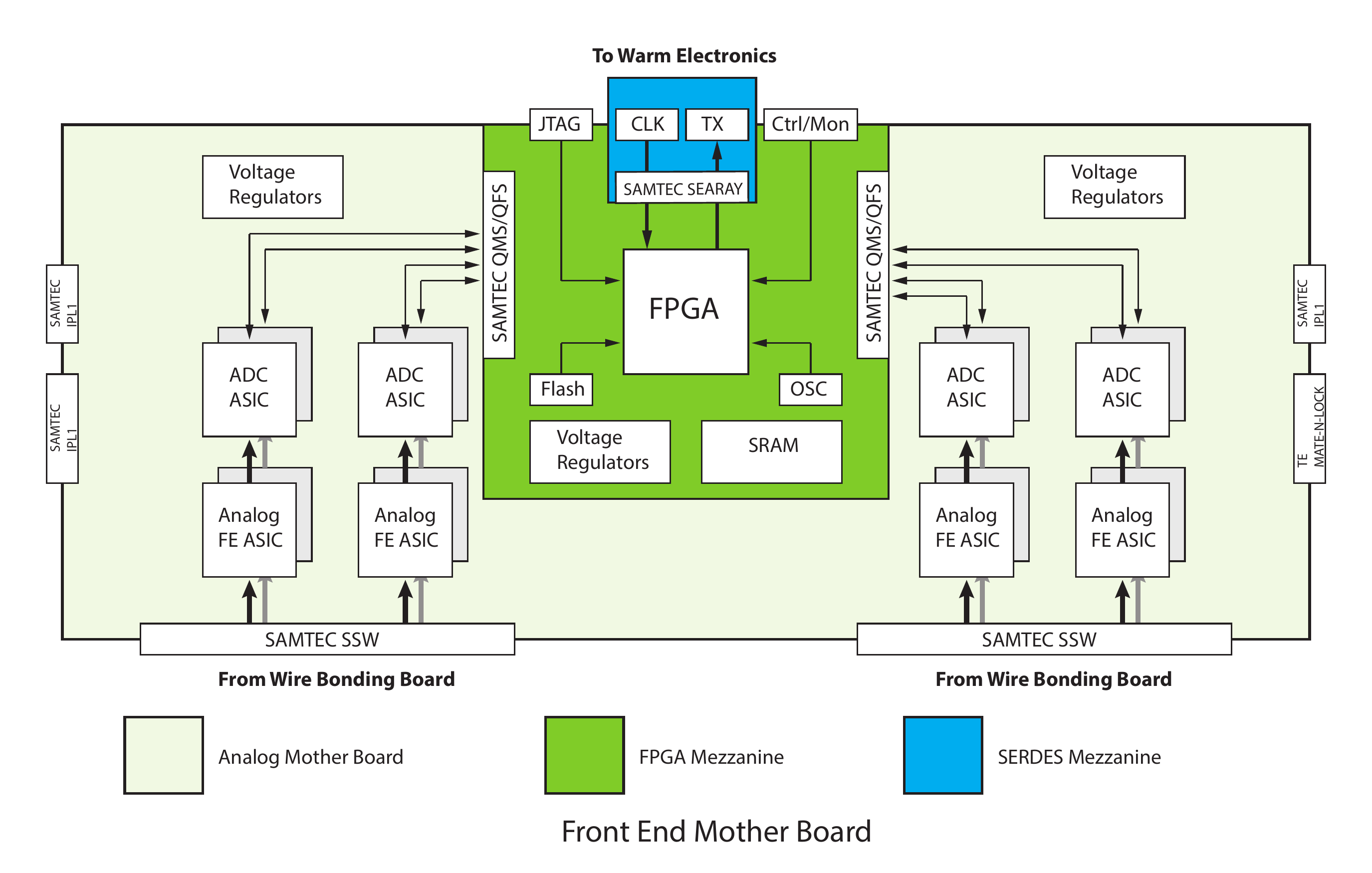}
\caption{\label{fig:FEMB}Block Diagram of Front End Mother Board Assembly.}
\end{figure}

Each side of the TPC has two APA modules, which are interconnected along the adjacent side. Each APA module will have 31 front end mother boards on two sides. 13 boards will be located on top of the TPC with each board processing 128 channels. 18 boards will be on the side without interconnection, with each board housing 64 channels. A block diagram of the 128-channel front end mother board is shown in Figure \ref{fig:FEMB}. Both analog FE ASIC and ADC ASIC have been designed and fabricated in a commercial CMOS process (0.18~$\mu$m and 1.8~V). This guarantees a high stability of the operating point over a wide range of temperatures, from room temperature to 77~K. The ASICs are packaged in a commercial, fully encapsulated plastic QFP 80 package. A minor revision of the analog front-end ASIC, to further improve the robustness and simplify the system design, including internal pulse generator, smart reset and improved input protection is being planned. It will greatly simplify the design of the electronic calibration system. 

The Cold FPGA will interface to analog FE ASICs and ADC ASICs on the analog mother board. It will control and monitor ASICs, and receive data from ADCs. Once data arrives at the FPGA, a second stage multiplexer will be implemented to further reduce the number of data links to outside of the cryostat. The design is to support transparent data readout without any compression over 2Gbit/s serial links. An efficient zero-suppression scheme can be implemented in the FPGA to greatly reduce the total data volume if proven to be feasible and necessary. Each mother board processes 128 or 64 detection channels. The clock will come in through RX links while data is sent out over TX links. Voltage regulators used on board have been qualified in liquid Nitrogen. On the board, SRAM is used to temporarily buffer events if a more sophisticated algorithm is used to process data. A commercial SRAM chip working in cryogenic temperature has also been identified. The estimated power dissipation is ~20 mW/channel. The ALTERA Cyclone IV GX FPGA will be used in the FPGA mezzanine design. It has been tested in LN2 successfully, both fabric logic and a high speed transmitter are working properly at cryogenic temperature.

\begin{figure}
\centering
\includegraphics[width=0.8\textwidth]{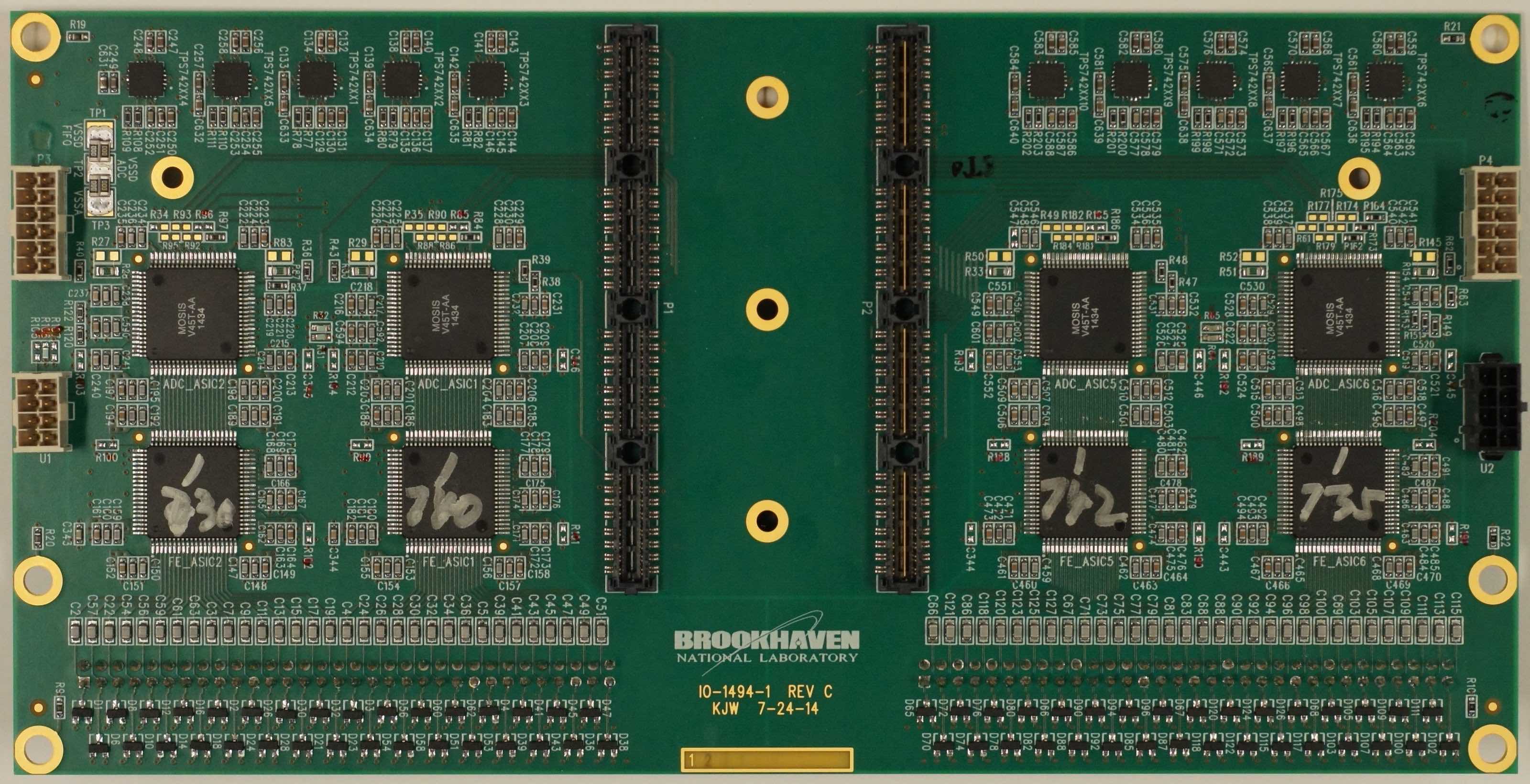}
\caption{\label{fig:AMB}Prototype 128-ch analog mother board designed for LBNE 35ton prototype TPC.}
\end{figure}

\begin{figure}
\centering
\includegraphics[width=0.8\textwidth]{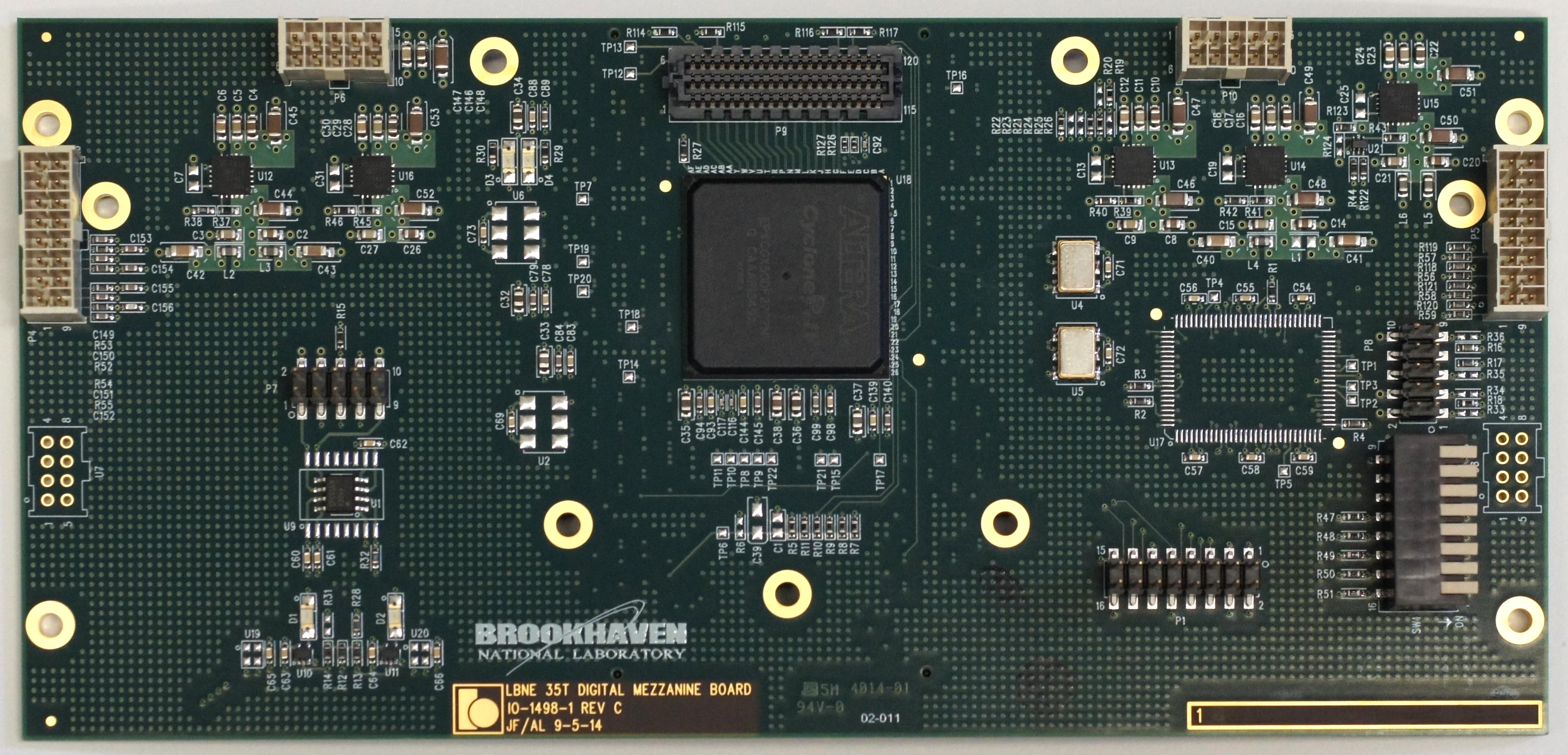}
\caption{\label{fig:FPGAMezz}Prototype FPGA mezzanine designed for LBNE 35ton prototype TPC.}
\end{figure}

The prototype front end mother board is being designed for the LBNE 35 ton prototype. The picture of the prototype 128-ch analog mother board is shown in Figure \ref{fig:AMB}  and prototype FPGA mezzanine is shown in Figure \ref{fig:FPGAMezz}. SERDES mezzanine is a passive adapter board, which will be plugged on the FPGA mezzanine to interface to cold cable. The candidate cold cable is made by Gore using twinax cable and ERNI hard metric connector. The 50ft Gore twinax cable has been tested with feed-through pin carrier running at 2 Gbit/s successfully.

\subsubsection*{Warm Interface Electronics}

The detector signal is digitized inside the cryostat. After multiplexing in FPGA, it is sent out of the cryostat over copper serial link to the warm interface electronics installed on the top of the signal feed-through. The warm interface board will be the bridge between the cold electronics and back end readout electronics and DAQ system.

The warm interface board will use ALTERA Cyclone V FPGA, which will receive high speed serial link coming out of cryostat, perform data preparation, then send the data to the readout and DAQ system through fiber optical links. On board FPGA has computing power to do further data processing if necessary before sending the data to downstream electronics. By default, data interface board will send data over a 2 Gbit/s link to the TPC readout module, which processes un-compressed 64-ch worth of data. The whole TPC will require 176 fiber optical links to carry data from \detch~ TPC channels. A 12-ch parallel fiber optical link will be used to minimize the volume required on board, requiring a total of 16 fiber bundles. This will ease the design of the electronics assembly on the top of the signal feed-through. 

Timing, control and monitoring are also functions of the warm interface electronics. The system clock and synchronization signal will be distributed from warm interface board to the front end mother board assembly. Slow control and monitoring information will also be communicated between the warm interface board and the front end mother board assembly, using an I2C like protocol. The front end mother board assembly can be remotely programmed and monitored through a control PC. The communication between the warm interface board and control PC is by Gigabit Ethernet over a fiber optical link.

The front end analog ASIC has a built in calibration capacitor, to facilitate the electronic calibration. With plans to revise the front end analog ASIC adding a built-in calibration pulse generator, there is no need to distribute the calibration pulse signal from outside the cryostat. This will greatly simplify the system design.

The link between warm interface electronics and back end readout electronics and control system is only through fiber optical links. This effectively eliminates the possibility to have ground loops between the detector and the DAQ system.

Both cold electronics and warm interface electronics will be powered by a floating low voltage power supply. It is planned to have a service board installed on the top of the signal feed-through. The service board will be responsible for power management and distribution.

\subsubsection*{Signal Feed-through}

\begin{figure}
\centering
\setlength{\fboxsep}{0pt}
\setlength{\fboxrule}{0.6pt}
\fbox{\includegraphics[width=0.8\textwidth]{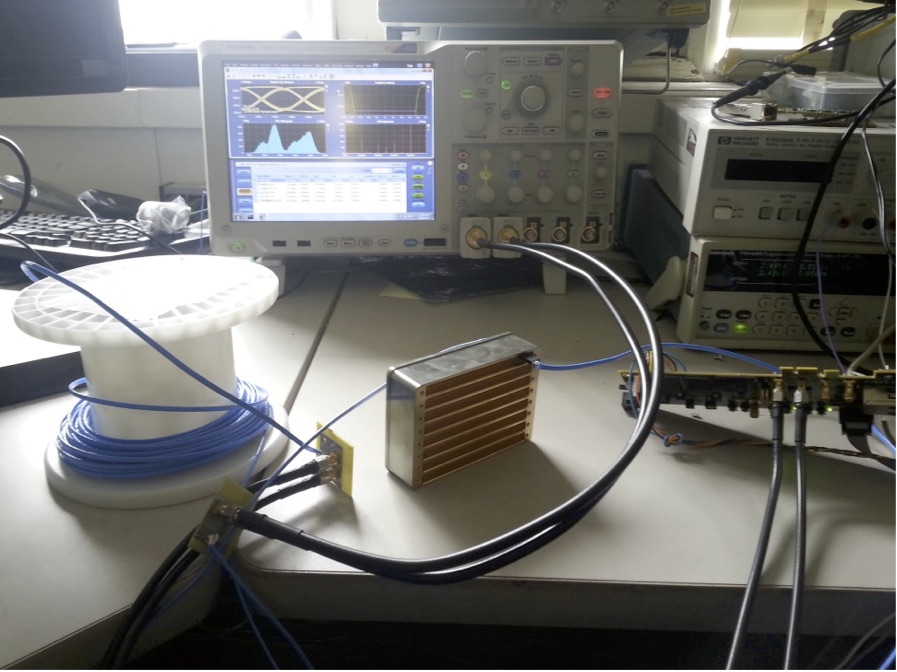}}
\caption{\label{fig:PinCarrier} ATLAS pin carrier is tested with Gore cable for 2 Gbit/s signal transmission.}
\end{figure}

The signal feed-through design has to consider two important factors: 100\% hermeticity and high speed signal transmission capability. The ATLAS style pin carrier~\cite{atlas_ft} shown in Figure \ref{fig:PinCarrier} is designed for a LAr Calorimeter and its hermeticity has been certified. It has also been used for the MicroBooNE signal feed-through design. Since the pin carrier design is available, no additional engineering design is needed. The pin carrier is suitable for both warm flange and cold flange if \larnd decides to use double flanges setup to improve the efficiency of the purification system. The manufacturer of the pin carrier has been contacted; they are still available and capable of building the same type of pin carriers with reasonable cost.

\begin{figure}
\centering
\setlength{\fboxsep}{0pt}
\setlength{\fboxrule}{0.6pt}
\fbox{\includegraphics[width=0.8\textwidth]{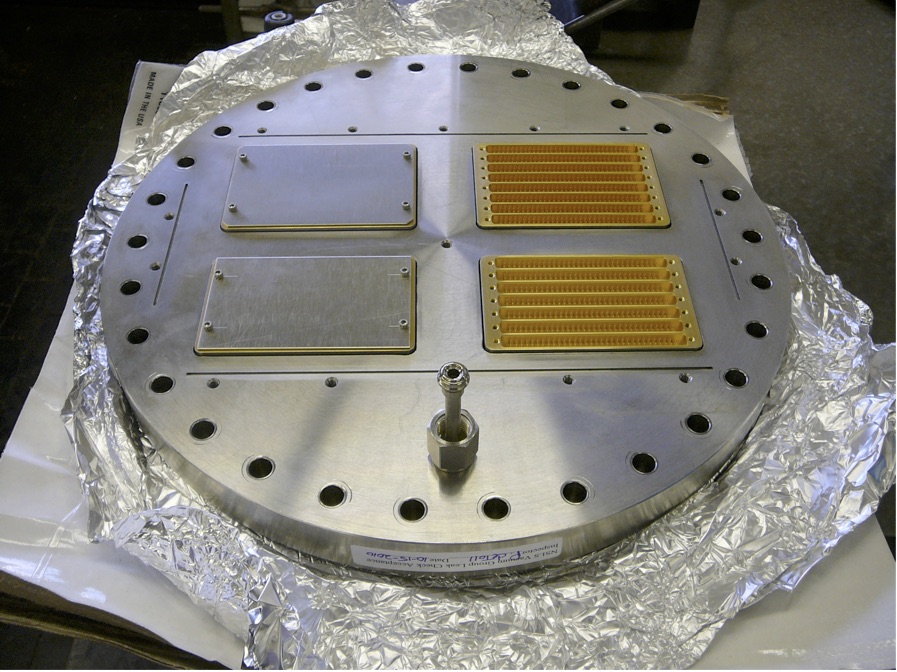}}
\caption{\label{fig:Flange}ATLAS pin carriers are welded on flange.}
\end{figure}

Two 8-row pin carriers and two 7-row pin carriers will be welded on a 14-inch conflat flange with a Faraday cage mounted on the top to provide the shielding for warm interface electronics. A picture of the flange with pin carriers welded on is shown in Figure \ref{fig:Flange}. There will be four signal feed-throughs for \larnd TPC readout. Each feed-through is used to read out 
%one anode plane assembly
one APA of 2,816 channels. A total of 44 serial links on each feed-through will come out of the cryostat running at 2 Gbit/s. Considering redundancy, ground pins and fiber organization, it will require 8 64-pin rows. The warm interface board will need proper cooling; it is envisioned that each board will occupy two pin carrier rows to allow sufficient air flow. The service board will occupy another two rows for power management and distribution. Therefore, the signal feed-through flange with 1920-pin, 30-row pin carriers is enough to handle one full anode plane assembly. It will be investigated if the readout of two APAs on one side of TPC could be fit on the one signal feed-through assembly.

\subsection {The TPC readout}

In MicroBooNE, the TPC signals are digitized outside of the cryostat in a board developed by BNL and joined to a Nevis board, the FEM, that provides the compression, storing and trigger application. In \larnd the signals will be digitized within the cryostat and after emerging from the feedthrough, will go through a copper to optical transceiver and arrive on optical fibers at an adaptation of the MicroBooNE FEM  Nevis boards. The FEMs will each receive signals from 64 wires  and will be modified to include an optical to copper converter AFBR-59R5LZ and a deserializer TLK2501IRCP. The signals will then be treated in exactly the same way as in MicroBooNE, thus capitalizing on the extensive Nevis hardware and firmware design and development performed for MicroBooNE 

A total of 176 FEMs will be required and will be housed in eleven 6U crates, sixteen 64-channel FEMs per crate. The crates each also house a fast data transmission module (XMIT) and a crate controller (CC) used for parameter transmission to the FEMs and for a slow debugging mode readout. A MicroBooNE crate layout  is shown in Figure \ref{fig:TPCCrate}.
\begin{figure}
\centering
\includegraphics[width=0.4\textwidth]{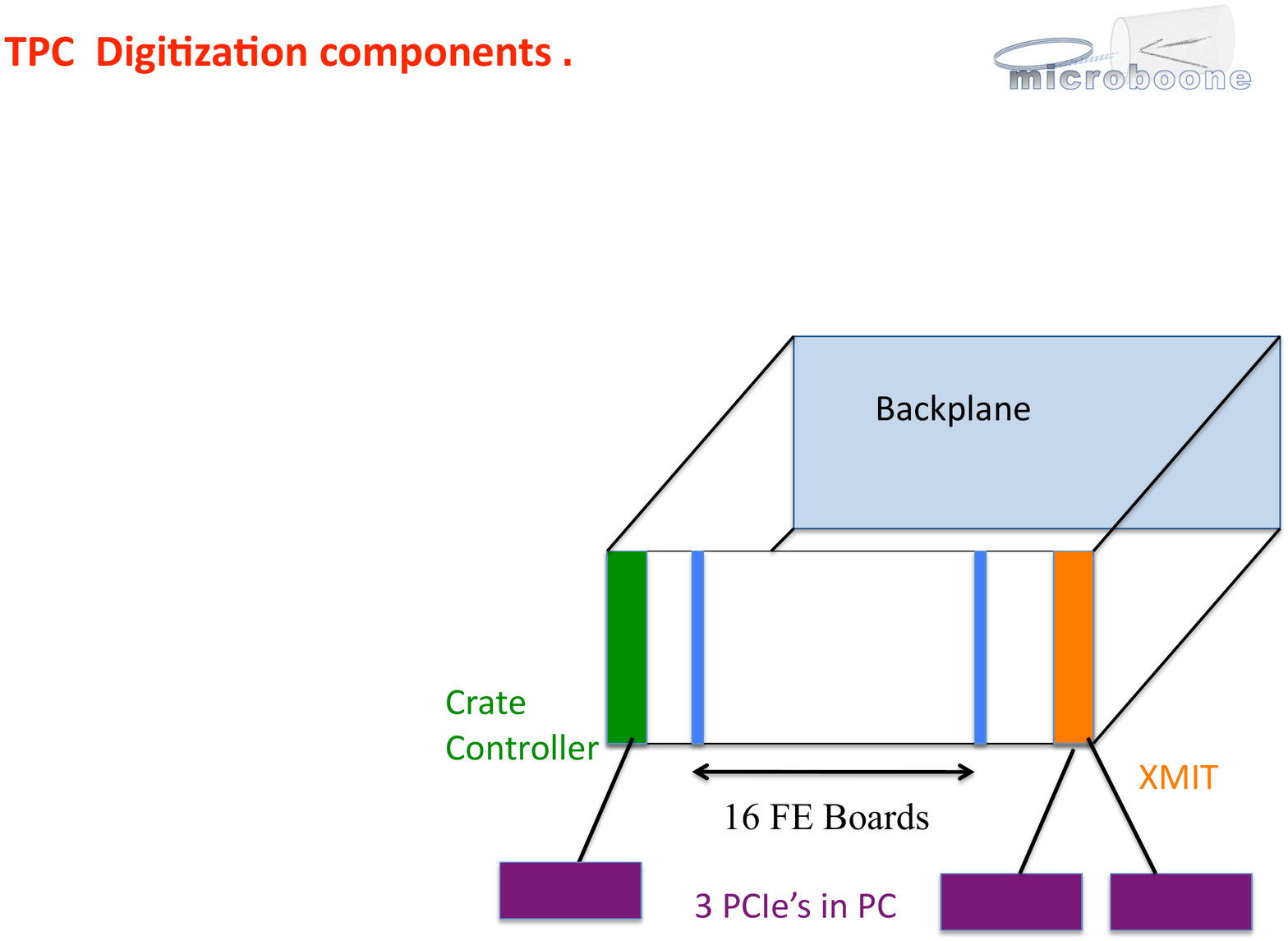}
\caption{\label{fig:TPCCrate}A readout crate showing, from left to right, the crate controller, up to 16 FEM modules, the XMIT and the 3 PCIe cards used for communicating with the crate and resident in a PC.}
\end{figure}

The digitized data stream is shown in Figure \ref{fig:Path}.  The FPGA stores the data from 64 wires sequentially in time in a 1M x 36 bit 128 MHz SRAM memory, grouping two ADC words together in each 36 bit memory word. This requires a data storage rate of (64/2) x 2 MHz = 64 MHz. Since data reduction and compaction algorithms rely on the sequential time information of a given wire, the data readout out from this SRAM memory takes place in wire order in alternate clock cycles, again at the rate of 64 MHz. The SRAM chip size and memory access speed permit continuous readout of the TPC data. The data is arranged in frames of 1.28~ms, the maximum drift time. Since the readout clock is not synchronous with the accelerator spill time, the 3.84 ms worth of data relevant to an accelerator neutrino event spans four 1.28 ms long frames. In order to reduce the amount of data being transmitted, the FPGA trims the four frames  to span the exact 3.84 ms required.
\begin{figure}
\centering
\includegraphics[width=0.5\textwidth]{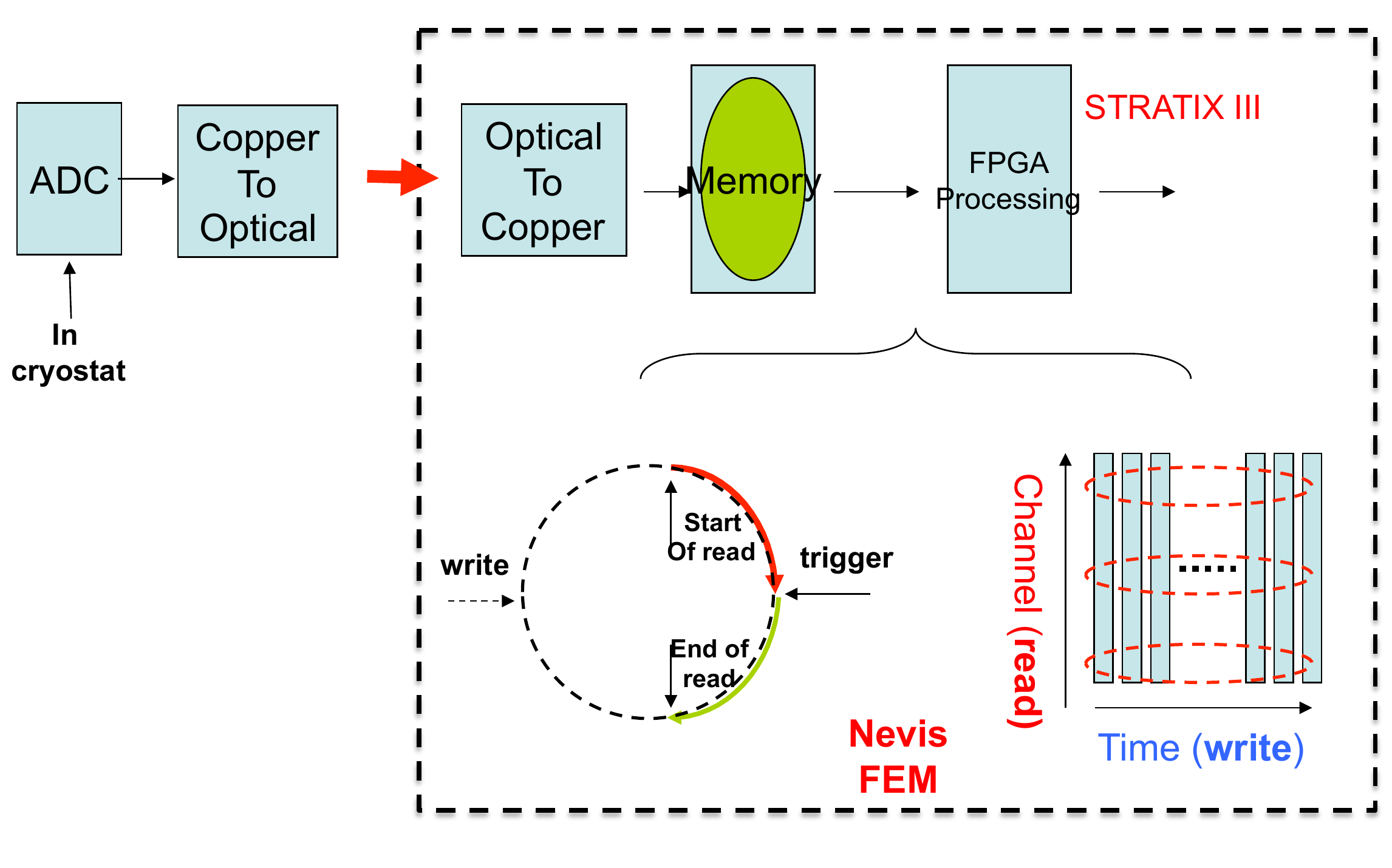}
\caption{\label{fig:Path}The Nevis FEM data flow, showing the circular buffer and the "write" in time order and "read" in channel order scheme.}
\end{figure}
Experience with a Fermilab test stand demonstrates that on any given wire, successive data samples vary relatively slowly in time. In most cases, two adjacent data samples either coincide or differ by one ADC count. Huffman coding provides for lossless data compression by taking advantage of this slow variation of the data stream. For accelerator neutrino events, lossless Huffman coding compression yields a compression factor of eight to ten and proves sufficient; but for the continuous supernova data, further compression by about an additional factor of ten, is necessary to limit the size of the data sample and to match disk writing speeds, resulting in unavoidable data loss. A method called dynamic decimation (DD) handles this case. The DD scheme relies on recognizing regions of interest (ROI) in the data stream that contain waveforms corresponding to drift charges. Parts of the data stream not containing ROI contribute to pedestal determination. In DD the FPGA samples the ROI data at the same rate as accelerator data, but reduces the pedestal sampling to a much lower rate (e.g. 1/16). The final data record of a wire contains full coverage in time, with or without drift signals from a charged track. A too-high threshold for ROI can result in loss of resolution for small signals, but the data still appear as pedestal, although sampled at a lower rate. An independent Huffman coding stage further reduces the data volume after dynamic decimation.
After going through their respective compression schemes, the beam and supernova data are stored in two separate DRAM buffers as shown in Figure \ref{fig:Components}. 
\begin{figure}
\centering
\includegraphics[width=0.5\textwidth]{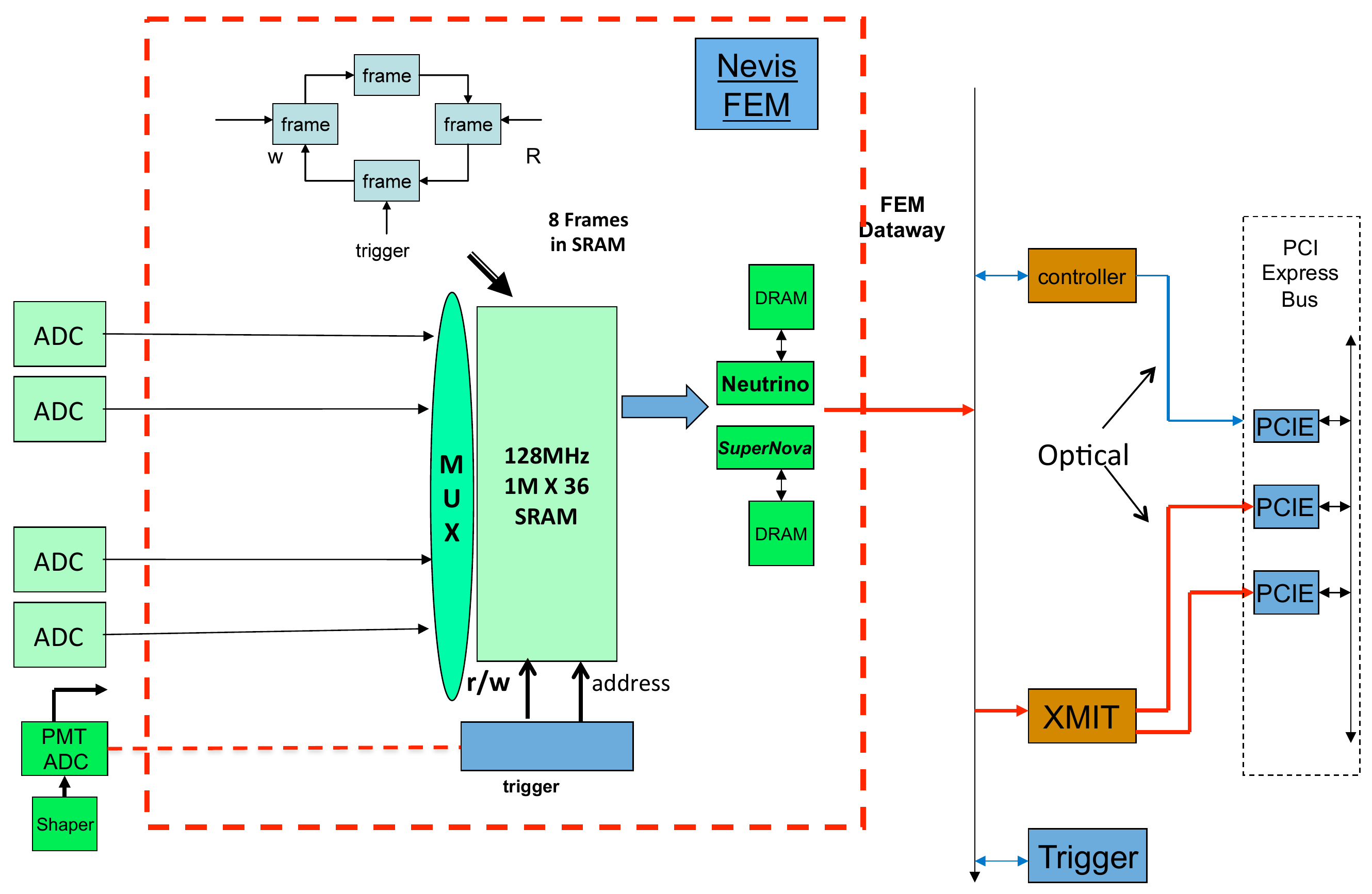}
\caption{\label{fig:Components}The Structure of the Nevis readout system: FEMs, dataway, XMIT, crate controller and three PCIe cards PC resident.}
\end{figure}
The data is then transmitted to the crate backplane dataway on connectors shown on the right of the schematic. A photograph of the Nevis FEM board currently in use in MicroBooNE is shown in Figure \ref{fig:FEMPix}. 
\begin{figure}
\centering
\includegraphics[width=0.5\textwidth]{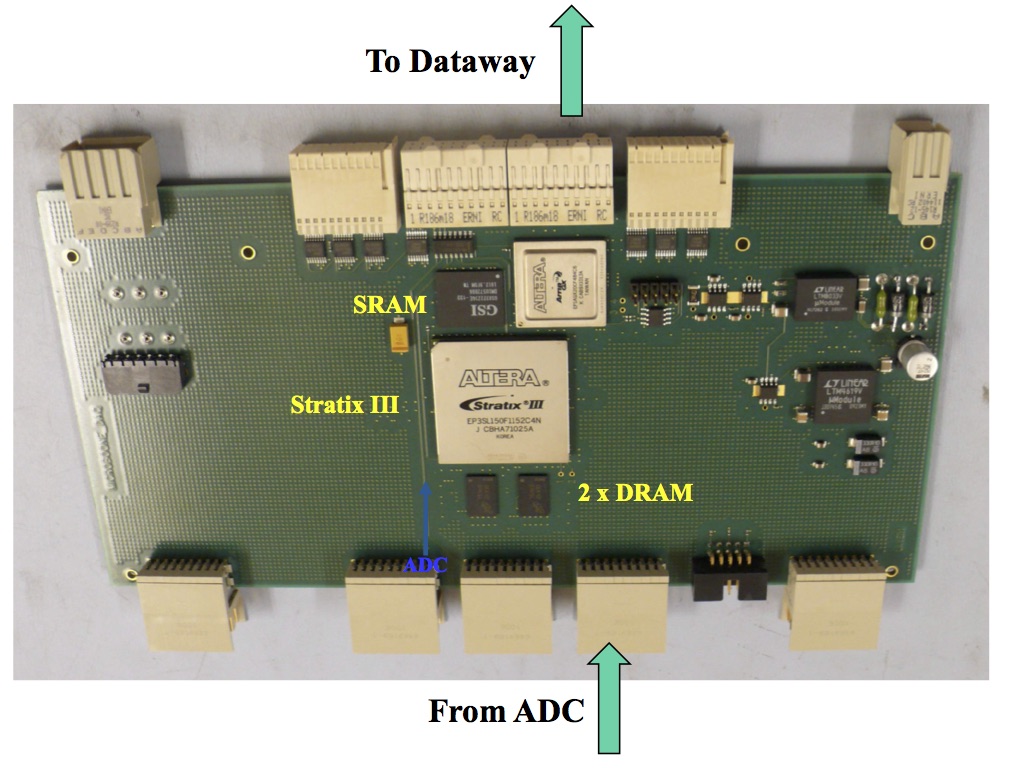}
\caption{\label{fig:FEMPix}The MicroBooNE Nevis FEM.}
\end{figure}
The transmission is controlled by the XMIT module. Each XMIT module includes two optical links, one used for the triggered data stream and the other for the continuous supernova data stream. These links connect to optical transceivers housed on two PCI Express interface cards developed at Nevis for use in ATLAS and resident in a PC. Each crate is connected to a dedicated PC (the sub-event PC). The two transceivers on the interface cards handle 6.4 GB/s traffic and connect to a 4-lane PCI Express bus, with each lane accommodating 2.5 GB/sec. The XMIT module transmits data to the PC based on the token passing technique. The XMIT module generates the token and passes it to the first FEM module. This module receives the token and transmits its data, if any. The data transits from one board to the next on the crate backplane until it reaches the XMIT module. The token passes on to the next FEM module when an active FEM finishes transferring its data. Since a module only drives the data to its neighbor, it forms a point-to-point short link. Data can flow at rates up to 512 MB/sec on this path, a factor of ten faster than the expected traffic on this dataway. Figure \ref{fig:BackPl} shows the layout of the crate backplane. In addition to the token passing dataway, a secondary bus serves to download the FPGA codes and initiate data/parameter readback via the crate controller and a third PCI Express card in the PC. This slower readback can be used to read out the FEM without the XMIT, a useful feature for system development, for operating a lightweight test stand, and for other purposes.

\begin{figure}
\centering
\includegraphics[width=0.5\textwidth]{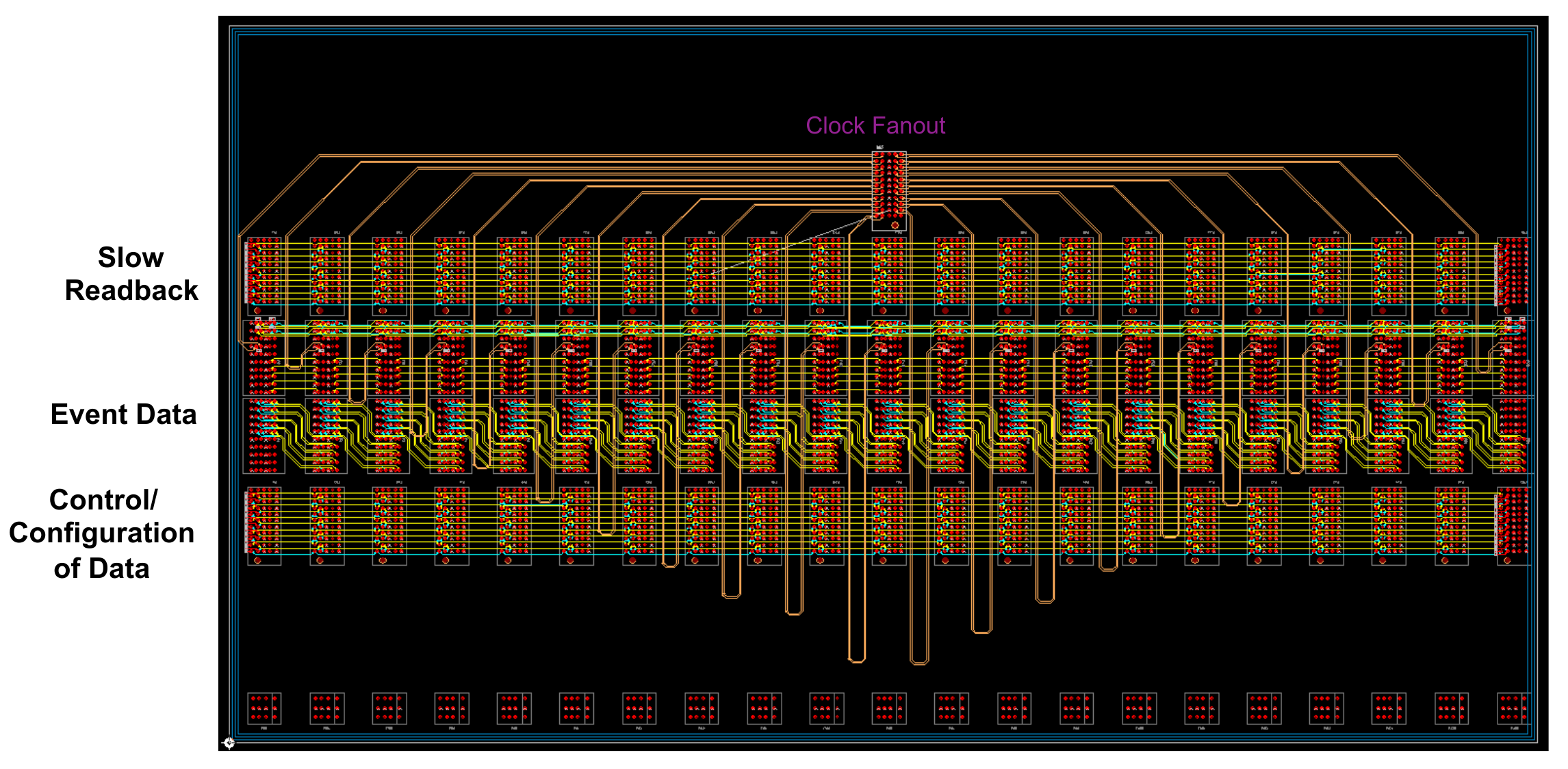}
\caption{\label{fig:BackPl}The Readout crate backplane.}
\end{figure}

An additional PC, the event-building PC, collects the triggered data sent to it by the eleven sub-event PCs and builds triggered events through a switch. Each sub-event PC also stores the supernova data in circular buffers large enough to accommodate a few hours of continuous data. The buffers are resident in disk drives that can accommodate 100 MB/sec writing speeds, but for which we have assumed a more conservative 50 MB/sec by providing an overall supernova compression factor of 80-100.

\subsection {The Trigger Board}
The trigger board (TB) flags time frames that must be treated differently than those for continuous reading of supernova events. The light detection system generates one or more triggers based on the detector signals. Example trigger conditions include: sum of all light detector pulse heights above a threshold, sums of groups of light detector pulse heights above a lower threshold, and number of light detectors above the threshold satisfying a multiplicity requirement. Each trigger condition receives a code ranging from 1 to n, with n likely not exceeding 7, and 0 meaning no trigger. This code is transmitted serially one bit at a time to the TB on one cable as shown in Figure \ref{fig:TrigB}. A second cable carries a marker to identify the first bit of a trigger code transmission. 
\begin{figure}
\centering
\includegraphics[width=0.5\textwidth]{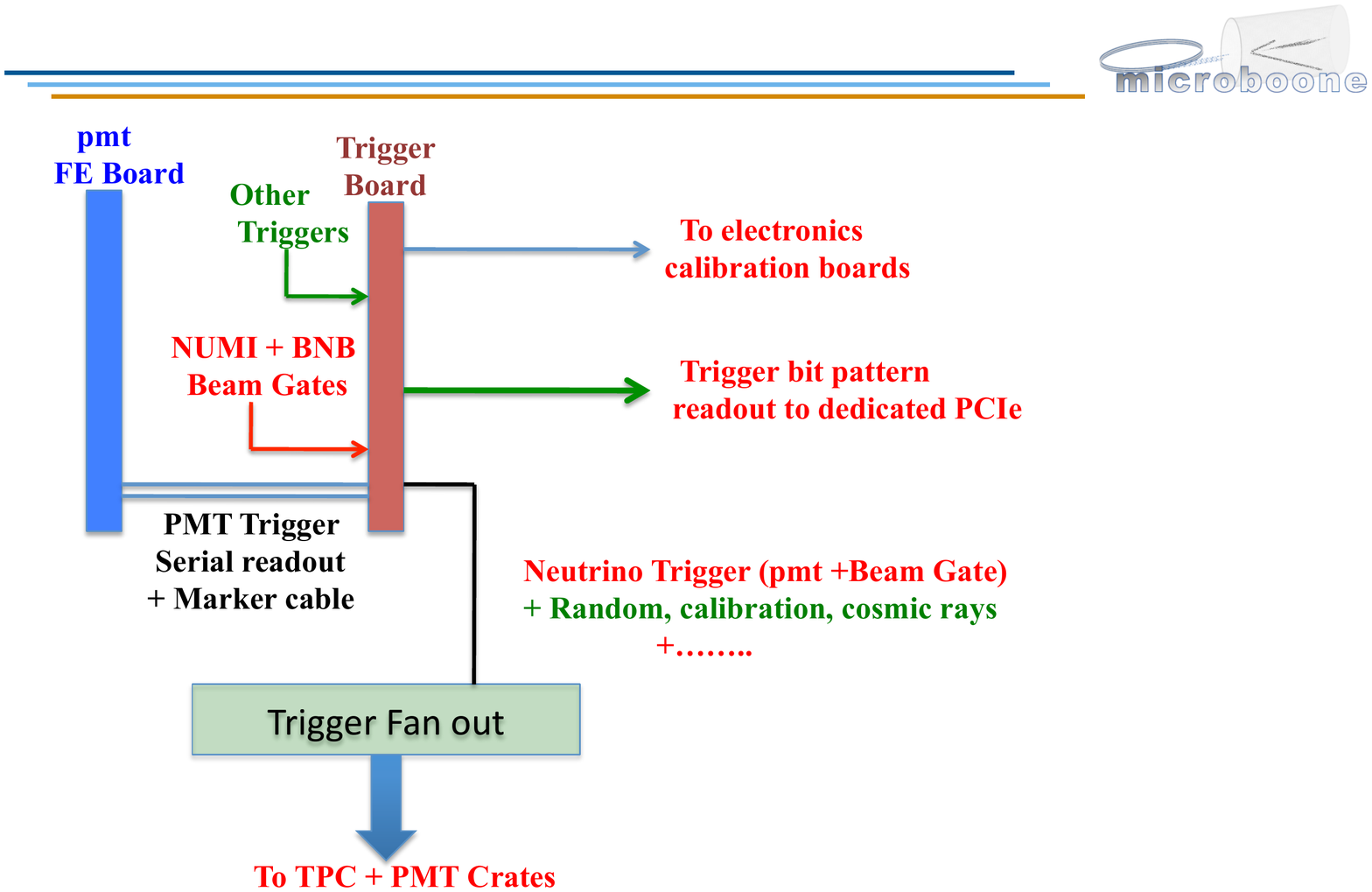}
\caption{\label{fig:TrigB}The MicroBooNE trigger scheme.}
\end{figure}

Booster and NUMI beam gates also input to the TB, where they can be placed in coincidence with light detector triggers to flag neutrino interaction candidates. Valid light detector triggers enter a logical OR with other utility triggers sent to the trigger board to form calibration triggers for the electronics, random triggers for noise measurement, off-beam triggers for cosmic ray response studies, and others as deemed necessary. 
All inputs to the trigger board will be via front panel LEMO connectors, as will the trigger output. An OR of all triggers passes to a fan-out module on a single cable, and from there is distributed to all crate controllers, and, through the crates backplane dataway, to the FEMs. Upon receiving a trigger, an FEM inhibits its supernova readout mode with its associated decimation, initiates the finer-grained readout scheme and directs the data to the appropriate readout path. Activation or masking of each of the trigger modes will be computer controlled as will the setting of the various trigger thresholds and conditions. The actual cause of the trigger will be available at the event building stage and off-line, as this information will be read out as a bit pattern from the TB via an optical fiber connected to a PCIe card resident in the sub-event PC connected to the crate housing the TB.

\subsection {An Alternative Scheme}

An alternative scheme of TPC front end electronics, readout and DAQ system is shown in Figure \ref{fig:LAr1-ND.Electronics.Alt}. The front end electronics could adopt the cold digital ASIC which will be developed for LBNE, if it becomes available in time. The only change is the FPGA mezzanine will be replaced by a cold digital ASIC mezzanine. It is also possible that a small section of TPC readout to be instrumented by a digital ASIC for R\&D purposes if the design is not sufficiently mature will be installed on the detector. The alternative TPC readout module will use a more advanced FPGA to process 256 channels of detector signals, it will greatly simply the system design and the number of readout crates will be reduced by a factor of 2-3. Also the modern USB 3 link could be used to replace the optical fiber link with a custom designed PCIe card, for data transmission to a DAQ PC farm.

\begin{figure}
\centering
\includegraphics[width=0.8\textwidth]{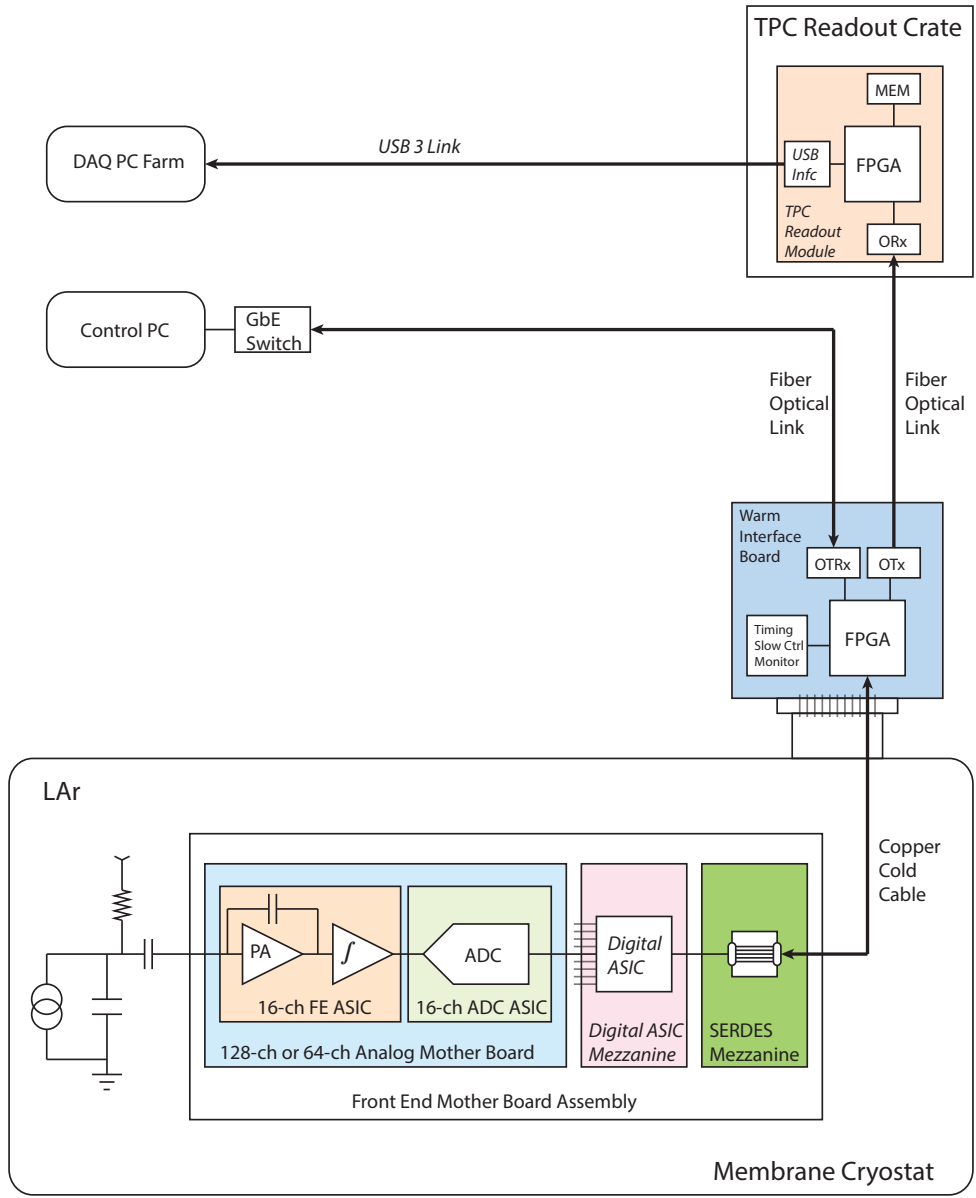}
\caption{\label{fig:LAr1-ND.Electronics.Alt}Information flow of a single TPC readout channel.}
\end{figure}

\section{UV Laser-Based Field Calibration System}
\label{Laser}

\subsection{Calibration of Drift Field by UV Laser Beam}

The knowledge of the electric field inside the drift volume of a TPC is a key aspect for performing subsequent event reconstruction. 
Since distortions of particle tracks due to field non-uniformities are indistinguishable from particle multiple scattering, they affect the accuracy of the particle momentum reconstruction based on track scattering angles.
Deviations of the field map from perfectly uniform in a \lartpc may arise due to accumulation of positive argon ions in the drift volume. 
Ions are created by ionizing particles produced in neutrino interactions as well as by cosmic rays. %  ionizing in the LAr volume.
While free electrons are quickly (within a few milliseconds) swept towards the readout system, ions have significantly lower mobility, and their drift velocity in a LAr detector at nominal drift field is of the order of 0.5 cm/s.
The rate of cosmic muons in the fiducial volume of the \larnd detector is estimated to be $\sim$2200~$\mu$/s (or roughly 110~$\mu$/m$^2$/s, considering only the top surface area). Positive ion charge is therefore produced by cosmic ray muons at a rate of $\sim$1.7~nC/s. These ions are continuously neutralized at a cathode. An example of positive ion charge distribution in equilibrium for a LAr1-ND-like geometry is shown in figure \ref{Ions}, Left. Such accumulated volume charge leads to distortion of the drift field and, consequently, deviation of reconstructed track coordinates from the true ones by up to 2.5~cm (see figure \ref{Ions}, Right).  Ion drift velocity is comparable to local argon flow velocities, produced by global argon re-circulation flow and thermal convection. Therefore, the resulting distribution of positive space charge inside the drift volume of the LAr1-ND TPC may show a sophisticated dynamic behavior.

\begin{figure}[ht]
\centering	
\includegraphics[width=0.99\linewidth]{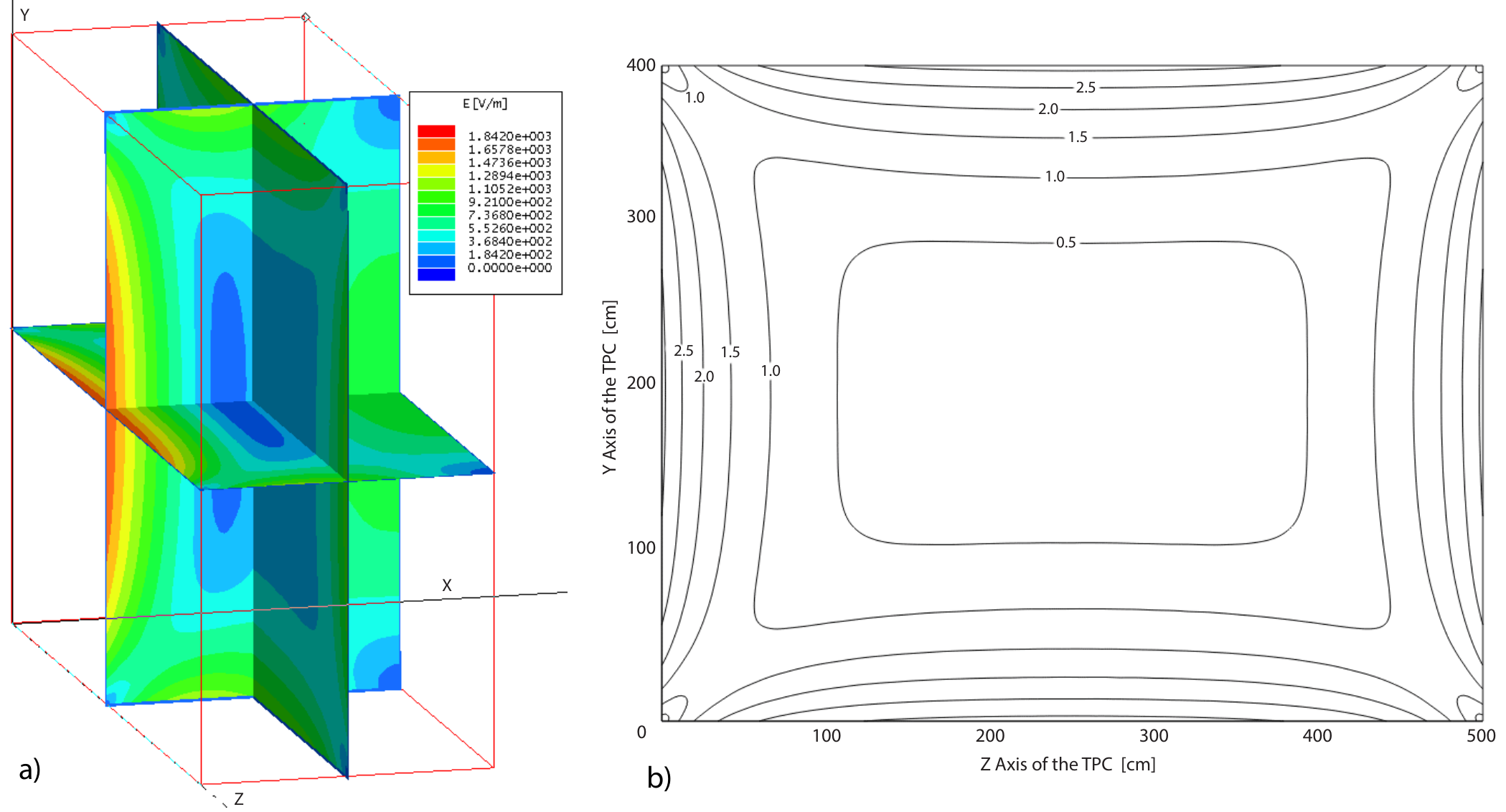}
\caption{(Left) Electric field strength of the accumulated space charge on 3 orthogonal cut planes inside a drift volume of the one half of \larnd (red outlined box). In this view the beam goes into the page (parallel to the z-xis), the cathode is on the left and anode is on the right.  The nominal 500V/cm drift field is not included.  This distorting field is high at the middle of the cathode, and middle of the wire planes, causing longitudinal distortions along the drift.  The field is also relatively high at the middle of the field cage walls, causing transverse distortions.   (Right) Maximum transverse distortion (lensing effect) in the TPC for electrons originating from the cathode surface.  The magnitude of the distortion decreases when the electron starting point is closer to the wire planes.}
\label{Ions}
\end{figure}

\begin{comment}
\begin{figure}[ht]
\centering	
\includegraphics[width=0.49\linewidth]{VETOCALIB/Potential.eps}
\includegraphics[width=0.49\linewidth]{VETOCALIB/Coordinate.eps}
\caption{Left: the distorting potential distribution due to positive space charge in equilibrium in a rectangular LAr TPC. Right: the deviation of a drifting track from its true coordinates due to positive ion space charge.}
\label{Ions}
\end{figure}
\end{comment}

Analyzing the curvature of initially straight ionization tracks allows to reconstruct the distribution of the drift field vector across the whole volume of the detector \cite{ATField}. This method was successfully exploited in the ARGONTUBE long drift TPC \cite{ARGONTUBE0,ARGONTUBE1,ARGONTUBE2} to derive non-uniformity of the electric field along its 5~m long drift volume.
The method to generate straight ionization tracks at defined locations in liquid argon is described in \cite{Badhrees:2010zz}. The thin photon beam from the pulsed UV laser with $\lambda$=266~nm ionizes argon via multi-photon absorption. The resulting ionization track is straight, characterized by low electron density, and therefore practically not subjected to charge recombination losses, unlike cosmic muon tracks. Those tracks are also free of $\delta$-electrons, which complicate track reconstruction in the case of muons. 
For the \uboone detector a set of such tracks are required in order to cover the whole sensitive volume to reconstruct field distortion. Such tracks are created one-by-one by steering pulsed laser beams with the use of a custom-designed opto-mechanical feed-through (see \cite{LaserFT}). The pulse rate of the laser generator is 10~Hz, capable of producing the minimum required set of 100 tracks within one minute (taking into account steering time).

\subsection{Laser Beam Arrangement and Beam Optics}

A typical scheme of producing a fan of straight ionization tracks from one laser source in a TPC is shown in Figure \ref{Arrangement}. A Nd:YAG laser (Surelite I-10) from Continuum, Inc. emitting light at a wavelength of 1024~nm is used as the primary light source. Inside the laser head nonlinear crystals are installed in the beam line for frequency doubling and summing, resulting in a wavelength of 266~nm, needed for ionization of liquid argon. The pulse maximum energy at this wavelength is 60~mJ and the pulse duration is $\approx$5~ns. The maximum repetition rate is 10~Hz. The beam has a divergence of 0.5~mrad and the diameter of about 5~mm.

The beam is delivered to the top of the rotating optical feed-through via a beam conditioning optics. This optics allows to set the beam attenuation and diameter and allows computer-controlled adjustment of the beam direction within few degrees. The feed-through brings the beam into the cryostat and provides a capability of steering it across the whole detector active volume. The arrangement  of the four optical rotating units inthe LAr1-ND cryostat is shown in Figure \ref{Arrangement}, Right. % and \ref{TPC_top}.

\begin{figure}[ht]
\centering	
\includegraphics[width=0.49\linewidth]{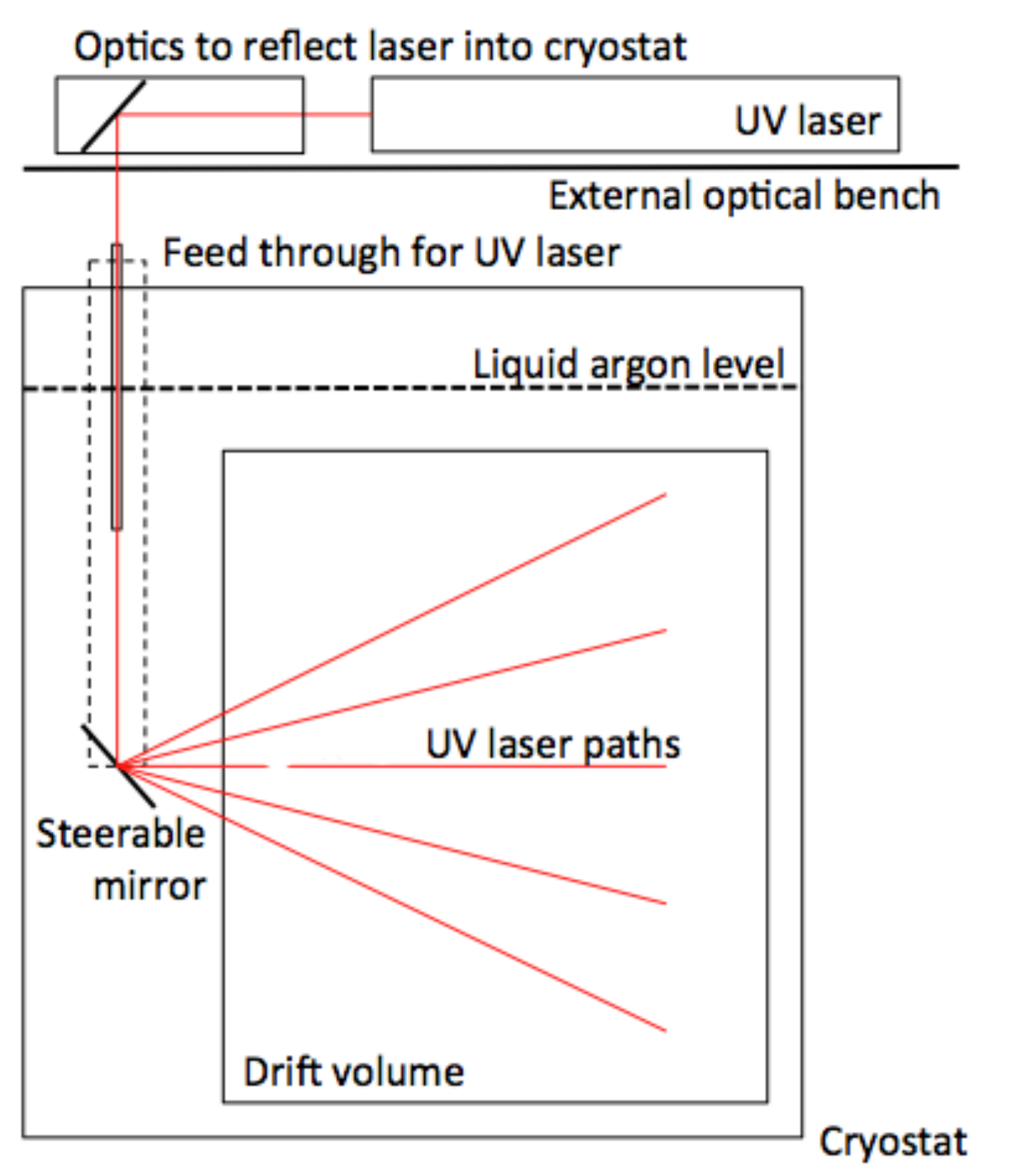}
\includegraphics[width=0.49\linewidth]{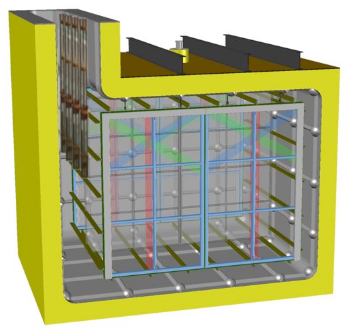}

\caption{Left: a typical scheme of producing a fan of straight ionization tracks from one laser source in a TPC. Right: arrangement of four rotating optical feed-through in the LAr1-ND cryostat.}
\label{Arrangement}
\end{figure}

%\begin{figure}[ht]
%\centering	
%\includegraphics[width=0.8\linewidth]{VETOCALIB/Top_View_LASER.png}
%\caption{The horizontal arrangement of the rotating optical feed-through in %the LAr1-ND cryostat. The red circles depict the locations of four rotating %optical feedthrough units.}
%\label{TPC_top}
%\end{figure}

%\begin{figure}[ht]
%\centering	
%\includegraphics[width=0.8\linewidth]{VETOCALIB/Cryostat_Laser.png}
%\caption{A typical scheme of producing a fan of straight ionization tracks from one laser source in a TPC.}
%\label{Arrangement2}
%\end{figure}

The four steerable mirrors are located at half-height of the TPC active volume in front of corresponding beam entry apertures in the TPC field-shaping structure, as shown in Figure \ref{fig:Apertures}.
Each aperture has a diameter of 50~mm in order to maximize the deflection angle for the laser beam.

\subsection{Rotating Feed-Through}

To deliver undistorted UV laser beam into the active volume of the LAr1-ND detector, a rotating optical feed-through with the steerable mirror has been designed. The mirror is mounted on a horizontally rotatable support structure. A rack and pinion construction, where the mirror is mounted on the front side of a half gear (pinion) provides the necessary freedom for the vertical movement. All movable components are motorized to allow for remote control and automation of the mirror movement. The mirror support structure was fabricated out of polyamide-imide (Duratron T4301 PAI), which has a very low outgassing rate, low thermal expansion coefficient and is certified for operation at 87~K. To minimize the probability of discharges due to the close location of the feedthrough to the field cage, no conductive parts were used in the support structure. The two principal parts of the feedthrough are shown in Figure \ref{fig:feedthrough}. At the left, the top rotating unit, operating at room temperature is seen. The steerable mirror, mounted at the bottom of the feedthrough support column, is shown at the right. In Figure \ref{fig:Apertures} steerable mirrors mounted on the TPC field-shaping cage are shown. These units operates in liquid argon at the temperature of 87~K.

\begin{figure}[h]
\centering
\setlength{\fboxsep}{0pt}
\setlength{\fboxrule}{0.6pt}
\fbox{\includegraphics[height=0.4\textheight]{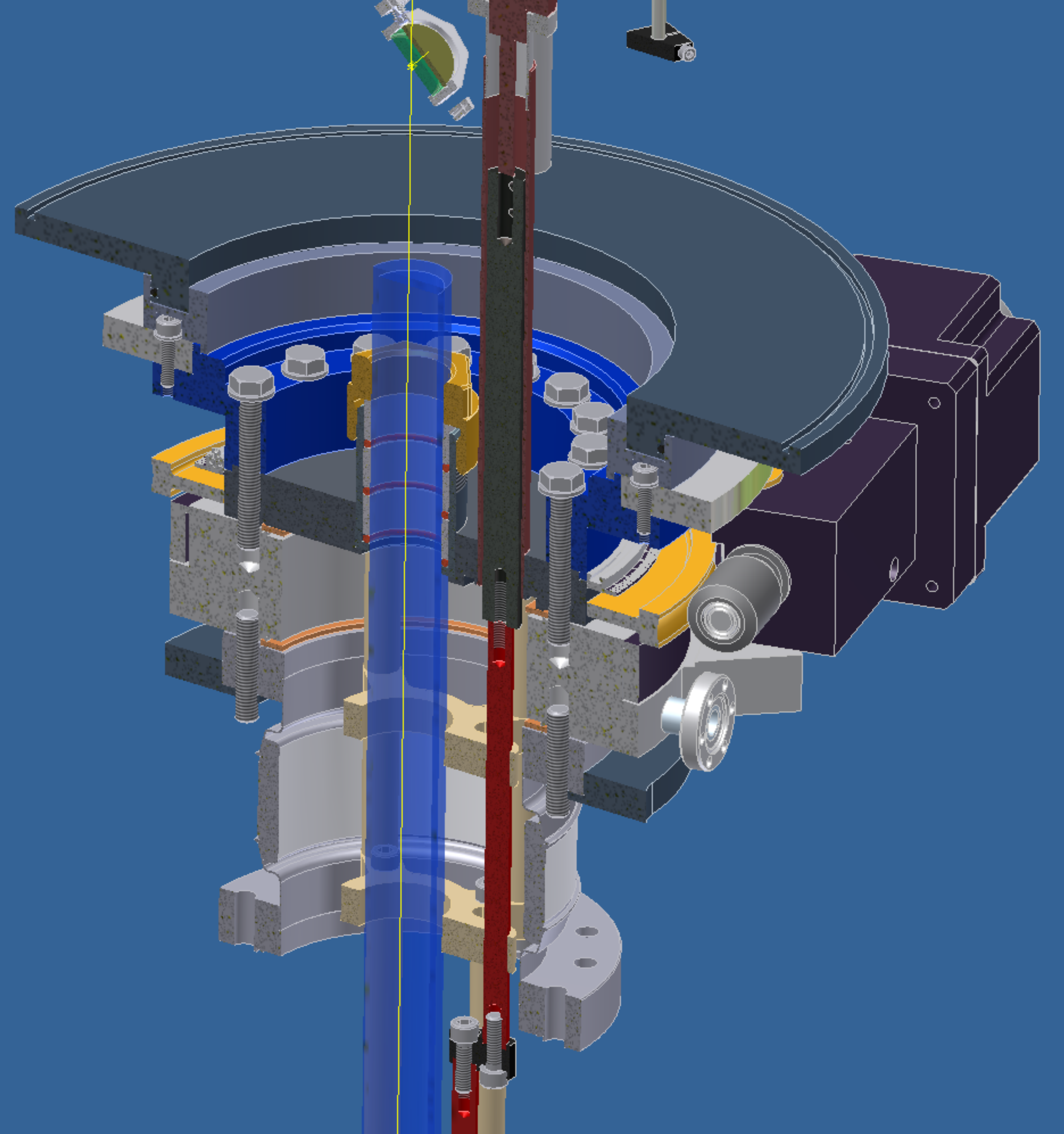}}
\fbox{\includegraphics[height=0.4\textheight]{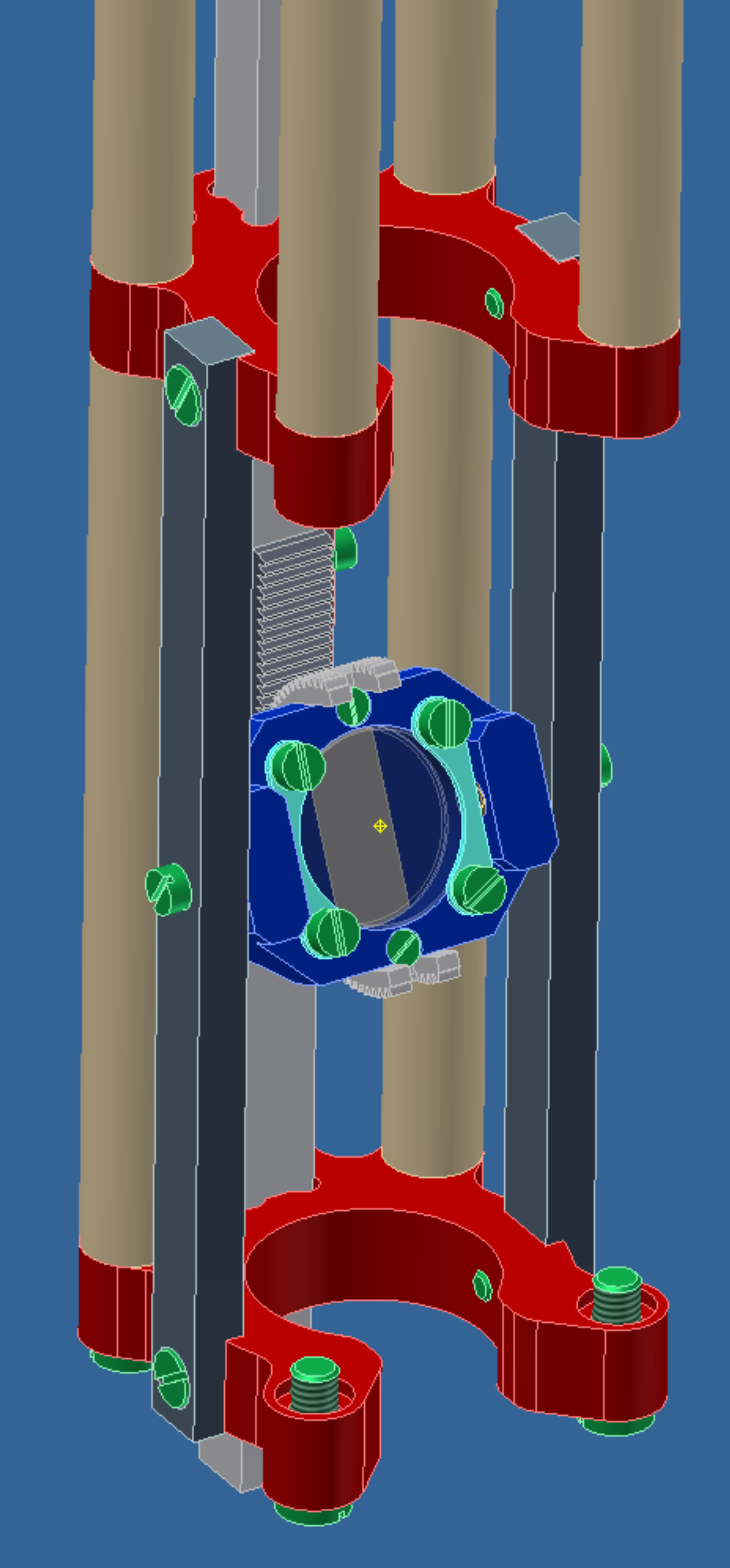}}
\caption{Left: CAD cutaway drawing of the feedthrough construction is shown. The yellow line indicates the path of the UV laser beam. Right: the cold mirror including the support structure.}
  \label{fig:feedthrough}
\end{figure}

\begin{figure}[ht]
\centering	
\setlength{\fboxsep}{0pt}
\setlength{\fboxrule}{0.6pt}
\fbox{\includegraphics[width=0.6\textwidth]{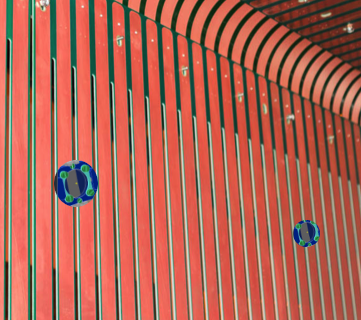}}
\caption{Steerable mirrors as seen through the 50~mm apertures in the TPC field-shaping cage.}
\label{fig:Apertures}
\end{figure}

\subsection{Expected Performance}

An algorithm of field calibration has as input an array of detector events with one straight ionization track in each. The result of the algorithm is the coordinate correction map, which converts apparently curved track images back to true coordinate system, where they are straight. The algorithm is iterative with optimizable iteration step and therefore reconstruction accuracy.  An example of simulated reconstruction in 2-D space is shown in Figure \ref{fig:Reco} for the case of a similar laser system that has been implemented in \uboone. The distortion magnitude is reduced down to a few millimeters in 99\% of the detector volume.  A similar performance is anticipated for the \larnd laser calibration system.

\begin{figure}[ht]
\centering	
\includegraphics[width=1.0\linewidth,trim=10mm 0mm 10mm 0mm,clip]{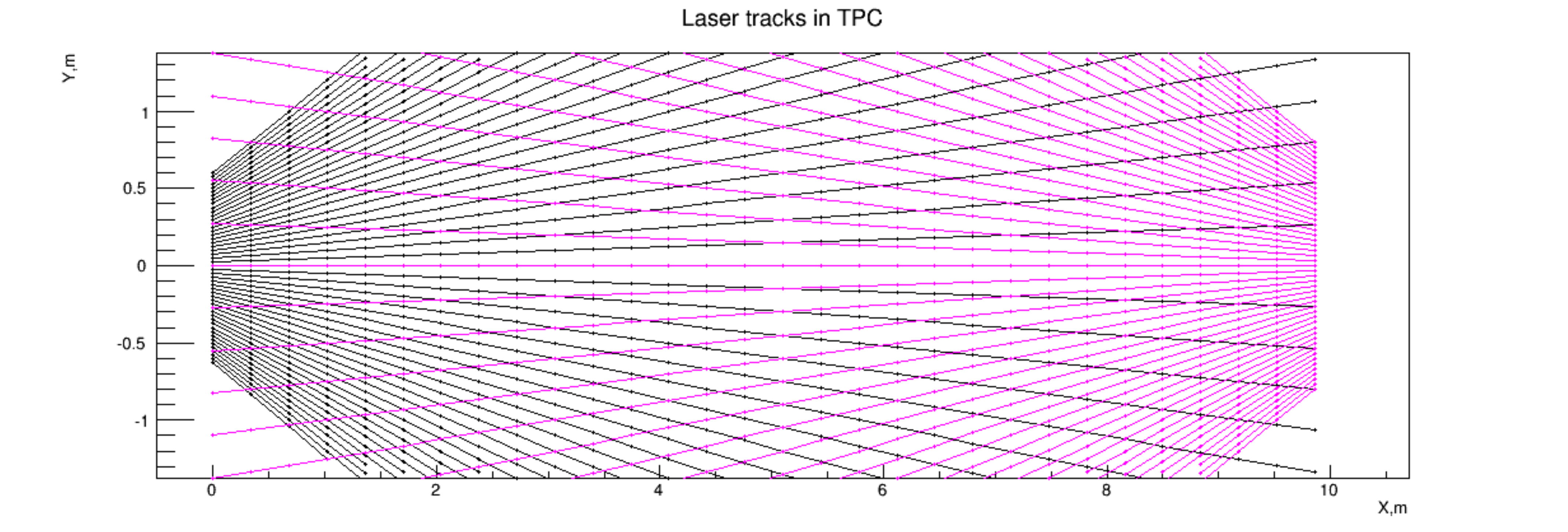}
\includegraphics[width=1.0\linewidth,trim=10mm 0mm 10mm 0mm,clip]{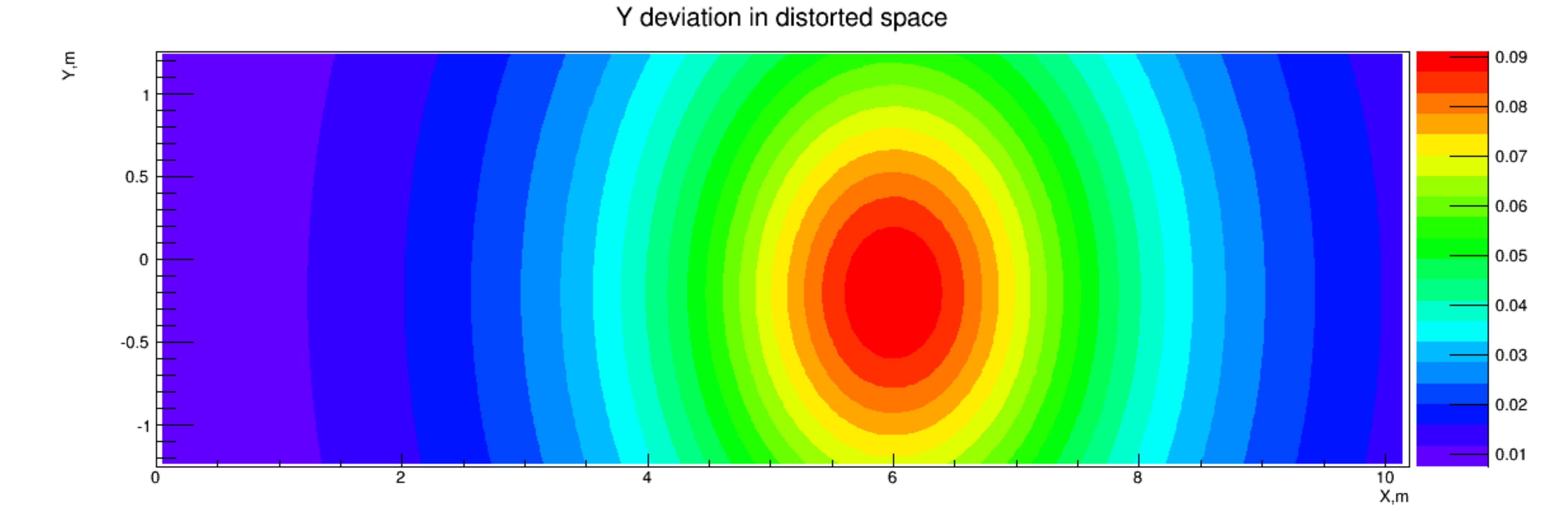}
\includegraphics[width=1.0\linewidth,trim=10mm 0mm 10mm 0mm,clip]{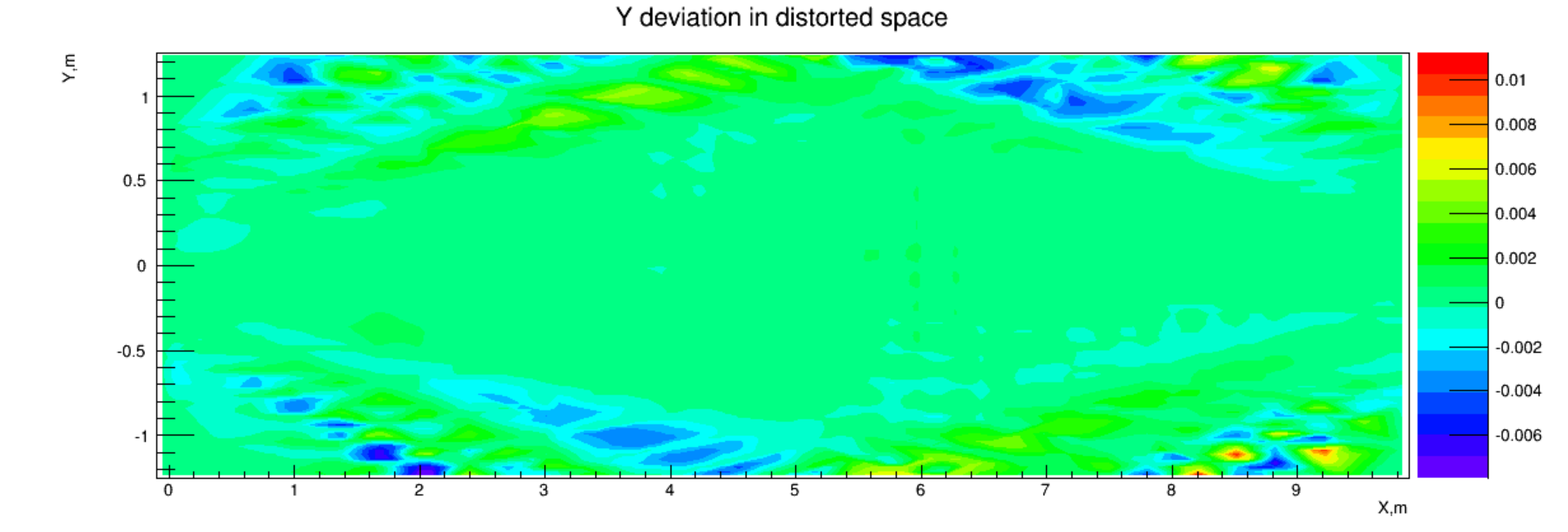}
\caption{Example simulation based on \uboone showing the performance of the laser calibration system.
(Top) True laser beam trajectories in the \uboone \lartpc.
(Middle) map of Y coordinate of track deviation under influence of an ad-hoc non-uniform electric field, which in this example is slightly offset from center.
(Bottom) map of the residual Y-coordinate deviation from the true ones after application of the reconstruction correction based on the laser tracks. The color scale in the lower two plots is in units of meters.} 
\label{fig:Reco}
\end{figure}

\FloatBarrier

\section{Light Detection System}
\label{LightDetectionSystem}

\subsection{Introduction and Motivation}
\label{LightIntro}

Ionized and excited argon molecular states in the LAr volume will produce 128~nm vacuum ultraviolet (VUV) scintillation photons through recombination and de-excitation processes. %\cite{scint1,scint2}.  
The scintillation light includes a nanosecond-scale fast component (from singlet $\rm{Ar}_2^*$ decay with a lifetime $\sim$6~ns)  as well as a microsecond-scale slow component (from triplet $\rm{Ar}_2^*$ decay with a lifetime $\sim$1.6~$\mu s$).   Since recombination rates are reduced with higher electric field strength, a change in the field strength has opposite impacts on the amount of free ionization and scintillation light available for detection.  At typical TPC field strengths, energy deposition by ionizing radiation is shared approximately equally between free ionization electrons and VUV scintillation photons, yielding approximately $2.9\times 10^4$ $e^-_{free}/{\rm MeV}$ and $2.4\times 10^4$ $\gamma/{\rm MeV}$ with a 500~V/cm electric field.  It thus appears very appealing and natural to further optimize \lartpc detector performance by combining information from the available scintillation light with that from the ionization charge.  

To be detected, scintillation photons are usually shifted from the vacuum ultraviolet to the visible to match the quantum efficiencies of available photodetectors, which typically peak around 430--450 nm.  In most systems, this has been achieved using a fluorescent material to downshift the direct scintillation light (such as Tetraphenyl Butadiene, TPB) either coated on the surface of cryogenic photomultiplier tubes (as in ICARUS) or on plates mounted in front of the PMTs (as in MicroBooNE).  
%The traditional method for detecting the scintillation light in LAr uses Tetraphenyl Butadiene (TPB) coatings on cryogenics photomultiplier tubes (PMTs) or coated plates placed in front of the PMTs.  Standard phototubes suffer from charge depletion at cryogenic temperatures; this problem is solved by a platinum undercoating of the photocathode.
PMTs must be located outside of electric field regions in the detector, thus in ICARUS %, which is the only neutrino \lartpc to have run with a light collection system, 
and in \uboone %, which is now under construction, 
the PMTs are located just outside of the wire planes which are held close to ground.  The resulting small photocathode area and limited solid angle coverage results in a relatively low light yield ($\sim$1 phe/MeV for the past ICARUS system and $\sim$2 phe/MeV for \uboone). %and $\sim$0.2-0.3 phe/MeV for the proposed LBNE system) 

Detection of scintillation light can play several important roles in \lartpcs, depending on the time, energy, and position resolution performance of the light detection system (LDS) that is implemented.  Increased collection efficiency could result in the improvement of all three performance metrics and enable improved background rejection and access to additional physics topics.    

For a surface detector in a beam, prompt light signals provide a `trigger', indicating when an interaction has occurred in coincidence with the neutrino beam. In \larnd, neutrino interactions in the active TPC volume are expected in about 5\% of the 1.6~\us long BNB beam spills. With a LDS time resolution of 1--2~ns, neutrino events could further be correlated with the 53 MHz Booster beam RF substructure (81 $\sigma$=1.15 ns wide pulses spaced 19 ns apart), leading to a potential 3-4$\times$ reduction in random cosmogenic backgrounds in the \nue event sample (see Part I: Oscillation Physics Program).  Figure \ref{RF} shows the BNB RF structure as measured by the \MB Cerenkov light detector.  
%There are 81 proton RF pulses during the 1.6 $\mu$sec beam spill, each proton RF pulse having $\sigma= 1.15$ nsec.  
Also shown is the \MB time reconstruction of CCQE muon neutrino events relative to the beam RF time of the first proton pulse as determined by the Resistive Wall Monitor (RWM) discriminator.  The CCQE muons exhibit the same time structure as the RF pulse, with a total time resolution of $\sigma_{t}= 1.75$~ns.  The extra spread is due to RWM timing jitter and event reconstruction time resolution.  %In addition, by looking outside of the RF pulse, beam-related backgrounds can be isolated and studied.

\begin{figure}[htb]
\centering
\setlength{\fboxsep}{0pt}
\setlength{\fboxrule}{0.6pt}
\mbox{\fbox{\includegraphics[width=0.5\textwidth,trim=2mm 0mm 0mm 1mm,clip]{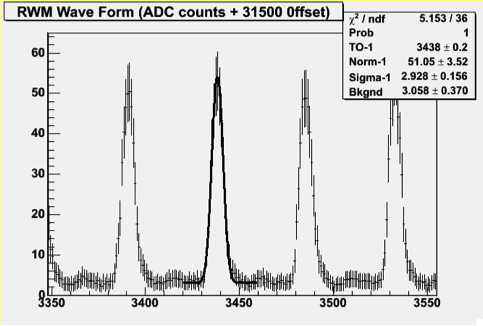}}
\fbox{\includegraphics[width=0.5\textwidth,trim=2mm 0mm 0mm 1mm,clip]{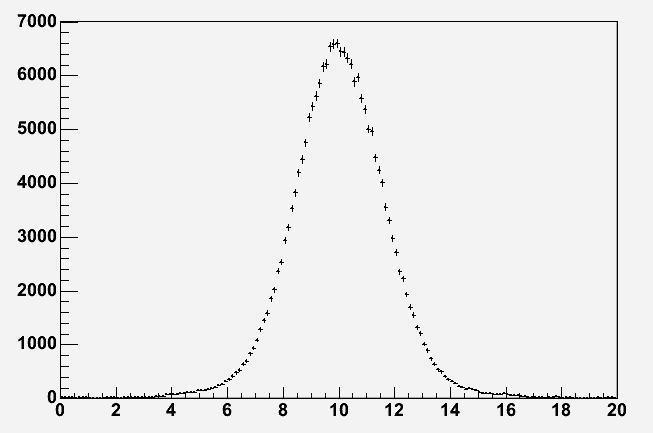}}}
\caption{Left: The Booster-BNB 53MHz RF beam structure (2.5 counts/nsec). Right: the absolute time (nsec) reconstruction for CCQE muon neutrino candidates in MiniBooNE with all 81 bunches overlayed on top of each other, demonstrating that with good track timing, the RF structure of the beam can be reconstructed.}
\label{RF}
\end{figure}

Good timing reconstruction resolution, in conjunction with the beam RF structure, can also be used to tag neutron events that are produced by neutrino interactions in the surrounding dirt.  Due to the extended interaction geometry and subluminal speed of the neutrons, they interact in the detector out of phase with the beam RF time structure.  This produces a flat time response in the reconstructed bunch time (right plot of Figure \ref{RF}), which can aid in background rejection and/or measurement of the neutron-dirt rate.  

Prompt light signals also provide the unknown event $t_0$ necessary to reconstruct non-beam related events such as cosmic rays or supernova neutrinos.  A system capable of associating multiple light pulses with their sources in the detector permits identification and 3D reconstruction of different events occurring throughout the 1.28~ms TPC readout window.  The $t_0$ of each interaction in the detector is necessary to determine the location of the ionization along the drift direction and get an accurate reconstruction of the energy deposited along a track (by accounting for attenuation along the drift).  The ability to identify the time of individual events will further contribute to rejecting cosmogenic backgrounds in the \nue analysis.  An average of 2.9 cosmic muons are expected in the TPC volume per readout window, distributed across the 20~m$^2$ area of the TPC (see Part I of this proposal).  A position resolution of the LDS of better than 1~m would be needed to enable association of light signals with different activity in the detector.  It should be noted, however, that the association with entering muons will be significantly aided by the external cosmic ray tagging system described in Section \ref{sec:CRTS}. 

A LDS with increased detection efficiency could improve the reconstruction threshold of argon neutrino detectors to as low as a few MeV, enabling access to whole new fields of study such as low-energy nuclear effects. This requires a more uniform light collection as well as at least a ten-fold improvement in the collection efficiency compared to existing LAr neutrino detectors.  In addition to lowering the threshold, using the scintillation light for calorimetric reconstruction allows the compensation of charge recombination effects, thereby increasing the linearity of the overall energy resolution in the detector. The improvement in resolution obtained by higher light collection has been demonstrated in simulation \cite{Szydagis, Sorel} and an example of the power of such a combined energy calculation in xenon has been shown by the EXO collaboration \cite{EXO}. 

An enhanced light readout system can also contribute to particle identification.  The PID will derive from pulse shape discrimination (PSD) methods used in dark matter noble liquid detectors which effectively distinguish nuclear recoils from minimally ionizing particles (MIPs) using the time structure of the scintillation light alone.  In neutrino interactions, this would aid in the separation of neutron interactions (which produce heavily ionizing recoil protons) from gamma scatters (that produce minimally ionizing electrons).

Sufficient light collection may also provide a way of determining the sign of the incoming neutrino without using a magnetic field through improved tagging of Michel electrons coming from stopping muons, including those at lower energies.  Efficient reconstruction of this very well known process is particularly helpful in argon because negatively-charged muons have a 75\% capture rate on argon atoms, in which case there is no Michel electron emitted at the end of a muon track, whereas this capture does not happen for positively-charged muons.  Ref.~\cite{Sorel} shows that light detection systems with efficiency $\sim 1 \times 10^{-3}$ can isolate a sample of $\mu^-$ events with relatively little $\mu^+$ contamination. This capability will be essential in raising the sensitivity in anti-neutrino running both in sterile neutrino searches and CP-violation searches in LBNF due to the large wrong-sign component of the anti-neutrino beam.

\subsection{A Light Collection System for \larnd}
\label{LightDesigns}

%The relatively small volume of \larnd provides an excellent test-bed for light detection systems (LSD) being designed and optimized for LAr neutrino detectors. 

The relatively small volume of \larnd makes it an excellent test-bed for new light detection system designs being considered for future LAr neutrino detectors, especially LBNF.  The collaboration is committed to taking full advantage of this opportunity without incurring undue risk to the science goals of the SBN program.  Hybrid systems that provide redundancy and side-by-side performance comparisons are also being considered and would fit well with the R\&D goals of the experiment.  LDS approaches currently being evaluated for \larnd include: a system based on acrylic light guide bars read out at the ends with SiPMs, a system based on TPB-coated reflector foils to increase collection efficiency without increasing the number of photodetectors, and a traditional TPB-coated PMT based system.

Detailed MC simulations of the light generation and detection are being developed to compare reconstruction performance criteria such as track time, calorimetric energy, and position resolution.  %The time scale for these studies is to have preliminary results by the first months of the new year.
These results will inform the final design choice and determine the feasibility of a hybrid system comprised of elements from more than one of the present concepts.  Studies are in progress and will be completed to enable a technology decision early in 2015.

%basis for deciding on exactly what type of system will need to be built.  A hybrid system with two different LDS can also be considered for \larnd to take advantage of the different strengths of each LDS.  
%This will drive the choice for the TPC cathode, that divides the LAr volume into two TPC regions in the beam direction. The cathode could be transparent or opaque (see Sect.~\ref{CPA}).  The default design is a transparent cathode, but an opaque cathode may be required if each of the two TPCs regions will be instrumented with a different LDS.

\subsubsection*{Light Guide Bar Light Detection System}

A light guide bar based light detection system has been initially proposed for \larnd~\cite{LAr1-NDPAC}, based on a design originally developed for LBNE.  In the LBNE design the system is  positioned inside the APAs with wrapped wires. Thin profile light guides, measuring 100 cm $\times$ 2.54 cm $\times$ 0.64 cm, with wavelength shifter (WLS) deposited on their surface are used to collect the 128~nm LAr scintillation light.  The WLS converts the VUV photons to $\sim$425 nm photons, some of which enter the bar.  The downshifted photons are internally reflected to the end where they are detected by 3 SensL SiPMs whose QE is well matched to the $\sim$425~nm photons (See Fig.~\ref{LightGuide}).  This design allows one to collect light over a larger area with a relatively small number of readout channels.  Four cast acrylic bars are assembled into ``paddles''.  Each LBNE APA measures 7 m $\times$ 2.5 m and contains 20 of these light guide paddles.  The solid angle subtended by an LBNE APA from a point displaced 2 m from its center is 15\% and the fraction of the APA surface covered by light guides is 12\%.  A total of 108 APAs are used in each of 2 cryostats, totaling $108 \times 2 \times 20 \times 4 = 17,280$ light guide bars in LBNE. 

%However, LBNE will be located deep underground and so its photon detection system will not have as stringent a requirement on cosmic background rejection as will \larnd's. 

%%%%%%%%%%%%%%%%%%%%%%%%%%%%%%
\begin{figure}[tb]
\begin{centering}
\includegraphics[width=0.9\textwidth]{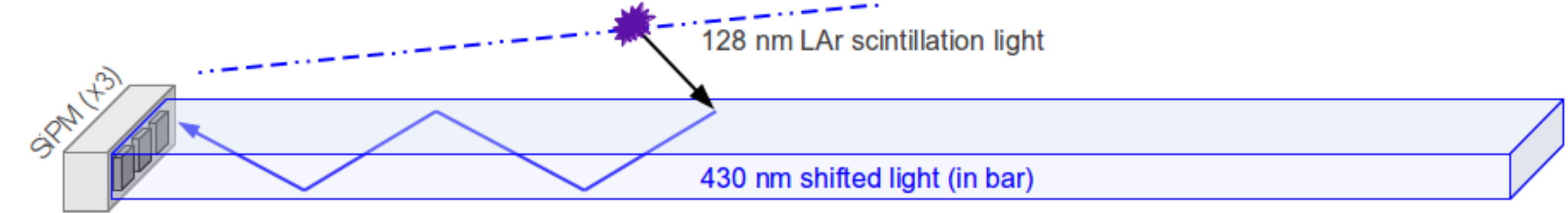}
\caption{\scriptsize \small Cartoon of LAr scintillation light detection with a light guide coated with WLS.}
\label{LightGuide}
\end{centering}
\end{figure}
%%%%%%%%%%%%%%%%%%%%%%%%%%%%%%%%%%%%%%%%%%%%

The light guide photon detection system being considered for \larnd aims to maximize the active area of the light guide bars  % as well as boosts their efficiency.
to have both a high photon detection efficiency and good granularity for timing and position resolution, which are especially valuable for a detector operating at the surface. The present design consists of 1,000 acrylic light guide bars coated with WLS embedded acrylic, each measuring 100 cm $\times$ 2.54 cm $\times$ 0.64 cm, mounted behind the \larnd wire planes, as shown in Fig.~\ref{LAr1ND_Design}.  The solid angle subtended by a \larnd APA from a point displaced 2 m from its center is 16\% and the fraction of the APA surface covered by light guides is 87\%.   Both ends of each bar will be read out by an array of 3 SiPMs, smoothing out the position-dependent response of each bar and improving the overall light collection efficiency of each bar by a factor of $\sim$2 relative to the LBNE design.  Furthermore, \larnd will use new low-noise SensL SiPMs (MicroFC-60035-SMT), which have an order of magnitude lower dark rate.   

%The dense coverage enables better position resolution for identifying cosmic rays in a surface detector.

%%%%%%%%%%%%%%%%%%%%%%%%%%%%%%
\begin{figure}[tb]
\centering
\mbox{\includegraphics[width=0.47\textwidth,trim=10mm 0mm 0mm 0mm,clip]{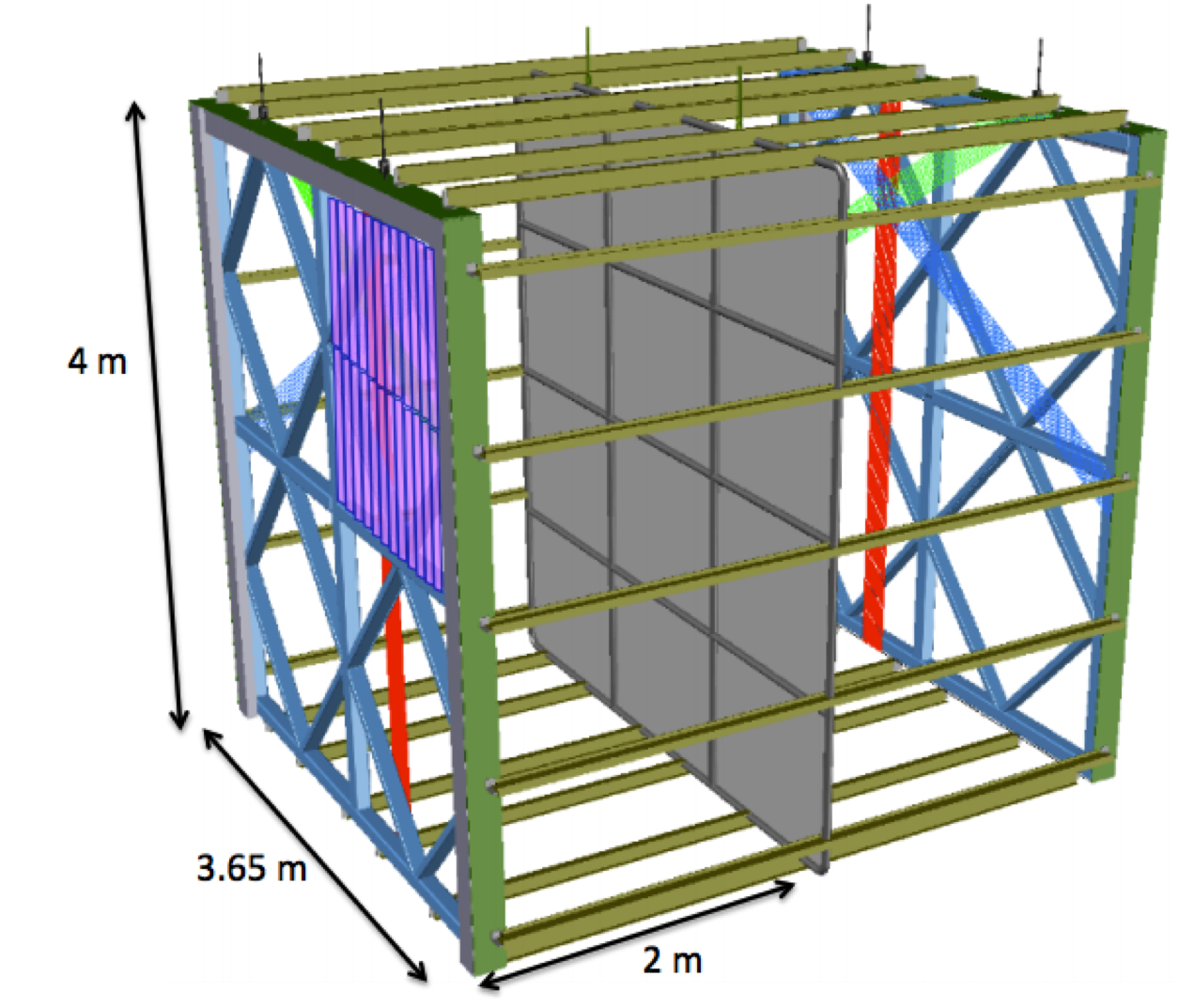}~~~~~
\includegraphics[width=0.47\textwidth]{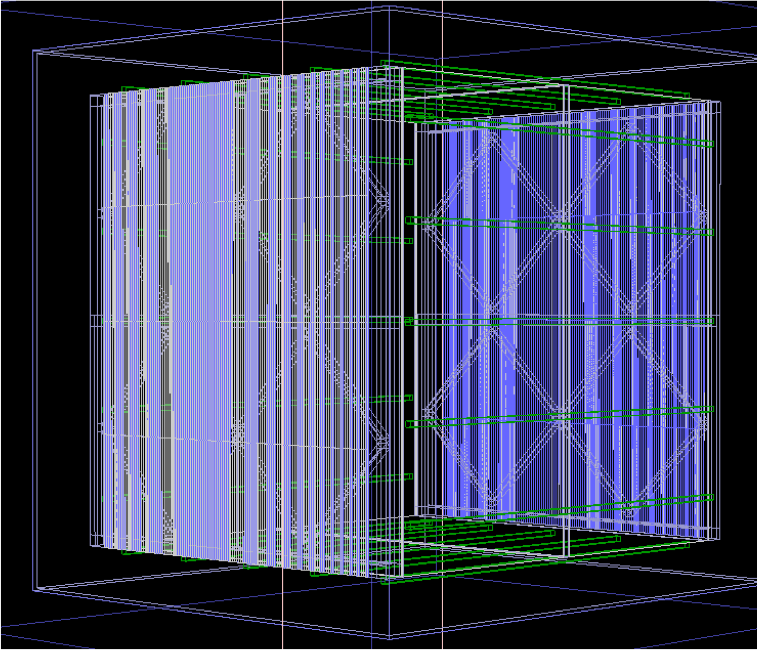}}
\caption{\scriptsize \small \textbf{Left:} Cartoon of the light guide design for \larnd. \textbf{Right:} Light guide design implemented in Geant4.}
\label{LAr1ND_Design}
\end{figure}
%%%%%%%%%%%%%%%%%%%%%%%%%%%%%%%%%%%%%%%%%%%%

The \larnd light guide photon detection system builds off the LBNE design and therefore benefits directly from LBNE-related R\&D efforts. Measurements of light guide bars in LAr have yielded valuable information on their performance and led to significant improvements in their design and quality.  In particular, as an improvement over hand-painting the bars, the WLS solution is now applied to the surface by dip-coating the acrylic bars and then allowing them to dry in a low humidity environment. Recent measurements of these improved light guides have shown that attenuation lengths of over 100 cm are routinely achievable.  Furthermore, preliminary data from studies of these light guides in pure LAr suggests that the global quantum efficiency (defined as the number of photoelectrons divided by the number of incident 128~nm photons) 50 cm from one end of the bar is $\sim$0.5\%.  Therefore, the global quantum efficiency when reading out both ends of the bar is expected to be $>1$\%. Preliminary estimates indicate that the light guide based photon detection system will collect 24 phe/MeV from a point displaced 2~m from its center and more than 24 phe/MeV when averaged over the entire TPC volume. A full MC simulation is in progress to confirm these performances.

\subsubsection*{TPB-coated Reflector Foil Light Detection System}

%An enhanced scintillation light collection system}
A second design, based on a concept adapted from liquid argon dark matter detectors, proposes the installation of TPB-coated reflector foils inside the TPC volume to enhance light collection.  The higher quantities of light collected would open the door to exploring physics topics not accessible to \lartpcs with standard readout schemes.
%like improvement in energy resolution up to a factor two when the scintillation light signal is combined with the TPC charge signal in event reconstruction \cite{szydagis}.  This technique will be especially powerful for lower energy events that are usually close to the threshold for standard LArTPC detectors. New PID techniques through pulse shape discrimination are also enabled with fast waveform processing. 
By trapping light within the volume using TPB-coated reflective foils, light yield is enhanced naturally with less need to increase the number of DAQ readout channels as compared to solutions that achieve similarly-high light yields by scaling up photocathode coverage.  A smaller number of readout channels allows for easier and more practical implementation of high-speed signal digitization, which would immediately improve the timing resolution of the system.
% and enable studies of beam exotica such as dark photons (see Sect.~\ref{LAr1-NDPhysics}).

A system of this type is currently implemented in the LArIAT experiment (\lartpc In A Test Beam). 
% The aim is to use scintillation light readout in combination with the response of the TPC to improve the calorimetric energy reconstruction with beam particles of pre-measured type and energy.  This will enable the development of new PID techniques for particles commonly observed in neutrino interactions ($\mu$, $\pi$, K, p) from pulse shape features of the scintillation light. 
LArIAT's light readout system collects many more scintillation photons than typical liquid argon neutrino experiments, with simulations estimating about 40 phe/MeV at zero field -- substantially higher than the reach of both current and planned liquid argon neutrino detectors ($\sim$1 phe/MeV for ICARUS, $\sim$2 phe/MeV for MicroBooNE, and $\sim$0.2-0.3 phe/MeV for the proposed LBNE system). The light yield simulation was cross-checked against available analytic predictions~\cite{Segreto} and has been recently validated through measurements on a small-scale prototype where two high-QE cryogenic PMTs %as well as three silicon photomultipliers (SiPMs) of different types 
were used to collect and read out scintillation light in argon.  Bench tests of SiPMs mounted to custom on-board preamps are also underway at FNAL and early results are encouraging.  The forthcoming LArIAT test beam run will provide information both on technical and physical aspects of a reflective foil enhanced efficiency light system for \lartpcs.  

The LArIAT optical system, with reflectors covering 60\% of the active volume's inner surfaces and a photocathode coverage of 0.35\%, appears scalable up to the \larnd dimensions with no need to transition through a dedicated R\&D phase.  A zero-field light yield on the order of 100 phe/MeV (corresponding to 0.5\% photon detection efficiency) is the target for the \larnd design.  Preliminary simulations indicate a photocathode coverage of approximately 0.4\% coupled with 70\% inner surface coverage of TPB-coated reflector foil is required to reach this level.  For a \larnd half-module, this translates to 1900 cm$^2$ of distributed photocathode area and about 50 m$^2$ of TPB-coated foil covering the inner surfaces of the field cage and cathode.  A full MC simulation is in progress to confirm these specifications.

%The LArIAT optical system, with reflectors covering 60\% of the active volume's inner surfaces and a photocathode coverage of about 0.3\%, appears scalable up to the \larnd dimensions with no need to transition through a dedicated R\&D phase.  A light collection efficiency enhanced to order 100 phe/MeV (corresponding to 0.5\% photon detection efficiency) is the target light yield for the \larnd design. A photocathode coverage of 0.1\% coupled with 70\% inner surface coverage of TPB-coated reflector foil is required to achieve this level~\cite{Segreto}. For a \larnd half-module, this translates to $\sim$750 cm$^2$ of distributed photocathode area and about 50 m$^2$ of TPB-coated foil covering the inner surfaces of the field cage and cathode. A full MC simulation is in progress to confirm these performances. 

%%%%%%%%%%% Figure 2 %%%%%%%%%%%%%%%%%%%%%%%%%%%%%%
\begin{figure}[t]
\begin{centering}
 \includegraphics[width=5.5in]{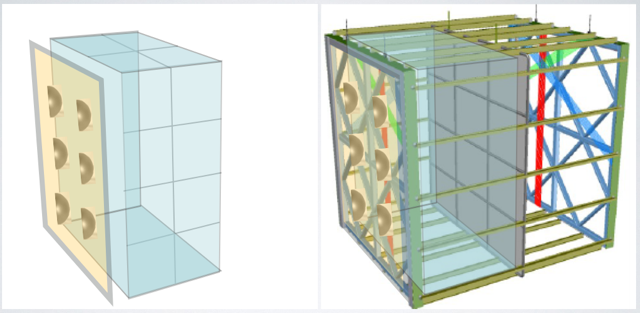}
 \caption{\scriptsize \small Schematic view of an enhanced efficiency light detection system in \larnd.  Shown on the left is the photosensor array behind the wire plane as well as modular TPB-coated reflector tiles lining the inner field cage of one half-module of the \larnd detector.}
\label{EES}
\end{centering}
\end{figure}
%%%%%%%%%%%%%%%%%%%%%%%%%%%%%%%%%%%%%%%%%%%%

The wavelength-shifting TPB film (200 $\mu$g/cm$^2$) is deposited  by vacuum evaporation on a substrate like Vikuiti ESR, a highly reflective and non-metallic foil made by multi-layer polymer technology.  Each foil is 65 $\mu$m thick, and can be fashioned in a variety of different sizes (for example, 50$\times$ 50 cm$^2$) and mounted to thin, rigid supports with comparable coefficients of thermal expansion to create modular TPB-coated reflector tiles.  These tiles can then be installed in an array to cover the inner surfaces of the field cage, and may also be easier to work with in the TPB deposition process.

% On the cathode, a double metal mesh structure can be built with the foil suspended in between to prevent charge build-up on the foil. Foils held between two layers of cathode mesh would have extended freedom of movement to accommodate any possible thermal expansion of the reflective substrate.  For foils mounted to tiles using plastic screws, using slightly oversized screw holes would 
% ensure a similar freedom of movement.

The choice of photosensor is restricted to high-QE cryogenic PMTs, SiPMs, or a combination of the two.  Adequate photocathode coverage could be provided by about 50 3-inch diameter PMTs positioned in an array behind the wire planes, or by about 1600 1.4 cm$^2$ SiPMs distributed behind the wire planes and possibly onto the field cage.  Silicon photomultipliers offer several notable advantages over PMTs, given their high QE coupled with small occupancy and low bias voltage.  The forthcoming LArIAT run will provide full characterization of the SiPM response to LAr scintillation light in operating conditions to compare directly to PMT performance.  Electronics for reading out SiPM arrays (groups of 16 SiPM channels) as well as single channel SiPM chips, which include on-board bias voltage filtering and preamp/shaping circuitry as shown in Figure 57, have been developed.  Further modifications to optimize the gain and timing response of the boards will be investigated.  The TPB-coated reflector tiles described above may provide the necessary support for mounting variations of these miniature SiPM boards onto the field cage walls if such a layout proves advantageous.

%The choice of photosensor is restricted to high-QE cryogenic PMTs, SiPMs, or a combination of the two.  Adequate photocathodic coverage could be provided by 20 3-inch diameter PMTs positioned in an array behind the wire planes, or by 600 1.2 cm$^2$ SiPMs distributed behind the wire planes and possibly onto the field cage. Silicon photomultipliers offer several notable advantages over PMTs, given their high QE coupled with small occupancy and low bias voltage.  The forthcoming LArIAT run will provide full characterization of the SiPM response to LAr scintillation light in operating conditions to compare directly to PMT performance. Electronics for reading out SiPM arrays (groups of 16 SiPM channels combined in a single readout channel) as well as single channel SiPM chips, which include on-board bias voltage filtering and preamp/shaping circuitry as shown in Figure~\ref{miniboards}, have been developed. Further modifications to optimize the gain and timing response of the boards will be investigated. The TPB-coated reflector tiles described above may provide the necessary support for mounting variations of these miniature SiPM boards onto the field cage walls if such a layout proves advantageous.

%%%%%%%%%%% Figure 3 %%%%%%%%%%%%%%%%%%%%%%%%%%%%%%
\begin{figure}[t]
\centering
\setlength{\fboxsep}{0pt}
\setlength{\fboxrule}{0.6pt}
\fbox{\includegraphics[width=4in]{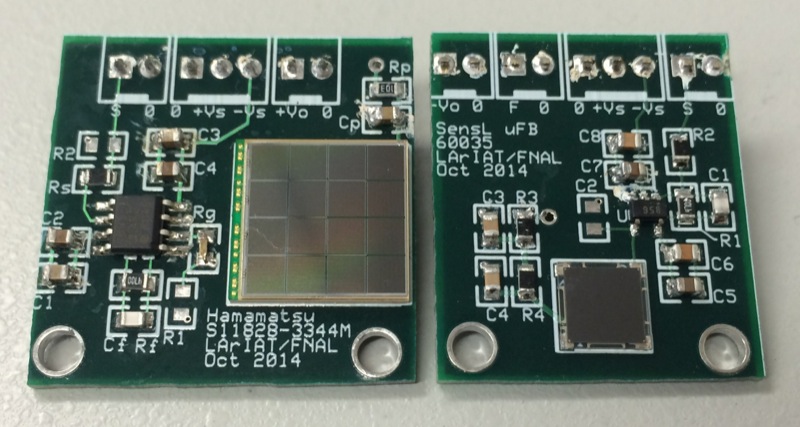}}
\caption{\scriptsize \small Custom readout electronics developed by LArIAT.  Boards include bias voltage filtering, preamp, and readout for the Hamamatsu S11828-3344M, a 4x4 SiPM array with a total active area of $1.2 \times 1.2$ cm$^2$ (left), and for the SensL MicroFB-60035 single SiPM channel with area $0.6 \times 0.6$ cm$^2$ (right).}
\label{miniboards}
\end{figure}
%%%%%%%%%%%%%%%%%%%%%%%%%%%%%%%%%%%%%%%%%%%%

The choice of the DAQ system for the TPB-coated reflector foil light detection system depends on the photosensor adopted, and in the case of PMTs, the limited number of readout channels and the fast signal formation enables the use of fast waveform digitizers like the CAEN V1751 (10 bit, 1 GS/sec ADC) currently used in LArIAT. However, their use would require full and detailed engineering and augmented DAQ performance due to the potential increase in readout channels. The Silicon Photomultiplier signal Processor module described in Sect.~\ref{sec:light_electronics} seems an appropriate solution for the SiPM option, offering 14-bit, 150 MS/sec ADC for digitizing signals. This module have full waveform-recording capabilities with flash ADC, but also allows for flexible FPGA data processing algorithms needed for real-time pulse height and area measurements. 

% A potential drawback of this system is the difficulty of applying direct ``flash-finding'' algorithms to tag cosmic events inside of the drift window and match tracks seen by the TPC to well-timed signals from the PMTs, since the directionality of the detected light is largely washed out by the reflections among multiple inner surfaces of the detector.  Studies are underway to investigate if the arrival time of this reflected light alone is enough to preserve some localization capabilities.  Another solution being investigated to mitigate this effect is a hybrid solution with separate detectors sensitive to either the direct light or the reflected light only.

\subsubsection*{PMT-based Light Detection System}

%design for Improved Timing and Track Matching.}
Light detection using TPB-coated cryogenic PMTs is known to work well in modest-sized \lartpcs, like the ICARUS T600 and \uboone.  This design is not easily scalable to big detectors, and different systems, like the ones reported in the previous sections, have been developed in view of larger \lartpcs for future applications.  

In a PMT-based system, by using fast PMTs and good pixelization, the goal is to detect as much prompt and delayed scintillation light information (time and charge) as possible.  This will then provide good event time and position reconstruction and track matching with the TPC. 
%The benefits of such a system include excellent cosmic/dirt background rejection and low risk as PMTs in LAr has been proven to work. %and minimal cabling and
%electronics system.  
The preference is to have a transparent cathode to maximize the prompt light collection.  A PMT-based system designed for \larnd would use Hamamatsu R11065 3" diameter PMTs with 25\% QE, and 6.5 ns transit time spread (full width at half max) at LAr temperatures.  
% Figure \ref{R11065} shows the R11065. 
The PMTs are flat face and would have a 3" TPB wavelength shifter affixed to the surface.  The combined PMT+TPB efficiency at 128~nm is expected to be 12.5\% .

The \larnd detector instrumented with 121 PMTs on each side arranged in an $11 \times 11$ grid over a $4~m \times 4~m$ frame would have 1\% total photocathode coverage, or 6.7\% on each of the instrumented sides. Light yield in this configuration is estimated to be 13~phe/MeV for a source 2~m away from the the center of an APA, where about 25\% (50\%) comes as prompt light for lightly (heavily) ionizing particles.  A full MC simulation is in progress to confirm these performances.  The option to instrument the outer veto region (LAr outside the active TPc boundaries) is being investigated as well.

\subsubsection*{Electronics for the Light Detection System}
\label{sec:light_electronics}

Several options exist for reading out SiPM or PMT signals, each of which has its own relative merits. These options include a direct signal digitization or the signal digitization after SiPM/PMT pulse shaping. Fast (direct) digitization of SiPM pulses has been implemented by the HEP Electronics Group at Argonne National Laboratory (ANL) to support the development of the photon detection systems for LBNE. The FEE developed at ANL is being adopted for SiPM readout, but the implementation is flexible and may be modified for use with PMTs, if desired.

Each SSP (SiPM Signal Processor) module (Fig. \ref{SIPMDiagram}) consists of 12 readout channels packaged in a self-contained 1U module.  Each channel contains a fully-differential voltage amplifier and a 14-bit, 150 MSPS analog-to-digital converter (ADC) that digitizes the waveforms received from the SiPMs.  The digitized data is then processed by a Xilinx Artix-7 Field-Programmable Gate Array (FPGA).  The FPGA implements an independent Data Processor (DP) for each channel.  The processing incorporates a leading edge discriminator for detecting events and a constant fraction discriminator (CFD) for sub clock timing resolution. In the simplest mode of operation, the module can perform waveform capture, using either an internal trigger or an external trigger.  
% Up to 2046 waveform samples may be read out for each event. When waveform readouts overlap the device can be configured to offset, truncate or completely suppress the overlapping waveform.  Pile-up events can also be suppressed. 
As an alternative to reading full waveforms, the Data Processors can be configured to perform a wide variety of data processing algorithms, including several techniques for measuring amplitude, and also timing of the event with respect to a reference clock. 
% All timing and amplitude values are reported in a compact event record.  Each data processing channel stores up to 340 events records when not storing waveforms. The SSP can be configured to trigger readout in several ways, including self-triggered, use of an external trigger, or use an external gate to readout all events within a time-window.  The digitization clock and timestamp across multiple SSP can be synchronized using an external clock source. A Xilinx Zynq FPGA, onboard the MicroZed system-on-module, handles the slow control and event data transfer.  The SSP has two parallel communication interfaces; USB 2.0 and 10/100/1000 Ethernet.  The 1 Gb/s Ethernet supports full TCP/IP.  The module includes a separate 12-bit high-voltage DAC for each channel to provide up to 30 V of bias to each SiPM.  The module also feature charge injection for performing diagnostics and linearity monitoring, and also voltage monitoring.
       
\begin{figure}[t]
\begin{centering}
 \includegraphics[width=5.0in]{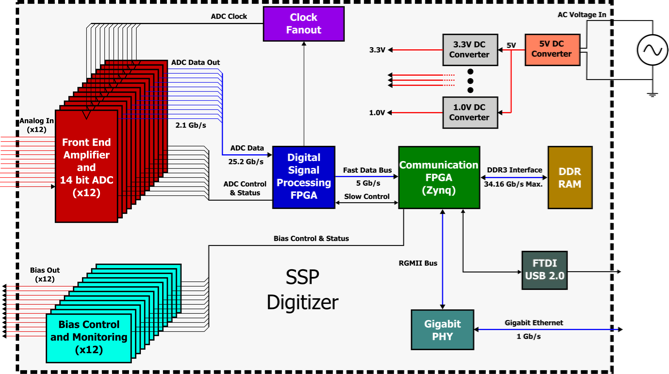}
 \caption{\scriptsize \small Block diagram and physical realization of the prototype SSP module.}
\label{SIPMDiagram}
\end{centering}
\end{figure}

% The system developed for the prototype studies was optimized for flexibility and stand-alone use.  
% The unit incorporates a high-performant Field Programmable Gate Array (FPGA), which is capable of executing algorithms on the digitized data.  The unit is stand-alone, containing all DC power supplies needed for the unit, and is packaged in a 1U chassis.  It has a USB interface, so that the data may be read directly by a PC or laptop.  It also has a GbE interface for use in a larger system.  
% In a large detector system, many of these features would not be needed, and there are aspects that might be eliminated or reconfigured to reduce cost.

If a faster timing response becomes one of the requirements of the photon detection system, a faster ADC with a shaper will be incorporated into the front-end electronics to achieve improved timing resolution.  The present system samples the waveforms at 150 MSPS, and achieves $\sim$2-3 ns resolution on single photo-electron signals.  Given that the time constant of the prompt light is 6-7 ns, it may become necessary to improve the timing resolution to the sub-nanosecond level. This would require the use of faster ADCs and possibly pulse shaping.  
%Such a system could enhance the performance to measure the time of a neutrino event with respect to the beam RF.  This would help with the rejection of cosmic rays, as well as identification of multiple beam interactions.  
This is not part of the system specifications currently, but would be an R\&D activity should the performance be required. 

\section{Cosmic Ray Tagging System}
\label{CosmicRayTagging}

\label{veto}
 As described in Part I of this Proposal, cosmic ray muons are the most abundant background in a \lartpc at surface. Although muons do not contribute significantly to the background for the sterile neutrino search, they produce $\delta$-rays, which in turn produce photons by Bremsstrahlung. These photons, via Compton scattering or pair production, could possibly mimic a $\nu_{e}$-like interaction signature.

The addition of a cosmic ray tagging system, that detects cosmic ray muons and measures their time and position relative to events internal to the TPC, is a way to mitigate the cosmic ray background.

From Monte Carlo simulations, the average number of cosmic muon tracks, seen in each of the \larnd TPC events is about 3. Each muon track is surrounded by tracks of electrons and positrons, originating from Bremsstrahlung of delta-electrons produced by muons, and, for rare very high energy muons, by muon Bremsstrahlung in the liquid argon.
The cosmic ray tracker system chosen for the \larnd detector is composed of scintillators external to the \larnd cryostat providing 4$\pi$ solid angle coverage.

\subsection{Scintillating Tracker Design and Operation}

The \larnd  cosmic ray tracker is composed of scintillating planes, each consisting of an array of 1x10x400~cm scintillating bars as shown in  Figure \ref{Strip}. One plane provides coordinate resolution in X - and the adjacent plane provides resolution in Y coordinate. In the following description each such sandwich is referred to as X-Y tracker plane. 

Each bar is made of BC-440 (or similar) plastic scintillator with the emission maximum at 434~nm and bulk attenuation length of $>$400~cm. The bar is wrapped in a diffuse reflector foil (Tyvek) to make the light field inside the bar more uniform and, additionally, in an aluminum-coated Mylar foil. The insensitive gap between two adjacent bars is about 0.5~mm.

In order to provide a more efficient and uniform collection of scintillation light along the bar, two wavelength-shifting fibers (1 mm diameter multi-clad Kuraray WLS Y11(200) S-type \cite{Y11}) are glued into the scintillating bar near its edges at both lateral sides. The light is transmitted by the fibers to the bar edge, where it is detected by the Hamamatsu S12825-050P multi-pixel Geiger avalanche photo-diodes (MAPDs, also known as SiPMs).
Matching of the Y-11 emission spectrum to the SiPM sensitivity is illustrated in Figure \ref{MppcY11}. 

In order to mitigate cosmic ray events, 
%that contribute to the background for $\nue$ and $\numu$ detected by LAr1ND TPC, 
a 4$\pi$ solid angle coverage of the TPC is %required.
highly desirable.
%In the direction of maximum flux of cosmic muons
On the top, where the flux of cosmic rays is maximum, an additional X-Y plane is installed at the distance of 2~m above the X-Y plane covering the top surface of the TPC as shown in Figure \ref{Veto}. These two planes in combination form a telescope, that provides coordinate resolution of $10/\sqrt{12}\approx2.9$~cm at the detector planes and angular resolution of $2\times 2.9/200\approx0.03$~rad.   

% The set of the cosmic ray events giving the largest contribution to the experiment background is shown in Figure \ref{Events} together with the scenarios of their identification and rejection with the proposed cosmic ray tracker. The average number of cosmic muon tracks, seen in each of the LAr1ND TPC events is about 3. Each muon track is surrounded by tracks of electrons and positrons, originating from delta-electrons produced by muons, and, for rare very high energy muons, by muon bremsstrahlung in the liquid argon.
% These electrons and positrons tracks, unconnected to the original muon track, are contributing to the background for $\nue$ CC interactions. An example of the simulated distribution of their distance from the parent muon track is shown in Figure \ref{PSala1}.
% Each muon track detected by the TPC is surrounded by the cylinder with the radius $R$ defined on the basis of the above distribution, as illustrated in Figure \ref{Events}, left top. All electromagnetic (EM) activity within this cylinder can relate to the muon, and therefore, must be vetoed.

In order to increase the rejection of the EM background and the detection efficiency of $\nue$ CC interactions the following algorithm is proposed. The time distribution of the muon-related signal from the scintillating tracker is compared to that of the internal TPC light detection system and to the beam gate. Muon tracks, detected in the TPC are extrapolated to the scintillating planes. The time distribution of the signal from the scintillating tracker channels, that are crossed by the extrapolated muon trajectories is used to build correspondence with the internal TPC light collection system signals. Those tracks, that are seen by the tracker and internal light collection system outside of the beam gate are unambiguously identified as cosmic muons, 
%the EM activity within the $R$ distance from the track is discarded. 
The remaining event with the signature of no signal in scintillating tracker and the  signal in the TPC light collection system in time with the beam gate can be identified as the beam-related. All 7 X-Y planes of the scintillating tracker are essential for this procedure.

\begin{figure}[t]
\centering	
\includegraphics[width=0.6\linewidth]{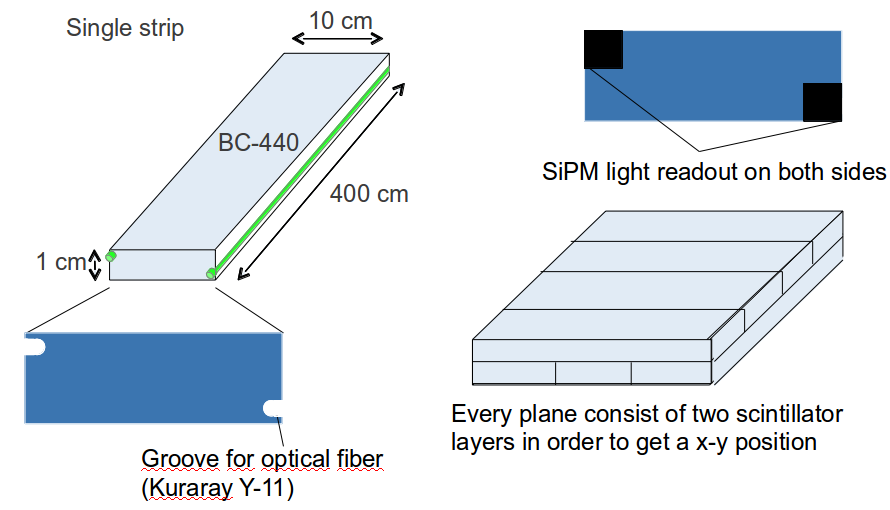}
\caption{Veto plane composition: scintillating bars with WLS fibers glued into grooves along strip sides. The material for the bars is BC-440 or similar plastic scintillator. Scintillation light is collected with Kuraray Y-11 WLS fibers and transmitted to the bar edge, where it is detected by Hamamatsu S12825-050P SiPMs.}
\label{Strip}
\end{figure}

\begin{figure}[t]
\centering	
\includegraphics[width=0.3\linewidth]{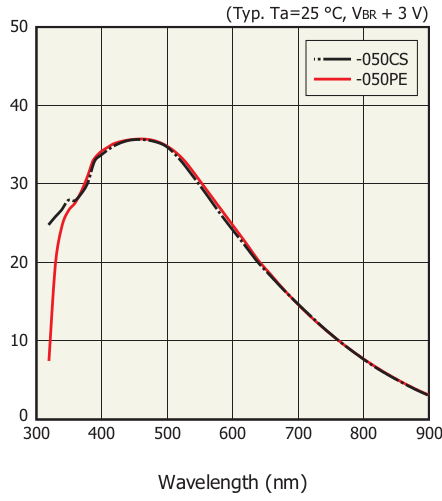}
\includegraphics[width=0.5\linewidth]{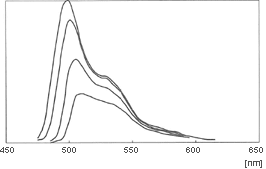}
\caption{SiPM spectral sensitivity (left), and emission spectrum from Y11 WLS fiber (right).}
\label{MppcY11}
\end{figure}

\begin{figure}[t]
\centering	
\includegraphics[width=0.6\linewidth]{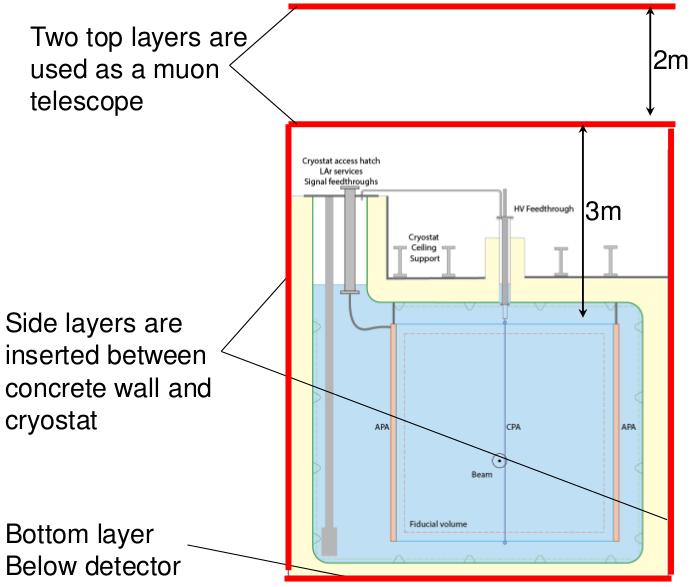}
\caption{Veto planes arrangement around the detector.}
\label{Veto}
\end{figure}

% \begin{figure}[t]
% \centering	
% \includegraphics[width=0.49\linewidth]{VETOCALIB/Veto_case1.png}
% \includegraphics[width=0.49\linewidth]{VETOCALIB/Veto_case2.png}
% \includegraphics[width=0.49\linewidth]{VETOCALIB/Veto_case3.png}
% \includegraphics[width=0.49\linewidth]{VETOCALIB/Veto_case4.png}
% \caption{Cosmic ray events, which give most significant contribution to the background in LAR1ND TPC and their mitigation with the proposed cosmic tagging system.}
% \label{Events}
% \end{figure}

\subsection{Electronic Readout System}
\label{sec:CRTS}

As mentioned above, in the proposed design the Hamamatsu S12825-050P photodiodes are used to convert wavelength-shifted scintillation light to the electronic analog signal.  The most relevant parameters of these diodes are shown in Figure \ref{Mppc} as a function of the over-voltage (the difference between the applied reverse-bias voltage and the breakdown voltage). At the over-voltage of 2.0~V the PDE at the maximum of the spectral sensitivity is about 32\% and the gain exceeds 10$^6$. The  pulse frequency at the threshold allowing to detect single photons is of the order of 100 kHz at 20$^{\circ}$C. 
The signal from photodiode is amplified and shaped by a CITIROC multi-channel front-end ASIC, designed by Omega. %reference! 
The analog signals are discriminated at the level of a 1.5 of single photo-electron response.
To suppress the dark current rate the logic coincidence is used at each end of the scintillating bar, between the signal from two diodes glued to two WLS fibers at each side of the bar.
The resulting logic signals together with digitized peak values of the shaped analog signals and time stamp are stored in a FIFO buffer and eventually transmitted via Ethernet link to the experiment DAQ system for event building and storage.

\begin{figure}[t]
\centering	
\setlength{\fboxsep}{0pt}
\setlength{\fboxrule}{0.6pt}
\raisebox{5mm}{\fbox{\includegraphics[width=0.35\linewidth]{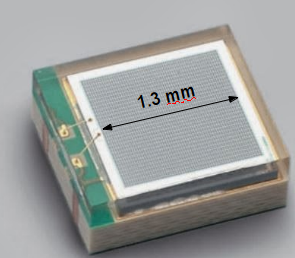}}} \quad \quad \quad
\includegraphics[width=0.45\linewidth]{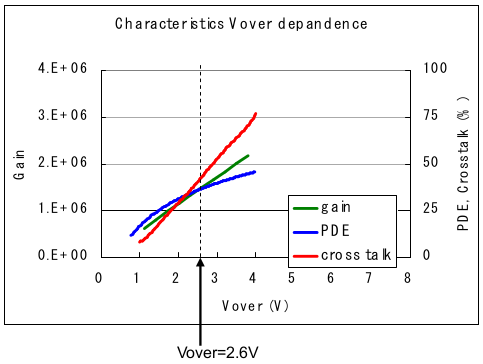}
\caption{Hamamatsu S12825-050P SiPM (left), photon detection efficiency (PDE), noise and pixel cross-talk as a function of over-voltage  at 408~nm (right).}
\label{Mppc}
\end{figure}

\subsection{Expected Performance}

The performance of the proposed tracker configuration is estimated on the base of the experimental data published in reference \cite{Musienko}. The authors studied the light yield and timing characteristics of a similar scintillating tracker with 16~m long scintillator bars, Kuraray Y-11(200)S WLS fiber and Hamamatsu S10362-13-050C photodiodes, very similar to S12825-050P used in the present proposal.
The thickness of the bars studied was 7~mm and the width from 1~cm to 4~cm.
The effective attenuation length was found to be $\approx$ 4.2~m. Extrapolating their data to the case of 1x10x400~cm bars yields the expected data on light yield, shown in the table \ref{LY}. The 
detection inefficiency in each channel is calculated as a Poisson probability to obtain a signal lower than 2 p.e. The single end trigger inefficiency is given by the logic AND between left and right channels
at each end of the bar.

\begin{table}[ht]
\caption{The summary on the expected MIP light yield and detection efficiency for the proposed cosmic ray tracker. The data is obtained by extrapolation of experimental results published in \cite{Musienko}.}
\vspace{1mm}
\centering	
\begin{tabular}{lcc}
%\toprule
%\hline
                      		&   Near end (1.0~m)   &   Far end (4.0~m) \\ %\hline
Light yield, p.e.     		&   22                   & 11             \\ %\hline
Detection inefficiency		&   6.4x10$^{-9}$          & 2.0x10$^{-4}$    \\ %\hline
Single end trigger inefficiency & 1.2x10$^{-8}$          & 4.0x10$^{-4}$    \\ %\hline
Single channel dark count rate  & 10 kHz  & 10 kHz \\ %\hline
Single end trigger  dark count rate  & 10 Hz   & 10 Hz \\ %\hline
%\bottomrule
\end{tabular}
\label{LY}
\end{table}

The velocity of the re-emitted light propagation in the Y11 fiber is found to be 16.00$\pm$0.08 cm/ns and the decay time of Y11 fiber 12$\pm$0.5 ns \cite{Musienko}. Therefore the coincidence window for left and right discriminated signals is conservatively chosen to be about 100~ns (defined by CITIROC FE ASIC timing characteristics). The rate of dark current pulses above the 1.5 p.e. threshold is about 10 times lower than that for single photo-electron \cite{Hamamatsu}, that is of the order of 10 kHz.
The resulting rate of accidental left-right coincidence is about 10 Hz. Such a low rate allows us to minimize the useless part of the data flow from the scintillating tracker and to use inexpensive  low-rate transmission channels from the tracker planes to the processing unit.
At the processing unit a more sophisticated logic of correlation of the signals from both ends of the bar, as well as with the signals from the perpendicular bars can be implemented.

The expected minimum sum of the signal from both ends of the bar is 36 p.e. on each of the left and right WLS fiber channels. The total sum per bar is therefore 72 p.e. The minimum total sum of the dark count pulses is 2x4=8 p.e. Setting a constraint of having the total sum per bar of at least 20 p.e will bring the rate of fake hits per bar down by more that ten orders of magnitude, making it negligible. The additional inefficiency introduced by such cut is of the order of 10$^{-13}$, therefore, also negligible. 

The probability of the multiple muons hitting the same scintillator channel is estimated from the Monte-Carlo simulation of cosmic muon flux at the detector surface by CRY simulation package.
The resulting multiplicities are shown in Figure \ref{Mult}. The graph at the left shows average multiplicity of muon tracks in the whole top 4x4~m$^2$ surface of the tracker X-Y plane, while the right plot shows the multiplicity in each single scintillating bar within 100~ns of the signal integration period.
The probability to have more than one hit per bar is of the order of 10$^{-5}$.

\begin{figure}[t]
\centering	
\includegraphics[width=0.45\linewidth]{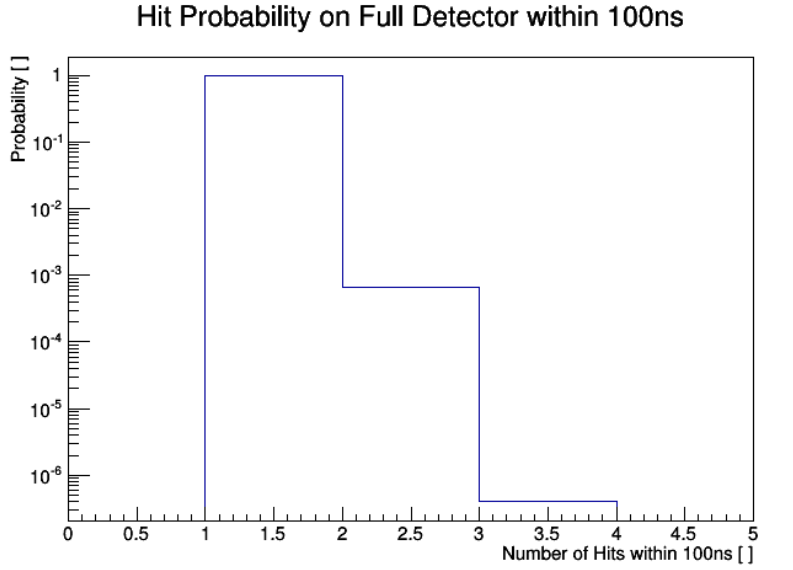}\quad
\includegraphics[width=0.47\linewidth]{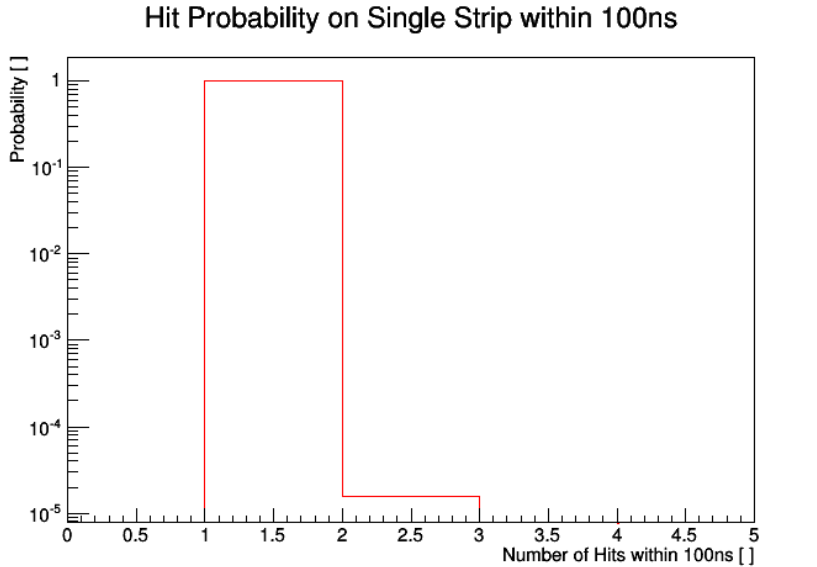}
\caption{The expected muon multiplicity over the whole top scintillator plane (left) and over one 1x10x400~cm scintillator bar.}
\label{Mult}
\end{figure}

Since the detection inefficiency of the proposed light readout system is negligible, the total detection efficiency of the tracker is limited by the 0.5~mm thick insensitive gaps between adjacent scintillating bars. The detection efficiency of one coordinate plane is therefore about 99.5\%.
If the coincidence of X and Y planes is required, the detection efficiency drops to 99\% per X-Y plane.
%Together with the information from chapter \ref{CosPhys} one can determine the background rejection efficiency and the related fiducial volume size, see Figure \ref{LAr1Fiducial}. The maximum fiducial volume of LAr1-ND is chosen to be 35.5~m$^{3}$.
%This allows rejection of the cosmic-muon related background events by a factor of 1000, significantly improving sensitivity of the LAr1ND detector to beam-related $\numu$ and $\nue$ CC interaction.

%\begin{figure}[ht]
%\centering
%\includegraphics[width=0.8\linewidth]{VETOCALIB/LAr1FiducialCut.png}
%\caption{Remaining fraction of background as a function of fiducial volume size of the LAr1-ND detector. This simulation with an energy cut of 50~MeV for electrons/positrons is made for LAr1-ND .}
%\label{LAr1Fiducial}
%\end{figure}
\section{The \larnd Detector: A Development Toward LBNF}
\label{sec:RD}

\larnd presents an excellent opportunity for the continued development of the \lartpc technology toward the LBNF program.  The design of the \larnd detector is largely based on current LBNF-type technology, but alternate solutions can also be pursued where it is valuable to inform final choices for the LBNF detector.  The designers of \larnd systems are, in many cases (i.e. cryostat, cryogenics, TPC, cold electronics), the same teams working on LBNF designs, ensuring good communication of ideas and lessons learned.  \larnd's location 110~m from the Booster Neutrino Beam target will provide a unique opportunity to test specific components and new concepts in a high-rate neutrino beam.  

%Common teams with LBNF have designed the cryostats and cryogenics, the TPC and the cold electronics for \larnd. %The \larnd design has many similarities with LBNF, but also implement some different concepts.  

Tables~\ref{tab:lar1ndvslbnf_cryo}, \ref{tab:lar1ndvslbnf_TPC}, and \ref{tab:lar1ndvslbnf_electronics} compare different systems of the \larnd and LBNF detector designs, highlighting key similarities and differences.  

A comparison of the \larnd and LBNF design of the cryostat and cryogenic systems is reported in Table~\ref{tab:lar1ndvslbnf_cryo}. The same membrane-style cryostat is used. One difference is the location of the cryogenic pumps.  Also, in \larnd, there is the possibility that the liquid completely fills the main volume, touching the top plate, in order to minimize the outgassing from the surface and cabling. This choice is currently under study. 

Table~\ref{tab:lar1ndvslbnf_TPC} compares the features of the \larnd and LBNF TPC designs and many similarities exist.  The main difference is
that the \larnd design does not wrap the readout wires around the APA frames since this is not needed with the drift volume only on one side. 

As reported in Table~\ref{tab:lar1ndvslbnf_electronics} the \larnd electronics is largely based on an already developed LBNF design. The choice of the cold FPGA for digital processing is due to the long lead time needed to develop a dedicated ASIC for this task.  This work is underway, and could be tested in the detector in a future phase of running.   

Finally, as discussed in Section \ref{LightDetectionSystem}, \larnd provides an excellent test-bed for light collection systems in a LAr detector.  The TPB coated acrylic light guide design is based on concepts developed for LBNF, and \larnd will be a direct test of this approach in a running neutrino experiment.  Other approaches are being developed in attempt to enhance light collection with increased collection efficiency and improved time resolution. \larnd provides an opportunity to test new approaches, possibly side-by-side, with the goal of informing an optimized design for LBNF in the future.  

%to tag and reconstruct low energy particles that are usually close to the threshold for standard \lartpc detectors, will be very valuable  in view of  studies of low energy events, i.e. supernova neutrino events in LBNF.  
%, which has a very similar design to one of the proposed light detection system for LBNF, will allow to perform a test in real experimental conditions. One of the other proposed system, the  LArIAT-style enhanced light collection system, designed to be able to tag and reconstruct up to low energy particles that are usually close to the threshold for standard \lartpc detectors, will be very valuable  in view of  studies of low energy events, i.e. supernova neutrino events in LBNF.

\begin{table}[h]
\centering
\captionsetup{justification=centering}
\label{tab:lar1ndvslbnf_cryo}
\includegraphics[width=1.\textwidth]{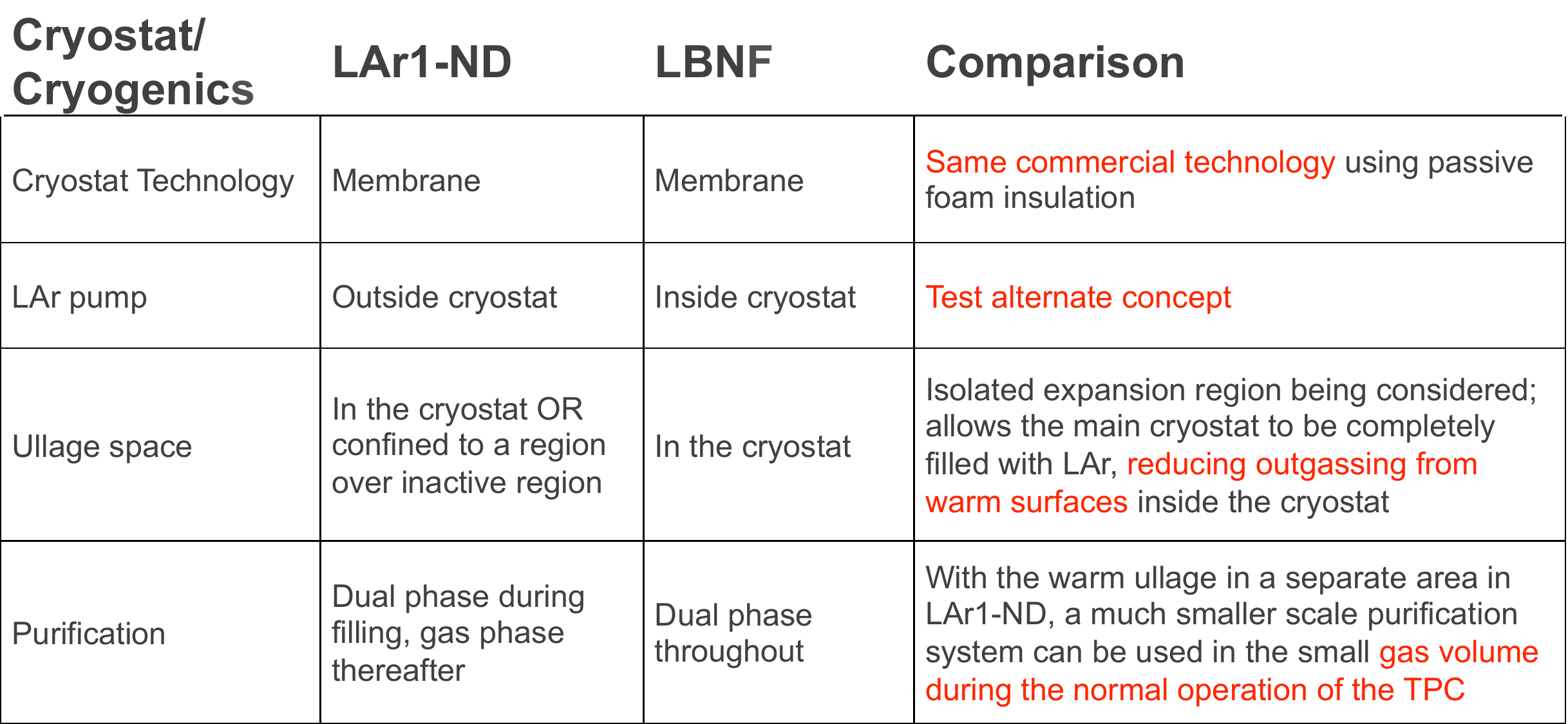}
\caption{\label{tab:lar1ndvslbnf_cryo}}
%\larnd vs LBNF: cryogenics and cryostat design}
\end{table}

\begin{table}[h]
\centering
\captionsetup{justification=centering}
\includegraphics[width=1.\textwidth]{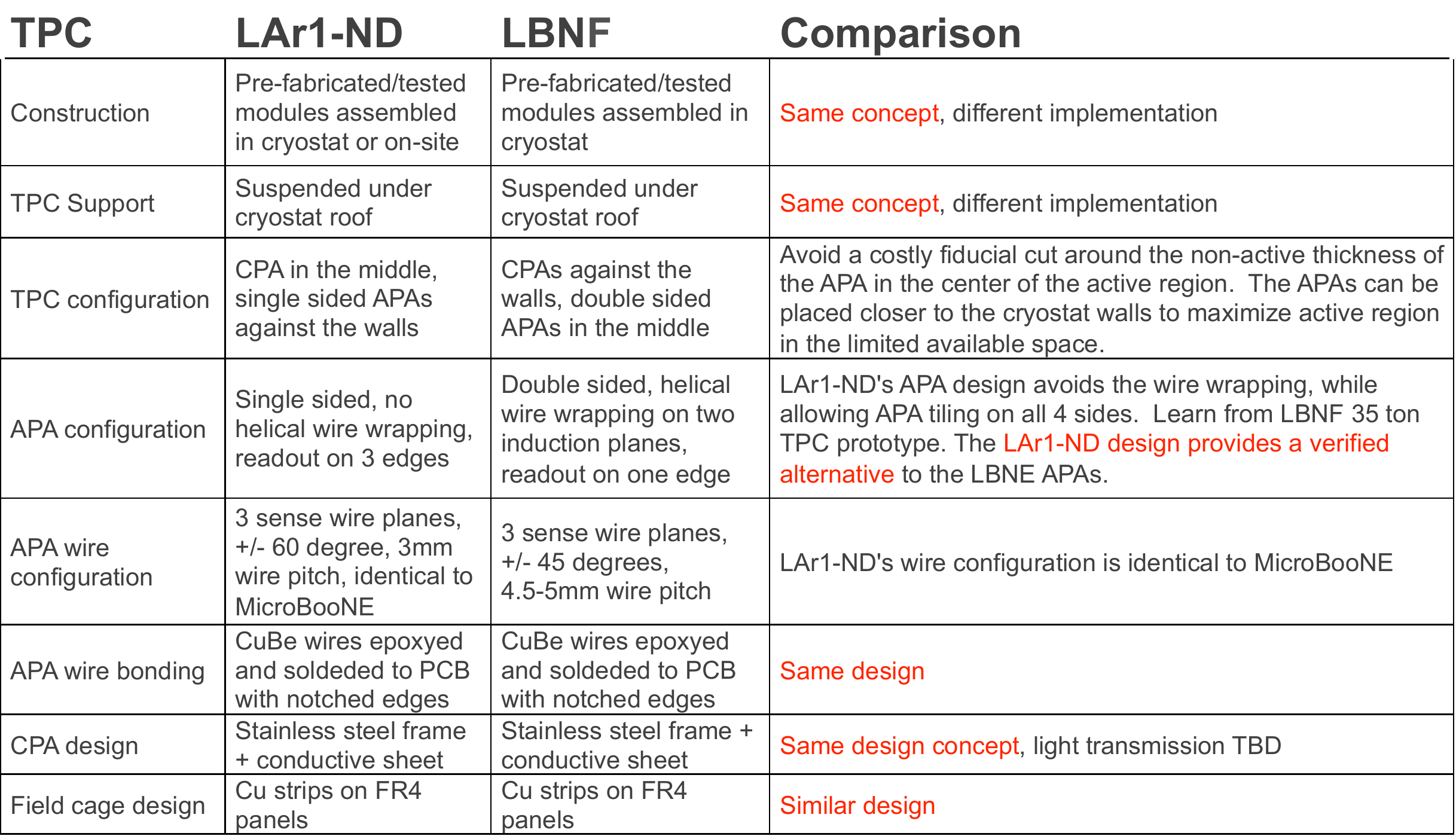}
\caption{\label{tab:lar1ndvslbnf_TPC}}
%LAr1-ND vs LBNF: TPC design}
\end{table}

\begin{table}[h]
\centering
\captionsetup{justification=centering}
\includegraphics[width=1.\textwidth]{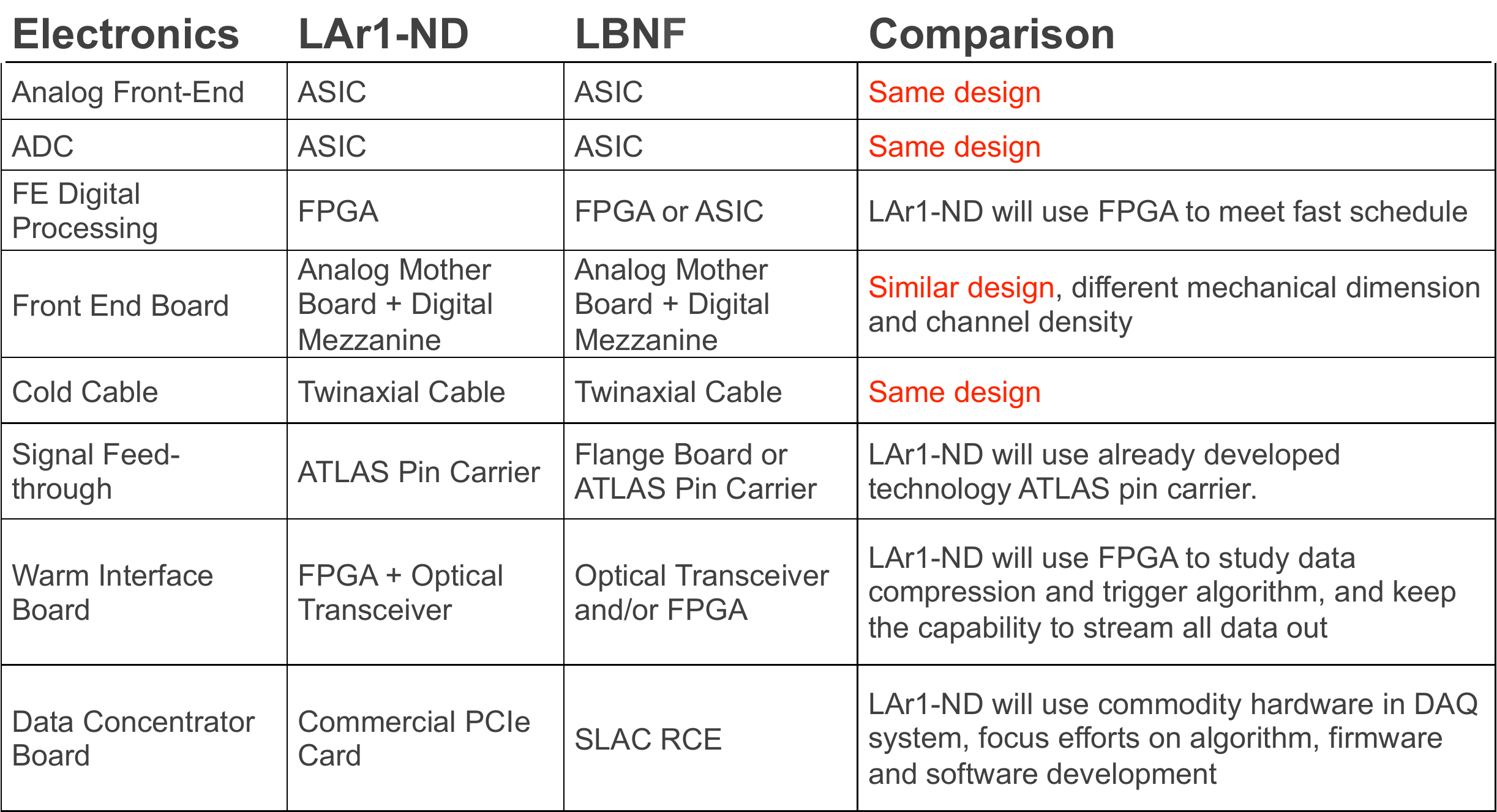}
\caption{\label{tab:lar1ndvslbnf_electronics}}
%\larnd vs LBNF: cold and warm electronics design}
\end{table}

% \begin{table}
% \centering
% \includegraphics[width=0.8\textwidth]{lar1ndvslbnf_light.png}
% \caption{\label{tab:lar1ndvslbnf_light}
% LAr1-ND vs LBNF: light collection system}
% \end{table}

\FloatBarrier
\clearpage \clearpage
\pagestyle{empty}

\proposalTitle 

\setcounter{part}{2}
\part{T600 Design and Refurbishing}

%\begin{center}
%{\large
%DRAFT \\
%\bigskip
%\today
%\proposaldate
%}\end{center}
 
\ifcombine
\clearpage
\else
\clearpage
\tableofcontents
\addtocontents{toc}{\protect\thispagestyle{empty}}
\clearpage
\setcounter{page}{0}
\fi

\pagestyle{fancy}
\rhead{III-\thepage}
\lhead{T600 Design and Refurbishing}
\cfoot{}

%%%%%%%%%%%%%%%%%%%%%%%%%%%%%%%%%%%%%%%%%%%%%%%%%%%%%%%%%%%%%%%%%%%%%%%%%%%%%%%%%%%%%%%%%%%%%%%%%%%%%%%%%%%%%%%%%%%%%%%%%%%%%%%%%%%%%%%%%%%%%%%%%%%%%%%%%%%%
\section{Introduction}

Imaging detectors have always played a crucial role in particle physics. In the past century successive generations of detectors realized new ways to visualize particle interactions, driving the advance of physical knowledge and the discovery of unpredicted phenomena, even on the basis of single fully reconstructed events. In particular, bubble chamber detectors were an incredibly fruitful tool, permitting to visualize and study particle interactions, providing fundamental contributions to particle physics discoveries. Gigantic bubble chambers, like Gargamelle~\cite{gargamelle,gargamelle2} (3 tons of mass), were extraordinary achievements, successfully employed in particular in neutrino physics. Two major limitations of bubble chambers in the search for rare phenomena are the impossibility to scale their size towards much larger masses, and their duty cycle which is intrinsically limited by the mechanics of the expansion system.

In 1977 C. Rubbia~\cite{rubbia} conceived the idea of a Liquid argon Time Projection Chamber (\lartpc), i.e. the calorimetric measurement of particle energy together with three-dimensional track reconstruction from the electrons drifting in an electric field in sufficiently pure liquid argon. The \lartpc successfully reproduces the extraordinary imaging features of the bubble chamber, its medium and its spatial resolution being similar to those of heavy liquid bubble chambers, with the further feature of being a fully electronic detector, potentially scalable to huge masses (several kton). In addition the \lartpc provides excellent calorimetric measurements and has the big advantage of being continuously sensitive and self-triggering. 

The \icarus cryogenic detector is the biggest \lartpc ever realized, with the cryostat containing 760 tons of LAr (476 tons active mass). Its construction finalized many years of R\&D studies by the ICARUS Collaboration~\cite{ica1, ica2, ica3, ica4, ica5}, with prototypes of growing mass developed both in laboratory and with industry involvement. Nowadays, it represents the state of the art of this technique and it marks a major milestone in the practical realization of large-scale LAr detectors. 

The pre-assembly of the ICARUS T600 detector began in 1999 in Pavia (Italy); one of its two 300-tons modules was brought into operation in 2001. A test run lasting three months was carried out with exposure to cosmic rays on the surface, allowing for the first time an extensive study of the main detector features~\cite{ICARUS_NIM}. After the test, the detector was de-commissioned and, in 2004, the two cryostats housing the internal detectors were transported to their final site, in the Hall B of the underground Gran Sasso National Laboratories (LNGS). A number of activities on the \icarus plant were then necessary for the completion of the detector assembly in its underground site. In the first months of 2010 the T600, see Fig.~\ref{ica_hallb}, was finally brought into operation~\cite{ICARUS_jinst} taking data with the CERN to Gran Sasso (CNGS) neutrino beam and with cosmic rays. The ICARUS experiment has operated with a remarkable detection efficiency
and it has successfully completed a three years physics program being exposed to the CNGS beam from October 2010 to December 2012. Neutrino events have been collected, corresponding to 8.6~$\times$~10$^{19}$ protons on target with an efficiency exceeding 93\%. Additional data were also collected with cosmic rays, to study atmospheric neutrinos and proton decay. From the technological point of view, the T600 run was a complete success, featuring a smooth operation, high live time, and high reliability. A total of about 3,000 CNGS neutrino events has been collected and is being actively analyzed. 

The successful operation of the \icarus \lartpc proves the enormous potential of this detection technique, addressing a wide physics program with the simultaneous exposure to the CNGS neutrino beam and cosmic-rays~\cite{ICARUS_EPJ, ICARUS_EPJ-2, ica_nuvel1, ica_nuvel2}. Moreover, the solutions adopted for the argon recirculation and purification systems permitted to reach an impressive result in terms of argon purity, which is one of the key issues for the superb detector performance. A corresponding free electron lifetime exceeding 15 ms has been obtained, a milestone for any future project involving \lartpcs. This result~\cite{LArPurity} demonstrates the effectiveness of the single phase \lartpc detectors~\cite{modular1, modular2}, paving the way to the construction of huge detectors with longer drift distances: for example, with the achieved purity level, at 5 m from the wire planes the maximum signal attenuation is only 23\%.

The T600 decommissioning process started in June 2013, with the cryostat emptying phase lasting less than one month in a safe and smooth way. A warming-up phase followed, that brought the cryostats to room temperature in about one month. The T600 dismantling started in September 2013 and globally lasted about 10 months. After it was concluded, the cryostats were opened, to recover the internal TPC detectors and the cryogenic plant and electronics to be re-used in future projects.

The movement of the two T600 modules to CERN has been already completed. The \icarus TPCs are ready for their complete overhauling (CERN WA104 project), preserving most of the existing operational equipment, while upgrading some components with up-to-date technology in view of the T600 future non-underground operation at FNAL. 

\begin{figure}[htbp]
\centering
\includegraphics[width=0.8\textwidth]{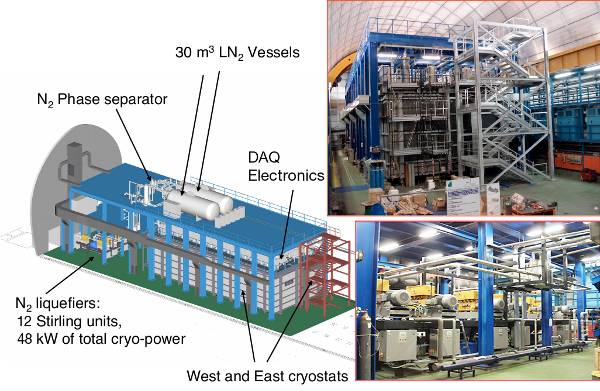}
\caption{Left: schematic view of the whole \icarus plant in Hall B at LNGS. Right-top: photo of the detector installation. Right-bottom: details of the cryo-cooler plant.}
\label{ica_hallb}
\end{figure}

This Design Report is organized as follows. Sec.~\ref{ica_physics},
after recalling the SBN experimental program, briefly reviews the potential of the \icarus standalone physics program with the NuMI beam. Requirements for the detector to operate at shallow depths are resumed in Sec.~\ref{ica_requirements}. The present \icarus detector configuration is described in Sec.~\ref{ica_design}, while the T600 overhauling activities foreseen at CERN as WA104 program are shown in Sec.~\ref{wa104}, with particular emphasis on the new Light Collection System, the new Electronics and the new Cryogenic and Purification systems. The possibility to complement the detector with an external cosmic ray tagging system is discussed in Sec.~\ref{ica_crts}. 
%The T600 infrastructures at FNAL are shortly presented in Sec.~\ref{ica_building}. Finally, Sec.~\ref{ica_funding} resumes the funding and the organization of the ICARUS T600 program.

%%%%%%%%%%%%%%%%%%%%%%%%%%%%%%%%%%%%%%%%%%%%%%%%%%%%%%%%%%%%%%%%%%%%%%%%%%%%%%%%%%%%%%%%%%%%%%%%%%%%%%%%%%%%%%%%%%%%%%%%%%%%%%%%%%%%%%%%%%%%%%%%%%%%%%%%%%%%%
\section{Physics of Far Detector}
\label{ica_physics}

%%%%%%%%%%%%%%%%%%%%%%%%%%%%%%%%%%%%%%%%%%%%%%%%%%%%%%%%%%%%%%%%%%%%%%%%%%%%%%%
\subsection{The SBN experimental program}

In recent years, several experimental ``anomalies" have been reported which, if confirmed, could be hinting at the presence of additional ``sterile" neutrino states with larger mass-squared differences participating in the mixing~\cite{reactor, Gallex, Sage, lsnd, mbosc1, mbosc2, mbcomb}.
An important contribution to the sterile neutrino search has been already given by the ICARUS Collaboration with the T600 detector running in the underground INFN-LNGS Laboratory and exposed to the CNGS neutrino beam~\cite{ICARUS_jinst, ICARUS_EPJ, ICARUS_EPJ-2}. 

As already described in details in the Part I of this Proposal, the future short-baseline experimental configuration is proposed to include three \lartpc detectors located on-axis along the Booster Neutrino Beam (BNB). 
The Near Detector (\larnd) will be located in a new building directly downstream of the existing \SB enclosure, 110~m from the BNB target.  The \uboone detector, which is currently in the final stages of installation, is located in the Liquid argon Test Facility (LArTF) at 470~m.  The Far Detector (the existing \icarus) will be located in a new building, 600~m from the target, between \MB and the NO$\nu$A Near Detector surface building. The challenge of predicting absolute neutrino fluxes in accelerator beam experiments, and the large uncertainties associated with neutrino-nucleus interactions, strongly motivate the use of multiple detectors at different baselines, to reduce systematic uncertainties in the search for oscillations.

The observed set of anomalous results in neutrino physics calls for a conclusive new experiment capable of exploring the parameter space in a definitive way and to clarify the possible existence of eV-scale sterile neutrinos. 
 
%%%%%%%%%%%%%%%%%%%%%%%%%%%%%%%%%%%%%%%%%%%%%%%%%%%%%%%%%%%%%%%%%%%%%%%%%%%%%%
\subsection{T600 Physics with the NuMI beam}

The physics outreach of the T600 detector as a stand alone detector can be enhanced with the study of neutrino cross-sections and interaction topologies at energies relevant to the Long Baseline Neutrino Facility (LBNF) program, exploiting the off-axis neutrinos from the NuMI beam.

The NuMI beam-line is fed by 120 GeV protons with 4~$\times$~10$^{13}$ protons per pulse. The secondary beam includes a double-horn focusing system which allows for different variable energy configurations producing a neutrino beam directed, towards the far MINOS detector, with a slope of $\sim$~50~mrad.

Given the NuMI repetition rate (0.53 Hz) and its spill duration (8.6~$\mu$s), one trigger every 12 s is expected in the T600, mainly due to cosmic rays occurring in the coincidence gate. About 1 neutrino event from the NuMI beam every 150 s is also foreseen.

The T600 will collect a large neutrino event statistics in the 0$\div$3 GeV energy range with an enriched component of electron neutrinos (several \%) from the dominant three body decay of secondary \textit{K}. A careful and detailed analysis of these events will be highly beneficial for the future LBNF LAr program, allowing to study very precisely detection efficiencies and kinematical cuts in all neutrino channels and event topologies.

A FLUKA-based Monte Carlo simulation~\cite{Fluka,Fluka2,Flukaneutrino} of the NuMI beam line has been set up according to the available technical drawings~\cite{numi} for the low energy beam configuration. The obtained neutrino fluxes have been compared with those published by the MINOS collaboration at the MINOS near detector position. Even if not all the geometry details were available and/or included in the simulation, our results agree with the MINOS ones within 20\%, indicating that reliable prediction of neutrino rates at off-axis positions is possible.

Muon neutrino event rates are comparable with the ones from the Booster beam, while the electron neutrino component is enhanced in the off-axis beam.
 
This amount of data would allow a detailed evaluation of detection efficiency and background reduction at the energy of the second oscillation maximum in the LBNF expected signal.

%%%%%%%%%%%%%%%%%%%%%%%%%%%%%%%%%%%%%%%%%%%%%%%%%%%%%%%%%%%%%%%%%%%%%%%%%%%%%%%%%%%%%%%%%%%%%%%%%%%%%%%%%%%%%%%%%%%%%%%%%%%%%%%%%%%%%%%%%%%%%%%
\section{Requirements for Detector Performance} 
\label{ica_requirements}

The \icarus detector in the present configuration is already well suited for sterile neutrino searches at FNAL. Nonetheless, it was designed for the low background, deep underground conditions of LNGS laboratory, where the single prompt “trigger” has always ensured the unique timing connection to the main “image” of the event. However, the situation will be substantially different for a detector of this magnitude if placed at shallow depths (a few meters deep), since several additional and uncorrelated “triggers” (due to cosmic rays) will be generally occurring continuously and at different times during the $\sim$~1~ms duration of the T600 readout window~\cite{arxiv_rubbia}. This represents a new problem since, to reconstruct the true position of the track, it is necessary to precisely associate the different timings of each element of the image to their own specific delay with respect to the trigger. The specific investigation of the oscillation anomalies at shallow depths is based on the search of a signal with the presence of a neutrino-induced, single ionizing electron (or positron). High energy cosmic muons creating secondary showers may also produce single ionizing background electrons or positrons with similar energies. At the neutrino energies of the FNAL Booster Beam, the intrinsic $\nu_e$CC contamination occurs at the very low rate of $\sim$~500 $\nu_e$CC/y, while a possible LSND-like oscillation signal will produce a few hundred $\nu_e$CC/y (e.g. $\sim$~170 ev/y for \dmsq = 0.43 \eVsq, \sinth = 0.013). On the other hand, as already described in details in Part I of this Proposal, the cosmic ray background is very prolific of events: in a pit covered by 3 m of concrete, cosmic muon rates in coincidence with the beam trigger window of 1.6~$\mu$s, will produce the huge rate of 0.83~$\times$~10$^6$ cosmics per year (c/y). Moreover, during the 1 ms long duration of each readout window, $\sim$~11 cosmic ray tracks are expected over the full T600, in agreement with the ICARUS measurements at surface carried out in 2001 test run~\cite{ICARUS_NIM}. It is concluded that in its original configuration the ICARUS \lartpc detector cannot perform a practical search for LSND-like anomalies at shallow depths, since the cosmic trigger events are too much frequent.
As already pointed out in Part I, depending on the background type, several reduction strategies can be applied. In addition to offline analysis techniques for background reduction, mostly based on electron/photon discrimination through dE/dx evaluation, the T600 detector will require the implementation of the following three features:

\begin{itemize}
 \item the realization of a new light collection system, to allow a more precise event timing and localization;
\item the exploitation of the BNB bunched beam structure, lasting 1.15 ns (FWHM $\sim$ 2.7~ns) every 19 ns, to reject cosmic events out of bunch as proposed in a SBN note~\cite{bb_rubbia}; 
\item the realization of a cosmic ray tagging system, external to the LAr fiducial volume, to
automatically identify entering charged tracks with position and timing information: this would greatly facilitate the reconstruction and identification of muon tracks. It has to be reminded that crossing muons can be identified by the 3D reconstruction software, however the 3D reconstruction itself needs the information on the track absolute time t$_0$.
\end{itemize}

%%%%%%%%%%%%%%%%%%%%%%%%%%%%%%%%%%%%%%%%%%%%%%%%%%%%%%%%%%%%%%%%%%%%%%%%%%%%%%%%%%%%%%%%%%%%%%%%%%%%%%%%%%%%%%%%%%%%%%%%%%%%%%%%%%%%%%%%%%%%%%%%%%%%%%%%%%%%%%
\section{The T600 Detector: present configuration}
\label{ica_design}

The \icarus detector consists of a large cryostat split into two identical, adjacent modules, with internal dimensions 3.6 $\times$ 3.9 $\times$ 19.6 m$^3$ each, filled with about 760 tons of ultra-pure LAr.
The modules will be referred in the text as \textit{West module} (the oldest one) and \textit{East module} (the newest one), with respect to CNGS beam coming from the North.
A uniform electric field (E$_{drift}$~=~500~V/cm) is applied to the LAr bulk: each module houses two TPCs separated by a common cathode.

Charged particles, generated for example by a neutrino interaction in LAr, produce ionization along their path. Thanks to the low transverse diffusion of charge in LAr, the images of the tracks (produced by ionization electron clouds) are preserved and, drifting along the electric field lines, are projected onto the anode, as illustrated in Fig.~\ref{ica_illustration}.
The TPC anode is made of three parallel planes of wires, 3 mm apart, facing the 1.5 m drift path. Globally, 53,248 wires with length up to 9 m are installed in the detector. By appropriate voltage biasing, the ionization charge induces signals in non-destructive way on the first two planes (Induction-1 and Induction-2), then it is finally collected by the last one (Collection plane).

\begin{figure}[htbp]
\centering
\includegraphics[scale=0.6]{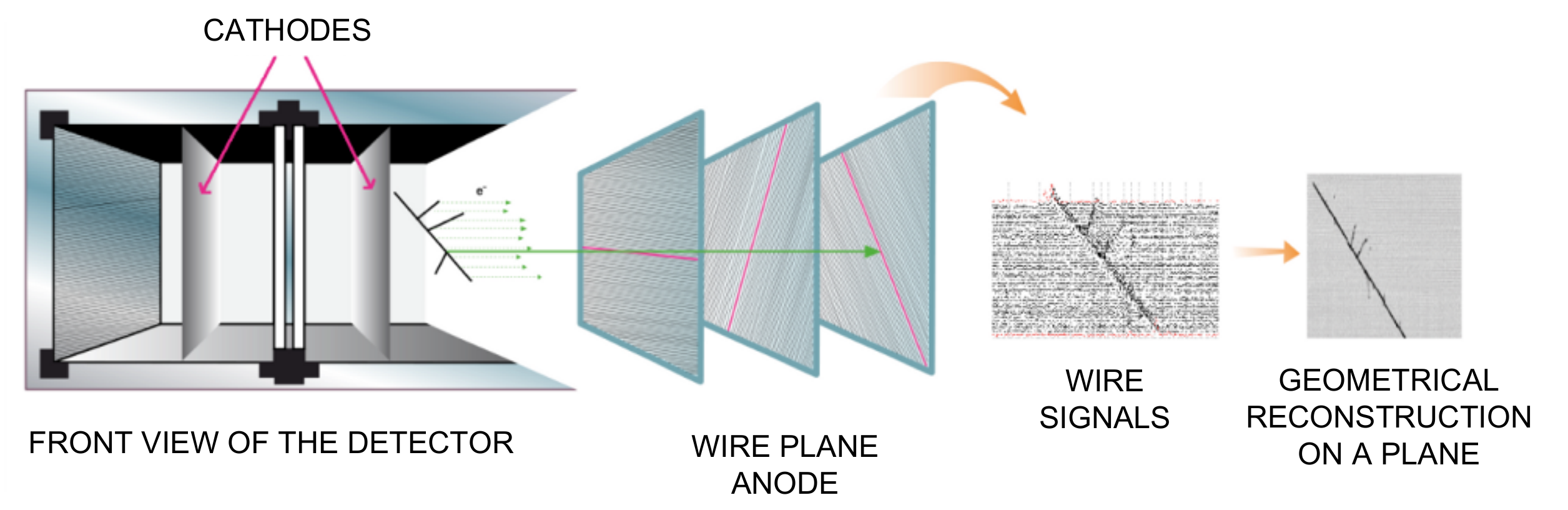}
\caption{Illustration of the \icarus working principle: a charged particle ionization path in LAr and its geometrical reconstruction.}
\label{ica_illustration}
\end{figure}

Wires are oriented on each plane at a different angle (0$^\circ$, +60$^\circ$, -60$^\circ$) with respect to the horizontal direction. Therefore, combining the wire/drift coordinates on each plane at a given drift time, a three-dimensional reconstruction of the ionizing event can be obtained. A remarkable resolution of about 1~mm$^3$ is uniformly achieved over the whole detector active volume (340~m$^3$ corresponding to 476~t).

The measurement of the absolute time of the ionizing event, combined with the electron drift velocity information (v$_{drift}$ $\sim$ 1.6 mm/$\mu$s at E$_{drift}$ = 500 V/cm), provides the absolute position of the track along the drift coordinate. The determination of the absolute time of the ionizing event is accomplished by the prompt detection of the scintillation light produced in LAr by charged particles. To this purpose, arrays of cryogenic Photo Multiplier Tubes (PMTs), coated with wavelength shifter to allow the detection of Vacuum Ultra-Violet (VUV) scintillation light ($\lambda$~=~128 nm), are installed behind the wire planes.

The electronics was designed to allow continuous read-out, digitization and independent waveform recording of signals from each wire of the TPC. The read-out chain is organized on a 32-channel modularity. A Decoupling Board receives the signals from the chamber and passes them on to an Analogue Board via decoupling capacitors; it also provides wire biasing voltage and the distribution of the test signals. Digitization is performed by 10-bit fast ADCs, which continuously read data and store them in circular buffers. Stored data are read out by the DAQ  when a trigger occurs. The trigger relies on the detection of scintillation light by the PMTs, in coincidence with the CERN-SPS proton extraction time for the CNGS beam.

This Section is organized as follows: Par.~\ref{TPC_design} describes in details the main component of the \icarus internal detectors: the mechanical structure, the wire planes, the cabling and the High Voltage system. Par.~\ref{ica_old_pmt} shows the present layout of the Light Collection System, while the Electronics and DAQ
in the LNGS configuration are described in Par.~\ref{ica_old_ele}. Finally, Cryogenics and Purification systems are presented in Par.~\ref{ica_old_cryo}.

%%%%%%%%%%%%%%%%%%%%%%%%%%%%%%%%%%%%%%%%%%%%%%%%%%%%%%%%%%%%%%%%%%%%%%%%%%%%%%
\subsection{TPC design}
\label{TPC_design}

%%%%%%%%%%%%%%%%%%%%%%%%%%%%%%%%%%%%%%%%%%%%%%%%%%%%%%%%%%%%%%%%%%%%%%%%%%%%%%
\subsubsection*{TPC mechanical structure}
\label{TPC_structure}

Each one of the two LAr cryostats hosts a mechanical structure that sustains the different internal detector subsystems and the control instrumentation, namely: (1) the TPC wire planes and the relative HV electrode system (cathode and field-shaping electrodes), %for the read out of the ionization charge,
(2) the PMT system for the scintillation light detection and (3) sensors and probes of the slow control system. Once the cryostat is filled, the structure is totally immersed in LAr.

All materials of the mechanical structure were chosen and treated to guarantee high LAr purity and minimal radioactive contamination: the main components (beams and pillars) have been built with AISI 304L stainless steel; other parts (supports, spacers, etc.) are made of PEEK\texttrademark~. The stainless steel structure has dimensions of 19.6~m in length, 3.6~m in width and 3.9~m in height, for a total weight of $\sim$ 20~tons, see Fig.~\ref{ica_structure_drawing} and ~\ref{ica_structure}.

\begin{figure}[htbp]
\centering
\includegraphics[scale=0.6]{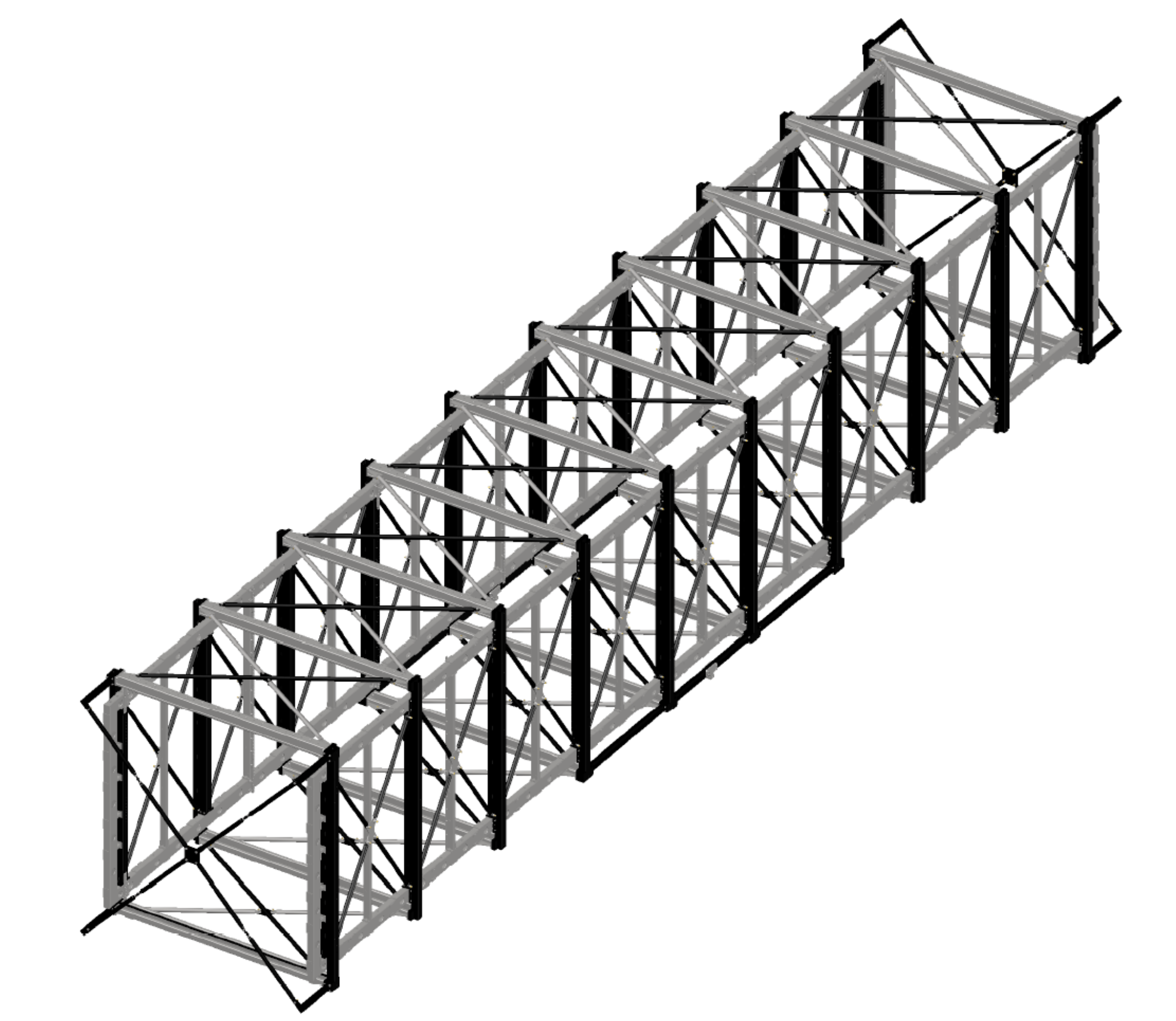}
\caption{Drawing of the bare inner mechanical structure of one T600 module, working as a support for all the internal detector subsystems.}
\label{ica_structure_drawing}
\end{figure}

\begin{figure}[htbp]
\centering
\setlength{\fboxsep}{0pt}
\setlength{\fboxrule}{0.6pt}
\fbox{\includegraphics[width=0.7\textwidth]{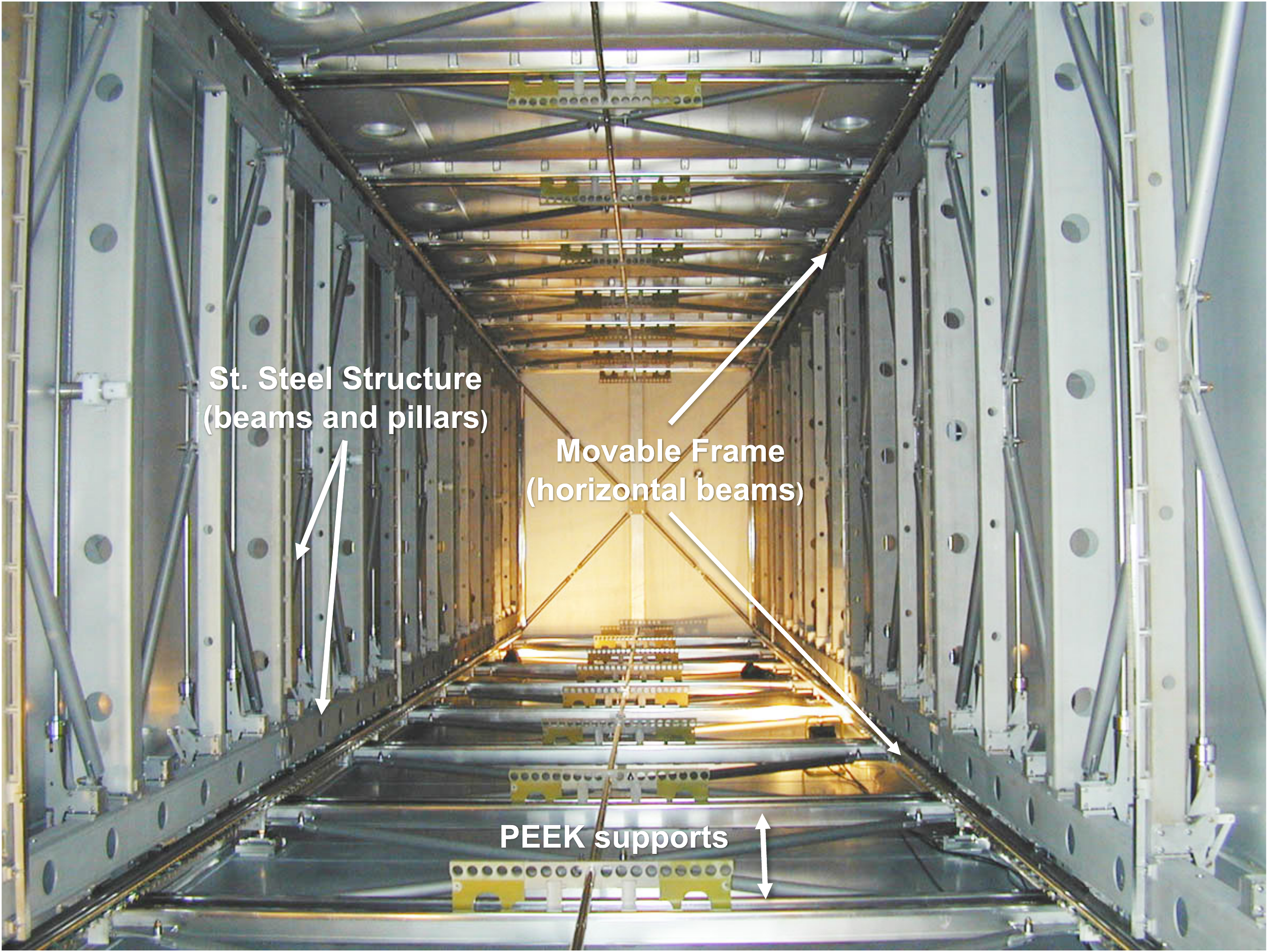}}
\caption{The internal sustaining structure of a T600 module.}
\label{ica_structure}
\end{figure}

To cope with the different thermal shrinking between aluminum and stainless steel, the stainless steel structure leans on the cryostat aluminum floor by means of 10 adjustable feet positioned on corresponding
reinforced pads, and rigidly linked to it only in two pads at half-module
length. In this way, the structure is practically independent from deformations of the cryostat induced by cooling and by the different operating conditions (vacuum and overpressure). Moreover, the sustaining structure is self-supporting and is rigid enough to allow for transportation.

Rocking frames to hold the TPC wires are positioned on the vertical long sides of the mechanical structure. The latter was dimensioned in such a way to sustain the total mechanical tension of the wires applied to the two wire frames, whose design is based on the concept of the variable geometry design (weight bridge). This is based on movable and spring loaded frames, to set the proper tension of the wires after installation, see Fig.~\ref{ica_springs} and Fig.~\ref{ica_corners}. This system allows for a precise detector geometry and planarity, compensating for any possible over-stress during the cool-down and LAr filling phases, and counteracting the flexibility of the structure. 

\begin{figure}[htbp]
\centering
\setlength{\fboxsep}{0pt}
\setlength{\fboxrule}{0.6pt}
\fbox{\includegraphics[width=0.6\textwidth]{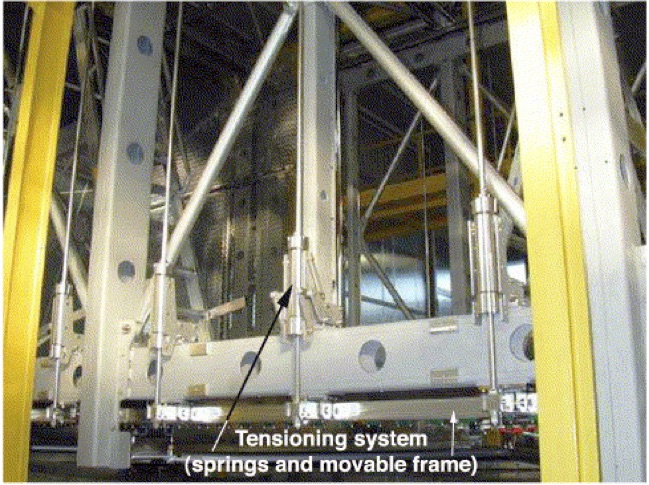}}
\caption{Detail of the wire tensioning mechanics: a 2 m long portion of the wire frame equipped with three tensioning springs.}
\label{ica_springs}
\end{figure}

\begin{figure}[htbp]
\centering
\setlength{\fboxsep}{0pt}
\setlength{\fboxrule}{0.6pt}
\fbox{\includegraphics[width=0.6\textwidth]{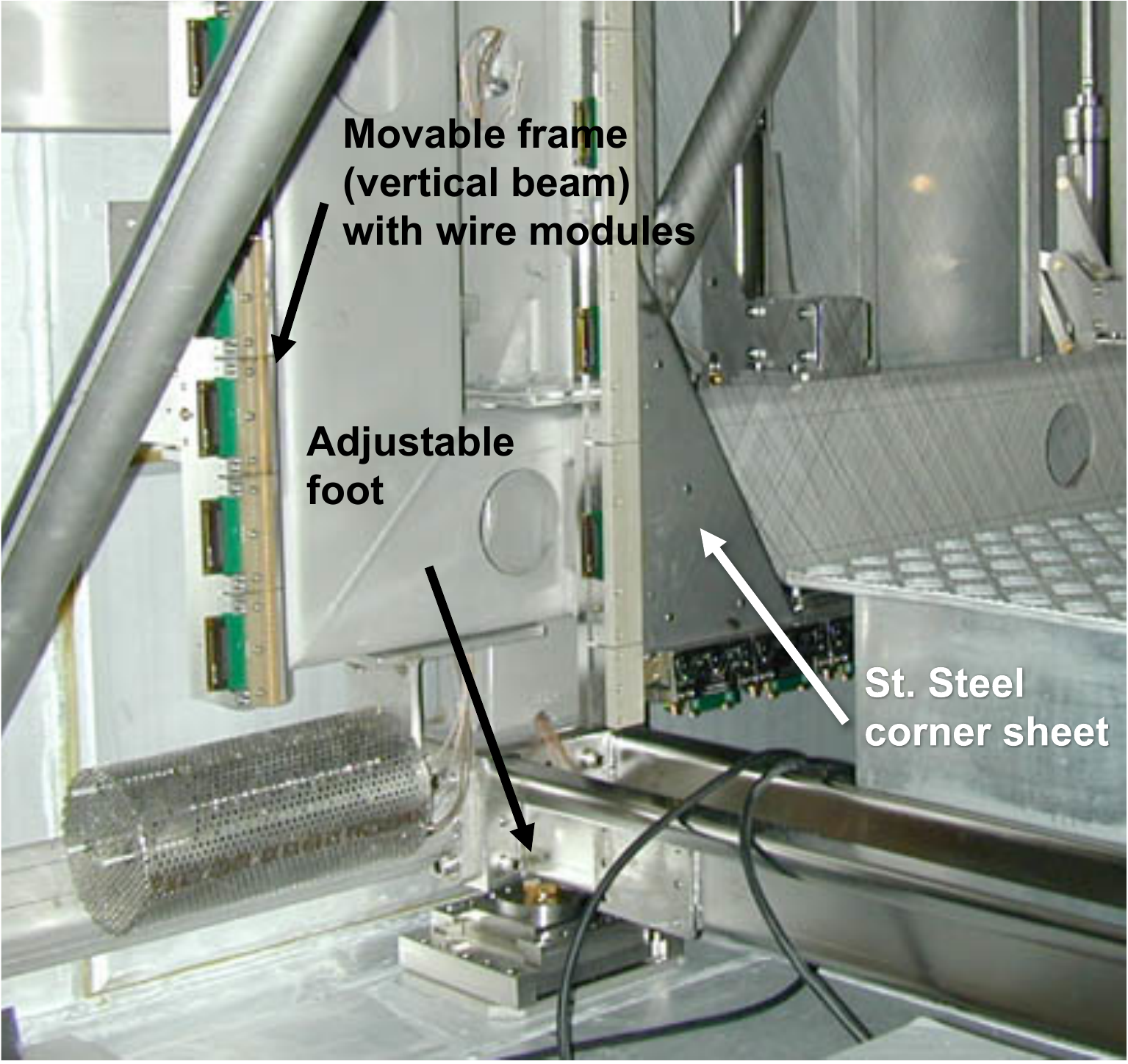}}
\caption{Detail of the internal structure showing an adjustable foot and the wires at the corner.}
\label{ica_corners}
\end{figure}

%%%%%%%%%%%%%%%%%%%%%%%%%%%%%%%%%%%%%%%%%%%%%%%%%%%%%%%%%%%%%%%%%%%%%%%%%%%%%%%
\subsubsection*{TPC wire planes}
\label{TPC_wires}

The anode of each TPC consists of a system of three parallel wire
planes (of 17.95~$\times$~3.16~m$^2$ surface), 3 mm apart from each other, for a total of 13312 wires/chamber, see Fig.~\ref{ica_tpc}; a total of 53,248 wires is mounted on the whole T600 detector (four chambers). Wires are made of AISI 304V stainless steel with a wire diameter of 150~$\mu$m; wire length ranges from 9.42 m to 0.49 m depending on the position of the wire in the plane itself. Thirteen windings (slipknots) around two gold-plated stainless steel ferrules at both ends of each wire are realized, in a "guitar chord" configuration, see Fig.~\ref{windings}. In this way a very safe holding is
guaranteed by the wire friction itself. From the mechanical point of view, wires are strung with a nominal tension of 12~N (5~N for the longest wires), which is high enough to limit sagittas (due to gravity and to electrostatic forces) to values negligible with respect to the distance between the wires. The wire elongation is still well below the elastic limit (39~N nominal value).

\begin{figure}[htbp]
\centering
\setlength{\fboxsep}{0pt}
\setlength{\fboxrule}{0.6pt}
\fbox{\includegraphics[height=0.4\textheight]{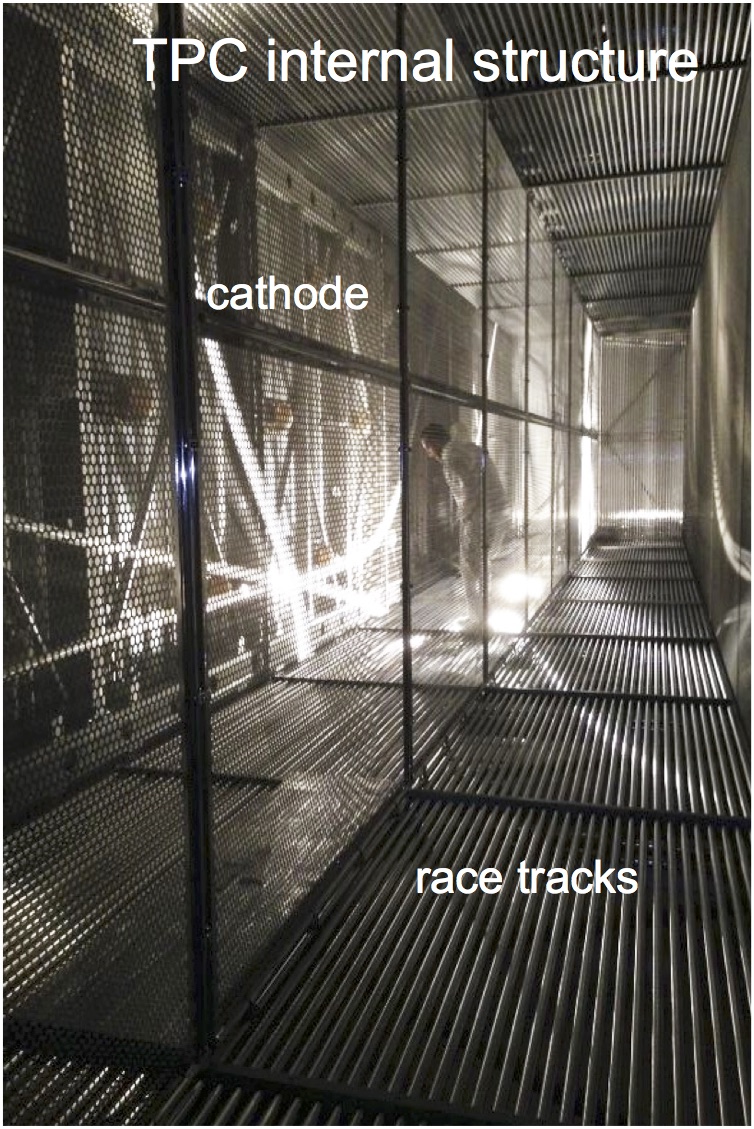}}
\setlength{\fboxsep}{0pt}
\setlength{\fboxrule}{0.6pt}
\fbox{\includegraphics[height=0.4\textheight]{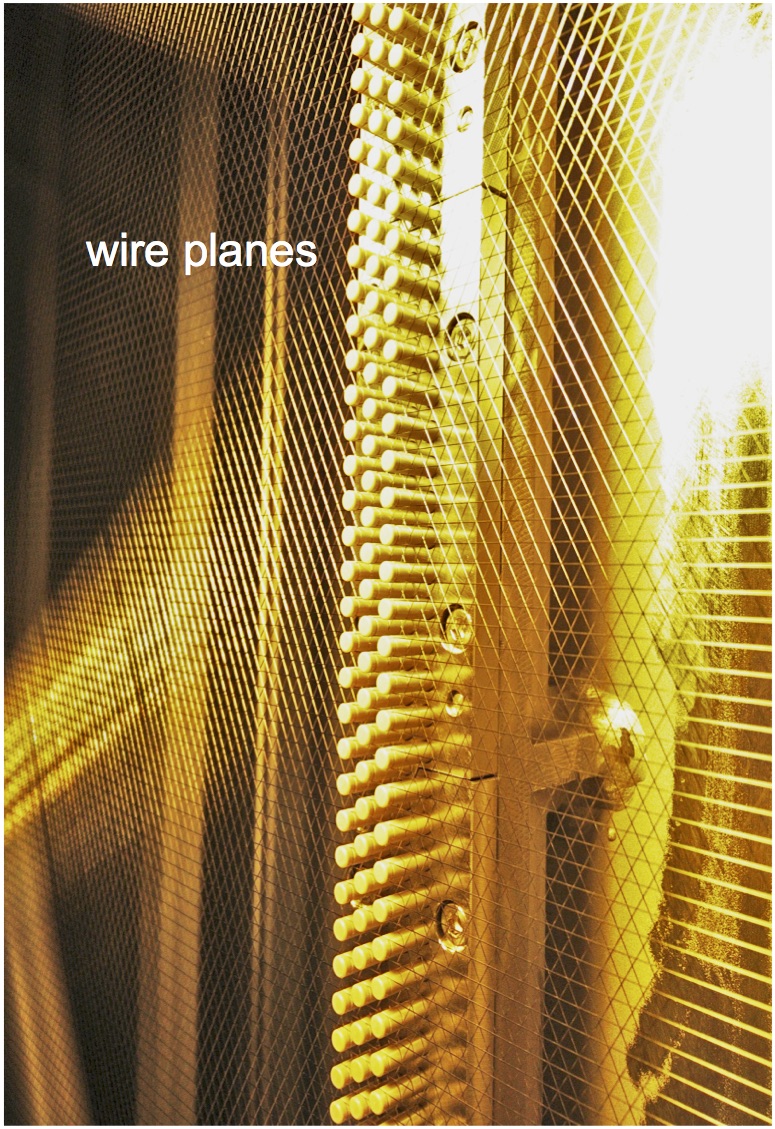}}
\caption{Left: internal TPC structure: cathode, race tracks and wire planes are highlighted. Right: detail of the three wire plane structure.}
\label{ica_tpc}
\end{figure}

\begin{figure}[htbp]
\centering
\includegraphics[scale=0.5]{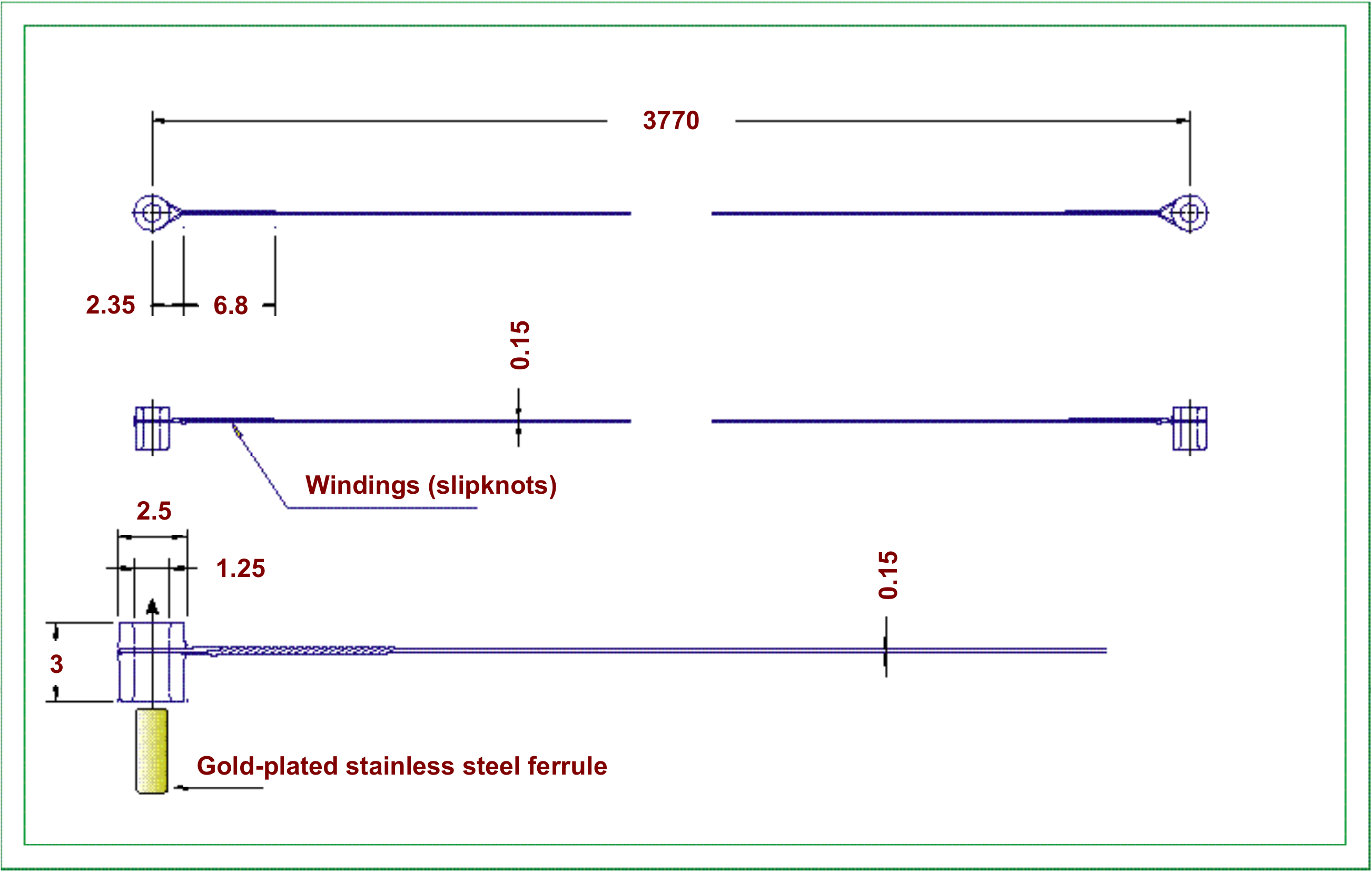}
\caption{Schematic of the Wire holding structure, with detail of the winding in a "guitar chord" concept.}
\label{windings}
\end{figure}

The variable geometry design demonstrated its reliability since none of the wire broke and no damages at the wire chamber structure occurred: in the 2001 Pavia test run; during the transport of the two modules from Pavia to LNGS; during all the installation movements on site; during the commissioning, run and de-commissioning at LNGS.
%
%By appropriate voltage biasing, the first two signal sensing planes (Induction-1 and Induction-2) provide induced signals in a non-destructive way, whereas the last plane (Collection) finally collects the ionization charge. On each chamber, the wire planes are oriented at 0$^{\circ}$, $\pm$60$^{\circ}$ angles with respect to the horizontal direction. \textbf{CUT??}
%
The wires are stretched in the elastic frame sustained by the mechanical structure, as described above. Two coplanar, adjacent sets of horizontal wires (1056 units), 9.42~m long, form the Induction-1 plane, stretched between the vertical beams of the wire frame and a central fixed beam. For both the Induction-2 and Collection planes (wires inclined at $\pm$60$^{\circ}$) the standard length of the wires stretched between the upper and lower beams of the frame is 3.77~m (4640 wires per plane), whereas wires of decreasing length (960 wires per plane) are used in the triangular-shaped portions, between one vertical and one horizontal beam, at the corners of the planes (Fig.~\ref{ica_corners}). The single wire capacitance in the various planes has been calculated to be 20 pF/m for the first (Induction-1) and third (Collection) plane, and 21 pF/m for the intermediate (Induction-2) plane.

The wires are anchored by special holders onto the wire frame in groups of 32 units (the \textit{wire modules}). Each holder is formed either by one or two (according to the different cases) PEEK\texttrademark~ combs contained in stainless steel supports which also embed one or two printed circuit boards. The wire ferrules, held by the PEEK\texttrademark~ shell at both ends of the wire module, are hung on the comb pins. The printed circuit board establishes the electric connection between the 32 pins of the comb and a single
connector also mounted onto the board. Fig.~\ref{ica_holder} shows the technical design of the mechanical system holding 2 %$\times$ 32-
wire modules for the wires at $\pm$60$^{\circ}$.

The wire modules are individually mounted onto the beams of the elastic frame (de-tensioned position). The elastic frame is schematically
subdivided into portions about 2 m long. Each portion comprises 18$\times$ 2 combs/connectors. After the installation of the wire modules was completed, wires were tensioned by loading the springs of the movable frame, see Fig.~\ref{ica_springs}.

\begin{figure}[htbp]
\centering
\includegraphics[scale=0.5]{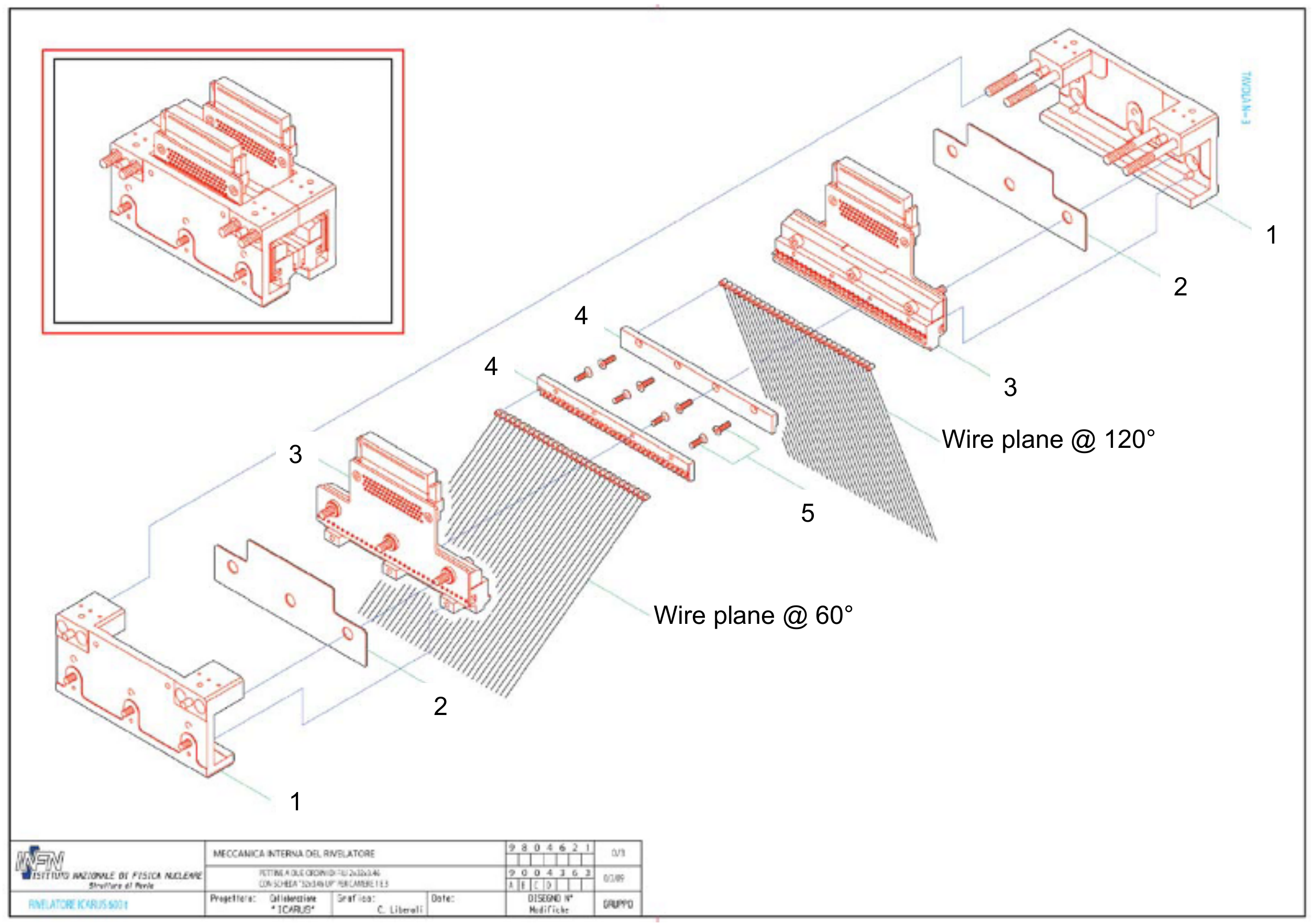}
\caption{Technical design of the mechanical system holding the 32-wire modules at $\pm$60$^{\circ}$: detail of the components (from left to center and reverse):
mechanical support, spacer, printed board with connector, PEEK\texttrademark~ shell with 32-wire ferrules, PEEK\texttrademark~ comb (and reverse).}
\label{ica_holder}
\end{figure}

%%%%%%%%%%%%%%%%%%%%%%%%%%%%%%%%%%%%%%%%%%%%%%%%%%%%%%%%%%%%%%%%%%%%%%%%%%%%%%%%%
\subsubsection*{Cabling}
\label{TPC_cabling}

The individual wire signal transfer to the read-out electronics outside the cryostat is provided by twisted-pair flat cables (34 pairs, flexible, halogen free). Thirty-two pair lines (pairs 1$\div$32) of the 68
available contacts are dedicated to wire signals, one to the test-pulse signal and the last pair, referred to ground, is used as a screen between
signal and calibration conductors. Inside the cryostat the flat cables, suitably terminated with male connectors at both ends, are plugged at one end to the female connectors (32-wire channels + 1 test pulse channel +  1 screen channel) mounted onto the printed boards of the wire modules. At the other end, the cables are plugged to similar connectors embedded (inner side) in specially designed vacuum tight feed-through flanges. Flanges are mounted on the top of the cryostat, at the end of the way-out chimneys. Each one of the 96 feed-through flanges installed in the T600 detector can provide signal transmission for 18 wire modules (576 wires) and for cables for test pulse calibration. The complete connection layout from wires to the read-out electronics is schematically displayed in Fig.~\ref{ica_cab}. A refurbishment of the internal detector cabling is foreseen during the T600 overhauling activities at CERN, see Par.~\ref{wa104}.

\begin{figure}[htbp]
\centering
\includegraphics[scale=0.5]{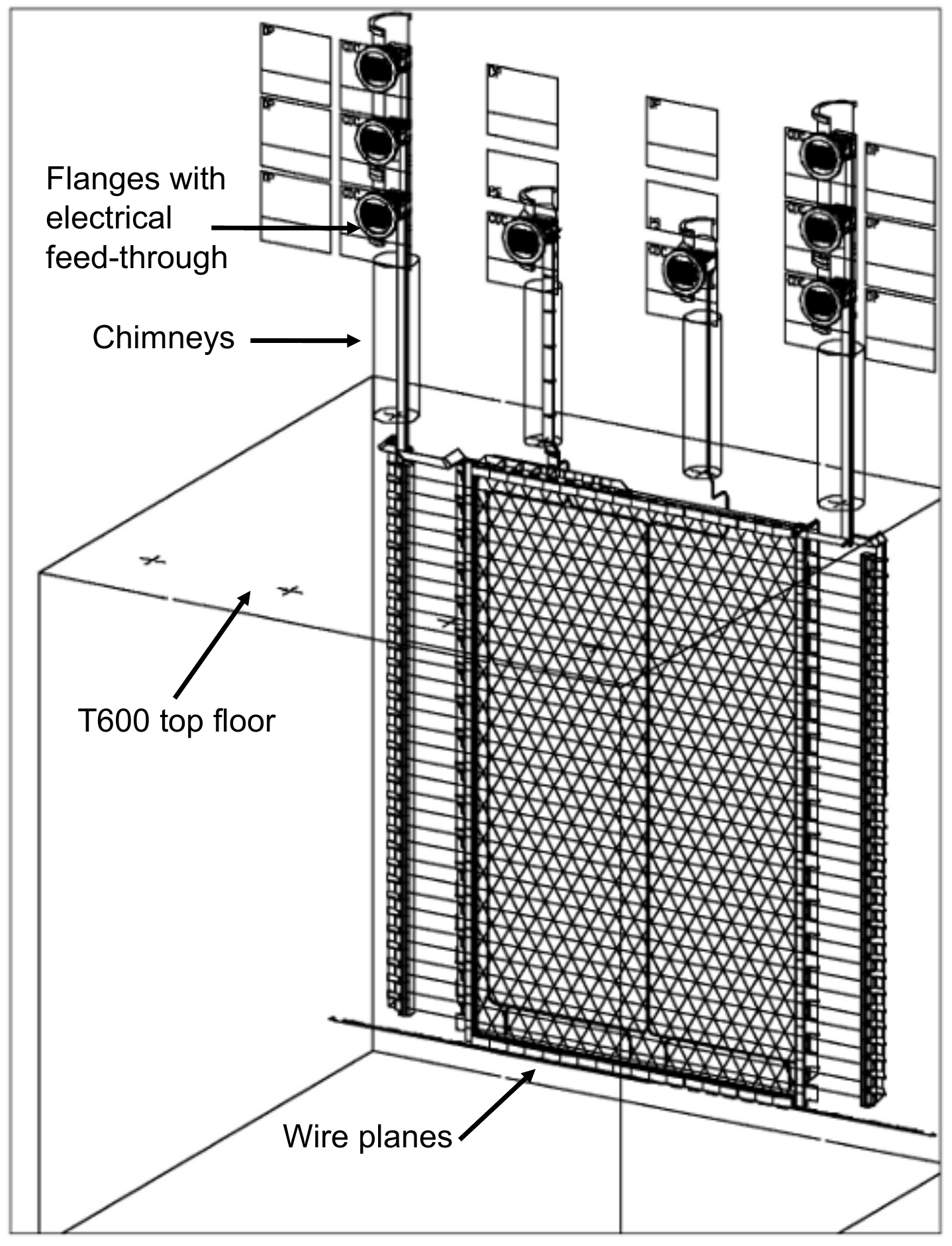}
\caption{Wires, cables, chimney and feed-through layout. The back side of the electronic racks (LNGS configuration) is indicated. Note that the wire chamber longitudinal dimension is not in scale, i.e. the figure includes only the non-repetitive portions of the TPC.}
\label{ica_cab}
\end{figure}

%%%%%%%%%%%%%%%%%%%%%%%%%%%%%%%%%%%%%%%%%%%%%%%%%%%%%%%%%%%%%%%%%%%%%%%%%%%%%%%%%
\subsubsection*{TPC HV system}
\label{TPC_HV}

The HV system has to produce a stable and uniform field over the $1.50 \times 3.16 \times 17.95$ m$^3$ entire drift volume. The system is made of several components, the most important of which are the cathodes, the field-shaping electrodes and the HV feed-throughs. 

In the present configuration the cathodes are built up by an array of nine panels (Fig.~\ref{oldcathode} and Fig.~\ref{oldcathode_dwg}) made of pierced stainless steel sheets. This solution implies an optical transparency between the two drift regions, which allows the detection of the scintillation light by means of all PMTs positioned between the wire chambers and the cryostat walls. 

\begin{figure}[tbp]
\centering
\setlength{\fboxsep}{0pt}
\setlength{\fboxrule}{0.6pt}
\fbox{\includegraphics[width=0.4\textwidth]{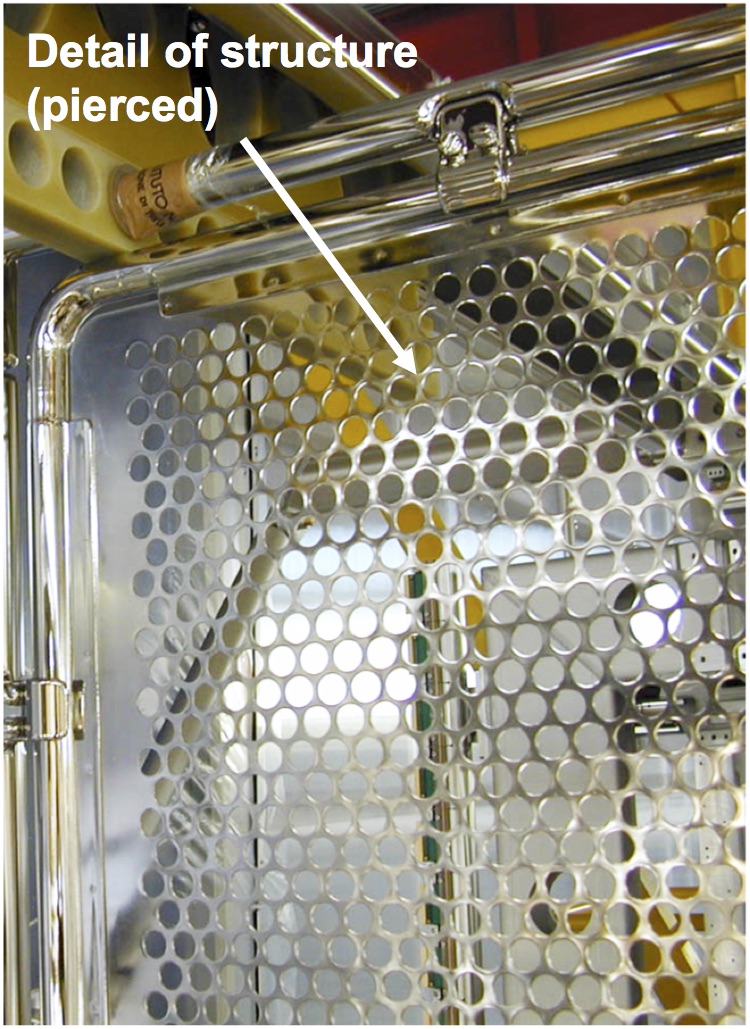}}
\caption{Detail of one cathode panel in the LNGS run configuration.}
\label{oldcathode}
\end{figure}

\begin{figure}[htbp]
\centering
\includegraphics[scale=0.85, angle=90]{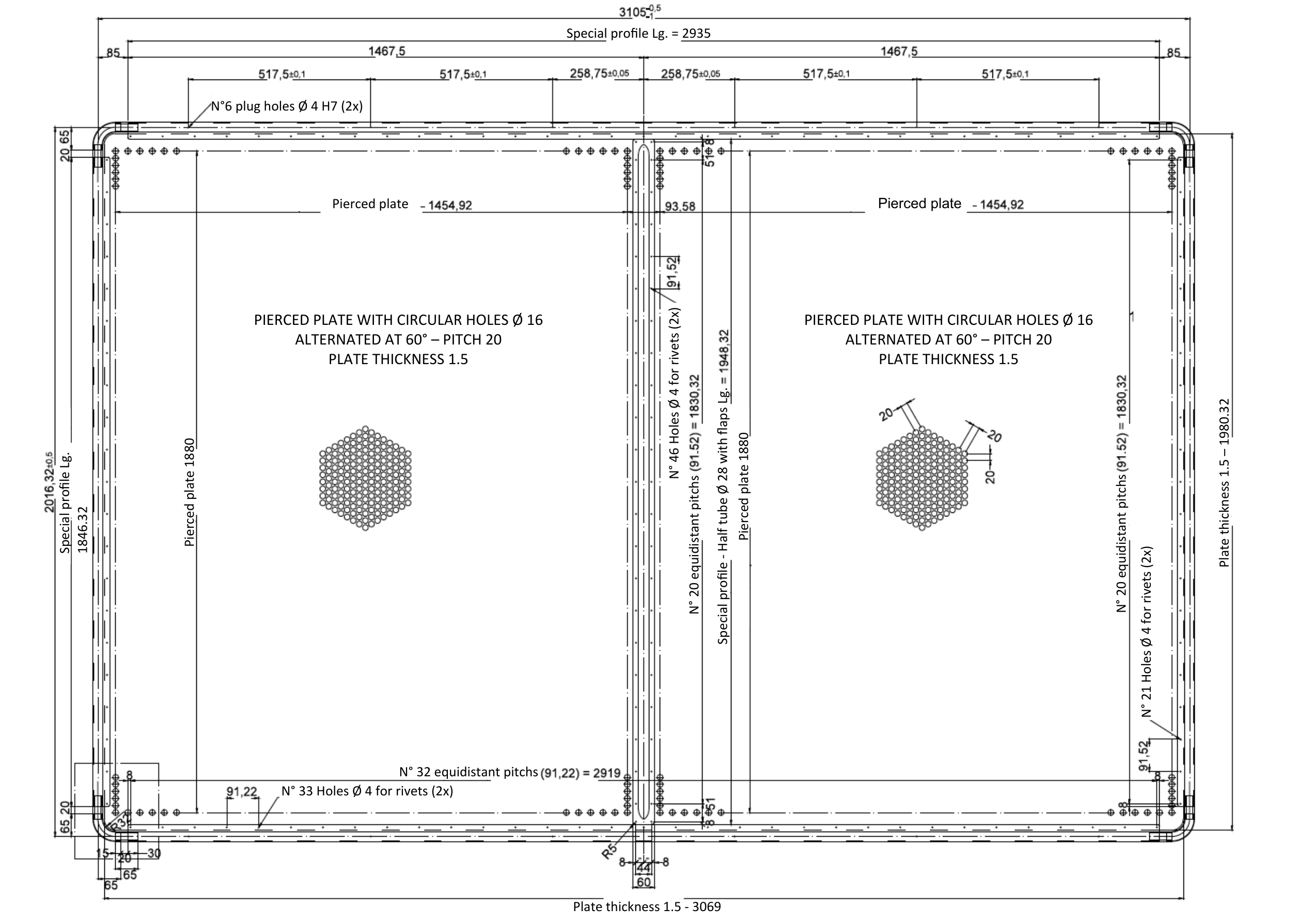}
\caption{Technical drawing of one cathode panel in the LNGS run configuration.}
\label{oldcathode_dwg}
\end{figure}

The electric field in each drift volume is kept uniform by means of the field-shaping electrodes (\textit{race tracks}, see Fig.~\ref{ica_rtracks} and Fig.~\ref{ica_rtdraw}). These consist of 29 rectangular rings (18.1 $\times$ 3.2 m$^2$) for each wire chamber, made of 2 m long stainless steel tubular elements (34 mm diameter, 0.8 mm thick) connected by two welded terminals. The distance between race tracks is 49.6 mm. In the upper part, between the race tracks and the gaseous Ar (GAr) phase, a grounded metallic shielding is interposed. The race tracks are set at a
potential linearly decreasing from the cathode value to the first wire plane voltage, to ensure uniform electric field and hence constant drift
velocity inside the volumes. The biasing potentials of the race tracks are
obtained through resistive voltage degraders. The HV degrader is based on four resistor chains, one for each drift volume, with the ``hot" end connected to the cathode and the ``cold" end set to ground. Resistor chains are made of 30 steps. Intermediate contacts are connected to the field-shaping electrodes. The resistance for each step is 25~M$\Omega$, obtained by connecting four 100~M$\Omega$ resistors in parallel. For a 0.5~kV/cm drift field the voltage across each resistor is 2.5~kV.

\begin{figure}[htbp]
\centering
\setlength{\fboxsep}{0pt}
\setlength{\fboxrule}{0.6pt}
\fbox{\includegraphics[width=0.5\textwidth]{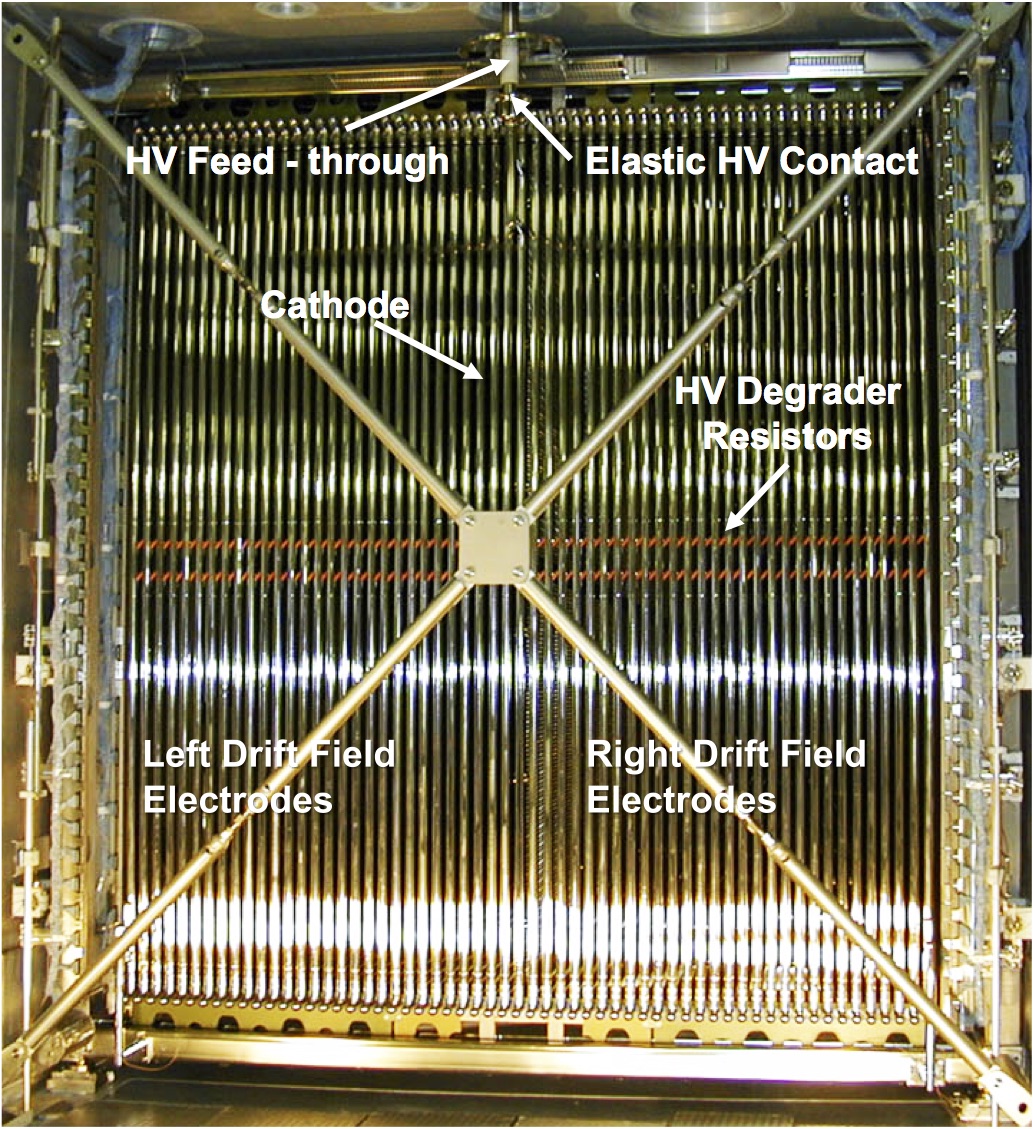}}
\caption{The first half-module of the T600 detector before closing. Some components of the HV system are visible: feed-through, cathode, field electrodes (race tracks), voltage divider.}
\label{ica_rtracks}
\end{figure}

\begin{figure}[htbp]
\centering
\includegraphics[scale=0.6]{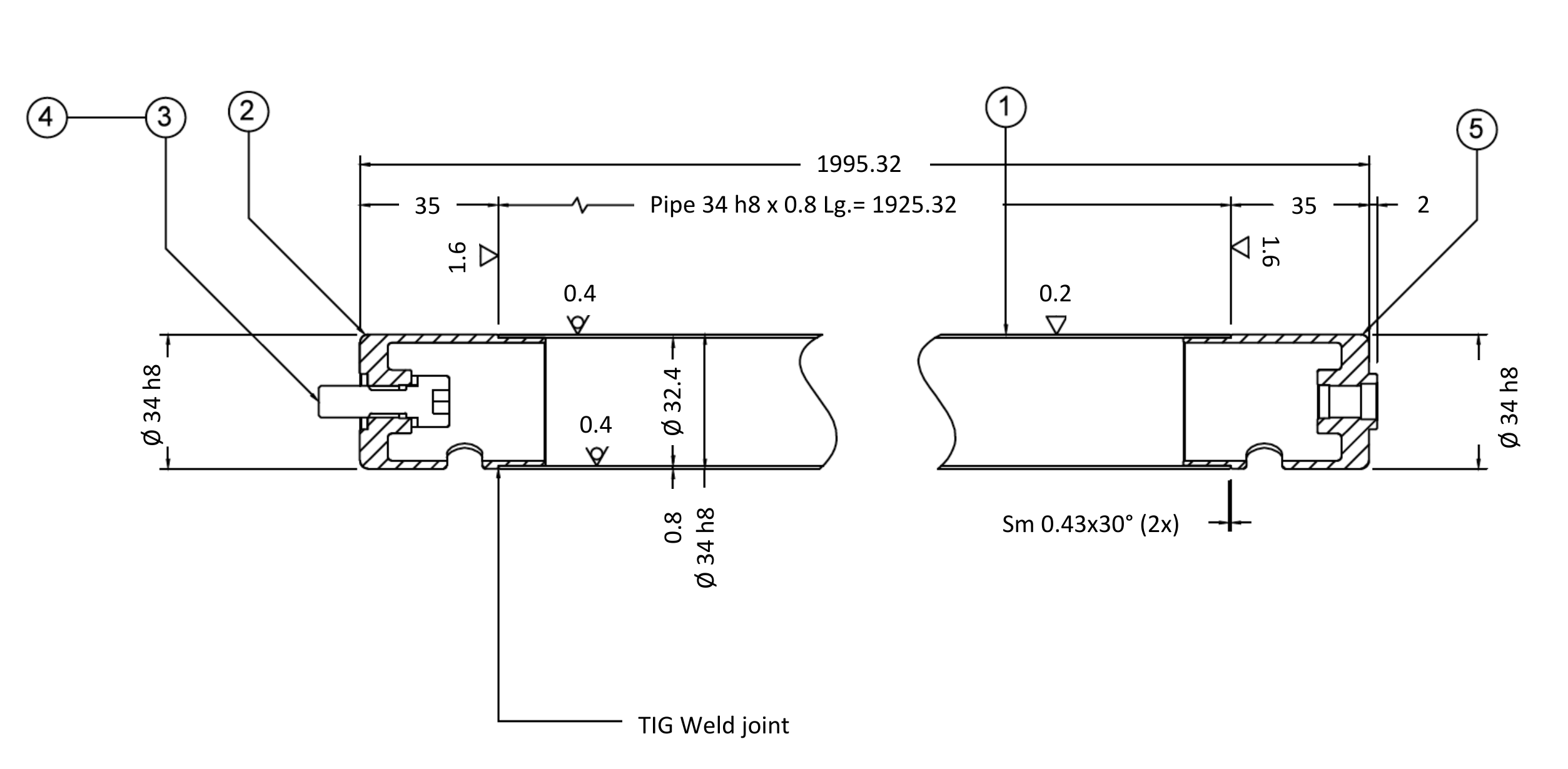}
\caption{Technical drawing of one race track element.}
\label{ica_rtdraw}
\end{figure}

The HV generated by an external power supply is brought to the internal cathode via an hermetic feed-through. A feed-through coaxial geometry has been adopted: the design is based on an inner conductor (HV) and an outer conductor (ground) insulated by UHMW PE (ultra-high-molecular-weight polyethylene) as shown in Fig.~\ref{ica_fthr}. The outer conductor, made of a stainless steel tube, surrounds the insulator extending inside the cryostat up to the LAr level. By such a geometry the electric field is always confined in regions occupied by high dielectric strength media (UHMW PE and LAr). The inner conductor is made of a thin wall stainless steel tube, to minimize the heat input and to avoid the creation of argon gas bubbles around the HV lower end. A female contact, welded at the upper end for the connection to the HV cable, and a round-shaped elastic contact for the connection to the cathode, screwed at the lower end, complete the inner electrode. Special care has been taken in the assembling to ensure the complete filling with the PE dielectric of the space between the inner and the outer conductors and to guarantee leak tightness at ultra-high-vacuum level.

The HV system of \icarus had no failures during the three years of run at LNGS, at the operating voltage of 76 kV. Moreover, in the last days of the LNGS run few tests were carried out with an operating voltage about twice its nominal value (150~kV, corresponding to E$_{drift} \simeq 1$ kV/cm), with no failures for five days. After this the HV system was switched off to allow the T600 decommissioning procedures to start.

\begin{figure}[htbp]
\centering
\setlength{\fboxsep}{0pt}
\setlength{\fboxrule}{0.6pt}
\fbox{\includegraphics[scale=0.5]{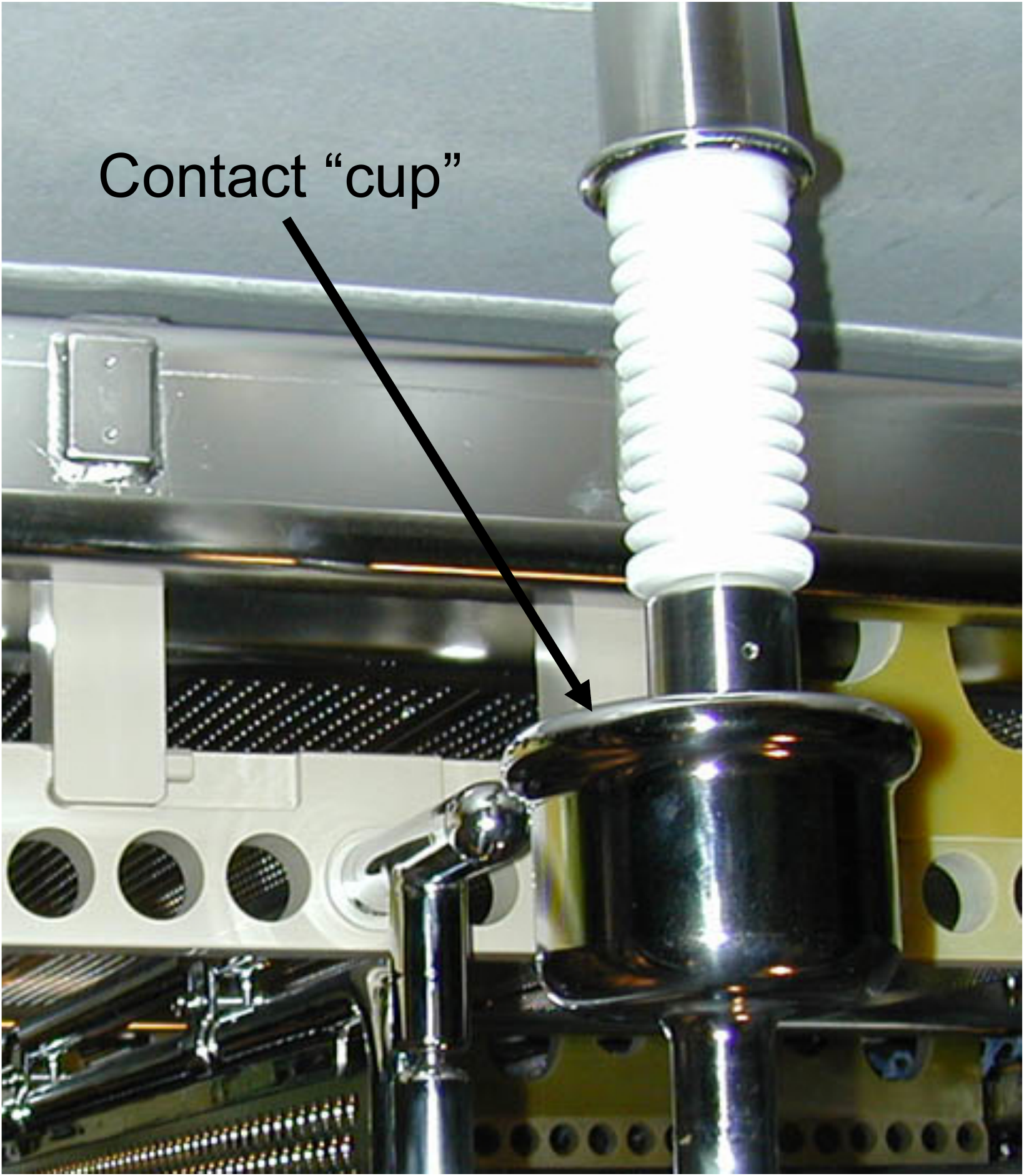}}
\caption{Picture of the HV feed-through. The ''cup'' for the contact to the cathode panels is also visible.}
\label{ica_fthr}
\end{figure}

The main characteristics of the present T600 TPC internal detector configuration are resumed in Tab.~\ref{ICA_tpc}.

\begin{table}[htbp]
\begin{tabular}{ll}\hline
Number of read-out chambers (TPC) in T600 & 4 \\
Number of wire planes per chamber & 3 \\
Distance between wire planes & 3 mm \\
Wire orientation with respect to horizontal & 0$^\circ$, $\pm$60$^\circ$ \\
Wire diameter & 150 mm \\
Wire length &  \\
\hspace{0.5cm}Horizontal wires & 9.42 m \\
\hspace{0.5cm}Wires at $\pm$60$^\circ$ & 3.77 m \\
\hspace{0.5cm}Wires at the corners ($\pm$60$^\circ$) & 3.81-0.49 m \\
 & \\
Wire pitch (normal to the wire direction) & 3 mm \\
Wire capacitance Ind.-1, Ind.-2, Coll. & 20, 21, 20 pF/m \\
Wire nominal tension & 12 N (5 N for hor. wires) \\
Number of wires/wire module & 32 \\
Number of wire modules/chamber &  \\
\hspace{0.5cm}Horizontal wires & 66 \\
\hspace{0.5cm}Wires at $\pm$60$^\circ$ & 2 $\times$ 145 \\
\hspace{0.5cm}Wires at the corners $\pm$60$^\circ$ & 2 $\times$ 30 \\
 & \\
Number of wires/chamber & \\
\hspace{0.5cm}Horizontal & 2112 \\
\hspace{0.5cm}At $\pm$60$^\circ$ & 2 $\times$ 4640 \\
\hspace{0.5cm}At the corners ($\pm$60$^\circ$) & 2 $\times$ 960 \\
Total & 13312 \\
Total number of wires in T600 & 53,248 \\
Wire plane voltage biasing (typical) & -220 V, 0 V, +280 V \\
Cathode HV (nominal) & 75 kV \\
Cathode to Collection plane distance & 1.50 m \\
Sensitive volume/chamber & 85 m$^3$ \\
\hspace{0.5cm}Length & 17.95 m \\
\hspace{0.5cm}Width & 1.50 m \\
\hspace{0.5cm}Height & 3.16 m \\
Maximum drift length in LAr & 1482 mm \\
Maximum drift time in LAr (at nominal field) & 950 $\mu$s \\
\hline
\end{tabular}
\caption{Main characteristics of the \icarus TPCs.}
\label{ICA_tpc}
\end{table}

%%%%%%%%%%%%%%%%%%%%%%%%%%%%%%%%%%%%%%%%%%%%%%%%%%%%%%%%%%%%%%%%%%%%%%%%%%%%%%%
\subsection{Light Collection system}
\label{ica_old_pmt}

Charged particles deposit energy in liquid argon mainly by excitation and ionization of Ar atoms, leading to scintillation light emission and free electron production, respectively. Additional scintillation light comes from the recombination of electron-ion pairs, which is inversely proportional to the strength of the electric field applied to the detector active volume. As a consequence, free-electron yield rises with the field value while photon yield decreases. In both cases saturation occurs, for minimum ionizing particles, at E$_{drift} >$ 10 kV/cm. At the nominal drift field applied in \icarus, approximately the same amount of photons ($\sim$ 4,000 $\gamma$/mm) and free electrons ($\sim$ 5,000 ion-electron pairs per mm) are produced for minimum ionizing particles (m.i.p.)~\cite{ica_reco}.

Scintillation light emission in LAr is due to the radiative decay of excited molecules (Ar$_2^*$) produced by ionizing particles, releasing monochromatic VUV photons ($\lambda$ $\sim$ 128 nm) in transitions
from the lowest excited molecular state to the dissociative ground state. A fast ($\tau$~$\sim$~6~ns decay time) and a slow ($\tau$~$\sim$~1.6~$\mu$s) components are emitted; their relative intensity depends on dE/dx, ranging from 1:3 in case of minimum ionizing particles up to 3:1 in case of $\alpha$-particles. This isotropic light signal propagates with negligible attenuation throughout each TPC volume. Indeed, LAr is fully transparent to its own scintillation light, with measured attenuation length in excess of several tens of meters and Rayleigh-scattering length of the order of 1 m. Because of their short wavelength the scintillation photons are absorbed by all materials inside the detector without reflection. 

The design of the T600 detector PMT system, in the LNGS configuration, resulted from dedicated R\&D activities on the LAr scintillation light detection, carried on during the second half of the 90's~\cite{ica4}. The adopted solution is based on the large surface Photo-Multiplier 9357FLA Electron Tube, a 12-stage dynode PMT with hemispherical glass window 200 mm (8") diameter, manufactured to work at cryogenic temperatures~\cite{ica_pmt}. The PMT sensitivity to VUV photons (128 nm) was achieved by coating the glass window with Tetra-Phenyl-Butadiene (TPB), which acts as fluorescent wavelength shifter from VUV wavelengths to the PMT sensitive spectrum. A TPB coating of thickness 0.2~mg/cm$^2$ on sand-blasted glass guarantees a conversion efficiency better than 90\% and good adhesion after immersion in LAr, resulting in a PMT response with 4\% overall quantum efficiency~\cite{ica_pmt2}.

PMTs are located in the 30~cm space behind the wire planes of each TPC, at 5~mm distance from the Collection wires, with a dedicated sustaining structure specially designed to compensate the thermal stresses occurring during the cooling of the T600 cryostat (Fig.~\ref{ica_pmt_old} Left). Three rows of 9 PMTs, spaced by 2~m, found place in the East module behind each wire chamber for a total amount of 27+27 photo-devices. In the West module only the two central rows were deployed; two additional PMTs were placed in the top and bottom positions in the Right chamber at the center of the longitudinal direction, for an overall amount of 20 PMTs (Fig.~\ref{ica_pmt_old} Right). Despite the small number of PMTs deployed inside the T600 detector in the LNGS configuration, the PMT system allowed to get a 100\% trigger efficiency for CNGS-induced events above 300 MeV of deposited energy, with a remarkable stability during the three years of data taking~\cite{ica_lngs_trigger}.

\begin{figure}[htbp]
\centering
\includegraphics[width=0.95\textwidth]{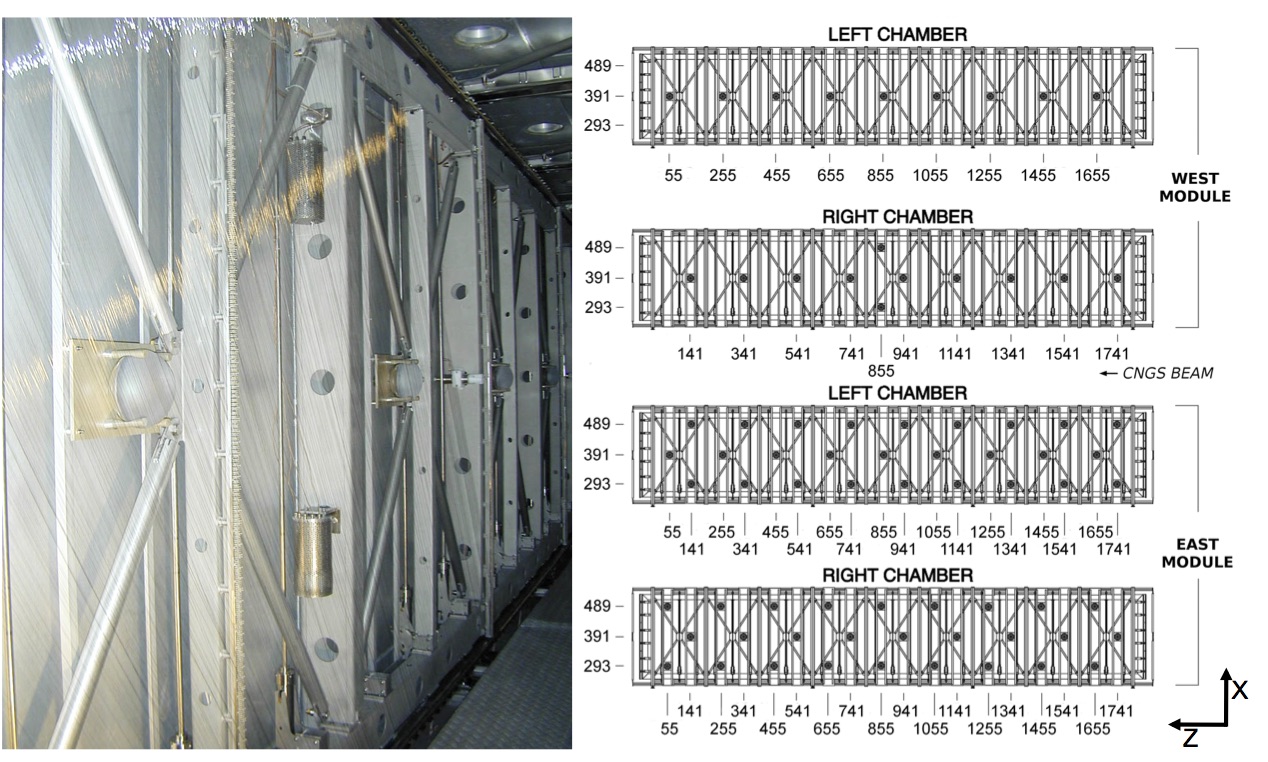}
\caption{Left: internal view of one TPC, with a few PMTs clearly visible together with their sustaining structure. Right: PMT’s deployment in the two \icarus cryostats. PMT coordinates in cm are related to the reference frame used in Hall B%whose origin is set at the ground floor (vertical axis, x), at the center of the two modules (drift coordinate, y) and at the downstream end of the wire chambers for the longitudinal direction (z) along the CNGS beam line
.}
\label{ica_pmt_old}
\end{figure}

%%%%%%%%%%%%%%%%%%%%%%%%%%%%%%%%%%%%%%%%%%%%%%%%%%%%%%%%%%%%%%%%%%%%%%%%%%%%%%%%
\subsection{Electronics, DAQ and Trigger}
\label{ica_old_ele}

\subsubsection*{Electronics and DAQ}

The present T600 electronics is designed to allow continuous read-out, digitization and independent waveform recording of signals from each wire of the TPC. The read-out chain is organized on a 32-channel modularity. A Decoupling Board receives the signals from the chamber and passes them to an Analogue Board via decoupling capacitors; it also provides wire biasing voltage and the distribution of the test signals. 

The Analogue Board hosts the front-end amplifiers and performs 16:1 channel multiplexing and 10-bit ADC digitization at 400 ns sampling time per channel. The overall gain is about 1,000 electrons per ADC count, setting the signal of minimum ionizing particles to 15 ADC counts, with a dynamic range of about 100 times the signal of one m.i.p. A 3~$\mu$s decay constant is used for the unipolar signals coming from the Collection and Induction-1 wires, while a 100~$\mu$s decay constant is used for the bipolar current signals (Induction-2 wires). A digital Board hosts a 10 bit waveform recorder, which continuously reads the data, stores them in multievent circular buffers, each covering a full drift distance. When a trigger signal occurs, the active buffer is frozen, following data are written to the next free buffer, and the stored data are read out by the DAQ. This configuration guarantees no dead time, until the maximum DAQ throughput (1 full-drift event per second) is reached. The average electronic noise achieved with the specially designed low noise front-end is well within expectations: 1,500 electrons r.m.s. to be compared with 15,000 free electrons produced by a minimum ionizing particle in 3~mm (S/N $\sim$ 10).

\subsubsection*{LNGS run Trigger}

Two different trigger systems based on the detection of scintillation light and ionization charge produced by charged particles in LAr have been realized for the \icarus detector LNGS run~\cite{ica_lngs_trigger}. They exploited the PMT system and the new S-Daedalus FPGA boards, spanning few orders of magnitude in event energy deposition. The main \icarus trigger for detecting CNGS beam related events required the coincidence of
the PMT \textit{local trigger} in at least one of the four TPC chambers with a 60~$\mu$s gate opened in correspondence of the proton spill extraction delayed for the $\sim$ 2.44 ms CNGS neutrino time-of-flight. The mentioned PMT local trigger is obtained, separately for each TPC, as the linear sum of the collected PMT signals, properly discriminated in order to account for the different number of devices deployed in the two modules.

The combined analysis of the performance of the PMT and S-Daedalus independent trigger systems demonstrated an almost full PMT trigger efficiency for CNGS neutrino events above 300 MeV energy deposition on the full T600 active volume. Efficiency remains at $\sim$ 98.5\% down to 100~MeV. The stability of the trigger system was verified within the measurement uncertainty, comparing different data sets collected during the CNGS run~\cite{ica_lngs_trigger}.

The T600 LNGS run Trigger Manager, built in a commercial National Instrument PXI crate, handled the different trigger sources: scintillation light collected by PMTs, timing synchronization with the
CNGS extractions, charge signal collected on wires and test pulses for calibration. The system consisted of a Real Time (RT) controller (PXIe-8130) and two FPGA boards (PXI-7813R and PXI-7833). The RT controller implemented all the features that implied communication with external devices, such as the DAQ process or the CNGS Early Warning reception. Communication with the DAQ was implemented in handshake between the DAQ main process and the trigger manager. The RT controller also monitored the number of available buffers in the digital boards and prevented the generation of new triggers in case all the buffers were full. The maximum number of buffers available for full drift recording was 8. The DAQ throughput, for full drift event recording, was limited to 0.8~Hz mainly because of the adopted VME architecture. The FPGA boards implemented time critical processes, like the synchronization with the LNGS time, the opening of CNGS gate and the time stamp of each trigger. They also kept record of the trigger source and the trigger mask, monitored trigger rates from each source and controlled the overall system stability.

In the architecture of the \icarus DAQ system adopted for data taking at LNGS~\cite{ica_lngs_trigger}, all the 96 readout units work autonomously pushing their own data to 4 receiving workstations, one per TPC chamber. This segmentation and parallelization of the data stream allowed reaching a $\sim$~1~Hz building rate on the whole T600, safely exploiting the data link at half of the $\sim$~50~MB/s available bandwidth.

%%%%%%%%%%%%%%%%%%%%%%%%%%%%%%%%%%%%%%%%%%%%%%%%%%%%%%%%%%%%%%%%%%%%%%%%
\subsection{Cryogenics and Purification systems}
\label{ica_old_cryo}

The \icarus cryogenic plant was mainly installed in the North end of Hall B of LNGS, i.e. behind the cryostat when entering the Hall.
The final design of the system was driven by compliance to strict requirements on efficiency, safety, anti-seismic constraints and reliability for several years of operations in a confined underground location. A schematic view of the apparatus has been shown in Fig.~\ref{ica_hallb} Left. The technical requirements for the plant, and its components, were developed requesting fulfillment of strict conditions in terms of mechanics, electronics, radiochemical and electronegative purity, and are summarized below:

\begin{itemize}
\item full cryogenic containment for safety needs;
\item extremely high LAr purity: residual contamination of electronegative molecules such as water and oxygen lower than 0.1 part per billion, to allow ionization electrons to drift over several meters;
\item extremely precise control of the components differential temperature during detector cool-down, in order to avoid stresses on the TPC precision mechanics. In particular the requests were of $\Delta$T~$<$~50~K on the wire-chamber structures, $\Delta$T~$<$~120~K on the cold vessels;
\item fast cooling to liquid argon temperature, to ensure good starting purity;
\item very high temperature uniformity in steady state conditions ($\Delta$T~$<$~1~K in the main volume) to guarantee uniform electron drift velocity;
\item thermal losses as low as possible, to reduce operation costs and minimize power consumption in emergency situations;
\item very high stability and operation reliability to fulfill the strict underground safety requirements (this point will be re-discussed to fit the different FNAL safety standards and rules);
\item full redundancy, to assure uninterrupted operation over several years.
\end{itemize}

The T600 detector is made by two adjacent aluminum LAr containers (parallelepipedal in shape), each with an inner volume of 275~m$^3$. The two modules are independent from the point of view of LAr containment and purification plants, while nitrogen cooling
system and thermal insulation are common to both. The main design and construction of the cryostats was carried on in collaboration with Air Liquide Italia Service (ALIS) Company~\cite{airliquide}.
The two cold vessels containing the TPCs are realized with 15~mm thick aluminum honeycomb panels, mechanically reinforced by extruded profiles. The external and internal skins work as double cryogenic containment. The external dimensions of the vessels are 4.2~$\times$~3.9~$\times$~19.9~m$^3$, i.e. the maximum size allowed for fitting the boxes into the LNGS underground laboratory. This solution, though unconventional, was preferred mainly for its lightness and rigidity to stand stresses during the emptying phase, and the overall LAr and detector weight.
A single thermal insulation vessel %(warm vessel)
surrounded the two modules. The insulation was designed to behave as an additional tight container in case of cryogenic liquid spillages. Every insulation wall is composed of separated metallic boxes, with outer skin made of stainless steel. The inner and side skins are instead of Pernifer\texttrademark~, to avoid thermal shrinking.
%Walls were welded together both internally and externally.
The boxes were filled with insulating honeycomb panels (0.4~m thick of Nomex\texttrademark~ or equivalent material) and super-insulation layers placed on the inner cold surface.

%During detector operation, each box was continuously evacuated to enhance insulation performance, with the exception of the roof panels. The nominal heat load in case of $10^{-4}$ mbar residual pressure was about 10 W/m$^2$ on average, in order to reduce the heat losses due to gas conduction and convection, which are limited by the honeycomb cell geometry but still present due to residual gas.
%Radiative losses are suppressed by the super-insulation layers.
%Holes for pipes and chimneys for cables of the detector signals are all located on the ceiling of the insulation vessel matching the position of all feed-throughs on the two module ports (about 100 in total). Tightness of the insulation vessel is ensured by special bellows mounted around all the crossing tubes.

A thermal shield was placed between the insulation vessel and the aluminum containers, to intercept the residual heat losses through the insulation walls, thus avoiding boiling of the LAr bulk. Boiling nitrogen was circulated in the thermal shield, with a gas/liquid ratio equal to 1:5. This solution guarantees a fast cooling-down phase with thermal gradients within specification, and it forces LAr de-stratification during normal operations, thus maintaining uniform and stable temperature in the LAr bulk.

The argon and nitrogen circulation lines will be now detailed in the following, while a comprehensive view of the cryogenic system is drawn in Fig.~\ref{fig:ica_cryoplant}.

\begin{figure}[htbp]
\centering
\includegraphics[scale=0.6]{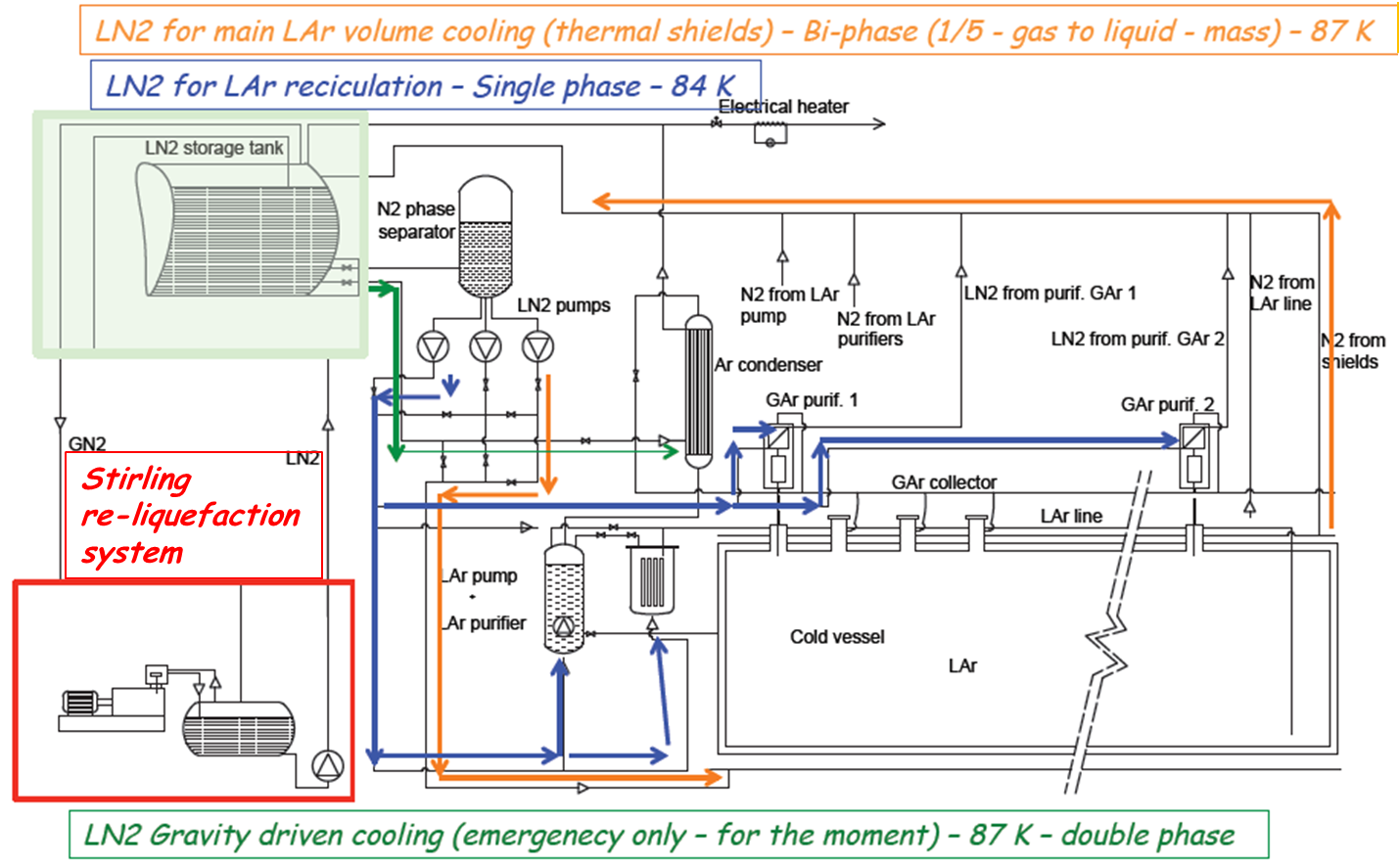}
\begin{center}
\caption{Drawing of the existing plant of the \icarus detector, containing all main elements of the system: LN$_2$ lines and storage tanks, GAr/LAr recirculation, Stirling re-liquefaction system. LN$_2$ lines are color-coded for clarity.}
\label{fig:ica_cryoplant}
\end{center}
\end{figure}

In order to maintain high LAr purity in the active volume, each T600 module is equipped with two GAr and one LAr recirculation units.
The gas recirculation system collects argon gas coming from the chimneys hosting the read-out cables and the feed-through flanges. The gas on the T600 top is warm and dirtier than the liquid from being in contact with hygroscopic plastic cables; moreover it could be further polluted by possible small leaks due to the presence of several joints on each chimney. The collected gas is re-condensed and then made to drop into a LN$_2$-cooled Oxysorb\texttrademark~ filter, placed after the re-condenser. Newly purified LAr is injected back into the main volume right below the liquid/gas interface. The condenser is normally fed with LN$_2$ at the temperature required for efficient re-condensation of the gas, by means of forced circulation.

During LNGS data taking, gas recirculation was usually kept at the maximum available rate of 25 GAr Nm$^3$/h/unit.
Gas recirculation is specially helpful during the filling phase, as it allows purification of the dirty warmer gas, while the outgassing rate decreases exponentially with temperature.
On the other hand the system acts as detector pressure stabilizer during steady state operations.

Continuous liquid recirculation is used to massively purify LAr and to reach and maintain purity levels as high as possible after initial filling. It can also be exploited to rapidly restore argon purity in case of accidental pollution during operations.
LAr is extracted from the main volume at about 2 m below the liquid surface on one of the 4 m long sides of the T600 modules (endcaps); it is then reinjected on the opposite side, 20 m apart, at the module floor level, through a horizontal pierced pipe that provides uniform distribution over the vessel width.
Recirculation is forced by means of immersed cryogenic pumps %(ACD CRYO AC-32 centrifugal pump)
placed inside independent dewars. Before reinjection, LAr is sent through a battery of four Oxysorb/Hydrosorb\texttrademark~filter cartridges, connected in parallel. Each set of filters has a nominal O$_2$ absorption capacity exceeding 200 normal liters, which is enough to purify an entire module, starting from standard commercial liquid argon (O$_2$ concentration $\sim 0.5$ ppm). The maximum recirculation rate of 2~m$^3$/h can be achieved, resulting from the pump throughput and the filter set impedance: with this value, one full-volume recirculation can be carried on in about six days. LN$_2$ is used to cool the pump vessel, purifier cartridges and all the Ar transfer lines.

Two-phase nitrogen coming from thermal screen, together with nitrogen employed in GAr/LAr recirculation, is sent back to a $\sim 1$~m$^3$ phase separator connected to two 30~m$^3$ liquid nitrogen storages, filled up to about 80\%. %In steady state conditions, these two reservoirs were fully dedicated to nitrogen storage, while during commissioning one was used for liquid argon.
All the residual nitrogen gas produced in the various processes was converted back into liquid by a dedicated re-liquefaction system. This was designed to work in closed loop for safe operation in confined place; however operation in open circuit, with liquid nitrogen delivery by trucks, was also foreseen in case of prolonged emergency stops of the apparatus.
The re-liquefaction system consists of twelve Stirling \cite{stirling} Cryogenics BV SPC-4 (4-cylinder) cryo-coolers, %based on the “reverse Stirling thermodynamic cycle”,
delivering 4.1~kW of cold power each at 84 K with a nominal efficiency of 10.4\%. The units operate independently, and automatically switch on/off to keep the nitrogen pressure at a fixed point, thus only delivering the actual cold power needed by the system.

Such required cold power can be determined by the consumption due to the insulation losses (heat input through joints, cryostat feet and
cables), the nitrogen screen cooling (circulation pump and distribution lines) and the GAr/LAr recirculation-purification systems.
During steady-state operations the cryogenic plant performed very well, successfully undergoing several safety, efficiency, redundancy tests and it demonstrated stability over the whole operating period. The average request of cold power was around 24~kW, mainly due to insulation losses. This was largely within the capability of the re-liquefaction system, as on average never more than 10 of the twelve Stirling units were found operational simultaneously.

The capability of the cryogenic plant allowed also performing a smooth detector commissioning: in order to ensure an acceptable initial LAr purity, the cryostats were evacuated to a pressure lower than $10^{-4}$~mbar, before commissioning. Vacuum phase lasted for three months, after which the cryostats could be cooled down to a temperature of 90~K within 7 days from start.
Finally filling phase took place, lasting about two weeks, with the use of commercial LAr, pre-purified in-situ before being injected in the detector, at a rate of $\sim 1$ m$^3$/hour/cryostat. During the whole period the GAr recirculation was operating at maximum speed to intercept outgassing from the inner walls. One month after filling, LAr recirculation and purification was started on both cryostats.

Operations at LNGS demonstrated the very high reliability of the existing cryogenic plant, in particular for what concerns argon purification and stability of the system. As a matter of fact it was clearly shown in~\cite{LArPurity} how, for most of the data taking period, argon purity could be kept at a level corresponding to a free electron life-time higher than 7~ms. Furthermore at the end of the LNGS run the value of 16~ms (increasing) was reached, thanks to the use of a new recirculation pump that was tested in one of the two modules (for details please refer to~\cite{LArPurity}, and see Fig.~\ref{fig:purity} below).

\begin{figure}[htbp]
\centering
\includegraphics[scale=0.6]{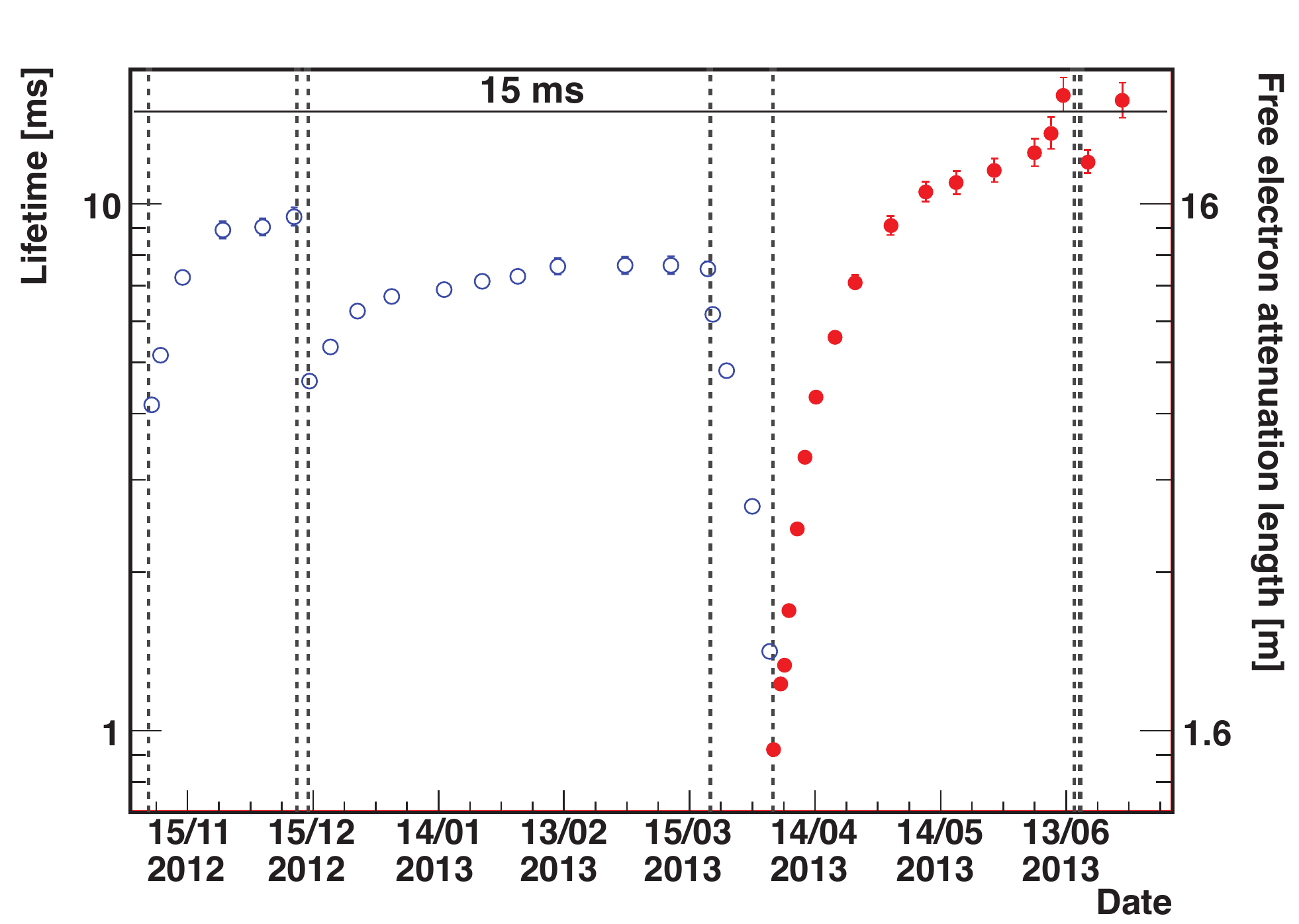}
\begin{center}
\caption{Detail of the LAr purity in the East module of the T600 detector, as measured in the last months of operation. The high value of the electron life-time can be appreciated, as well as the ever-increasing trend in the last days of data taking achieved with a new model of circulation pump. Drops in purity correspond to stops of the recirculation for pump maintenance/substitution.}
\label{fig:purity}
\end{center}
\end{figure}

Other cryogenic parameters affecting the \lartpc performance were also accurately monitored during steady-state detector operations.
In particular, the internal temperature in the two modules, directly connected to the electron drift velocity, was found to be stable and uniform at a level better than 0.25~K, well within the requirements (see Fig.~\ref{fig:tempLNGS}).
The same behavior was confirmed by the data on the internal absolute pressure (see Fig.~\ref{fig:pressLNGS}): a very high stability was measured, with variations contained within around 10~mbar, far lower than those of ambient pressure, despite various stops of the recirculation system due to pump maintenance.

\begin{figure}[htbp]
\centering
\includegraphics[scale=0.6]{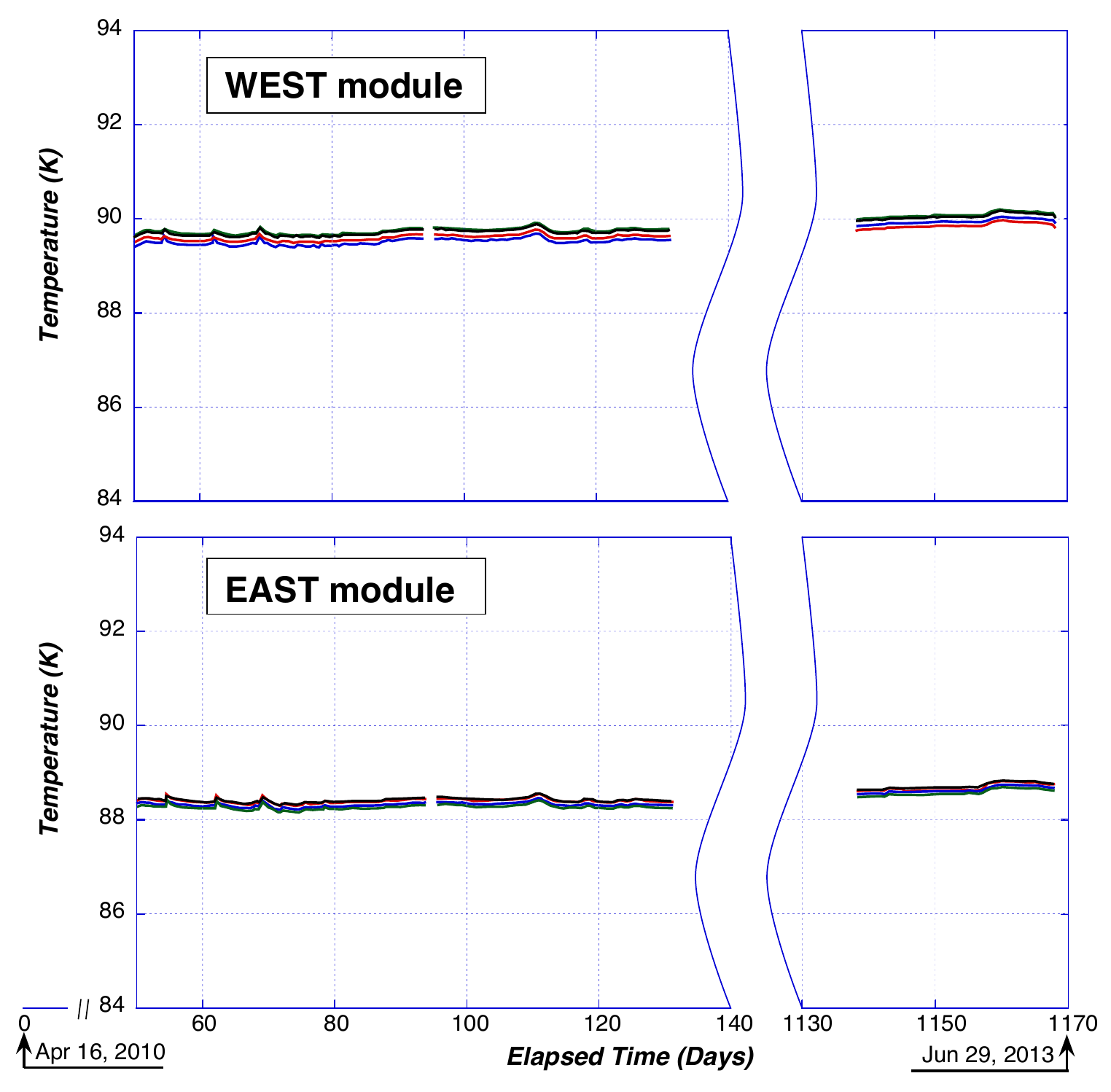}
\begin{center}
\caption{Trend of the temperature in the two modules, measured in three different positions (bottom, middle height, top) during two periods of the LNGS run, one at its beginning in 2010, and the second in 2013, close to the end of data taking.}
\label{fig:tempLNGS}
\end{center}
\end{figure}

\begin{figure}[htbp]
\centering
\includegraphics[scale=0.4]{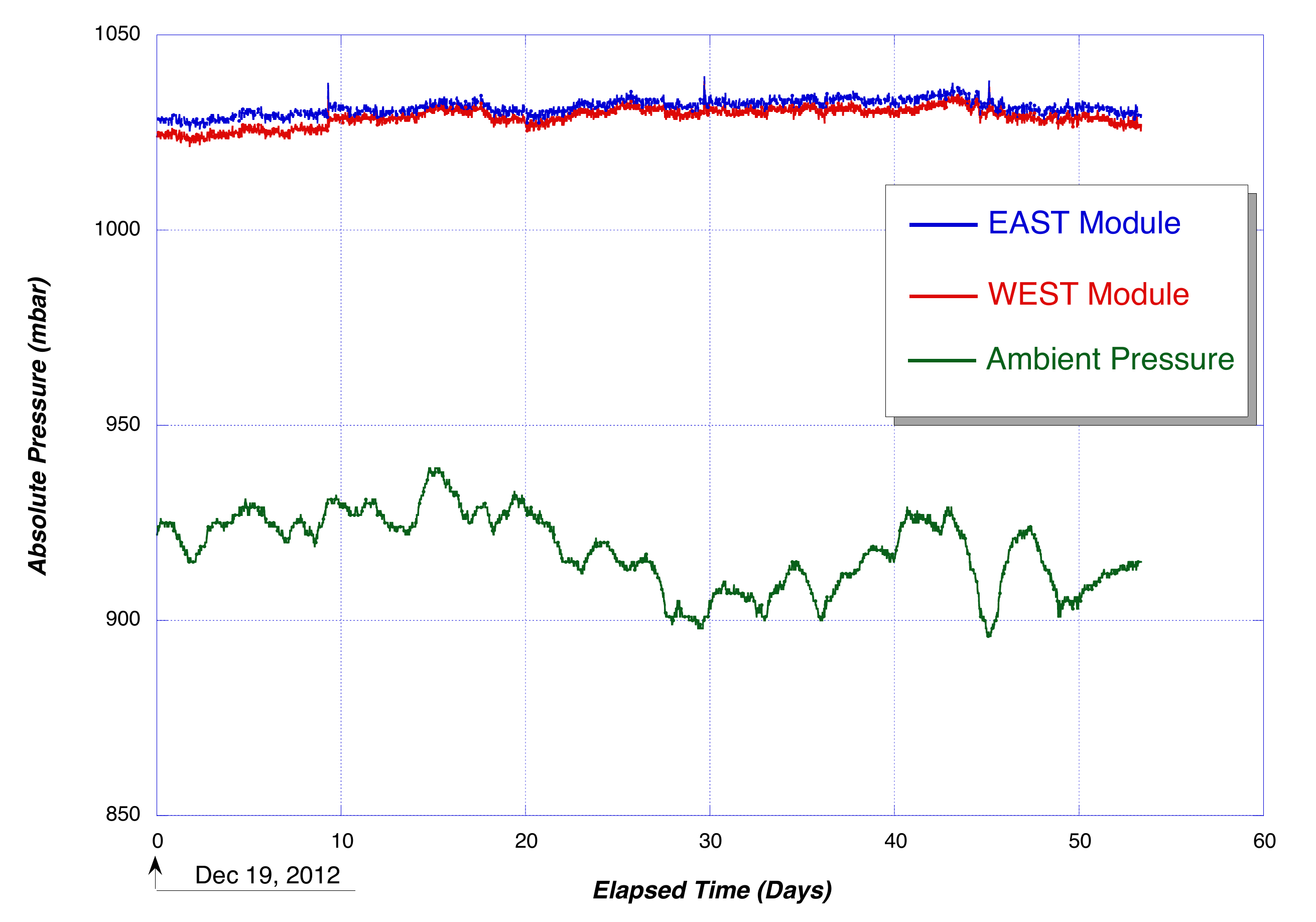}
\begin{center}
\caption{Absolute internal pressure in the two modules during a two-month period between the end of 2012 and the beginning of 2013. Data can be compared with the trend of ambient pressure.}
\label{fig:pressLNGS}
\end{center}
\end{figure}

In general, the cryogenic system of the T600, repeatedly pre-tested against different types of emergencies, performed very well during operations in limiting conditions (deep underground location), and it allowed obtaining unprecedented results on argon purification. This result largely justifies the decision of carrying most of the plant and its design on to the next stage of the detector life within the Fermilab SBN program (see details in Section~\ref{ica_new_cryo}).

%%%%%%%%%%%%%%%%%%%%%%%%%%%%%%%%%%%%%%%%%%%%%%%%%%%%%%%%%%%%%%%%%%%%%%%%%%
%%%%%%%%%%%%%%%%%%%%%%%%%%%%%%%%%%%%%%%%%%%%%%%%%%%%%%%%%%%%%%%%%%%%%%%%
\section{Overhauling of the T600 Detector: WA104}
\label{wa104}

The \icarus detector has been moved to CERN for a complete overhauling, preserving most of the existing operational equipment, while upgrading some components with up-to-date technology in view of its future non-underground operation. The refurbishing (CERN WA104 project) will include the following main activities:

\begin{itemize}
 \item substitution of the present cathodes with new ones of improved planarity;
 \item implementation of a new light collection system, to allow a more precise event localization and disentangle beam events from the background induced by cosmic rays;
 \item implementation of new readout electronics;
 \item other internal TPC updating: slow control system and cabling;
 \item realization of new vessels for LAr containment and new thermal insulation, based on a similar technology, as foreseen for LBNF and the SBN Near Detector;
 \item complete review and maintenance of the cryogenics and purification systems.
\end{itemize}

The transfer of the two T600 TPCs to CERN has been already completed:
after the positioning of the first module into its transport vessel (see Fig.~\ref{ica_box1}, Fig.~\ref{ica_box2} and Fig.~\ref{ica_box3}), the first cargo has arrived to CERN at the middle of November 2014 (see Fig.~\ref{ica_box4}). Movement operations at LNGS proceeded smoothly, with 4 to 6 people continuously involved for three weeks during October 2014. After the arrival of the second transport vessel at LNGS in November 2014, also the second T600 TPC was moved to CERN, being on site at the middle of December 2014.

\begin{figure}[htbp]
\centering
\setlength{\fboxsep}{0pt}
\setlength{\fboxrule}{0.6pt}
\fbox{\includegraphics[width=0.8\textwidth]{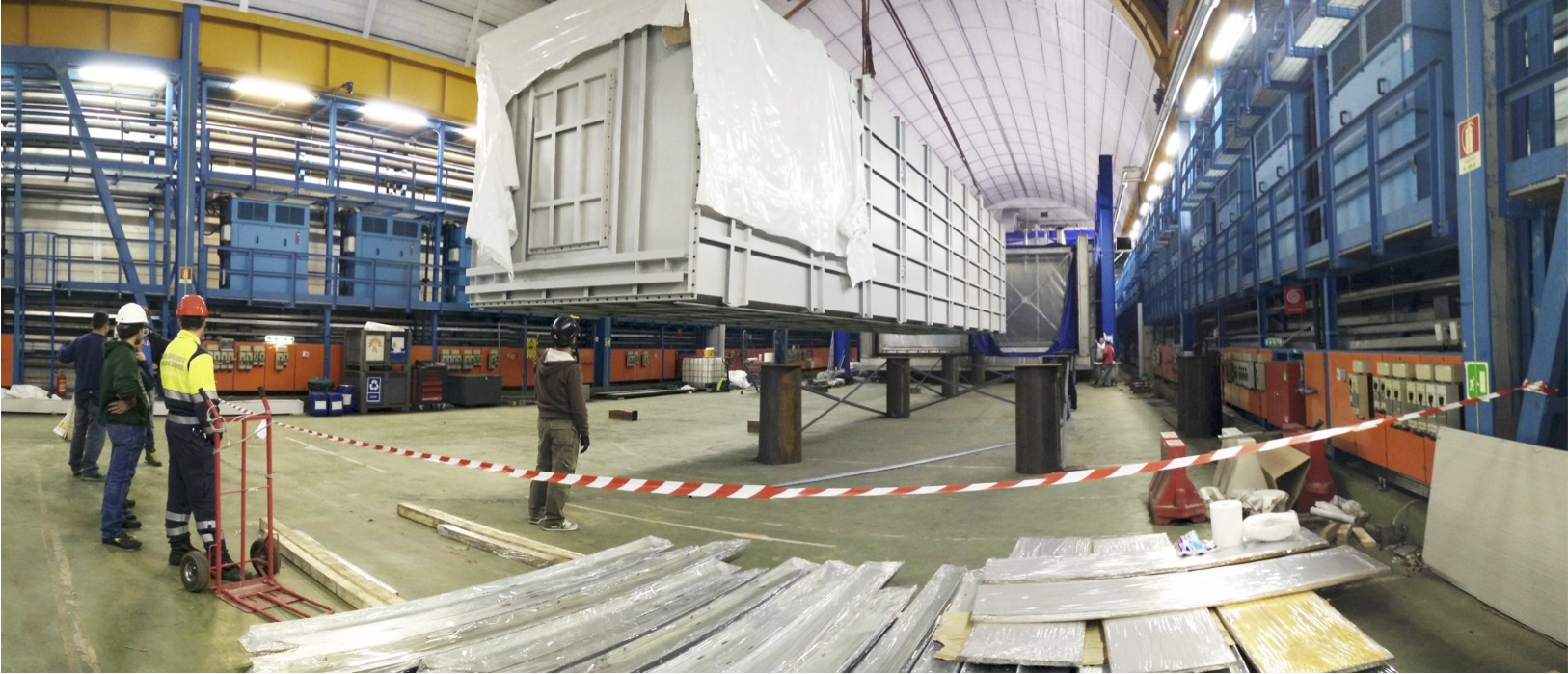}}
\caption{Transport vessel positioning in front of the TPC at LNGS.}
\label{ica_box1}
\end{figure}

\begin{figure}[htbp]
\centering
\setlength{\fboxsep}{0pt}
\setlength{\fboxrule}{0.6pt}
\fbox{\includegraphics[width=0.7\textwidth]{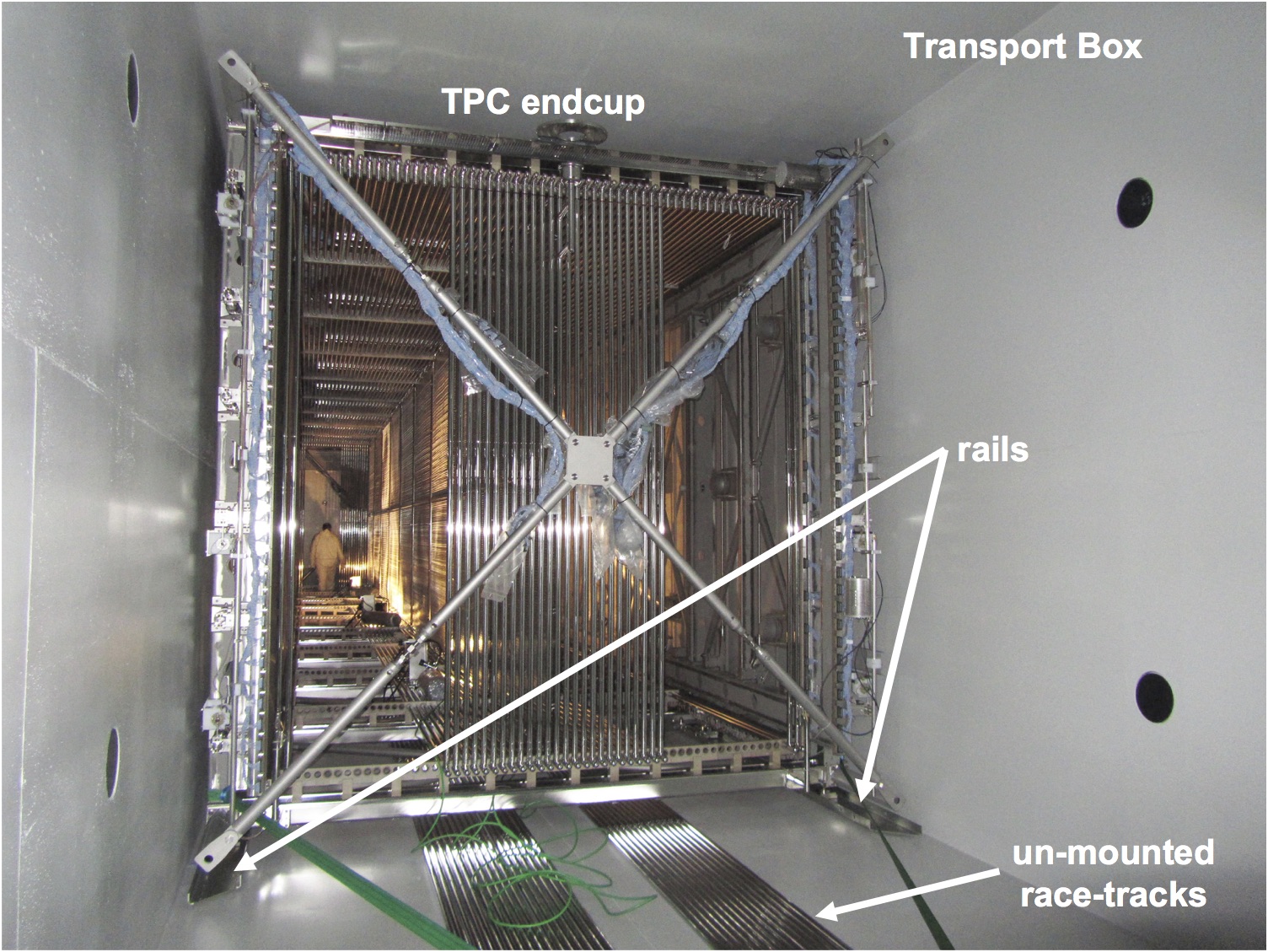}}
\caption{Start of movement of the TPC inside the transport vessel.}
\label{ica_box2}
\end{figure}

\begin{figure}[htbp]
\centering
\setlength{\fboxsep}{0pt}
\setlength{\fboxrule}{0.6pt}
\fbox{\includegraphics[width=0.7\textwidth]{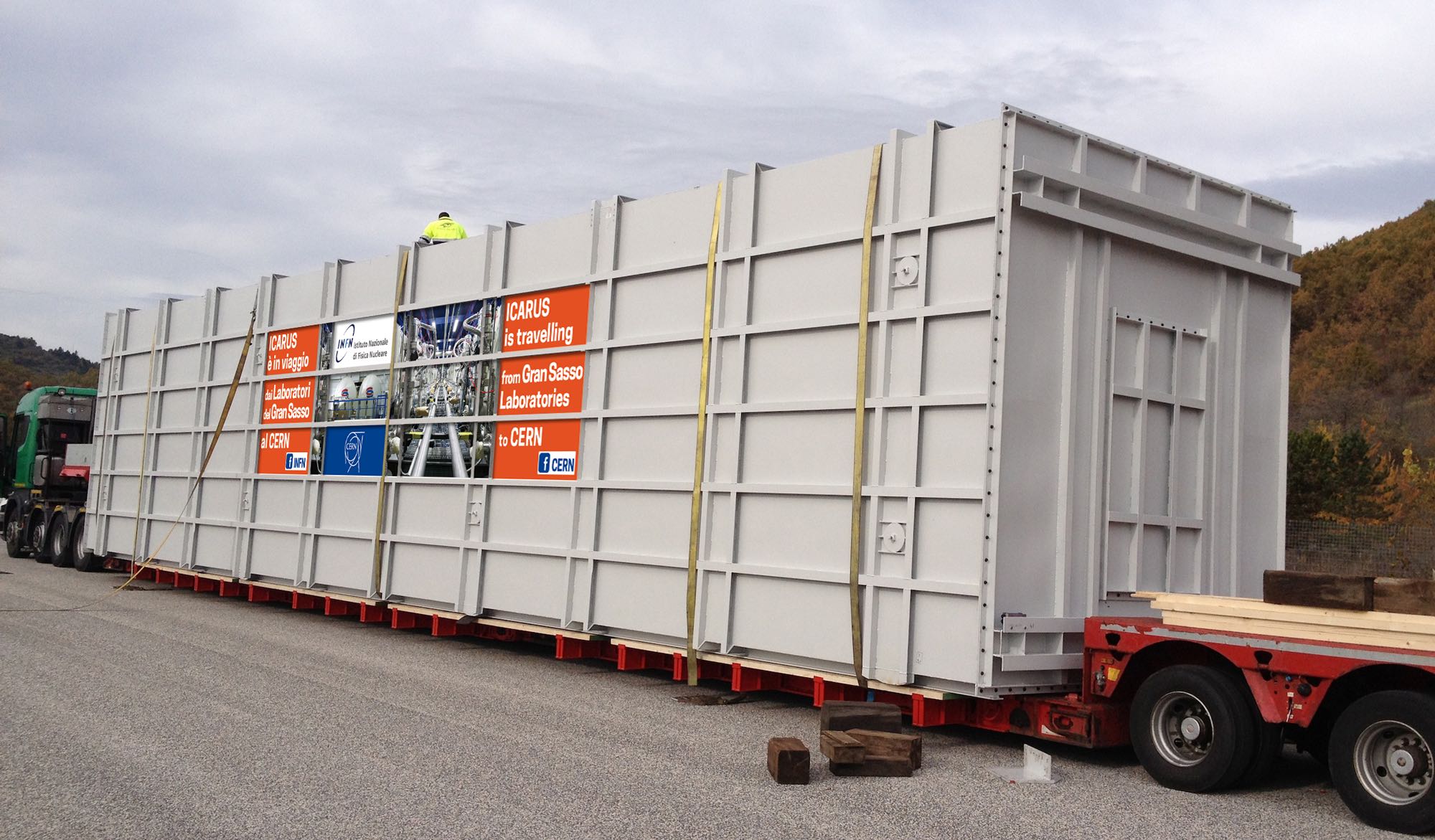}}
\caption{The transport vessel on its way to CERN.}
\label{ica_box3}
\end{figure}

\begin{figure}[htbp]
\centering
\setlength{\fboxsep}{0pt}
\setlength{\fboxrule}{0.6pt}
\fbox{\includegraphics[width=0.7\textwidth]{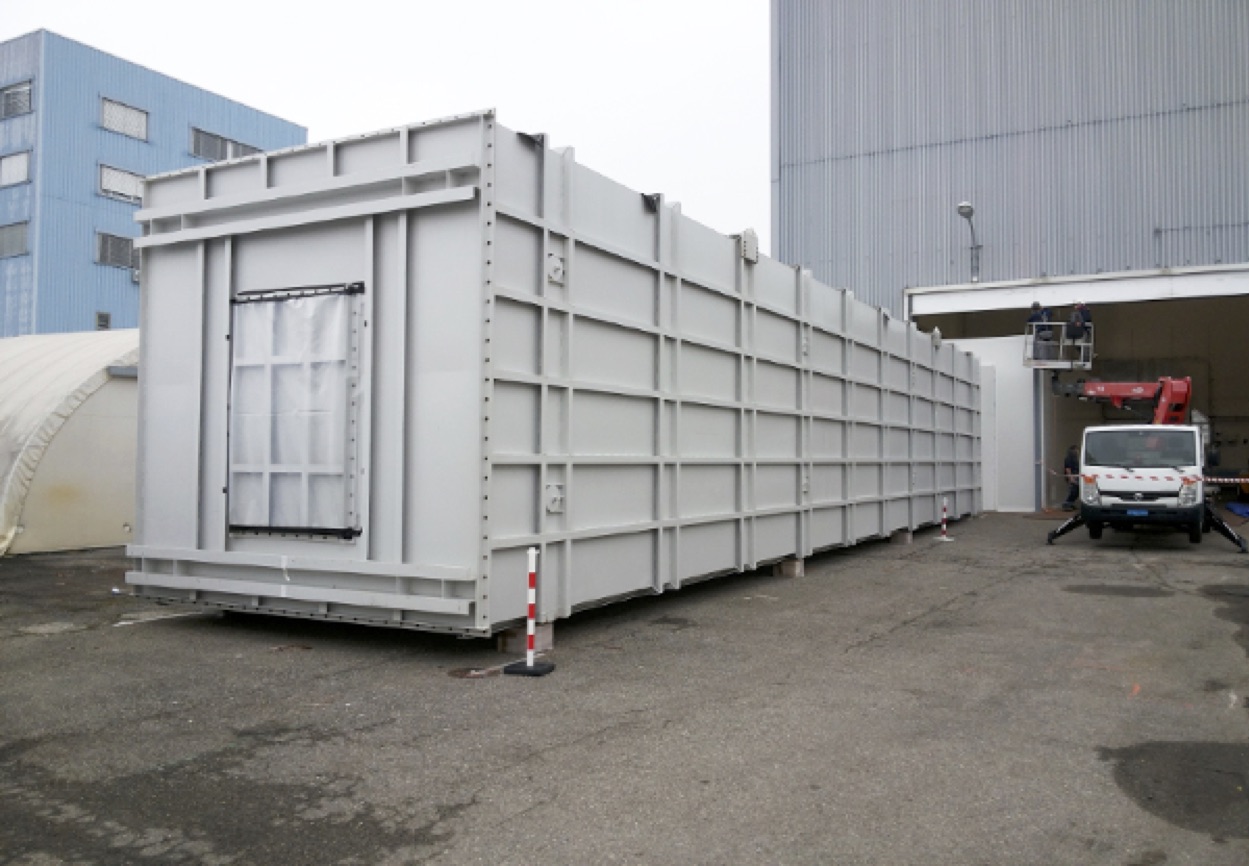}}
\caption{The transport vessel in front of building 185 at CERN.}
\label{ica_box4}
\end{figure}

All overhauling activities will be carried on at CERN building 185, which has been outfitted accordingly, with all necessary services (e.g. electrical, ventilation, heating, air recirculation in clean room). A dedicated clean room, to house the TPCs during operations, has been already completed (see Fig.~\ref{ica_croom}). Since December 19$^{th}$ 2014, the first transported module is inside the clean room, while the second one has been stored inside the building (see Fig.~\ref{tpc_cleanroom}) 

\begin{figure}[htbp]
\centering
\setlength{\fboxsep}{0pt}
\setlength{\fboxrule}{0.6pt}
\mbox{\fbox{\includegraphics[height=0.3\textheight]{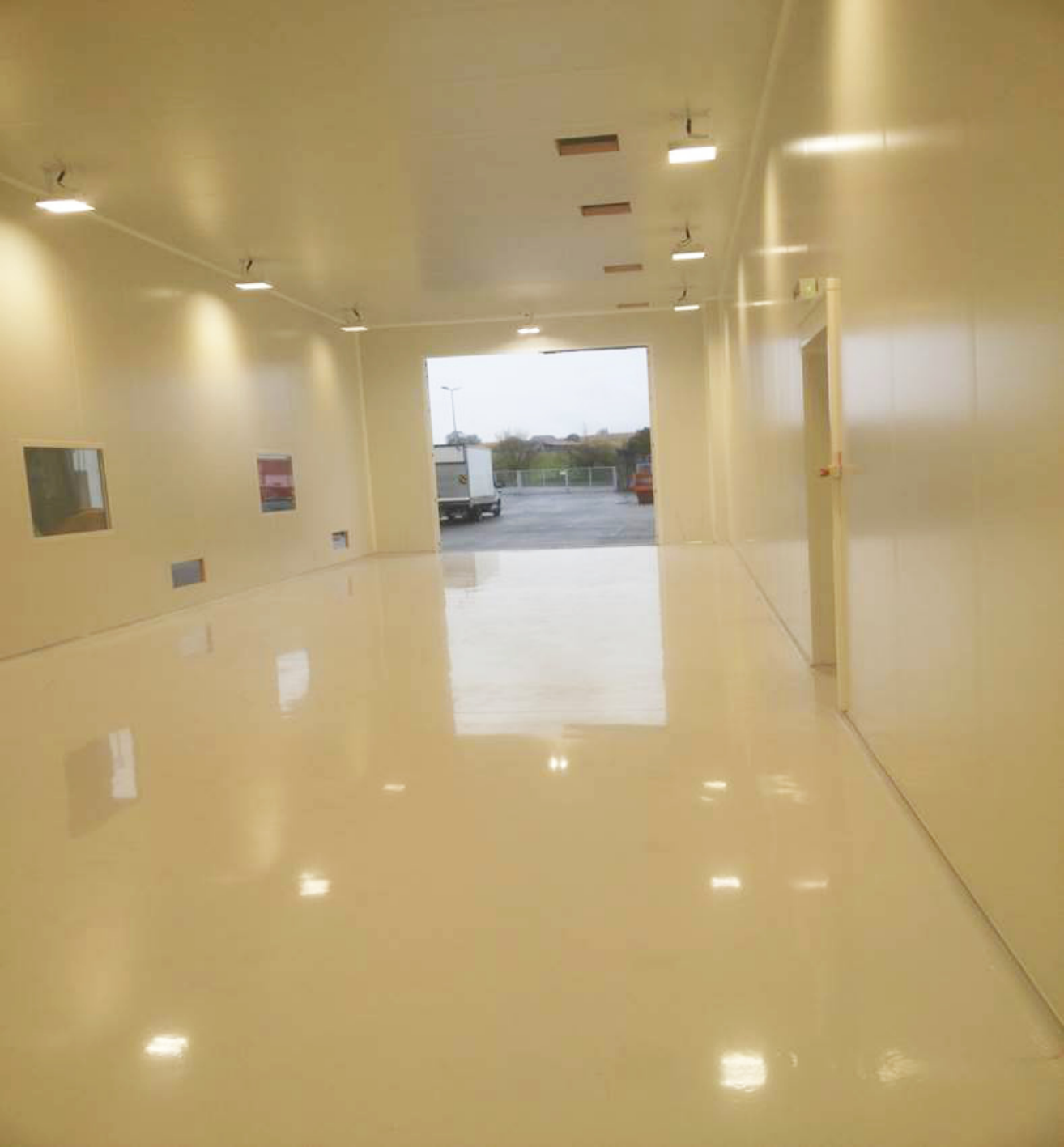}}\qquad
\fbox{\includegraphics[height=0.3\textheight]{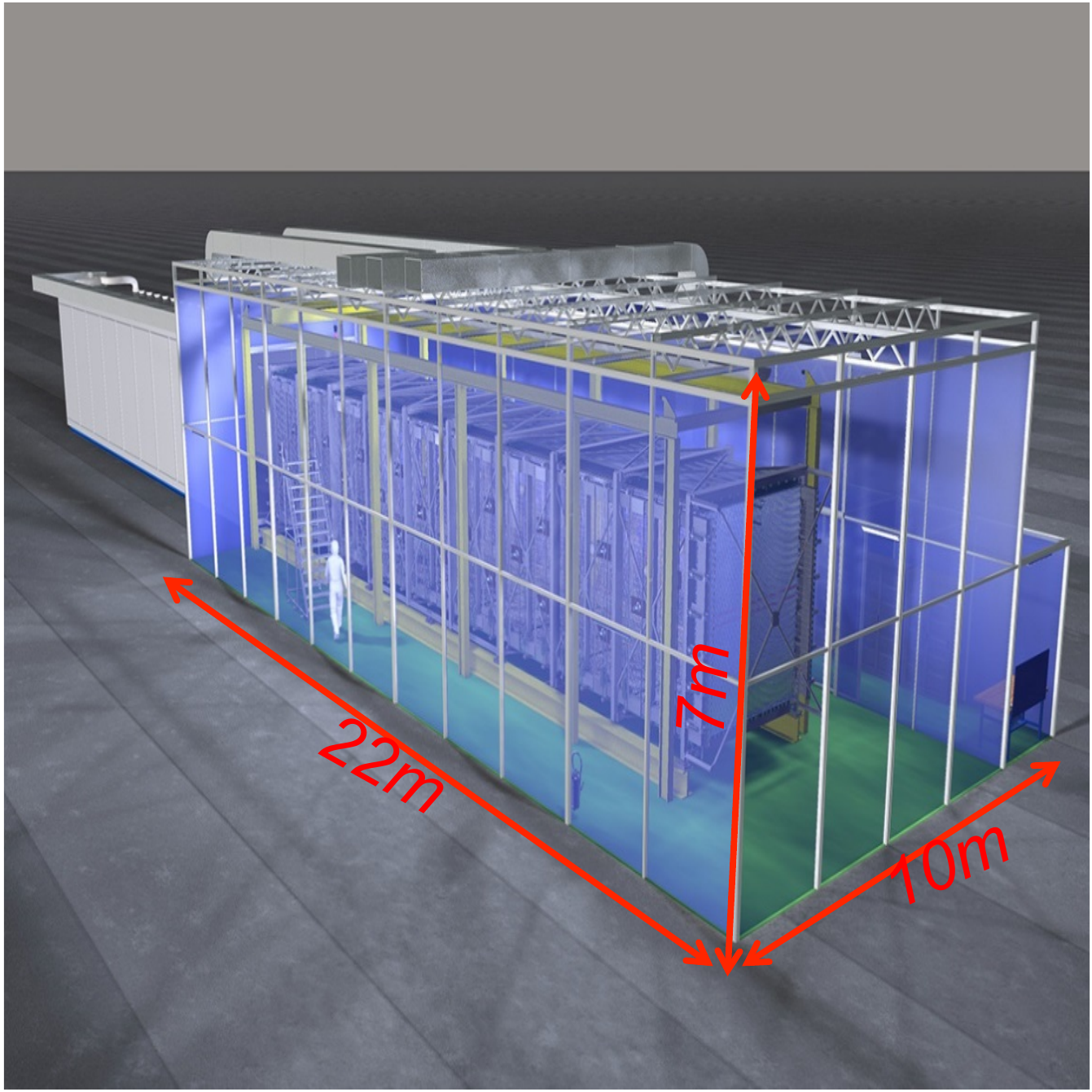}}}
\caption{Left: Clean room implemented inside CERN building 185, adapted for the T600 overhauling. Right: Sketch of the clean room housing the T600 TPCs.}
\label{ica_croom}
\end{figure}

\begin{figure}[htbp]
\centering
\setlength{\fboxsep}{0pt}
\setlength{\fboxrule}{0.6pt}
\fbox{\includegraphics[width=0.7\textwidth]{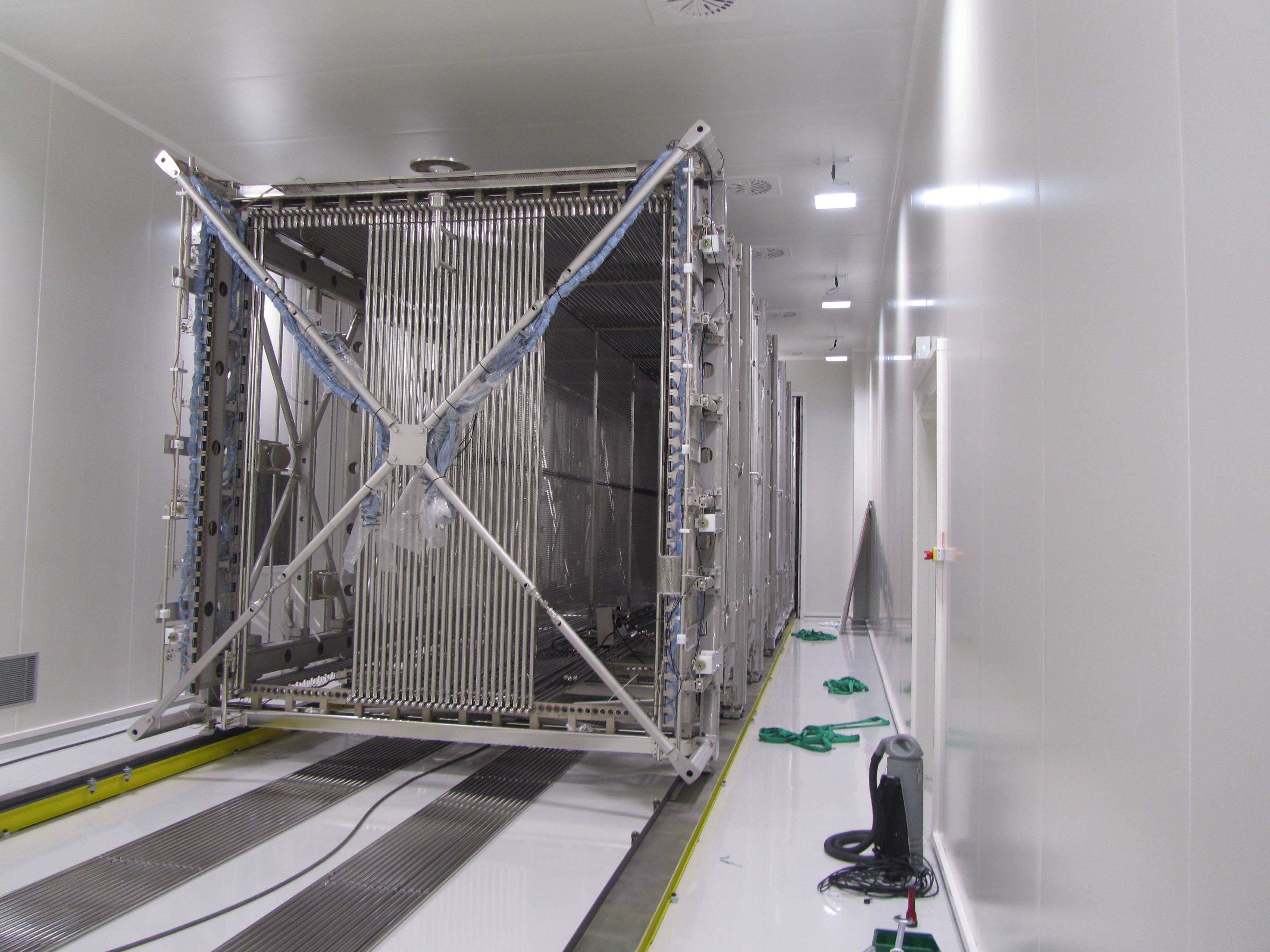}}
\caption{The first transported module inside the clean room at CERN.}
\label{tpc_cleanroom}
\end{figure}

This Section is organized as follows: Par.~\ref{TPC_mod} describes the main modifications of the \icarus internal detectors. Par.~\ref{ica_new_pmt} shows the new layout of the Light Collection System, while the Electronics and DAQ
in the FNAL configuration are described in Par.~\ref{ica_new_ele}. Finally, Cryogenics and Purification new systems are presented in Par.~\ref{ica_new_cryo}.

%%%%%%%%%%%%%%%%%%%%%%%%%%%%%%%%%%%%%%%%%%%%%%%%%%%%%%%%%%%%%%%%%%%%%%%%%%%%%%%%%
\subsection{TPC modifications}
\label{TPC_mod}

Minor changes are expected to be implemented, for what concerns the T600 TPC internal structure, with respect to the present configuration.

Distortions in the uniformity of the electric field in the drift volume of the T600, due to positive ion charge accumulation induced by cosmic rays, have been investigated within the ICARUS collaboration. According to our estimates for ICARUS at shallow depths, the effect could be at most of a few mm, in agreement with the data collected in the 2001 technical run on surface in Pavia. To correct these distortions by making the  electric field more uniform, some additional widely spaced shaping wire planes could be installed  inside the sensitive volume, at the voltage of the potentials of the field cage electrodes. For instance, two arrays of wires (with a   pitch of the $\sim 10$ cm)  at 50~cm and 100~cm from the HV plane, anchored to the corresponding two field cage electrodes, could reduce the field distortion by almost an order of magnitude.

Moreover, small deviations from the linearity of the drift field have been found in the region close to the cathode plane on both modules. This is due to the not perfect planarity of the cathodes, owing to their pierced structure. This was confirmed by visual inspection after the first cryostat opening in October 2014, where displacements from planarity of the order of 5 mm were found. Thus, in view of the SBN experiment at FNAL, it has been decided to change the present cathodes with  new ones of improved planarity. 
The cathode surface could be either opaque or transparent to the scintillation light, depending on the request to perform the coincidence of the light signals from the two PMT arrays at either sides of the cathode. The detailed design of the new cathodes is currently under completion. Other activities on the T600 TPCs concern the updating of the slow control system for temperature, pressure and cryostat wall deformation monitors, as well as the design of new cabling for internal wires, PMTs and slow control sensors.

%%%%%%%%%%%%%%%%%%%%%%%%%%%%%%%%%%%%%%%%%%%%%%%%%%%%%%%%%%%%%%%%%%%%%%%%%%%%%%%
\subsection{New Light Collection System Layout and Implementation}
\label{ica_new_pmt}

The future operation of the ICARUS \lartpc at the FNAL BNB at shallow depths  requires an improved light collection system, able to detect with full efficiency the prompt scintillation light from events with energy depositions down to $\sim 100$~MeV. %, to trigger on the full neutrino beam spectrum.

The renovated T600 photo-detector arrangement should again collect the VUV scintillation signal which is present in the LAr simultaneously to the ionization, converting it to visible light.

The detection process in the \lartpc is initiated by the trigger signal opening a long ``imaging" readout window, in which tracks are recorded in a time sequence, collected serially by the readout planes, while the electrons travel towards the end of the drift path. The full image of the event is therefore progressively extracted from the drift time distributions and from the many readout wires.

As already mentioned in Sec.~\ref{ica_requirements}, the fast scintillation light signal in LAr must be put in coincidence with the 1.6~$\mu$s beam trigger gate, giving the huge rate of $\sim$ 0.83 $\times$ 10$^6$~c/y. Moreover, during the long duration of each readout window, there will be on average 44 cosmic ray-induced scintillation light signals spread over the whole T600, four times the number of the cosmic tracks in a single TPC (accounting for the cathode transparency and because the time interval during which a light signal can be linked to a charge deposition is twice the maximum drift time).

The new light collection system has to be able to localize the track associated with every light pulse %, not only along the drift direction but also 
 along the 20~m of the longitudinal detector direction, with an accuracy better than 1~m, which is smaller than the expected average spacing between cosmic muons in each TPC image.

In this way, the light collection system would be able to provide unambiguously the absolute timing %t$_0$
for each track; and to identify, among the several tracks in the \lartpc image, the event in coincidence with the neutrino beam spill. The time accuracy of the incoming event with the new light collection system is expected to be at 1 ns level, allowing the exploitation of the bunched beam structure, lasting 1.15 ns (FWHM $\sim$ 2.7 ns) every 19~ns, to reject cosmic events out of bunch as described in a SBN note~\cite{bb_rubbia}. An overall time resolution of 1.3~ns would then allow a background reduction of a factor $\sim$ 4 by rejecting cosmic events occurring outside the RF buckets with a 2$\sigma$ accuracy.

\subsubsection*{Tests on new cryogenic PMT models}

The baseline solution for the T600 photo-detection system will rely on large surface Photo-Multiplier Tubes with hemispherical glass window of 200~mm (8") diameter, manufactured to work at cryogenic temperature. First tests were carried out to choose the most suitable PMT model. Three new large area PMTs, Hamamatsu R5912 Mod and R5912-02 Mod, and ETL 9357 KFLB, have been characterized both at room and at cryogenic temperature~\cite{pv_pmt}.

Tested PMTs have a borosilicate glass window and a bialkali photo-cathode (K$_{2}$CsSb) with platinum undercoating, to restore the photo-cathode conductivity at low temperature. Hamamatsu R5912 Mod and R5912-02 Mod PMTs have 10 and 14 dynodes, respectively, while the ETL 9357 KFLB has 12 dynodes. Photo-cathode uniformity, gain, linearity, dark count rate and Quantum Efficiency (QE) for LAr scintillation light have been measured.

For gain, linearity and uniformity measurements, PMTs were illuminated with a 405~nm NICHIA NDV1413 laser diode, using an Avtech AVO-9A-C-P2-LARB pulse generator and an optical fiber (7~$\mu$m core diameter, 3~m long). An appropriate support was used to maintain the fiber in a fixed orientation, normal to the PMT window, while allowing to move it in various positions on the window itself. A CANBERRA 2005 pre-amplifier and an ORTEC-570 amplifier  were used to form PMT signals, then acquired with an ORTEC-Easy-8k,12 bit Multi Channel Analyzer. PMTs dark count rate has been measured with a different acquisition system, i.e. with a CAEN V812 discriminator and a CAEN V560 counter. The discrimination threshold was gradually increased from 1 to 255~mV, with 1~mV steps.

To test them at cryogenic temperature, the PMTs were directly immersed in liquid nitrogen (T~=~77~K), to simulate real experimental conditions. Measurements were carried out after a couple of days of rest in the cryogenic bath. The same setup and acquisition system described above were used, with the fiber and the other cables allowed to enter by a proper feed-through, used to preserve darkness conditions and thermal insulation (see Fig.~\ref{fig:setup}).

\begin{figure}[ht]
\centering
\includegraphics[scale=0.4]{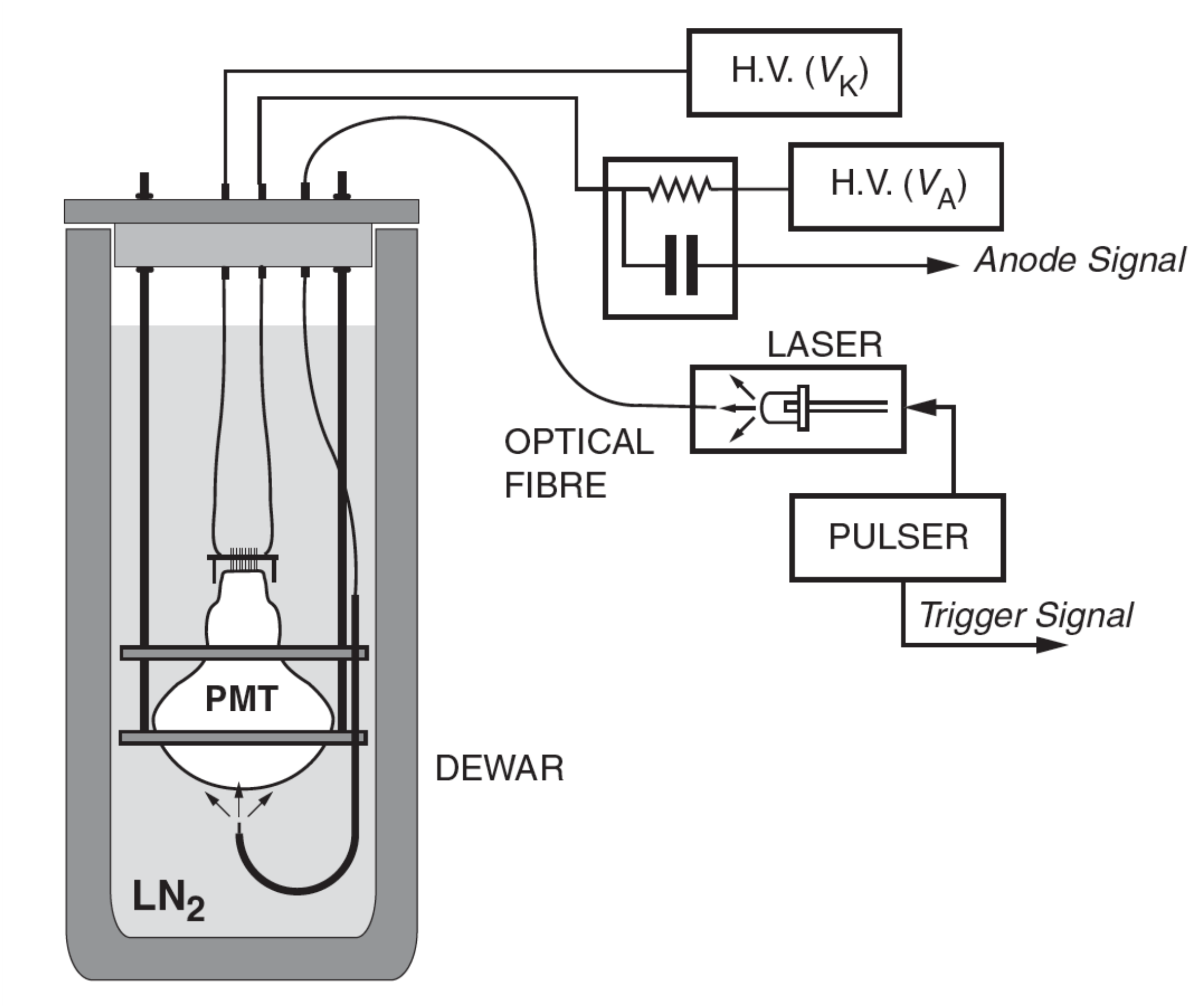}
\caption{Experimental setup for the characterization of PMTs immersed in liquid nitrogen.}
\label{fig:setup}
\end{figure}

Hamamatsu PMTs showed a good uniformity, within 10\%, up to 10 cm from the tube axis, where a gain reduction occurs, probably due to the electric field non-uniformity in the peripheral region of the tube. While this behavior does not occur with ETL 9357 KFLB, lower uniformity is measured, $\sim$20\%; furthermore a very low signal has been measured in a specific region of the photo-cathode for this model: this can be explained by a degradation of the photo-cathode for the specimen under test. Due to this problem and after
discussions with the manufacturer, it was decided not to test the ETL 9357 KFLB at cryogenic temperature, pending further tests by the producer. 

The gain of the devices was estimated from Single Electron Response (SER)  as a function of the applied voltage and operating temperature. The gain reduction occurring at 77~K is evident for both Hamamatsu devices, being $\sim$70\% in the R5912 and $\sim$35\% in the R5912-02 with respect to room temperature data (see Fig.~\ref{ica_pmt_qe} Left). Hamamatsu R5912 MOD remains linear up to 400 phe, while R5912-02 MOD reaches again the saturation regime after a few photo-electrons, about only 10 phe. 

%R5912-02, at room temperature, has a dark count rate lower than 0.4~kHz for all thresholds; the increase at T = 77~K is very pronounced, but the rate remains lower than 2~kHz. For R5912 dark count rate is higher, both at room ($\geq$0.4~kHz up to 0.9~phe of threshold) and cryogenic ($\geq$2~kHz up to 0.6~phe of threshold) temperature. This increase should be lower in LAr, being the temperature higher than the LN$_{2}$ one. However, the relatively high dark noise rate does not hinder the use of the devices in real experimental conditions.
In general both photo-detectors showed a good behavior and are suitable for cryogenic application.

A different experimental setup was used to measure the QE of the photo-cathodes in the VUV light region. To make them sensitive to VUV light, the PMT sand-blasted glass windows were deposited with a TPB coating of $\sim$~0.2~mg/cm$^2$.

The measured QE accounts then for: the shifting efficiency of the TPB, a geometrical factor (on average half of the photons will be re-emitted in the opposite direction with respect to the photo-cathode) and the QE of the PMT for blue light. As shown in Fig.~\ref{fig:sistem}, the PMT under test was placed inside a steel chamber optically connected to a McPHERSON 234/302 VUV monochromator. 

The experimental setup included a McPHERSON 789A-3 scanner, a McPHERSON 632 Deuterium lamp, a rotating Al+MgF$_{2}$ mirror, an AXUV-100 reference photo-diode and collimating optics. The whole system was kept under vacuum, down to $10^{-4}$ mbar, to prevent ultraviolet light absorption. Thanks to the rotating mirror, the light spot was directed alternatively on the PMT surface or on the reference photo-diode. The QE was obtained by comparing the current measured with the PMT and the same collected with the reference diode, keeping the light constant. Measurements were carried out by means of a picoammeter. Results for LAr (128 nm) and LXe (165 nm) emission peaks are reported in Fig.~\ref{ica_pmt_qe} Right. ETL 9357 KFLB has, at 128 nm, a QE = $4.7\%\pm0.7\%$, while Hamamatsu PMTs present a higher value, QE = $7.0\%\pm0.6\%$. 

\begin{figure}[ht]
\centering
\includegraphics[scale=0.5]{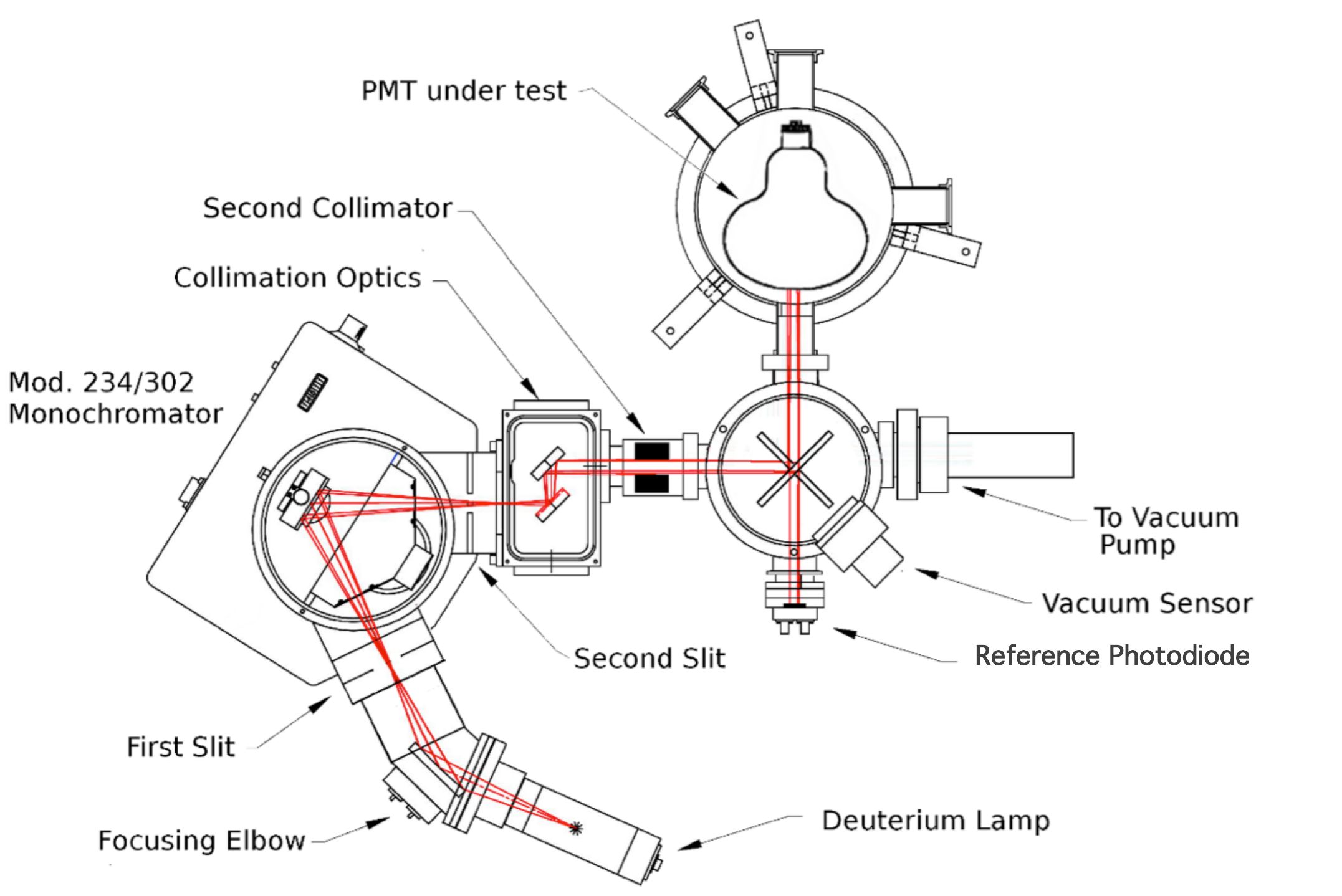}
\caption{Experimental setup for the evaluation of the response of PMTs to the VUV light.}
\label{fig:sistem}
\end{figure}

\begin{figure}[ht]
\centering
\includegraphics[scale=0.4]{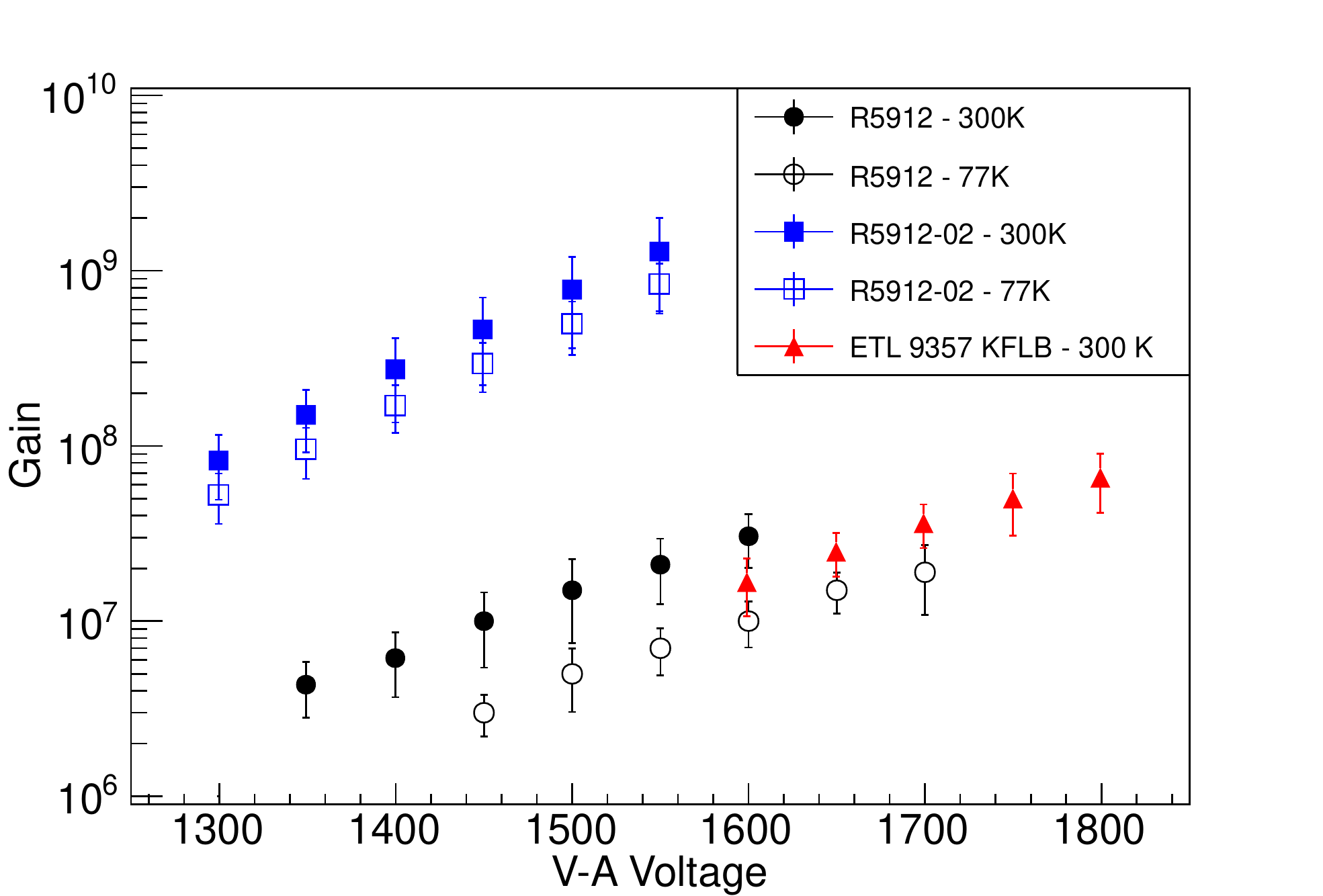}
\includegraphics[scale=0.4]{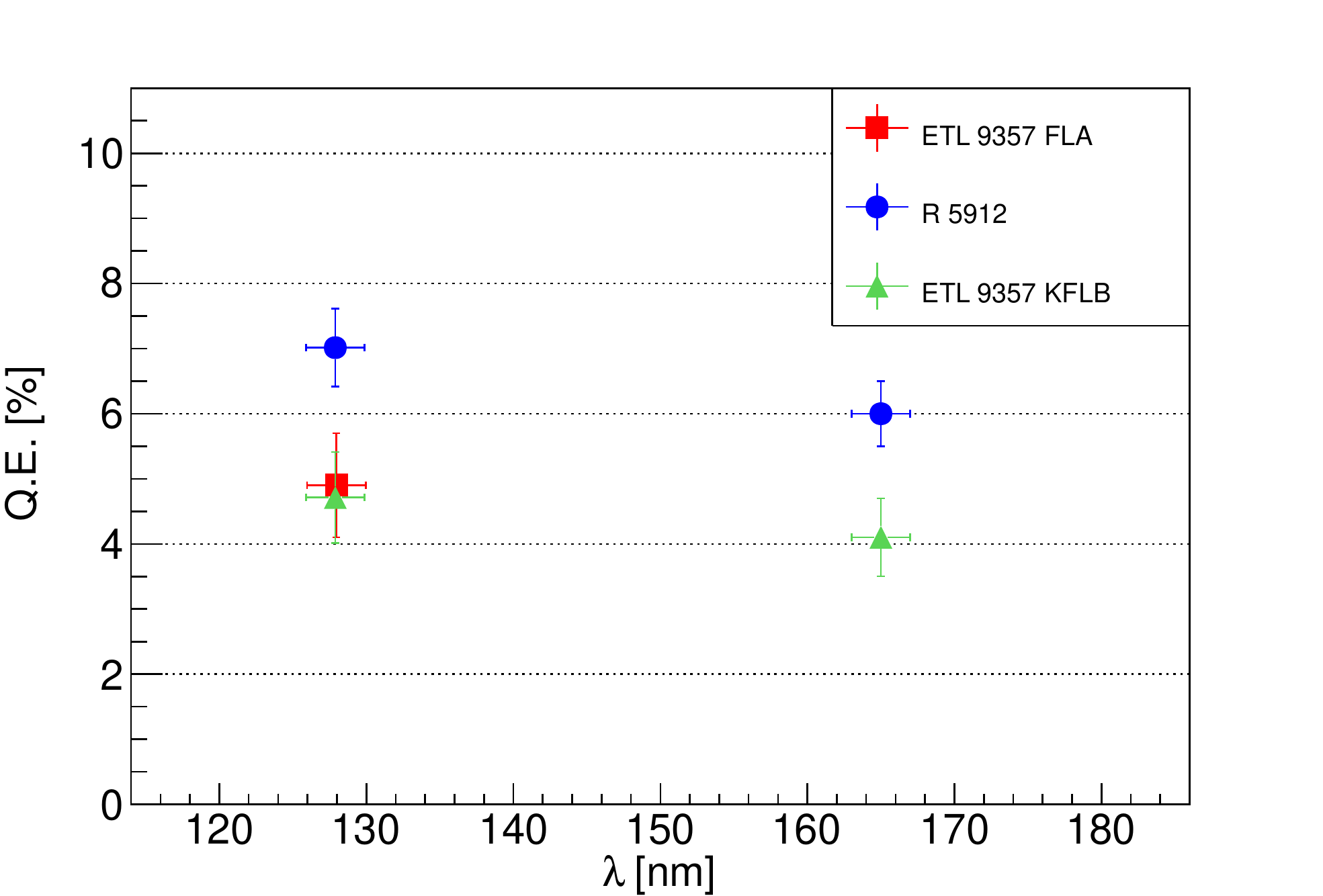}
\caption{Left: gain trends for tested PMTs at room (full spots) and at cryogenic (empty spots) temperature as a function of the voltage between anode and cathode. Right: Quantum Efficiency of the tested devices at LAr emission peak (128 nm) and LXe emission peak (165 nm).}
\label{ica_pmt_qe}
\end{figure}

\subsubsection*{New Light Collection System Layout}

After the choice of the PMT model and the purchase of the devices, a careful evaluation of the performance of each PMT before its final mounting inside the T600 detector will be carried out. Measurements will be focused on the main PMT parameters which are temperature dependent, to isolate possible defects in the devices. In particular, the following features will be characterized: shape of the anode pulse, SER of the anode pulse, gain, single-electron transit time (spread, pre- and late-pulsing), after-pulses, dark-count rate and spectrum. To this purpose a test facility will be setup in a dedicated INFN laboratory.

PMTs will be located in the 30~cm space behind the wire planes with sustaining structures: the sustaining system will allow the PMT positioning behind the Collection wire planes according to the geometrical planned disposition. PMTs induce spurious signal on wire planes: %which contribute to the noise of the detection system.
to reduce this drawback, still under investigation, each PMT will be surrounded by an electrostatic shield. A drawing to show a 90-PMTs layout behind the wire planes is presented in Fig.~\ref{PMTlayout}.

\begin{figure}[htbp]
\centering
\includegraphics[width=0.8\textwidth]{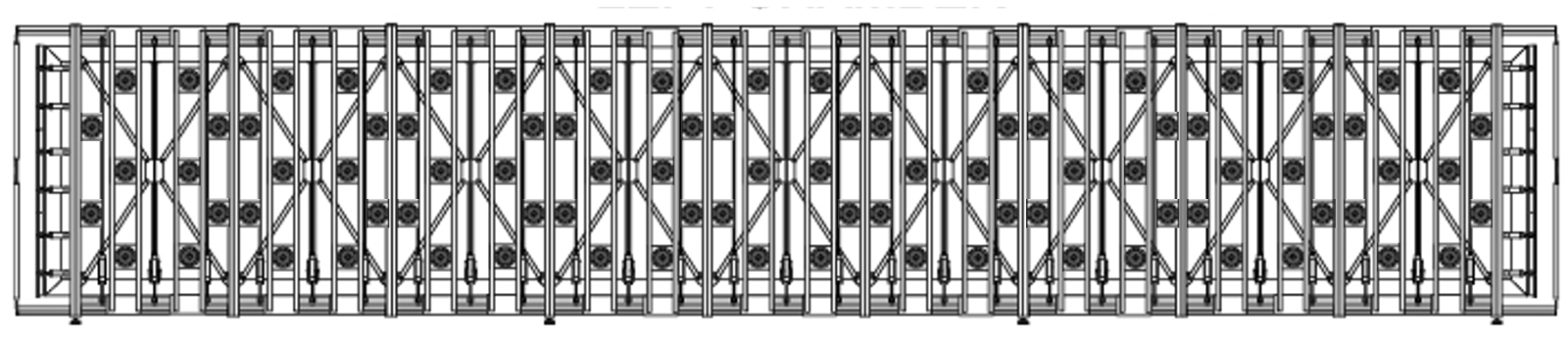}
\caption{Drawing showing a 90-PMTs layout behind the wire planes. PMTs are depicted as dark circles. This design yields a 5\% photo-cathode coverage.}
\label{PMTlayout}
\end{figure}

To prevent any impedance mismatch in the signal lines, PMT negative biasing will be adopted. %This will also lead to the advantage of a low-voltage capacitor in the PMT voltage divider final stages.
This scheme will adopt two cable for each device, one for the DC HV power supply and one for the signal. The main disadvantage is the operation with photo-cathodes at HV voltage close to the collecting wire planes. 
%Considering a $\sim$ 3 cm gap, 
Despite electric fields of about 600~V/cm are expected, the TPC electric field and the electron collection on the wire planes will be not altered. Moreover, the %possible
adoption of a fine-mesh grid in front of each PMT should avoid any interference between charge collection and light detection. 

%The standard readout of the PMTs is based on waveform 12-bit ADC digitizers with circular buffering. A sampling rate around 20 MHz will allow for accurate event start time determination, once the PMT signals are shaped with $\sim 100$~ns decay time pre-amplifiers.
Fast waveform digitizers are required to exploit the bunched beam structure. 1~GHz waveform digitizers with zero suppression will be adopted, preventing the use of shaping pre-amplifiers; the input dynamics must permit the recording of the scintillation light fast component pulses and, at the same time, of the single photons arriving from the slow component de-excitation. A PMT timing calibration/monitoring system will also be implemented. The baseline solution consists of pulsed laser diodes and optical diffusors installed at various locations in front of the PMTs, possibly on the TPC cathodes. Short pulses ($<1$~ns) of laser light will be transmitted inside each TPC, by means of optical fibers and feed-throughs, to each of the diffusors. This will allow the PMTs lighting with approximately uniform intensity. A dedicated R\&D activity is foreseen to evaluate the timing performance of this system and to optimize the optical fiber feed-through implementation.

Dedicated calculations have been set up to evaluate the performance of the upgraded ICARUS light collection system, in terms of event localization for both cosmic muons and electromagnetic showers. 
Cosmic muon tracks in the active liquid argon volume are simulated as straight lines with directions distributed as $\cos^{2}\theta$ around the vertical axis, where $\theta$ is the zenith angle. About 40,000 VUV photons/cm (corresponding to the amount of scintillation light from 2.1~MeV/cm energy deposition of a minimum ionizing particle in a field of 500~V/cm) are produced uniformly along the muon tracks and are emitted isotropically. Rayleigh scattering and the presence of delta rays in a 15 cm radius cylinder surrounding the muon path are also included in the simulation.

Electromagnetic showers are simulated as clusters containing single 1~MeV points ($\sim 21,000$ photons) up to the deposited energy. No reflection and diffusion on walls have been simulated, since the LAr VUV scintillation photons are absorbed by all materials. The number of detected photons is derived in terms of solid angle calculation.

To estimate the localization capability of the light collection system, a Monte Carlo simulation has been carried out. For each simulated event 
the barycentre of the light emission has been calculated, by averaging on the coordinates of the PMTs and weighing on the different signal intensities. This was done both for the vertical and horizontal coordinates. The spatial resolution of the system, evaluated as the difference between the simulated track barycenter and the same quantity reconstructed as described above, is found to be better than half a meter, as shown in Fig.~\ref{MC_result}. This will allow to strongly narrow the LAr region in which to search for neutrino beam-induced events.

\begin{figure}[htbp]
\centering
\includegraphics[width=8cm]{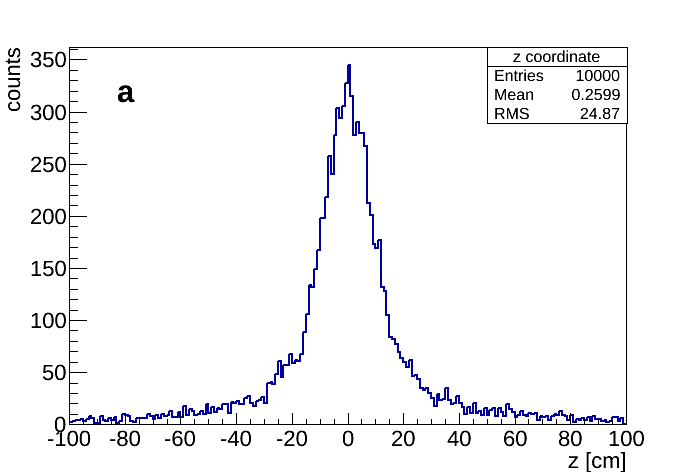}
\includegraphics[width=8cm]{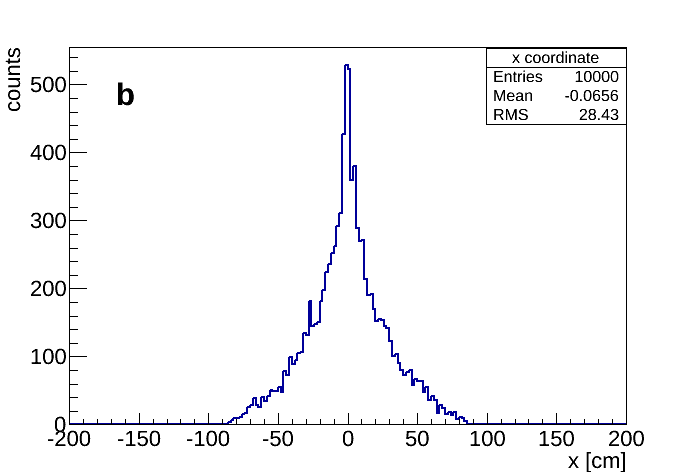}
\includegraphics[width=8cm]{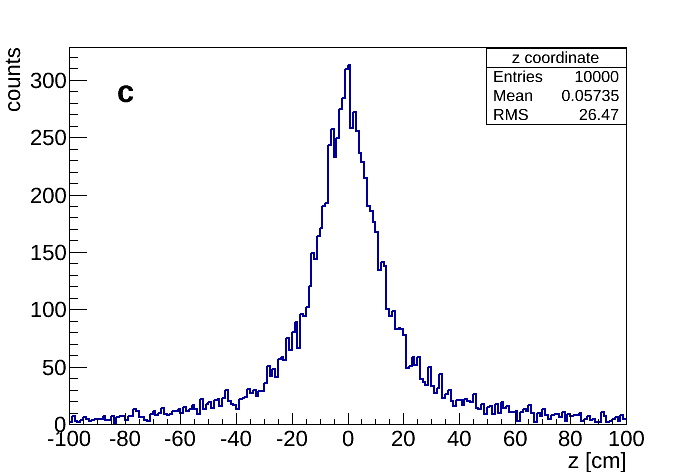}
\includegraphics[width=8cm]{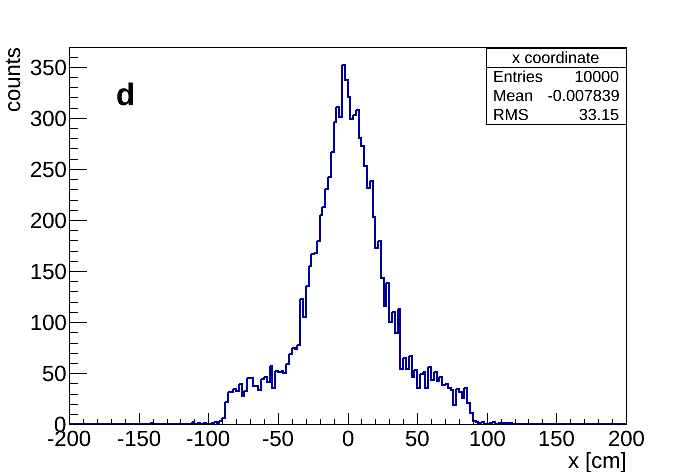}
\caption{Monte Carlo evaluation of the residuals from the correct position, both along the beam direction $z$ and along the vertical coordinate $x$, for electromagnetic showers of about 200~MeV (a,b) and cosmic rays (c,d). In plot (d) the residual distribution is affected by the detector boundaries along the $x$ coordinate.}
\label{MC_result}
\end{figure}

%%%%%%%%%%%%%%%%%%%%%%%%%%%%%%%%%%%%%%%%%%%%%%%%%%%%%%%%%%%%%%%%%%%%%%%%%%%%%%
\subsection{New Electronics, DAQ and Trigger}
\label{ica_new_ele}

\subsubsection*{Electronics}

The \icarus electronics was designed starting from an analogue low noise “warm” front-end amplifier followed by a multiplexed (16 to 1) 10-bit AD converter and by a digital VME module that provides local storage, data compression, and trigger information. The overall architecture, based on VME standard, was appropriate for the experiment, taking into account that the T600 electronics design started in 1999. The first production was carried out in 2000 and tests, at surface on the first T600 module, were successfully performed in 2001.

The present architecture (essentially a waveform recorder followed by circular buffers switched by trigger logic) is still valid, however possible improvements are now conceivable, taking advantage of new more performing and compact electronic devices.

\begin{figure}[htbp]
\centering
\includegraphics[width=0.3\textwidth]{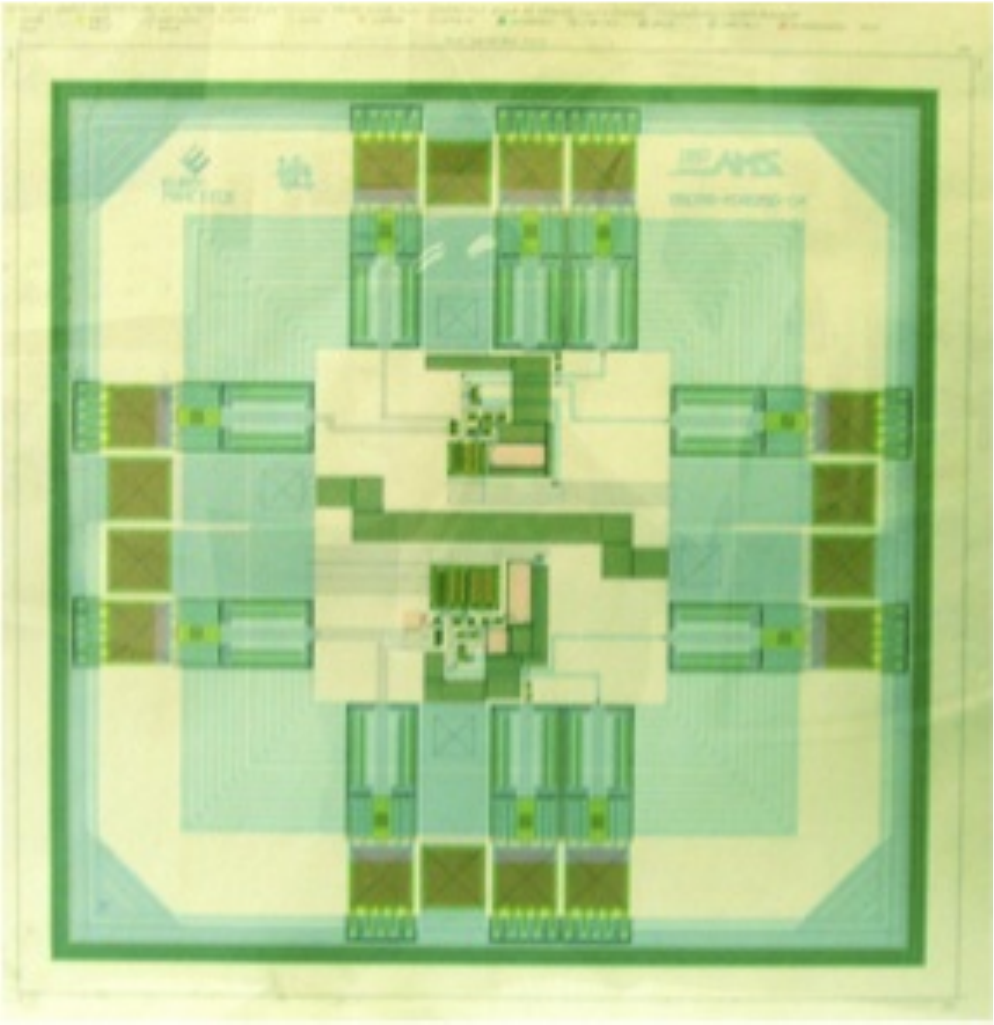}
\caption{BiCMOS dual channel custom analog pre-amplifier.}
\label{bicmos}
\end{figure}

\begin{figure}[htbp]
\centering
\includegraphics[width=0.5\textwidth]{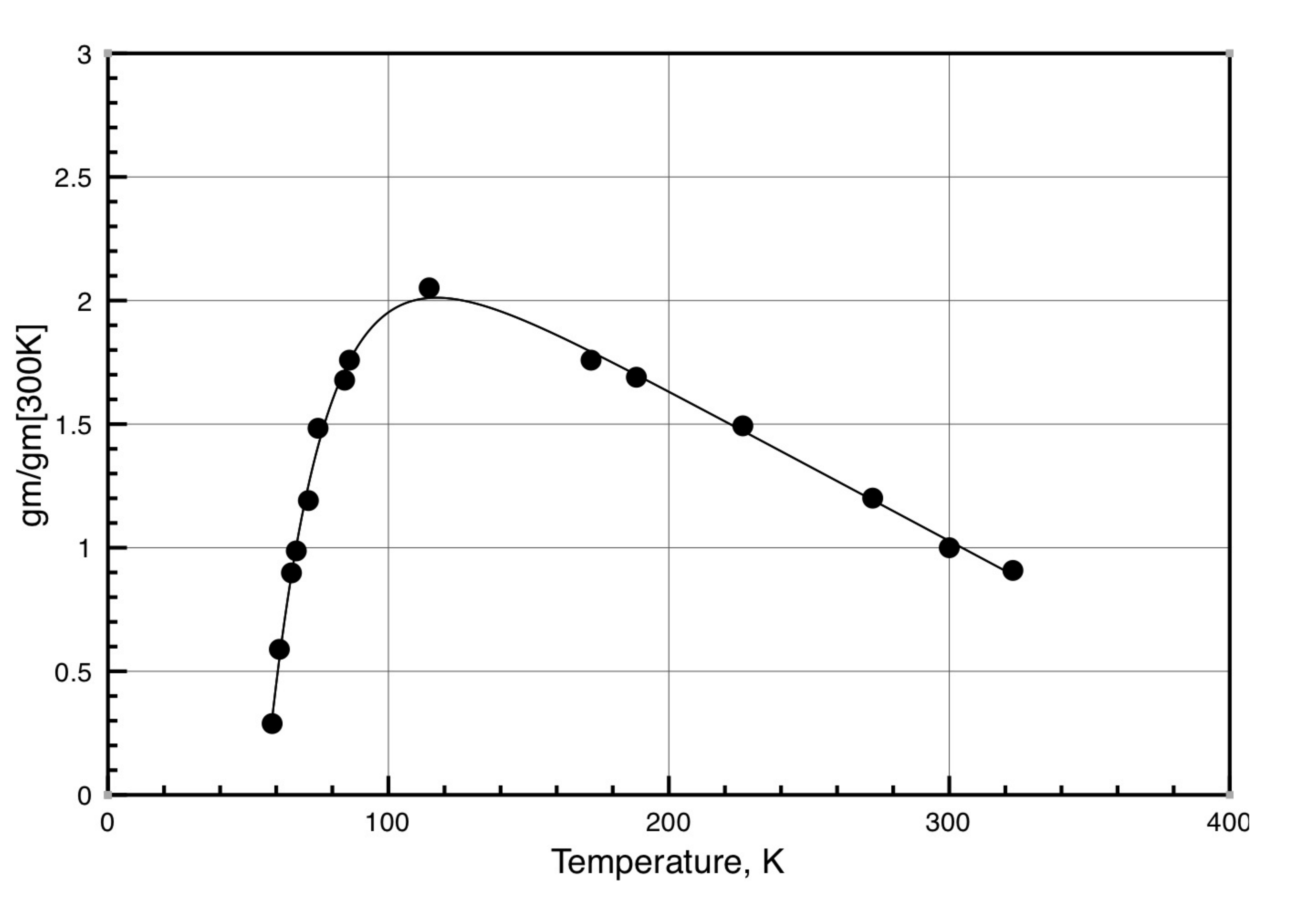}
\caption{Typical variation of the input stage transconductance $g_m$ with temperature for junction Fet.}
\label{transc} 
\end{figure}

The analogue front-end amplifier, used in the T600 LNGS configuration, is perfectly adequate: the only proposed change is the adoption of a smaller package for the BiCMOS (see Fig.~\ref{bicmos}) custom amplifier, dual channel, which is already available. The possibility  to have the front-end in LAr is also considered.

The amplifier serial input noise, \textit{e}, linearly increases with detector and cable capacitance, $C_d$, and decreases with input stage transconductance, $g_m$:

\begin{equation}
e^2 = \frac{C_d^2}{g_m} .
\end{equation}

Transconductance is 26\% higher at LAr temperature (86~K), see Fig.~\ref{transc}, and, together with the reduction of cable length, an improvement of S/N is expected with cold amplifiers. However, in the case of large mass \lartpcs, a detector lifetime in the order of tens of years is expected. In this period, it is natural to foresee improvement programs in the electronics, because of its constant evolution and progress. An architecture allowing for major and easy upgrading with an accessible electronics has been then chosen.

The gain of the front-end amplifier and filter is $\frac{1 V}{164 fC}$. The 10 bit ADC input range is 1 V, therefore the least count is equivalent to 1,000 electrons. This value matches with the amplifier noise of $\sim$ 2,000 electrons, given a detector capacitance of 450 pF (signal wires plus cables).

The T600 run at LNGS on the CNGS neutrino beam confirmed a S/N better than 10 for m.i.p. on about 53,000 channels. 
A relevant change, in the new electronics design, concerns the adoption of serial ADCs (one per channel) in place of the multiplexed ones used at LNGS. The main advantage is the synchronous sampling time (400 ns) of all channels of the whole detector, not to mention compactness and price.
A very reliable and cost effective new flange (CF200) has been developed for the T600 future operations, see Fig.~\ref{fig:ICA_Flange} Left. The internal structure of a via in this CF200 flange is shown in Fig.~\ref{inflange}. The external contacts, on both sides, not visible in Fig.~\ref{inflange}, are on a different plane respect to the via, and allow for SMD connectors use. The white squares are brass disks that reinforce the flange structure, in order to stand atmospheric pressure without deformation in case of use in vacuum vessels.

\begin{figure}[htbp]
\centering
\setlength{\fboxsep}{0pt}
\setlength{\fboxrule}{0.6pt}
\fbox{\includegraphics[width=0.5\textwidth]{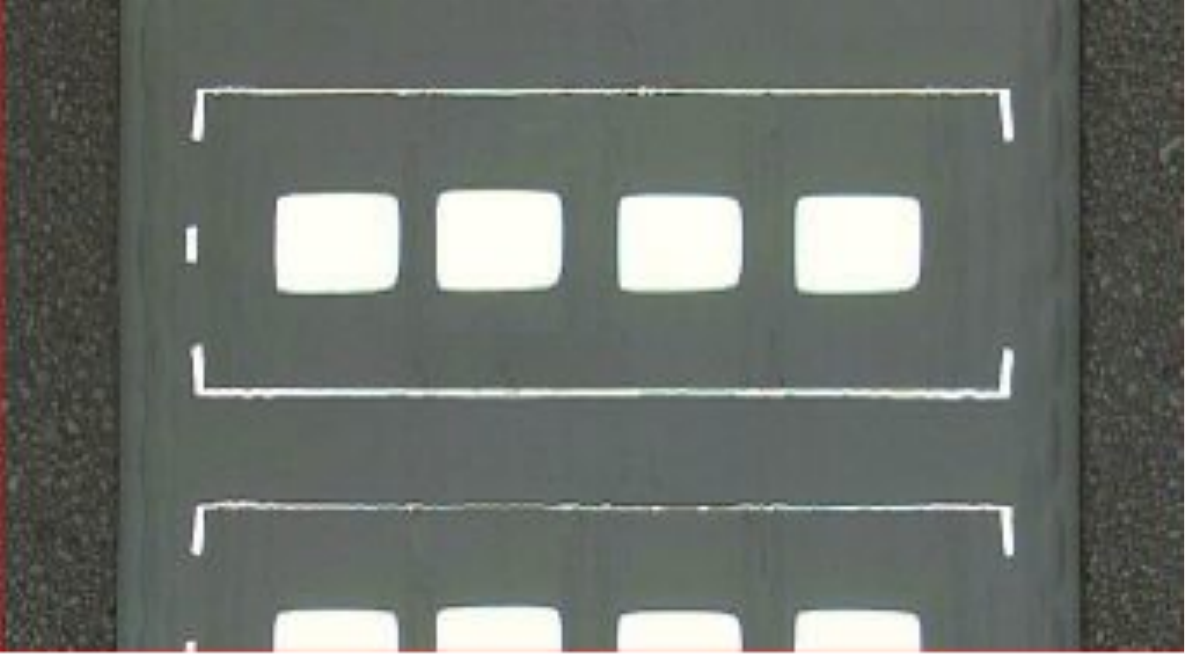}}
\caption{Section of signal flange (5 mm): the copper via contact (white lines in the photo) is fully embedded in solid G10.}
\label{inflange} 
\end{figure}

This flange allows the connection of 16 cables (512 channels), to exploit the external side of the flange as an electronics cards backplane in a special crate (Fig.~\ref{fig:ICA_Flange} Right). The connectors on the external side allow for direct insertion of electronics boards, where both analogue and digital electronics, with a compact design, are housed. 

In Fig.~\ref{fig:ICA_Board} the first working prototype board is shown. It serves 64 channels and uses serial optical links. The 8 boards of one flange may use the same serial optical link as it is shown in the block diagram of Fig.~\ref{boards}.
The digital part is fully contained in a high performance FPGA (Altera Cyclone V) that allows easy firmware upgrading. Behind the FPGA, sockets for front-end amplifiers are visible together with the direct insertion connectors that convey wire signals. In Fig.~\ref{preamps} the pre-amplifiers are shown. The board is “scored” so the amplifiers will be “snapped” in eight sets of eight pre-amplifiers.

Performance, in terms of throughput of the read-out system, has been improved replacing the VME (8 - 10 MB/s) and the sequential order single board access mode inherent to the shared bus architecture, with a modern switched I/O. Such I/O transaction can be carried over low cost optical Gigabit/s serial links. 

As mentioned before, the cold electronics option could be considered, provided a suitable design is found. The ICARUS collaboration does not have a ready-to-use cold pre-amplifier, even if in the past many tests, eventually abandoned, were carried out for this solution \cite{coldelect}. If such cold pre-amplifier is found, one should adapt a pre-amplifier board inside the cold vessel, close to the wire support, modulo 32 or 64, and substitute the piggy back 8-channel amplifier boards, described before, with suitable receivers, maintaining the digital part as it is and the 400~ns sampling time.

\begin{figure}[htbp]
\centering
\setlength{\fboxsep}{0pt}
\setlength{\fboxrule}{0.6pt}
\mbox{\fbox{\includegraphics[height=0.25\textheight]{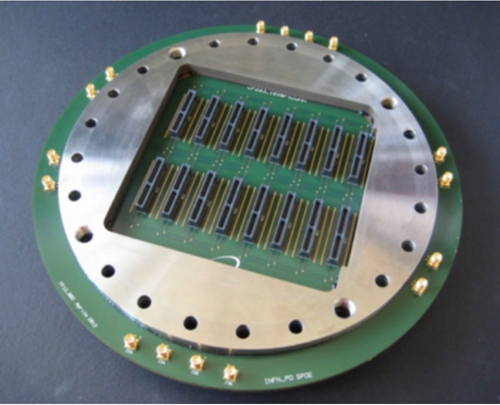}}\quad
\fbox{\includegraphics[height=0.25\textheight]{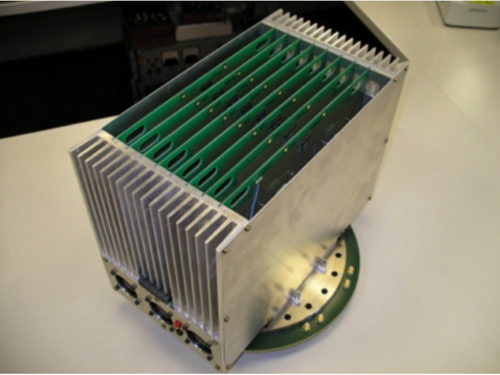}}}
\caption{\label{fig:ICA_Flange}Left: the new T600 flange without the cards cage. Right: the new T600 flange with the cards cage and 8 boards inserted without front panels.}
\end{figure}

\begin{figure}[htbp]
\centering
\setlength{\fboxsep}{0pt}
\setlength{\fboxrule}{0.6pt}
\fbox{\includegraphics[width=0.5\textwidth,trim=0mm 70mm 0mm 70mm]{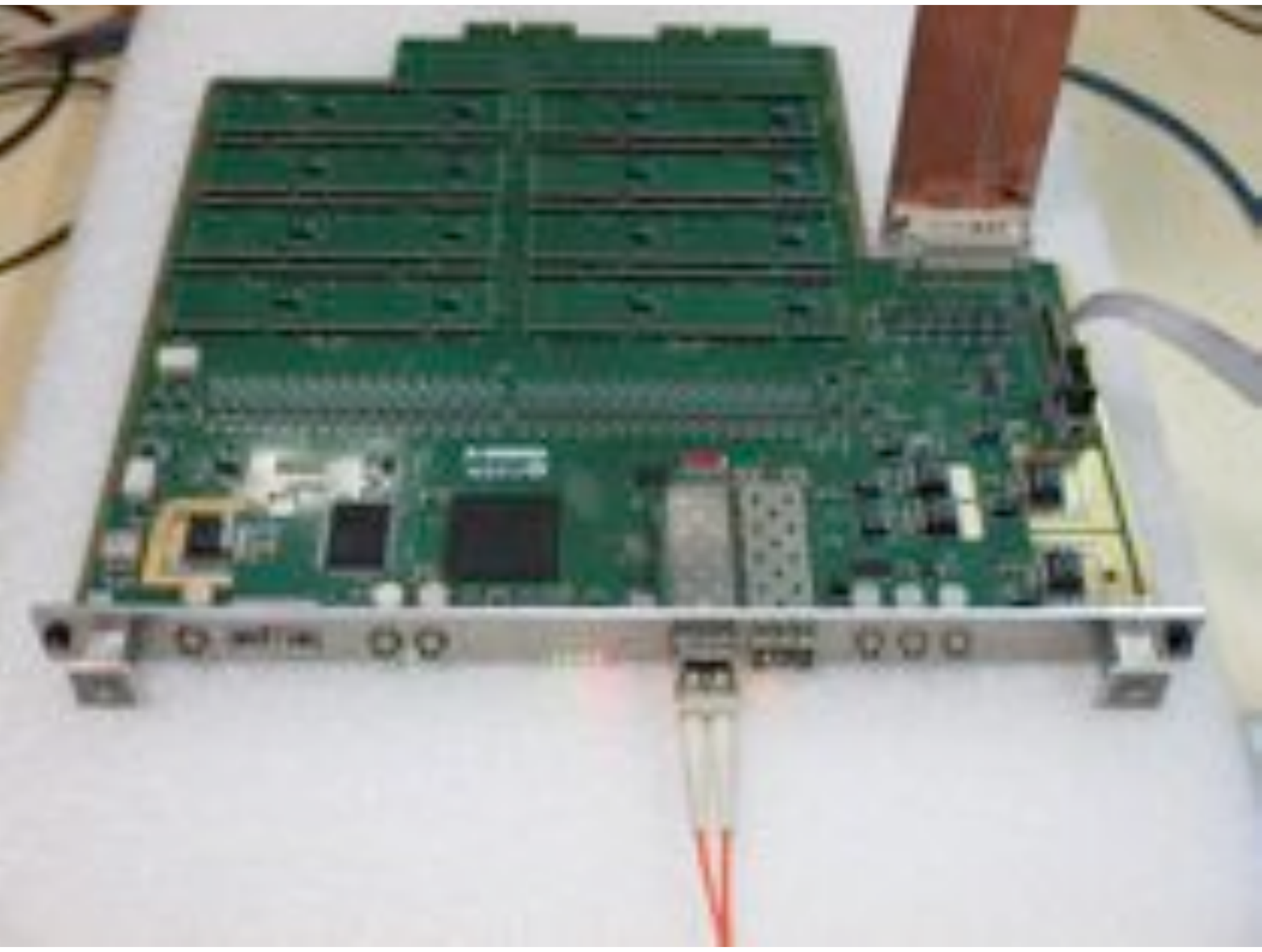}}
\caption{\label{fig:ICA_Board} First working new electronics prototype board.}
\end{figure}

\begin{figure}[htbp]
\centering
\includegraphics[width=0.5\textwidth]{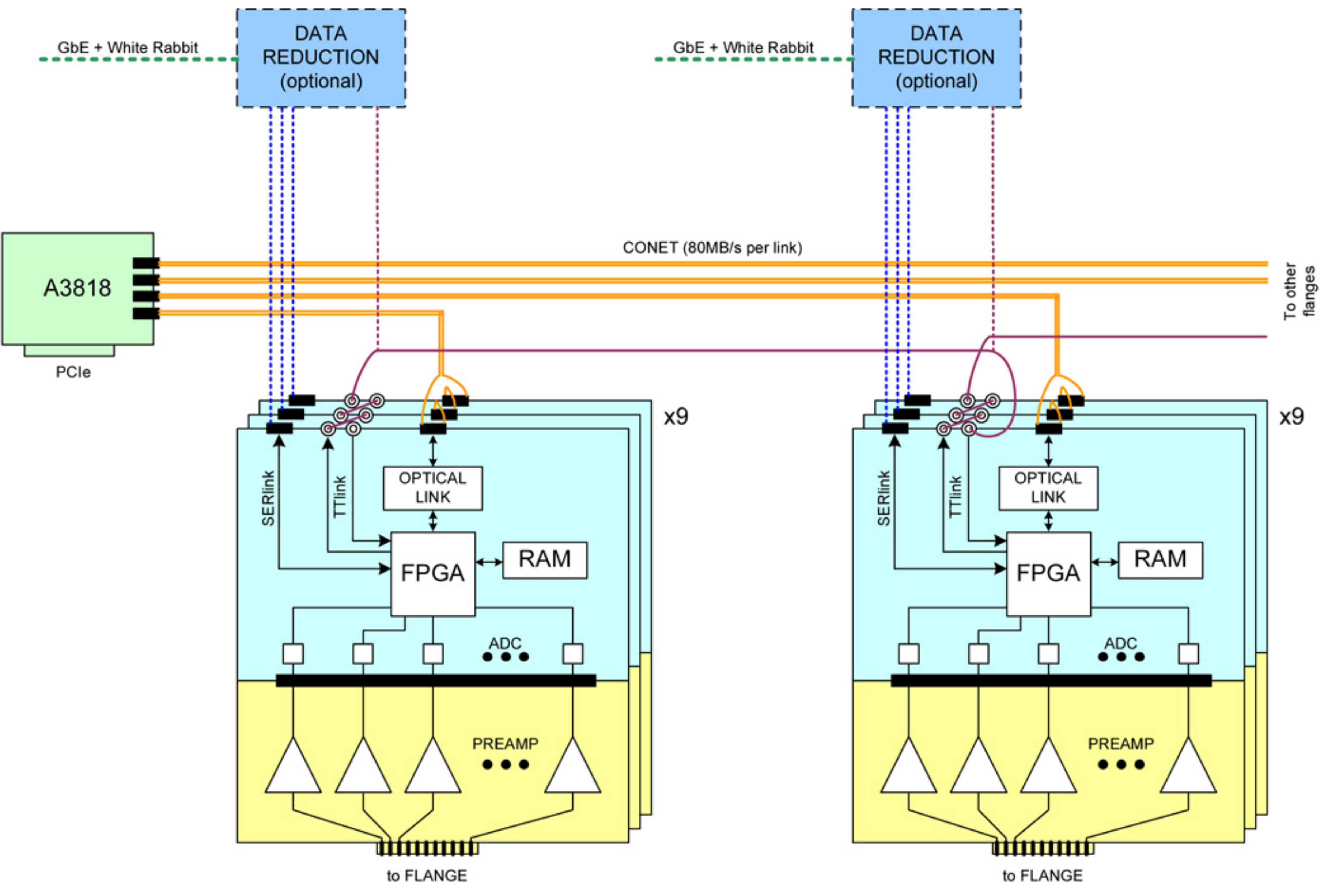}
\caption{\label{boards} Arrangement of the read-out boards on different flanges.}
\end{figure}

\begin{figure}[htbp]
\centering
\setlength{\fboxsep}{0pt}
\setlength{\fboxrule}{0.6pt}
\fbox{\includegraphics[width=0.5\textwidth]{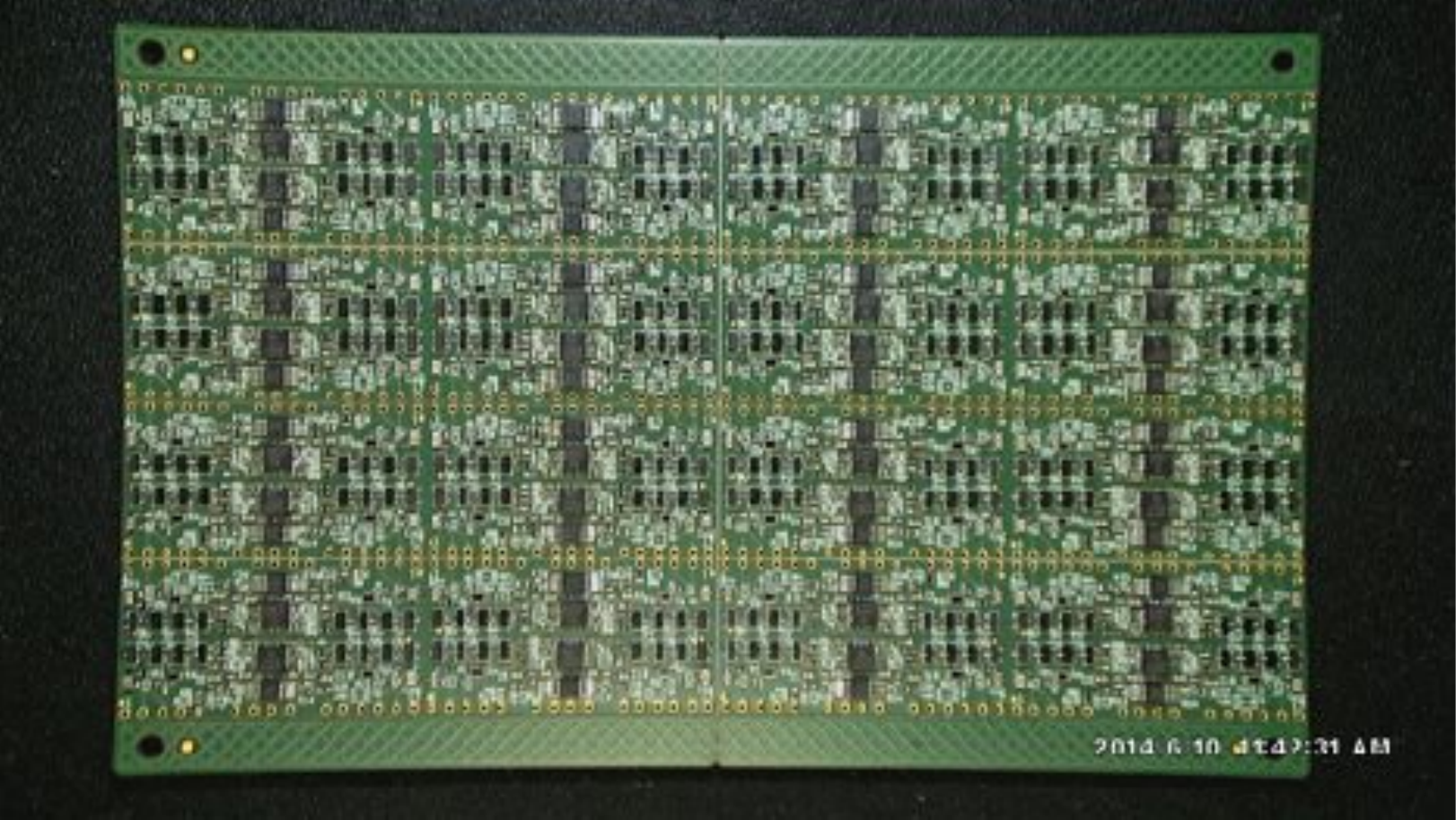}}
\caption{\label{preamps} 64 front-end amplifiers before “snapping” in 8 sets of 8 amps.}
\end{figure}

\subsubsection*{Trigger and DAQ}

The trigger system of the T600 detector will exploit the coincidence of the prompt signals from the scintillation light in the \lartpc, recorded by the PMT system, with the proton spill extraction of the BNB within a 1.6~$\mu$s gate. 

PMT digitized pulses are sent to a front end dedicated board to be processed by  FPGA modules, requiring a logic on multiple PMT signals for the generation of the trigger. Discrimination thresholds have to be set to guarantee the detection of all the event associated to each neutrino interaction with energy E$>$100~MeV.

The PMT trigger signal will be then sent to the T600 Trigger Manager, where it will be combined with the time information from the beam spill to initiate the readout of all the TPCs. A multi-buffer event recording will be adopted with a 3-level veto, as for the CNGS beam exploitation, able to give different priorities to different trigger sources, thus minimizing DAQ dead-time. The system, similar to the LNGS run one described in Sec.~\ref{ica_old_ele}, will consist of a Real Time (RT) controller and FPGA boards, communicating with the DAQ in handshake mode. The RT controller will monitor the number of available buffers in the digital boards, preventing the generation of new triggers in case they are full. The FPGA boards will implement time critical processes, like the opening of Booster Beam gate and the time stamp of each trigger. FPGA boards are also expected to record the trigger source and mask, to monitor the trigger rates and to control the overall system stability.
The T600 Trigger Manager will also allow combine in the trigger logic the signals coming from the new cosmic ray tagging system (see Sec.~\ref{ica_crts}).

At the nominal BNB intensity of 5~$\times$~10$^{12}$~pot/spill, $\sim$~1~neutrino interaction, either charged or neutral current, every 180 spills is expected to trigger the T600 detector at the far position with vertex in the \lartpcs (1 neutrino every 240 spills considering charged currents only). A slightly lower trigger rate, one every 210 spills, will come from beam-associated events; the dominant trigger source, 1 over 55 spills, is expected from cosmic rays. Globally, about 1 event every 10 s is foreseen in the T600 \lartpc at the standard 4 Hz repetition rate of the Booster Neutrino Beamline. 
This $\sim$~0.1~Hz trigger rate is well within the 50 MB/s DAQ throughput already realized for the CNGS data taking at LNGS. Actually, the significant improvement in the readout throughput achievable with the new serial optical links will ensure an even better performance. Therefore, the present DAQ is largely adequate to operate without dead-time not just with the standard 4 Hz repetition rate of the Booster Beam, but also up to the $\sim$15 Hz maximum repetition rate.

Provided that a precise beam extraction signal is available, the trigger and DAQ system is also well suited to the exploitation of the bunched beam structure, accounting for the 1~GHz sampling of the PMT waveforms and the precise spatial reconstruction of the neutrino interaction vertex in the TPCs.
The excellent performance in event timing achievable with the \icarus detector has been proven by the precision measurement of the neutrino velocity on the CNGS beam~\cite{ica_nuvel1, ica_nuvel2}. This result relied mainly on:

\begin{itemize}
\item the waveform of the extracted proton beam time-structure signal, recorded at CERN with a 1~GHz sampling triggered by the kicker magnet signal;
\item an absolute GPS-based timing signal, distributed by LNGS laboratory to the \icarus detector via a $\sim$8~km optical fiber, synchronized with the CERN absolute timing within few ns;
\item the waveform of the PMT trigger signal recorded with a 1~GHz sampling;
\item the evaluation, with $\sim$1~ns accuracy, of the time corrections corresponding to the distance of the event from the closest PMT and the position of the interaction vertex along the $\sim$18~m of the detector length. Note that the time corrections also include the contribution of the PMT transit time, different for each device.
\end{itemize}

A similar strategy could be adopted at FNAL as well, if the waveform of the fine bunched structure of the Booster beam will be provided with $\sim$1~ns resolution. As a further simplification, a precise matching of the neutrino interaction in the T600 active volume with the corresponding bunch could be obtained without need for an absolute timing, if the beam extraction signal will be delivered directly to the T600 detector exploiting the recent successfully developed White Rabbit timing protocol~\cite{whiterabbit}.

%%%%%%%%%%%%%%%%%%%%%%%%%%%%%%%%%%%%%%%%%%%%%%%%%%%%%%%%%%%%%%%%%%%%%%%%%%%%%%
\subsection{New Cryogenic and Purification systems}
\label{ica_new_cryo}

The SBN program provides a first opportunity for the CERN and FNAL engineering groups to collaborate on the design of \lartpc infrastructure. Once established, this collaboration could have a significant impact on designs for other short and mid-term projects leading to a long-baseline neutrino facility. While most of the components of the T600 detector will be reused after overhauling at CERN, the LN$_2$ delivery system is expected to be replaced and a completely new cryostat and cryogenic layout will need to be developed. The following describes the cryostat and cryogenic needs for the Far Detector.

New cryostats will host the refurbished T600 detector.
LAr will be contained in two mechanically independent vessels, of about 270~m$^3$ each. According to the past experience, to efficiently outgas the internal surfaces and obtain an appropriate LAr purity, the cold vessels must be evacuated to less than $10^{-3}$~mbar. Therefore the vessels need to be tight to better than $10^{-5}$ mbar $l$ s$^{-1}$.
%(0.45 bar relief valve settings in addition to 0.55 bar hydrostatic pressure).
The new T600 vessels will be parallelepipedal in shape with internal dimensions 3.6 (w) $\times$ 3.9 (h) $\times$ 19.6 (l) m$^3$. Aluminum welded extruded profiles (see Fig.~\ref{fig:AluProfiles}) will be employed, designed in collaboration with industries and Milano Politecnico (Italy): they are requested to be super clean, vacuum-tight and to stand a 1.5~bar maximal operating internal overpressure. Executive design for both the profiles and welding (mounting) procedures has already been procured: further details are shown in Fig.~\ref{fig:frontdoor}, Fig.~\ref{fig:exploded} at the end of the section.

\begin{figure}[htbp]
\centering
\includegraphics[scale=0.6]{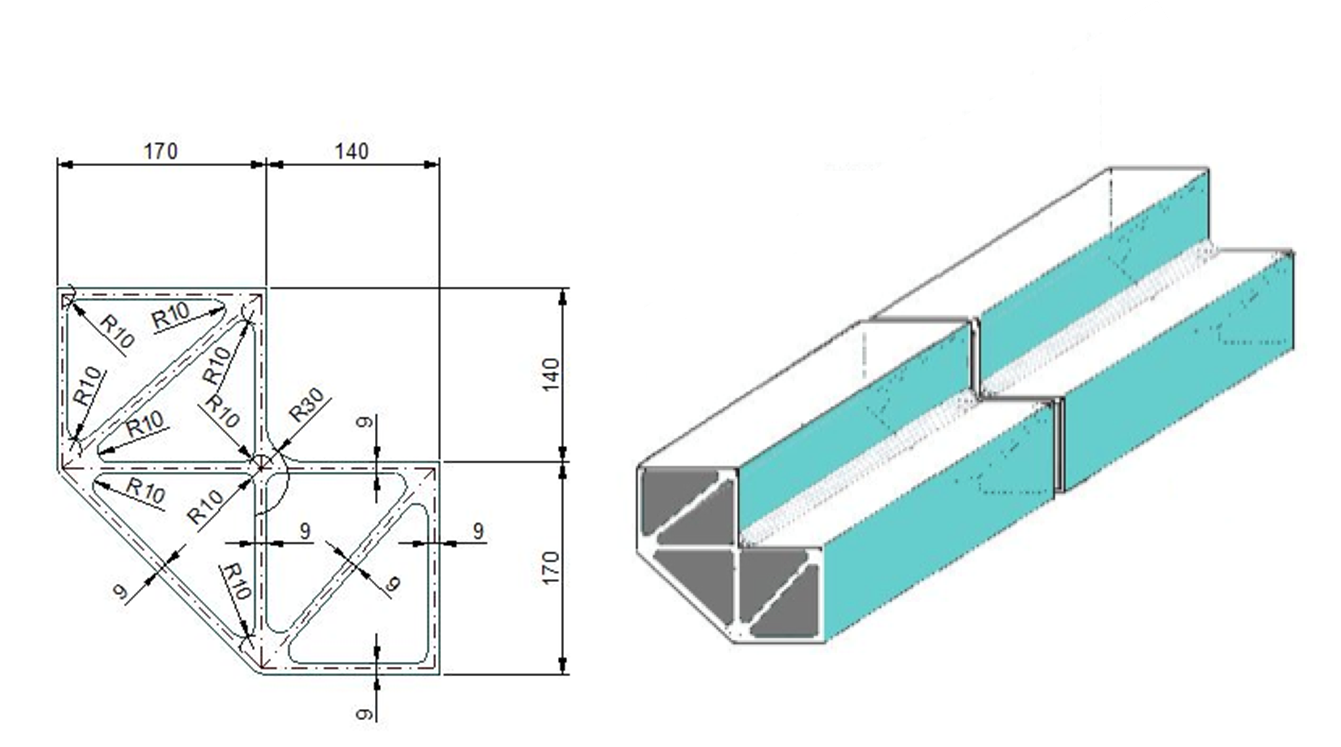}
\begin{center}
\caption{Detail of an aluminum extruded profile corner, 2D and 3D. More drawings are shown at the end of the section.}
\label{fig:AluProfiles}
\end{center}
\end{figure}

Such new solution represents a significant simplification with respect to the aluminum honeycomb used in the LNGS run, whereas it implies a slight increase in the cryostat weight, 30 t each. Use of aluminum LAr vessels is also particularly attractive as it offers very good shielding against external electronic noises, and it provides large thermal conductivity that improves the temperature uniformity inside the LAr.
As in the LNGS run, walls are double-layered and can be evacuated, leading to efficient leak detection and repair.

The cold vessels will be enclosed inside a common heat exchanger (thermal shield) in which two-phase (gas+liquid) nitrogen is circulated. As in the past run, a mass ratio less than 5:1 will be kept between the liquid and the gas phases, which ensures temperature uniformity all along the shield.
%This shield is essential in minimizing the heat load to the bulk of the sensitive LAr volume to: (1) suppress any risk of LAr boiling that could affect the operation of the HV system; (2) establish very good temperature uniformity ($\Delta$T $<$ 1 K) through LAr de-stratification, as required to have uniform electron drift velocity and LAr purity; (3) reduce internal temperature gradients during the cool down, making it faster. %The critical period during which the argon purifier is switched-off is kept to a minimum in order to preserve an acceptable initial LAr purity.

A purely passive polyurethane foam is chosen for insulation, based on the “membrane tanks” technology. This technique has been developed for 50 years and is widely used for large industrial storage vessels and ships for liquefied natural gas \cite{gttsite,ihisite}.
The solution has been adapted to the ICARUS design by the GTT firm, and it is similar to the one to be used for the membrane cryostat of the Near Detector. In Fig.~\ref{fig:insulation} and Fig.~\ref{fig:insulTempGrad} details of the insulation elements and expected thermal gradients respectively are shown, as an example. In Tab.~\ref{tab:insulFlux} the thermal flux through the various elements is listed.
An insulation thickness of 600~mm will be used for the bottom and lateral sides; for the top-side a maximum thickness of about 400~mm will be used. With this configuration, the expected average thermal losses will be of around 10~W/m$^2$, resulting in a heat loss through the insulation of $\sim 6.6$~kW.
All the external heat contributions (cables, pumps, transfer lines, etc.) can be accounted for a value not exceeding 5.4~kW, leading to a total heat load of about 12~kW.

\begin{figure}[htbp]
\centering
\includegraphics[scale=0.7]{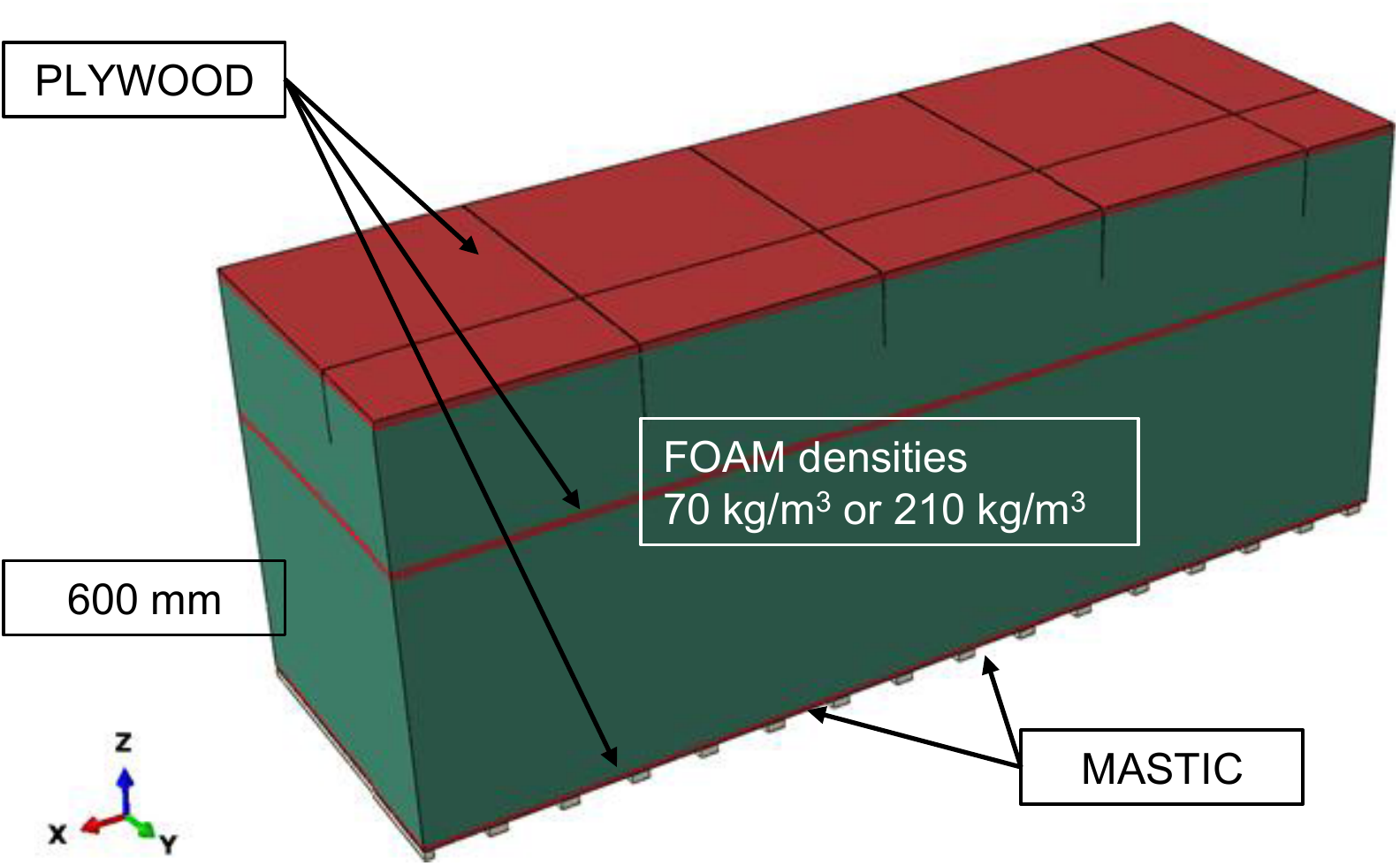}
\begin{center}
\caption{3D model of the newly-proposed T600 insulation. 600 mm element displayed}
\label{fig:insulation}
\end{center}
\end{figure}

\begin{figure}[htbp]
\centering
\includegraphics[width=0.8\textwidth]{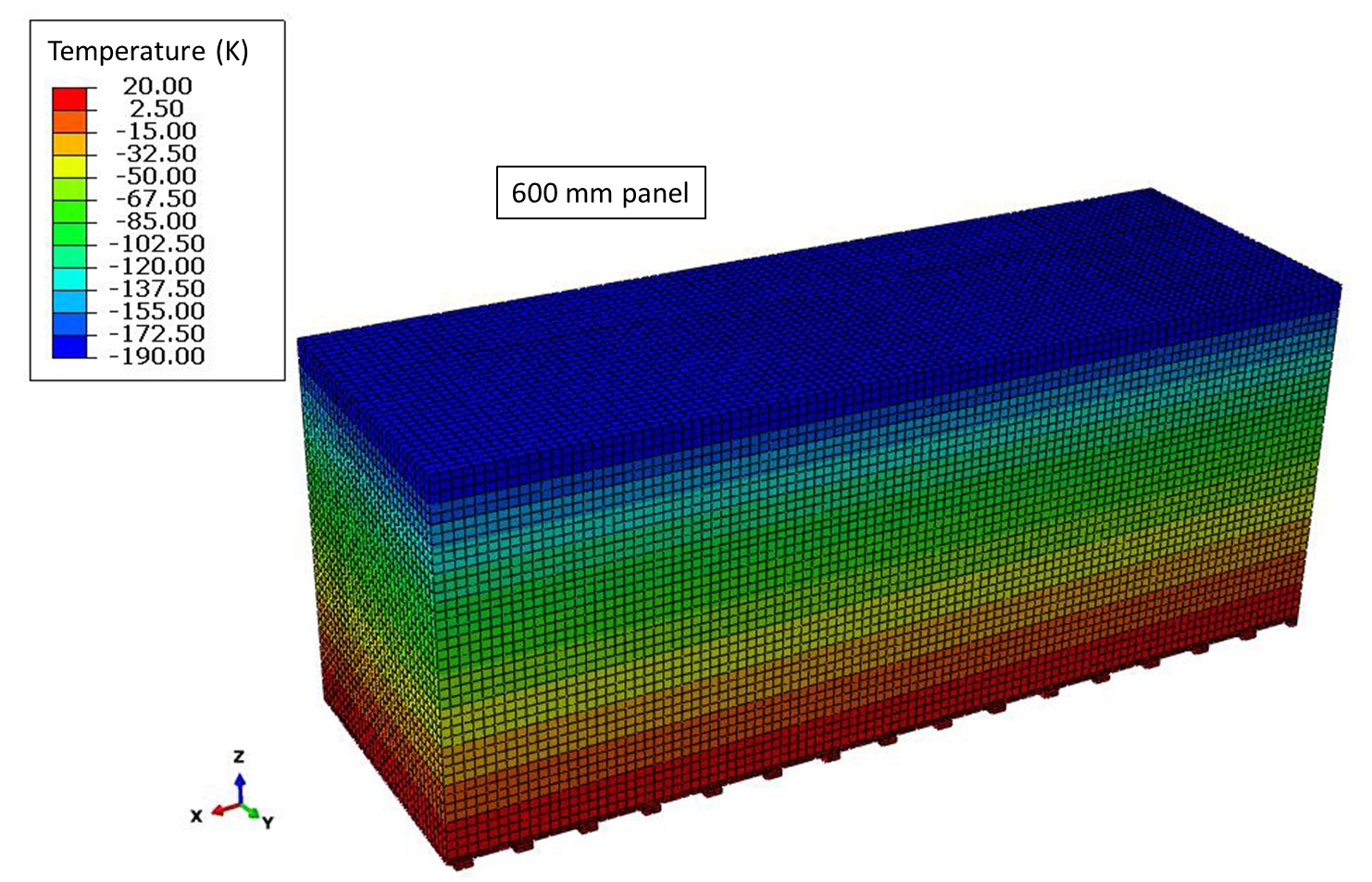}
\begin{center}
\caption{Finite elements study of the thermal gradient in the 600 mm thick element. Note that on this picture ambient temperature is on the lower side of the element} %ref to GTT doc needed?
\label{fig:insulTempGrad}
\end{center}
\end{figure}

\begin{table}[htbp]
\begin{center}
\begin{tabular}{|c|c|c|c|c|}
\hline
\hline
Panel thickness (mm) & Foam density (kg/m$^3$) & Total heat (W) & Surface (m$^2$) & Thermal flux (W/m$^2$) \\
\hline
\hline
400 & 70  &  7.800 & 0.750 & 10.40 \\
\hline
400 & 210 & 13.530 & 0.750 & 18.04 \\
\hline
600 & 70  &  5.200 & 0.743 &  7.00 \\
\hline
600 & 210 &  9.026 & 0.743 & 12.15 \\
\hline
\hline
\end{tabular}
\end{center}
\caption{Thermal flux through the insulation elements, as a function of thickness and foam density, from GTT study. The foam density will be usually of 70~kg/m$^3$, while amounting to 210~kg/m$^3$ in correspondence to the feet of the detector.}
\label{tab:insulFlux}
\end{table}

The Far Detector renovated cryogenic design is being developed, wherever possible, with a focus on commonalities with the Near Detector one, to be used across both experiments and also as a stepping stone for LBNF collaborative efforts. With this idea in mind, this system is expected to be modular in design, able to be enlarged for future projects, and portable, i.e. constructed on skids that can be tested separately, prior to delivery to FNAL for installation.

The original scheme of \icarus cryogenics and LAr purification systems will be preserved, and most of the present plant, INFN property, will be reused. In the same way, the requirements will remain the same, with respect to the past, as described in Sec.~\ref{ica_old_cryo}. 

One main difference, with respect to the previous LNGS run, will be in the logistics, due to the different location of the detector (at shallow depths). %Apart from that, the requirements on the plant will remain the same.
If further updates were needed, they would be carried on by the hosting laboratories, following the specification given by the ICARUS Collaboration.
On the other hand, during the different stages of the program (overhauling and, later, commissioning and data taking) it will be responsibility of the hosting laboratory to conceive and take care of the necessary maintenance of circuitry and control systems. The same goes for what concerns the plants-related logistics.

At Fermilab, the cooling circuit will be operated in open loop: the Stirling re-liquification will not be used.  
Fig.~\ref{fig:LArSystem} shows once again the existing cryogenic system on the \icarus detector, this time highlighting the Stirling re-liquefaction system that will be discarded with the implementation of the open-loop LN$_2$ delivery system.

Further discussions on specific technical aspects of the cryogenic system are underway, mainly regarding the purification system (filters), best re-condensation strategy, ullage conditions. However, given the already discussed very successful operation of the existing plant (see Sec.~\ref{ica_old_cryo}), the ICARUS Collaboration intends to maintain the choices made in its previous experience, with the exceptions described above, and carry them on to the coming SBN program at FNAL.

The envisioned schedule for the development of the ND/FD cryogenic systems is related to the request to have the Near and Far Detectors ready for commissioning in fall 2017, with data taking starting in April 2018.%: it is reported below in Tab.~\ref{tab:CryoSchedule}. 
LAr Cryogenics groups are being formed both at CERN and FNAL, and are set to collaborate to meet the goals.

Fig.~\ref{fig:LArSystem} shows once again the existing cryogenic system on the \icarus detector, this time highlighting the Stirling re-liquefaction system that will be discarded with the implementation of the open-loop LN$_2$ delivery system.

\begin{comment}
Removed since redundant with Part 4 (PJW 12/15/14)
\begin{figure}[htbp]
\centering
\includegraphics[height=0.6\textwidth, trim=0mm 0mm 0mm 0mm, clip]{T600_LN2_Rev_4.pdf}
%\includegraphics[scale=1.0]{LN2.png}
\begin{center}
\caption{Schematic diagram of the new design proposed for the LN$_2$ delivery system for the \icarus detector (open loop circuit).}
\label{fig:newLN2}
\end{center}
\end{figure}
\end{comment}

\begin{figure}[htbp]
\centering
\includegraphics[scale=0.8]{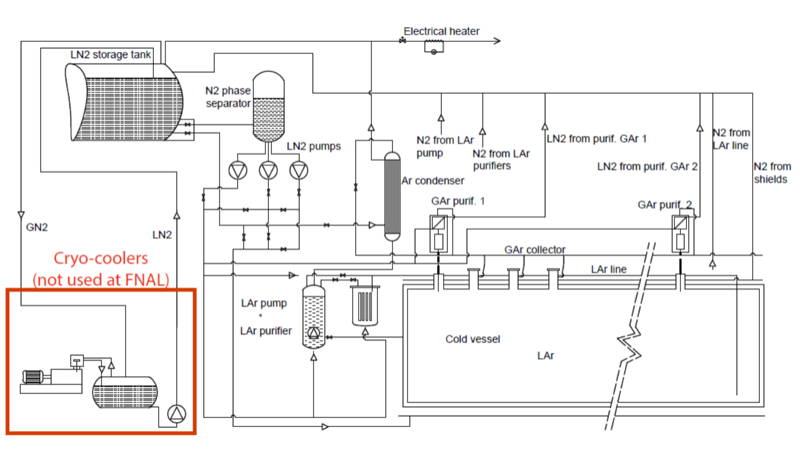}
\begin{center}
\caption{Schematic diagram of the existing T600 cryogenic system, which will be maintained during the FNAL operations as well. The highlighted section in the lower left represents the old Stirling re-liquefaction units, which will be replaced by a new open-loop LN$_2$ delivery system.
%, shown in Fig.~\ref{fig:newLN2} and described in text.
}
\label{fig:LArSystem}
\end{center}
\end{figure}

The \icarus detector is expected to be delivered to FNAL in the first half of 2017, with about 6 months needed for installation. Commissioning can take place during the second half of 2017. It will require from 3 to 5 months, based to the experience gained at LNGS. In Gran Sasso 5 months were needed, including about 3 of vacuum pumping.
The consumption of LN$_2$ and LAr during this commissioning phase can be estimated, based on the fact that the total expected heat loss through the new insulation, as mentioned above, is of the order of 10~kW, and that larger cold power consumption during the first cooling down can be assumed. Approximately 100,000 liters of LN$_2$ should be needed for cooling down, along with 273,000 liters of LAr per module of the T600. Assuming present pricing for the cryogenic liquids, Tab.~\ref{tab:commCosts} can be constructed, summarizing the estimated costs of LN$_2$ and LAr for the commissioning phase.

\begin{table}[htbp]
\begin{center}
\begin{tabular}{|l|c|c|c|}
\hline
\hline
Material & Quantity (l) & Price(US\$)/liter & Total cost (US\$) \\
\hline
\hline
Liquid nitrogen & 100,000 & 0.07 & 7,000 \\
\hline
Liquid argon (per module) & 273,000 & 1.5 & 409,500 \\
\hline
\textbf{Total$^*$} & & & \textbf{826,000} \\
\hline
\hline
\end{tabular}
\end{center}
\caption{Summary of costs of the commissioning phase, for what concerns usage of LN$_2$ and LAr during cooling down. $^*$Note that Liquid argon cost is detailed for one module of the T600, while the grand total accounts for both modules ($2 \times 409,500 \$ $). }
\label{tab:commCosts}
\end{table}

%Further discussions on specific technical aspects of the cryogenic system are underway, mainly regarding the purification system (filters), best re-condensation strategy, ullage conditions. However, given the already discussed very successful operation of the existing plant (see Sec.~\ref{ica_old_cryo}), the ICARUS Collaboration intends to maintain the choices made in its previous experience, with the exceptions described above, and carry them on to the coming SBN program at FNAL.

%In general, the cryogenic system of the T600 detector, repeatedly pre-tested against different types of emergencies, performed very well during operations in difficult conditions (deep underground location), and it allowed obtaining unprecedented results on argon purification. This result largely justifies the decision of carrying the plant and its design on to the next stage of the detector life within the Fermilab SBN program.

Before closing this section, further drawings of the new cold vessels are reported, referring as an example to one of the endcaps of the vessel, exploded (Fig.~\ref{fig:frontdoor}), and to the whole exploded view (Fig.~\ref{fig:exploded}), respectively. 

\begin{figure}[htbp]
\centering
\includegraphics[scale=0.4]{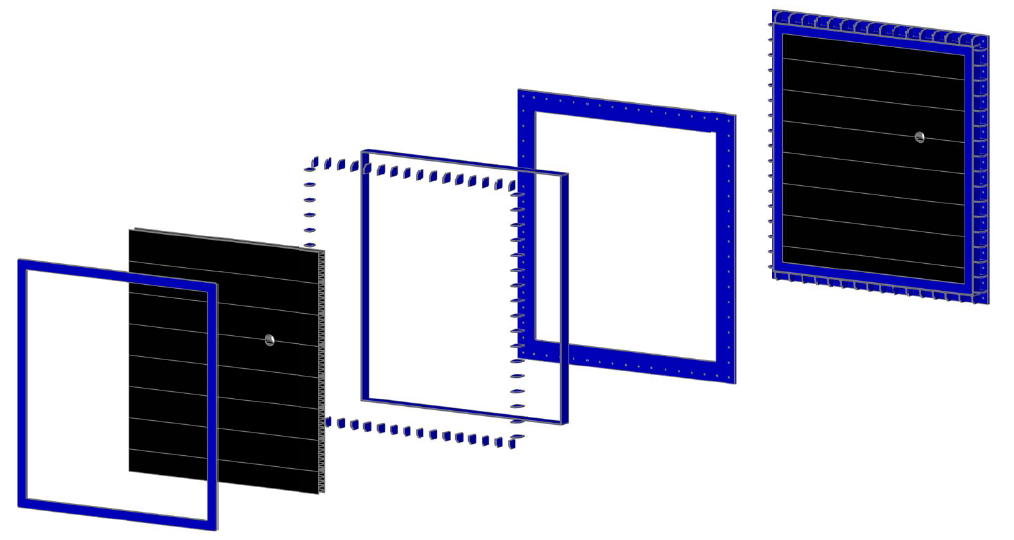}
\begin{center}
\caption{Exploded detail of one of the endcaps of the new aluminum vessel. The hole corresponding to the LAr extraction line (for liquid recirculation) is visible.}
\label{fig:frontdoor}
\end{center}
\end{figure}

\begin{figure}[htbp]
\centering
\includegraphics[scale=0.4]{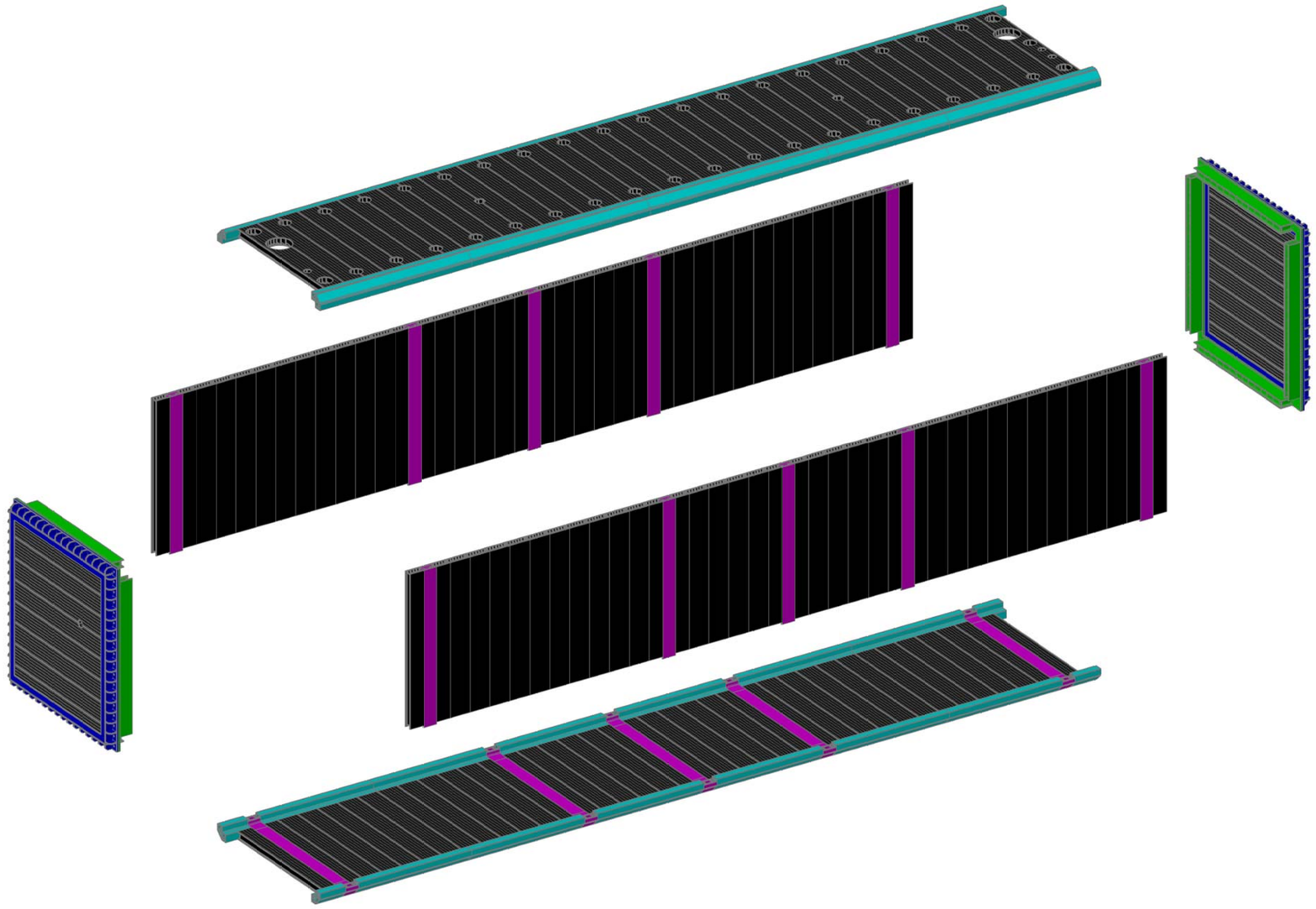}
\begin{center}
\caption{Exploded view of the whole cold vessel.}
\label{fig:exploded}
\end{center}
\end{figure}

%%%%%%%%%%%%%%%%%%%%%%%%%%%%%%%%%%%%%%%%%%%%%%%%%%%%%%%%%%%%%%%%%%%%%%%%%%%%%%%%%
%%%%%%%%%%%%%%%%%%%%%%%%%%%%%%%%%%%%%%%%%%%%%%%%%%%%%%%%%%%%%%%%%%%%%%%%%%%%%%%%%
\section{Cosmic Ray Tagging System}
\label{ica_crts}

As already mentioned in Sec.~\ref{ica_requirements}, very effective new methods must be introduced to reduce the cosmic ray related signals~\cite{arxiv_rubbia}. For example, a segmented, fast anti-coincidence with 4$\pi$ coverage detector (Cosmic Ray Tagging System, CRTS), may record each charged particle crossing the outer boundaries of the LAr containers. 
At the nominal BNB intensity of 5~$\times$~10$^{12}$~pot/spill, only 1 neutrino CC interaction every  240 spills is expected to trigger the T600, with vertex in the \lartpcs. This rate has to be compared with the expected cosmic rays rate of 1 every 55 beam spills. The CRTS detector could be used to deplete by a significant factor the spurious cosmic ray induced interactions, by tagging events in the beam spill without any crossing cosmic ray. According to the expected time resolution of the PMT detection system and of the CRTS, a tagging window $<$100 ns will reduce spurious coincidences generated by CRTS.

The positions and the timings of all random muon tracks crossing the walls of the CRTS during the T600 imaging window will be recorded. Each muon track reconstructed in the TPC may be then correctly determined by associating the charge image with the corresponding absolute drift time t$_0$ coming both from the CRTS and from the internal light collection system, matching the track geometry with the CRTS recorded positions.

This would be achieved by means of a system which provides signals, independently from the \lartpc and the light collection system, that indicate the passage of charged particles through the surface of the LAr sensitive volume. These signals would be used as anti-coincidence to identify and recognize the interactions generated by external particles.

\subsection{CRTS efficiency}

The performance of the CRTS system will depend both on the intrinsic efficiency of each CRTS detector unit and on the coverage of the adopted layout, which must approximate the ideal limit of a complete 4$\pi$ solid angle.

According to the overall size of the T600 cryostat, a shell made of particle detectors with a large surface area (order of 1,000~m$^2$) will be required. The CRTS spatial granularity has to allow the unambiguous association between the reconstructed tracks in the TPC and the position where the cosmic rays cross the CRTS. 

Since the TPC mixes drift time with space coordinates, a relation between the absolute time $t_{true}$ and position $y_{true}$ on the CRTS along the drift coordinate $y$ is determined, as:

\begin{equation}
y_{true} = y_{img} + v_{drift} \cdot (t_{true} - t_0) ,
\label{ytrue}
\end{equation}

\noindent
where $y_{img}$ is the position on CRTS extrapolated by the recorded image on TPC, v$_{drift}$ is the drift velocity and $t_0$ is the absolute trigger time opening the acquisition window. Among all the possible pairs ($t$, $y$)$_{CRTS}$ only the one satisfying Eq.~\ref{ytrue} is considered to correctly tag the muon track crossing the CRTS. The other CRTS coordinates $x$ and $z$ are instead determined unambiguously.

The absolute time $t_0$ is expected to be obtained from the PMT system, with resolution of the order of 1 ns, in order to exploit the bunch structure of the beam. Therefore, the time resolution requested for the detectors comprising the CRTS must be at least of the same order of magnitude. As a consequence, the uncertainty in localizing the track in space, due to the mentioned time resolution, is not less than 30 cm.

These considerations must be taken into account, when choosing the detector technology to be employed for the CRTS and its geometrical segmentation, as discussed in the next section.

\subsection{CRTS layout}

There are few well consolidated technologies to realize large-area detectors with high space and time resolution. Among these, the following may be taken into account: Resistive Plate Chambers (RPC) and plastic scintillator slabs coupled to PMTs or SiPMs. The RPCs allow building large panels (up to 3~m$^2$ of area) that can be easily arranged in large walls, but their implementation involves the use of gas mixtures and HV operation which may pose special safety issues. 

A very promising detector technology, based on the use of LAr readout plates, has been proposed~\cite{arxiv_rubbia}. These plates, deployed directly inside the cryostats few centimeters away from its edges, could detect the presence of the dE/dx signals generated by the cosmic particles in the LAr with a relatively modest electric field ($\sim$ 1~kV/cm) added between the readout plates and the cryostat grounded walls. The performance of these detectors in terms of efficiency, rates, stability and noise must be carefully investigated.

The choice of the adopted detector solution must also take into account the size and shape of the T600 detector and its mechanical structure. As already discussed, the CRTS should completely surround the T600 volume, and for each side it is necessary to evaluate the most suitable technical solution, either internal or external.

The design of the top side of CRTS, which is interested by the largest amount of incoming cosmic particles flow, is certainly the most critical. Due to the presence of many dead spaces inside the TPCs, including flange feed-throughs, signal wire and HV cables, the deployment of such a detector inside the T600 will limit its effective geometrical coverage. It should be rather positioned at a suitable distance (about 3~m) from the T600 upper floor, conveniently above the readout electronics racks and the GAr recirculation units. 

Similarly, vertical CRTS side walls could be more easily placed outside, positioned close to the cryostat walls, since the internal space behind the wire chambers is occupied by the PMTs, the slow control sensors and the TPC mechanical supports.

On the contrary, the bottom side CRTS detectors should be more conveniently placed into the LAr cryostat, due to the presence of the external mechanical supporting structure placed below it.

The presence of the electronics racks inside the CRTS envelope could prevent adopting a completely hermetic structure, to allow the necessary access to the TPC upper floor for maintenance and inspection purposes. For this reason, the CRTS top plane has to be kept separated from the CRTS vertical walls surrounding the T600,
and the cooling system of the electronics must be carefully designed taking into account a reduced heat dissipation by natural convection. 

For what concerns the future steps for a CRTS design finalization, first of all a detailed Monte Carlo simulations have to be carried out to study how to disentangle the genuine neutrino events from cosmogenic-induced backgrounds. The combined action of the CRTS and of the internal PMT system has to be studied in detail to define a possible analysis strategy for the neutrino event selection and reconstruction.

The probability of an autoveto signal by charged particles generated in the
neutrino interactions and escaping the \lartpcs has to be carefully
evaluated. According to results of a preliminary evaluation, considering the CTRS system installed inside the cryostats and surrounding
the LAr volume, $\sim$ 45\% of $\nu_\mu$CC and $\sim$ 10\% of $\nu_e$
events are expected to provide a signal. Alternatively $\sim$ 10\% of
$\nu_\mu$ and few percent of $\nu_e$ events are expected to give a signal in an external
muon tagging system.

The T600 CRTS must be realized in a common framework with the \larnd CRTS, being mandatory for the two detectors to adopt the same design of the cosmic ray detection efficiency with identical sensitivities and systematics.

\FloatBarrier

 \clearpage
\pagestyle{empty}

\proposalTitle

\setcounter{part}{3}
\part{Infrastructure and Civil Construction}

%\begin{center}
%{\large
%DRAFT \\
%\bigskip
%\today
%\proposaldate
%}\end{center}

\ifcombine
\clearpage
\else
\clearpage
\tableofcontents
\addtocontents{toc}{\protect\thispagestyle{empty}}
\clearpage
\setcounter{page}{0}
\fi

\pagestyle{fancy}
\rhead{IV-\thepage}
\lhead{SBN Infrastructure and Civil Construction}
\cfoot{}

%%%%%%%%%%%%%%%%%%%%%%%%%%%%%%%%%%%%%%%%%%%%%%%%%%%%%%%%%%%%%%%%%%%%%%%%%%%%%%%%%%

\section{Introduction}

The Short Baseline Neutrino program is proposed to include three Liquid Argon Time Projection Chamber detectors (\lartpcs) located on-axis in the Booster Neutrino Beam (BNB) as shown in Figure~\ref{fig:SBN_overview}.  The near detector (\larnd) will be located in a new building directly downstream of the existing \SB enclosure 110~m from the BNB target as also shown in Figure~\ref{fig:buildings}~(right).  The \uboone detector, which is currently in the final stages of installation, is located in the Liquid Argon Test Facility (LArTF) at 470~m.  The far detector (the existing \icarus) will be located in a new building, 600~m from the target between \MB and the \nova near detector surface building as shown in Figure~\ref{fig:buildings}~(left).

The following sections address the new infrastructure required to support these detectors: 
\begin{itemize}
\item cryostats for the near and far detectors,
\item cryogenic systems for the near and far detectors, 
\item buildings for the near and far detectors, and 
\item common computing and software systems.  
\end{itemize}
The infrastructure required for the the \uboone detector is not described here since the detector installation will have been completed by early 2015. However, development of common computing and software systems for the SBN program can benefit significantly by the participation of \uboone in the development and experience from the use of these tools on \uboone data.

\begin{figure}[htbp]
\centering
\setlength{\fboxsep}{0pt}%
\setlength{\fboxrule}{0.5pt}%
\fbox{\includegraphics[width=1\textwidth,trim=20mm 0mm 0mm 0mm,clip]{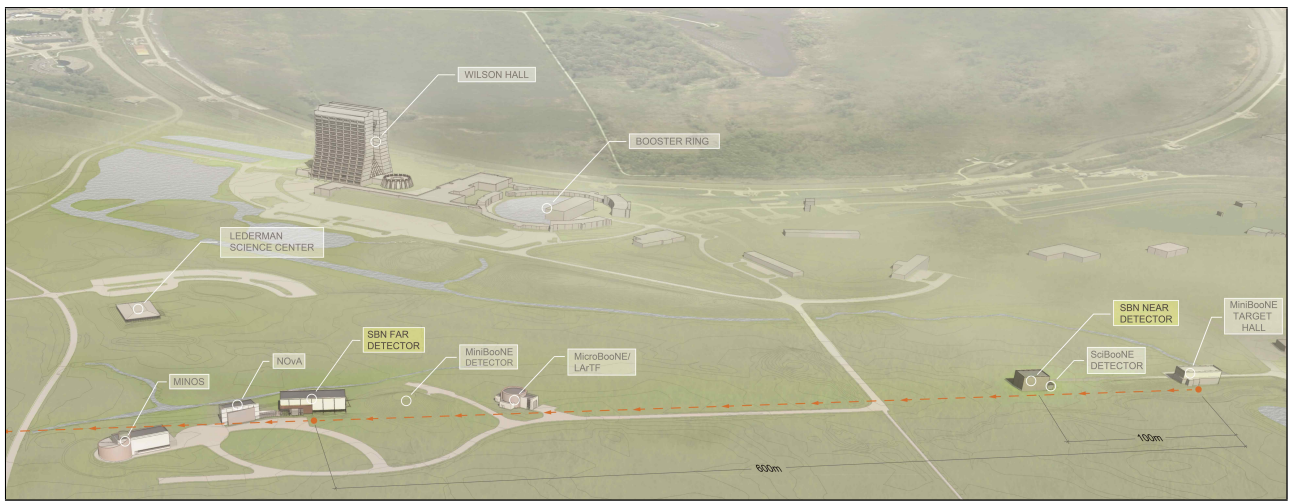}}
\caption{Diagram of the Fermilab neutrino beamline area (looking east) showing the axis of the BNB (red dashed line) and approximate locations of the SBN detectors at 110~m, 470~m, and 600~m. }
\label{fig:SBN_overview}
\end{figure}

\begin{figure}[htbp]
\centering
\setlength{\fboxsep}{0pt}
\setlength{\fboxrule}{0.6pt}
\mbox{\fbox{\includegraphics[height=0.23\textwidth, trim=0mm 30mm 30mm 30mm, clip]{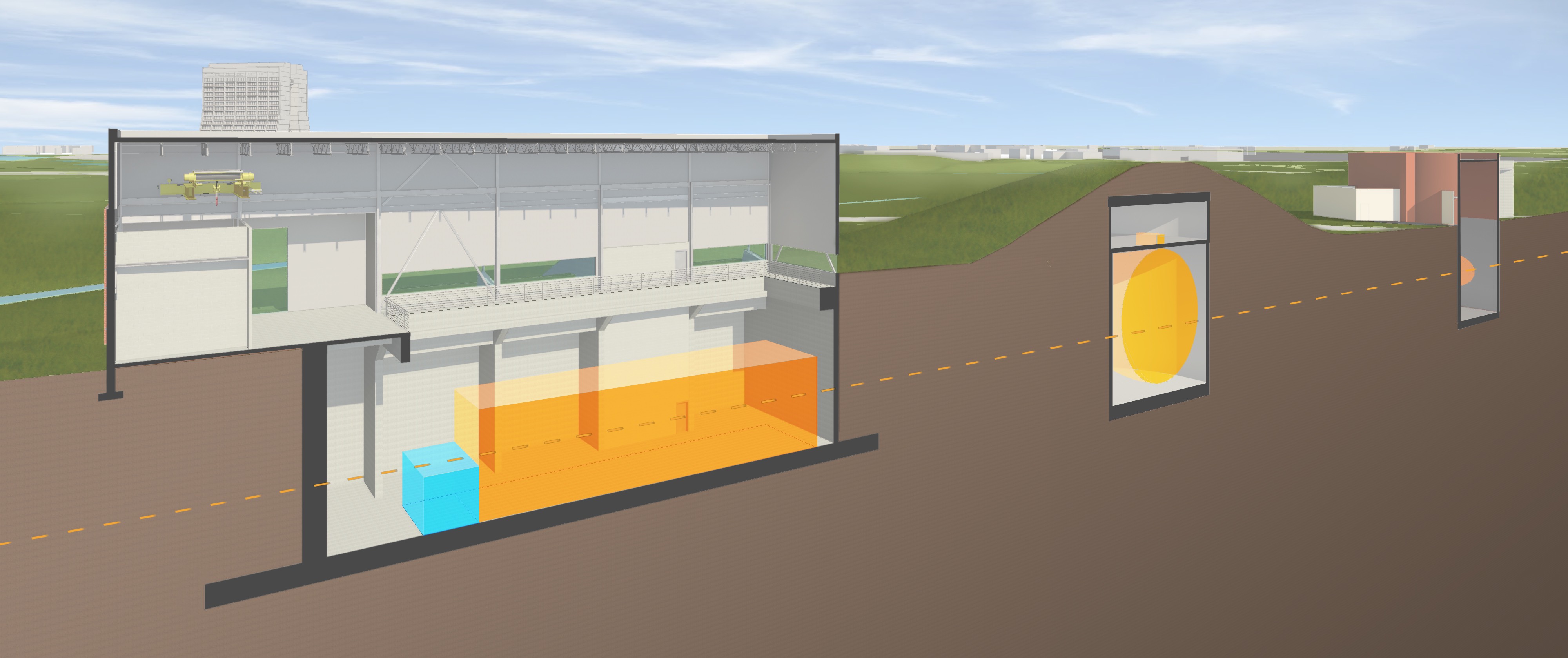}}
\fbox{\includegraphics[height=0.23\textwidth, trim=30mm 20mm 30mm 20mm, clip]{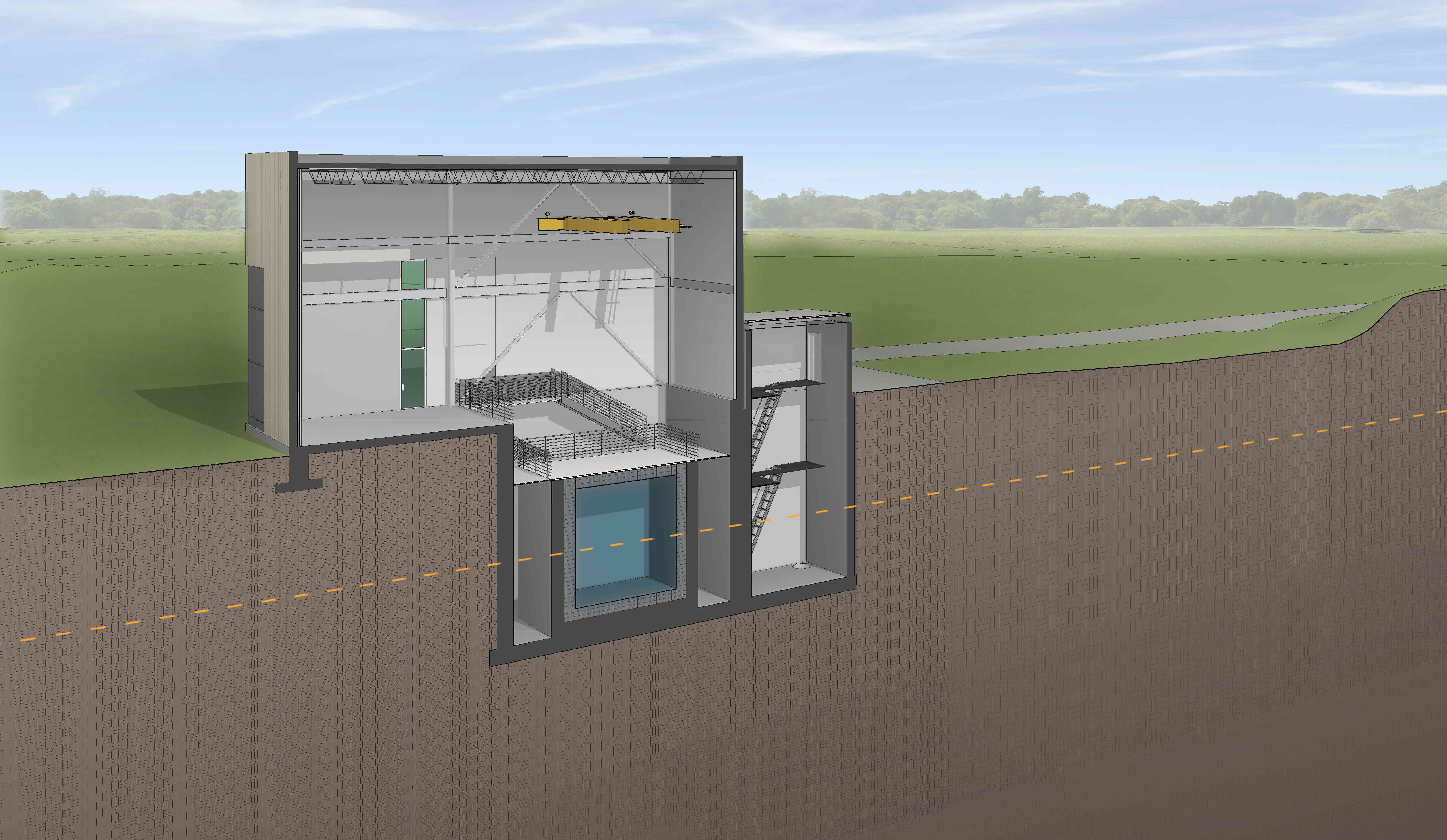}}}
  	\caption{Cross-sectional views of a design concept for the far detector building (left) and near detector building (right).  In the left view the existing enclosures for the MiniBooNE and MicroBooNE detectors are also seen.}
\label{fig:buildings}
\end{figure}

The detectors of the SBN program will need external detectors for tagging/vetoing cosmic ray muons as explained in Part~1 of this proposal.  As described in Parts~2~and~3 the conceptual designs for cosmic taggers the near and far detectors are being developed.  While these concepts were developed independently and described separately, it is likely that the realization of these systems will be managed as a joint project with a common underlying design.  The Collaborations have made provisions in their funding requests for the necessary resources.  

The SBN program provides an excellent opportunity for a collaborative effort on the design of \lartpc infrastructure between the recently formed LAr Cryogenic engineering groups at CERN and Fermilab along with engineering resources within INFN.  These teams are also collaborating on developments for other short and mid-term projects leading to a long-baseline neutrino facility. The following two sections describe the cryostat and cryogenic needs for the near and far detectors.     

\FloatBarrier

\section{Cryostats}

\subsection{Near Detector Cryostat\label{sec:ND_cryostat}}

The near detector will use a membrane tank technology to contain the base design of 220 tons of LAr equivalent to about 158 m$^3$. The design is based on a scaled up version of the LBNE 35 Ton Prototype. The cryostat will be housed in a dedicated building next to the existing SciBooNE hall where the cryogenic system components will be located. The two buildings will be connected with an underground tunnel spanning about 9 feet. The cryostat will use a steel outer supporting structure with a metal liner inside to isolate the insulation volume. An alternative that was considered was a concrete supporting structure with vapor barrier and heating elements embedded in the concrete to control the temperature.

The scope of the Near Detector cryostat subsystem includes the design, procurement, fabrication, testing, delivery and oversight of a cryostat to contain the liquid argon and the TPC. This section describes a reference design, whose scope encompasses the following components:

\begin{itemize}
\item steel outer supporting structure,
\item main body of the membrane cryostat (sides and floor),
\item top cap of the membrane cryostat. 
\end{itemize}

A membrane cryostat design commonly used for liquefied natural gas (LNG) storage and transportation will be used. In this vessel a stainless steel membrane contains the liquid cryogen. The pressure loading of the liquid cryogen is transmitted through rigid foam insulation to the surrounding outer support structure, which provides external support. The membrane is corrugated to provide strain relief resulting from temperature related expansion and contraction. The vessel is completed with a top cap that uses the same technology.

Two membrane cryostat vendors are known: GTT (Gaztransport \& Technigaz) from France and IHI (Ishikawajima-Harima Heavy Industries) from Japan. Each one is technically capable of delivering a membrane cryostat that meets the design requirements for the Near Detector. To provide clarity, only one vendor is represented in this document, GTT; this is for informational purposes only. Figure~\ref{fig:insulation} shows a 3D model of the GTT membrane and insulation design.

\begin{figure}[htbp]
\centering
\includegraphics[scale=0.5]{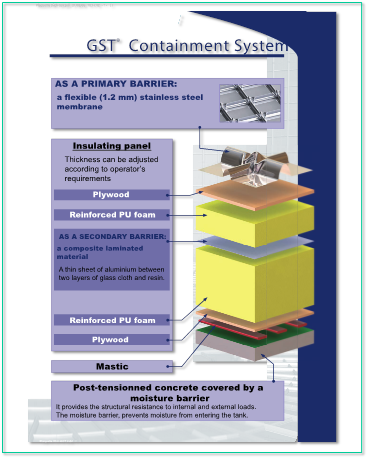}
\begin{center}
\caption{Exploded view of the membrane cryostat technology.}
\label{fig:insulation}
\end{center}
\end{figure}

\subsubsection*{Cryostat Design}

The conceptual reference design for the Near Detector cryostat is a rectangular vessel measuring 6.38 m in length (parallel to the beam direction), 5.17 m in width, and 4.80 m in height; containing a total mass of 220 tons of liquid argon. Figure~\ref{fig:LAr1-ND_158m3} shows a 3D view of the Near Detector cryostat with a “neck” and two main plates constituting the top: plate A and plate B. Two cold penetrations are located on plate A; all the other penetrations are located on plate B. To minimize the contamination from warm surfaces, the liquid argon level touches the membrane underneath top plate A. The gas is all contained in the neck region underneath plate B.  An alternative design is being considered with a single removable plate for the top.  The TPC could be directly hung from underneath this top plate.
In this case the gas ullage will all be contained over the liquid argon bath.

\begin{figure}[ht]
  \begin{center}	\includegraphics[width=1.0\textwidth, trim=0mm 0mm 0mm -20mm, clip] {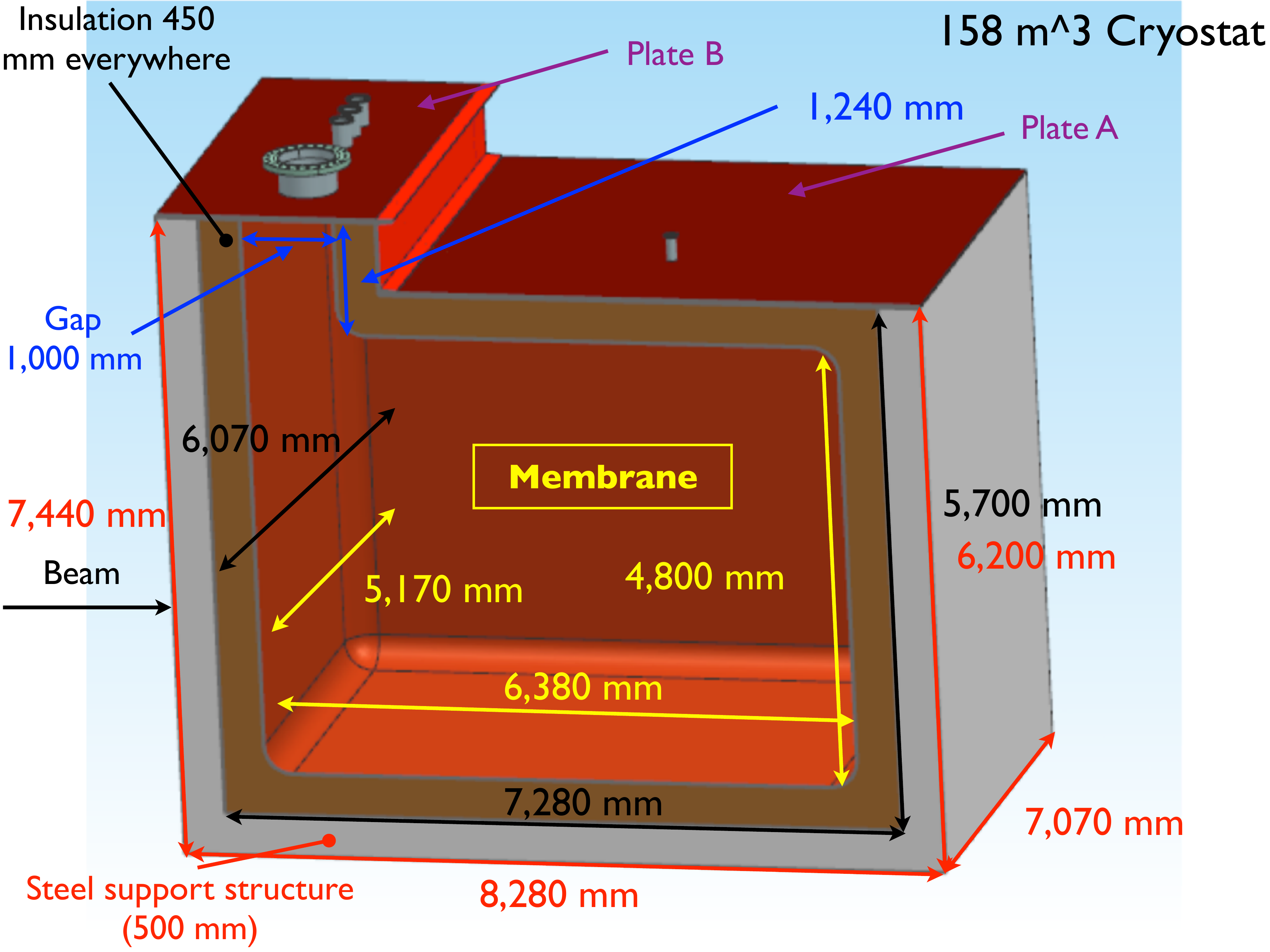}
  	\caption{A 3D view of the \larnd cryostat using membrane technology.}
  	\label{fig:LAr1-ND_158m3}
  \end{center}
\end{figure}

\subsubsection*{Design parameters}

%The requirements for the cryostat are listed in Section xxx (from Near Detector chapter when document is completed). 
This design is meant also to test technical solutions that may be of interest for the Long Baseline Neutrino program as well. The use of a cold ullage ($<100$~K) to lower the impurities in the gas region, and of a LAr pump outside the cryostat to minimize the effect of noise, vibration and microphonics to the TPC inside the LAr are Value Engineering studies for the Long Baseline program performed in synergy with the LBNF cryostat and cryogenics team.

The design parameters for the Near Detector cryostat are listed in Table~\ref{tab:near_cryostat}.

\begin{table}[htbp]
\centering
\begin{tabular}{|l|c|}
\hline
\multicolumn{1}{|c|}{Design Parameter}	& Value  \\ 
\hline \hline
Total Cryostat volume 				&	166 m$^3$  		\\ \hline
Total LAr volume					&	158 m$^3$  		\\ \hline
Liquid argon total mass				&	220,000 kg		\\ \hline
Inner dimensions of the cryostat	& 6.38 m (L) x 5.17 m (W) x 4.80 m (H)
														\\ \hline
Depth of liquid argon				& 4.80 m (5\% ullage all in the neck
region)													\\ \hline
Insulation							& 0.45 m Polyurethane foam
														\\ \hline
Primary membrane 					& 1.2 mm thick SS 304L corrugated stainless steel											\\ \hline
									& 0.07 mm thick aluminum between fiberglass cloth. \\ 	
Secondary barrier system			&  Overall thickness is ~1 mm located\\ 
									& between insulation layers.			\\ \hline
Outer support structure				& Steel enclosure with metal liner to isolate \\ 
									& the outside from the insulation space	\\ \hline
Liquid argon temperature			& $88\pm 1 ^{\circ}$K	\\ \hline
Operating gas pressure				& Positive pressure. Nominally 70 mbar ($\sim$1psig)											\\ \hline
Vacuum								& No vacuum			\\ \hline
Design pressure						& 350 mbar ($\sim$5~psig) + LAr head \\ \hline
Design temperature					& $77^{\circ}$K (liquid nitrogen temperature for flexibility)								\\ \hline
Temperature of all surfaces in 		&    \\
the ullage during operation			& $100^{\circ}$K	\\ \hline
Minimize noise/vibration/microphonics & 				\\ 
inside cryostat						& LAr pump preferably outside the cryostat													\\ \hline
Leak tightness						& $1 \times 10^{-6}$ mbar~$\ell$~s$^{-1}$		\\ \hline
Heat leak							& $< 15$\ W/m$^2$		\\ \hline
Lifetime							& 10 years (5 years of run + 5 years of potential upgrade)											\\ \hline
Thermal cycles						& 20 complete cycles (cool down and total warm up)												\\ \hline

\end{tabular}
\caption{Design parameters for the Near Detector cryostat.}
\label{tab:near_cryostat}
\end{table}

\subsubsection*{Insulation system and secondary membrane}

The membrane cryostat requires insulation applied to all internal surfaces of the outer support structure and roof in order to control the heat ingress and hence required refrigeration heat load. Choosing a reasonable, maximum insulation thickness of 0.45 m and given an average thermal conductivity coefficient for the insulation material of 0.0283 W/(m$\cdot$K), the heat input from the surrounding steel is expected to be about 3 kW total. It assumes that plates A and B are both foam insulated. This is shown in Table~\ref{tab:heat_load}. The overall heat leak is then about 13 W/m$^2$.

The insulation material is a solid reinforced polyurethane foam manufactured as composite panels. The panels get laid out in a grid with 3 cm gaps between them (that will be filled with fiberglass) and fixed onto anchor bolts anchored to the support structure. The composite panels contain the two layers of insulation with the secondary barrier in between. After positioning adjacent composite panels and filling the 3 cm gap, the secondary membrane is spliced together by epoxying an additional overlapping layer of secondary membrane over the joint. All seams are covered so that the secondary membrane is a continuous liner.

The secondary membrane is comprised of a thin aluminum sheet and fiberglass cloth. The fiberglass-aluminum-fiberglass composite is very durable and flexible with an overall thickness of about 1 mm. The secondary membrane is placed within the insulation space. It surrounds the bottom and sides. In the unlikely event of an internal leak from the primary membrane of the cryostat into the insulation space, it will prevent the liquid cryogen from migrating all the way through to the steel support structure where it would degrade the insulation thermal performance and could possibly cause excessive thermal stress in the support structure. The liquid cryogen, in case of leakage through the inner (primary) membrane will escape to the insulation volume, which is purged with GAr at the rate of one volume exchange per day.

\begin{table}[htbp]
\centering
\begin{tabular}{|l|c|c|c|c|}
\hline
\multicolumn{1}{|c|}{Element}	
		& Area (m$^2$)	& k (W/m$\cdot$K)	& $\Delta$T (K) & Heat Input (W) 	\\ \hline \hline														
Base	&	38			&	0.0283	& 205			& 495				\\ \hline 
End Walls	&	75		&	0.0283	& 205			& 973				\\ \hline  
Side Walls	&	77		&	0.0283	& 205			& 987				\\ \hline 
Roof	&	38			&	0.0283	& 205			& 495				\\ \hline
Total	&				&			&				& 2,945				\\ \hline

\end{tabular}
\caption{Heat load calculation for the near detector cryostat (insulation thickness = 0.45 m for all)}
\label{tab:heat_load}
\end{table}

\subsubsection*{Cryostat Configuration}

This section describes the configuration of the cryostat only. The TPC is described in Part~II, the \larnd CDR.  
With the intent to minimize the contamination in the gas region, the ullage will be kept cold ($<100$~K). A possible way to achieve this requirement is to spray a mist of clean liquid and gaseous argon to the metal surfaces in the ullage and keep them cold, similar to the strategy that was developed for the cool down of the LBNE 35 Ton prototype.

\subsubsection*{Outer Support Structure}

Two types of outer support structures have been evaluated: steel and concrete. With the current cryostat dimensions, the two are similar in cost, but the steel one presents some advantages. The current reference design is a steel support structure with a metal liner on the inside to isolate the insulation region and keep the moisture out. This choice allows natural and forced ventilation to maintain the temperature of the steel within acceptable limits, without the need of heating elements and temperature sensors, otherwise embedded within the concrete. It reduces the time needed for the construction: the structure will be prefabricated in pieces of dimensions appropriate for transportation, shipped to the destination and only assembled in place. Fabrication will take place at the vendor’s facility for the most part. This shortens the construction of the outer structure on the detector site, leaving more time for completion of the building infrastructure. If properly designed, a steel structure may allow the cryostat to be moved, should that be desired later in the future.

\subsubsection*{Main body of the membrane cryostat}

The sides and bottom of the vessel constitute the main body of the membrane cryostat. They consist of several layers. From the inside to the outside the layers are stainless steel primary membrane, insulation, thin aluminum secondary membrane, more insulation, metal vapor barrier, and steel outer support structure. The secondary membrane contains the LAr in case of any primary membrane leaks and the vapor barrier prevents water ingress into the insulation. The main body does not have side openings for construction. The access is only from the top. There is a side penetration for the liquid argon pump for the purification of the cryogen.

\subsubsection*{Top cap}

In the current reference design two plates constitute the top cap: plate A and plate B. The stainless steel primary membrane, intermediate insulation layers and vapor barrier continue across the top of the detector, providing a leak tight seal. The secondary barrier is not used nor required at the top. The cryostat roof is a removable steel truss structure that bridges the detector. Stiffened steel plates are welded to the underside of the truss to form a flat vapor barrier surface onto which the roof insulation attaches directly. Depending on the number and size of the penetrations, Plate B may be the primary container for the gaseous argon itself. In that case there will be radiation shields only and no membrane underneath instead of the same polyurethane and membrane configuration as plate A. The truss structure rests on the top of the supporting structure where a positive structural connection between the two is made to resist the upward force caused by the slightly pressurized argon in the ullage space. The hydro-static load of the LAr in the cryostat is carried by the floor and the sidewalls. Everything else within the cryostat (TPC planes, electronics, sensors, cryogenic and gas plumbing connections) is supported by the steel plates under the truss structure. All piping and electrical penetration into the interior of the cryostat are made through this top plate, primarily through Plate B to minimize the potential for leaks.
Studs are welded to the underside of plate A to bolt the insulation panels. Insulation plugs are inserted into the bolt-access holes after panels are mounted. The primary membrane panels are first tack-welded then fully welded to complete the inner cryostat volume.

Table~\ref{tab:cryostat_top} presents the list of the design parameters for the top of the cryostat.

\begin{table}[htbp]
\centering
\begin{tabular}{|p{2.0in}|p{4.7in}|}
\hline
\multicolumn{1}{|c|}{Design Parameter}	& \multicolumn{1}{|c|}{Requirement} 	\\ \hline \hline						
Configuration			
	& Removable metal plate reinforced with trusses  anchored to the membrane cryostat support structure. Contains multiple penetrations of various sizes and a  manhole. Number, location and size of the penetrations TBD. Provisions shall be made to allow for removal and\ re-welding six (6) times.								\\ \hline
Plate/Trusses non-wet	& Steel if room temperature.				\\
material				& SS 304/304L or equivalent if at cryogenic temperature. \\ \hline
Wet Material			& SS 304/304L, 316/316L or equivalent. 	\\ \hline
Fluid					& Liquid argon (LAr)									\\ \hline
Design Pressure			& 5.0 psig ($\sim$350 mbar)								\\ \hline
Design Temperature		& 77 K (liquid nitrogen temperature for flexibility)	\\ \hline
Inner Dimensions		& To match the cryostat									\\ \hline
Maximum allowable roof deflection		
						& 0.018 m								\\ \hline	
Maximum static heat leak& $< 20$ W/m$^2$												\\ \hline
Temperatures of all surfaces in the ullage during operation
						& $<$ 100 K	 \\ \hline
Additional design loads & Top self-weight										\\ 
						& TPC ($\sim 2,300$ kg total, to be distributed over all anchors) \\
                        & TPC anchors (TBD) \\
						& Live load (488 kg/m$^2$) \\
						& Electronics racks (400 kg in the vicinity of the feedthroughs \\
						& Services (150 kg on every feed through) \\ \hline
TPC anchors				& Capacity: to be determined by the number of anchors (1,000 kg each anchor, if six). Number and location TBD. Minimum 6. \\ \hline
Grounding plate			& 1.6 mm thick copper sheet brazed to the bottom of the top plate \\ \hline
Lifting fixtures		& Appropriate for positioning the top at the different parts that constitute it. \\ \hline
Cold penetrations		& Minimum 2. Location and design TBD. \\ \hline
Lifetime				& 10 years (5 years of run + 5 years of potential upgrade) \\ \hline
Thermal cycles			& 20 complete cycles (cool down and total warm up) \\ \hline

\end{tabular}
\caption{Near detector cryostat top requirements}
\label{tab:cryostat_top}
\end{table}

\subsubsection*{Cryostat grounding and isolation requirements}

The cryostat has to be grounded and electrically isolated from the building. Table~\ref{tab:cryostat_grounding} presents the list of the current grounding and isolation requirements for the cryostat. Figure~\ref{fig:top_plate_ground} shows the layout of the top plate grounding.

\begin{table}[htbp]
\centering
\begin{tabular}{|l|p{5.0in}|}
\hline
\multicolumn{1}{|c}{~~~Parameter~~~}	& \multicolumn{1}{|c|}{Requirement} 	\\ \hline \hline						
\multirow{3}{*}{Isolation}	
	& 1) The cryostat membrane and any supporting structure, whether it is a steel structure or a concrete and rebar pour,  shall be isolated from any building metal or building rebar with a DC impedance greater than 300~k$\Omega$.\\
	& { 2) The outer support structure shall be electrically isolated from  the building.}	\\					
	& 3) All conductive piping penetrations through the cryostat shall have dielectric breaks prior to entering the cryostat and the top plate.												\\
						 \hline
\multirow{3}{*}{Grounding}	
	& 1) The cryostat, or “detector” ground, shall be separated from the “building” ground.	\\
	& 2) A safety ground network consisting of saturated inductors shall be used between detector ground and building ground.   \\
	& 3) Parameters TBD.						\\ \hline
\multirow{7}{*}{Top Plate Grounding}				
	& 1) The top plate shall be electrically connected to the outer support structure. Parameters TBD.\\
	& 2) The top grounding plate shall be electrically connected to the cryostat membrane by means of copper braid connections.\\
	& ~~a) Each connection shall be at least 1.6 mm thick and 63.5 mm wide. \\
	& ~~b) The length of each connection is required to be as short as possible. \\
	& ~~c) The distance between one connection and the next one shall be no more than 1.25 m.\\
	& ~~d) The layout can follow the profile of several pieces of  insulation, but it shall be continuous. \\
	& ~~e) The DC impedance of the membrane to the top plate shall be less than 1~$\Omega$. 
	\\ \hline
\end{tabular}
\caption{Near detector cryostat grounding and isolation requirements.}
\label{tab:cryostat_grounding}
\end{table}

\begin{figure}[ht]
  \begin{center}	\includegraphics[width=1.0\textwidth, trim=0mm 0mm 0mm -20mm, clip] {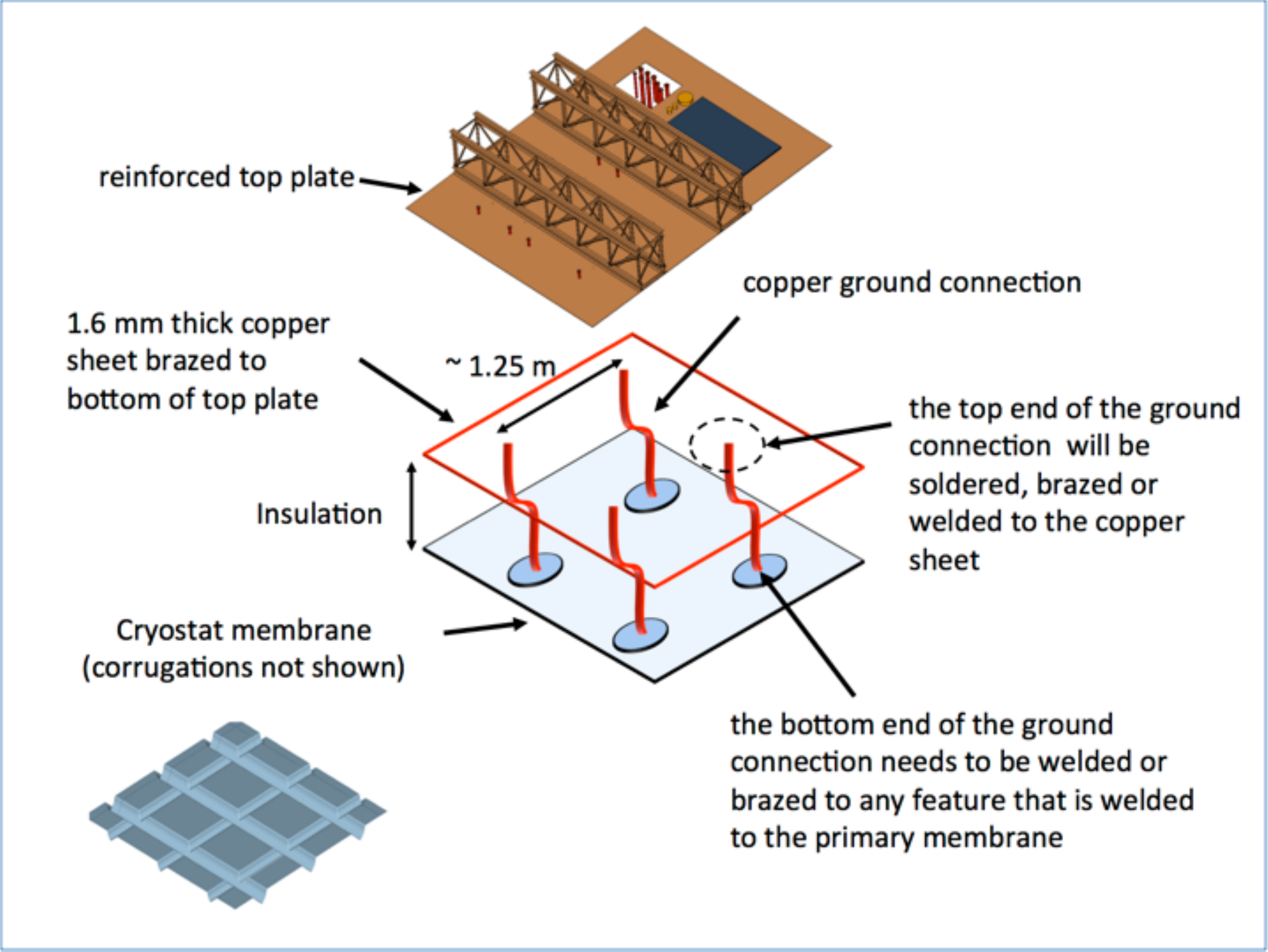}
  	\caption{Top plate grounding layout for the near detector cryostat.}
  	\label{fig:top_plate_ground}
  \end{center}
\end{figure} 

\subsubsection*{Leak prevention}

The primary membrane will be subjected to several leak tests and weld remediation, as necessary. All (100\%) of the welds will be tested by an Ammonia colorimetric leak test (ASTM E1066-95) in which welds are painted with a reactive yellow paint before injecting a Nitrogen-Ammonia mixture into the insulation space of the tank. Wherever the paint turns purple or blue, a leak is present. The developer is removed, the weld fixed and the test is performed another time. Any and all leaks will be repaired. The test lasts a minimum of 20 hours and is sensitive enough to detect defects down to 0.003 mm in size and to a 10${-7}$ std-cm$^3$/s leak rate (equivalent leak rate at standard pressure and temperature, 1~bar and 273 K).
To prevent infiltration of water vapor or oxygen through microscopic membrane leaks (below detection level) the insulation spaces will be continuously purged with gaseous argon to provide one volume exchange per day.
The insulation space will be maintained at 30 mbar, slightly above atmospheric pressure. This space will be monitored for changes that might indicate a leak from the primary membrane. Pressure control devices and safety relief valves will be installed on the insulation space to ensure that the pressure does not exceed the operating pressure inside the tank.
The purge gas will be recirculated by a blower, purified, and reused as purge gas. The purge system is not safety-critical; an outage of the purge blower would have negligible impact on LAr purity.

\FloatBarrier

\subsection{Far Detector Cryostat\label{sec:FD_cryostat}}

The features of the Far Detector new cryostats and insulation were already described in Part III Section V~D. In the following, only a brief summary of the specifications is reported.

New cryostats using a passive polyurethane foam insulation, similar to that 
used for the membrane cryostat of the near detector, have been designed to house the refurbished T600 detector as shown in Figure~\ref{fig:T600_Cryostat}.  
The inner cryostats will consist of Aluminum vessels constructed from welded extruded profiles designed by a  collaboration between industries and Milano Politecnico (Italy).  The vessels are required to be super clean, vacuum-tight and to stand a 1.5 bar maximal operating internal over-pressure. Figure~\ref{fig:AluProfiles} shows a 3D model of the vessel assembly.

\begin{figure}[htbp]
\centering
\includegraphics[height=0.4\textwidth, trim=0mm 0mm 0mm 0mm, clip]{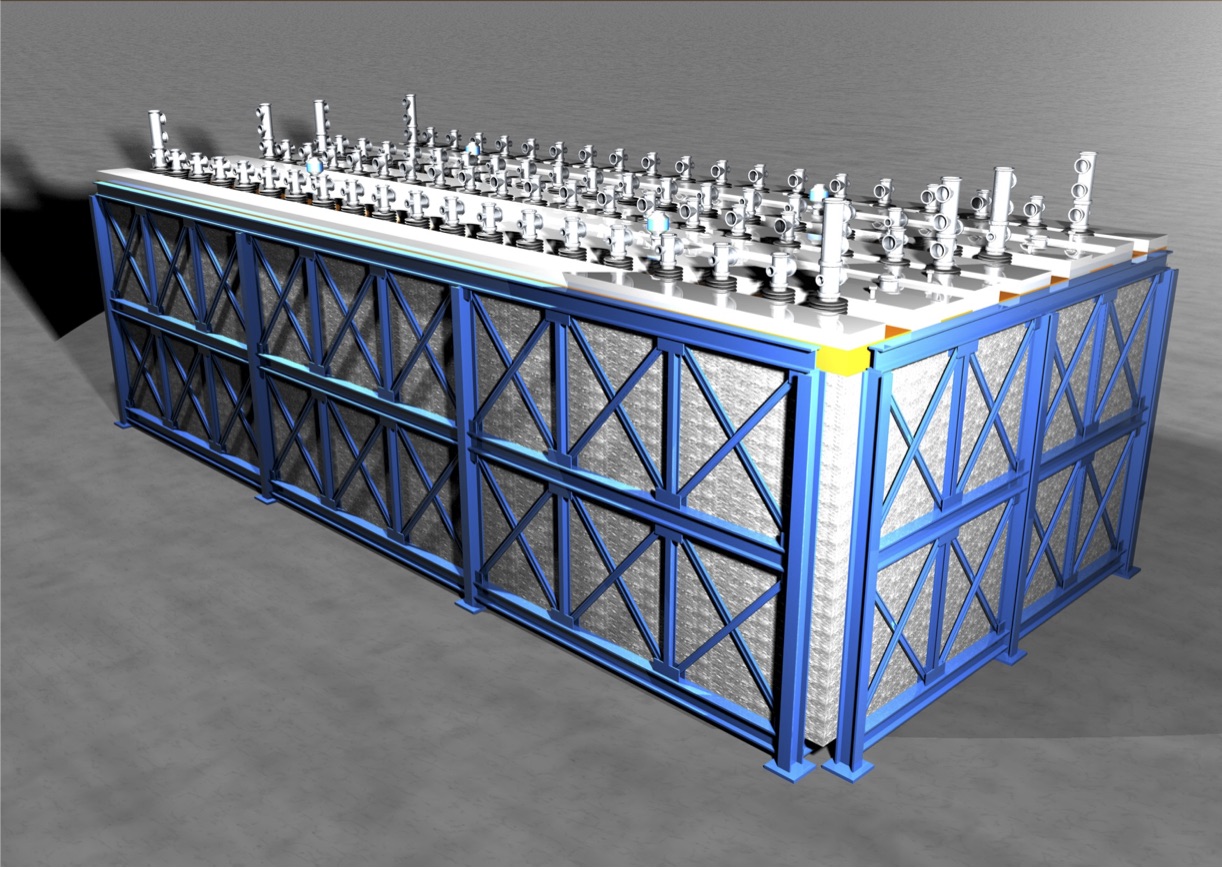}
\begin{center}
\caption{A 3D model of the T600 detector in new cryostat consisting of new Aluminum inner vessels, polyurethane insulation and outer cryostat.}
\label{fig:T600_Cryostat}
\end{center}
\end{figure}

\begin{figure}[htbp]
\centering
\includegraphics[height=0.35\textwidth, trim=0mm 0mm 0mm 0mm, clip] {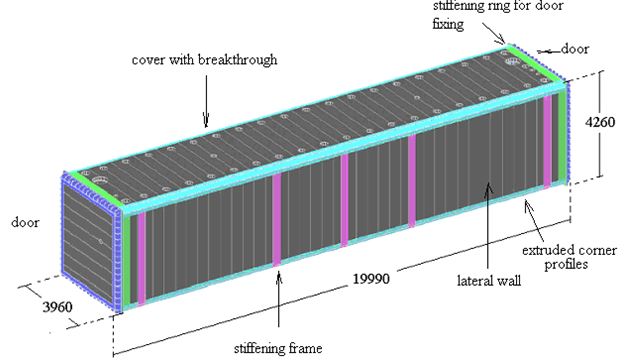}
\vspace*{5mm}
\includegraphics[height=0.35\textwidth, trim=0mm 0mm 0mm 0mm, clip] {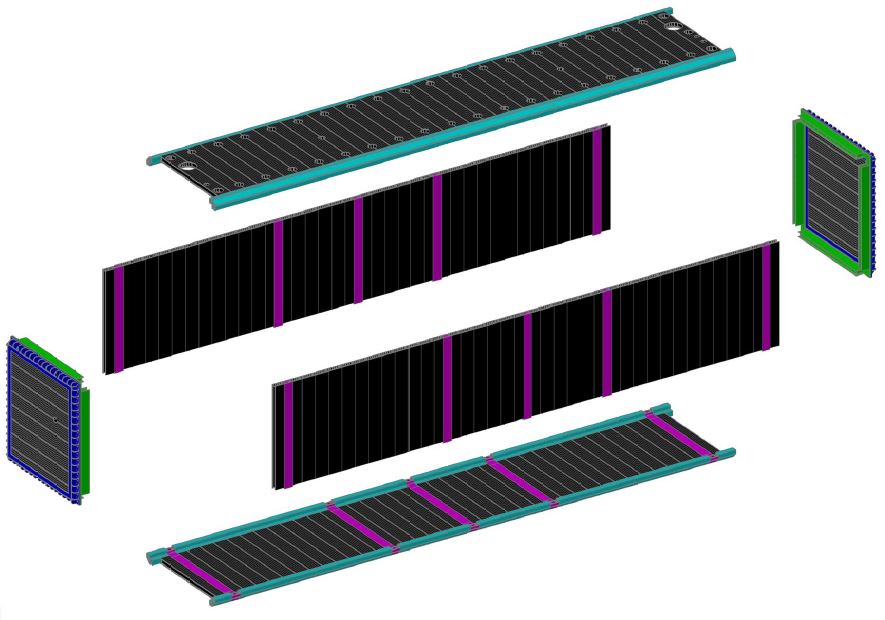}
\begin{center}
\caption{3D model of the proposed new Aluminum vessels for the T600 far detector.}
\label{fig:AluProfiles}
\end{center}
\end{figure}

\begin{figure}[htbp]
\centering
\includegraphics[height=0.35\textwidth, trim=0mm 0mm 0mm 0mm, clip]{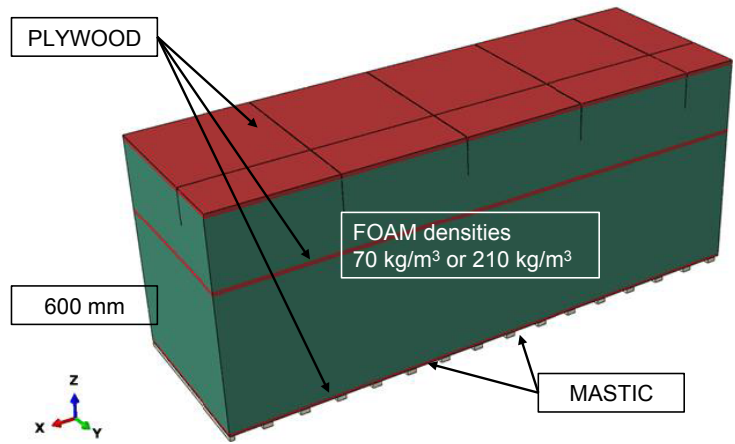}
\begin{center}
\caption{A 3D model of the insulation for proposed new T600 cryostat. A 600 mm thick element is displayed}
\label{fig:T600_insulation}
\end{center}
\end{figure}

The inner cryostats will be enclosed in a passive polyurethane foam insulation developed by GTT, similar to that 
used for the membrane cryostat of the Near Detector, as shown in Figure~\ref{fig:T600_insulation}. The foam insulation will be contained in a new outer frame and coupled to boiling-\lntwo cooling shields, used for heat interception. %A preliminary design contract was awarded to the GTT firm in January 2013. 
Expected heat loss through the insulation is estimated at approximately 6.6 kW. %No membrane technology is expected to be employed in this configuration for the far detector.

Although not described in detail here, the grounding and isolation for the far detector cryostat will need to be handled with the same care as described above for the near detector.  The grounding and isolation for the T600 will abide by all Fermilab safety standards. 

\FloatBarrier

\section{Cryogenic Systems}
\label{sec:cryogenics}

The near and far detector cryogenic designs are being developed with a focus on commonalities which can be used across both detectors and also as a stepping stone for LBNF collaborative efforts.  These systems will be modular in design and constructed on skids that can be tested separately prior to delivery to Fermilab for installation.  Figure~\ref{fig:ln2_System} outlines the basic \lntwo supply system which is proposed by CERN and agreed as an appropriate solution for both detectors.  Each experiment will rely on \lntwo tankers for regular deliveries to local dewar storage. Storage dewars will be sized to provide several days of cooling capacity in the event of a delivery interruption. Note that for the far detector, the original Stirling machines used at Gran Sasso will not be used in the \lntwo cycle.  The lower estimated heat leak of the newly designed vessels allows for use of an open loop system typical of other \lartpc vessels operated at Fermilab (LAPD, LBNE 35 Ton proto., and \uboone). 

\begin{figure}[ht]
  \begin{center}	
  	\fbox{\includegraphics[height=0.55\textwidth, trim=0mm 0mm 0mm 0mm, clip] {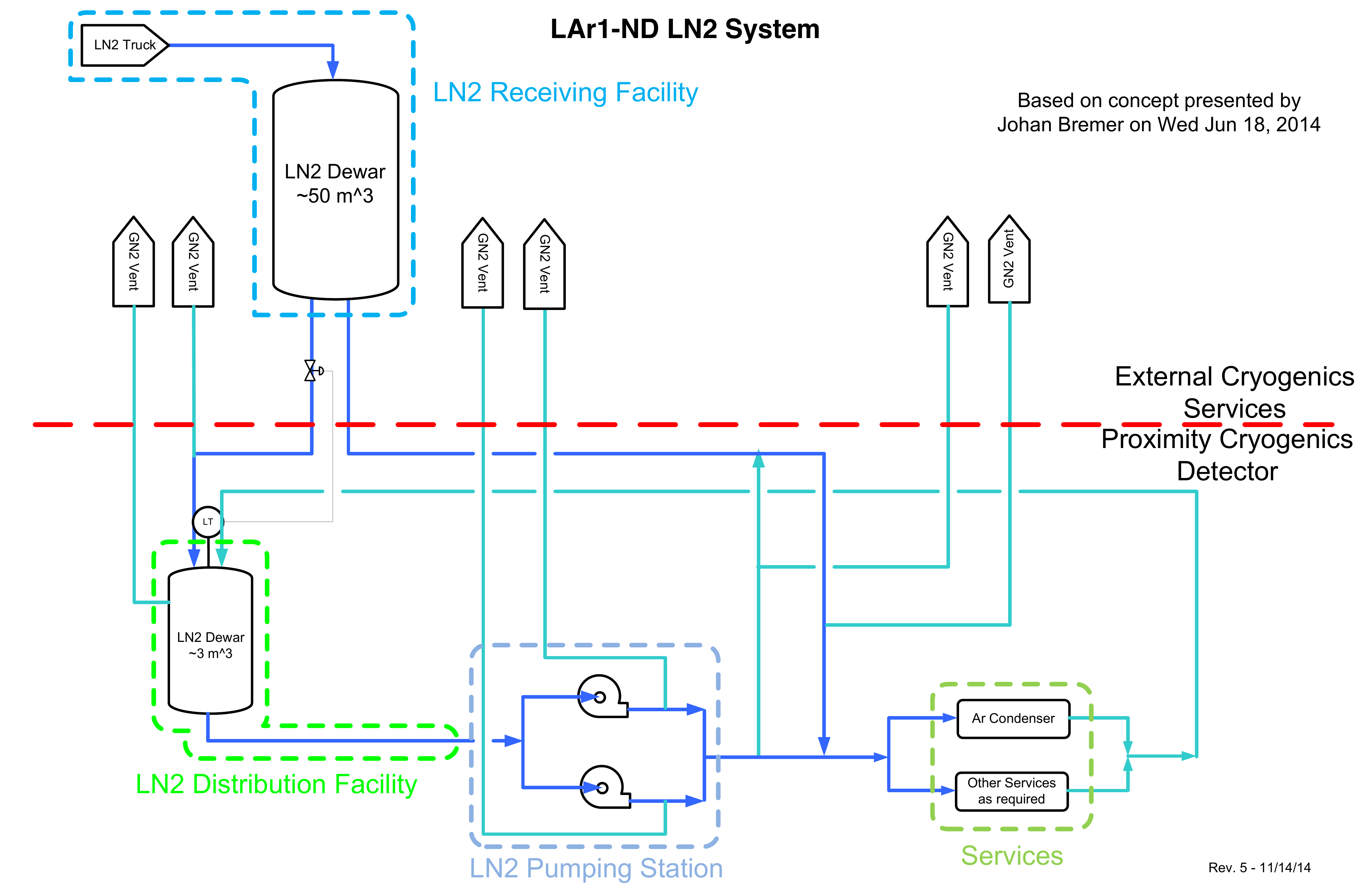}}	\fbox{\includegraphics[height=0.55\textwidth, trim=0mm 0mm 0mm 0mm, clip] {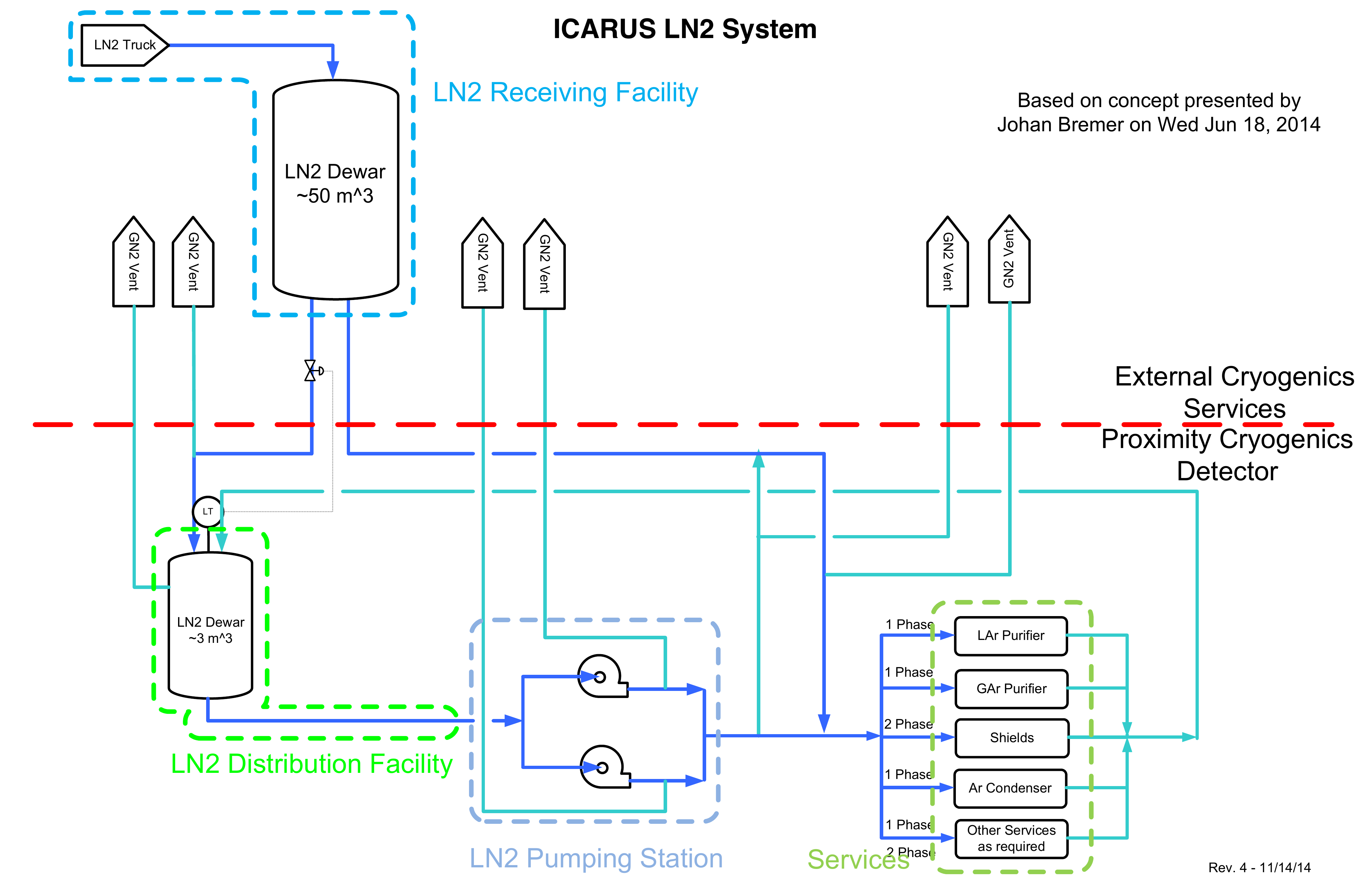}}
  	\caption{Schematic diagrams for the proposed \lntwo systems for \larnd (top) and T600 (bottom) detectors.}
  	\label{fig:ln2_System}
  \end{center}
\end{figure}

Figure~\ref{fig:LAr1-ND_Cryo} shows a schematic diagram of the proposed \larnd liquid argon system.  It is based on experience in the design of the LBNE 35 ton prototype and the MicroBooNE detector systems.

\begin{figure}[ht]
  \begin{center}	
  	\fbox{\includegraphics[width=0.9\textwidth, trim=0mm 0mm 0mm 0mm, clip] {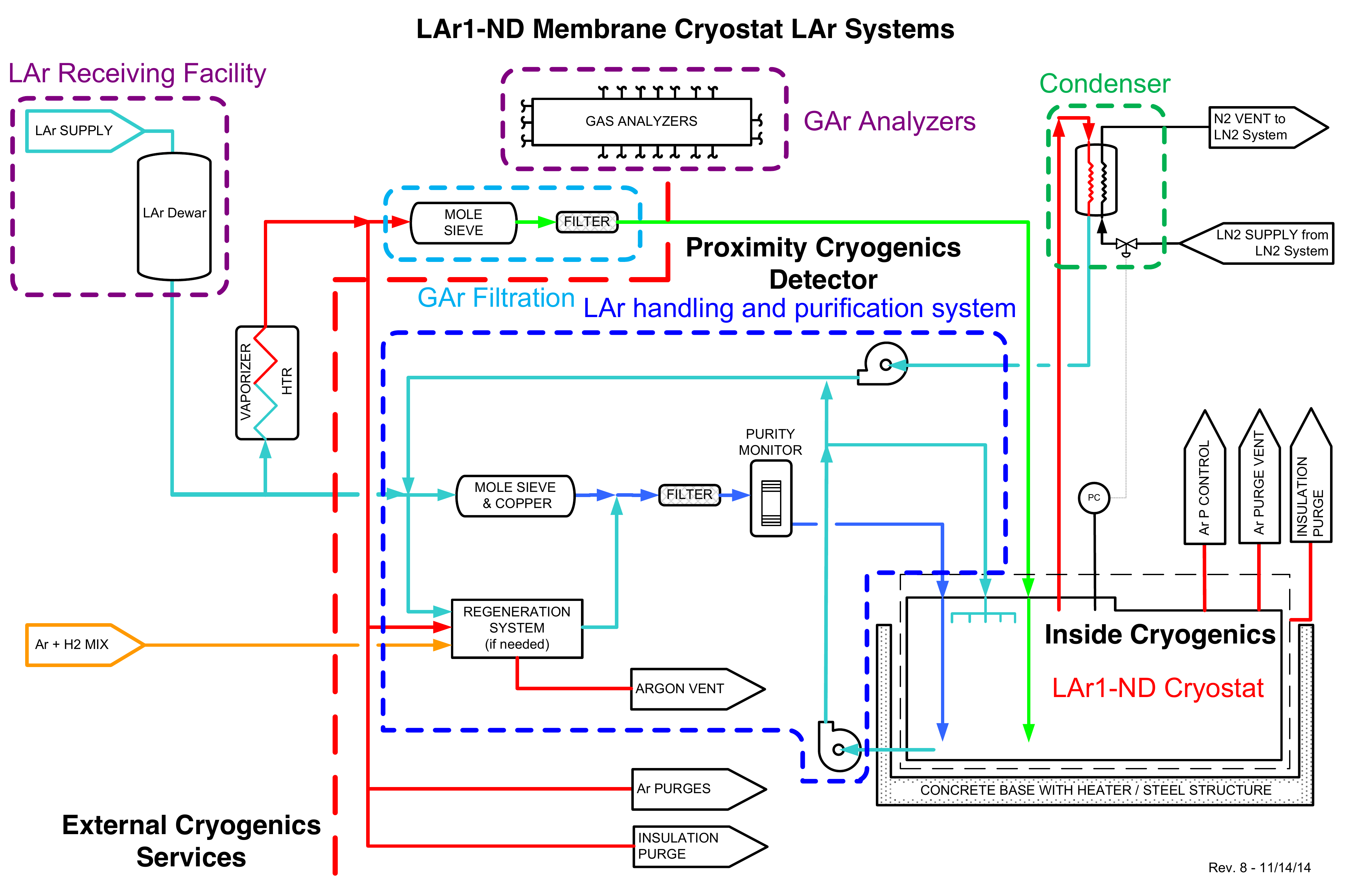}}
  	\caption{Schematic diagram for the proposed \larnd liquid argon systems.}
  	\label{fig:LAr1-ND_Cryo}
  \end{center}
\end{figure}

Preliminary discussions on the requirements and development of the T600 cryogenic system are ongoing. These discussions include the purification system, best re-condensation strategy, and ullage conditions. It is not expected that these aspects will change significantly from previous experience, where the systems performed well enabling the experiment to achieve very high levels of Argon purity with electron lifetime exceeding 15 ms. A description of the existing T600 cryogenic and purification systems can be found in \cite{ICARUS_jinst} and the latest results on Argon purity are detailed in \cite{LArPurity}.

The existing cryogenic system on the \icarus detector is therefore meant to be kept as is, apart from the implementation of the open-loop LN$_2$ delivery system.  Figure~\ref{fig:cryoT600} shows a schematic diagram of the T600 argon system including the existing \lntwo refrigerators.  These refrigerators would be replaced by a system like that shown in Figure~\ref{fig:ln2_System} (bottom).

\begin{figure}[ht]
  \begin{center}	\includegraphics[width=0.9\textwidth, trim=0mm 0mm 0mm 0mm, clip] {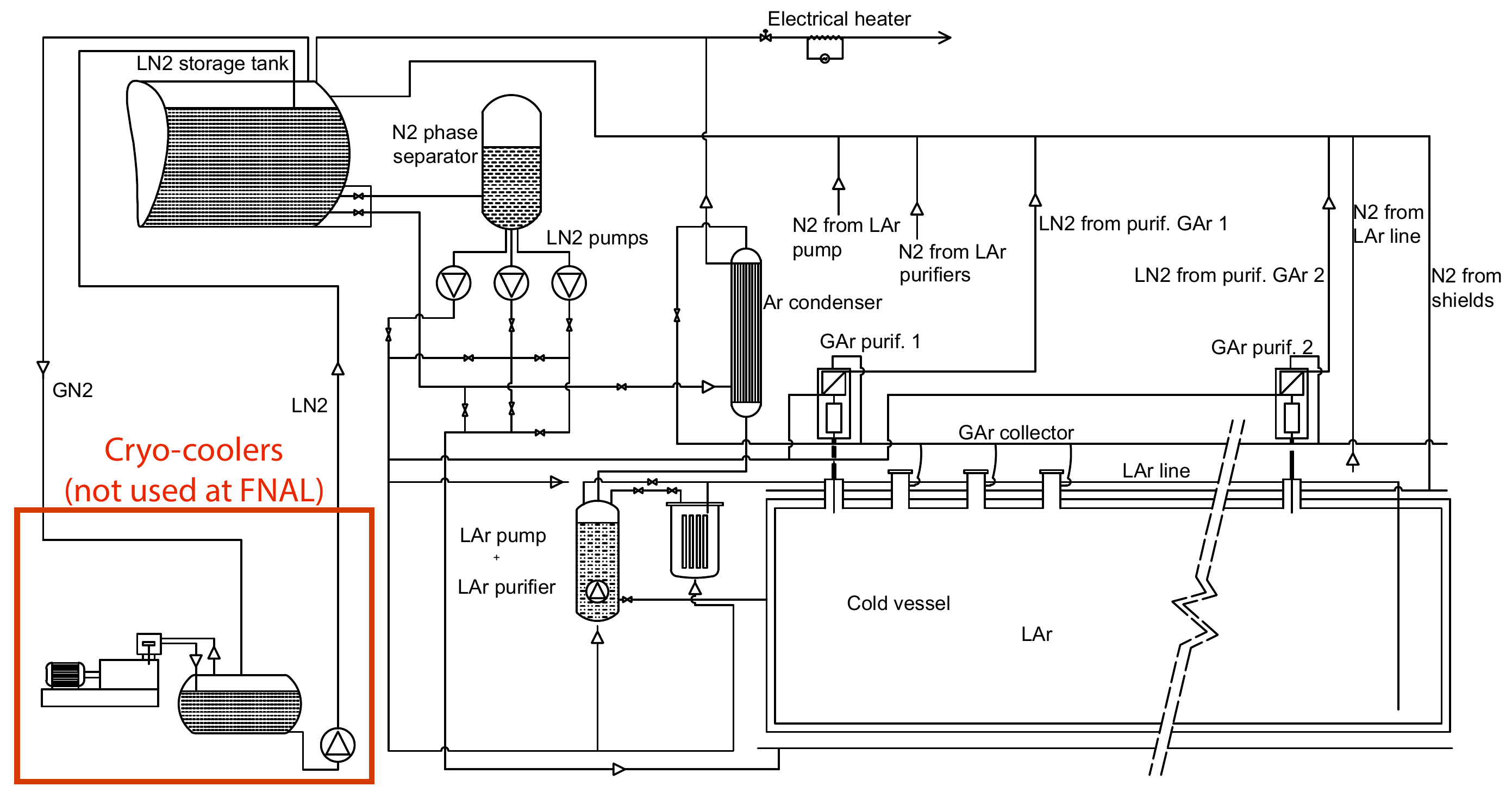}
  	\caption{Schematic diagram of  T600 cryogenic system.  The section in the lower left would be replaced by the new liquid nitrogen delivery system show in Figure~\ref{fig:ln2_System}.}
  	\label{fig:cryoT600}
  \end{center}
\end{figure}

The responsibility for the design and construction of the cryogenic systems will be shared between CERN and Fermilab, along with INFN, by Liquid Argon Cryogenics groups that have been formed at each laboratory.  A preliminary division of the responsibilities by deliverable is outlined in Table~\ref{tab:CryoNeeds}.  The schedule for the development of the ND/FD cryogenic systems, cryostats, and detectors is described in Part~VI of the proposal.
%driven by the plan to have both the near and far detectors ready for commissioning in fall 2017 to ensure to prepare for data taking starting in spring 2018. 
%A draft set of cryogenic system milestones is reported in Table~\ref{tab:CryoSchedule}. 

\begin{comment}
Replaced with latex table copied from Part III

\begin{table}[htbp]
\begin{center}
%\includegraphics[scale=0.6]{tabLAr.png}
%\includegraphics[scale=0.6]{tabLN2.png}
%\includegraphics[scale=0.6]{tabAncill.png}
\end{center}
\caption{Draft proposal for CERN/FNAL responsibilities in the various items related to the cryogenic systems.}
\label{tab:CryoNeeds}
\end{table}
\end{comment}

\begin{table}[htbp]
\begin{center}
\begin{tabular}{|l|c|c|}
\hline
\hline
LAr/GAr System & Service Type & Responsible \\
\hline
\hline
LAr Receiving Facility & Cryo & FNAL \\
\hline
LAr/GAr Transfer Lines & Cryo/Non Cryo & FNAL \\
\hline
GAr/H2 Supply and Transfer Lines & Non Cryo & FNAL \\
\hline
GAr Filtration & Non Cryo & shared \\
\hline
GAr Analyzers & Non Cryo & shared \\
\hline
Condenser & Cryo & shared \\
\hline
LAr handling and purification System & Cryo & shared \\
\hline
Inside piping & Cryo/Non Cryo & shared \\
\hline
GAr handling system & Non Cryo & shared \\
\hline
\hline
LN2 System & Service Type & Responsible \\
\hline
\hline
LN2 Receiving Facility & Cryo & FNAL \\
\hline
LN2 Transfer Lines & Cryo & FNAL \\
\hline
GN2 returns & Non Cryo & INFN/CERN \\
\hline
LN2/GN2 handling system & Cryo/Non Cryo & INFN/CERN \\
\hline
LN2 Distribution Facility & Cryo & INFN/CERN \\
\hline
LN2 Pumping Station & Cryo & INFN/CERN \\
\hline
Services & Cryo & shared \\
\hline
\hline
Ancillary Items & Service Type & Responsible \\
\hline
\hline
Process Controls & Non Cryo & FNAL \\
\hline
Design/Drafting & Non Cryo & shared \\
\hline
Smart P$\&$IDs & Cryo/Non Cryo & shared \\
\hline
Safety aspects of cryogenic installation at Fermilab & Cryo & FNAL \\
\hline
\hline
\end{tabular}
\end{center}
\caption{Draft proposal for CERN, FNAL and INFN responsibilities, for what concerns the management of the cryogenic system maintenance and on-site logistics. The keyword 'shared' refers to tasks to be undertaken jointly by all groups.}
\label{tab:CryoNeeds}
\end{table}

\begin{comment}
\begin{table}[htbp]
\begin{center}
\begin{tabular}{|l|c|}
\hline
\hline
Milestone & Date \\
\hline
\hline
Cryogenic plant proposal submitted for peer review & 6/2015 \\
\hline
%\larnd technical proposal submitted for peer review & 3/2015 \\
%\hline
Cryogenics procurement plans released and active & 3/2016 \\
\hline
Start cryogenic plant commissioning & 9/2017 \\
\hline
Start detectors cooling and commissioning & 1/2018 \\
\hline
Start data taking with beam & 4/2018 \\
\hline
\hline
\end{tabular}
\end{center}
\caption{Milestones for development of the cryogenic system of the far and near detectors.}
\label{tab:CryoSchedule}
\end{table}

\end{comment}

%The \icarus detector is expected to be delivered to Fermilab in the first half of 2017, with about 6 months needed for installation. Commissioning will start in the second half of 2017 requiring from 3 to 5 months, based on the experience gained at LNGS, as already mentioned in Part III, Section VD. 

%%%%%%%%%%%%%

\FloatBarrier

\section{Requirements for Near and Far Detector Buildings}

We present preliminary lists of requirements for the far and near detector buildings for support of the cryostats, cryogenics and detector systems.  These are shown for example only, the complete list of requirements are contained in separate documents maintained by the Fermilab Engineering Support Services personnel responsible for the building design.

The following is a list of required infrastructure  for the cryostat and TPC installation in the near detector building:
\begin{itemize}
\item minimum free area around perimeter of cryostat of 0.92~m (per FESHM);

\item lay down space equivalent to one cryostat footprint for assembly staging;

\item a crane with capacity of 5~ton for cryostat assembly and TPC installation, higher capacity may be needed if TPC is installed with the cryostat top;

\item full crane coverage over the cryostat and lay down space;

\item minimum hook height above the cryostat 5.75~m (6.0~m if top cap installed with TPC already mounted).   
Detailed requirements for power and cooling are under development.

%The minimum hook height needed to install the top cap with the TPC already anchored underneath is 6.0 m over the sidewalls of the support structure.

\end{itemize}

\begin{comment}
\begin{table}[htbp]
\centering
\begin{tabular}{|l|c|}
\hline
\multicolumn{1}{|c|}{Design Parameter}	& Value  \\ 
\hline \hline
Free area around perimeter of cryostat  				
		&	Minimum 0.92 m per FESHM 		\\ \hline
Lay down space					
		&	Equivalent to one cryostat footprint		\\ \hline
Crane coverage				
		&	Over the cryostat and lay down space	\\ \hline
Crane capacity for TPC installation (Not including 	
		& 5 ton	\\ 
the installation of the top of the cryostat)				
		& 					\\ \hline
Minimum hook height above the cryostat 			
		& 5.75 m  \\ 
neck for TPC installation 			
	&  				\\ \hline

\end{tabular}
\caption{Near Detector Cryostat infrastructure requirements}
\label{tab:near_cryostat_infrastructure}
\end{table}
\end{comment}

The following is a list of key infrastructure requirements to support the installation and operation of the T600 detector in the far detector building:
\begin{itemize}
\item a  crane with 5/10~ton capacity;
\item 300 kW power is needed, to be divided among read-out electronics and cryogenic plant. This evaluation does not include general services as light, ventilation, heating. An UPS will be needed for control and monitoring systems;
\item a closed-circuit water cooling system, with flow rate of 5~m$^3$/h, and a pressure/temperature drop of 1.5 bar and 10$^{\circ}$C, respectively;
\item  for safety, separation walls to surround the T600 and cryogenic areas (minimum height 3-4~m), safety sensors (oxygen, smoke, temperature), emergency light, audio alarms;
\item for other general services, as no specific requests are needed, the FNAL Standards and Rules will apply, as in the case of Safety Ventilation: two flow rates systems are foreseen, one always running, the other to be started in case of emergency (e.g.: low Oxygen).
\end{itemize}

For both the near and far detector buildings, the requirement to place concrete blocks for the required overburden will most likely set a higher requirement for the crane capacity. 

\FloatBarrier

\section{Near Detector Siting and Construction}
\label{sec:ND_building}

The location of the near detector building is approximately 110~meters from the existing BNB target located in the MI-12 Building.  The new building incorporates conventional facilities to provide the spatial and infrastructure requirements needed to install and operate the components that comprise the near detector.  Figure~\ref{fig:ND_Building} shows a concept for the near detector building in cross section. In general, the construction will consist of a 1,300 square foot (120~m$^2$), below-grade enclosure centered on the existing Booster Neutrino Beam that will house the \lartpc and related electronics while the 2,300 square foot (215~m$^2$) above-grade portion will provide a means for staging and installing the detector components as well as personnel access.  %The below grade construction will be designed to bear the load up of up to 3~m of concrete shielding placed over the pit inside the building.  This shielding can be added after detector installation if needed to reduce cosmogenic backgrounds.  
The site work will include utility extensions from MI-12, cryogenic storage tanks, gravel staging areas and vehicle access to the near detector building.  Figure~\ref{fig:ND-Building2} shows an aerial view of the building with the SciBooNE and MI-12 target buildings to the left.

\begin{figure}[t]
  \begin{center}
	\includegraphics[width=0.85\textwidth, trim=30mm 20mm 30mm 20mm, clip]{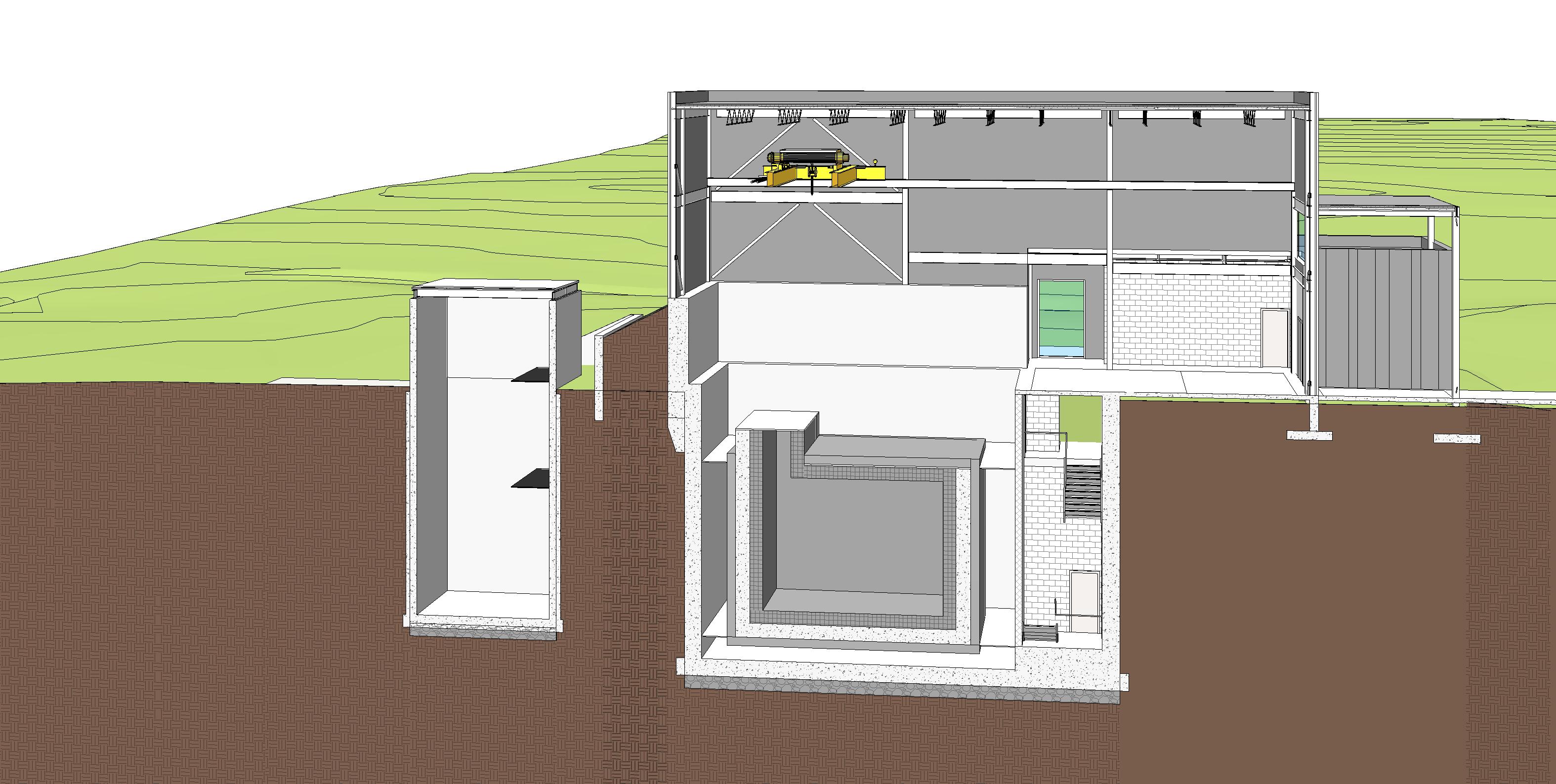}
  	\caption{Cross-sectional view of a design concept for the building that will house the near detector below-grade. The design includes a surface building. The existing SciBooNE enclosure to the left will be used for the cryogenic system.  The beam enters from the left in this view. }
  	\label{fig:ND_Building}
  \end{center}
\end{figure}

\begin{figure}[h]
  \begin{center}
	\includegraphics[width=0.85\textwidth, trim=0mm 30mm 30mm 60mm, clip]{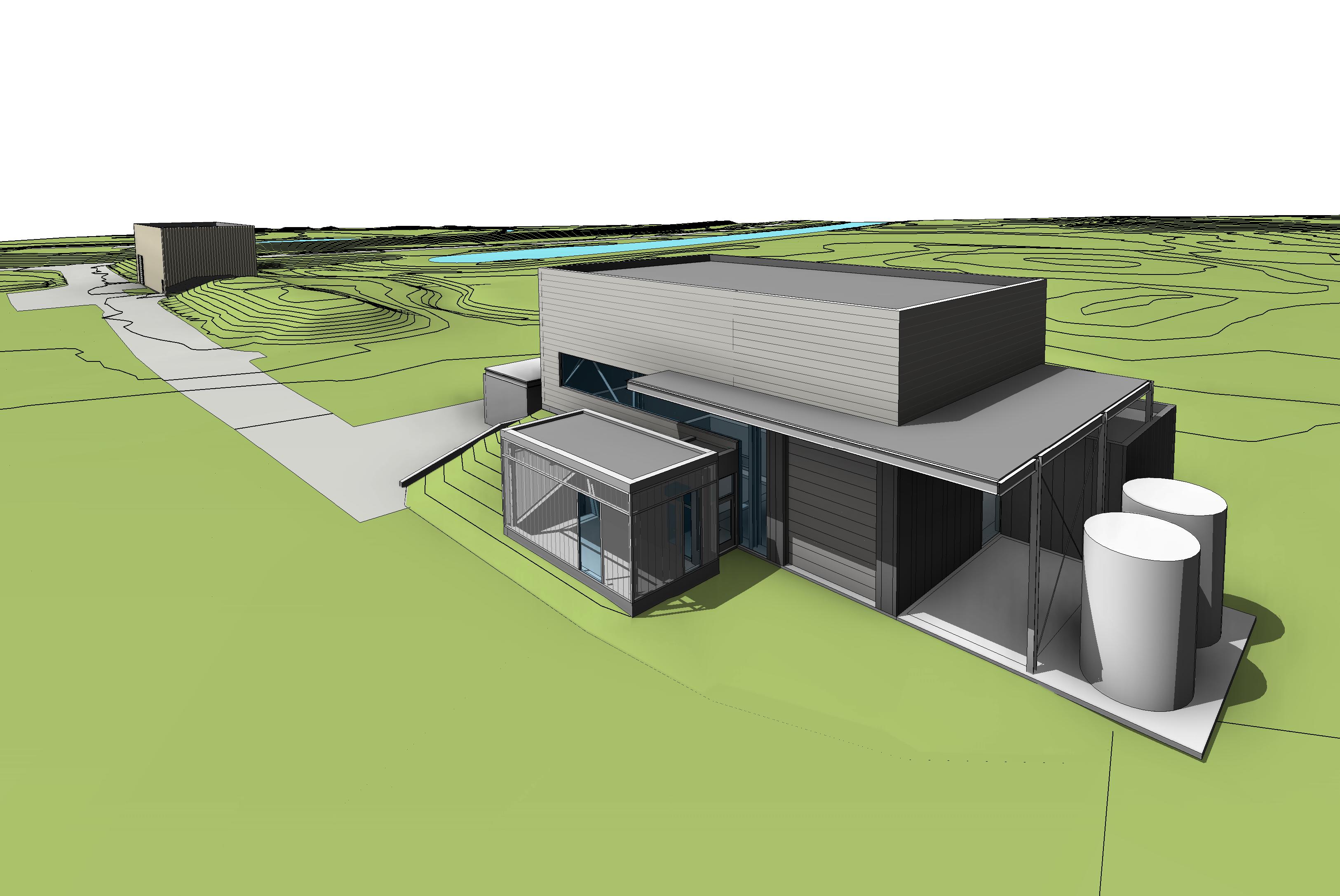}
  	\caption{Aerial view of design concept of the  near detector building.}
  	\label{fig:ND-Building2}
  \end{center}
\end{figure}

%\FloatBarrier

  This near detector location, north of the existing SciBooNE Detector Enclosure allows the existing SciBooNE enclosure to be re-purposed to provide support space for the cryogenic equipment required to operate the detector.  The SciBooNE enclosure can be seen on the left in Figure~\ref{fig:ND_Building}.  An alternate solution with the cryogenic equipment located on the surface is under study.

The lower level of the near detector building will house the 220 ton near detector.  The detector will be located to align slightly off center horizontally (to the east) of the existing Booster Neutrino Beam, placing the floor of the lower level at elevation 713 feet (217.3~m), or approximately 30 feet (9.1~m) below existing grade.  The floor plan of the lower level includes access around the near detector and a stairway to grade.  The stairs to grade will include a landing at the top of the detector to provide access to the cryostat and supporting equipment.  An opening will be cut into the existing SciBooNE Detector Enclosure at the lower level to allow piping and communication access between the two spaces.  The below grade walls and floors will be constructed of cast-in-place concrete and will include a groundwater underdrain system connected to the existing SciBooNE Detector Enclosure sump pump.

The upper level of the near detector building will provide unloading/loading, staging and support space for the construction, assembly and operation of the near detector.  The structure will be designed to accommodate a 5 ton capacity overhead bridge crane to unload and transport detector components from the grade level loading dock to the below grade  detector enclosure.  While not installed initially, the structure will be designed to accommodate the installation of up to 9.84 feet (3 meters) of removable precast shield blocks over the detector.  This shielding can be added after detector installation if needed to reduce cosmogenic backgrounds. The surface building will be a steel framed, metal sided building with a cast-in-place concrete foundation.

\section{Far Detector Siting and Construction}
\label{sec:FD_building}

The new far detector building incorporates the conventional facilities to provide the spatial and infrastructure required to assemble, install, and operate the physics components that comprise the T600 far detector.  The location of the far detector building is approximately 600~meters from the existing BNB target just downstream of the existing MiniBooNE experiment building. In general, the construction will consist of a 7,100 square foot (660~m$^2$) below-grade enclosure housing the relocated T600 detector as well as related electronics while the 4,000 square foot (370~m$^2$) above-grade portion will provide a means for staging and installing the detector components as well as personnel access.  Figure~\ref{fig:FD-Building} shows a cross-section view the design concept for the far detector building at the early stages of final design.  Figure~\ref{fig:FD-Building2} shows an aerial view of the building with the MiniBooNE hill visible to the right.

The site work for the far detector building will include utility extensions from existing utility corridors, storage tanks, gravel staging areas, and vehicle access to the far detector building.  

The lower level of the far detector building is sized to house the T600 detector.  The detector will be located to align both horizontally and vertically with the existing BNB, placing the floor of the lower level at elevation 713 feet (217.4~m), or 32~feet (9.7~m) below existing grade.  The floor plan of the lower level includes code required access space around the detector as well as space to the north end of the detector for detector support equipment.    Surrounding the detector enclosure are several alcoves that will house electronics and support equipment required for detector operations.

The structure will be designed to accommodate a 10 ton capacity overhead bridge crane to unload and transport detector components from the grade level loading dock to the below grade  detector enclosure.  The lower level of the building will be designed to accommodate up to 9.84 feet (3 meters) of earth equivalent shielding over the below grade detector enclosure if this is found to be needed to reduce cosmogenic backgrounds.   Prior to the installation of the concrete shield blocks, the lower level will be open to the crane bay above.  Once the shielding blocks are in place and crane access is not available, a 5,000 pound (2,200~kg) capacity material hoist will be used to transport equipment between the upper and lower levels. The lower level will include two (2) code-compliant exit stairs to grade as well as a duplex underdrain system which will collect groundwater.

The upper level of the far detector building will provide unloading/loading, staging and support space for the construction, assembly and operation of the far detector in addition to the mechanical, electrical and toilet facilities required to operate the building. The surface building will be a steel framed, metal sided building with a cast-in-place concrete foundation.  The building will have exposed finishes.   

The SBN Far Detector Building will be designed to allow the detector to be installed through removable roof sections.

\begin{figure}[ht]
  \begin{center}
	\includegraphics[width=1\textwidth, trim=0mm 30mm 30mm 30mm, clip]{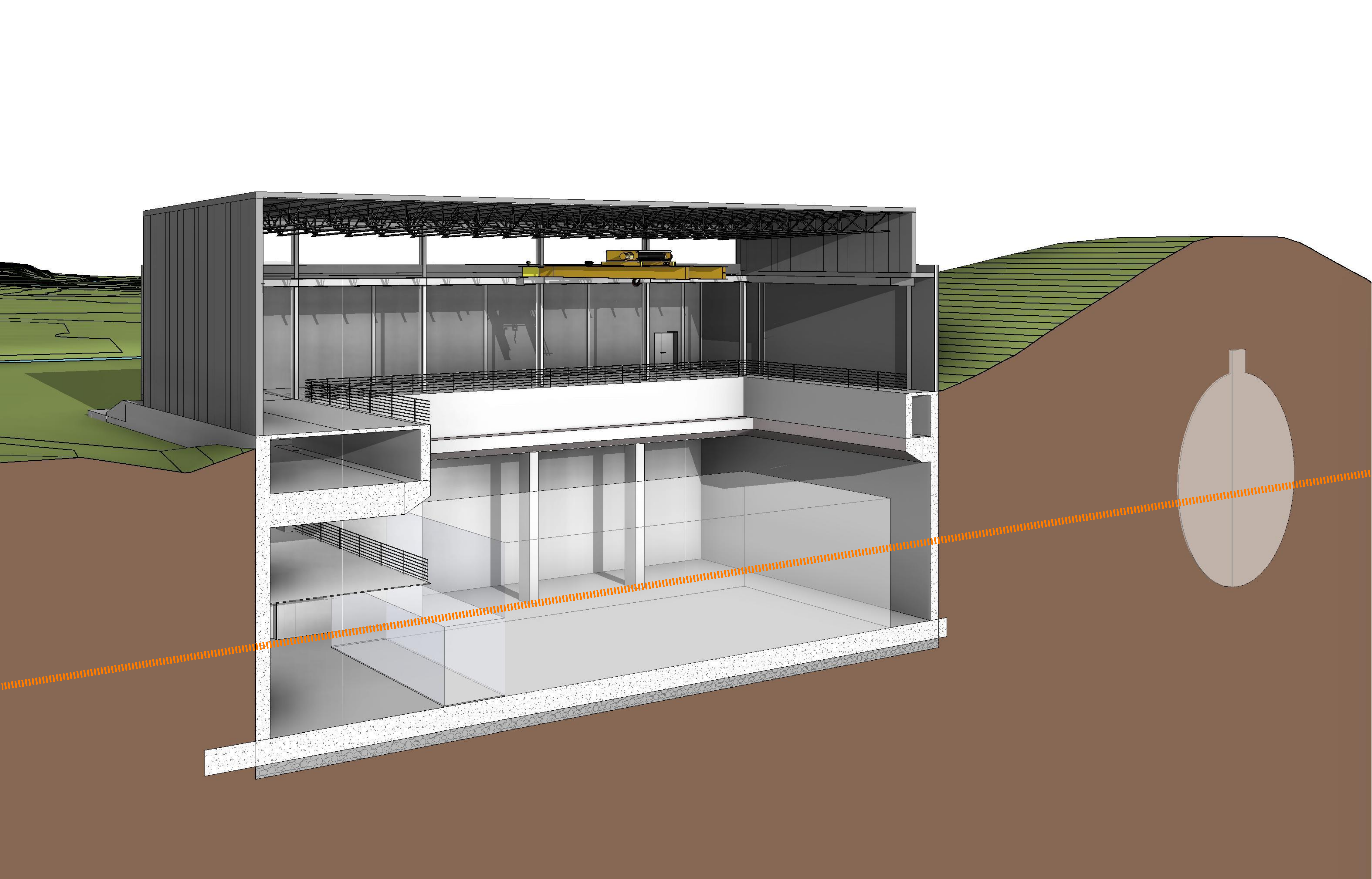}
  	\caption{Cross-sectional view of a design for the far detector building.  The T600 cryostat will be housed in the below-grade portion.  Equipment can be lowered into the below-grade area from the surface building using the internal overhead crane.  The two T300 TPC modules will be installed through a removal roof section.  The beam enters from the right in this view. The MiniBooNE hill is to the right.}
  	\label{fig:FD-Building}
  \end{center}
\end{figure}

\begin{figure}[ht]
  \begin{center}
	\includegraphics[width=0.9\textwidth, trim=0mm 30mm 30mm 30mm, clip]{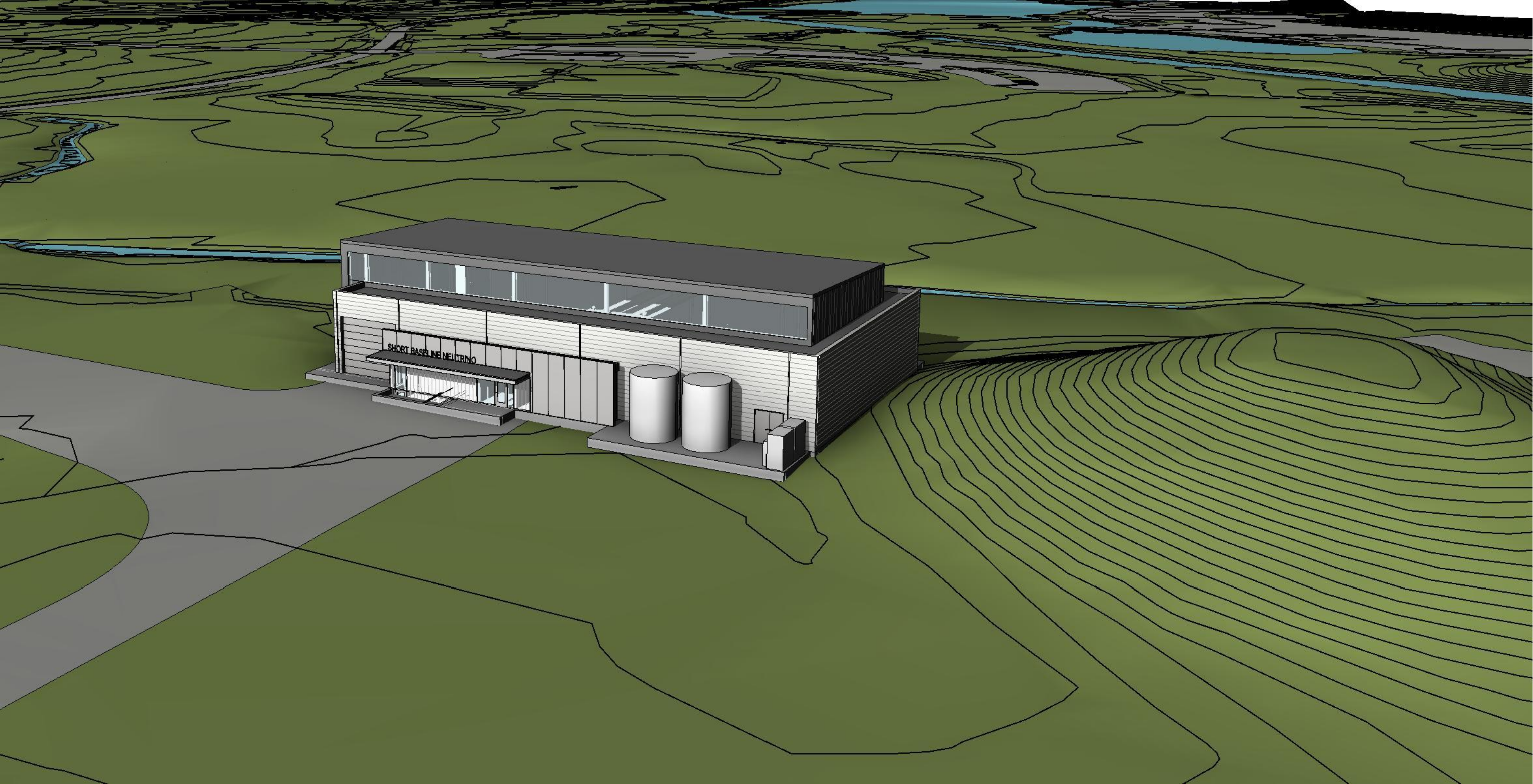}
  	\caption{Aerial view of design for the  detector building.}
  	\label{fig:FD-Building2}
  \end{center}
\end{figure}

\FloatBarrier

\section{Computing Infrastructure and Software}
\label{sec:Software_computing}

The details of the computing needs for the near and far SBN detectors have not yet been fully specified, but it is recognized that the use of similar or identical applications, software libraries, and user interfaces would simplify development, maintenance, and operation.  Developing common solutions will be valuable through the full data processing stream from data acquisition software to event reconstruction and analysis.  It will not be possible to have completely identical software due to physical differences in the design of the detectors and electronics.  However, common frameworks can be used where differences are handled through geometry databases and detector specific versions of some modules. Not only will common solutions result in a more efficient use of scarce programming resources, in a number of places it will be critical to maximizing the sensitivity of the measurements.  For example, common reconstruction tools will be needed to ensure that systematic effects between the different detectors can be carefully studied and minimized.  

In this section, we outline where common solutions could most benefit the program: data acquisition, data quality monitoring, analysis framework, and reconstruction tools.  The description uses several Fermilab support products as examples: the {\em art} analysis framework, the {\em artdaq} data acquisition package, and the LArSoft tools interface.   We then discuss the current state of automated event reconstruction within the collaborations and plans to advance this critical area in common.   

\subsection{Data Acquisition and Data Quality Monitoring}

Common data acquisition infrastructure would allow developers and shift crews to more easily switch between the DAQ systems on the different detectors.  Of course, the DAQ software can not be completely identical because of differences in the detectors, the readout electronics, and any  online analysis needs.  However, the use of a common DAQ software framework would provide the benefit of common infrastructure and functionality while supporting experiment-specific customizations.

There are
 several data acquisition frameworks in use in high energy physics today, including {\em artdaq} which has been developed by the Scientific Computing Division at Fermilab.  As an example of what such a framework provides, {\em artdaq} includes core functionality in the areas of data transfer, event building, process management, system and process state behavior, control messaging, message logging, and configuration of software and hardware.  It also provides online processing and data quality monitoring functionality using the {\em art}framework which is used in the offline environments of many of the current experiments at Fermilab.  

With frameworks such as this, experimenters can focus on the development of the software components that are particular to their experiment.  In the case of {\em artdaq}, this includes the modules that read out and configure the electronics that are used by the experiment.  It also includes the reconstruction, filtering, and compression modules that are run online, and the software modules that monitor the quality of the data as it is being acquired.  The framework model that is implemented in {\em artdaq} is shown graphically in Figure~\ref{fig:artdaq}.

\begin{figure}[ht]
  \begin{center}
	\includegraphics[width=0.8\textwidth, trim=0mm 0mm 0mm 0mm, clip]{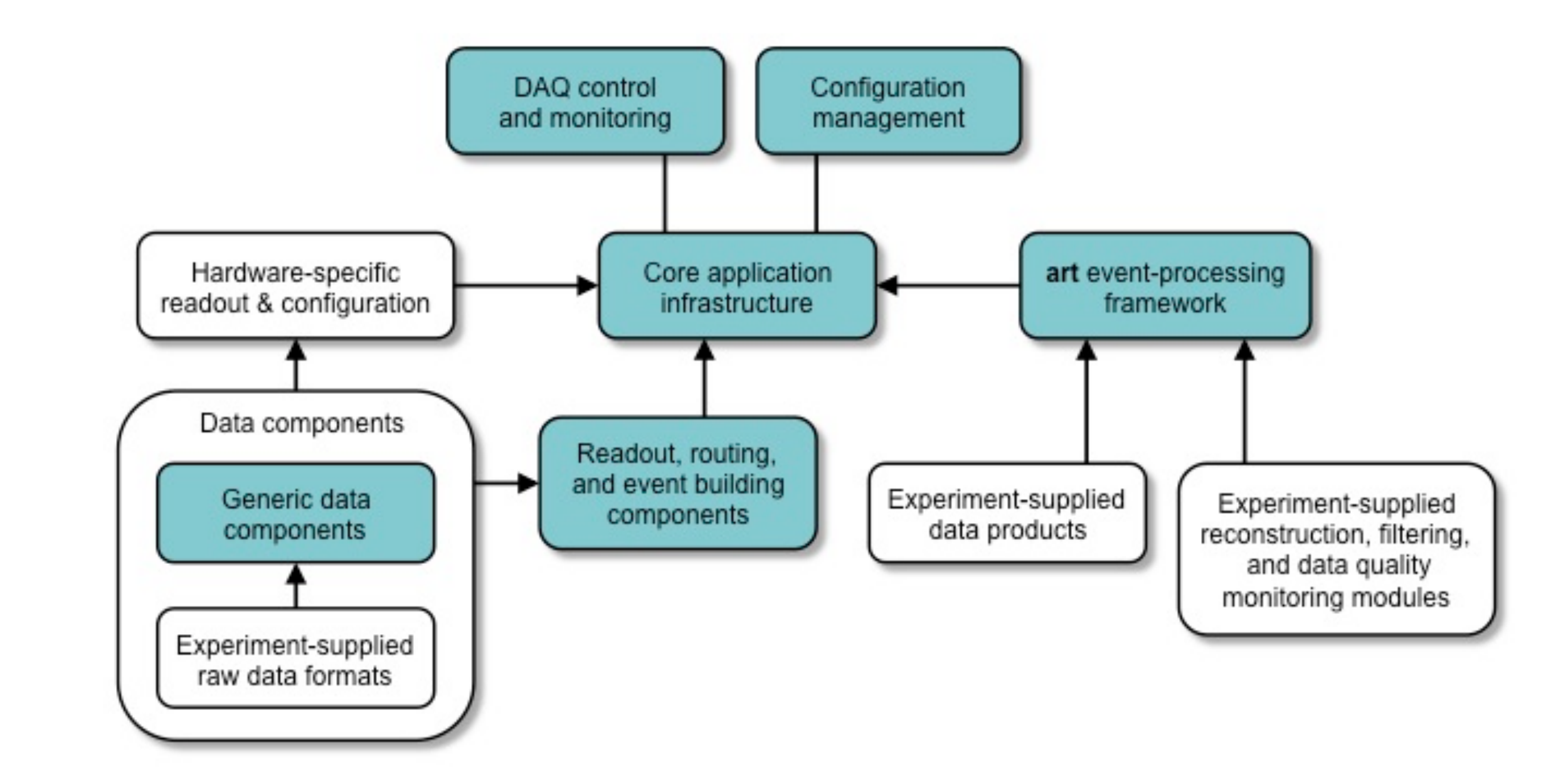}
  	\caption{Sample data acquisition framework architecture, as demonstrated by the {\em artdaq} toolkit.  The blue-shaded boxes show functions that are provided by the framework.  The white boxes show the components that are developed by each experiment.}
  	\label{fig:artdaq}
  \end{center}
\end{figure}

The {\em artdaq} framework is currently used or will be used by several experiments including the DarkSide-50 experiment, the LArIAT testbeam experiment, the LBNE 35-ton prototype detector, and Mu2e.  The benefits of using such a framework has been demonstrated in each of these, most notably the LBNE 35t DAQ in which experimenters from the UK developed the software interfaces to the custom DAQ hardware.

In addition to making maintenance and operation easier, the use of a common DAQ system or framework would allow for coordinated operation, if that would be useful.  For example, an umbrella run control application could allow shift operators to st{\em art}and stop data taking runs for the SBN detectors simultaneously, even if the internal timing structure of the data and the analysis and storage of the data are independent.

Lastly, the use of a common DAQ framework may provide benefits for software organization.  If were useful for the software components that provide convenient access to the raw data for each of the three detectors to be grouped into a single software package, that could be facilitated by the use of a single framework.

The implementation of \larnd experiment-specific items in Figure~\ref{fig:artdaq}, including low-level hardware access for configuration, control and readout, would benefit greatly from experience with MicroBooNE.  Although MicroBooNE does not use {\em artdaq}, as it was designed prior to its existence, many of the design and control features are common.  Also, the \larnd front-end electronics are quite similar to that of MicroBooNE, as shown in Part~3 of this proposal.  The FEM (Front End Module) readout boards are quite similar (with the exception of “cold digitization” in the \larnd case).  The Nevis PCIe interfaces to the sub-event computers are identical, allowing much of the lowest level Linux software to be recycled.  Since the channel counts and expected neutrino rates at both MicroBooNE and \larnd (11,262 channels) are quite similar, the data throughput requirements and the design of the event builder need not be improved or changed.  If {\em artdaq} is chosen, the lower level readout and control code will have to be integrated into that framework; experience with the successful integration in LArIAT will help in this regard.  In {\em artdaq} nomenclature, there would be one BoardReader process per sub-event computer plus FEM crate.

The \icarus detector requires readout  of  significantly more channels (53,248) which makes a common front-end solution more difficult.  However, with the downstream solution shown in Figure~36 of Part~3 of this proposal, with a CAEN A3818 PCIe interface, ICARUS will share a common computer-side sub-event readout bus.  One {\em artdaq} BoardReader per A3818 would match with similar function as with \larnd, one in each sub-event computer.  With a cluster of identical sub-event Linux computers forming the first stage of ICARUS event building, we can make the data flow elements downstream of the PCIe bus look essentially identical.  Any readout control functionalities will of necessity be distinct from \larnd.  Extensive experience with the CAEN A3818 PCIe interface on LArIAT, MINER$\nu$A and CMS experiments would greatly benefit work in this direction.  In any choice of framework, whether {\em artdaq} or other solution, having the frameworks at both ICARUS and \larnd identical will speed development and ease maintenance enormously.

To control all aspects of the items in Figure~\ref{fig:artdaq} in a convenient manner for the shift-crew to understand and operate, we will need a Run Control graphical user interfaces  (GUI).  The MicroBooNE experiment has successfully ported the \nova GUIs based on the QT graphics framework for overall Run Control, application control and resource management.  As even porting a GUI is difficult and time consuming, ICARUS and \larnd should come up with a common solution to avoid duplication of effort.  Behind the scenes of these applications we need a run configuration database to store configuration options and actual configuration used during run time.  The {\em artdaq} group plans to implement such functionalities, and we tag along or follow the MicroBooNE example.  In any case, having a similar DAQ architecture at both far and near detectors will simplify all of these issues.

The final element of online software spans the boundary between online and offline analysis: the online data quality monitor (DQM).  The DQM process should have fast access to the data as they arrive, in real-time in order that DQM can find problems with the experiment quickly to avoid beam time loss.  Depending on the data rates and the processing speed of the DQM process(es), it may not be possible for the DQM process to analyze all of the data in real-time, so a variable prescaling system is needed to, say, “process one out of three events”, where three can be adjusted so that DQM in no way impedes the main data flow.  The DQM input mechanism can be via data files on disk, via a shared memory or via a network event service.  The latter option allows the DQM to run anywhere and to not need to access the disk files or shared memory directly.  In any scenario, the DQM must never be allowed to cause the data acquisition to slow or stop, so communications between the DAQ and DQM must be flexible and robust.  The DQM process normally produces a set of histograms or other ROOT objects that let the shift-crew and detector experts see the status quickly.   To display the histograms, we envision a light-weight interface to the ROOT objects via a web server like http://if-wbm.fnal.gov which hosts histogram displays for MINOS, MINER$\nu$A and LArIAT.

\subsection{Data Storage and Processing}

At this time, it is difficult to precisely evaluate data sizes for the SBN detectors. We assume that data from the detector will be staged to disk and then to tape for archival storage. New data are assumed to remain on disk until an initial pass of production processing can be performed. Subsequent reprocessing of the data, which can be assumed to occur at least once or twice per year, will draw data from tape with a cache disk front end. Production data will, under ideal circumstances, remain on disk for as long as needed for analysis purposes. At least two complete production passes must be resident on disk at any given time. Additional storage will be needed for testing, staging intermediate results, user analysis and management of production processing. 

Monte Carlo data production will consume additional storage. We assume that each pass of data production is accompanied by a complete generation and reconstruction of the production Monte Carlo samples. The statistics in each of the Monte Carlo samples will be approximately ten times that. 

The following data model is based upon experience from other neutrino experiments and broad assumptions about data sizes and rates. Given the large uncertainties at this time about the details of the data and the data model required, we attempt to provide an upper bound on the storage that may be required. Even within this estimate, there are large uncertainties that could change the bound by large factors. Nonetheless, it is instructive to work through some reasonable set of assumptions for the data model to see the order of magnitude of the answer. We should also note that storage technologies change. Any reference to a specific technology in the following should be regarded in that light. The important points will be the total data sizes and rates.
 
Assuming the new DAQ and beamline operates at a maximum rate of 5 Hz with an addition 5 Hz of simultaneous cosmic ray data taking, the expected upper bound of the data rate from \icarus is expected to be or order 500 MB/sec. The \larnd detector will contribute an additional 100 MB/sec in this limit. The raw data is assumed to be compressed, but not zero suppressed. The upper bound on the total raw data size over the course of the SBN run would then be about 1 PB.

The signal samples are approximately 10\% of the total, so about 100 TB. Since the raw data is compressed, each pass of reconstructed data will be larger than 100 TB. Low level data, however, could be dropped for the production output. Here, we assume that the production data output is 150 TB per full re-processing pass on the final dataset, which is still small on the scale of the total storage needs for the experiment. We assume that the dataset will be re-processed at least twice per year for the three years of data taking, and twice more in the year after data taking, which results in about 6 x 150 TB, or about 1 PB of production output over the course of the experiments.

For Monte Carlo data, we assume that we will require 10 times the signal statistics, and that Monte Carlo data is twice as large as raw data in order to accommodate truth information. This yields a total size of 2 PB for a full pass of Monte Carlo on the final dataset. If we assume the same number of re-processings, we get approximately 10 PB for the full Monte Carlo sample by the end of the experiment, which obviously dominates the estimated data size. 

Summing over all of these datasets, we arrive at an upper bound on the total production data size that is on the order of 10 PB. 

Additional disk storage or the equivalent will be required for the following uses: 
 
\begin{itemize} 
\item	Staging raw data from the detector before writing the archiving the data to tape.  Ideally, this data will remain on disk until the first pass of production processing is completed, and large enough to weather latencies and downtimes in the production and archival apparatus.
\item	Staging of data from tape for production data re-processing
\item	Staging production output data. Ideally, production output will remain on disk until it is no longer needed for analysis. We should assume at that at least two full production passes must be on disk at once.
\item	Storage associated with production processing, which includes both special purpose staging and large scale production testing
\item	Storage for smaller analysis datasets and user analysis
\item	Storage associated with a DOE mandated data management plan
\end{itemize}

It is difficult to estimates the needs for these categories, but the asymptotic bound assuming the data size estimates above put the size in the area of a few PB.

Existing technologies deployed at Fermilab, such as dCache and Enstore, already operate at the capacities and throughput scales implied by the above data estimates, so should not introduce unknown costs.

CPU:

Given that the development of simulation and reconstruction algorithms are at a very early stage with only limited exposure to real data (only that from ArgoNeuT), it is premature to make estimates of the CPU capacity required to process and analyze the SBN data. Suffice it to say that there must be sufficient processing available to perform the first pass of production processing in real time with data taking, while at the same time that a full production pass on the data is being performed. In order to provide rapid turn around for analysis, a full reconstruction re-processing of the data, including the accompanying Monte Carlo production, should be possible on the time scale of one to three months. 

Even knowing the production processing requirements, it is often difficult to estimate the analysis CPU in advance of actual analysis. In many experiments, the analysis CPU dominates the total consumed, while in others, it is comparable to or smaller than the production CPU. In order to enable the construction of sensible CPU demand models, it is important that the data processing and job submission infrastructure used by the SBN experiments provides the capability to track the specific use for any given job. Fermilab supports several systems that provide this capability. 

\subsection{Data Analysis Framework and Tools}
The {\em art}product supported by Fermilab is a general-purpose data processing framework for offline data production and analysis that is well-suited for use in neutrino and related experiments. A number of experiments have adopted the {\em art}framework, including \nova, g-2 and Mu2e, as well as the LAr-based experiments ArgoNeuT, MicroBooNE, the former LBNE, LArIAT,  and LAr-1ND. The question of using {\em art} for \icarus will be addressed separately later in this section.

The choice of \lartpc technology and the consequent similarities in readout geometries across experiments offers a unique opportunity for developing common solutions for the simulation, reconstruction and analysis of data for the experiments. The LArSoft project, a joint venture between the experiments, software providers and Fermilab, supports the development and maintenance of an integrated, art-based, experiment-agnostic software suite for simulation, reconstruction and analysis of \lartpc data. All of the current \lartpc-based experiments at Fermilab are members of LArSoft; Fermilab manages the project. Participating experiments contribute algorithms to the LArSoft suite. By using generic interfaces to the services that provide otherwise detector-specific information, the algorithms can be decoupled from the details of the experiment for which code was originally written. In participating in LArSoft, the member experiments also gain access to the algorithms and tools contributed by other experiments.  

Some elements of the offline software necessarily require specific knowledge of the particular experiment. In some cases, the required functionality can be hidden behind interfaces that are sufficiently general to be used for all experiments. In either case, each experiment must develop and maintain this software, which includes the geometry description for the detector, readout electronics simulation and digitization algorithms, interfaces to calibration data, etc. Some fraction of the photon reconstruction software may also need to be experiment specific. 
%Given the existing common software base, these elements represent a significant reduction in the amount of software that must be developed in order to obtain a full simulation and reconstruction chain. \larnd, for instance, completed the necessary detector-specific elements and ran a complete simulation and analysis chain in significantly less than a month using only two part-time developers. 

At the present time, \uboone contributes code more quickly than do the other experiments, due largely to the proximity to data taking. As a result, most of the reconstruction and simulation algorithms were developed either by ArgoNeuT, the original LArSoft user, or MicroBooNE. While MicroBooNE has successfully completed 5 Monte Carlo challenges using the LArSoft suite over the past few years, the remaining active experiments,  LBNE, LArIAT and LAr-1ND have successfully used the same simulation and reconstruction code, but with their respective detector geometries. Notably, LBNE found that a track merging algorithm designed to fix certain reconstruction pathologies in MicroBooNE was able to fully reconstruct tracks that crossed TPC boundaries. While it is clear that each experiment may require algorithms that are only used in their respective experiments or for specific analyses (something  which may be particularly true for the SBN analyses), it is equally clear that the existing LArSoft software provides a strong foundation from which to begin development.

The case of the \icarus detector is special in that it comes with a legacy data processing and analysis framework, plus simulation and reconstruction algorithms. The choices from this point are to either port the ICARUS code into LArSoft, port the LArSoft code into the ICARUS framework, or to allow each set of code to evolve independently. Oscillation experiments will benefit from using the same or very similar algorithms across all detectors by offering better control of systematics.  We argue that 
porting the ICARUS code into LArSoft is preferred because it allows the most direct and complete sharing of the experience gained by ICARUS for the benefit of the other experiments including eventually the long baseline program. Porting LArSoft into the ICARUS framework might increase the cost of sharing from ICARUS back to the other LArSoft participants. Leaving each independent raises the cost of sharing in either direction, so would be preferred only in the case that porting is prohibitively expensive.

The problem of porting code from one framework into another presents a number of issues that need to be addressed. In no particular order, these issues include the compatibility between data structures and their relationships to each other and the framework; the dependence of algorithm code on framework-specific features; the compatibility of the analysis-level plug-ins between the frameworks; and the feature sets available within each of the frameworks. Issues such as build systems and external dependencies need to be resolved as well, but those typically do not present major impediments.

LArSoft has adopted the strategy of framework independence for all of its algorithms and data structures, although implementation is far from complete. If successful, the task of porting the code into another framework reduces to the relatively simple task of writing an interface layer between the framework and the algorithms and data structures. In LArSoft, this layer does little more than interact with the framework to access data structure and other services managed by the framework and hand them in framework-independent forms to the algorithms.

\begin{comment}
In approaching \icarus code, the cost of performing a similar exercise on the ICARUS code base  would need to be evaluated, where the worst case is a complete re-implementation in the LArSoft framework. One could reduce the cost of a complete re-implementation by porting only the data structures and using only the algorithms that exist in the target framework, but this loses the experience embedded in the algorithms in the legacy framework. 
\end{comment}

\subsection{Toward an Automated Reconstruction}

Reconstruction of events in \lartpcs is challenging since the fine-grained tracking and calorimetric aspects of \lartpcs provide a large amount of information on each neutrino event. Taking full advantage of this information requires a precise, efficient, and automated event selection and reconstruction package. 

Reconstruction algorithms are being developed by all three collaborations.  MicroBooNE and \larnd simulation and reconstruction software is in the LArSoft framework, while the ICARUS collaboration has a legacy data analysis and reconstruction framework.

The raw data from LAr TPC detectors consists of signal waveforms from each wire in each of the views (three for all the SBN detectors). The reconstruction of events takes these waveforms as input and proceeds in six steps as outlined below:
\begin{enumerate}
\item Hit finding: each of the signal waveforms is analyzed to find pulses that are returned as “hits,” each representing a time and deposited charge on the analyzed wire. The hits are the basic building blocks for the remaining steps.
\item 2D clustering: a clustering algorithm associates the hits in each view into logical groups called clusters.
\item 3D reconstruction: pattern-recognition algorithms match these two-dimensional (2D) clusters across views to create collections of hits associated with the same three-dimensional (3D) objects. This step can involve splitting or merging 2D clusters to better match candidate 3D groupings.
\item Tracks/showers spatial/calorimetric reconstruction: Candidate 3D objects are passed to track- and shower-reconstruction algorithms to produce collections of tracks and showers. 
\item Identification of primary and secondary vertices: vertex reconstruction associates tracks (and showers) to produce the full event. The event is represented as a hierarchical collection of the tracks and showers, starting with the event interaction point, arranged according to the logical structure of the interaction. 
\item Particle identification: this is performed via $dE/dx$ versus residual range measurement or decay/interaction topologies (stopping particles identification, photon/electron discrimination). 

\end{enumerate}

% Schematically, the reconstruction procedure of \lartpc data goes through:
% \begin{itemize}
% \item hit finding,
% \item identification of 2D clusters with track/shower distinction,
% \item 3D  spatial/calorimetric reconstruction of objects (tracks, showers..),
% \item identification of primary and secondary vertices,
% \item determination of the full event   topology, and
% \item particle identification via $dE/dx$ or decay/interaction  topologies: (photon/electron, stopping particles...). 
% \end{itemize}

Individual pieces of the reconstruction chain are already at a good level of development and have been used in the analysis of \icarus data in the CNGS beam at 
Gran Sasso laboratory and ArgoNeuT data in the NuMI beam at Fermilab. It should be noted that due to the bubble-chamber like quality of
\lartpc data, visual scanning is a very powerful tool that provides an understanding of many features of neutrino interactions that was possible with other technologies and existing experiments. As described in the following, some of the analyses of the ArgoNeuT data have taken advantage of visual scanning.  The relatively small size of the ArgoNeuT data sample has made this possible.

A big effort is ongoing with the goal of optimizing the most challenging parts of the reconstruction, namely  merging two-dimensional  (2D) views into a three dimensional (3D) picture and reconstructing the interaction vertex.   The full chain of event selection and reconstruction is being automated to be ready for the analysis of much larger data samples from MicroBooNE and future \lartpc experiments. Monte Carlo simulated events, with the inclusion of detector effects, are of course an essential tool in order to develop and test the algorithms.

\subsubsection*{Scale of necessary reconstruction}
The huge number of beam and cosmogenic events that will be recorded in
the three SBN detectors (shown in Table~\ref{tab:trigrates}) demonstrates the necessity of developing automatic reconstruction tools. It is nonetheless foreseen that subsamples of
all event topologies and all electron neutrino candidate events will
be visually examined.
% Table ~\ref{tab:trigrates} collects all possible trigger sources,
% prior to application of any  energy or fiducial volume cut, which will constitute
% the initial data sample. Differently from Table~VI of Part I, %\ref{tab:bgrates},
% here  cosmogenic triggers include
% all event topologies in coincidence with the beam spill, independently
% on the presence of photons. The last two rows are related to the
% ``dirt events'' described in Section~II.F of Part I. %\ref{sec:Dirt}. 
% Here again, all possible triggering events are included, specifically muons or other charged/neutral particles that reach the detector active volume but produced externally by
% beam neutrino interactions. 
Table ~\ref{tab:trigrates} contains the expected trigger rates in the active volume,
prior to application of any  energy or fiducial volume cut.  These triggers together with a significant rate of crossing muons and other particles from neutrino interactions outside the active volume will constitute the initial data sample. 
Differently from Table~VI of Part I, %\ref{tab:bgrates},
here  cosmogenic triggers include all event topologies in coincidence with the beam spill, independently from the presence of photons.

\begin{table}[tbh]
\caption{Expected trigger rates in the active volume for a $6.6 \times 10^{20}$ protons on target (POT) exposure, delivered in $1.32 \times
10^{8}$ beam spills for \larnd and T600, and for a $1.32 \times 10^{21}$ POT exposure for \uboone. A significant rate of crossing muons and other particles from neutrino interactions outside the active volume is also expected.}
\label{tab:trigrates}
\begin{tabular}{|p{0.4\linewidth}|c|c|c|}
\hline
&\larnd & \uboone & ICARUS \\  \hline \hline
%&&&\\
\numu CC in active volume (AV)& $5.2\times10^6$ &$3.4\times10^5$ &$5.6\times10^5$\\
\nue CC in AV &$3.7\times10^4$&$2.2\times10^3$  & $3.5\times10^3$\\
$\nu$ NC in AV&$2.0\times10^6$&$1.3\times10^5$ & $2.1\times10^5$ \\
Cosmogenic triggers &$3.0\times10^6$&$1.8\times10^6$&$2.5\times10^6$ \\
%$\mu$ from $\nu$ events outside AV& & &$5.0\times10^5$\\
%remnants from   $\nu$ events outside AV& & &$3.0\times10^5$\\
\hline
\end{tabular}
\end{table}

The aim of the automatic reconstruction algorithms will be:
\begin{itemize}
\item rejection of cosmogenic events and ``in drift time''  cosmogenic tracks,

\item identification of the neutrino interaction and its classification (CC, NC..), and
\item estimation of the neutrino energy.
\end {itemize}
For the first item, the coupling with the light detection and muon tagging systems will be certainly beneficial. 

Reconstruction that enables $\nu$ interaction identification involves:
\begin{itemize}
\item electron/photon  discrimination  via initial part of the cascade,
\item other discriminating features, e.g. energy in the primary vertex region,
\item discrimination between pions and  muons.
\end {itemize}
Identification and reconstruction of the neutrino primary interaction vertex is the first prerequisite for the above tasks.
Work in ICARUS software framework has shown that reconstructing the primary vertex requires that the full net of tracks must be reconstructed first (where tracks meet at interaction vertices). From this stage one can search for the primary vertex e.g. by looking for the track directions.
Electromagnetic shower reconstruction and identification needs primary vertex reconstruction as well. To identify electron neutrino events it is required to 
select showers pointing to the primary vertex, validate if the shower is not separated from the vertex and if it is possible to identify the shower as related to single electrons.

%In addition, the ICARUS T600 data add realistic checks, although the event topology is different from the one expected on the Booster beam due to the much higher energy of the CNGS beam.

\subsubsection*{Current reconstruction capabilities and planned near term work}
Reconstruction algorithms of \lartpc data  have been largely developed on real data from ArgoNeuT and \icarus experiments. The \uboone collaboration is giving a very important contribution continuing the development of most of the reconstruction and simulation algorithms in the  LArSoft  framework.

%\subsubsection*{ArgoNeuT}

\subsubsection*{ICARUS}
Many pieces of the reconstruction are ready in the ICARUS  framework. A novel approach is followed for 3D reconstruction~\cite{ICARUS-3D}, that outclasses the standard approach of hit-by-hit matching through common timing. Hit-by-hit is prone to ambiguities (especially for objects developing in a direction perpendicular to the drift direction), to missing matches (i.e. in case of objects developing in a direction parallel to one of the wire orientation)  and to position quantization (wire pitch). The ICARUS approach instead  performs a global fit of all the three independent views starting from pre-identified 2D ``objects'' or ``clusters'', allowing to overcome the above difficulties. 
Particle identification via dE/dx is also ready, and tested both on Monte Carlo and real events. It is based on a Bayesian neural network algorithm as described in~\cite{NNpid}. 

The automatic identification of 2D ``objects'' or ``clusters'' has
also been developed and nearly finalized.  It is based on a
segmentation algorithm that proceeds from the event periphery towards the
center,  building segments and vertices. The 2D segmentation is fed to
the 3D reconstruction to solve ambiguities. Several 2D-3D iteration
steps can be envisaged.

Further work is ongoing for the reconstruction of showers and the automatic identification of the primary vertex.

The ICARUS automatic reconstruction has been tested on a large sample of real CNGS events containing muons produced in the materials upstream of the detector. The reconstruction efficiency was checked against visual scanning and found to be about 90\% (this sample contains many tracks parallel to the wire planes, thus with almost constant t$_0$ along the track, which is a  very adverse condition for the reconstruction).
Tests on low energy  real CNGS neutrino events provided  encouraging results.

Reconstruction of fully simulated Monte Carlo event samples at the BNB energy  has also been performed and shown to give the same identification efficiency as visual scanning. %, about 50\% 
 In this case, the position of the primary vertex has been assumed as given from the simulation.  

Full automated reconstruction, including the identification of the primary vertex, has been attempted on a sample of simulated \numu CC Quasi-elastic events at 1~GeV. The position of the primary vertex was reconstructed within 3~cm in 92\% of the events, out of which 72\% resulted to have the correct track multiplicity (zero, one, or more protons depending on final state effects in the interaction). The muon initial direction was reconstructed within $20^\circ$ in 90\% of events, and within  $4^\circ$ in 83\% of cases. 

The currently available reconstruction algorithms were used to prepare methods for automatic background rejection. Two approaches were studied:
\begin{itemize}
\item 3D reconstruction of tracks in order to identify muon tracks crossing the detector. The efficiency was checked on a generation of cosmic muons with the results of 95\% tracks correctly reconstructed from the full sample. Improvements are under investigation. Known t$_0$ was assumed in this study.
\item Identification of muon signal and its EM activity in 2D projection of the Collection view. Since the Collection view is the source of the calorimetric and dE/dx measurements, the region in the close proximity to cosmic muons and induced EM activity should be excluded from the signal analysis. Clustering algorithms were used to study the potential of such an approach, obtaining an almost complete background reduction with an estimated few percent loss in fiducial volume.
\end{itemize}

\subsubsection*{LArSoft}
ArgoNeuT was the first experiment using the LArSoft package to simulate and reconstruct neutrino events collected during a run in the Fermilab's NuMI LE beam at the MINOS near detector hall in 2009-2010.
The ArgoNeuT experiment \cite{ArgoNeuT_detector}, a 240 kg active volume \lartpc, collected several thousand $\nu$ and anti-$\nu$ interactions.  Automated event selection has been used to extract different samples of events: 1) CC (anti-)$\nu_{\mu}$ events, combining TPC tracking information  with the downstream MINOS near detector and 2) events with electromagnetic shower activity in the TPC. Fully automated reconstruction has been used  for some analyses, requiring the reconstruction of simple of inclusive topologies, while semi-automated reconstruction procedures, guided by visual scanning have been used  for detailed reconstruction of final state event topologies.  Individual events have been categorized in terms of exclusive topologies observed in the final state and semi-automated geometrical reconstruction has  allowed to reconstruct low energy hadrons (protons with 21 MeV kinetic energy threshold) at the vertex of $\nu$ events. 

Automated geometrical and calorimetric reconstruction of  a high statistics sample of minimum ionizing tracks, through-going muons produced by neutrino interactions upstream the
detector, has demonstrated the reliability of the geometric and calorimetric reconstruction in the ArgoNeuT detector \cite{ArgoNeuT_TM}. Analyses $\nu_{\mu}$ and anti-$\nu_{\mu}$ CC inclusive events~\cite{ArgoNeuT_CCnumode},~\cite{ArgoNeuT_CCantinumode}, coherent charged pion production on argon~\cite{ArgoNeuT_coherent} and  highly ionizing tracks ~\cite{ArgoNeuT_recombination} have been performed through fully automated geometrical and calorimetric reconstruction and particle identification (PID). Analyses requiring the complete reconstruction of the final state kinematics~\cite{ArgoNeuT_2p} have been performed through semi-automated geometrical reconstruction of protons at the vertex followed by fully automated calorimetric reconstruction and PID. Analyses of the selected samples of events with electromagnetic shower activity in the TPC using semi-automated reconstruction procedures to study of NC $\pi_0$ events and electron-gamma separation  are expected to be finalized soon.

% The raw data for MicroBooNE consists of signal waveforms from each wire in each of the three views (U, V and W). The reconstruction of events takes these waveforms as input and proceeds in five steps as outlined below:
% \begin{enumerate}
% \item Each of the signal waveforms is analyzed to find pulses that are returned as “hits,” each representing a time and deposited charge on the analyzed wire. The hits are the basic building blocks for the remaining steps.
% \item A clustering algorithm associates the hits in each view into logical groups called clusters.
% \item A pattern-recognition algorithm matches these two-dimensional (2D) clusters across views to create collections of hits associated with the same three-dimensional (3D) objects. This step can involve splitting or merging 2D clusters to better match candidate 3D groupings.
% \item Candidate 3D objects are passed to track- and shower-reconstruction algorithms to produce collections of tracks and showers. 
% \item Finally, vertex reconstruction associates tracks (and showers) to produce the full event. This event is represented as a hierarchal collection of the tracks and showers, starting with the event interaction point, arranged according to the logical structure of the interaction. 

% \end{enumerate}

In LArSoft multiple modules for performing a task using different methods may exist. LArSoft currently contains multiple hit-finding, hit-cluster, and charged-particle-track-finding algorithms. 
Given the existing common software base, these elements represent a significant reduction in the amount of software that must be developed in order to obtain a full simulation and reconstruction chain. \larnd, for instance, completed the necessary detector-specific elements and ran a complete simulation and analysis chain in significantly less than a month using only two part-time developers. This modularized structure allows for fast, independent development of new algorithms, and seamless incorporation of new and different algorithms into the reconstruction chain. Development and optimization of the different algorithms, test and studies of the performances of automated reconstruction chains are in progress on big samples of \uboone Monte Carlo events. 

Because the \uboone TPC is situated near the surface of the earth it experiences significant exposure to cosmic rays, which must be removed during the reconstruction phase. This is facilitated using a two-pass reconstruction. The first pass proceeds through the reconstruction steps 1--3 described above and the results are passed to track reconstruction. The resulting tracks are then analyzed for consistency with the cosmic ray background (aided by the PMT system). Hits associated to tracks identified as coming from cosmic rays are removed from the event. The second pass then runs through all six reconstruction steps using the remaining hits, which are taken to belong to a beam-induced event. Data from \uboone
will be very important in order to evaluate the performance of this procedure.

% The reconstruction software is part of LArSoft, a common simulation, reconstruction, and analysis software package for \lartpcs. The backbone behind LArSoft is Fermilab’s {\em art} event-processing framework. Users write code modules to do a stage of the reconstruction or analysis, and those modules are scheduled and run through an event-processing state machine in {\em art}, with configurations provided by a user-specified file.

\subsubsection*{Steps to a joint reconstruction effort}
 It is recognized that teams from the different experiments need to collaborate on the development of common solutions for the simulation, reconstruction and analysis of data for the combined analysis of the three detector data.  The details of this reconstruction software have not yet been fully specified. A  workshop dedicated to \lartpc event reconstruction to define a common strategy will be organi
 zed. %in the first months of 2015. 
 As a first step to test the current performance of the already existing LArSoft and ICARUS tools it is proposed that  ICARUS atmospheric neutrino data, which are in the energy region  of the BNB  events are reconstructed with LArSoft tools and ArgoNeuT data are reconstructed with the ICARUS tools. 
 
There is an ongoing effort to merge ICARUS reconstruction algorithms into the common framework of LArSoft. Currently the ICARUS algorithms for clusters and 3D track reconstruction are being adapted to the structure of LArSoft. 
The reconstruction in the LArSoft framework is implemented as modules that allow to perform various stages of reconstruction. Such a modular approach would allow to use any configuration of algorithms already available in the LArSoft together with algorithms developed in the ICARUS framework.
This would allow to verify and test algorithms on data from different liquid argon detectors, as well as give possibility to compare with each other independently developed algorithms. 
%All of them could be run on any data.

The development of common computing and software systems for the SBN program will benefit significantly of the use and development of these tools on the soon to come MicroBooNE data.

It is the natural role of Fermilab as the host laboratory to provide and support software infrastructure such as the  art, {\em artdaq}, and LArSoft.  For the SBN program to successfully take advantage of these tools,  it will be essential that sufficient resources are available from Fermilab to assist in code development, code porting, and user support. This support will be needed in parallel with the construction and refurbishing of the physical detectors. 
 \clearpage
\pagestyle{empty}

\proposalTitle

\setcounter{part}{4}
\part{Booster Neutrino Beam}

%\begin{center}
%{\large
%DRAFT \\
%\bigskip
%\today
%\proposaldate
%}\end{center}
 
\ifcombine
\clearpage
\else
\clearpage
\tableofcontents
\addtocontents{toc}{\protect\thispagestyle{empty}}
\clearpage
\setcounter{page}{0}
\fi

\pagestyle{fancy}
\rhead{V-\thepage}
\lhead{Booster Neutrino Beam}
\cfoot{}

%%%%%%%%%%%%%%%%%%%%%%%%%%%%%%%%%%%%%%%%%%%%%%%%%%%%%%%%%%%%%%%%%%%%%%%%%%%%%%%%%%
\section{Overview}

The  short-baseline neutrino program described in this proposal makes use of the existing Booster Neutrino Beamline (BNB). The BNB is a conventional horn focused neutrino beam, fed with 8~GeV protons from Fermilab's Booster accelerator. The beamline was originally optimized for the MiniBooNE detector, the primary user of the beamline over the last decade. 
One of the considerations when designing the beamline was to have as large a flux as possible at 500~MeV, while keeping the flux at higher energies as low as possible. The higher energy neutrinos produce $\pi^0$s in the MiniBooNE detector through Neutral Current interactions and these present significant background for the $\nu_e$ appearance measurement. The \lartpc technology provides much better background rejection and so the constraint of reduced high energy neutrino flux can be relaxed. Maximizing flux at all energies should be generally beneficial.

In the existing beamline configuration the 8~GeV protons from the Booster are guided through the transport line to the target hall as shown in Figure~\ref{fig:tgthall}. The primary beamline ends with a quadrupole triplet that focuses the beam on the target. The target is embedded within the 1.8~m long horn, and the target horn assembly lies just downstream of the final triplet. A 2.14~m long collimator about 3~m downstream of the target shields the entrance to the decay pipe region.  
\begin{figure}[!h]
    \begin{centering}
      \includegraphics[width=0.6\textwidth]{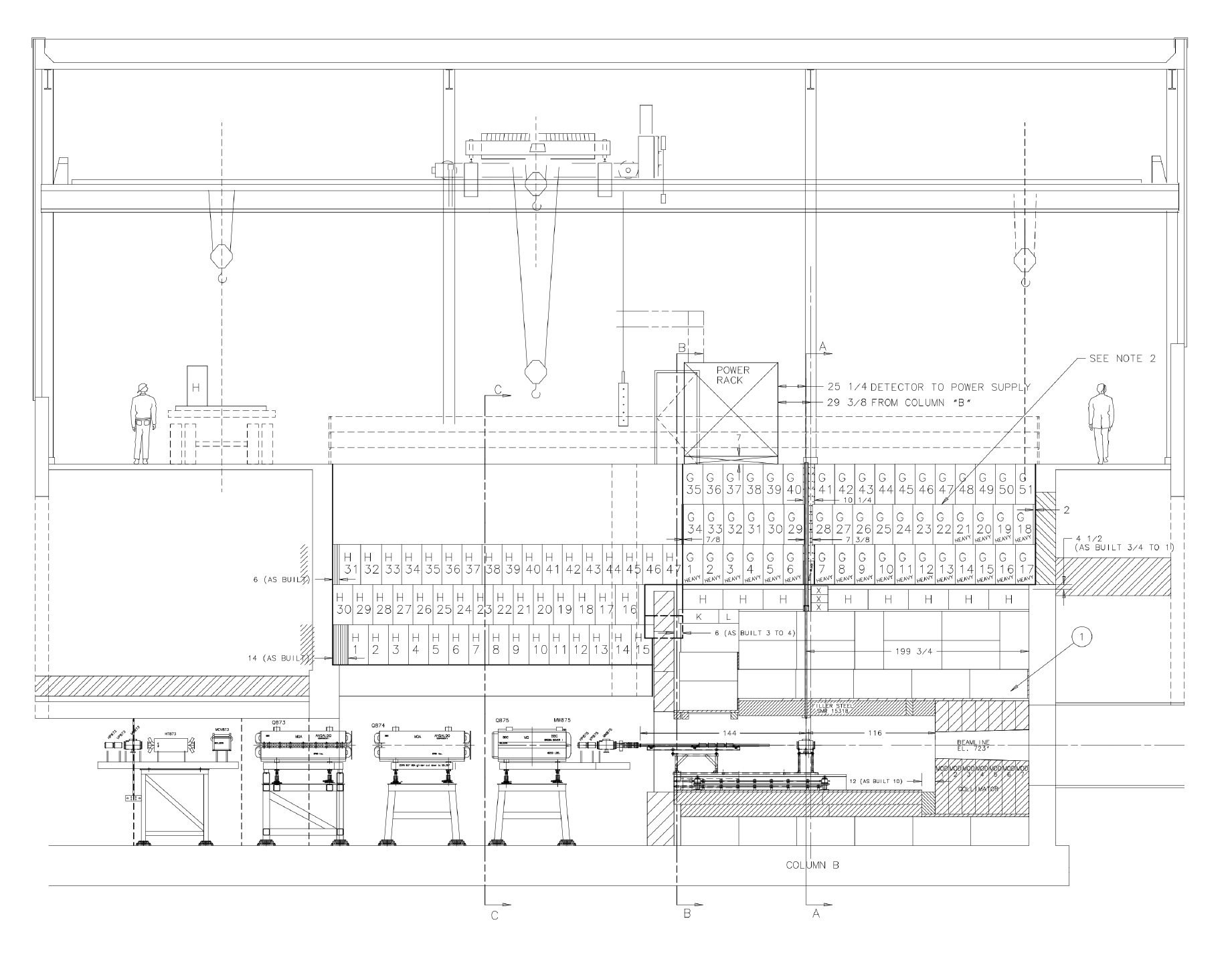}
      \includegraphics[width=0.35\textwidth]{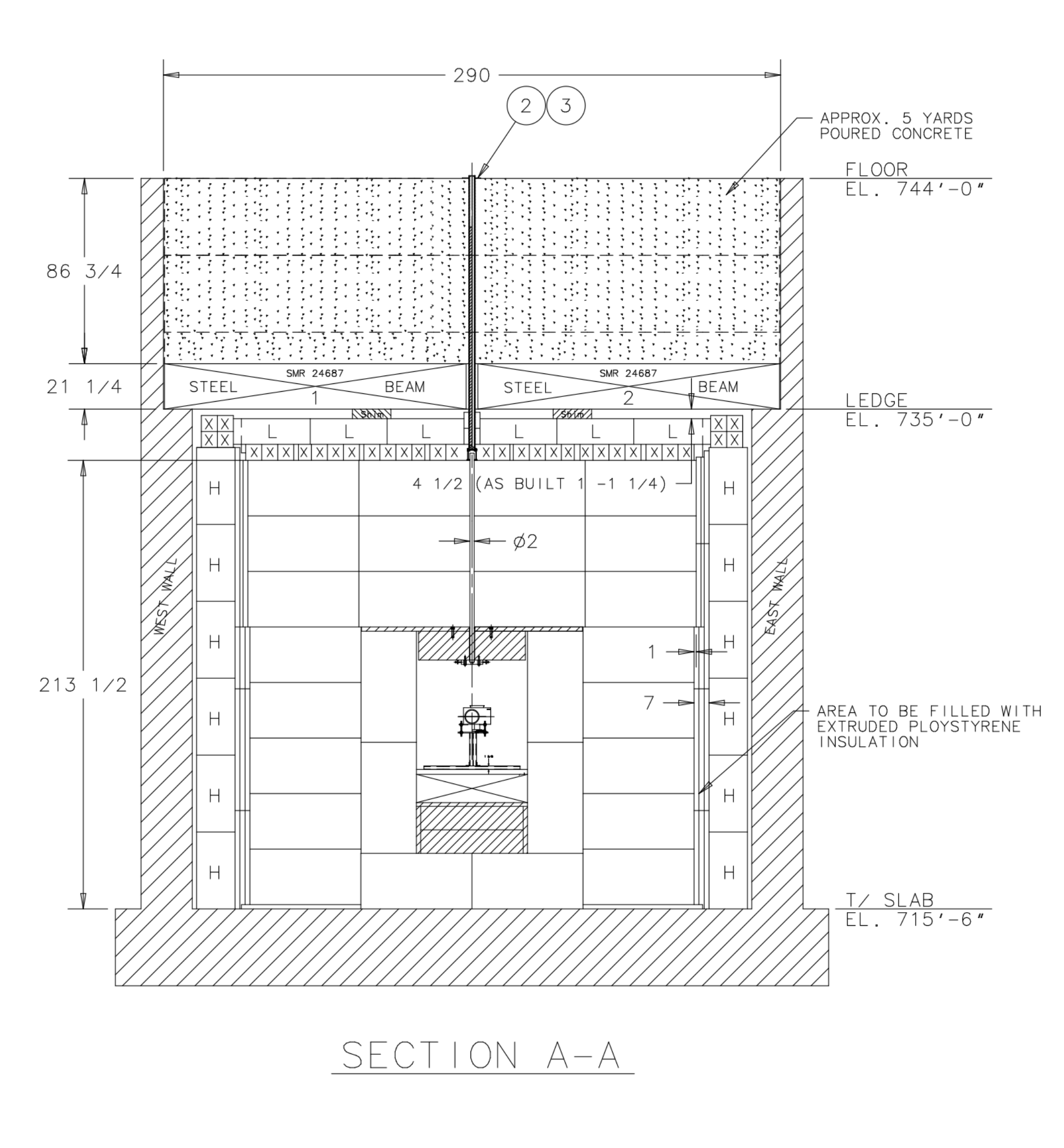}
      \caption{\label{fig:tgthall} Side (left) and beam (right) view of the target hall region. Final focusing triplet (Q873, Q874 and Q875) can be seen in the side view. The MiniBooNE horn is inserted into the target pile just upstream of the collimator noted as section A in the drawing. This region is 2m high and 1.4m wide.~\cite{TGTHall} }
    \end{centering}
  \end{figure}
 
The Booster operates at the 15~Hz repetition rate with up to 5~Hz average rate delivered to BNB.  The intensity per spill is typically about $4.5 \times 10^{12}$ protons. The time structure of individual beam spills is determined by Booster parameters. The harmonic number for Booster is 84 (81 buckets are filled with beam) and the RF frequency is 53~MHz.  This results in 1.6$\mu$s long spill comprised of a train of 81, roughly 1~ns wide buckets mutually separated by $\sim$19~ns.

The next few sections describe how the neutrino interaction rate in the detectors can be doubled by replacing the existing single horn system with a re-optimized two horn system. The additional space needed for this larger system can be made available in the BNB target building without any need for civil construction by condensing the final components of the proton beamline immediately upstream of the target.

%%%%%%%%%%%%%%%%%%%%%%%%%%%%%%%%%%%%%%%%%%%%%%%%%%%%%%%%%%%%%%%%%%%%%%%%%%%%%%%%%%
\section{A Re-optimized Horn Configuration}

This section discusses the a reoptimization of the target and horn system to better match the capabilities of the \lartpc detectors, the future users of the beamline. In addition to the reoptimization motivated by the change in detector technology there is also a push to reoptimize that comes from better knowledge of the system components that is now available. Since the MiniBooNE horn was originally designed, precise measurements of pion production in the beryllium target have been made by the HARP experiment~\cite{HARP} and the kinematic distributions are much better known. These data additionally allow for better optimization of the shape of the inner conductor and the focusing system.

Preliminary studies have been made to estimate possible gains with a reoptimized focusing system. A fast Monte Carlo was developed and used to optimize the horn current, shape of inner conductor of horn 1 (and horn 2), horn position(s), and target position in order to provide focusing of pions that produces the most neutrino events in the on-axis detector(s). The geometry of the optimal design was then simulated using full GEANT4 based Monte Carlo (MC) used by MiniBooNE and other BNB experiments to calculate the neutrino flux. The detailed beam simulation was tuned to match HARP hadron production measurements. Comparing the predicted flux using the full beam MC enables a realistic comparison of the optimized system to the existing MiniBooNE horn focusing.

Figure~\ref{fig:horn_design} shows the shapes and locations of the current single horn system and the re-optimized two horn system. Figure~\ref{fig:twohorn} shows the fluxes that result from the same proton delivery to the current and re-optimized systems.

\begin{figure}[!h]
    \begin{centering}
      \includegraphics[width=0.6\textwidth]{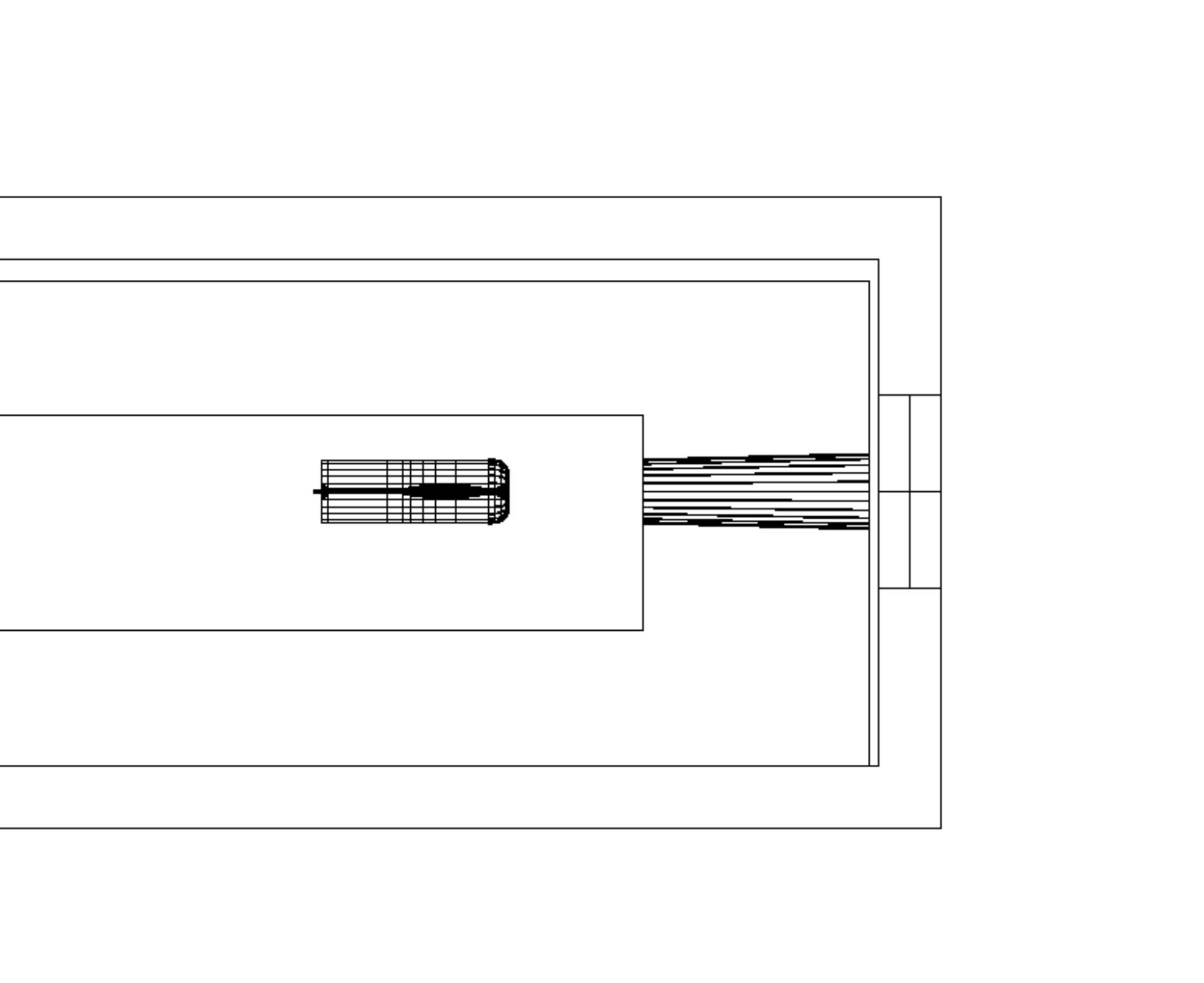}
      \includegraphics[width=0.6\textwidth]{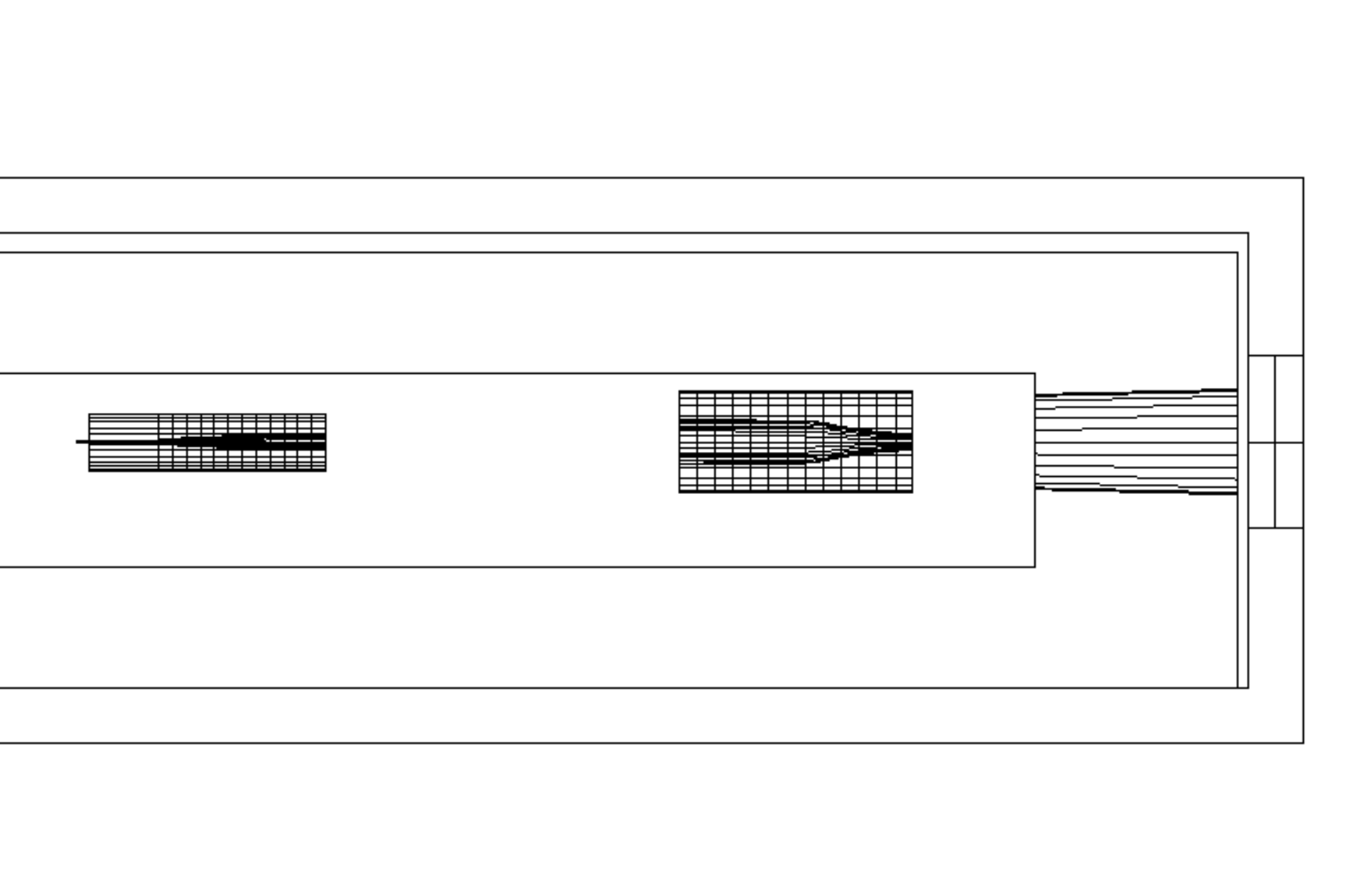}
      \caption{\label{fig:horn_design} Two plan views of the target chase showing the shape and location of the current single horn (top) and optimized two horn (bottom) systems.}
    \end{centering}
  \end{figure}

\begin{figure}[!h]
\begin{centering}
\includegraphics[width=0.49\textwidth]{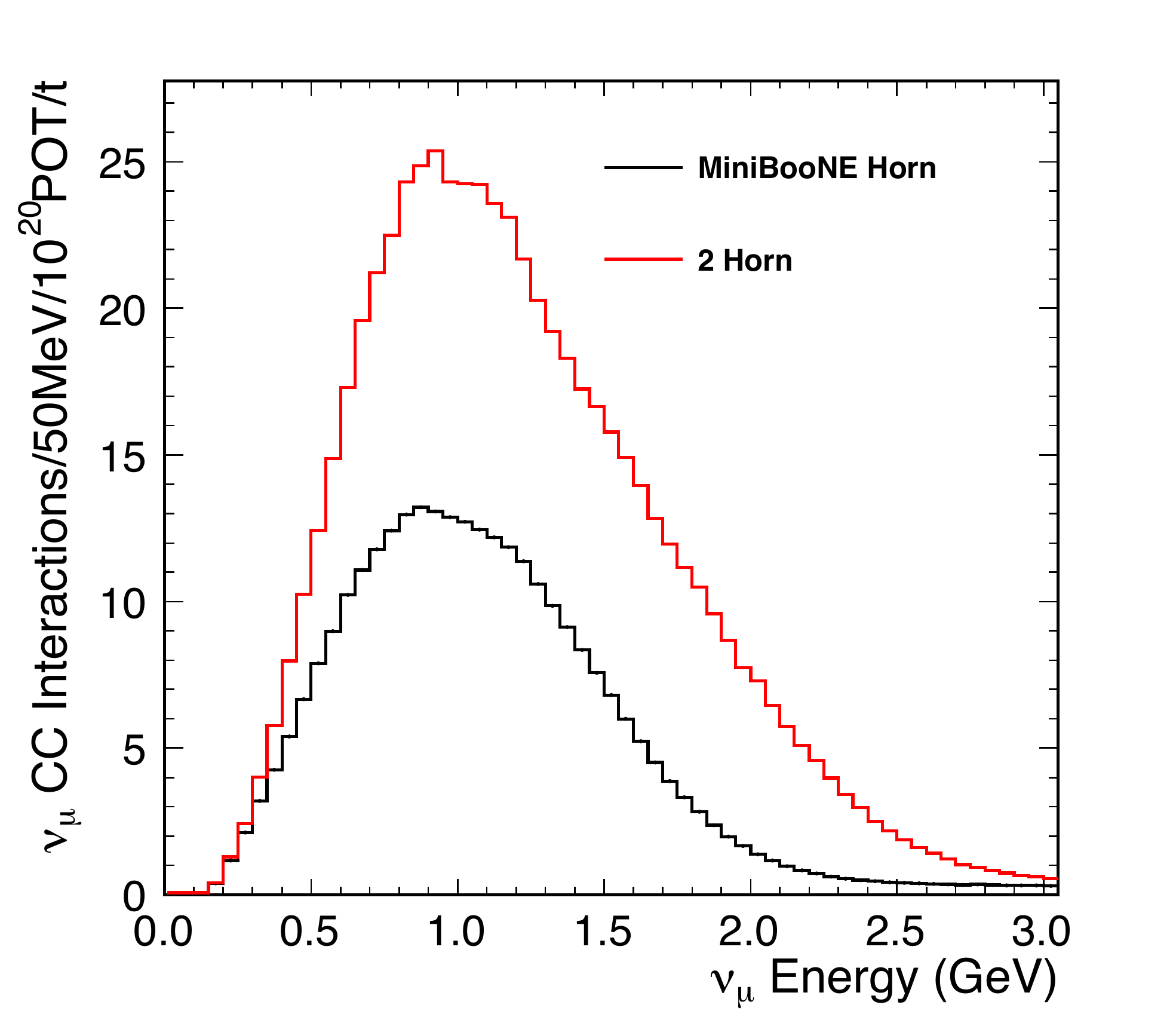}
\includegraphics[width=0.49\textwidth]{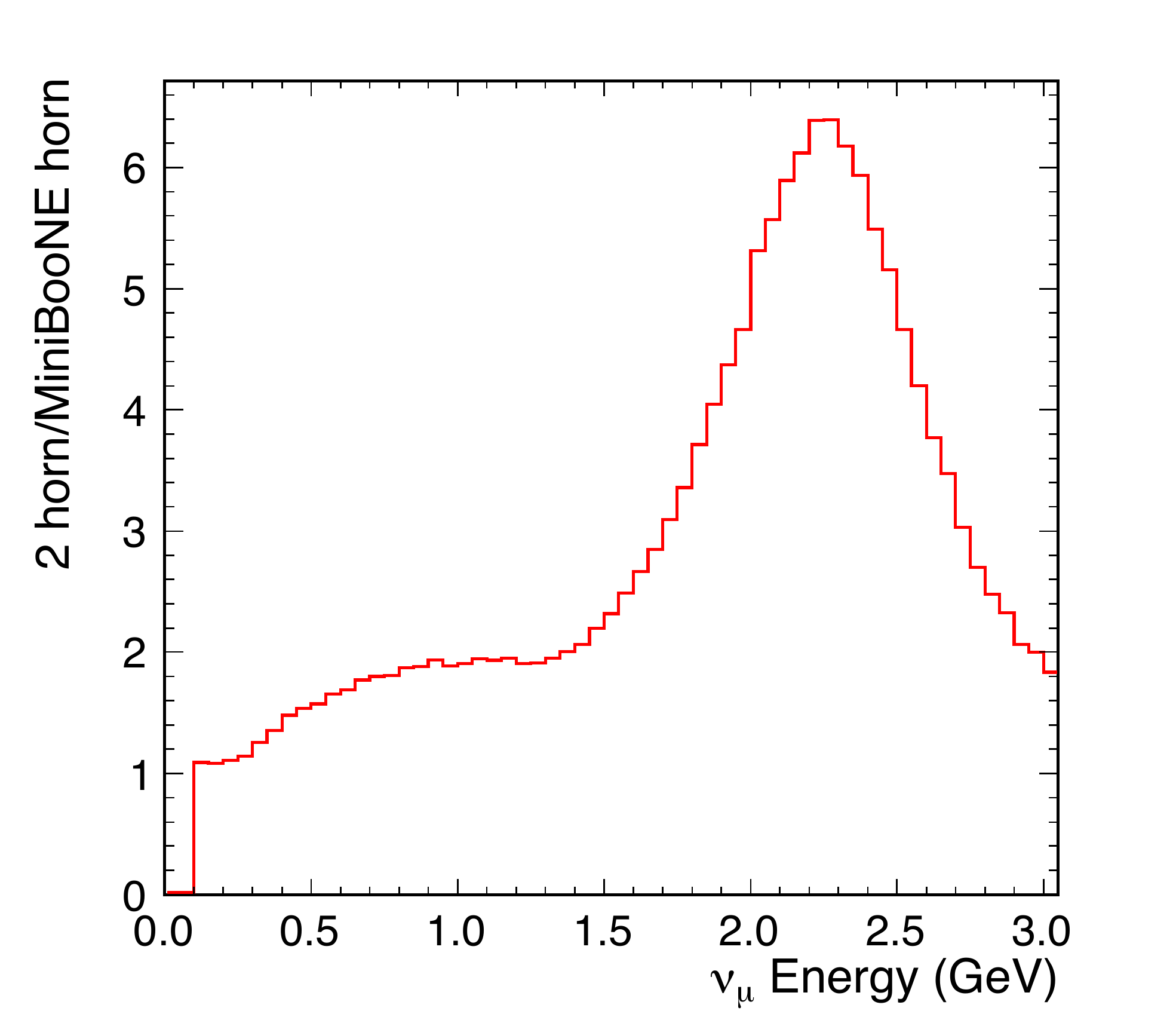}
      \caption{\label{fig:twohorn} Comparison of the expected neutrino flux with a two horn system to the present MiniBooNE horn focusing. The two horns were optimized to give the most neutrino events, while fitting the constraints of the existing target hall.   }
    \end{centering}
  \end{figure}
  
  Table~\ref{tab:twohorn_eventrate} shows the expected event rates with two horn system. It is important to note that the intrinsic \nue component which presents irreducible background for $\nu_{e}$ appearance measurement remains fractionally the same.
  
\begin{table}
 \begin{center}
    \begin{tabular}{|l|c|c|c|c|}
    
    \hline
                                       & \multicolumn{2}{c|}{CC ($Events/t/10^{20}POT$)} & \multicolumn{2}{c|}{Flux ($\nu/m^{2}/10^{6}POT$)} \\
                                    & MiniBooNE Horn & ~~2 Horn~~  & MiniBooNE Horn & ~~2 Horn~~ \\
       \hline \hline
       $\nu_{\mu}$          &    302.0                &  636.6     &    7.02  &  12.6    \\
       $\bar{\nu}_{\mu}$ &         2.6               &     2.9       &   0.44  &  0.41    \\
       $\nu_{e}$              &        2                   &     3.8      &   0.039 & 0.067    \\
       $\bar{\nu}_{e}$     &       0.06               &  0.06        &  0.004 & 0.004  \\ \hline
       \end{tabular}
    \caption{\label{tab:twohorn_eventrate}Predicted neutrino event rates with a two horn system compared to the present BNB configuration with MiniBooNE horn. The rates were calculated using CC inclusive cross section on Ar. Significant increase in the event rate is expected with reoptimized 2 horn system.}
\end{center}
\end{table}

%Finally, the choice of target material is being studied. The target material provides a very useful handle in optimizing the neutrino spectrum. The BNB uses a 71.1~cm long beryllium target. The target material was chosen for practical reasons; to allow efficient air-cooling and produce less residual radioactivity. Over the last decade, a lot of experience handling heavier targets has been developed at Fermilab. With heavier materials it is possible to build shorter targets while maintaining the same interaction length. Since a horn focal length depends on particle momentum, relatively higher momentum particles are focused from the upstream end of the target compared to the downstream end. This results in a broader spectrum for long targets, and more peaked spectrum for shorter targets.  Preliminary studies of the yields were done comparing beryllium, carbon, inconel and tungsten targets indicating that inconel provides the most flux in the peak.

It can also be seen from Table~\ref{tab:twohorn_eventrate} that the optimized two horn system has a much smaller wrong sign component compared to the original MiniBooNE horn configuration. Both the longer first horn, and the additional second horn further defocus wrong sign (WS) mesons. In neutrino mode this results in a reduction of the WS component by a factor of $\sim$2. While this is not an important feature in the neutrino mode, similar reduction is expected in the antineutrino mode where the WS component is significant. Hence, the two horn system would provide a much cleaner measurement of antineutrino oscillations as well as cross sections because the statistical and systematic uncertainties associated with subtracting the wrong sign component would be greatly reduced,

Further optimization of the system is possible. About 20\% of the neutrino flux in the MiniBooNE configuration is lost due to pion interactions within the horn conductor. The thickness of conductors in these preliminary studies was taken to be the same as for MiniBooNE horn. Thinner inner and outer conductor could be used, further reducing the losses. The transverse size of the first horn was kept the same as the original MiniBooNE horn. The horn current was limited 250kA, the upper limit of the present MiniBooNE power supply, and both horns were pulsed with same current. All of these parameters could be modified to fine tune the system. The possibility of movable target and horn longitudinal positions will also be explored. This would allow the beam to be tuned to higher or lower energies. Future information from running of MicroBooNE or other experiments might make that a useful capability to have, just as it was for the NuMI target and horn system in the pre-NOvA era.
 
Physical constraints of the existing target hall were taken into account in these preliminary studies. The two horn system requires more room along the beam axis than is presently available (see Figure~\ref{fig:tgthall}). 
Some modifications of the primary beamline and shielding within the target hall region would be necessary to accommodate the modified design as discussed in Section~\ref{sec:BNBreconf}. 

The total length of the optimized system was limited to a realistically achievable size. The transverse size of the second horn was limited to the dimensions of the chase. To fully take advantage of the larger second horn, the opening of the collimator at the entrance of the decay pipe was enlarged from 30 to 50cm. 

%There are two options that would provide more space along the beamline. The final focusing triplet could be replaced with much more compact system, and/or the bending of the beamline upstream of the target hall could be modified allowing the usage of shorter dipole magnets. These modifications would provide several meters of space, enough to fit the second horn.
 
The preliminary studies demonstrate that it is feasible to build a new focusing system that would increase event rate by a factor of 2 or more. This system would provide a huge improvement in the statistics as it doubles the count rate of every detector in the beamline. Further optimizations are possible as well as fine tuning of the horn focusing to shape the spectrum and maximize the physics potential of the experiment. 

The new system should be designed to take advantage of present and future accelerator upgrades. The present target/horn system and target hall shielding limits operations to 5~Hz average beam rate with up to $5 \times 10^{12}$ per spill. The first phase of Proton Improvement Plan (PIP) is presently underway and will allow Booster to deliver beam at a rate of up to 15~Hz starting in FY2016. Future improvements planned for PIP II will allow increasing booster rate to 20~Hz and spill intensity $6.5\times 10^{12}$ protons. The design of an  upgrade to horn system components should be made capable of handling the higher repetition rate and spill intensity.

%%%%%%%%%%%%%%%%%%%%%%%%%%%%%%%%%%%%%%%%%%%%%%%%%%%%%%%%%%%%%%%%%%%%%%%%%%%%%%%%%%
\section{Making Space for the New Horn Configuration}
\label{sec:BNBreconf}

In order to accommodate the two-horn system, an additional 5m of space is needed in the Booster Neutrino Beamline.

The Booster Neutrino Beamline (BNB) has three sections.
The first section is in the Main Injector tunnel, the second section is a carrier pipe transporting the beam under a road, and the third section points the beam toward the detectors and focuses the beam on the target.
This third section is composed of optics to capture the beam from the carrier pipe, a regular lattice to transport the beam through the arc, a vertical dogleg to raise the beam to the height of the target, and a final focusing triplet.
The third section is located in the MI12 tunnel and the target hall.
The MI12 tunnel is ten feet wide, and the height is eight feet or nine feet, six inches.
The tunnel changes height at the approximate center of the first magnet of the dogleg.
The target hall, when all shielding is in place, is 23 feet wide, 24 feet deep, and 13 feet high ($7.0 \times 7.3 \times 4.0$~meters).

In order to gain the additional 5m the dogleg can be moved upstream, beginning in the last cell of the arc and ending at the transition to the higher enclosure.
A slight adjustment of position of the quadrupole matching the lattice to the final focusing triplet is also required.
A calculation using TRANSPORT~\cite{Carey:1998nu} shows that a 1mm round beam can be focused at the center of the target with the quadrupole at acceptable currents.

In addition to changes in the beamline, the target pile must also be reconfigured.
The present target pile consists of steel blocks filling the downstream half of the target hall.
The steel is covered by concrete blocks above.
The pile has concrete stacked in front, with an opening large enough to accommodate the horn.
Adding an additional 5m of shielding upstream of the existing target pile should be possible.

Figure~\ref{fig:beamline_reconfig} shows the present and proposed beamline configurations indicating how the space needed for a two horn system can be recovered by adjusting the beamline components.

\begin{figure}[!h]
    \begin{centering}
      \includegraphics[width=\textwidth]{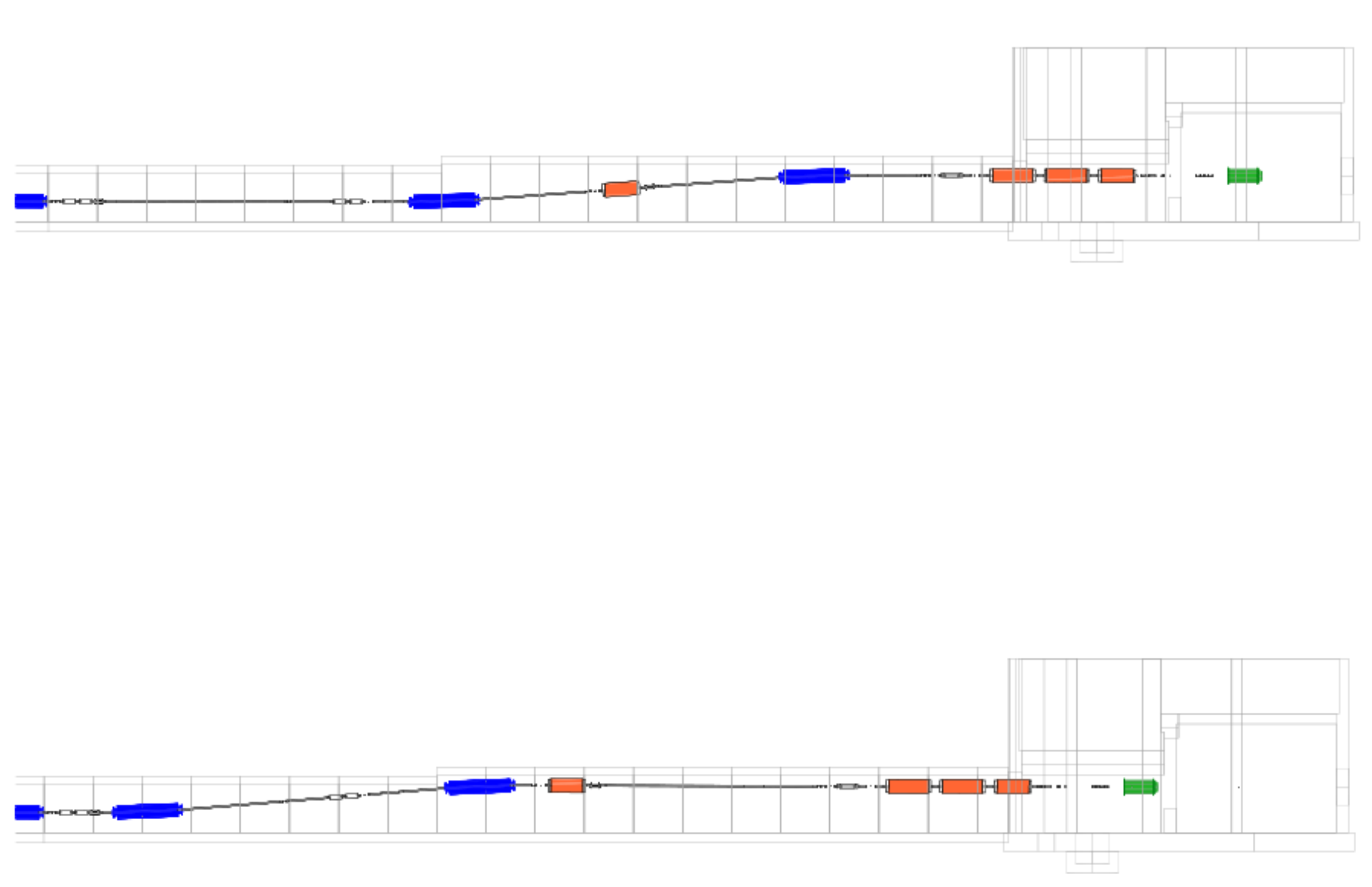}
      \caption{\label{fig:beamline_reconfig} Side elevations of the Booster Neutrino Beamline (BNB). The present (top) and proposed (bottom) beamline configurations are shown.  The dogleg dipoles are shown in blue, the triplet and matching quadrupoles in orange, and the horn in green.  The new dogleg is initiated at the beginning of the last cell of the lattice and completed where the enclosure roof rises.  The triplet and horn are moved five meters upstream.  The location of the matching quadrupole is adjusted slightly. }
    \end{centering}
  \end{figure}

Figure~\ref{fig:tgthall_reconfig} shows the line 5m upstream of the existing target pile.
The new target pile would not occlude the door, although it would cover the sump.
Shielding would have to be configured such that the pumps in the sump can be replaced.
Existing utilities, such as the cooling skid for the horn, would have to be relocated, perhaps upstream, under the raised beamline.

Another option would be to reconfigure the target pile to allow for more space downstream of the target.
This would entail removing all equipment from the MI12 service building, removing the existing shielding blocks, and handling the radioactive steel.
However, enough space exists so as not to cover the sump.

\begin{figure}[!h]
    \begin{centering}
      \includegraphics[width=0.6\textwidth]{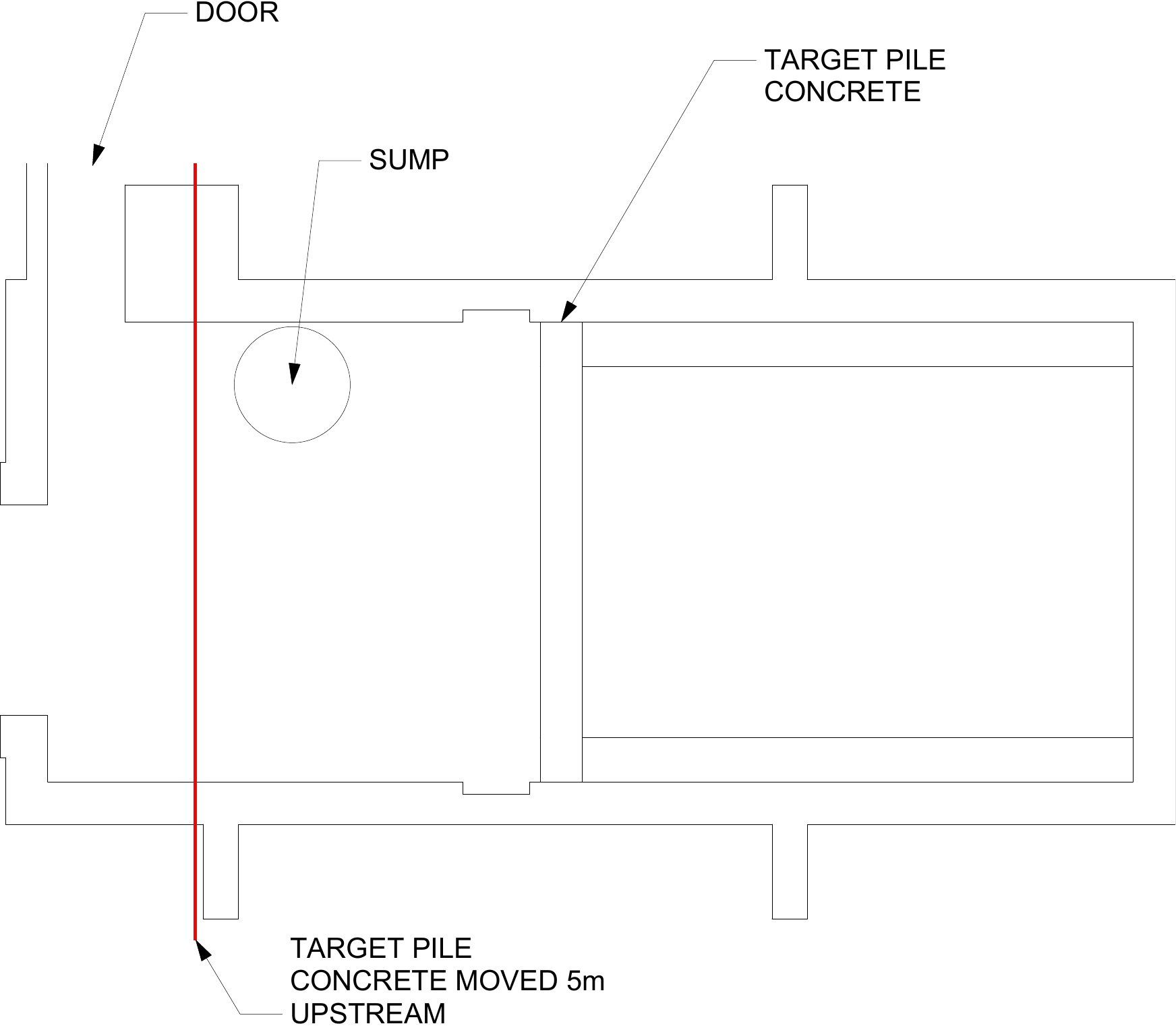}
      \caption{\label{fig:tgthall_reconfig} Plan view of target hall showing front face of existing target pile and location 5m upstream.  The door is not blocked, but the sump would be covered. }
    \end{centering}
  \end{figure}

%%%%%%%%%%%%%%%%%%%%%%%%%%%%%%%%%%%%%%%%%%%%%%%%%%%%%%%%%%%%%%%%%%%%%%%%%%%%%%%%%%
\section{Secondary Beamline Instrumentation}

In this section the current secondary beamline monitoring is described along with some possible upgrades. It should be noted that these monitoring upgrades are completely independent of the horn system upgrades of the previous three sections.

The present secondary beam has minimal instrumentation, consisting of a cross formed by 22 loss monitors located behind the 50m absorber.
Twelve loss monitors are placed vertically, approximately six inches apart; and ten loss monitors placed horizontally, five on each side of the vertical column, also spaced approximately six inches apart.
The loss monitors are read out through a segmented wire ionization chamber (SWIC) scanner, allowing one to see horizontal and vertical profiles.

The fifty meter absorber consists of 24 blocks of steel stacked roughly into a cube, ten foot on side.
The steel is rough cut.
This is followed by a ten foot square by three foot deep concrete block.
The secondary monitor follows.
A stack of steel, eight foot square by two feet deep, ends the absorber.
The absorber is buried directly in the ground -- no enclosure exists -- eliminating the possibility of easily repairing the muon monitor.
A steel pipe carries the signal wires to the surface.

Ideally, one would replace the 50m absorber and provide better instrumentation.
A hadron monitor would be placed at the upstream end and a muon monitor at the downstream end.
The existing steel would be removed and replaced with more uniform steel plates, eliminating any transverse gaps.
An enclosure would be provided to allow for the repair or replacement the hadron or muon monitor.

Constructing such a feature would entail digging into the berm and removing the present 50m absorber.
Controls would be in place to manage the irradiated aggregate and steel.
The existing water barrier would be breached and resealed around the new enclosure.
Power would be run to the new enclosure.
Adequate shielding would be placed between the absorber and enclosure, and a means of removing it thought about.
Rebuilding the 50m absorber would require significant engineering.

Another option would be a retractable profile monitor at the 25m absorber. This absorber consists of a series of steel and concrete plates that can be lowered into the secondary beamline halfway down the 50m absorber.

In the autumn of 2014, the 25m absorber hatch was opened and the modules adjusted longitudinally to provide a 3/4 inch gap.
A profile monitor, 5/8 inch thick, was inserted through this gap to nominal beam center.
The monitor consisted of 48 horizontal wires and 48 vertical wire, each plane having a 2mm pitch.
The primary beam was steered around the target and observed on the monitor.

With moderate engineering effort, one could design a profile monitor which would be remotely inserted for alignment runs and retraced for normal runs.
The monitor could be of adequate size to see both the primary and secondary beams.
By appropriate choice of gain one may be able to distinguish between primary and secondary beam.

%%%%%%%%%%%%%%%%%%%%%%%%%%%%%%%%%%%%%%%%%%%%%%%%%%%%%%%%%%%%%%%%%%%%%%%%%%%%%%%%%%
\section{Request}

Based on the preliminary studies outlined above we make the following requests
\begin{itemize}

\item A detailed study of the cost and schedule for conversion
to a two horn system should be initiated immediately. This should include the
cost of new horns, new or refurnished power supplies, and the necessary work for reconfiguration of the incoming beamline and of the collimator. The system should be capable of (or readily upgradeable to) operation up to 20Hz and of taking the beam intensities anticipated once the PIP II project is complete.

\item A detailed study of the cost, schedule, impacts, and benefit of improving the secondary beamline instrumentation of the BNB should be initiated immediately. This should include studies of what instrumentation might be placed near the horn(s), at the 25m absorber, and in the 50m absorber. The instrumentation should be capable of (or readily upgradeable to) operation up to 20Hz and of taking the beam intensities anticipated once the PIP II project is complete.

\end{itemize}

%%%%%%%%%%%%%%%%%%%%%%%%%%%%%%%%%%%%%%%%%%%%%%%%%%%%%%%%%%%%%%%%%%%%%%%%%%%%%%%%%%

 \clearpage
\pagestyle{empty}

\proposalTitle 

\setcounter{part}{5}
\part{Coordination and Schedule}

%\begin{center}
%{\large
%DRAFT \\
%\bigskip
%\today
%\proposaldate
%}\end{center}

\ifcombine
\clearpage
\else
\clearpage
\tableofcontents
\addtocontents{toc}{\protect\thispagestyle{empty}}
\clearpage
\setcounter{page}{0}
\fi

\pagestyle{fancy}
\rhead{VI-\thepage}
\lhead{SBN Program Coordination and Schedule}
\cfoot{}

%%%%%%%%%%%%%%%%%%%%%%%%%%%%%%%%%%%%%%%%%%%%%%%%%%%%%%%%%%%%%%%%%%%%%%

\section{Introduction}

This proposal is being submitted jointly by three separate scientific collaborations: ICARUS-WA104, \larnd, and \uboone.  Each has an existing organizational structure with an existing scientific mission.  The preparation of the proposal is the result of a collaborative effort between the collaborations guided by the SBN task force and the Fermilab SBN coordinator.  The mandate of the task force and the five associated working groups ends with the submission of the proposal.  In the desired event that this proposal is accepted, a new organizational structure will be necessary to ensure a successful program.  The form of the future SBN organization has been the topic of discussions between the leadership of the three collaborations, the management of Fermilab, the management of INFN, and the management of CERN. Since these discussions have not concluded, we only present here some general observations.  The new structure will need to ensure the following will happen:
\begin{enumerate}
\item the \uboone experiment (E-974) carries out the already  approved physics program as an independent collaboration; 
\item the \icarus detector is refurbished, transported to Fermilab, installed and commissioned;
\item the \larnd detector is designed, constructed, installed and commissioned;
\item the required infrastructure (e.g. buildings, cryogenics, computing) are constructed or purchased and installed; 
\item necessary common activities, like cosmic taggers, are designed, constructed and installed;  
\item necessary reconstruction and analysis software for a multi-detector oscillation analysis is designed, developed and tested;
\item necessary improvements of the booster neutrino beam for an increased neutrino intensity are studied in detail to provide implementation before the starting of data taking.

\end{enumerate}
All of these activities must take place simultaneously in a very short time-frame.

We expect that the three collaborations will continue to exist in whatever new structure is created for some time to come.  %this organizational structure at least through the commissioning of the near and far detectors in early 2018.  
The collaborations each have a clear unique role to play in delivering the first three of the above items.  The fourth is the responsibility of Fermilab, as host laboratory, in collaboration with international partners such as CERN and INFN. The collaborations have a vested interest in delivery of this infrastructure so the organization must account for that.  The last items are clearly of interest to members of all three collaborations.     

The three scientific collaborations have worked together through the task force and working groups for the past eight months. As discussed in Part~1 of this proposal, the successful analysis of data from all three detectors for the oscillation analyses will require a coordinated effort on common reconstruction and analysis tools.  This effort should start well in advance of start of operations.  Joint reconstruction and oscillation analysis group(s) could form the nucleus of the future scientific organization.  It will be natural for members of the SBN physics working groups and other relevant people from the collaborations to initiate this effort. 

%In the following sections, we describe the proposed organization for oversight of the SBN program that will ensure that the detectors and necessary infrastructure are completed.  

In the following sections we describe the schedule and funding for completion of the components listed above.

\FloatBarrier

\section{Schedule}

Initial data-taking with all three detectors operational is foreseen in spring 2018. By this time, the MicroBooNE detector will already have been operational with beam for several years.  All steps to prepare the near and far detectors must be accomplished by this time, including design and construction of the near and far buildings, construction of cryostats and cryogenics, preparation of the \icarus detector, construction of the near detector, detector installation and commissioning. The proposed schedule is very tight, but with a good level of coordination, it is judged to be feasible.
Table~\ref{tab:milestones} shows the high level set of milestones.     

\begin{table}[h]
\centering
\begin{tabular}{|l|c|}
\hline
 Milestone & Date \\

\hline \hline

Far Detector building: final CE requirements and start final design
  & Nov 2014	 \\ \hline
Far Detector: T600 at CERN and refurbishing starts  &Dec 2014	\\ \hline
Submit SBN proposal to PAC & Dec 2014 \\ \hline
Near Detector: Start preliminary design of TPC and installation & Dec 2014 \\ \hline
PAC Review of SBN Proposal & Jan 2015 \\ \hline
Near Detector cryostat: Start preliminary design 	&  Jan 2015  \\ \hline
Cryogenic plants: Start preliminary design & Jan 2015  \\ \hline 
Draft MOUs: e.g. Fermilab-INFN, Fermilab-CERN, Fermilab-CH-NSF   & Feb 2015 \\ \hline
Near Detector building: final CE requirements and start final design
  & Feb 2015	 \\ \hline
Independent Review of Near Detector, ND Cryostat and ND Cryogenics (CD-1/2 like) & May 2015 \\ \hline
Independent Review of Far Detector refurbishing, cryogenics and installation planning  & May 2015 \\ \hline  
Far Detector building: ground breaking & May 2015 \\ \hline
Near Detector building: ground breaking & Aug 2015 \\ \hline
Independent review of near detector production readiness & Nov 2015 \\ \hline
Independent review of far detector production readiness & Nov 2015 \\ \hline
Near Detector building: beneficial Occupancy & Sept 2016 \\ \hline
Independent review of installation readiness for near and far detectors & Oct 2016 \\ \hline
Far Detector building: beneficial Occupancy & Nov 2016 \\ \hline
Near Detector cryostat: start installation  & Nov 2016 \\ \hline
Far Detector cryostat: start installation  & Dec 2016 \\ \hline
Far Detector: \icarus ready at CERN for transport &Dec 2016 \\ \hline
Near and Far Detector Buildings Complete & Jan 2017 \\ \hline
Far Detector: start T600 installation &Mar 2017 \\ \hline
Near Detector: Start LAr1-ND installation & April 2017 \\ \hline
Far Detector: T600 Installed & May 2017 \\ \hline
Near Detector: LAr1-ND Installed & July 2017 \\ \hline
Far Detector: Cryogenic plant complete & Aug 2017 \\ \hline
Near Detector: Cryogenic plant complete & Oct 2017 \\ \hline
Start detectors cooling and commissioning 	&Nov 2017 	 \\ \hline
Start data taking with beam &Apr 2018 \\ \hline

\end{tabular}
\caption{Overall Milestones for construction, installation and initial commissioning of the Short Baseline Neutrino Program}
\label{tab:milestones}
\end{table}

Figure~\ref{fig:Schedule} shows the schedule in a summary format for near and far detector, pointing to the start of data taking with beam to April 2018.

\subsection{Near Detector Schedule}
\label{sec:ND_Schedule}

The schedule for design, fabrication, assembly and installation of the components of near detector is shown in Table~\ref{tab:ND_Schedule}.
%~\&~\ref{tab:ND_Schedule2}. 
A detailed resource loaded schedule for these tasks is under development and will be presented at the Independent review in the Spring of 2015.

\begin{table}[h]
\begin{center}
\begin{tabular}{|l|c|c|}
\hline
\hline

{\bf TPC}						& 	Start	&	End	\\ \hline \hline
Requirements Documents			&			& Feb 2015	\\ \hline
Preliminary Design				& Nov 2014	& Jan 2015	\\ \hline
Design Review					& 			& Mar 2015	\\ \hline
Final Design 					& Jan 2015	& Jun 2015	\\ \hline
Production Readiness Review
								&			& Jul 2015	\\ \hline
Fabrication	(APAs, CPAs, FCA)	& Jul 2015	& Apr 2016	\\ \hline
QA (Cold Tests)					& Apr 2016	& Jun 2016	\\ \hline
Delivery to Fermilab by			&       	& Sep 2016	\\ \hline
Assembly						& Sep 2016	& Feb 2017	\\ \hline
Installation					& Apr 2017	& Jul 2017	\\ \hline \hline

{\bf Cold Electronics}			& 			&		\\ \hline \hline
Requirements Documents			&			& Mar 2015	\\ \hline
Preliminary Design				& Jan 2015	& Sep 2015	\\ \hline
Design Review 					& 			& Oct 2015	\\ \hline
Final Design 					& Oct 2015	& Mar 2016	\\ \hline
Production Readiness Review
								&			& Mar 2016	\\ \hline
Fabrication						& Apr 2016	& Jun 2016	\\ \hline
Assembly						& Jul 2016	& Sep 2016	\\ \hline
Delivery to Fermilab by			& 	& Oct 2016	\\ \hline
Installation					& Nov 2016	& Jul 2017	\\ \hline \hline

{\bf Light Detector}			& 			& 	\\ \hline \hline

Preliminary Design				& Dec 2014	& Mar 2015	\\
\hline
Choice of technology			&			& Mar 2015			\\ \hline
Final Design 					& Apr 2015	& Sep 2015	\\ \hline
Production Readiness Review
								&			& Oct 2015	\\ \hline
Fabrication						& Nov 2015	& May 2016	\\ \hline
Assembly						& Jun 2016	& Nov 2016	\\ \hline
Delivery to Fermilab by			& Dec 2016	& Dec 2016	\\ \hline
Installation					& Jan 2017	& Jul 2017	\\ \hline 

{\bf Laser Calibration}			& Start		& End	\\ \hline \hline
Requirements Documents			&			& Jan 2015 	\\ \hline
Preliminary Design				& 			& Complete	\\ \hline
Final Design 					& Mar 2015	& Jul 2015	\\ \hline
Production Readiness Review
								&			& Jul 2015	\\ \hline
Fabrication						& Jul 2015	& Jan 2016	\\ \hline
Assembly						& Jan 2016	& Jun 2016	\\ \hline
Delivery to Fermilab			& Jun 2016	& Aug 2016	\\ \hline
Installation					& Jul 2017	& Sep 2017	\\ \hline

%\hline

\end{tabular}
\end{center}
\caption{Summary of \larnd detector component schedules}
\label{tab:ND_Schedule}
\end{table}

\begin{comment}
\begin{table}[h]
\begin{center}
\begin{tabular}{|l|c|c|}
\hline
\hline

{\bf Laser Calibration}			& Start		& End	\\ \hline \hline
Requirements Documents			&			& Jan 2015 	\\ \hline
Preliminary Design				& 			& Complete	\\ \hline
Final Design 					& Mar 2015	& Jul 2015	\\ \hline
Production Readiness Review
								&			& Jul 2015	\\ \hline
Fabrication						& Jul 2015	& Jan 2016	\\ \hline
Assembly						& Jan 2016	& Jun 2016	\\ \hline
Delivery to Fermilab			& Jun 2016	& Aug 2016	\\ \hline
Installation					& Jul 2017	& Sep 2017	\\

%{\bf Integration and Installation}			
%								& 			&		\\ \hline %\hline
%Requirements Documents			& Dec 2014	& Jan 2015 	\\ \hline
%Preliminary Design				& Feb 2015	& May 2015	\\ \hline
%Preliminary Design Review		& Jun 2015	& Jun 2015	\\ \hline
%Final Design 					& Jul 2015	& Sep 2016	\\ \hline
%Installation Readiness Review
%								&			& Oct 2016	\\ \hline
%Fabrication					& Nov 2016	& Dec 2016	\\ \hline
%Delivery to Fermilab			& Jan 2017	& Jan 2017	\\ \hline
%Installation					& Feb 2017	& Jul 2017	\\ \hline %\hline
\hline 
\hline

\end{tabular}
\end{center}
\caption{Summary of LAr1-ND detector component schedules}
\label{tab:ND_Schedule2}
\end{table}
\end{comment}

\FloatBarrier
\subsection{Far Detector Schedule}
\label{sec:T600_Schedule}

As for the Near Detector, all steps to prepare \icarus must be accomplished by this time, from the submission of the initial design report to start-up of the civil engineering work, construction of cryostats and cryogenics, transportation to site, and overall detector preparation and commissioning. 

\begin{table}[h]
\begin{center}
\begin{tabular}{|l|c|c|}
\hline
\hline

{\bf  movement to CERN}		& Start		& End	\\ \hline \hline
Laser survey					&			& Complete 	\\ \hline
Transportation Frames   		& 			& Complete	\\ \hline
Disassembly work 				&  	        & Complete	\\ \hline
TPC transport to CERN           &  	        & Complete	\\ \hline
Equipment transport to CERN		&  	        & Complete	\\ \hline
Building 185 preparation		&  	        & Complete	\\ \hline
\hline

{\bf T600 cryostats}		    & 			& 	\\ \hline \hline
Engineering cold vessel + production & Jan-13  & Sep-16	\\ \hline
Engineering cold vessel supports &  	    & complete	\\ \hline
Engineering warm vessel 		&  Dec14  & May15	    \\ \hline
GTT preliminary study    		&  	      & complete	\\ \hline
Procurement extruded Aluminum  		&  Dec14  & May15	\\ \hline
Assembly cold vessels  			&  Jun15   & Dec16		\\ \hline
Procurement Warm Vessel + supports 	&  Dec14  & Mar16	\\ \hline
Procurement Insulation  		&  Jan15  & Dec16		\\ \hline
Warm vessel assembly 			&  Oct15  & Jun16		\\ \hline
Insulation installation 		&  Sep16  & Dec16		\\ \hline
Cold shields		 			&  Jan15  & Dec16		\\ \hline
\hline

\end{tabular}
\end{center}
\caption{Summary of ICARUS-WA104 detector overhauling schedules}
\label{tab:FDschedule}
\end{table}

The proposed schedule is illustrated in tables ~\ref{tab:FDschedule} and ~\ref{tab:FDschedule2}. It is a very tight schedule, but with a good level of coordination, it is judged to be feasible, as it is based on the previous experience gained with the Pavia test run in 2001 \cite{ICARUS_NIM} and the Gran Sasso Physics Run, 2010-2013 \cite{ICARUS_jinst}. Moreover most of the milestones are related to operations at CERN in the frame of the WA104 program, therefore a periodic verification of the project development will be assured.
The \icarus detector has already be transported to CERN.

\begin{table}[h]
\begin{center}
\begin{tabular}{|l|c|c|}
\hline
\hline

{\bf T600 cryogenics} 			& 		   &	\\ \hline \hline
Reorganization and packaging    & Jan-15   & Dec-16     \\ \hline
New hardware 				    & May-15   & Oct-16     \\ \hline 
Tests and maintenance 			& Jan-15   & Mar-16     \\ \hline
Cryo group at CERN 				& Mar-15   & Oct-16 	\\ \hline
Vacuum test 					& Aug-16   & Nov-16     \\ \hline
Cold gas test (if possible) 	& Nov-16   & Dec-16     \\ \hline
\hline

{\bf T600 refurbishing}		& 			&	\\ \hline \hline
TPC-1 Cabling TPC internal  & Dec-14   &	Aug-15 \\\hline
TPC-1 New PMT procurement   & 	Oct-14 &	Aug-15 \\\hline
TPC-1 New PMT installation  & Mar-15 &	Sep-15 \\\hline
TPC-1 installation in cold vessel  & Nov-15 &	Jan-16 \\\hline
TPC-2 Cabling TPC internal  & Jan-16 &	Oct-16 \\\hline
TPC-2 New PMT procurement   & Oct-15 &	Jul-16 \\\hline
TPC-2 New PMT installation  & Feb-16 &	Oct-16 \\\hline
TPC-2 installation in cold vessel & Nov-16 &	Dec-16 \\\hline
New electronics     		&	Mar-14 &	Dec-16 \\\hline
Final assembly into the warm vessel & Dec-16 & Jan-17 \\ \hline
\hline

{\bf T600 controls and tests} 			& 	Start	   & End	\\ \hline \hline
 Slow controls hardware    & Jan-16 &	Nov-16 \\\hline
Slow controls software     & Jan-16 &	Nov-16 \\\hline
DAQ system 				   & Jan-16 &	Nov-16 \\\hline
Tests of Slow controls 	   & Jan-16 &	Nov-16 \\\hline
Tests of DAQ 			   & Jan-16 &	Nov-16 \\\hline

\hline

\end{tabular}
\end{center}
\caption{Summary of ICARUS-WA104 detector overhauling schedules}
\label{tab:FDschedule2}
\end{table}

%\clearpage
\subsection{Infrastructure Schedule}

Table~\ref{tab:cryogenics_Schedule} shows the schedule for the design, procurement and installation of the cryogenic systems for the near and far detectors.  Table~\ref{tab:ND_Cryostat_Schedule} shows the schedule for the design, procurement and installation of the near detector cryostat. The schedule for the far detector cryostat is included in the schedule for the far detector (Table~\ref{tab:FDschedule}).
The schedule for the construction and installation of Cosmic taggers for both near and far detector are presented in Tab. ~\ref{tab:Cosmics-tagger}.

\begin{table}[h]
\begin{center}
\begin{tabular}{|l|c|c|}
\hline
\hline

{\bf Near Detector Cryostat}			
								& 	Start		&	End	\\ \hline \hline
Requirements Documents			&			& Dec 2014 	\\ \hline
Preliminary Design				& Jan 2015	& May 2015	\\ \hline
Production Readiness Review		&			& Jun 2015	\\ \hline
%Procurement Documents			& n/a		& n/a 		\\ \hline
Final Design					& Jul 2015	& Nov 2015	\\ \hline
Procure membrane cryostat materials 															& Dec 2015	& Jun 2016	\\ \hline  
Procure support structure  		& Dec 2015	& Apr 2016	\\ \hline  
Procure top plate				& Feb 2016	& Oct 2016	\\ \hline
Delivery to Fermilab (structure)& May 2016	& Jun 2016	\\ \hline
Delivery to Fermilab (top plate)& Oct 2016	& Nov 2016	\\ \hline
Assembly						& Jul 2016	& Nov 2016	\\ \hline
Installation					& Dec 2016	& Jan 2017	\\ \hline
Safety Review Complete			& 			& Nov 2017	\\ \hline \hline

\hline
\hline
\end{tabular}
\end{center}
\caption{Summary of schedules for design, procurement and installation of the near detector cryostat.}
\label{tab:ND_Cryostat_Schedule}
\end{table}

\begin{table}[h]
\begin{center}
\begin{tabular}{|l|c|c|}
\hline
\hline

{\bf Near Detector LAr Cryogenics}			
								& 	Start		&	End	\\ \hline \hline
Requirements Documents			&			& Dec 2014 	\\ \hline
Preliminary Design				& Jan 2015	& Jun 2015	\\ \hline
Production Readiness Review		&			& Aug 2015	\\ \hline
Procurement Documents			& Sep 2015	& Mar 2016	\\ \hline
Final Design					& Apr 2016	& Jun 2016	\\ \hline
Fabrication 					& Jul 2016	& Apr 2017	\\ \hline  
Delivery of skids to Fermilab	& Jun 2017	& Jul 2017	\\ \hline
Installation					& May 2017	& Oct 2017	\\ \hline
Safety Review Complete			& 			& Nov 2017	\\ \hline \hline

{\bf Near Detector LN$_2$ Cryogenics}			
								& 			&		\\ \hline \hline
Requirements Documents			&			& Dec 2014 	\\ \hline
Preliminary Design				& Jan 2015	& Jun 2015	\\ \hline
Production Readiness Review		&			& Aug 2015	\\ \hline
Procurement Documents			& Sep 2015	& Nov 2015	\\ \hline
Final Design					& Dec 2015	& Mar 2016	\\ \hline
Fabrication 					& Apr 2016	& Dec 2016	\\ \hline  
Delivery to Fermilab			& Jan 2017	& Feb 2017	\\ \hline
Installation					& Mar 2017	& Jul 2017	\\ \hline
Safety Review Complete			& 			& Nov 2017	\\ \hline \hline

{\bf Far Detector LAr Cryogenics}			
								& 			&		\\ \hline \hline
Existing Cryogenics at CERN		&			& Nov 2014 	\\ \hline
Reorganization and Packaging (Part 1)				
								& Jan 2015	& Jun 2015	\\ \hline
Production Readiness Review		&			& Aug 2015	\\ \hline
Reorganization and Packaging (Part 2)			
								& Sep 2015	& Dec 2016	\\ \hline
New Hardware					& May 2015	& Oct 2016	\\ \hline
Test and Maintenance 			& Jan 2015	& Mar 2016	\\ \hline  
Vacuum Test						& Aug 2016	& Nov 2016	\\ \hline 
Cold Test						& Nov 2016	& Dec 2016	\\ \hline
Delivery to Fermilab			& Jan 2017	& Feb 2017	\\ \hline
Installation					& Mar 2017	& Aug 2017	\\ \hline
Safety Review Complete			& 			& Sep 2017	\\ \hline \hline

{\bf Far Detector LN$_2$ Cryogenics}			
								& 			&		\\ \hline \hline
Requirements Documents			&			& Dec 2014 	\\ \hline
Preliminary Design				& Jan 2015	& Jun 2015	\\ \hline
Production Readiness Review		&			& Aug 2015	\\ \hline
Final Design					& Sep 2015	& Nov 2015	\\ \hline
Procurement Documents			& Dec 2015	& Mar 2016	\\ \hline
Fabrication 					& Apr 2016	& Dec 2016	\\ \hline  
Delivery to Fermilab			& Jan 2017	& Feb 2017	\\ \hline
Installation					& Mar 2017	& Jul 2017	\\ \hline
Safety Review Complete			& 			& Sep 2017	\\ \hline \hline

\hline
\hline
\end{tabular}
\end{center}
\caption{Summary of schedules for design, procurement and installation of the cryogenic systems for the near and far detectors.   }
\label{tab:cryogenics_Schedule}
\end{table}

Table~\ref{tab:Civil_Schedule} shows the schedule for design and construction of the far detector and near detector buildings.  Also shown is the schedule for the associated site preparation work.   The site preparation includes elements that will be completed after construction of the buildings is completed.

\begin{table}[h]
\begin{center}
\begin{tabular}{|l|c|c|}
\hline
\hline

{\bf Site Preparation}			& 	Start	& End		\\ \hline \hline
Preliminary Design				& Oct 2014	& Nov 2014	\\ \hline
Requirements Documents			&			& Nov 2014 	\\ \hline
Final Design 					& Jan 2015	& Mar 2015	\\ \hline
Bidding	and Procurement			& Apr 2015	& May 2015	\\ \hline
Construction					& May 2015	& Jan 2017	\\ \hline \hline

{\bf Far Detector Building}		& 			&		\\ \hline \hline
Preliminary Design				& Jun 2014	& Sep 2014	\\ \hline
Requirements Documents			&			& Nov 2014 	\\ \hline
Final Design 					& Nov 2014	& Mar 2015	\\ \hline
Final Design Review				&			& Mar 2015  \\ \hline
Bidding	and Procurement			& Apr 2015	& May 2015	\\ \hline
Construction					& May 2015	& Jan 2017	\\ \hline
Beneficial Occupancy 			& 			& Nov 2016	\\ \hline \hline

{\bf Near Detector Building}	& 			&		\\ \hline \hline
Preliminary Design				& Jun 2014	& Sep 2014	\\ \hline
Requirements Documents			&			& Feb 2015 	\\ \hline
Final Design 					& Feb 2015	& May 2015	\\ \hline
Final Design Review				&			& May 2015	\\ \hline
Bidding	and Procurement			& Jun 2015	& Aug 2015	\\ \hline
Construction					& Aug 2015	& Nov 2016	\\ \hline
Beneficial Occupancy 			& 			& Sep 2016	\\ \hline \hline

\end{tabular}
\end{center}
\caption{Summary of civil construction schedules for the site preparation, far detector building and near detector building }
\label{tab:Civil_Schedule}
\end{table}

\begin{table}[h]
\begin{center}
\begin{tabular}{|l|c|c|}
\hline
\hline

{\bf Cosmic Tagger}				& 			&		\\ \hline \hline
Requirements Documents			&			& Jan 2015 	\\ \hline
Preliminary Design				& Nov 2014	& Mar 2015	\\ \hline
Final Design 					& Mar 2015	& Jul 2015	\\ \hline
Production Readiness Review
								&			& Jul 2015	\\ \hline
Fabrication	of strips			& Jul 2015	& Jun 2016	\\ \hline
Assembly bottom plane			& Mar 2016	& Jun 2016	\\ \hline
Assembly side planes			& Jun 2016	& Dec 2016	\\ \hline
Assembly top planes				& Mar 2016	& Apr 2017	\\ \hline
Delivery of bottom to Fermilab 	& Jun 2016	& Aug 2016	\\ \hline
Delivery of sides to Fermilab 	& Dec 2016	& Feb 2017	\\ \hline
Delivery of top to Fermilab 	& Apr 2017	& Jun 2017	\\ \hline
Installation bottom				& Nov 2016	& Nov 2016	\\ \hline
Installation sides				& Sep 2017	& Sep 2017	\\ \hline
Installation top				& Dec 2017	& Dec 2017	\\ \hline \hline

\end{tabular}
\end{center}
\caption{Summary of construction and installation of the Cosmics tagger}
\label{tab:Cosmics-tagger}
\end{table}

\begin{figure}[h]
\centering
\includegraphics [width=1.05\textwidth, trim=0mm 15mm 0mm 25mm, clip]{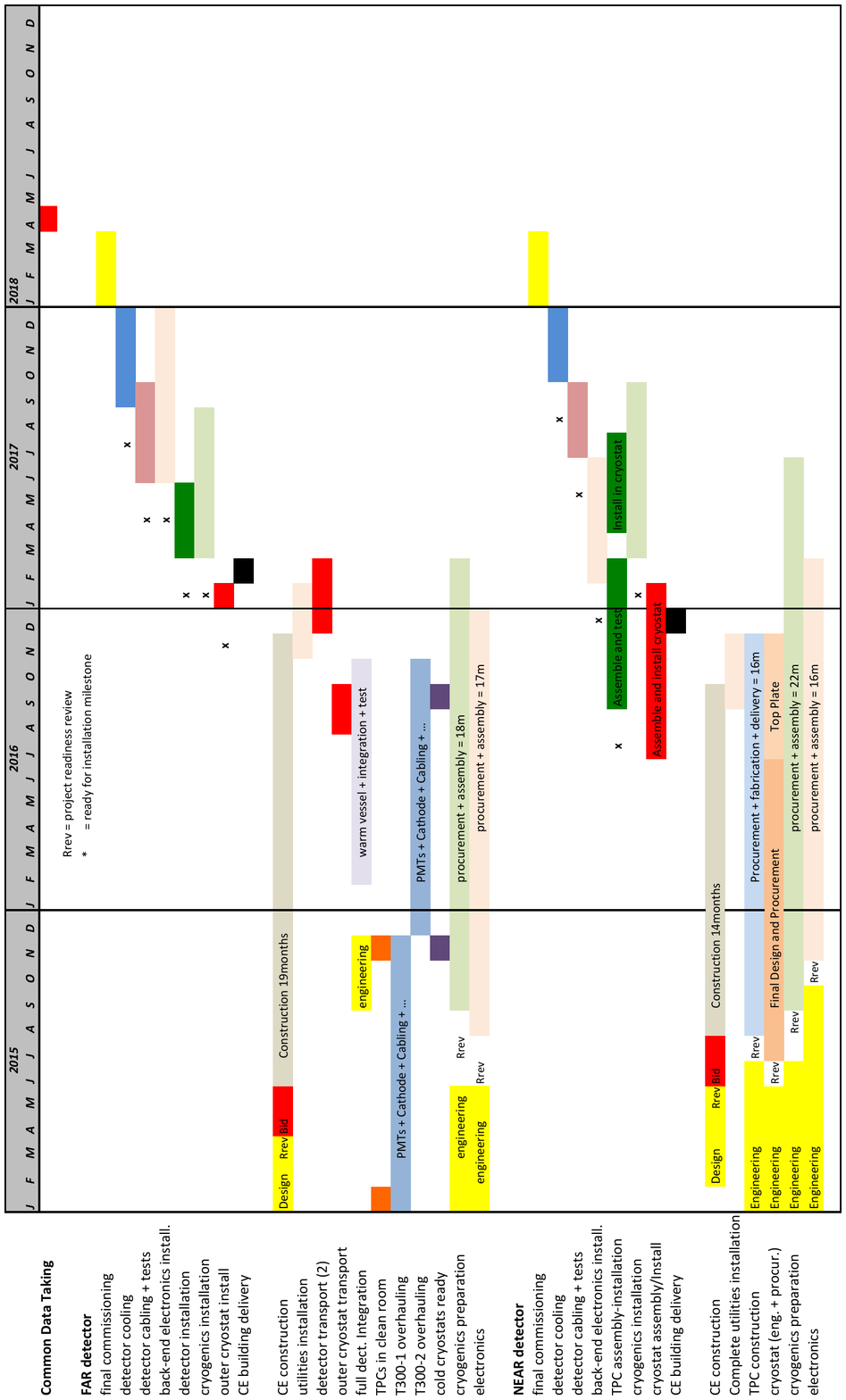}
\begin{center}
\caption{Overall summary schedule for the far and near detector construction}
\label{fig:Schedule}
\end{center}
\end{figure}

\ifarxiv
\else

\FloatBarrier
\section{Program Funding}

The funding of the overall program will have several sources from the U.S. and Europe.  As the host laboratory, Fermilab will be responsible for the design, construction and outfitting of the buildings for the near and far detectors.  This construction will be funded as three separate DOE General Plant Projects (GPPs):
\begin{itemize}
\item Site preparation - extension of utilities, preparation of roads etc
\item Far detector building
\item Near detector building

\end{itemize}
The maximum cost of a GPP (\$10M) is a significant constraint for the design of the far detector building, however a it has been shown that at least two different solutions fit within the constraint, provided external funding for ancillary items (such as shielding) is found.    The design and construction starts of the far and near detector facilities are staged to match expected funding profiles yet provide beneficial occupancy in time for detector installation. A total of \$14.8M has been allocated in the Fermilab budget plan in FY2014-2016 for the design and construction of the buildings.  This is sufficient for construction based on estimates of the conceptual designs.

Costs for cryostats and cryogenics infrastructure will be shared between Fermilab, CERN, and INFN. CERN and Fermilab will share in the design and construction of the near detector cryostat and cryogenics. CERN and INFN will have primary responsibility for the far detector cryostat and cryogenics.  Fermilab's role in the far detector will focus primarily on the LN$_2$ delivery and support of the safety documentation process.  CERN and INFN will act with in-kind contributions of engineering and procurements which can be done in Europe (e.g. near detector cryostat material, LAr cryogenic plants components, and existing \icarus infrastructure components).  A draft of the sharing of these responsibilities is shown in Table~\ref{tab:Cryo_resources}. 
The overhauling of the \icarus at CERN is financed by the WA104 collaboration through a Memorandum Of Understanding between CERN and INFN. Transport of components to FNAL will be financed in the same way.  
Figures~\ref{fig:wa104MOU}~and~\ref{fig:wa104MOU2} are extracted from the MOU signed between INFN and CERN for the overhauling work. A summary is also given in Table~\ref{tab:T600CERNcostsSum}, the costs are quoted in swiss francs (CHF).

\begin{table}[h]
\begin{center}
\begin{tabular}{|l|c|c|}
\hline
\hline
LAr/GAr System & Service Type & Responsible \\
\hline
\hline
LAr Receiving Facility & Cryo & FNAL \\
\hline
LAr/GAr Transfer Lines & Cryo/Non Cryo & FNAL \\
\hline
GAr/H2 Supply and Transfer Lines & Non Cryo & FNAL \\
\hline
GAr Filtration & Non Cryo & shared \\
\hline
GAr Analyzers & Non Cryo & shared \\
\hline
Condenser & Cryo & shared \\
\hline
LAr handling and purification System & Cryo & shared \\
\hline
Inside piping & Cryo/Non Cryo & shared \\
\hline
GAr handling system & Non Cryo & shared \\
\hline
\hline
LN2 System & Service Type & Responsible \\
\hline
\hline
LN2 Receiving Facility & Cryo & FNAL \\
\hline
LN2 Transfer Lines & Cryo & FNAL \\
\hline
GN2 returns & Non Cryo & INFN/CERN \\
\hline
LN2/GN2 handling system & Cryo/Non Cryo & INFN/CERN \\
\hline
LN2 Distribution Facility & Cryo & INFN/CERN \\
\hline
LN2 Pumping Station & Cryo & INFN/CERN \\
\hline
Services & Cryo & shared \\
\hline
\hline
Ancillary Items & Service Type & Responsible \\
\hline
\hline
Process Controls & Non Cryo & FNAL \\
\hline
Design/Drafting & Non Cryo & shared \\
\hline
Smart P$\&$IDs & Cryo/Non Cryo & shared \\
\hline
Safety aspects of cryogenic installation at Fermilab & Cryo & FNAL \\
\hline
\hline
\end{tabular}
\end{center}
\caption{Draft proposal for CERN, FNAL and INFN responsibilities, for what concerns the management of the cryogenic system maintenance and on-site logistics. The keyword 'shared' refers to tasks to be undertaken jointly by all groups.}
\label{tab:Cryo_resources}
\end{table}

\begin{figure}[h]
\centering
\includegraphics[width=1\textwidth, trim=20mm 45mm 20mm 45mm, clip]{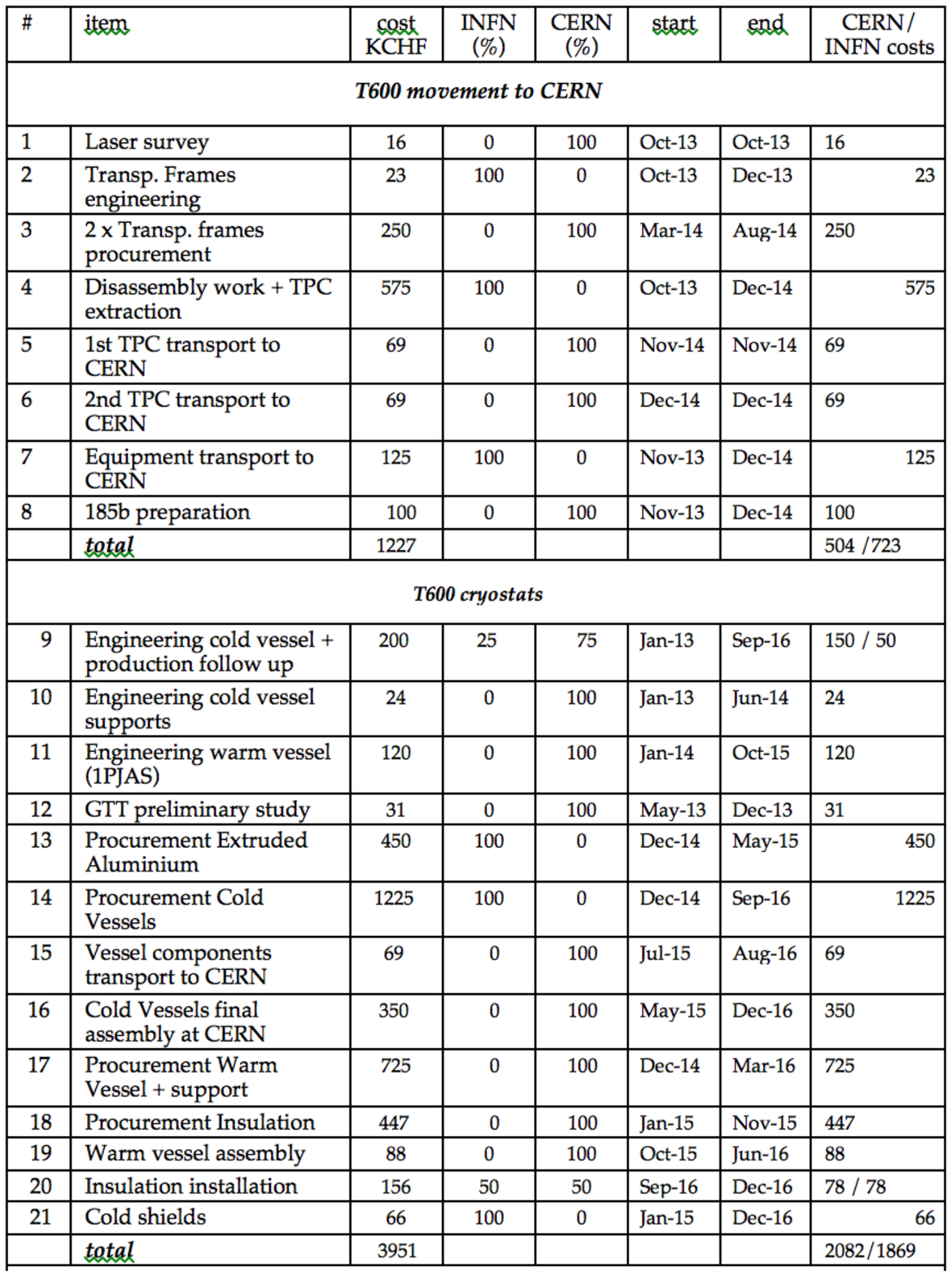}
\begin{center}
\caption{Signed MOU table between CERN and INFN for the ICARUS detector overhauling at CERN}
\label{fig:wa104MOU}
\end{center}
\end{figure}

\begin{figure}[h]
\centering
\includegraphics[width=0.9\textwidth, trim=0mm 5mm 0mm 5mm, clip]{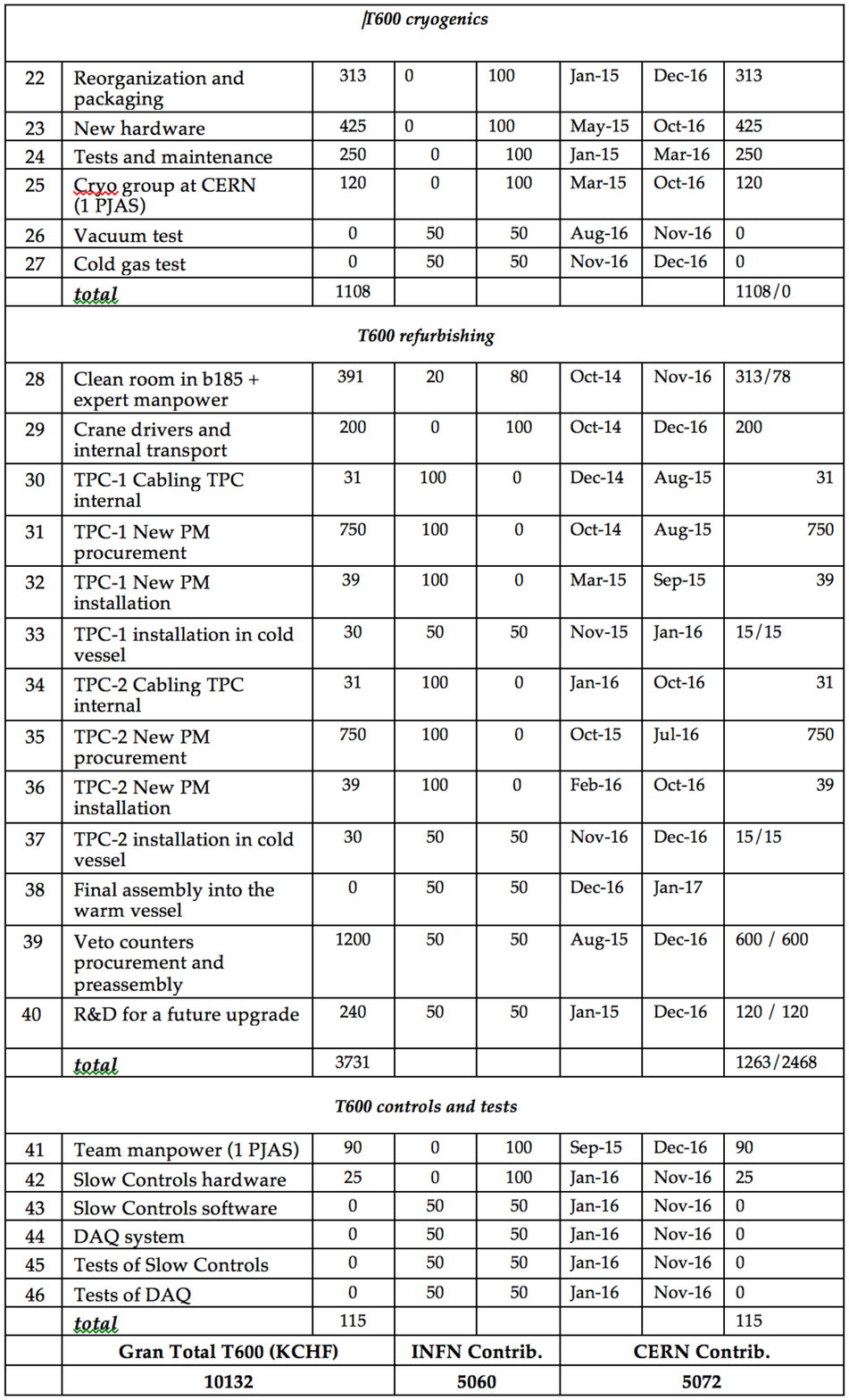}
\begin{center}
\caption{Signed MOU table between CERN and INFN for the ICARUS detector overhauling at CERN}
\label{fig:wa104MOU2}
\end{center}
\end{figure}

\begin{table}[h]
\begin{center}
\begin{tabular}{|l|c|c|c|}
\hline
\hline
Main Operation & Total(kCHF) & INFN & CERN \\
\hline
Movement to CERN & 	1227 & 723 & 504 \\
\hline
Cryostats & 3951 & 1869 & 2082 \\
\hline
Cryogenics & 1108 & 0 & 1108 \\ 
\hline
Refurbishing\footnote{The new electronics %refurbishing 
will be separately addressed by INFN} & 6731 & 4263 & 2468 \\
\hline
Controls and tests & 115 & 0 & 115 \\
\hline
Total & 13132 & 8060 & 5072 \\
\hline
\hline
\end{tabular}
\end{center}
\caption{Overall summary of INFN/CERN funding for \icarus overhauling operations at CERN.}
\label{tab:T600CERNcostsSum}
\end{table}

%\FloatBarrier

The construction of the new near detector (e.g. TPC, light detection) will be financed by several funding agencies including the DOE, US-NSF, CERN, UK-STFC, and  Switzerland.   Table~\ref{tab:LAr1-ND_funding} shows the expected breakdown of responsibilities by funding source and responsible institutions.  The exact sharing of responsibilities will be defined in a dedicated MOU or MOUs, which will be monitored by Fermilab. Funding has already been secured for design and construction of the LAr1-ND TPC through a US-NSF MRI grant to Univ.\ of Chicago, Syracuse Univ., and Yale Univ.\ and through a UK-STFC grant to Lancaster Univ., Univ.\ of Liverpool, Univ. of Manchester, Unv.\ of Sheffield, and Univ.\ College London\footnote{Part of a grant to multiple UK institutions for LBNF and SBN activities}.  Funding from the DOE through Fermilab started with fiscal year 2015 (Oct 2015).  Additional funding proposals have been submitted by the Univ.\ of Bern (CH-NSF) and Columbia Univ.\ (US-NSF). As described in Part~2 of this proposal, there are several alternatives under consideration for the light detection system. The institutional responsibilities for this system will be determined when a technology choice is made. 

\begin{table}[h]
\begin{center}
\begin{tabular}{|l|c|c|c|}
\hline
\hline
Detector System & Funding Source & Institutions \\
\hline
TPC Integrated Design	& UK-STFC 	& Lancaster, Liverpool, Manchester, Sheffield, \\
						& 	\& 	US-NSF			& BNL, Chicago, Syracuse, Yale  \\
\hline
TPC CPAs 				& UK-STFC 		 	& Liverpool \\
\hline
TPC APAs				& UK-STFC   & Manchester, Sheffield, \\
						&		\& US-NSF			& Chicago, Syracuse, Yale \\ 
\hline 
TPC Field Cage  		& US-NSF   			& Yale \\
\hline
Cold Electronics		& DOE				& BNL  \\
\hline
Warm Electronics		& US-NSF\footnote{Grant proposal submitted}									
											& Columbia  \\
\hline
Light Detection			& US-NSF \& DOE		& TBD		\\
\hline
Laser Calibration		& CH-NSF\footnotemark[\value{footnote}]  
											& Bern		\\
\hline
Cosmic Tagger			& CH-NSF
\footnotemark[\value{footnote}]$^,$\footnote{plan to combine with far detector effort}
											& Bern      \\ 
\hline
Detector Integration and Installation			
						& DOE 				& FNAL lead      \\ 
\hline
\hline
\end{tabular}
\end{center}
\caption{Overall summary of funding sources and institutional responsibilities for construction of the near detector.}
\label{tab:LAr1-ND_funding}
\end{table}

The sharing of operation costs for the SBN detector will be defined in due time in a dedicated agreement, shared by all partners.  The boundary between construction and operations will need to be defined in the initial agreements for the construction.  The typical Fermilab boundary includes detector commissioning costs as part of operations.

\FloatBarrier

Table~\ref{tab:ica_fte} details the ICARUS-WA104 contribution to the SBN program in terms of personnel for a three-year activity plan both from CERN and INFN. This includes scientific as well as technical resources. The temporary manpower needed at CERN for several aspects of the overhauling, like crane drivers and experts manpower (Project Associates: PJAs), is already quoted in the MOU figures in CHF and it is in addition to the FTE quoted figures. 

\begin{table}[h]
\begin{center}
\begin{tabular}{|c|c|}
\hline
\hline
ICARUS-WA104 T600 group & Manpower (FTE) \\
\hline
\hline
 Gran Sasso Science Insitute (GSSI) & 1 \\
\hline
 INFN, Sezione di Catania & 2.5 \\
\hline
 INFN, Sezione di Milano & 1.5 \\
\hline 
 INFN, Sezione di Milano Bicocca & 0.5 \\
\hline  
 INFN, Laboratori Nazionali del Gran Sasso (LNGS) & 1 \\
\hline
 INFN, Sezione di Napoli & 0.5 \\
\hline
 INFN, Sezione di Padova & 7 \\
\hline
 INFN, Sezione di Pavia & 6 \\
\hline
 CERN Neutrino group and services & 9 \\
 \hline
 \hline
Total & 30 \\
\hline
%\hline
%\hline
\end{tabular}
\end{center}
\caption{Details of ICARUS-WA104 T600 contribution to SBN activities in terms of manpower (FTE) for a three-year activity plan.}
\label{tab:ica_fte}
\end{table}

The contributions of personnel for near detector and infrastructure (by Fermilab) are being accumulated as part of the development of the resource loaded schedule.  A preliminary accounting should be available prior to the PAC meeting.

Table~\ref{tab:lar1nd_fte} provides a preliminary list of the personnel contributions for the near detector in average FTE/year for construction and installation period. Both scientific and technical resources are included but not labor provided through contracted services. Contributions from some institutions are still pending (tbc). 
The Fermilab effort includes work on infrastructure such as cryogenics but excludes design and construction of the buildings.  
Effort on the \uboone experiment is not included.

The last row lists common support effort by Fermilab including support for far detector infrastructure.  
The Fermilab effort on beamline improvements have not been included at this time.   Needs for common support of data acquisition software and reconstruction software will be developed in the coming months. 

\begin{table}[h]
\begin{center}
\begin{tabular}{|l|c|}
\hline
\hline
\larnd Group 				& Average FTE/yr \\
\hline
\hline
Univ.\ of Bern 				& 4 	\\ \hline
Univ.\ of Chicago 			& 2 	\\ \hline
Columbia Univ. 				& 2.3 	\\ \hline
Indiana Univ.				& 1		\\ \hline
MIT 						& 2 	\\ \hline
Syracuse Univ.  			& 1.5 	\\ \hline
Yale Univ.  				& 1.5 	\\ \hline
Cambridge, Lancaster, Liverpool,   	
							&   \\
Manchester, Sheffield, UCL  & 5.9   \\ \hline
ANL							& 1.5	\\ \hline
BNL							& 7		\\ \hline
FNAL						& 12.5	\\ \hline
LANL						& 2.7 	\\ \hline
 \hline
\larnd Total 				& 43.9 	\\
\hline

  \multicolumn{2}{l}{} \\ \hline
General FNAL SBN Effort		& 5.5	\\ \hline
%\hline

\end{tabular}
\end{center}
\caption{Contributions to near detector development by institution in FTE/year averaged over three years.  The last row shows effort by Fermilab in areas of common support or infrastructure support for the far detector.}
\label{tab:lar1nd_fte}
\end{table}

%A draft proposal for the sharing of responsibility in the construction of the cryogenics systems for both far and near detectors is presented in table ~\ref{tab:Cryo_resources}

\begin{comment}

\FloatBarrier
\section{Environmental Safety and Health}
\label{sec:esh}

The design, construction, commissioning, and operation of the SBN detectors will be performed in compliance with the standards in the Fermilab ES\&H Manual (FESHM) and all applicable ES\&H standards in the Laboratory’s “Work Smart Standards” set. In addition, all related work, including work performed off-site, will be performed in compliance with applicable federal, state and local regulations. All work done at collaborating universities will be performed in accordance with existing university policies and applicable national, state, or local regulations.

No significant environmental, regulatory or political sensitivities are affected by this project.  All work done at Fermilab will be in compliance with Federal, state, and local regulations.  Environmental protection is integral to all aspects of work performed at Fermilab in accordance with Fermilab’s DOE-approved Environmental Management System, which is part of the Laboratory’s Integrated ES\&H Management Plan.  Compliance with the National Environmental Protection Act (NEPA) is part of this system. 

\end{comment}

\FloatBarrier
\fi

 \clearpage

\appendix

\section{Detector Volumes and Masses}
\label{sec:fv}

\begin{table}[h]
\begin{center}
\caption{\label{tab:fv}TPC active and fiducial volumes in each SBN detector used in these analyses.}
\begin{tabular}{|l|c|c|c||c||c|}\hline
Detector volume						& W (cm)			& H (cm)& L (cm)	& volume ($m^3$)	& argon mass (tons) \\ \hline\hline
\larnd Active						&  2$\times$200 	& 400 	& 500 		& 80.0 				& 112   \\
\larnd Fiducial (\numu analysis)	&  2$\times$183.5	& 370 	& 405		& 55.0				& 77.0  \\
\larnd Fiducial (\nue analysis)		&  2$\times$173.5 	& 350 	& 420 		& 51.0				& 71.4  \\ \hline
\uboone Active						&  256 				& 233 	& 1037 		& 61.9 				& 86.6  \\
\uboone Fiducial (\numu analysis)	&  226				& 203	& 942		& 43.2				& 60.5	\\
\uboone Fiducial (\nue analysis)	&  206 				& 183 	& 957 		& 34.2				& 47.9  \\ \hline
\icarus Active						& 4$\times$150 		& 316 	& 1795		& 340.3 			& 476   \\
\icarus Fiducial (\numu analysis)	& 4$\times$133.5	& 286  	& 1700      & 259.6				& 363  	\\
\icarus Fiducial (\nue analysis)	& 4$\times$123.5 	& 266 	& 1715  	& 225.4 			& 315	\\ \hline\hline
\end{tabular}
\end{center}
\end{table}

\FloatBarrier
\clearpage

\bibliography{main.bib}

\end{document}